\documentclass[12pt,a4paper]{book}
\usepackage{amsmath}
\usepackage{amssymb}
\usepackage{mathtools}
\usepackage{amsfonts}
\usepackage{a4wide}
\usepackage{fancyhdr}
\usepackage{graphicx}
\usepackage[english]{babel}
\usepackage[utf8]{inputenc}
\usepackage[T1]{fontenc}
\usepackage{lmodern}
\usepackage{adjustbox}
\usepackage{booktabs}
\usepackage{xcolor}
\usepackage{array}
\usepackage{multirow}
\usepackage{etoolbox}
\usepackage{graphicx}
\usepackage[bookmarks, colorlinks=true, allcolors=blue]{hyperref}
\usepackage[switch]{lineno}
\usepackage{silence}
\usepackage{siunitx}
\usepackage{subcaption}
\usepackage{tabularx}
\usepackage{tikz}
\usepackage{xcolor}
\usepackage{titlesec}
\usepackage{csquotes}
\usepackage{tikz}
\usepackage{tikz-cd}
\usepackage{bm}
\usepackage{algorithm}
\usepackage{algpseudocode}
\usepackage{url}
\usepackage{enumitem}
\usepackage[nomargin,inline,draft]{fixme}
\usepackage{adjustbox}
\usepackage{amsthm}
\usepackage[backend=biber, sorting=none, backref=true]{biblatex}
\usepackage{placeins}
\usepackage[capitalize,noabbrev]{cleveref}
\usepackage{microtype}

\DeclareFieldFormat{urldate}{}
\titleformat{\paragraph}
{\normalfont\normalsize\bfseries}{\theparagraph}{1em}{}
\titlespacing*{\paragraph}
{0pt}{3.25ex plus 1ex minus .2ex}{1.5ex plus .2ex}

\DeclareMathAlphabet{\pazocal}{OMS}{zplm}{m}{n}

\newcommand{\Lb}{\pazocal{L}}

\newcommand{\norm}[1]{\left\lVert#1\right\rVert}
\newcolumntype{C}[1]{>{\centering\arraybackslash}p{#1}}

\newcommand{\cosines}{\vc}

\newcommand{\idx}{\bm s}

\newcommand{\dlat}{{l}}
\newcommand{\dhid}{{c}}
\newcommand{\dang}{{a}}
\newcommand{\nref}{{n_0}}

\theoremstyle{definition}
\newtheorem{definition}{Definition}

\def\eqref#1{equation~\ref{#1}}

\def\1{\bm{1}}

\def\va{{\bm{a}}}

\def\vc{{\bm{c}}}

\def\vn{{\bm{n}}}

\def\vr{{\bm{r}}}

\def\vx{{\bm{x}}}

\def\vz{{\bm{z}}}

\def\mR{{\bm{R}}}

\def\mW{{\bm{W}}}
\def\mX{{\bm{X}}}

\DeclareMathAlphabet{\mathsfit}{\encodingdefault}{\sfdefault}{m}{sl}
\SetMathAlphabet{\mathsfit}{bold}{\encodingdefault}{\sfdefault}{bx}{n}

\newcommand{\E}{\mathbb{E}}

\newcommand{\R}{\mathbb{R}}

\renewcommand{\chaptermark}[1]%
         {\markboth{\thechapter.\ #1}{}}
\renewcommand{\sectionmark}[1]%
         {\markright{\thesection\ #1}}
\newcommand{\getTitle}{Deep Generative Models for Ultra-High Granularity Particle Physics Detector Simulation: A Voyage From Emulation to Extrapolation}

\newcommand{\getSubtitle}{Tiefe Generative Modelle für die Simulation von Teilchenphysik-Detektoren mit ultrahoher Granularität: Eine Reise von der Emulation zur Extrapolation}
\newcommand*{\getAuthor}{Baran~(Hosein)~Hashemi}
\newcommand*{\getPrintLocation}{München}       
\newcommand*{\getPrintYear}{2023}         

\newcommand*{\getPlaceOfBirth}{Rasht, Iran}
\newcommand*{\getSubmissionDate}{12. October 2023}
\newcommand*{\getExpertOne}{Prof. Dr. Thomas Kuhr}
\newcommand*{\getExpertTwo}{Prof. Dr. Lukas Heinrich}
\newcommand*{\getExamDate}{22.11.2023}

\newcommand*{\lang}{de-DE}

\hypersetup {
pdfpagemode = {UseNone},
pdftitle = {\getTitle},
pdfauthor = {\getAuthor},
pdflang = {\lang},
}

\begin{document}

  \frontmatter

  \begin{titlepage}

   {\sffamily
    \vspace*{\stretch{1}}
    {\parindent0cm
    \rule{\linewidth}{.7ex}}  
  \begin{flushright}
    \vspace*{\stretch{1}}
    {\LARGE \bfseries \baselineskip=40pt \getTitle{}}
    
    
    \vspace*{\stretch{1}}
    {\large\bfseries \getAuthor{}}
    
    \vspace*{\stretch{1}}
  \end{flushright}
    \rule{\linewidth}{.7ex}
    
    \vspace*{\stretch{5}}  
  \begin{center}
    \includegraphics[width=2in]{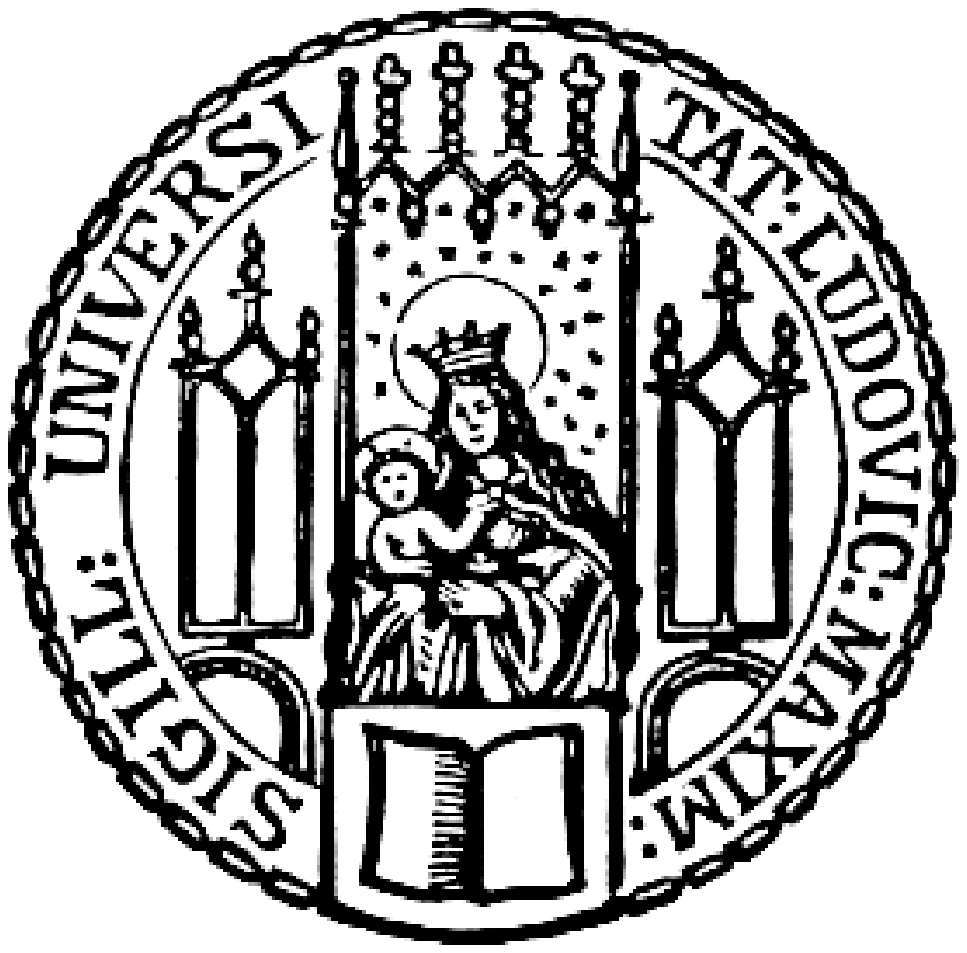}
    
    \vspace*{\stretch{1}} 
    {\Large \getPrintLocation{} \getPrintYear{}}
  \end{center}
    }

\end{titlepage}

\thispagestyle{empty}
\cleardoublepage

  \begin{titlepage}   
   {\sffamily
    \vspace*{\stretch{1}}
    {\parindent0cm
    \rule{\linewidth}{.7ex}}  
  \begin{flushright}
    \vspace*{\stretch{1}}
    {\LARGE \bfseries \baselineskip=40pt \getTitle{}}
    
    \vspace*{2ex}
    {\Large \bfseries \baselineskip=40pt \getSubtitle{}}
    
    \vspace*{\stretch{1}}
    {\large\bfseries \getAuthor{}}
    
    \vspace*{\stretch{1}}
  \end{flushright}
    \rule{\linewidth}{.7ex}
    \vspace*{\stretch{3}}
    
  \begin{center}
    {\large
     Dissertation\\
     der Fakultät für Physik\\
     der Ludwig-Maximilians-Universität\\
     München
     
    \vspace*{\stretch{1}}    
     vorgelegt von\\
     \getAuthor{}\\
     aus \getPlaceOfBirth{}
     
    \vspace*{\stretch{2}}   
     \getPrintLocation{}, den \getSubmissionDate{}
     }
  \end{center}
   }
\end{titlepage}

  \thispagestyle{empty}

   \vspace*{\stretch{1}}
  \begin{flushleft}
    {\large\sffamily
     Erstgutachter:  \getExpertOne{} \\[1mm]
     Zweitgutachter: \getExpertTwo{} \\[1mm]
     Tag der mündlichen Prüfung: \getExamDate{}\\
     }
  \end{flushleft}

\cleardoublepage{}




\chapter*{Zusammenfassung}
Die Simulation von Detektorantworten mit ultrahoher Granularität in der Teilchenphysik stellt eine kritische, jedoch rechenintensive Aufgabe dar.
Diese Arbeit zielt darauf ab, diese Herausforderungen zu bewältigen, indem sie sich auf den Pixel-Vertex-Detektor~(PXD) im Belle~II-Experiment konzentriert, der über 7,5 Millionen Pixelkanäle verfügt – den höchstaufgelösten Detektorsimulationsdatensatz, der jemals mit tiefen generativen Modellen analysiert wurde.
Als erste Beitrag führe ich das Intra-Event-Aware Generative Adversarial Network~(IEA-GAN) ein, ein Modell, das relationales Denken und selbstüberwachtes Lernen integriert, um ein „Ereignis“ im Detektor zu simulieren. Diese Studie etabliert PXD-Daten als feinkörnigen Datensatz und unterstreicht die Bedeutung der intra-eventuellen Korrelation für nachgelagerte physikalische Analysen. IEA-GAN emuliert PXD-Signaturen auf eine geometriebewusste und korrelierte Weise.

Aufbauend darauf führt diese Arbeit YonedaVAE ein, ein fortschrittliches generatives Modell, das das offene Problem der Out-of-Distribution~(OOD) Simulation von realen Daten angeht.  Inspiriert durch die Kategorientheorie verwendet YonedaVAE ein lernbares Yoneda-Embedding, um die Gesamtheit eines Ereignisses anhand seiner Sensorbeziehungen zu erfassen, und formuliert eine formale Sprache für intra-eventuelles relationales Denken. Dies wird ergänzt durch einen selbstdestillierten Mengengenerator und einen adaptiven Top-q Abtastmechanismus, der es dem Modell ermöglicht, Punktewolken mit variablen intra- und inter-eventuellen Kardinalitäten weit über die Kardinalität der Trainingsdaten hinaus zu sampeln. Eine variable intra-event Kardinalität wurde zuvor noch nicht erreicht und ist von entscheidender Bedeutung, wenn man sich mit unregelmäßigen Detektorgeometrien und Treffermustern durch den Detektor beschäftigt.

Diese Studie präsentiert die ersten Ergebnisse für ein generatives Modell, das auf realen Daten in der Teilchenphysik mit ultrahoher Granularität trainiert wurde. 
Sie zeigt, dass das YonedaVAE-Modell, trainiert auf frühzeitigen Hintergrunddaten eines Experiments, eine vernünftige Simulation eines späteren Experiments mit fast doppelter Leuchtkraft erreichen kann, während es gleichzeitig eine signifikante Speicherentlastung bietet. Bemerkenswert ist, dass YonedaVAE diese Extrapolation ohne vorherige Exposition gegenüber Hochleuchtkraftdaten erreicht, was seine Robustheit und Generalisierbarkeit unterstreicht. Insgesamt reduzieren diese Modelle nicht nur erheblich den Rechenaufwand, sondern erreichen auch eine noch nie dagewesene Präzision in Detektorsimulationen mit ultrahoher Granularität. Dies eröffnet neue Wege sowohl für die Recheneffizienz als auch für die Genauigkeit in der Teilchenphysik.
  \addcontentsline{toc}{chapter}{Zusammenfassung}
  \markboth{Zusammenfassung}{Zusammenfassung}
  

\chapter*{Abstract}
Simulating ultra-high-granularity detector responses in Particle Physics represents a critical yet computationally demanding task. 
This thesis aims to overcome these challenges by focusing on the Pixel Vertex Detector~(PXD) at the Belle~II experiment, which features over 7.5M pixel channels—the highest spatial resolution detector simulation dataset ever analyzed with deep generative models. 
As the first contribution, I introduce the Intra-Event Aware Generative Adversarial Network~(IEA-GAN), a model incorporating relational reasoning and self-supervised learning to simulate an ``event'' in the detector. This study establishes PXD data as a fine-grained dataset and underscores the importance of intra-event correlation for downstream physics analyses. IEA-GAN emulates PXD signatures in a geometry-aware and correlated manner.

Building upon this, this thesis introduces YonedaVAE, a Category Theory-inspired generative model that tackles the open problem of Out-of-Distribution~(OOD) simulation. Inspired by Category Theory, YonedaVAE introduces a learnable Yoneda Embedding to capture the entirety of an event based on its sensor relationships, formulating a formal language for intra-event relational reasoning. This is complemented by a self-supervised set generator and an adaptive Top-q sampling mechanism, enabling the model to sample point clouds with variable intra-event and inter-event cardinalities far beyond the training data cardinality. Intra-event variable cardinality has not been done before and is vital when one is dealing with irregular detector geometries and hit patterns through the detector.

This study presents the first results for a generative model trained on real data in ultra-high granularity particle physics. 
It shows that the YonedaVAE model, trained on an early experiment background data, can reach a reasonable simulation of a later experiment with almost double luminosity while providing a significant storage release. Remarkably, YonedaVAE achieves this extrapolation without previous exposure to high-luminosity data, showcasing its robustness and generalizability.

Collectively, these models not only substantially reduce computational overhead but also achieve unprecedented precision in ultra-high-granularity detector simulations, opening new avenues for both computational efficiency and accuracy in particle physics.
  \addcontentsline{toc}{chapter}{Abstract}
  \markboth{Abstract}{Abstract}

  \tableofcontents
  \addcontentsline{toc}{chapter}{Contents}
  \markboth{Contents}{Contents}

  \mainmatter\setcounter{page}{1}
  \chapter{Introduction}
\label{chap:1}
The current consensus to model the origin of our universe centers on an event known as the Big Bang~\cite{bennett_nine-year_2013,planck_collaboration_planck_2014}, which occurred 13.8 billion years ago. 
This explosive instant was characterized by extreme heat and dense energy, giving rise to elementary particles and their corresponding anti-particles. Mere fractions of a second later, cosmic inflation dramatically expanded the universe. What transpired next remains a fundamental enigma, serving as one of the motivating forces behind the creation of the Belle II experiment~\cite{abe_belle_2010}. This crucial and enigmatic process, termed ``Baryogenesis,'' violated baryon number conservation, resulting in the observed predominance of baryons over anti-baryons.

In mainstream cosmology, the Universe underwent an initial inflationary phase, swiftly transitioning to a period of radiation domination. This was followed by an era of matter domination, ultimately giving way to the dark energy-dominated epoch we find ourselves in today~\cite{peebles_cosmological_2003}. While we can somewhat simulate baryon formation in particle colliders, the puzzle of the universe's ``missing'' antimatter remains unsolved. To generate matter and antimatter at differing rates, a baryon-generating interaction must meet three critical conditions~\cite{sakharov_violation_1967}: violation of the baryon number, breaking of C-symmetry and CP-symmetry, and a deviation from thermal equilibrium.

The Belle~II experiment focuses particularly on CP-symmetry violation, which results in an imbalance between the number of left-handed and right-handed baryons and anti-baryons. This adds another layer of complexity to the baryogenesis conundrum. Current explanations within the Standard Model fail to account for the observed matter-antimatter disparity, signaling the need to delve into unexplored areas of physics. To navigate this uncharted territory, researchers employ two primary approaches. The first, known as the ``high energy frontier,'' is utilized by the Large Hadron Collider~(LHC) and aims to directly create and analyze new particles through high-energy collisions. 
SuperKEKB (Belle~II) takes a different tack and operates on the ``intensity frontier,'' focusing on high-precision experiments to identify deviations from the Standard Model and search for new, weakly-coupled mediators in the dark sector. 
Thus, the quest endures, the mysteries dating back to the universe's earliest moments to today's cutting-edge experiments, each discovery bringing us closer to understanding the nature of our existence.\vspace{1ex}

The key to high-precision measurements at SuperKEKB is the collider's ability to produce ``clean'' events. This enables precise measurements, particularly for events where particles like neutrinos escape undetected. 
High-precision measurement at Belle~II is achieved by recording a large number of collisions to offer a statistically robust sample for analyses. Additionally, the excellent reconstruction of particle trajectories, particularly their decay vertices, is crucial. SuperKEKB has attained an unprecedented instantaneous luminosity of \(4.7 \times 10^{34} \, \text{cm}^{-2} \text{s}^{-1}\) and aims for an even higher rate of \(8 \times 10^{35} \, \text{cm}^{-2} \text{s}^{-1}\). Luminosity is a measure of how many particles pass through a given area in a specific time period. It's essentially a measure of the ``brightness'' of the collider, and higher luminosity implies that more particle collisions are likely to occur, increasing the odds of observing rare processes.

Discoveries of New Physics at Belle~II are heavily dependent on the precise reconstruction of particle decay vertices. To achieve this level of precision, the Belle~II detector is equipped with a state-of-the-art PiXel vertex Detector~(PXD) located very close to the interaction point. This serves as the innermost sub-detector of Belle~II. The PXD, with its ``ultra-high granular'' matrix of sensors, records the passage of charged particles with a total readout of \(7.68 \times 10^6\) channels per event. However, its position as the innermost sub-detector results in a high level of ``background'' noise. ``background'' refers to any detector hit that is not of interest but cannot be fully eliminated from the data. These are usually events that look similar to the desired signal but are caused by different processes. These backgrounds, as artifacts in the PXD, directly influence the precision of particle trajectory finding and reconstruction.

To validate our understanding of theoretical models in physical processes, Belle~II, like other particle physics experiments, heavily relies on simulation for various tasks. These include data selection, statistical inference, and design optimization for new experiments. Consequently, emulating the PXD background responses is critical for studying experimental conditions. However, this emulation is both storage-intensive and computationally demanding. The current landscape for PXD background simulation includes three primary approaches: First is the somewhat accurate but storage-consuming and computationally intense ``Geant4 simulation.''
Second is the highly accurate but storage-intensive ``random trigger'' method. 
Lastly, there's the ``surrogate model,'' known for its fast and efficient simulation capabilities.\vspace{1ex}

In this thesis, my primary objective is to develop specialized Deep Generative Models~(DGM) as surrogates to enhance computational efficiency and data storage for Belle~II PXD detector simulation. 
To address the challenges presented by the ultra-high resolution of PXD data, I introduce a fresh perspective through ``event-based relational reasoning.'' 
In an event-based relational reasoning approach, each sensor in an event is studied in relation to the other sensors.
The central premise is that understanding each PXD sensor solely based on its intrinsic features is insufficient for accurate simulation. Instead, how each sensor interacts with other sensors within the event provides a contextualized representation, very much like like transitioning from ``syntax'' to ``semantics'' in linguistics. 
Through this thesis, I demonstrate that incorporating this relational inductive bias leads to the creation of more accurate surrogate models for detector simulation.

Within the ensuing  chapter,~\cref{chap:2}, I introduce the Belle~II detector and its software framework. Then, there will be an in-depth discussion on the PXD background and various ways of simulating its detector signatures.

\cref{chap:3} lays the foundation for deep generative models, relational reasoning, and self-supervised learning, the three Machine Learning~(ML) pillars of this thesis, by providing an overview of the prerequisite technologies and methodologies. 

In~\cref{chap:4}, I present an exhaustive and taxonomically organized review of the use of deep generative models in Experimental Particle Physics, with a special focus on detector simulation. By the end of this chapter, I construct a holistic view to demonstrate that current state-of-the-art deep generative models are incomplete for tackling ultra-high granular PXD background generation.

Moving on to~\cref{chap:5}, I delve into the specifics of PXD background generation and offer a task-categorization perspective for it. I also introduce two new models that I have developed: the Prior Embedding GAN~\cite{hashemi_pixel_2021}, founded on Contrastive Learning, and the Intra-Event Aware GAN~(IEA-GAN)~\cite{hashemi_ultra-high-resolution_2023}, which relies on Self-Supervised Learning and relational reasoning. These models are trained and tested on Geant4-simulated PXD background data. 
During this process, novel technologies in deep generative models tailored for fine-grained images will be introduced. For evaluation, I initially motivate the significance of ``intra-event correlation'' for the first time in the fast simulation domain via a study focused on Helix Parameter Resolution. Subsequently, I conduct a comprehensive evaluation of various figures of merit. As a result of this analysis, IEA-GAN has successfully integrated into the basf2 software~\cite{kuhr_computing_2011} suite as a surrogate module for emulating PXD background on the fly for analysis.

In~\cref{chap:6}, I engage with the ``real'' PXD detector background data coming from the random trigger. This chapter explores the challenges and motivations behind the Out-Of-Distribution~(OOD) simulation of PXD background data and extrapolation beyond current experimental limits, particularly concerning luminosity.
I then seek to unify relational reasoning concepts through the lens of Category Theory, the abstract study of mathematical objects and their interrelations. As a result, I introduce YonedaVAE, a zero-shot point cloud deep generative model capable of generating PXD background hits with an unprecedented cardinality of \num{100700} hits per event, despite being trained only on \num{7600} PXD hits per event from Experiment~\num{12}. Remarkably, YonedaVAE performs robustly when tested on data from Experiment~\num{26}, which has nearly double the luminosity of Experiment~\num{12}. 
The chapter presents results in two primary tasks: ``length extrapolation,'' where the model has access to an individual sensor condition, and ``context extrapolation,'' where it has access only to a global event-level condition during inference and has to solve an Inverse Problem. 
For evaluation, after an in-depth evaluation of low-level marginal distributions and NN-based metrics, I introduce a new diversity measure for detector simulation, the Vendi Score, previously used in Ecology and Protein Design. Then, I examine the geometrical and topological properties of the generated PXD background data via clustering analyses and Topological Data Analysis~(TDA). 
Finally, I study the tracking analysis of the generated PXD background in comparison to the real PXD detector background data. 
Consequently, I demonstrate the efficacy of YonedaVAE, which excels not only in the OOD simulation of PXD background but also effectively manages to condition its output based on both sensor locations and background levels.

In the final chapter,~\cref{chap:7}, a summary of the study's key findings will be presented, along with a thoughtful discussion regarding the limitations. 
These limitations offer crucial insights for potential refinements in the current methodologies and lay the groundwork for the next phase of investigations in this domain.

  \chapter{Belle II experiment: The PXD Saga}
\label{chap:2}

The Belle~II experiment, a precision marvel of modern particle physics located at the SuperKEKB accelerator in Tsukuba, Japan, initiated data collection~\cite{noauthor_superkekb_2019} from electron-positron collisions in March 2019. It has achieved an unprecedented instantaneous luminosity, reaching up to \(4.7 \times 10^{34} \, \text{cm}^{-2} \text{s}^{-1}\), and accumulated data equivalent to \(424 \, \text{fb}^{-1}\)~\cite{noauthor_belle_nodate}. 
A pivotal component of the Belle~II apparatus is its Vertex Detector~(VXD), which is instrumental for the precise reconstruction of both primary and secondary decay vertex locations. The VXD is comprised of an outer section with four layers of Silicon Vertex Detector~(SVD). The inner section features a two-layer PiXel Detector~(PXD), the protagonist of this thesis.

In the ensuing sections of this chapter, I will guide you through the experimental setup of the Belle~II experiment and the PXD sub-detector. As we traverse this journey, we will delve into the nuances of the responses recorded by the PXD and their origin.  
Eventually, I will discuss the simulation of PXD data signatures and its challenges. 
By the end of the chapter, I discuss the plan for this thesis in the following chapters.

\section{Belle~II Experiment}

The Belle~II experiment~\cite{abe_belle_2010} represents the next chapter in a rich legacy of collider experiments focused on B meson physics, collectively known as B factories. Situated at KEK in Tsukuba, Japan, Belle~II is the successor to the original Belle experiment and operates in conjunction with the SuperKEKB accelerator. This state-of-the-art facility is an asymmetric electron-positron collider featuring a High Energy Ring~(HER) for electrons at 7 GeV and a Low Energy Ring~(LER) for positrons at 4 GeV. Designed to achieve a peak luminosity of \(6.5 \times 10^{35} \, \text{cm}^{-2} \text{s}^{-1}\), it aims to outperform its predecessor, KEKB, by a factor of \num{40}.

\begin{figure}[!htb]
    \centering
    \includegraphics[width=0.75\textwidth]{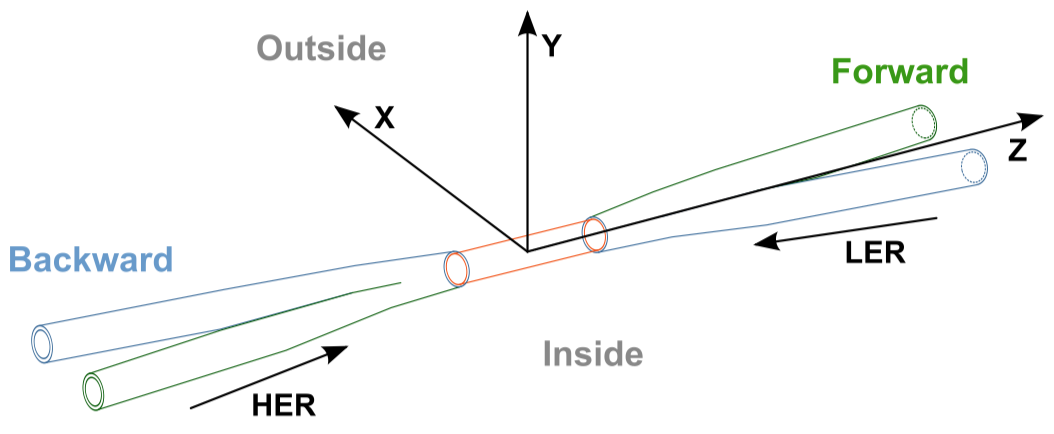}
    \caption{The Belle II coordinate system.}
    \label{fig:belle2_coord}
\end{figure}

The Belle~II detector employs a Cartesian coordinate system that is right-handed for its spatial description~(see~\cref{fig:belle2_coord}). The origin of this coordinate framework is situated at the point where particle interactions are expected to occur, commonly known as the nominal interaction point~(IP). At this point, both beams are crossing at an angle of 11 mrad and collisions occur. The orientation of the axes is established as follows:

\begin{itemize}
\item \( z\text{-axis} \): This axis is aligned with the magnetic field generated by the solenoid, and parallel to the $e^+$ beam. It essentially provides a longitudinal perspective, running parallel to the primary axis of the detector.

\item \( y\text{-axis} \): Oriented in the vertical direction, this axis points towards the ceiling of the detector hall. It serves as the upward vertical reference for the system.

\item \( x\text{-axis} \): This axis extends radially, pointing outward in a direction perpendicular to the plane formed by the \( z \) and \( y \) axes. It is oriented towards the exterior of the accelerator ring.
\end{itemize}

In this coordinate system, the positive \( z \)-axis is directed towards the ``forward'' orientation while the negative \( z \)-axis indicates the ``backward'' direction. Unless stated otherwise, all projections of the Belle II detector onto the \( xy \)-plane are designed to be viewed from the forward end to the backward end of the detector. 

The collision dynamics are carefully controlled to yield a center-of-mass energy of \(E_{\text{CMS}} = \sqrt{4 E_{\text{HER}} E_{\text{LER}}} \approx 10.58 \, \text{GeV}\), aligning closely with the mass of \(\Upsilon(4S)\) resonance. In this setup, B and $\bar{\text{B}}$ mesons are produced and almost immediately decay, initiating complex chains of subsequent decays. These chains are critical to the study of B meson physics, as they often yield a final set of stable particles whose properties can be thoroughly analyzed.

The asymmetry along the \( z \)-axis in Belle~II's detector design is a response to the unique collision dynamics, where the center-of-mass frame and the flight trajectories of produced B mesons are both directed towards the forward part of the detector. This forward focus introduces a Lorentz boost to the B mesons, effectively increasing their flight length within the detector. Thus, longer flight length yields a better resolution on lifetimes, enabling more precise measurements.
This leads to a division of the angular acceptance into three separate regions in terms of polar angles: the Forward, Barrel, and Backward regions.
The polar angle $\theta$ is the angle in the (x, z)-plane with respect to the z-axis. Specifically, the Forward region encompasses angles \( 17^\circ < \theta < 30^\circ \), the Barrel region is defined by \( 30^\circ < \theta < 125^\circ \), and the Backward region spans \( 125^\circ < \theta < 150^\circ \).

All but the outermost component of the Belle~II detector operates within a uniform solenoidal magnetic field of magnitude 1.5 T, oriented parallel to the detector's main axis. A recorded collision, namely an ``event'', involves mostly \( e^+ e^- \rightarrow \Upsilon(S) \rightarrow \bar{B} B \) reaction that typically yields around \num{12} tracks. The particles most commonly detected in such events are electrons, positrons, photons, muons, \( \pi^\pm \) mesons, \( K^\pm \) mesons, protons, and deuterons.

\begin{figure}[!htb]
    \centering
    \includegraphics[width=0.88\textwidth,clip]{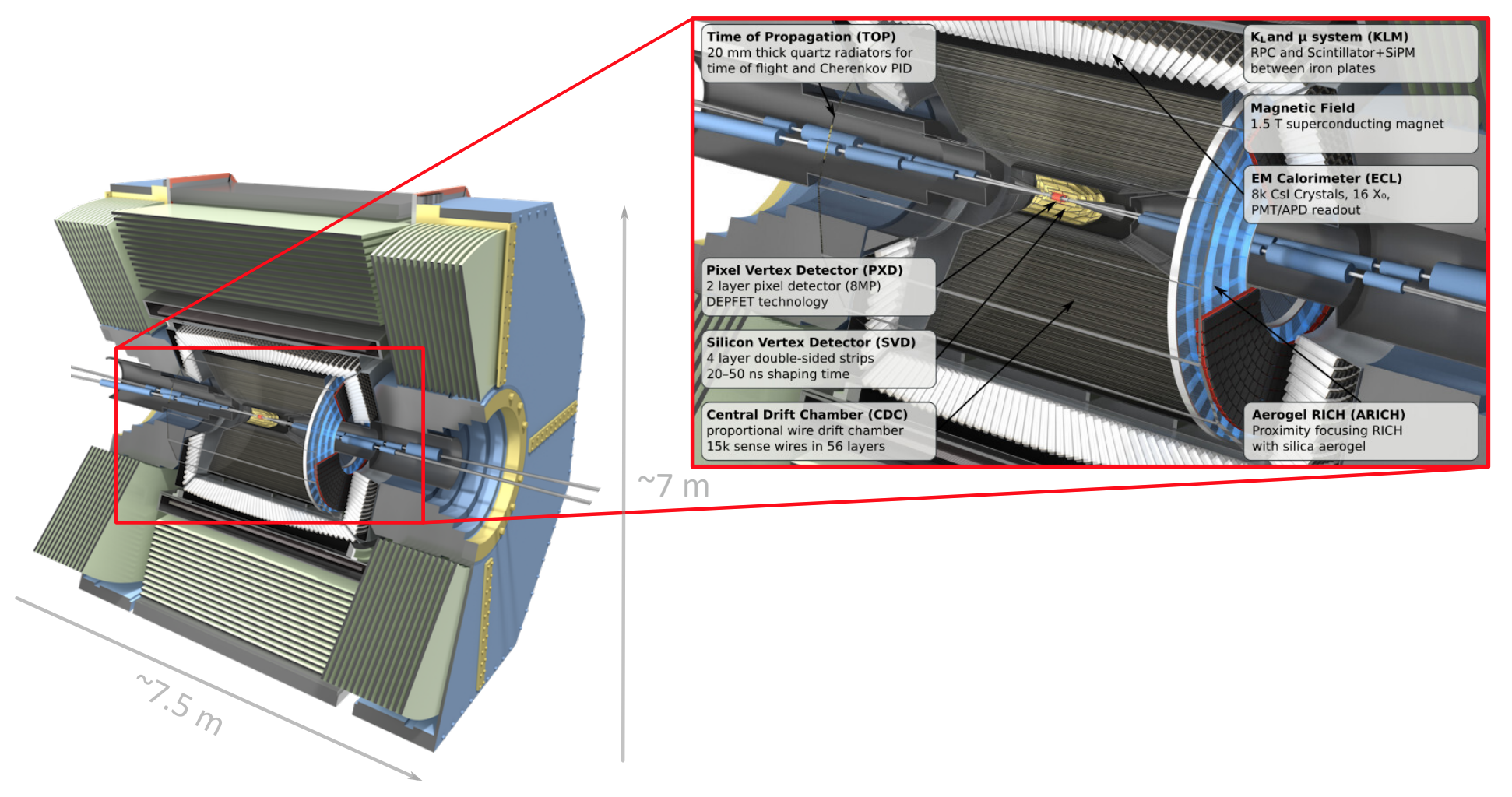}
    \caption{The Belle~II detector with a zoomed-in view of the inner part adopted from~\cite{giakoustidis_status_2023}}
    \label{fig:belle2}
\end{figure}

Belle~II is engineered to focus on capturing decay products of B mesons due to their extremely short lifetimes. It features a cylindrical, multi-layered structure that envelopes the Interaction Point~(IP), where the particle collisions take place. The sub-components of the Belle II detector are specialized for measuring various particle properties. Among these are the tracking sub-detectors, responsible for gauging particle momentum and locating decay vertices. Additionally, there are particle identification sub-detectors to classify the type of particle and a calorimeter tasked with reconstructing the energies of photons and electrons.
The detector consists of a beam pipe and the following sub-detectors from inside out, as shown in~\cref{fig:belle2} and~\cref{tab:belle2_sum} , the Vertex Detector~(VXD)~(the Pixel Vertex Detector~(PXD) and Silicon Vertex Detector~(SVD)), the Central Drift Chamber~(CDC), The Time of Propagation Counter~(TOP), The Aerogel Rich Detector~(ARICH), The Electromagnetic calorimeter~(ECL), and the KLong and Muon detector~(KLM). Each plays a specific role, and together, they provide a comprehensive dataset essential for the reconstruction of B meson decays.

The original Belle experiment and its contemporary BaBar set remarkable milestones, including the groundbreaking discovery of CP violation in the B meson system~\cite{belle_collaboration_observation_2002}. This discovery led to the validation of the Kobayashi-Maskawa model and earned a Nobel Prize in 2008~\cite{kobayashi_cp-violation_1973,noauthor_nobel_nodate}. By combining advancements in technology and insights from previous experiments like Belle and BaBar, the Belle~II experiment aims to push the frontiers of our understanding of particle physics in the luminosity front.


\begin{table}[!htb]
\label{tab:belle2_sum}
\centering
\caption{Summary of Belle~II detector components, adopted from~\cite{kou_belle_2019}}
\resizebox{\textwidth}{!}{%
\Large
\begin{tabular}{ccccc}
\toprule
\textbf{Purpose} & \textbf{Name} & \textbf{Component} & \textbf{Configuration} & \textbf{Readout channels} \\
\midrule
\multirow{2}{*}{Beam pipe} & \multirow{2}{*}{Beryllium} & \multirow{2}{*}{Cylindrical, inner radius} & 10 mm, 10 $\mu$m Au, 0.6 mm Be, 1 mm paraffin, 0.4 mm Be & \multirow{2}{*}{-} \\
\\
\addlinespace
\multirow{3}{*}{Tracking} & PXD & Silicon Pixel (DEPFET) & Sensor size: 15$\times$(L1 136, L2 170) mm$^2$, Pixel size: 50$\times$(L1a 50, L1b 60, L2a 75, L2b 85) $\mu$m$^2$ & 10M  \\
& SVD & Silicon Strip & Rectangular and trapezoidal, strip pitch: 50(p)/160(n) - 75(p)/240(n) $\mu$m & 245k  \\
& CDC & Drift Chamber & 14336 wires in 56 layers, inner radius of 160mm outer radius of 1130 mm & 14k  \\
\addlinespace
\multirow{2}{*}{Particle ID} & TOP & RICH with quartz radiator & 16 segments in $\phi$ at r $\sim$ 120 cm, 275 cm long, 2 cm thick quartz bars & 8k  \\
& ARICH & RICH with aerogel radiator & 2$\times$2 cm thick focusing radiators, HAPD photodetectors & 78k  \\
\addlinespace
Calorimetry & ECL & CsI(Tl) & Barrel: r = 125 - 162cm, end-cap: z = -102 - +196cm & 6624 (Barrel), 1152 (FWD), 960 (BWD)  \\
\addlinespace
\multirow{2}{*}{Muon ID} & KLM barrel & RPCs and scintillator strips & 2 layers with scintillator strips and 12 layers with 2 RPCs & $\theta$ 16k, $\phi$ 16k \\
& KLM end-cap & scintillator strips & 12 layers of (7-10)$\times$40 mm$^2$ strips & 17k  \\
\bottomrule
\end{tabular}%
}
\end{table}

\subsection{Belle~II Software Framework}
 Alongside hardware improvements, the experiment's analysis software has also been revamped to enhance both performance and user experience, culminating in a new software, basf2~\cite{kuhr_computing_2011,kuhr_belle_2019}. 
 The methods described in this thesis are also implemented in the Belle II software, basf2. In the following, I briefly skim through the bird-eye view of the underlying processes in Belle~II data-taking and the software pipeline based on~\cite{noauthor_belle_nodate}. Later, we go through the most relevant parts: simulation and reconstruction.
 
\begin{enumerate}
    \item \textbf{Data Collection and Event Triggers:} In the Belle~II experiment, data acquisition occurs in specialized sessions characterized by a constant flux of electron-positron collisions. These collisions happen at an elevated frequency, and the detector system continuously scans the collision outcomes in real time. The focus is on identifying unique markers that signify the creation of B \(\bar{\text{B}}\) meson pairs. 
    An autonomous component within the Belle~II online system~(shown in~\cref{fig:belle2_dataflow}), known as the ``trigger,'' oversees this detection process. The online system consists of the Data Acquisition~(DAQ), Level 1 Trigger~(L1), and the High-Level Trigger~(HLT). 
    When the trigger discerns that B \(\bar{\text{B}}\) mesons have been produced, it activates a comprehensive readout of the entire detector, converting this information into a format suitable for downstream analysis and reducing the amount of data as much as possible before they reach the first storage. The DAQ system is there to make sure that all trigger signals are synchronously delivered to all sub-detectors, and provides the high-speed data links to read out the full detector data for each event and forward it to the HLT system.
    The Belle~II trigger, responsible for starting the data readout of the whole detector for interesting events, is engineered to handle collision rates up to 30 kHz at full SuperKEKB design luminosity.

    \begin{figure}[ht]
    \centering
    \includegraphics[width=0.7\textwidth,clip]{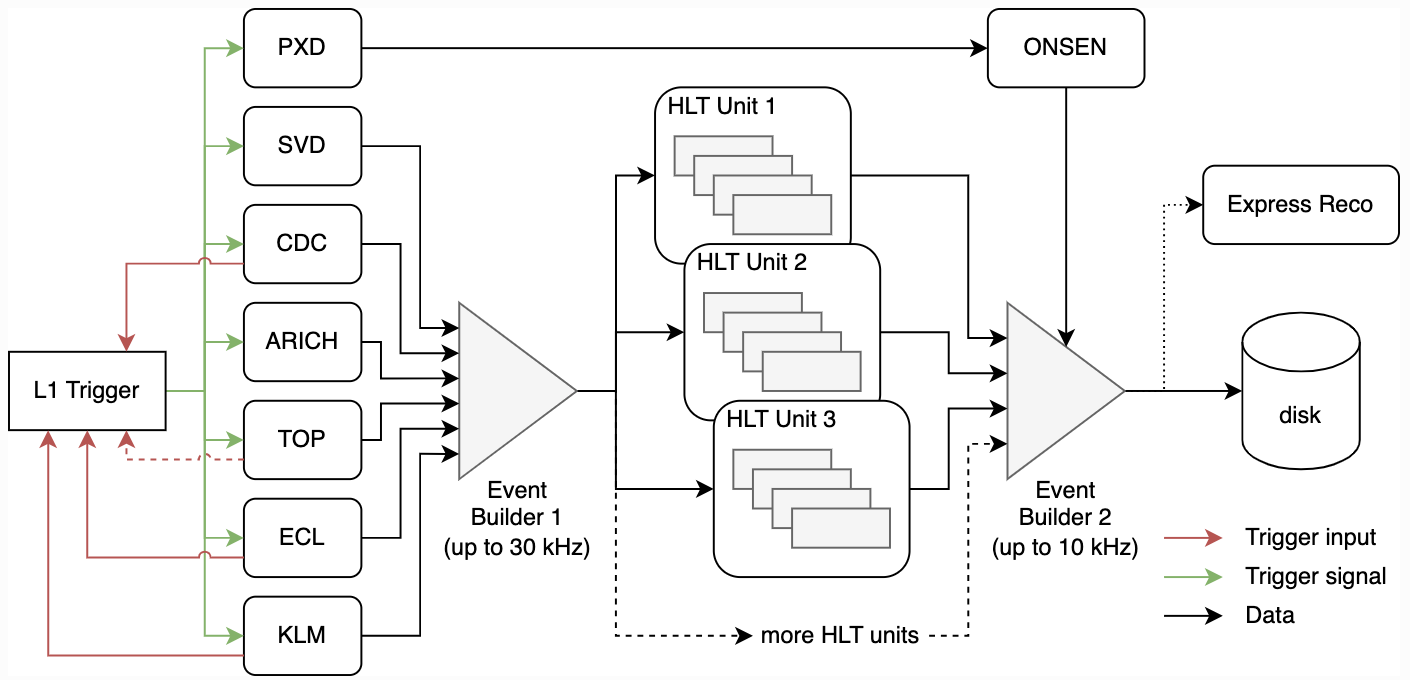}
    \caption{A schematic and simplified diagram of the Belle~II data flow, taken from~\cite{kou_belle_2019}}
    \label{fig:belle2_dataflow}
    \end{figure}

    \item \textbf{Event Classification and Data Segmentation:} In particle physics, the term ``event'' represents a specific, time-bound physical process, commonly a particle decay. Within the context of Belle~II, a ``signal event'' refers to, e.g., the unique occurrence tied to B \(\bar{\text{B}}\) meson pair and the subsequent decay products generated by each meson. 
    The organized data captured by the full detector readout, instigated by the trigger, exemplifies one such signal event. Collections of these signal events are then grouped into datasets that serve as the primary material for scientific discovery. It is crucial to distinguish these signal events from ``background noise,'' which includes extraneous physical processes that can interfere with the experimental results. They will be discussed in detail during the next section.
    As the luminosity of SuperKEKB escalates through a reduction of transverse beam size and a doubling of the beam currents, compared to its predecessor KEKB, the background noise at the Belle~II detector is substantially raised. Generally, this background interference originates from two primary sources: beam-induced processes and luminosity backgrounds. Beam-induced backgrounds are a consequence of beam particles interacting with elements like residual gas inside the beam pipe, bending magnets, or other particles within the same bunch. On the other hand, luminosity backgrounds result from beam collisions that yield inconsequential physics phenomena, such as Bhabha scattering or two-photon interactions. Identifying and discarding these background processes is of paramount importance, especially for low-multiplicity analyses.
    
    \item \textbf{Event generation and Simulation:} The ability to compare the observed data from the detector with theoretical expectations is paramount for the validation of results in particle physics. To fulfill this requirement, simulated events are generated that mimic the behavior of real detector events as closely as possible. This simulation process relies heavily on the Monte Carlo~(MC) method, which involves the repeated sampling of pseudo-random numbers. Simulated data enables the estimation of the statistical uncertainties associated with the recorded data and the estimation of background noise. It is also used to validate the detector and its operation, estimate the detector's eﬃciency, and perform calibration purposes. Simulation of data involves two steps, namely, event generation and the simulation of the detector response of each generated event. \emph{Event generation} consists of simulating the physics interactions under study. Based on the initial conditions set for colliding electrons and positrons, a range of particles are generated in accordance with the specific physics model being examined—be it advanced theories. These generated events are often categorized into different samples for specific analyses: ``signal MC'' pertains to the decay process being studied, whereas ``generic MC'' refers to basic standard model processes. Occasionally, additional ``background MC'' samples may be generated for processes that need to be excluded from the analysis. For Belle~II, this phase is efficient and tailored for various study objectives. 
    
    \item \textbf{Detector Simulation:}
    Upon generating the essential 4-momentum vectors and taking the detector geometry and the magnetic ﬁeld into account, the next task is to simulate their interactions with the detector material, capturing processes such as ionization, scintillation, Bremsstrahlung, pair production, and Cherenkov radiation, among others. The simulation of the detector response is called ``digitization.''
    This part of simulation has been the subject of extensive research and development, culminating in the simulation software Geant4~\cite{agostinelli_geant4simulation_2003}. Geant4 takes the generated particles and replicates their interactions within a virtual Belle~II environment. Subsequently, the energy deposition and particle interactions within each sub-detector are computed. Geant4 performs the simulation by tracking the particles, one at a time, through the geometry. It takes into account the eﬀect of the magnetic ﬁeld on the particle, the energy loss, and multiple scattering the particle experiences while traversing material and various other electromagnetic and hadronic eﬀects. 
    If a particle decays, the decay products are added to the list of particles. The initial four-momentum vectors given to Geant4 are called \emph{primary particles}, while all particles created from interactions or decay during the simulation are called \emph{secondary particles}.
    Additional specialized software modules then convert these Geant4 outputs into realistic detector signals. The digitization is the last step in the processing chain used solely for Monte Carlo data. Post-digitization, the workflow for both Monte Carlo and real-world data aligns in terms of data processing steps. For instance, the pixel detector software translates the energy deposits into pixel activations, \emph{PXD digits}. This part of the simulation is the computational bottleneck both from the time and space~(storage) complexity perspective. 
    
    For example, the PXD background data has the highest storage consumption among all sub-detectors, with an utterly infeasible hypothetical online simulation time. A solution to alleviate the detector simulation problem is to use surrogate models under the topic of ``Fast Simulation,'' the main focus of this thesis.
    
    \item \textbf{Reconstruction Procedures:} The term ``reconstruction'' in this setting refers to the act of methodically determining the attributes of a decaying \(B\) meson and its decay products. These reconstructed signal events serve as the foundational data for the downstream physical analyses. The reconstruction workflow adopts a bottom-up strategy, starting with the final state particles and their detector responses and moving backward to recreate each decay event until the originating B \(\bar{\text{B}}\) pair~(or one B) is reached. However, it’s never possible to uniquely identify all the particles in the interaction because of not only hadronic interactions and background processes but also because we a priori don't know which decay products correspond to which mother particles. So, all we can do is look at the detector response, find a set of most likely particles, and then leave it to the analyses to do a proper statistical analysis of the events. The first stages of reconstruction consist of two main procedures:
    
    \begin{itemize}
        \item \textbf{Clustering:} One of the first steps in reconstruction is ``clustering'', where one needs to modulate the detector responses in each sub-detector if they are related. This is done by means of a clusteriser, which groups together adjacent detector signals~(nearest neighbor) into clusters, performed for both real data and Monte Carlo data in the same way. One can then calculate geometric properties of these clusters like size, shape or center.
        For PXD, either a center of gravity or an analog head-tail algorithm~(for cluster multiplicities larger than two hit pixels), is used. These algorithms aggregate and interpret data from pixelated detectors such as PXD that results in ``PXD Clusters.'' For real detector data, the simulation stage is skipped, and the input is the ﬁred pixels, ``PXD Digits.''
        
        The ``Center of Gravity''~(cog) method is widely used for localizing the point where a particle has interacted with a detector, especially in situations where the interaction affects more than one pixel. The idea is to take a weighted average of the positions of all the pixels in a cluster, with the weights being the ionization strengths~(e.g., charge, energy deposition, etc.) in those pixels, as: 
        \[
            \text{cog} = \frac{\sum_{i=1}^{N} w_i \cdot x_i}{\sum_{i=1}^{N} w_i}
        \]
        Where \( w_i \) is the weight~(often the energy deposition) of the \( i^{th} \) pixel, and \( x_i \) is the position of the \( i^{th} \) pixel. \( N \) is the total number of pixels in the cluster. 
        The ``Analog Head-Tail'' algorithm is used primarily when the cluster multiplicity is larger than two hit pixels. The algorithm aims to identify the 'head' and 'tail' of a cluster, which correspond to the entry and exit points of the particle in the detector plane. This is particularly useful for distinguishing between particles that might have similar total charge deposition but different directions.

        The clustering of the PXD pixels is done in a row-wise manner, with increasing values for the column index and the row index. Starting with the pixel in the upper left corner. The clusteriser of basf2 checks each pixel to make sure that the ratio of its charge over a common noise level is above a given threshold. If it is, the left neighbor in the same row and the direct neighbors in the previous row are investigated. If one or more clusters have already been found in those neighboring pixels, the clusters are merged and the pixel is assigned to this cluster. Otherwise, a new cluster is created, and the pixel becomes its first member. The clusteriser proceeds with the pixel to the right of the current pixel or, if the pixel is the last pixel in the current row, with the ﬁrst pixel of the next row. The procedure is repeated until the last pixel in the last row has been processed. This clustering scheme investigates each pixel only once and requires only the current and the previous pixel row to be stored in memory, making it an efficient and memory-saving pixel clustering method. After having grouped all pixels into clusters, the position of each cluster is determined: if the size of the cluster, deﬁned by the number of pixels belonging to the cluster, exceeds a given threshold, the head-tail algorithm is used to calculate its position. The default threshold is set to 3 pixels. The Analog Head-Tail algorithm calculates the position of the cluster by using the outermost pixels of a cluster~(head and tail).

        \item \textbf{Track Reconstruction:} The Tracking~(Track Reconstruction) of Belle~II is designed to capture the spatial location of charged particles as they move through the detector. Reconstruction involves the inference of
        the trajectories of particles through the tracking detectors, called tracks. Multiple sensor layers work in tandem to collect this positional data, enabling the accurate reconstruction of particle paths, known as ``tracking.'' Belle~II incorporates two primary tracking steps to ensure precise trajectory identification and thereby contribute to accurate decay vertex estimates, the track ﬁnding and the track ﬁtting step. During the track ﬁnding step, the clusters from the PXD/SVD and the ﬁred wires in the CDC that belong to the same track are identiﬁed. In other words, it finds patterns in the hits or hit clusters in the tracking detectors.
        Then, the identiﬁed clusters and wires are passed to the track ﬁtting, which estimates the optimal track parameters. Here, the goal is to determine the best estimate of the kinematic variables describing the particle trajectories corresponding to each found hit/cluster pattern to obtain the particle position and momentum close to the interaction region as precisely as possible. 
        
       %
        
    \end{itemize}
\end{enumerate}

\section{Pixel Vertex Detector~(PXD)}
\begin{figure}[htb]
    \centering
    \includegraphics[width=0.65\textwidth]{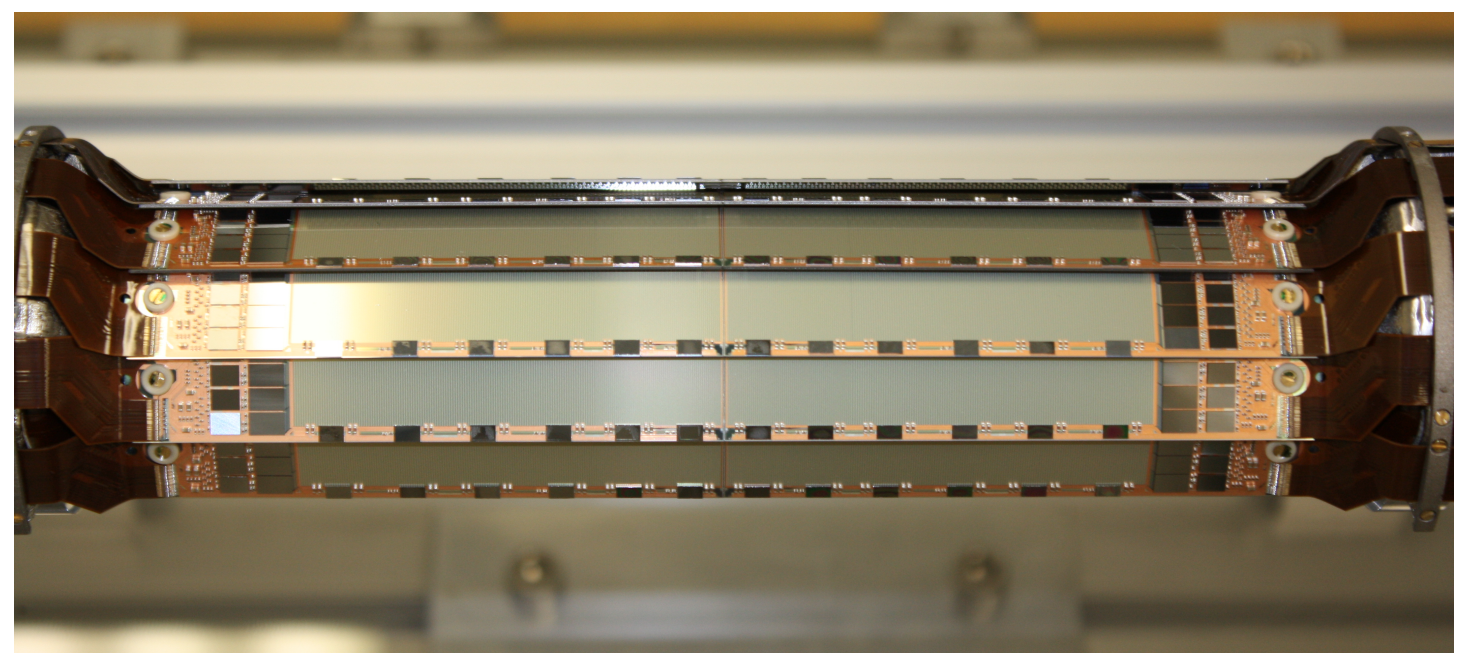}
    \caption{The first picture of the full PXD during the commissioning phase in DESY Hamburg.}
    \label{fig:pxd_img}
\end{figure}

\begin{figure}[!htb]
    \centering
    \begin{subfigure}[t]{0.6\textwidth}
        \centering
        \includegraphics[width=\textwidth]{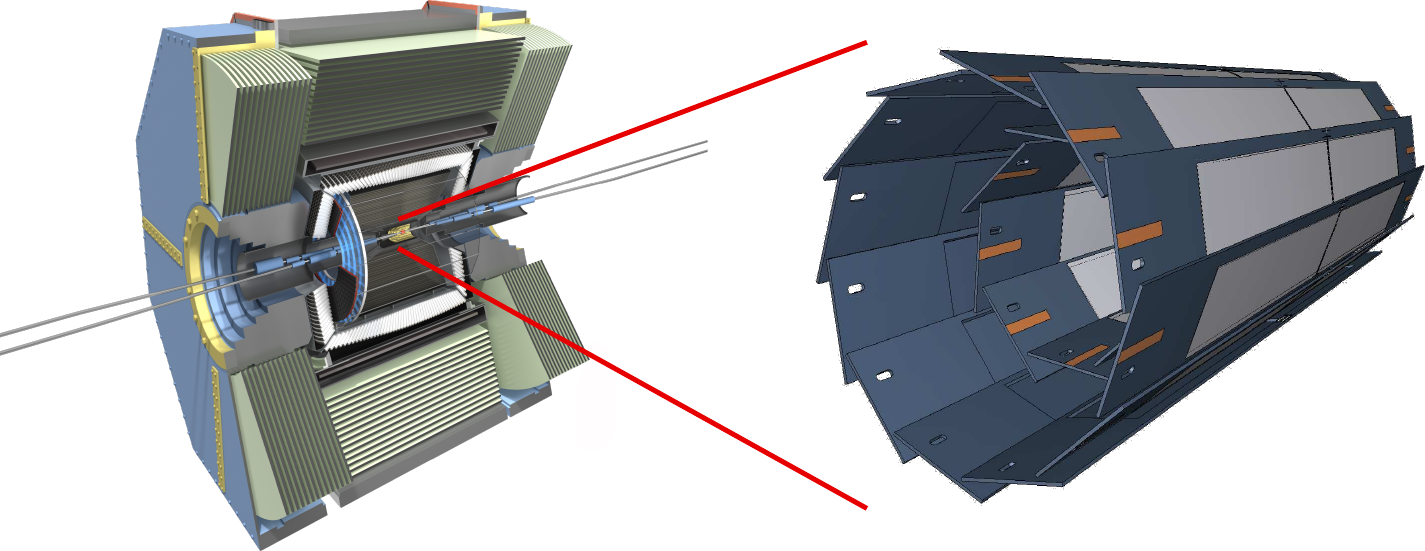}
        \caption{}
        \label{fig:pxd_location}
    \end{subfigure}%
    \hfill%
    \begin{subfigure}[t]{0.3\textwidth}
        \centering
        \includegraphics[width=\textwidth]{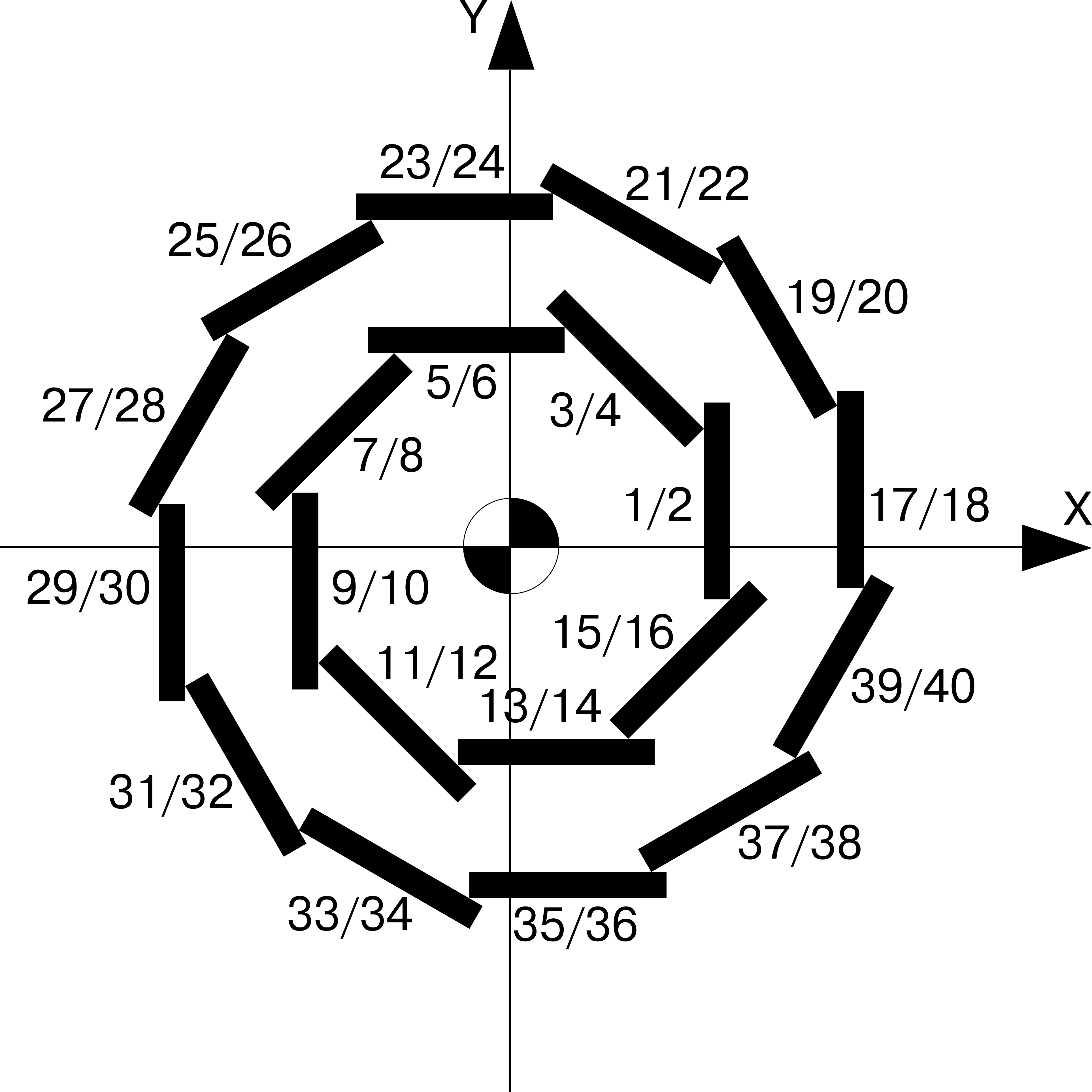}
        \caption{}
        \label{fig:pxd_layout}
    \end{subfigure}
    
    \begin{subfigure}[t]{\textwidth}
        \centering
        \includegraphics[width=\textwidth]{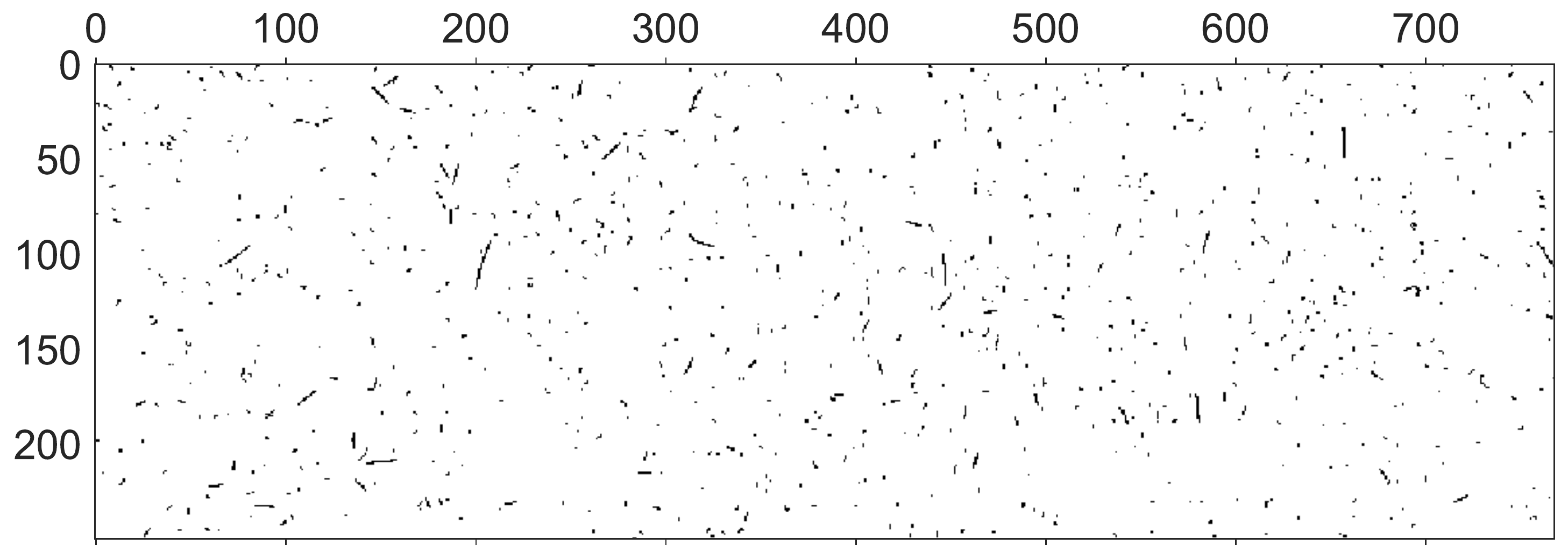}
        \vfill
        \includegraphics[width=\textwidth]{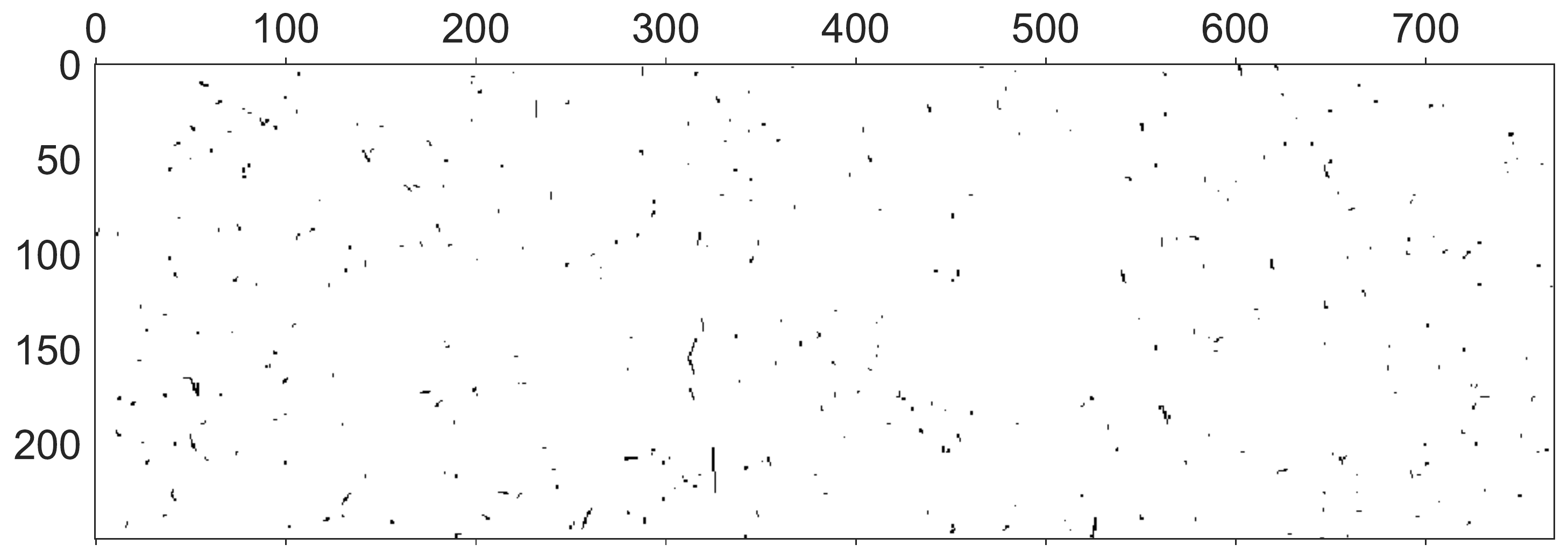}
        \caption{}
        \label{fig:pxd_sensor_examples}
    \end{subfigure}%
    \caption{
        \textbf{Pixel detector layout} The pixel detector~(PXD) is the inner-most sub-detector of the Belle~II experiment (\textbf{a}) and is configured in a two-layered overlapping sensor structure (\textbf{b}).
        Geant4 Simulated PXD Background image examples~(\textbf{c}) for sensors \num{7} (top) and \num{25} (down).
    }
    \label{fig:PXD_overview}
\end{figure}

The Pixel Vertex Detector~(PXD)~\cite{giakoustidis_status_2023} serves as the innermost layer of the Belle~II detector complex, shown in ~\cref{fig:pxd_img} and~\cref{fig:PXD_overview}. This semiconductor-based apparatus employs DePFET~(Depleted P-channel Field Effect Transistor) technology, a semi-conductor sensor that combines the detection of the passage of charged particles and the ampliﬁcation of their deposited energy within one device. The PXD is the first full-scale detector employing the DePFET technology in High Energy Physics.
The primary function of the PXD is to determine the positions of charged particles that emerge from collision events. A key objective is the highly accurate determination of decay vertices, a task for which the PXD is strategically located in close proximity to the interaction point~(IP). The device spans a comprehensive polar angle range, specifically \( 17^\circ < \theta < 150^\circ \), within the Belle~II setup. The PXD incorporates \num{40} pixel sensors, depicted in~\cref{fig:pxd_layout}, which are organized into two concentric circles~(annulus) around the IP, termed the inner and outer layers. These pixel sensors are paired into structures known as ``ladders,'' aligned along the \(z\)-axis. The geometrical arrangement of these ladders in the \(x\)-\(y\) plane resembles an octagon for the inner layer and a dodecagon for the outer layer. 

\begin{figure}[!htb]
  \centering
  \begin{subfigure}[b]{0.85\textwidth}
    \includegraphics[width=\textwidth]{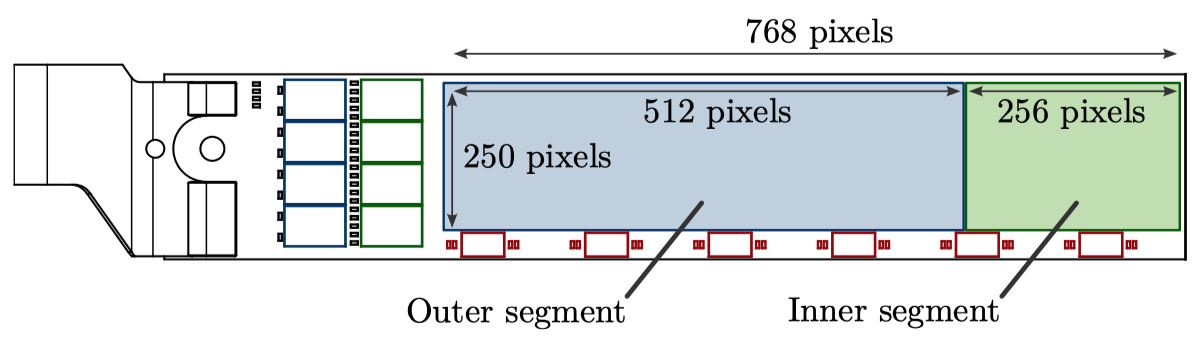}
  \end{subfigure}
  \\
  \begin{subfigure}[b]{0.85\textwidth}
    \includegraphics[width=\textwidth]{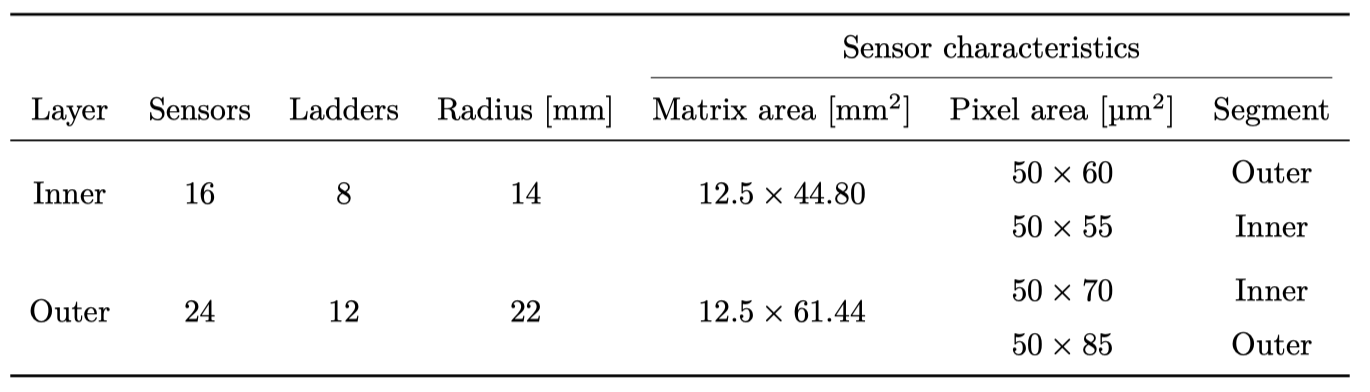}
  \end{subfigure}
  \caption{PXD specifications and detailed information adopted from~\cite{moll_comprehensive_2015}}
  \label{fig:pxd_detail}
\end{figure}

Each of these pixel sensors integrates a \(250 \times 768\) matrix of silicon pixels that are engineered to interact with charged particles, as shown in~\cref{fig:pxd_detail}. The activation of a pixel is closely linked with the ionization of underlying material within the silicon medium and producing electron-hole pairs. Data relating to the spatial coordinates of the particle-pixel interaction and the number of ionized electrons~(accumulated charge) is then collected as ``hits''~(see~\cref{fig:pxd_sensor_examples}). 

The output from the pixel sensor is essentially a map detailing the hits and their corresponding levels of charge displacement, measured during the readout interval. This raw data undergoes noise filtering and digitization, converting it into 8-bit integers that range from~\num{0} to~\num{255}.
Given the technical constraints of silicon pixel technology, physical necessities, and the need for rapid readout, the pixel sensors are calibrated for optimal spatial resolution, thus ensuring adequate tracking precision~\cite{moll_comprehensive_2015}.

Remarkably, the PXD can generate an ``ultra-high granular'' amount of raw data, owing to its intricate pixel configuration comprising a total of \(7.68 \times 10^6\) pixel channels. The data production rate can reach up to \(20 \, \text{GB/s}\), outpacing the combined data rates of all other components in the detector by over a factor of ten. The PXD necessitates a parallel readout time of \num{20} \( \mu s \) from the individual pixel sensors which is considerably longer than the approximately \num{10} \( \mu s \) required for a beam particle to complete one circuit of the collider.

\subsection{PXD Background: Types and Levels}
Due to its proximity to the IP and the incoming particle beams, the PXD is susceptible to a huge amount of radiation and background levels. PXD hits generated by background shower particles deteriorate the detector’s physics performance. The rates of these background processes are correlated with multiple factors e.g. beam size, beam current, luminosity, accelerator status, and vacuum conditions~\cite{kou_belle_2019}.
Generally, the processes contributing to background in the detector can be classified into five main categories:

\begin{itemize}
\item \textbf{Touschek Background:} At Belle~II, Touschek scattering is a predominant source of background. This phenomenon involves Coulomb interactions between two particles within the same beam bunch. A ``bunch'' refers to a group of particles that are packed together and travel around the accelerator ring almost as a single unit. These bunches are created to maximize the chances of interaction between particles when two such bunches cross paths in the detector. As a result of such interactions, the energy levels of the two participating particles diverge from the nominal beam energy; one particle experiences an energy increase while the other undergoes an energy loss. The rate of Touschek scattering is directly related to the square of the beam current and inversely related to both the number of bunches in the accelerator ring and the size of the beam~\cite{natochii_measured_2023}.

\item \textbf{Beam-Gas Background:} Scattering between the beam and residual gas molecules within the beam tube stands as another significant source of background noise at Belle~II. Two types of beam-gas scatterings are prevalent: the elastic Coulomb scattering, which alters the trajectory of the beam particles, and the inelastic Bremsstrahlung, which diminishes their energy. The rate of such scatterings is directly linked to the residual gas pressure inside the tube and the current of the beam. 

\item \textbf{Luminosity-dependent Background:} Background arising from beam interactions at the Interaction Point~(IP) is referred to as luminosity-dependent background, and its intensity is directly proportional to the luminosity, as shown in~\cref{fig:pxd_occ}~\footnote{Adopted from the PXD Analysis meeting 2.11.2022. by D.Pitzl}. 
In the case of SuperKEKB, the target luminosity is approximately 40 times greater than KEKB's peak luminosity, making the luminosity-dependent background a significant concern. 
A critical form of this background stems from radiative Bhabha scattering $(e^+ e^- \rightarrow e^+ e^- \gamma)$. In this process, beam particles emit photons and deviate from their nominal paths. At a high luminosity regime, this background dominates over other Belle~II backgrounds.
Two-photon processes $(e^+e^- \rightarrow e^+e^-e^+e^-)$ represent another source of background. The emitting low-momentum electron-positron pairs cause the beam particles to lose energy, similar to the radiative Bhabha process. Moreover, when possessing low transverse momentum, the emitted electron and positron curl in the Belle~II solenoid field, making multiple hits on the PXD.

\begin{figure}[!htb]
    \centering
    \includegraphics[width=0.6\textwidth]{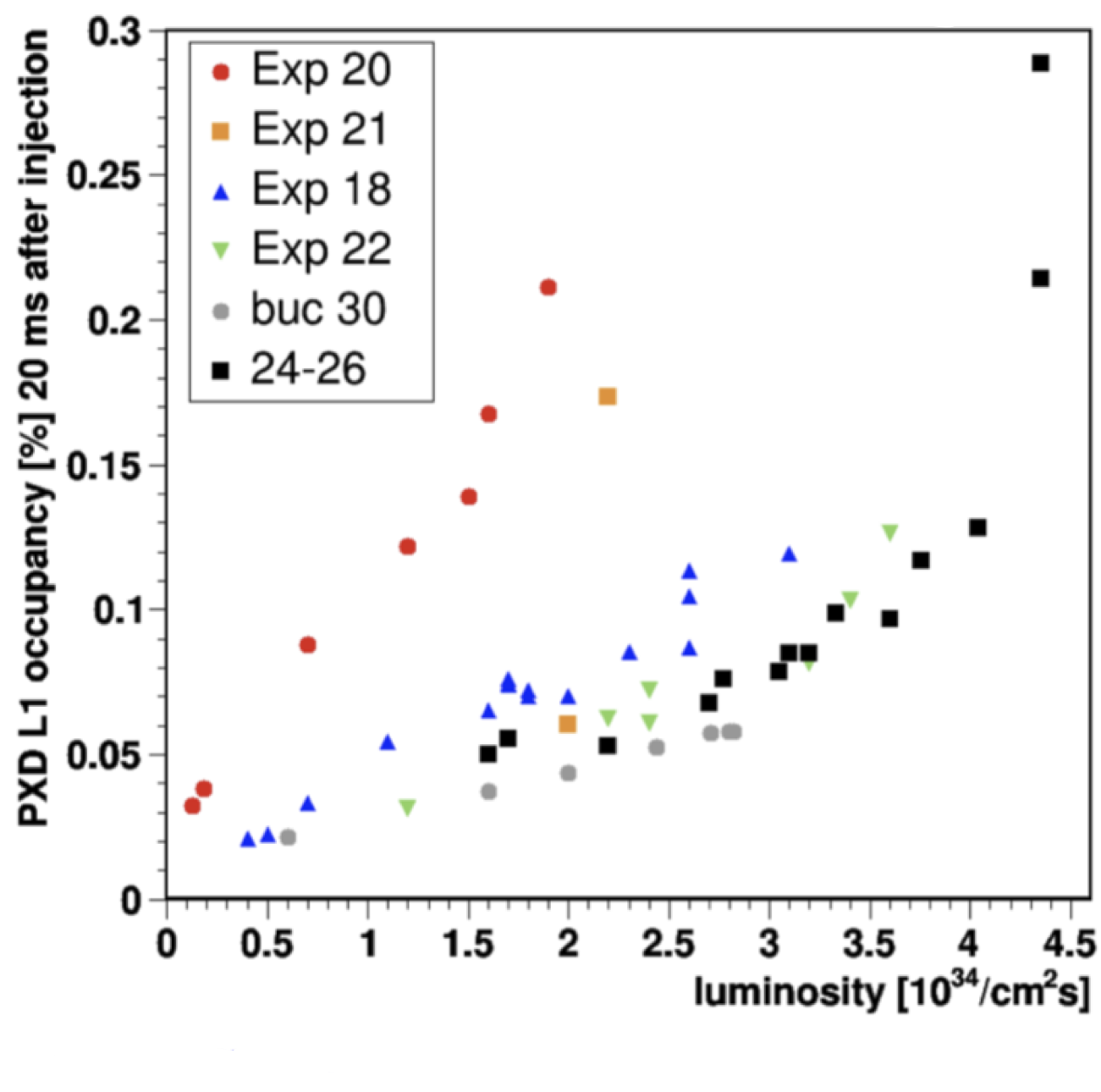}
    \caption{PXD occupancy~(layer 1 here) increases with beam currents and luminosity, coming from raw data. The occupancy of PXD is defined as the ratio of the number of hits to the total number of PXD pixels. The Experiment numbering corresponds to the Belle~II run periods.}
    \label{fig:pxd_occ}
\end{figure}

\item \textbf{Background from Synchrotron Radiation:} Another background component affecting the inner detectors of Belle~II arises from Synchrotron Radiation. The intensity of this radiation is governed by the square of both the beam energy and the magnetic field strength. Thus, the HER electron beam predominantly contributes to the Synchrotron Radiation background. The energy of Synchrotron Radiation photons impacting the PXD and the SVD ranges from a few kiloelectronvolts~(keV) to multiple tens of keV. 

\item \textbf{Background Due to Injection:} The beam lifetime in SuperKEKB is notably brief, necessitating frequent top-up injections through a betatron injection scheme~\cite{noauthor_accelerator_nodate}. When the overall beam current drops below 99\% of the nominal current, additional charge is injected into low bunch-current buckets at varying frequencies~(between 1--25 Hz). These new bunches oscillate horizontally around the main beam, causing temporary spikes in the background levels of the Belle II detector. These spikes last for milliseconds each time the newly-injected bunch crosses the interaction point.
\end{itemize}

Thus, these traversing background processes can deposit considerable energy, thereby reducing the lifespan of the detector and creating malfunctioning pixels or zones of reduced efficiency. Moreover, due to its proximity to the IP, PXD is also vulnerable to back-scattered low-energy Synchrotron Radiation photons. 
This huge amount of background manifests itself primarily in a sensor-level observable called the ``occupancy.'' The occupancy of a pixel sensor, defined as the ratio of the number of hits to the total pixel count in the matrix, is directly influenced by the level and type of background noise. 
Currently, the average PXD occupancy is below 0.3\%.

In reality, there exists a constraint due to bandwidth limitations. For a \num{30} kHz trigger rate, data loss starts to become a factor when the average occupancy of the inner PXD layer exceeds 3\%. Beyond this occupancy threshold, offline performance experiences a marked decline due to issues like cluster merging and a higher likelihood of incorrectly associating hits to particle tracks. It should be noted that performance degradation can begin even below this specified level. 

\section{PXD Background Simulation: Ideas and Challenges}
To accurately replicate experimental conditions, emulating background processes is essential. Two methods are currently used, as depicted in\cref{fig:bkg_opt}, for implementing background simulation: ``background simulation'' and ``random trigger'' Both techniques have their pros and cons, and the choice between them depends on the specific requirements of the simulation and the resources available.

\subsection{Background Simulation}
\begin{figure}[!htb]
    \centering
    \includegraphics[width=0.75\textwidth,clip]{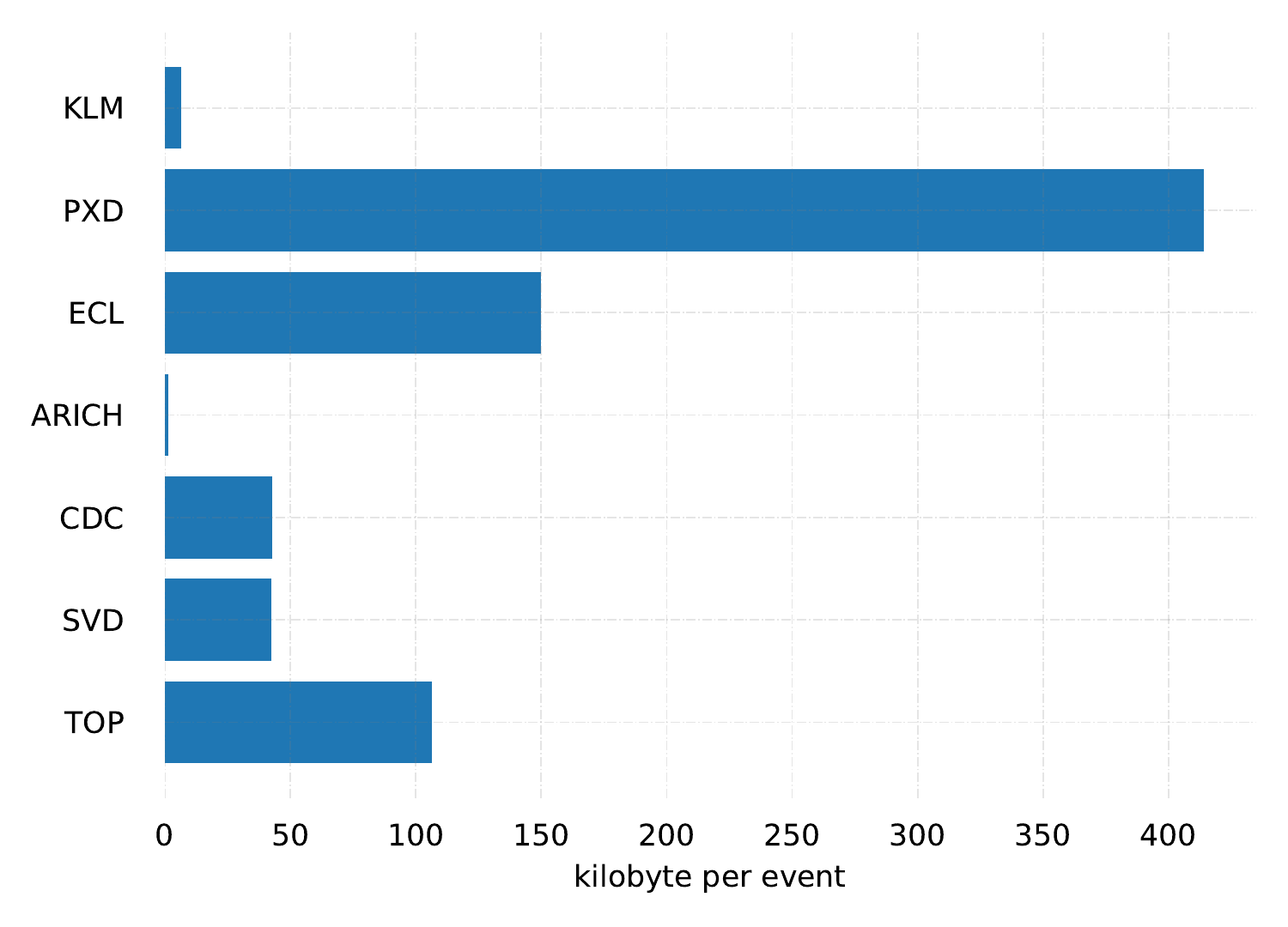}
    \caption{Geant4 simulated Background data volume comparison of PXD data per event compared to other sub-detectors at Belle~II. The PXD background data is nearly as much as the combined background data for all other sub-detectors.}
    \label{fig:data}
\end{figure}

In the ``background simulation'' approach, a collection of synthetically emulated background events serves as a resource. These backgrounds are simulated with the software framework Strategic Accelerator Design~(SAD)~\cite{noauthor_sad_nodate}. SAD, initialized with beam optics parameters and detector elements, tracks scattered particles through the sequence of detector elements. The tracking simulation initiates by establishing uniformly spaced scattering zones around each ring, generating particle bunches in these regions. These particles are randomly created within a three-dimensional Gaussian distribution. Their momentum and statistical weight are calculated using established scattering theories, specifically Coulomb, Bremsstrahlung, and Touschek scattering. These background events are essentially the particle outcomes of specific background processes. Using Geant4 for detector response simulation, the PXD digits are then stored in a dedicated file. This file is then utilized to overlay background particles onto those from a signal or physics event, as shown in~\cref{fig:bkg_sim}. This method improves the quality of digitization by combining energy deposits to generate sensor responses. However, it neither fully reconciles the discrepancy between simulated and real background data nor is it computationally efficient. 
The reason is that it has to be pre-generated and stored, thus creating a high ``space-complexity'' from the computational point of view, as shown in~\cref{fig:data}. On the other hand, it is extremely inefficient to run this on the fly and creates a high ``time-complexity'' issue.

\subsection{Random Trigger}
In the ``random trigger'' technique, an alternative to direct simulation, real experimental sensor responses to background noise are directly utilized randomly. This approach operates on the premise that each sensor's response can be associated with a unique background event. The procedure begins with a focus on simulating the energy deposits of only the signal particles. Subsequently, sensor-specific responses from actual recorded background events—coming from random trigger events—are overlaid during the digitization stage, as shown in~\cref{fig:bkg_rand}. The random trigger activates data collection separate from the event signature. Three kinds of random triggers exist~\cite{moll_comprehensive_2015}: a periodic one synced with SuperKEKB's bunch crossing signal, a pseudo-random trigger based on an independent local clock, and a delayed Bhabha trigger that fires shortly after a specific bunch passes. This results in a more faithful simulation of real PXD conditions than the former technique. This method even allows for the incorporation of noise from detector electronics. However, it also comes with challenges, such as the potential loss of certain background properties due to threshold effects, the lack of real detector background response for higher~(undetected) luminosity, and as always, the requirement for storage and transfer of a large volume of extra data. Thus, this also creates a high ``space-complexity'' issue from the computational point of view. 

\section{PXD Background Simulation: Surrogate models}
A strong solution to the above challenges and issues for PXD simulation is ``Surrogate models''. Surrogate models for fast detector simulation are simplified, computationally efficient approximations that emulate the behavior of more complex, detailed simulations of particle detectors on the fly of the analysis pipeline. 
Traditional detector simulation methods, as discussed above, are very computationally intensive and storage costly. 
A surrogate model aims to replicate the essential features of a full detector simulation but at a fraction of the computational cost. These models are constructed using Deep Generative Models~(DGM), where the surrogate model is trained on a dataset either generated by the Geant4 simulation or the real detector data. Once trained, the model can generate samples that are statistically similar to their training data and even generalize to the data beyond the training data. 
There are some key requirements that the model should meet to be effective and efficient:

\begin{enumerate}
    \item \textbf{Low time-complexity}: The surrogate model must be computationally efficient in order to facilitate fast simulations. Computational speed is crucial when performing large-scale simulations or when needing to iterate the model many times for optimization or fine-tuning. Thus, the model should take advantage of parallel processing capabilities and have to be optimized for the hardware it is expected to run on, whether it's a CPU or GPU.
    \item \textbf{Low space-complexity}: It should come with a minimal storage cost. Hence, the underlying compression technique has to reduce the storage footprint without sacrificing too much in terms of the downstream physics analysis.
    \item \textbf{Realistic and Diverse:} It has to generate samples as diverse as possible from the downstream physics analysis point of view. Thus, the sampling techniques should be capable of employing the nuanced behaviors and symmetries of the PXD to ensure that the model captures the diversity inherent in the real data.
    \item \textbf{Extrapolation:} 
    It has to be able to extrapolate to background levels beyond the current beam parameters and luminosities in order to analyze the PXD operation at higher luminosity and to do physics analysis beyond the current experimental limits. Therefore, the model should be robust against overfitting and incorporate a measure of control when extrapolating to give a range of plausible outcomes.
\end{enumerate}

\begin{figure}[!htbp]
    \centering
    \begin{subfigure}[b]{0.4\textwidth}
        \centering
        \includegraphics[width=\textwidth]{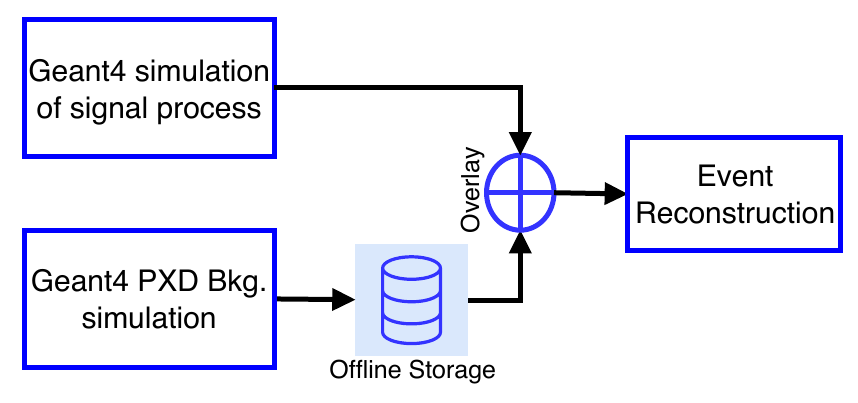}
        \caption{Background overlay with Geant4 simulation}
        \label{fig:bkg_sim}
    \end{subfigure}
    \hfill
    \begin{subfigure}[b]{0.4\textwidth}
        \centering
        \includegraphics[width=\textwidth]{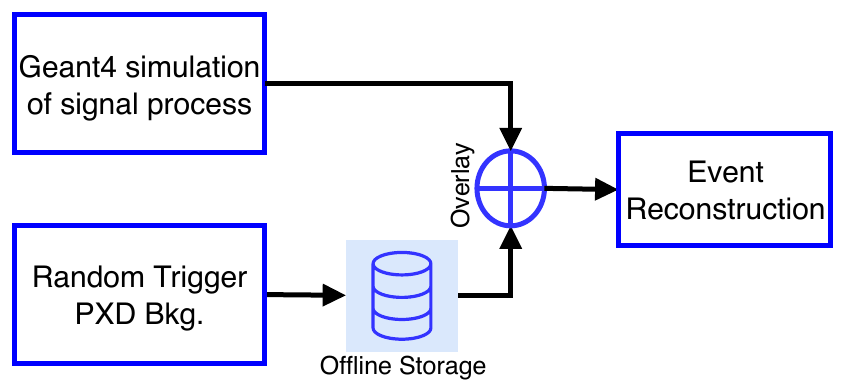}
        \caption{Background overlay with random trigger}
        \label{fig:bkg_rand}
    \end{subfigure}
    
    \vskip\baselineskip
    \begin{subfigure}[b]{0.4\textwidth}
        \centering
        \includegraphics[width=\textwidth]{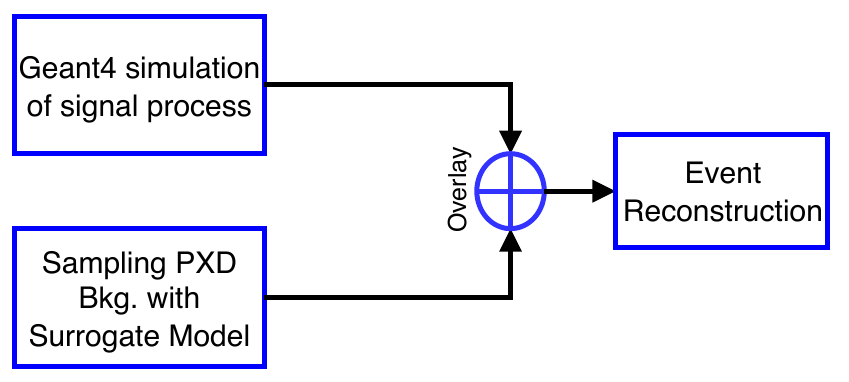}
        \caption{Background overlay with Surrogate model}
        \label{fig:bkg_surr}
    \end{subfigure}
    
    \caption{PXD Background simulation options.}
    \label{fig:bkg_opt}
\end{figure}

In this thesis, I propose the surrogate modeling solution as an amortized simulator for the PXD background simulation as a fast and efficient simulation method, as shown in~\cref{fig:bkg_surr}. 
First, a DGM is developed to emulate Geant4 simulated background data as an approximation of the real data with much lower data complexity and implement this approach into the Belle~II software, basf2. 
Then, for the final challenge of real PXD detector data, I introduce a much more efficient DGM as a solution that successfully satisfies all the above conditions. 

\section{PXD Background Simulation: Figure of Merits}
For evaluation, we have three main categories of metrics: Low-level Metrics, Neural Network-based Metrics, and Physics-Level Metrics. 
These metrics evaluate different aspects of the generated PXD background. Through the next chapters, I elaborate on and incorporate them to analyze and compare the results with each other. Some of them are being introduced for the first time in the Fast Detector Simulation domain. In the following, I briefly introduce them.

\subsection{Low-Level Metrics}
For Low-Level metrics, we are interested in analyzing the histogram projections of the data.
Therefore, the Low-Level observables as marginal distributions include charge distribution and average charge values per sensor, occupancy distribution and average occupancy per sensor, and the correlation between the average occupancies.

\subsection{Neural Network-based Metrics}
The necessity for incorporating Neural Network~(NN)-based metrics in the evaluation process of PXD background simulation arises from two main factors:

\begin{itemize}
  \item \textbf{Complexity of Data}: PXD background data is intrinsically complex and is influenced by numerous variables originating from the underlying Physics processes. Moreover, they exist in a high-dimensional space, often making evaluating their quality and diversity challenging using traditional metrics.
  Thus, simple 1D and 2D histogram-based Low-Level observables might not fully capture this complexity. NN-based metrics can distill this high dimensionality into a more manageable form by focusing on relevant feature spaces. This is particularly useful for understanding what aspects of the data are being captured~(or missed) by the DGM, such as the data modality~(mode collapse).
  
  \item \textbf{Quantitative Assessment}: While Low-Level metrics provide a direct evaluation of specific properties, they might lack the capacity for an overall quantitative assessment of data similarity. NN-based metrics offer a single quantitative score that can be compared across different models or iterations.
\end{itemize}

Thus, NN-based metrics offer a multifaceted evaluation approach that \emph{complements} the Low-Level metrics. They enable a more comprehensive, nuanced, and rigorous analysis of the performance and reliability of PXD background simulations. 
The NN-based metrics that I incorporate in this thesis are the Frechet Inception Distance~(FID)~(see~\cref{chap:5})~\cite{heusel_gans_2018}, Kernel Inception Distance~(KID)~(see~\cref{chap:5})~\cite{binkowski_demystifying_2021}, and Vendi Score~(see~\cref{chap:6})~\cite{friedman_vendi_2023} and show how they stand out as particularly insightful and interpretable evaluation tools.
FID and KID are used to measure the similarity between the distributions of generated data and real data in the feature space of an NN model that is pre-trained on the PXD background dataset with a very high precision. Vendi score, on the other hand, is to measure mode collapse~(see~\cref{chap:3}) and the diversity evaluation problem~(see~\cref{chap:6}) of the generated data.

\subsection{Clustering and Topological Data Analysis}

In PXD background simulation, a fundamental question arises concerning the spatial organization of PXD hits. Specifically, how are points in a PXD point cloud clustered? What is the complexity of these clusters, and what topological and geometrical features are discernible? Or, as we adjust our observation window, what shifts can be observed in the geometric representation of the PXD hits?

Topological Data Analysis~(TDA) serves as an instrumental tool for investigating these questions. It allows us to delve into the intrinsic shape of the data by formalizing the notions of proximity and continuity of data. TDA focuses on identifying topological features in data sets, such as connected components, loops, and higher-dimensional cycles~(see~\cref{chap:6}). These features offer invaluable insights into the structural properties of PXD hits across multiple scales, which is especially useful for identifying underlying patterns and anomalies for both the real PXD data and the generated ones. This is the very first time that detector signatures are being analyzed through the lens of TDA.

Crucially, the reconstruction process in Belle~II relies on clusters of PXD hits to perform its analyses. As such, it becomes imperative to accurately reproduce the background on this cluster level. To address this, I further complement the TDA perspective with clustering analysis implemented in the Belle~II software. Integrating these two approaches aims to capture a more complete understanding of the structural, geometrical, and topological properties inherent in the PXD background data. Clustering analysis in Belle II offers a different set of evaluative tools that focus on grouping data points based on similarity measures such as distance or density. The complementary nature of these methods allows for a more robust validation mechanism, ensuring that the background is faithfully represented at the cluster level. This integrated approach provides an interpretable framework for the TDA and the clustering analysis, fulfilling the requirements for a precise and comprehensive background simulation.

\subsection{Physics-Level Metrics}
\begin{figure}[!hbt]
    \centering
    \includegraphics[width=0.45\textwidth]{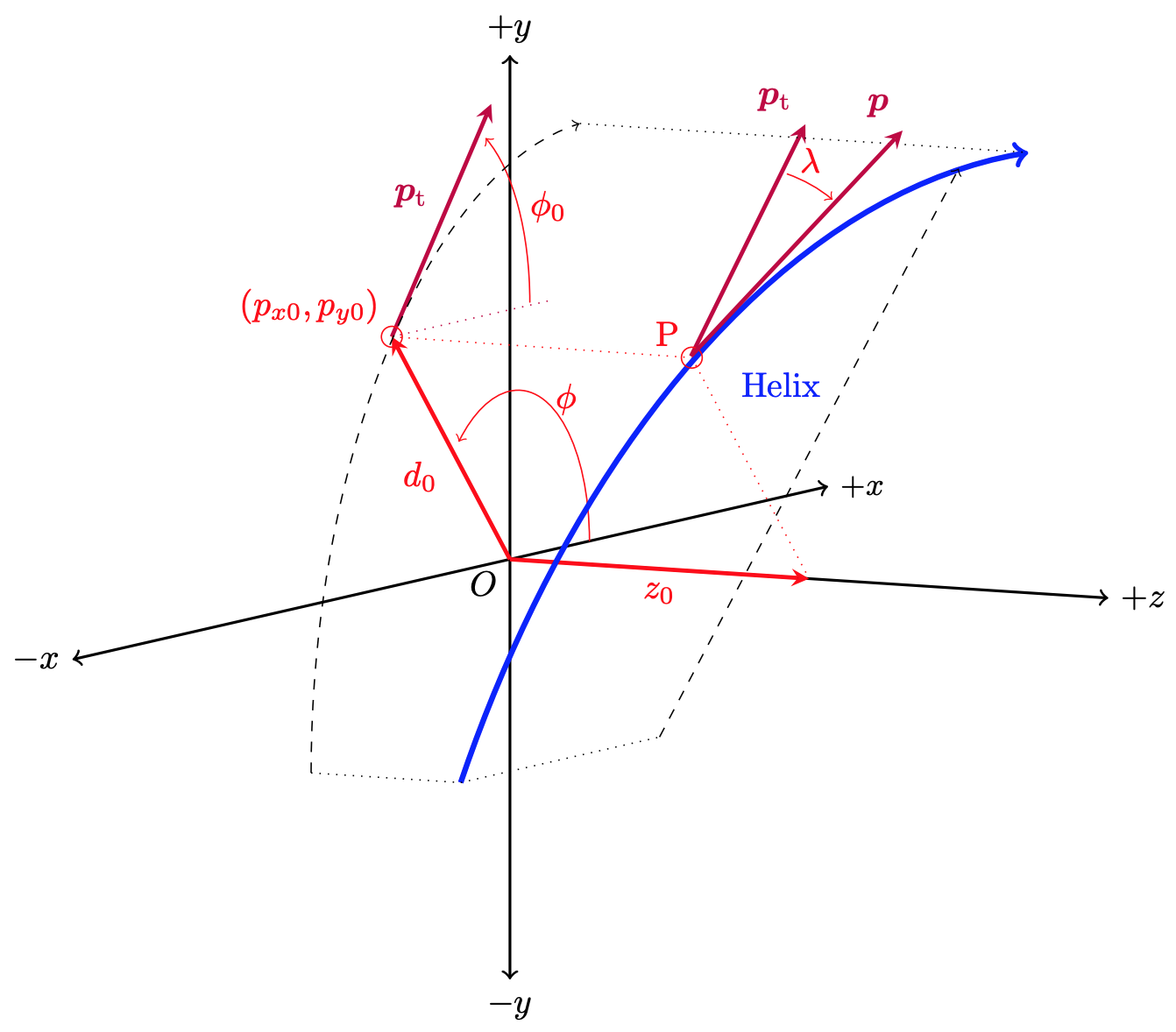}
    \caption{The perigee parametrization of the track helix adopted from~\cite{krohn_global_2020}}
    \label{fig:helix}
\end{figure}

Eventually, within this study, I also analyze the effect of generated and simulated PXD backgrounds on high-precision charged track reconstruction and compare them. This is indeed crucial for both decay vertex measurement, quantifying and correcting systematic uncertainties, and any subsequent physics analyses. The Helix parameters quantify the relative distance to the Point of Closest Approach~(POCA), a reference point on the particle track closest to the center of the coordinate system. The POCA is determined by extrapolating a particle track to the global detector z-axis.
In the Belle~II experiment, the trajectory of charged particles in a uniform magnetic field can be encapsulated by five helix parameters \(\{d_0, z_0, \phi_0, \omega, \tan\lambda\}\) relative to POCA, depicted in~\cref{fig:helix}. 

\begin{enumerate}
    \item \(\mathbf{d_0}\): Refers to the impact parameter, indicating the shortest perpendicular distance from POCA. This provides insight into the spatial offset of the helix trajectory from the coordinate system's center.
    \item \(\mathbf{z_0}\): Denotes the z-coordinate at the point where the helix is closest to the POCA, revealing the helix's vertical positioning relative to the central z-axis.
    \item \(\mathbf{\phi_0}\): Represents the azimuthal angle of the particle's momentum at its nearest proximity to the POCA, specifying the helix's initial rotational direction on the xy-plane.
    \item \(\mathbf{\omega}\): Characterizes the curvature of the helix, which is inversely related to the radius of curvature \(R\). The curvature provides a sense of the helix's tightness and is influenced by particle momentum and magnetic field intensity.
    \item \(\mathbf{\tan\lambda}\): Corresponds to the tangent of the dip angle \(\lambda\), determining the helix's inclination with respect to the xy-plane. A flat helix exhibits \(\tan\lambda = 0\), while a steeper helix is indicated by a larger \(\tan\lambda\).
\end{enumerate}

These helix parameters collectively offer a detailed geometric and kinematic description of a charged particle's trajectory. The resolution of these helix parameters serves as a compelling physics-level metric and comparative assessment for studying the effects of PXD background on charged track reconstruction~(see \cref{chap:5}).

  \chapter{On the Shoulder of Giants: 
The Modern Machine Learning Tools}
\label{chap:3}

\section{Introduction}

The arena of Artificial Intelligence has been evolving at a relentless pace, fueled by an incessant pursuit to replicate and augment human cognitive abilities. In this quest, the development of advanced machine learning techniques has been instrumental, providing the bedrock for sophisticated AI systems. This chapter embarks on a journey to explore these techniques, focusing on three pivotal advancements that have revolutionized the landscape: Transformers and Attention Mechanism, Deep Generative Models, and Self-Supervised Learning.

The first section, ``Transformers and Attention Mechanism'', dives deep into understanding these models that have redefined the field of natural language processing. With their ability to handle long-range dependencies in sequences, they form a key pillar in this study, providing us with the tools to extend and adapt these mechanisms to the context of detector simulation.

In the ``Deep Generative Models'' section, I venture into the realm of generative models, the vanguard of modern machine learning that strives to capture and mimic reality. By providing a deep dive into the two main latent variable models, Generative Adversarial Networks and Variational Auto Encoders, I aim to elucidate how these models serve as a launching pad for the innovative techniques and methods that I will introduce in the following chapters.

Finally, the ``Self-Supervised Learning'' section leads us to the frontier of machine learning research, the Dark Matter of Artificial Intelligence. By harnessing the power of SSL and enabling models to generate their own supervision, it has the potential to significantly improve the efficiency of learning algorithms. This concept will play a central role as I build upon it to devise novel learning strategies.

This chapter is designed to provide a comprehensive understanding of these three pillars of modern machine learning. By dissecting their inner workings, highlighting their strengths, and understanding their limitations, I lay the groundwork for the novel mechanisms and strategies I will introduce in the coming chapters. 
As we navigate this journey, we are indeed standing on the shoulders of giants, harnessing their wisdom and insights to forge new paths in the exciting world of AI research.

\section{Attention Mechanism and Transformers}
The concept of attention in neural networks is a powerful mechanism that allows a model to enhance its predictive ability by selectively focusing on specific subsets of data. 
This idea, inspired by the human cognitive function, assigns learned weights to quantify the degree of attention, thereby forming the output as a weighted average. In essence, humans do not process all incoming information simultaneously; instead, attention is selectively allocated to relevant information when necessary.
Initially implemented in the realm of computer vision to alleviate the computational load of image processing, attention was designed to concentrate on specific regions of images rather than the entire picture, effectively mimicking human perceptual tendencies. 
However, the inception of the attention mechanisms we recognize today is primarily traced back to the field of natural language processing. 
Bahdanau et al.~\cite{bahdanau_neural_2014}. employed attention in a machine translation model to rectify the structural issues inherent in recurrent neural networks, subsequently highlighting the benefits of attention. 
This endorsement paved the way for the refinement and popularization of attention techniques across a multitude of tasks.

\subsection{Attention Mechanism and Self-Attention}
\label{sec:self_dis}
Self-attention represents a unique form of attention mechanism that allows a model to make inferences about a specific section of a data sample by leveraging information from other portions of the same sample in a permutation-invariant way. This concept echoes the principles of non-local means~\cite{buades_non-local_2011}, an image denoising technique, where each pixel in the output image is a function of all pixels in the input image.

Formally, an attention mechanism, is the data of $(\mathbf{K},\mathbf{Q},\mathbf{V},\mathbf{s},\mathbf{A})$. The vector spaces $\mathbf{K}\in \mathbb{R}^{N\times d_k}$, $\mathbf{Q}\in \mathbb{R}^{N\times d_k}$ and $\mathbf{V}\in \mathbb{R}^{N\times d_v}$ are the set of Keys, Queries, and Values.
The bilinear map $\mathbf{s}:\mathbf{K}\times \mathbf{Q}\rightarrow \mathbb{R}^{N\times N}$ is a score function between the key and the query where the codomain is an attention score. 
The query serves as a request for information, and the corresponding attention score quantifies the relevance of the data encapsulated in the key vector with respect to the query.
In the case of self-attention, the keys, queries, and values all come from the same sequence, allowing each element to attend to all others in that sequence.
Different score functions are the additive~\cite{bahdanau_neural_2014}, simple Dot-Product~\cite{luong_effective_2015}, Scaled Multiplicative~\cite{vaswani_attention_2017}, General~\cite{sordoni_iterative_2016}, Activated General~\cite{ma_interactive_2017}, and similarity-based~\cite{graves_neural_2014}.
Then, the attention, $\textbf{Att}$, is defined as,

\begin{equation*}
\mathbf{Att}(\mathbf{K},\mathbf{Q},\mathbf{V}):= \mathbf{Align}(\mathbf{s}(\mathbf{K},\mathbf{Q}))\mathbf{V} ~,
\end{equation*}

where $d_k$ and $d_v$ are the dimensions of the corresponding vector spaces. The goal of the attention module is to produce a weighted average of the value vectors. Thus, the scores are redistributed via an alignment operation $\textbf{Align(.)}: \mathbb{R}^d \rightarrow [a, b]$ that maps the non-compact distribution to a compact one between $[a, b]$. Different alignment functions are the soft or global alignment, like the Softmax function~\cite{luong_effective_2015}, that also introduces a probabilistic interpretation to the input vectors. Another variant, the hard alignment~\cite{xu_show_2015}, forces the attention model to focus on exactly one feature vector in a deterministic sense.

For instance, the attention mapping used in the vanilla Transformer~\cite{vaswani_attention_2017} adopts the scaled dot-product as the bilinear map between keys and queries as

\begin{equation}
\mathbf{Att}(\mathbf{K},\mathbf{Q},\mathbf{V}):= \mathrm{softmax}(\frac{\mathbf{K}\mathbf{Q}^T}{\sqrt{d_k}})\mathbf{V} ~.
\label{eq:attention_map}
\end{equation}

As the dimensionality of the key, represented by $d_k$, increases, the dot product of the query and key can potentially escalate in magnitude. If these large values are then passed through the softmax function during the alignment stage, the resulting gradient can diminish significantly, impeding the model's ability to converge effectively. The incorporation of the normalization factor $\frac{1}{\sqrt{d_k}}$ serves as a countermeasure to this issue, effectively preventing the occurrence of vanishing gradients even when dealing with large inputs.

Rather than simply computing the attention once, the multi-head mechanism runs through the scaled dot-product attention of linearly transformed versions of keys, queries, and values multiple times in parallel via learnable maps $W_i^k$, $W_i^q$ and $W_i^v$.
The independent attention outputs over $h$ number of heads are then aggregated and projected back into the desired number of dimensions via $W^p$, 

\begin{equation}
\mathrm{MultiHead}(\mathbf{K},\mathbf{Q},\mathbf{V}):= [\biguplus_{i=1}^h \mathrm{H}_i]W^p ~,
\label{eq:attention_multi_head}
\end{equation}

where $\mathrm{H}_i$ is given by $ \mathbf{Att}(\mathbf{K}W_i^k,\mathbf{Q}W_i^q,\mathbf{V}W_i^v)$ and $\biguplus_{i=1}^h$ is the concatenation operation. 
Each query essentially asks for a different form of relevant information, allowing the attention model to incorporate more information into the computation of the context vector. Ergo, each head has the capability to learn and concentrate on distinct segments of the inputs, thus allowing the model to engage with a broader spectrum of information.

When used for processing feature vectors, the self-attention mechanism allows the model to summarize the information in the feature vectors that is important to the query. 
For instance, within the Natural Language Processing~(NLP) domain, self-attention facilitates the extraction of inter-word relationships, such as the connections between verbs and their corresponding nouns or the associations between pronouns and the nouns they represent. 
For images, self-attention aids in identifying the interrelations between different regions of the image manifold. For multi-modal learning, self-attention can create links between different representations of the data.

\subsection{Transformer}
\begin{figure}[!htb]
    \centering
    \includegraphics[width=0.65\textwidth]{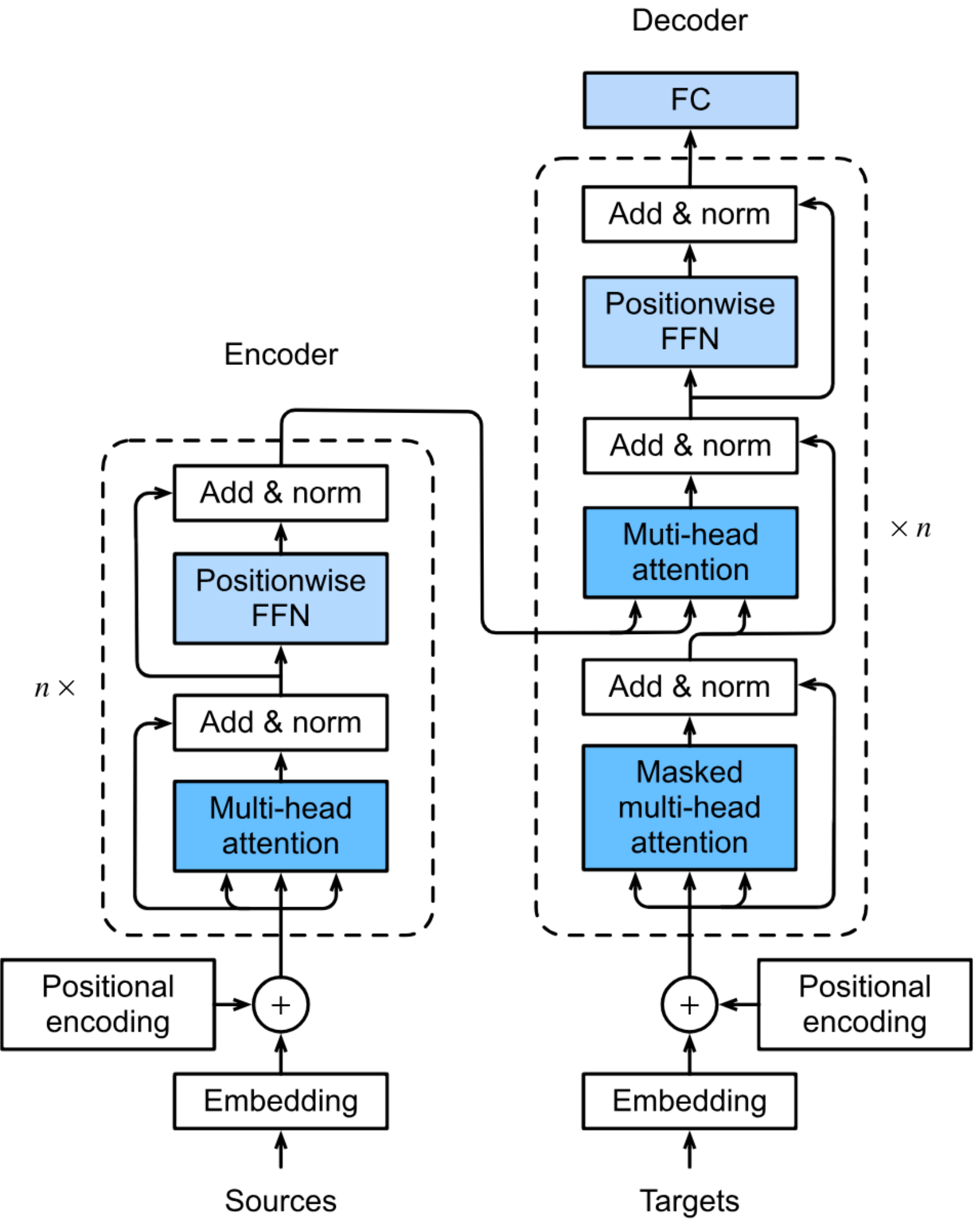}
    \caption{Transformer Encoder and Decoder Architecture, taken from~\cite{zhang_dive_2021}}
    \label{fig:transformer}
\end{figure}

The original Transformer model, as proposed by Vaswani et al.~\cite{vaswani_attention_2017}, operates as a sequence-to-sequence model and encompasses an encoder and a decoder~(shown in\cref{fig:transformer}), as commonly used in many Natural Machine Translation~(NMT) models both of which are assembled from $\mathrm{L}$ identical layers. Each layer in the encoder primarily consists of a multi-head self-attention mechanism and a position-wise feed-forward network~(FFN). To facilitate the construction of a deeper model, a residual connection~\cite{he_deep_2015}, is integrated around each component, followed by the Layer Normalization~\cite{noauthor_160706450_nodate}. Decoder layers, in contrast to encoder layers, incorporate cross-attention modules as well, situated between the multi-head self-attention mechanisms and the position-wise FFN. In cross-attention, the queries are projected from the outputs of the previous decoder layer, whereas the keys and values are projected using the outputs of the encoder. Additionally, the self-attention mechanisms within the decoder are modified to prevent each position from focusing on the positions that follow to create causal reasoning.

\textbf{Position-wise FFN} The position-wise FFN is a fully connected feed-forward module that operates separately and identically on each position. This can also be viewed as a convolutional layer with kernel size \num{1}.
\[
\text{FFN}(x) = \text{ReLU}(xW_1 + b_1)W_2 + b_2,
\]
where $x$ is the outputs of previous layer, and $W_1, W_2, b_1, b_2$ are trainable parameters.

\textbf{Residual connection and Layer Normalization}
Proposed by He et al.~\cite{he_deep_2015}, the residual connection is a shortcut connection that skips one or more layers. Formally, a residual block computes the output as \( \text{FFN}(x) + x \), where \( \text{FFN}(x) \) is the output of the neural network layers that are being skipped. The addition operation between \( \text{FFN}(x) \) and \( x \) is element-wise. By doing so, residual connections alleviate the vanishing and exploding gradient problems, thus making it feasible to train much deeper models effectively.

Layer Normalization, introduced by~\cite{noauthor_160706450_nodate}, aims to mitigate the issue of internal covariate shift by normalizing the activations across a layer for each data point in a mini-batch. The normalization is carried out as follows,

\[
\text{Norm}(x) = \frac{x - \mu}{\sqrt{\sigma^2 + \epsilon}},
\]

where \( \mu \) and \( \sigma^2 \) are the mean and variance of the activations across the layer for each data point, and \( \epsilon \) is a small constant for numerical stability.

In Transformers, both residual connections and layer normalization are extensively utilized to enhance the capabilities of the encoder and decoder structures, thereby facilitating the training of more robust and more complex models.

The Transformer architecture can be leveraged in three distinct manners:

\begin{itemize}

\item \textbf{Encoder–Decoder Configuration:} The complete Transformer architecture is employed. This is commonly utilized in sequence-to-sequence modeling tasks, such as neural machine translation, speech recognition, and video captioning. The main inductive bias here is the assumption that the structure of the input and output sequences can be different and that a mapping between the two can be learned. This assumption comes into play in tasks like machine translation, where the length and structure of the source and target sentences can vary greatly. The self-attention mechanism in both the encoder and decoder allows the model to focus on different parts of the input sequence when generating each element of the output sequence. This configuration also assumes that information flow from any part of the input sequence to any part of the output sequence is possible.
As the successor of Recurrent Neural Networks~\cite{rumelhart_learning_1987} and Long Short-Term Memory~\cite{noauthor_long_nodate} architectures, Transformer models have several benefits such as parallel processing that prompts to performance and scalability increase and bidirectionality which allows understanding of ambiguous words and complex contexts.

\item \textbf{Encoder-Only Configuration:} Only the encoder is utilized and the outputs of the encoder serve as a representation of the input sequence, such as the Bert~\cite{devlin_bert_2019} family. BERT, Bidirectional Encoder Representations from Transformers, utilizes only the encoder part of the Transformers. In BERT, the input sequence is transformed into contextualized word embeddings. Unlike previous models that analyzed sentences from either left-to-right or right-to-left, BERT is able to analyze the context of a word in relation to all other words in the sentence by reading the input sequence in both directions. Bert utilizes a 2-stage training. During pre-training, it is trained on either of two self-supervised tasks: Masked Language Model~(MLM) and Next Sentence Prediction~(NSP). In the MLM task, some percentage of the input tokens are masked randomly, and the model must predict those masked tokens based only on their context. In the NSP task, the model learns to predict whether a sentence follows another sentence. This pre-training step allows BERT to learn a robust and general representation of language. Then, for fine-tuning on specific tasks, an additional output layer is added to the pre-trained BERT model, and all the parameters are fine-tuned on the task-specific data. This is frequently employed in Natural Language Understanding~(NLU) tasks, such as text classification, sequence labeling, sentiment analysis, and named entity recognition. 
\item \textbf{Decoder-Only Configuration:} Only the decoder is used, with the encoder-decoder cross-attention module also being removed, such as the GPT~\cite{noauthor_improving_nodate} family of models. GPT, short for Generative Pretrained Transformer, employs a stack of Transformer decoders to generate a probability distribution over the vocabulary for the next token in a sequence, given all the previous tokens. 
Each token in the input sequence is processed in order autoregressively using masked attention, allowing each token to consider prior tokens in the same sequence. Importantly, this attention mechanism is ``causal,'' meaning that it only allows each token to attend to prior tokens in the sequence, preventing ``future'' tokens from influencing the output at the current position. This design enables GPT to generate coherent and contextually relevant text one token at a time. This is typically used for sequence generation tasks, such as language modeling, text generation, and music generation.
\end{itemize}

Compared with convolutional and recurrent networks, the Transformers carry a different inductive bias. Convolutional networks are recognized for their inductive biases of translation invariance and locality, facilitated through the use of shared local kernel functions. 
In a similar vein, Recurrent networks uphold the inductive biases of temporal invariance and locality via their Markovian structure~\cite{battaglia_relational_2018}. 
In contrast, the Transformer architecture, without the positional encoding, makes minimal presumptions about the data, thereby establishing it as a versatile and universal model. 
However, this absence of structural bias can lead the vanilla Transformer to be susceptible to overfitting. 
The Transformer can also be perceived as a Graph Neural Network GNN~\cite{wu_comprehensive_2021} with message passing, designed over a fully connected graph, with each input serving as a node in the graph. 
A significant distinction between Transformers and GNNs lies in the fact that the former does not introduce any preconceived notions about the structure of the input data — the message-passing process in the Transformer is solely dictated by similarity measures over the content.

\subsection{Problems with Transformer Models}
While transformers have proven to be a powerful tool in many areas, they also have their limitations. Some of the key issues with transformer models include:

\textbf{Temporal Information Loss:} Research~\cite{zeng_are_2022} has shown that despite their success in sequence modeling, transformers may not be the most ideal architecture for Long-Term Time Series Forecasting~(TSF) problems. It was shown that self-attention mechanisms, even with positional encoding, despite their semantic correlation-capturing ability, can result in temporal information loss. When the transformers were evaluated on various datasets, they failed to capture the scale and bias of future data and had difficulty predicting trends on aperiodic data.

\textbf{Limited Access to Higher Level Representations:} In Transformers, a process of incremental abstraction is carried out, layer upon layer, to generate increasingly complex interpretations of the input sequence. This processing method involves treating the representations for the input sequence in a parallel manner across each layer. However, this parallel processing approach presents a drawback~\cite{fan_addressing_2021}. A significant feature that Transformers lacks is the utilization of previously computed top-tier representations to calculate the present representation. These top-tier representations refer to the most abstract and intricate interpretations of the input sequence, which have already been derived in the context of autoregressive models.

\textbf{Complexity and Overfitting:} Transformers often require larger training data sets to perform well due to their complexity. However, an experiment~\cite{hassani_escaping_2022} showed that data set size was not a limiting factor for LTSF transformers, with models trained on a smaller training set performing marginally better. In another experiment~\cite{young_dateformer_2023}, the authors discovered that the performance of the transformers only dropped slightly when the look-back window started at different time steps, suggesting that transformers may be overfitting to the provided data.

\textbf{Destructive Bias from Improper Positional Embedding:} Transformers also can exhibit a destructive bias when a proper positional embedding is not used~\cite{hashemi_ultra-high-resolution_2023}.
Positional embeddings are crucial in transformers as they provide a sense of order/symmetry to the input data, which is inherently absent in the architecture due to its permutation equivariance~\cite{liu_learning_2020}. However, the use of static positional embeddings can lead to limitations. For instance, these embeddings are fixed after training, regardless of the task or the word ordering system of the source or target language~\cite{zheng_dynamic_2022}. This can lead to a destructive bias, where the model fails to generalize well to unseen data. Furthermore, the lack of proper positional encoding can lead to inconsistencies in predictions under small shift perturbations, demonstrating a lack of shift-equivariance~\cite{ding_reviving_2023}. This destructive bias can significantly affect the model's performance, especially in tasks that heavily rely on the order or position of the input data.

In this thesis, we widely incorporate Transformer-based models and their inductive bias for both Event approximation~(IEA-GAN) and point cloud generation~(YonedaVAE). We also introduce tricks and methods to overcome these issues within this path.

\section{Deep Generative Models and Simulation-Based Inference}
Given a classification problem, the neural network is trained so that it can classify a proper class or condition $y$ with a high probability $p(y|x)$, given the dataset $x$.

There are two main methods the model can utilize to reach decisions. One approach involves the explicit formulation of a classifier by modeling the conditional distribution $p(y|x)$. The problem is that $p(y|x)$ lacks the capability to deeply comprehend the data as it has no basis for understanding the uncertainties involved in decision-making. 
In other words, these models cannot simply acquire decision-making abilities without quantifying their beliefs about their environment in a probabilistic way~\cite{noauthor_probabilistic_nodate}. 

Alternatively, one can opt for a joint distribution $p(x, y)$, that can be decomposed into $p(x, y) = p(y | x)~p(x)$
To accomplish this, the estimation of the marginal distribution over objects, $p(x)$, becomes pivotal~\cite{noauthor_probabilistic_nodate,lasserre_principled_2006}.
The objective of ``Deep Generative Models''~(DGM) is to express and find the density of the data, $p(x)$, either implicitly or explicitly. 

This chapter places an emphasis on latent Variable models as our formulated results, including PE-GAN, IEA-GAN, and YoendaVAE, all fall under this category of models. The fundamental concept of latent variable models involves postulating a latent manifold and the subsequent density estimation process:

\[
z \sim p(z),~~x \sim p(x | z).
\]

In essence, the latent variables are associated with concealed factors within the data, and the conditional distribution $p(x | z)$ can be viewed as a \textbf{generator}. As a result, the joint distribution is factorized as $p(x, z) = p(x | z)p(z)$.
However, since during training, one only has access to \textbf{x}, the unknown \textbf{z} should be marginalized out to get rid of all unobserved random variables. This leads us to the definition of the~(marginal) likelihood function as follows:
\[
p(x) = \int p(x | z)p(z) dz.
\]
The pressing question then arises: \emph{how can we compute this integral? }
The integral over the latent manifold cannot be computed in practice efficiently, so we cannot directly evaluate the density of the observed data with a proper uncertainty evaluation. 
This means that we cannot directly ﬁnd the maximum-likelihood estimators that best ﬁt the given observations. 
The task of performing statistical inference when the data generating process does not have a tractable likelihood is known as \textbf{simulation-based inference} or \textbf{likelihood-free inference}. It is worth noting that the non-tractability amounts to not knowing the functional form of $p(x|z)$ but only being able to produce samples following the marginal distribution.

In the context of this thesis, there are two potential paths to tackle this issue with surrogate models as amortized simulators: Variational Autoencoders~(VAEs)~\cite{kingma_improved_2016,rezende_variational_2016} and Generative Adversarial Networks~(GANs)~\cite{goodfellow_generative_2014}.

In the case of VAEs, the tractability of inference is facilitated by using variational inference to approximate the posterior $p(z | x)$. This approach employs neural networks to parameterize the distributions, aiming to maximize the log-likelihood function, which measures the similarity between the data distribution and the model distribution in order to find a lower bound to the likelihood.

On the other hand, GANs operate based on a different principle. They utilize an adversarial loss function, where a discriminator, denoted as $D(.)$, is used to differentiate between real data and synthetic data generated by the model. This generator operates in an implicit manner, defined as $p(x | z) = \delta (x - G(z))$, where $\delta(.)$ represents the Dirac delta function. This adversarial mechanism encourages the generator to produce synthetic data that the discriminator cannot distinguish from real data, improving the model's ability to capture the data distribution. Although GANs do not learn the density directly, they do it implicitly, where one can indeed incorporate it for simulation-based inference~\cite{ramesh_gatsbi_2022}.

\subsection{Variational Autoencoders}

Considering that we possess a latent-based model $p(x | z)$, with a prior $p(z)$ and a posterior $p(z | x)$, density estimation through maximum likelihood is intractable due to the integral, $p(x) = \int_{z} p(x | z)p(z)dz$. A naive approach would be to use the Monte Carlo approximation:

\[
p(x) = \int p(x | z) p(z) dz
\]

\[
= E_{z\sim p(z)} [p(x | z)]
\]

\[
\approx \frac{1}{K} \sum_{k} p(x | z_k ),
\]

where $z_k \sim p(z)$ are samples from the latent prior. This method is simple and computationally feasible. However, if the manifold $z$ is multi-dimensional, one encounters the curse of dimensionality immediately.
This means that the number of samples needed to properly cover the space grows exponentially with respect to the dimension of the prior manifold.
Alternatively, one can apply \textbf{variational inference}~\cite{noauthor_introduction_nodate}. 
Variational inference reframes this challenge as an optimization problem by introducing an approximate version of the true, yet intractable, posterior which enables the creation of a tractable bound on $p(x)$. 
In other words, VAEs amortize the inference process.

Considering a family of known variational distributions parameterized by $\phi$, represented as $\{ q_{\phi}(z) \}_{\phi}$, we assume that they allocate non-zero probability mass to all latent priors, $z \in Z^M$. 
Subsequently, the logarithm of the marginal distribution can be approximated as follows:

\[
\ln p(x) = \ln \int p(x | z)p(z) dz
\]

\[
= \ln \int \frac{p(x | z)p(z)}{q_{\phi}(z)} q_{\phi}(z) dz
\]

\[
= \ln E_{z \sim q_{\phi}(z)} \left[ \frac{p(x | z)p(z)}{q_{\phi}(z)} \right]
\]

\[
\geq E_{z \sim q_{\phi}(z)} \left[ \ln \frac{p(x | z)p(z)}{q_{\phi}(z)} \right]
\]

\[
= E_{z \sim q_{\phi}(z)} [\ln p(x | z)] - E_{z \sim q_{\phi}(z)} [\ln \frac{q_{\phi}(z)}{p(z)}] .
\]

Here, the inequality arises from Jensen's inequality. Jensen's inequality states that the expectation of the convex function of the variable, $E[f(X)]$, is always greater than or equal to the function of the expectation of the variable, $f(E[X])$, $E[f(X)] \geq f(E[X])$.

By considering an amortized variational posterior, denoted as $q_{\phi}(z | x)$ instead of $q_{\phi}(z)$, we can rewrite our equation to:

\[
\ln p(x) \geq \mathbb{E}_{z \sim q_{\phi}(z | x)}[\ln p(x | z)] - \mathbb{E}_{z \sim q_{\phi}(z | x)}[\ln q_{\phi}(z | x) - \ln p(z)].
\]
In Variational Bayesian methods, the lower bound of the log-likelihood function is known as the Evidence Lower BOund~(ELBO). $p(z)$ is the marginal distribution or the prior.
In amortization of the inference, one trains a singular model, like a neural network with specific weights, and it produces the parameters of a distribution for a given input. 
What emerges from this is a model similar to an Autoencoder, characterized by a probabilistic encoder, $q_{\phi}(z | x)$, and a probabilistic decoder, $p(x | z)$. 
An ``Autoencoder''~\cite{hinton_reducing_2006} is a neural network-based lossy compression model designed to learn an identity function in an unsupervised way to reconstruct the original input while compressing the data in a representation-efficient way.
Thus, equipping it with the amortized variational posterior is a ``Variational Auto-Encoder''. 
Two essential components form the ELBO: the first part, $\mathbb{E}_{z \sim q_{\phi}(z | x)}[\ln p(x | z)]$, is the negative reconstruction error, as $x$ is encoded to $z$ and then decoded back. The second part, $\mathbb{E}_{z \sim q_{\phi}(z | x)}[\ln q_{\phi}(z | x) - \ln p(z)]$, can be viewed as a regularization term. It is interesting that it looks suspiciously like a Kullback-Leibler~(KL) divergence term.

From another point of view, in order to have a perfect reconstruction, one has to close the amortization gap, the gap between the ELBO and the true log-likelihood. In other words, the estimated posterior $q_{\phi}(z | x)$ should be very close to the intractable real posterior. One can use Kullback-Leibler divergence to quantify the distance between these two distributions. KL divergence $KL(X\parallel Y)$ measures how much information is lost if the distribution $Y$ is used to represent $X$.

\begin{align*}
\ln p(x) &= \mathbb{E}_{z \sim q_{\phi}(z | x)} [\ln p(x)]  \\
&= \mathbb{E}_{z \sim q_{\phi}(z | x)} \left[\ln \frac{ p(z | x) p(x)}{p(z | x)}\right]  \\
&= \mathbb{E}_{z \sim q_{\phi}(z | x)} \left[\ln \frac{p(x | z)p(z)}{p(z | x)}\right]  \\
&\begin{aligned}
    = \mathbb{E}_{z \sim q_{\phi}(z | x)} \Big[ \ln &\frac{p(x | z)p(z)}{p(z | x)}\frac{q_{\phi}(z | x)}{q_{\phi}(z | x)} \Big]
\end{aligned}  \\
&\begin{aligned}
    = \mathbb{E}_{z \sim q_{\phi}(z | x)} \Big[ \ln p(x | z) &- \ln \frac{q_{\phi}(z | x)}{p(z)} \\
    &+ \ln \frac{q_{\phi}(z | x)}{p(z|x)} \Big]
\end{aligned}  \\
&\begin{aligned}
    \Longrightarrow \ln p(x) - \text{KL} \left[ q_{\phi}(z | x) \parallel p(z | x) \right] &= \mathbb{E}_{z \sim q_{\phi}(z | x)} \left[ \ln p(x | z) \right] \\
    &- \text{KL} \left[ q_{\phi}(z | x) \parallel p(z) \right].
\end{aligned}
\end{align*}

The term, $\text{KL} [ q_{\phi} (z | x) \parallel p(z | x) ]$, measures the discrepancy between the variational posterior and the real posterior, which is unknown. The LHS is exactly what we want to maximize when learning the true distributions.

\textbf{Reparameterization Trick}:
The expectation term in the loss function invokes generating samples from the variational posterior, $q_{\phi} (z | x)$. The variational posterior typically assigns more probability mass to a smaller region than the prior. Nevertheless, the variational posteriors are almost deterministic, whereas the sampling process should be a stochastic process. A possible solution is the reparameterization trick~\cite{noauthor_random_nodate,kingma_auto-encoding_2022,rezende_stochastic_2014}. 
With this trick, a random variable can be represented as a sequence of transformations of an independent random variable that has a simple distribution with known parameters.
Incorporating the reparametrization of the Gaussian distribution can significantly decrease the variance of the gradient. This is because the source of randomness originates from the independent variable, and the gradient calculation is performed with respect to a deterministic function, not random entities.

\subsubsection{Likelihood Estimation:}
The log-likelihood is a critical stage in simulation-based inference. However, it's important to note that the ELBO is merely a lower bound for the log-likelihood and thus doesn't serve as an optimal approximation of it. 
There are several methods for likelihood estimation in VAEs, and we touch upon some of the most powerful ones here:

\begin{itemize}
    \item \textbf{Importance Sampling:} As underscored in various studies~\cite{rezende_stochastic_2014,burda_importance_2016}, an alternative approach is importance sampling that often suggested as a more effective means of approximating the log-likelihood. Formally 
    \[ \ln p(x) \approx \frac{1}{K} \sum_{k=1}^{K} \ln p(x | z_k) \] 
    where each \( z_k \) is sampled from the distribution \( q_\phi(z_k | x) \). The key point to observe here is that the logarithm operation is applied outside the expected value. Given a sufficiently large sample space, the importance of weighting yields a better estimate of the log-likelihood.

    \item \textbf{Annealed Importance Sampling~(AIS):} AIS is an advanced method of estimating the log-likelihood for VAEs~\cite{neal_annealed_2001,zhang_differentiable_2021}. It gradually transforms and ``anneals'' the simple initial distribution into the complex target distribution through a series of intermediate distributions. This is done by introducing a temperature parameter \( \beta \), varying from 0 to 1. At \( \beta = 0 \), the distribution is the prior \( p(z) \) and at \( \beta = 1 \), it becomes the posterior \( q_\phi(z | x) \). Formally, the unnormalized target distribution is defined as
    
    \[ \tilde{p}_\beta(z) = p(z)^{1-\beta} q_\phi(z | x)^\beta \]
    
    AIS computes the importance weights along this temperature annealing path, and the log-likelihood can be approximated as
    
    \[ \ln p(x) \approx  \frac{1}{K} \sum_{k=1}^{K} \ln \left[ \frac{p(x, z_k)}{q_\phi(z_k | x)}\right] \]
    
    where each \( z_k \) is sampled from the annealed distribution \( \tilde{p}_\beta(z) \). This method provides a tighter estimate of the log-likelihood compared to the standard importance sampling, especially for complex data distributions.
    
    \item \textbf{Bidirectional Monte Carlo~(BDMC):} BDMC~\cite{grosse_sandwiching_2015,hoffman_learning_2017,li_approximate_2017} is another approach to improve the estimate of the log-likelihood in the context of VAEs. 
    BDMC combines both forward and backward Markov chain Monte Carlo~(MCMC) transitions for a more refined approximation of the log-likelihood. This method has been shown to provide more accurate log-likelihood estimates compared to methods like importance sampling or AIS, especially when the proposal distribution differs significantly from the target distribution. The idea is to correct the bias induced by the proposal distribution by considering both the forward and backward transitions. BDMC can be particularly beneficial in settings where the posterior distribution is multi-modal or when the variational approximation is not flexible enough. Formally, the log-likelihood estimate can be written as:

    \[
    \ln p(x) \approx \frac{1}{K} \sum_{k=1}^{K} \ln \left[ \frac{1}{2} \left( \frac{p(x|z_k^{(fwd)})}{q_\phi(z_k^{(fwd)}|x)} + \frac{p(x|z_k^{(bwd)})}{q_\phi(z_k^{(bwd)}|x)} \right) \right]
    \]
    where each \( z_k^{(fwd)} \) and \( z_k^{(bwd)} \) is a sample from the forward and backward MCMC chains, respectively, initiated from the approximate posterior \( q_\phi(z_k | x) \). As before, the logarithm operation is applied outside the expected value, and the average is taken over a large number of samples \( K \). This approach provides a more balanced and accurate estimate of the log-likelihood when compared to unidirectional methods. 

    \item \textbf{Nested Variational Inference (NVI):} NVI~\cite{zimmermann_nested_2021} is a unique approach that seeks to improve the estimation of the likelihood in the context of models with latent variables. The methodology introduces a sequence of forward and reverse densities, bridging the gap between an initial easy-to-sample density and the ultimate target density. This bidirectional approach aids in generating more sophisticated proposals for complex densities, deriving from simpler ones. NVI has demonstrated improved performance, particularly when the discrepancy between the proposal and target distributions is large. The approach becomes especially advantageous in scenarios where the posterior distribution is intricate or the variational approximation lacks sufficient flexibility. Formally, each intermediate joint density $\hat{p}(x, z_k)$ can be estimated using the self-normalized importance sampling method given by 
    \[
    \hat{p}(x, z_k) = \frac{w_k p(x, z_k)}{\hat{Z}}
    \]
    where $w_k = p(x, z_k) / q_{\phi}(z_k|x)$, $z_k$ are samples from the posterior $q_{\phi}(z|x)$, $\hat{Z}$ is an estimate of the normalizer in the self-normalized importance sampling method,$\hat{Z} = \frac{1}{K} \sum_{k=1}^{K} w_k$, and $K$ is the number of samples used. 
    The log-likelihood is then approximated using weighted samples like in the AIS method. This framework offers a more refined and accurate estimate of the log-likelihood compared to the other methods.

\end{itemize}

\subsubsection{Problems with VAEs:}
Variational Autoencoders (VAEs) face a few challenges. A potential issue is the ``posterior collapse''~\cite{bowman_generating_2016}, which occurs when a powerful decoder treats the latent variable $z$ as mere noise, leading to the regularization term being minimized for priors like the standard Gaussian. 
Another problem, known as the ``hole problem''~\cite{rezende_taming_2018}, arises from a mismatch between the aggregated posterior and the prior. If there are regions where the prior assigns high probability, but the aggregated posterior assigns low, sampling from these regions can result in low-quality output. 
In the next chapter~(\cref{chap:4}), we touch upon technologies that try to extend VAEs to approach these issues. Furthermore, in this study (see \Cref{chap:6}), with YonedaVAE, I introduce several tricks and methods to alleviate these issues.

\subsection{Generative Adversarial Networks}
Given the Monte Carlo approximation of the log-likelihood, one can turn the problem of calculating the integral into a problem of sampling from a known prior. 
In other words, one needs only to model a parameterized $p_{\theta}(x | z)$ to generate samples.
To gauge the difference between the generated samples $x \in p_g$ and real samples $x \in p_x$, an adversarial loss can be utilized to implicitly represent the empirical distribution. In such a setup, the generator in a single dimension behaves like a Dirac delta function, represented as $\delta (x-G(z))$. Consequently, the marginal distribution transpires as an infinite mixture of delta peaks in the observable space~~\cite{tomczak_deep_2022}.

Given the generator $G$, a function $G:\mathbb{R}^d\rightarrow \mathbb{R}^n$, that maps a latent variable $z\in \mathbb{R}^d$ sampled from a distribution to the data manifold, $x \sim p(x|z) \in p_g$, and the discriminator $D$, a functional $D:\mathbb{R}^n\rightarrow [0,1]$, that takes an image $x \sim p(x) \in p_x$ and assigns a probability to it, they are the players of the following two-player minimax game with value function $V(D, G)$~\cite{goodfellow_generative_2014},

\begin{equation*}
\mathop{min}_{G}\mathop{max}_{D}~V(D,G) = \mathbb{E}_{x\in \mathbb{R}^n}[\log D(x)]+\mathbb{E}_{z\in \mathbb{R}^d}[\log(1-D(G(z))].
\end{equation*}

For a ﬁxed $G$, the objective for $D$ can be reformulated as:

\begin{equation}
\begin{split}
     \mathop{max}_{D} V(D,G) = \E_{x \sim p_x}\bigg[ \ln \frac{p(x)}{p(x)) + p(x|z)} \bigg] + \E_{x \sim p_g}\bigg[ \ln \frac{p(x|z)}{p(x) + p(x|z)} \bigg]
\end{split}
\end{equation}
\begin{equation*}
    = KL(p_d || \tfrac{1}{2}(p_d + p_g)) + KL(p_g ||  \tfrac{1}{2}(p_d + p_g)) + C.
\end{equation*}

This loss is fundamentally linked to the Jensen-Shannon divergence, which measures the similarity between the generative distribution $p_g$ and the data distribution $p_d$. Given ample capacity, the generator can implicitly reconstruct the data distribution.

Jensen-Shannon divergence has a particularly advantageous feature in that it behaves well when the magnitudes of both $p_g$ and $p_d$ are small. This attribute is an improvement over the asymmetric KL divergence employed in maximum likelihood models, which may not function well under these circumstances. However, it's worth mentioning that even the Jensen-Shannon divergence is not devoid of limitations. A notable instance arises when no probability mass is assigned to a data sample in a maximum likelihood model. While in such a case, the KL-divergence spikes to infinity, a GAN would experience no repercussions. This discrepancy highlights the unique properties and potential pitfalls of different divergence metrics in the context of generative models. As a result, a vast amount of research has been undertaken to improve its convergence and stability.

\subsubsection{Problems with GANs}
In general, training GANs is a highly brittle and painful task. It requires a significant amount of patience and hyperparameter tuning for domain-specific tasks.

\textbf{Vanishing gradient:}
Should the discriminator reach an optimal state, the corresponding loss function diminishes to zero. Consequently, the gradients effectively vanish. This absence of a gradient halts the learning process, as the lack of an error signal hinders the ability to make adjustments based on the model's performance. The conundrum deepens when considering the opposite situation. If the discriminator performs poorly, the generator lacks accurate feedback to guide its progress. In this scenario, the loss function fails to accurately mirror the reality of the data distribution. 
This double-edged sword scenario indeed poses a significant challenge to the successful training of GANs. The solution to this problem normally lies in balancing the complexity between the two models.

\textbf{Mode collapse:}
A significant challenge that may arise during the training process of a GAN is a phenomenon known as 'Mode Collapse.' In this scenario, the generator model converges to a state where it consistently generates nearly identical outputs, despite the stochasticity of input noise vectors.
While the generator, in this collapsed state, might succeed in deceiving the discriminator into believing that its outputs are real, it falls short in its primary goal: learning to accurately emulate the intricate and diverse $p(x)$. Formally speaking, if the generator's output is denoted by $G(z)$ for a random noise input $z$, the collapse can be seen when $G(z_1) \approx G(z_2)$ for $z_1 \neq z_2$.
In this thesis, I introduce methods originating from Contrastive and Self-Supervised Learning as powerful solutions to prevent mode collapse.

\textbf{Evading Nash equilibrium}
The training of a GAN is essentially a two-player non-cooperative game involving the generator and the discriminator, each aiming to optimize its own objective function. The goal of this game is to reach a Nash equilibrium, a state where no player can unilaterally improve their position by deviating from their current strategy while the other player's strategy remains fixed. 
GANs are typically trained using gradient descent techniques that are designed to find a low value of a cost function rather than to find the Nash equilibrium of a game~\cite{salimans_improved_2016}. However, this process does not explicitly account for the inherently interactive nature of the game being played. Each model updates its cost function independently, essentially ignoring the simultaneous adaptations of the other player.
This strategy of concurrent gradient updates does not guarantee convergence to a Nash equilibrium. The models may continue to oscillate without settling into a stable state, or they may converge to a non-optimal solution. 
Several strategies, like Geometric GANs~\cite{lim_geometric_2017}, Coulomb GANs~\cite{unterthiner_coulomb_2018}, Energy-based GANs~\cite{zhao_energy-based_2017}, moment matching GANs~\cite{li_mmd_2017}, f-GANs~\cite{arjovsky_wasserstein_2017}, contrastive GANs~\cite{kang_contragan_2021} have been proposed to address the problem of reaching Nash equilibrium in GAN training.

\textbf{Low dimensional supports}
A manifold is a space that, at each point, locally is an Euclidean space. To be specific, if the dimension of this Euclidean space is denoted as $n$, we refer to the manifold as an $n$-manifold.
For a given real-valued function, $f$, the ''support'' is the subset of the domain, $n$-manifold, that includes elements that are not mapped to zero. Arjovsky et al.~\cite{arjovsky_towards_2017} focuses on the issue of the supports of data being situated on low dimensional manifolds and the implications it has on the stability of GAN training. When the generator is tasked with creating a larger image based on a small dimensional random noise variable, $z$, it becomes challenging for it to entirely populate the high-dimensional space. The issue is further exacerbated when the intrinsic dimensions of the data are much lower than those of natural images, which can introduce biased information and lead to overfitting~\cite{su_unified_2023}. To address this, some approaches propose training the GAN using a low-dimensional representation of the dataset with the latent space of a pre-trained Autoencoder~\cite{martinez_ld-gan_2023}. Other methods propose to let the prior match the embedding distribution rather than imposing the latent variables to fit the prior, which can help preserve the geometric structure of the data manifold~\cite{geng_generative_2020}. Despite these advances, the problem of low-dimensional supports in GANs remains a significant challenge in the field.

\subsection{BigGAN and ContraGAN}
BigGAN~\cite{brock_large_2019} is a model that has been trained at an unprecedented scale, with modifications introduced to improve both conditioning and scalability. These modifications have resulted in a model that sets a new state-of-the-art in class-conditional high-resolution image synthesis. In this subsection, we go through some of the most important methods they combined to reach such a high performance in conditional image generation.

\textbf{Hinge Loss:}
BigGAN incorporates the hinge-loss variation of the adversarial loss~\cite{lim_geometric_2017},

\begin{align}
\Lb_D^{\mathrm{hinge}} &= -\mathbb{E}_{x\in \mathbb{R}^n}[\min(0,-1+D(x))]-\mathbb{E}_{z\in \mathbb{R}^d}[\min(0, -1-D(G(z))] ~, \\
\Lb_G^{\mathrm{hinge}} &= -\mathbb{E}_{z\in \mathbb{R}^d}[D(G(z))].
\end{align}
The intuition behind the Hinge loss is to create a ``margin'' of separation in the feature space. It tries to ensure that positive and negative examples are not just correctly classified but are also separated by a wide margin. The Hinge loss function encourages the model to correctly classify examples and penalizes miss classifications. However, unlike other loss functions, such as cross-entropy loss, Hinge loss does not just care about correct classification. It also cares about the confidence of the classification.

For the discriminator, the Hinge loss tries to maximize the difference between the average output for real images and the average output for generated images. This encourages the discriminator to be confident in its decisions, which in turn provides stronger gradients for training the generator. On the other hand, for the generator, the Hinge loss tries to maximize the discriminator's output for generated images. This encourages the generator to produce images that the discriminator is likely to classify as real.

\textbf{Class Conditioning:}
For a Deep Generative Model~(DGM) in order to capture and generate class-conditional samples, $p(x|z,y)$, many schemes for capturing the class conditions have been proposed since conditional GANs over input labels have been introduced~\cite{mirza_conditional_2014}.
The main idea is to minimize a specific measure between a class identification output of the discriminator and the actual labels after injecting an embedding of the conditional prior information into the generator.
For example, ACGAN~\cite{odena_conditional_2017} tries to capture ``data-to-class'' relations by introducing an auxiliary classifier.
However, BigGAN uses the ``projection discrimination''~(ProjGAN) mechanism~\cite{miyato_cgans_2018} to generate conditional samples. ProjGAN tries to capture these data-to-class relations by projecting the class embeddings onto the output of the discriminator via an inner product that contributes to the adversarial loss. 
It is implemented by taking an inner product between the class embedding and the global mean-pooled feature map and then adding this to the output of the discriminator, as follows, 

\begin{equation}
D(x, y) = D^{'}(x) + e(y) \cdot D^{'}D(x)
\end{equation}

Where $D^{'}()$ is the global mean-pooled feature map of the image $x$, $e(y)$ is the class label embedding.
BigGAN also uses class-conditional BatchNorm~\cite{de_vries_modulating_2017}~(CBN) to inject class labels into each of the subsequent layers. 
In standard conditional GANs, the class label information is typically incorporated into the model by simply concatenating the class label (or an embedding of the class label) with the noise vector that is input to the generator. However, this approach might not allow the class label information to sufficiently influence the generation process, especially in deeper layers of the generator. CBN addresses this issue by allowing the class label information to modulate the normalization parameters (i.e., the scale and shift parameters) of the BatchNorm layers in the generator. This means that the class label information can directly influence the activations of the neurons in each layer of the generator, allowing the model to more effectively condition the generation process on the class labels.
In a general form, CBN can be formulated as follows

\[
BN(x_i | y_i) = \gamma(y_i) \cdot \frac{x_i - \mu(X)}{\sqrt{\sigma^2(X) + \epsilon}} + \beta(y_i)
\]

where, \(x_i\) is the input to the BatchNorm layer, \(y_i\) is the class label, \(\mu(X)\) and \(\sigma^2(X)\) are the mean and variance of the batch \(X\), and \(\gamma(y_i)\) and \(\beta(y_i)\) are the scale and shift parameters, which are functions of the class label \(y_i\).

\textbf{Spectral Normalization:}
In order to stabilize the training, BigGAN employs Spectral Normalization~\cite{miyato_spectral_2018} in all layers except the class label embedding layer for the generator.
Spectral Normalization is a technique used to stabilize the training of the discriminator in GANs. It works by constraining the Lipschitz constant of a function, which helps to control the gradient norm and prevent the exploding gradient problem. The Spectral Normalization operation normalizes the weight matrix of a layer by the spectral norm, which is the largest singular value of the matrix. Formally, it can be written as

\begin{equation}
W_{SN} = \frac{W}{\sigma_1(W)}
\end{equation}

Where, \(W\) is the weight matrix of a layer in $G$ or $D$, \(\sigma_1(W)\) is the spectral norm of the weight matrix, which is the largest singular value of \(W\), and \(W_{SN}\) is the spectrally normalized weight matrix. This normalization operation ensures that the spectral norm (i.e., the largest singular value) of the weight matrix is 1, which helps to stabilize the training of the model.

\textbf{Orthogonal Initialization:}
Orthogonal Initialization is a method used to initialize the weights of both $G$ and $D$. Orthogonal Initialization has the property that the dot product of any pair of different rows or any pair of different columns is zero, and the dot product of a row with itself (or a column with itself) is \num{1}. When the initial weight matrices are drawn from the orthogonal group, the width needed to guarantee efficient convergence of the model becomes independent of the depth of the layers~\cite{hu_provable_2020}. 
This property can help to prevent vanishing and exploding gradients, which are common problems in the training GANs.

\textbf{Skip-z connections:}
The generator is enhanced with this mechanism that directly links the input latent vector to the deeper layers within the network. So, instead of solely connecting to the initial layer, they introduced direct skip connections from the noise vector $z$ to multiple layers within the generator. The underlying rationale for this approach is to enable the generator to directly manipulate features at varying resolutions and hierarchical levels using the latent space. The implementation of skip-z connections results in a moderate enhancement in performance.

\textbf{Orthogonal Regularization:}
Orthogonal regularization~\cite{brock_neural_2017} is a technique used to encourage the weights of the network to be orthogonal. This is done by adding a regularization term to the loss function that penalizes the network when its weights are not orthogonal. This can help to reduce overfitting and improve the generalization ability of the network, as it encourages the network to learn a set of diverse features that are not correlated with each other. Formally, the regularization term \(R(W)\) is added to the generator's loss function during training

\begin{equation}
R(W) = ||W^T W - I||_F^2
\end{equation}

Where, \(W\) is the weight matrix of a layer in the neural network, \(I\) is the identity matrix of the same size as \(W^T W\), \(||\cdot||_F\) denotes the Frobenius norm (which is a measure of the total magnitude of all the elements in the matrix), \(R(W)\) is the orthogonal regularization term.

\begin{figure}[!htb]
    \centering
    \includegraphics[width=0.85\textwidth]{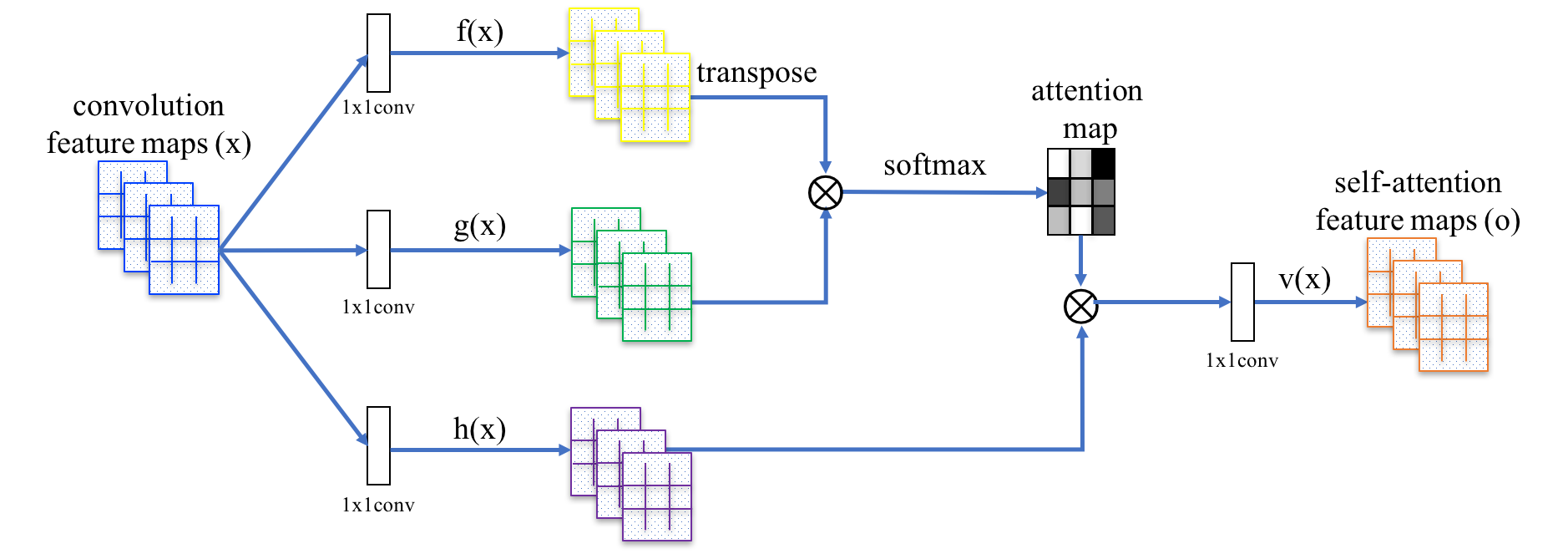}
    \caption{SAGAN architecture, taken from~\cite{zhang_self-attention_2019}}
    \label{fig:SAGAN}
\end{figure}

\textbf{Architecture: Self-Attention non-local Blocks.} 
The self-attention block used in BigGAN is borrowed from the Self-Attention GAN~(SAGAN)~\cite{zhang_self-attention_2019}, depicted in\cref{fig:SAGAN}, allows the model to focus on global relations and long-range dependencies in the image manifold. 
It works by calculating the attention score for each pair of positions in the input. The attention mechanism computes a weighted sum of features, where the weight assigned to each feature is determined by the attention score, as follows
\[
o_i = \sum_{j} \frac{\exp(f_i^T g_j)}{\sum_{k}\exp(f_k^T g_k)} h_j.
\]

Where $o_i$ is the output of the self-attention block for position $i$. $f_i$ and $g_j$ are the feature vectors at positions $i$ and $j$, respectively. These vectors are obtained by applying linear transformations~($1 \times 1$ convolutions) to the input feature vectors. $h_j$ is another transformed version of the input features at position $j$. The term $\exp(f_i^T g_j)$ computes the attention score. The softmax function, represented by $\frac{\exp(f_i^T g_j)}{\sum_{k}\exp(f_k^T g_k)}$, normalizes these scores across all positions $k$, ensuring the alignment.

Since traditional convolutional layers in a neural network are local and translation invariant, meaning they perform the same computation for every region of the input. However, they have a limited ``field of view'' (determined by the kernel size), and therefore, the long-range dependencies could be washed out over high-resolution images. The self-attention mechanism, on the other hand, allows the model to consider all positions in the input at once and learn to weigh their influence accordingly. This is particularly useful in scenarios where the relevant information is spread out and not confined to local neighborhoods, such as in images where the context of a pixel can be informed by distant pixels.

\begin{figure}[!htb]
    \centering
    \begin{subfigure}[b]{0.3\textwidth}
        \includegraphics[width=\textwidth, height=8cm, keepaspectratio]{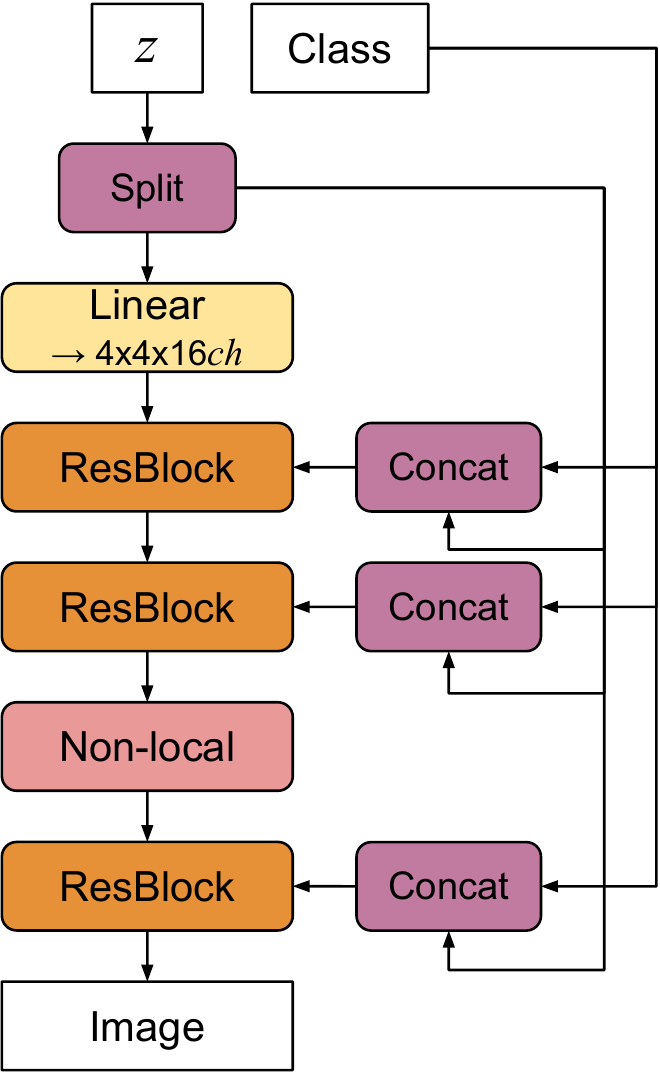}
        \caption{The Generator}
    \end{subfigure}
    \hfill
    \begin{subfigure}[b]{0.3\textwidth}
        \includegraphics[width=\textwidth, height=8cm, keepaspectratio]{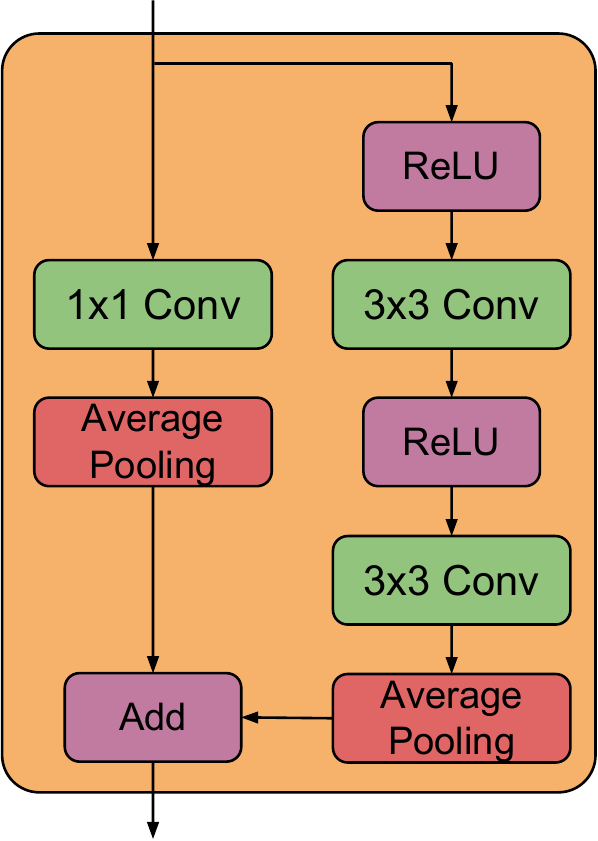}
        \caption{Residual blocks in the discriminator}
    \end{subfigure}
    \hfill
    \begin{subfigure}[b]{0.3\textwidth}
        \includegraphics[width=\textwidth, height=8cm, keepaspectratio]{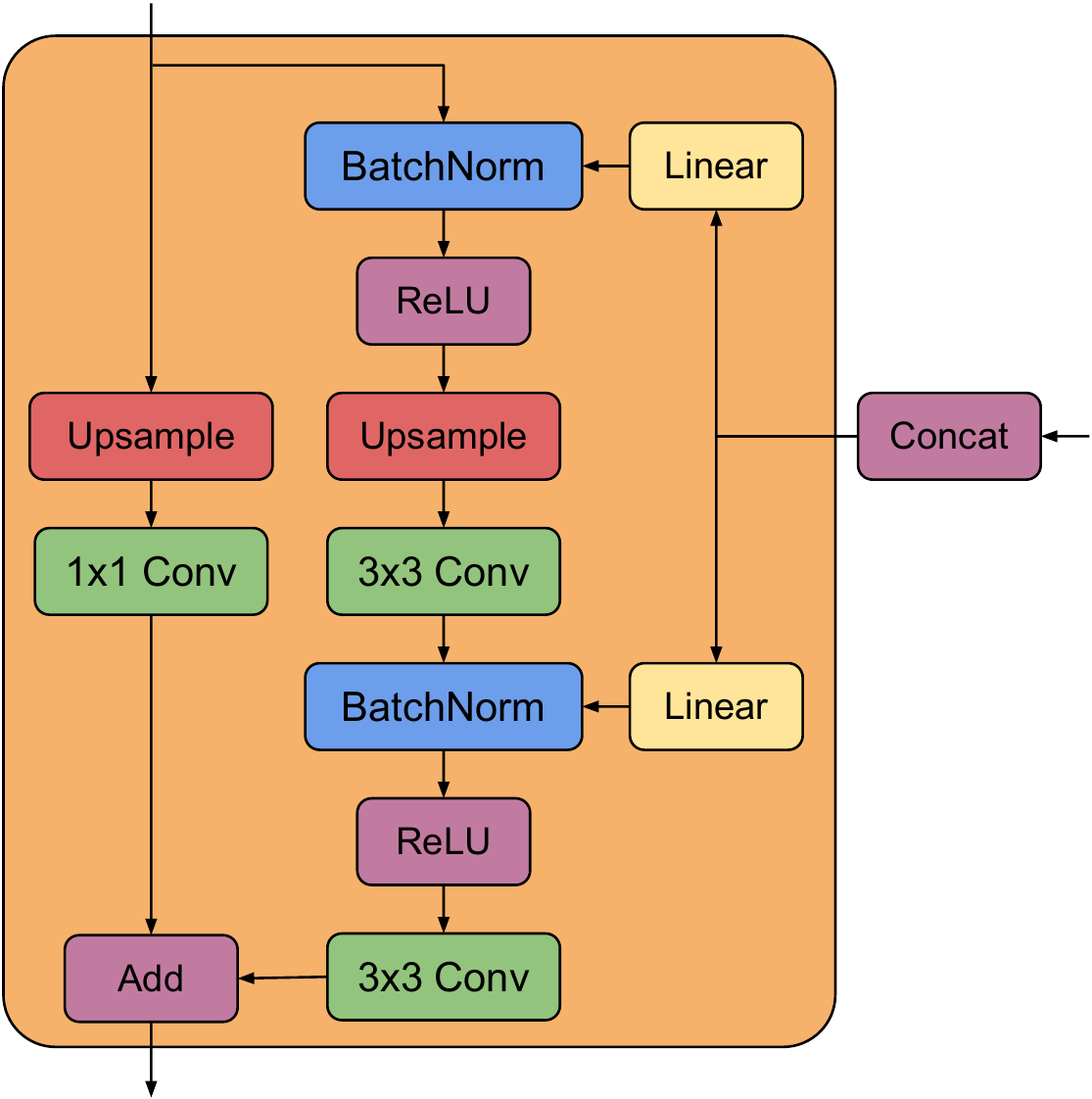}
        \caption{Residual blocks in the generator}
    \end{subfigure}
    \caption{BigGAN's main building blocks, taken from~\cite{brock_large_2019}}
    \label{fig:biggan}
\end{figure}

\textbf{Architecture: BigGAN.}
The BigGAN model, shown in \cref{fig:biggan}, employs the ResNet GAN architecture~\cite{zhang_self-attention_2019,miyato_spectral_2018}, but with a modified channel pattern in the discriminator. 
The first convolutional layer in each block of the discriminator has the same number of filters as the output filters. 
The model uses a shared class embedding in the generator and skip connections for the latent vector. The latent vector is divided into equal-sized chunks, each of which is combined with the shared class embedding and passed to a corresponding residual block. 
Each block's conditioning is linearly projected to create per-sample gains and biases for the CBN layers.

\textbf{Architecture: BigGAN-deep.}
The BigGAN-deep model differs from the original BigGAN in its structure and complexity. In the generator, It employs simplified skip-z conditioning where the entire $z$ is concatenated over the class embedding and passed to each residual block in the CBN blocks. 
The model uses bottleneck residual blocks~\cite{he_deep_2015}, which include two extra $1 \times 1$ convolutions to manage the number of channels before and after the $3 \times 3$ convolutions. 
In the generator, for channel reduction, they retain the first group of channels and drop the rest to produce the required number of channels. However, in the discriminator, they pass the input channels unperturbed and concatenate them with the remaining channels produced by a $1 \times 1$ convolution.
At each resolution, there exist two blocks, whereas BigGAN uses one, making BigGAN-deep four times deeper than BigGAN. Despite this, BigGAN-deep has fewer parameters due to the bottleneck structure in its residual blocks. 

\begin{figure}[!htb]
    \centering
    \begin{subfigure}[b]{0.3\textwidth}
        \includegraphics[width=\textwidth, height=8cm, keepaspectratio]{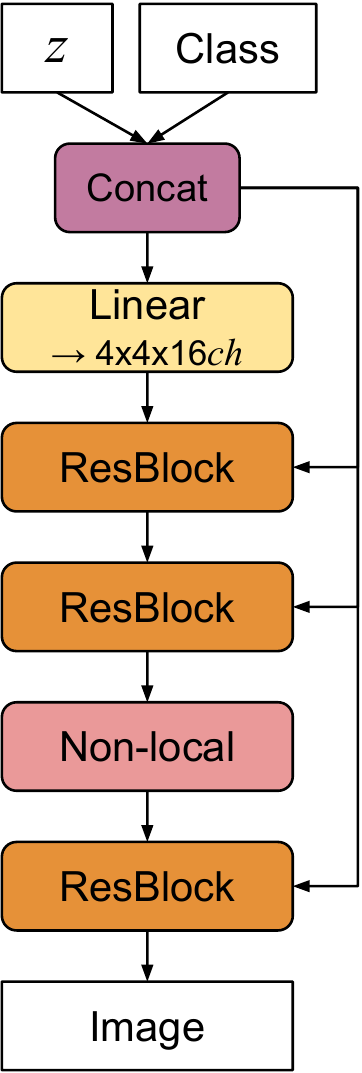}
        \caption{The Generator}
    \end{subfigure}
    \hfill
    \begin{subfigure}[b]{0.3\textwidth}
        \includegraphics[width=\textwidth, height=8cm, keepaspectratio]{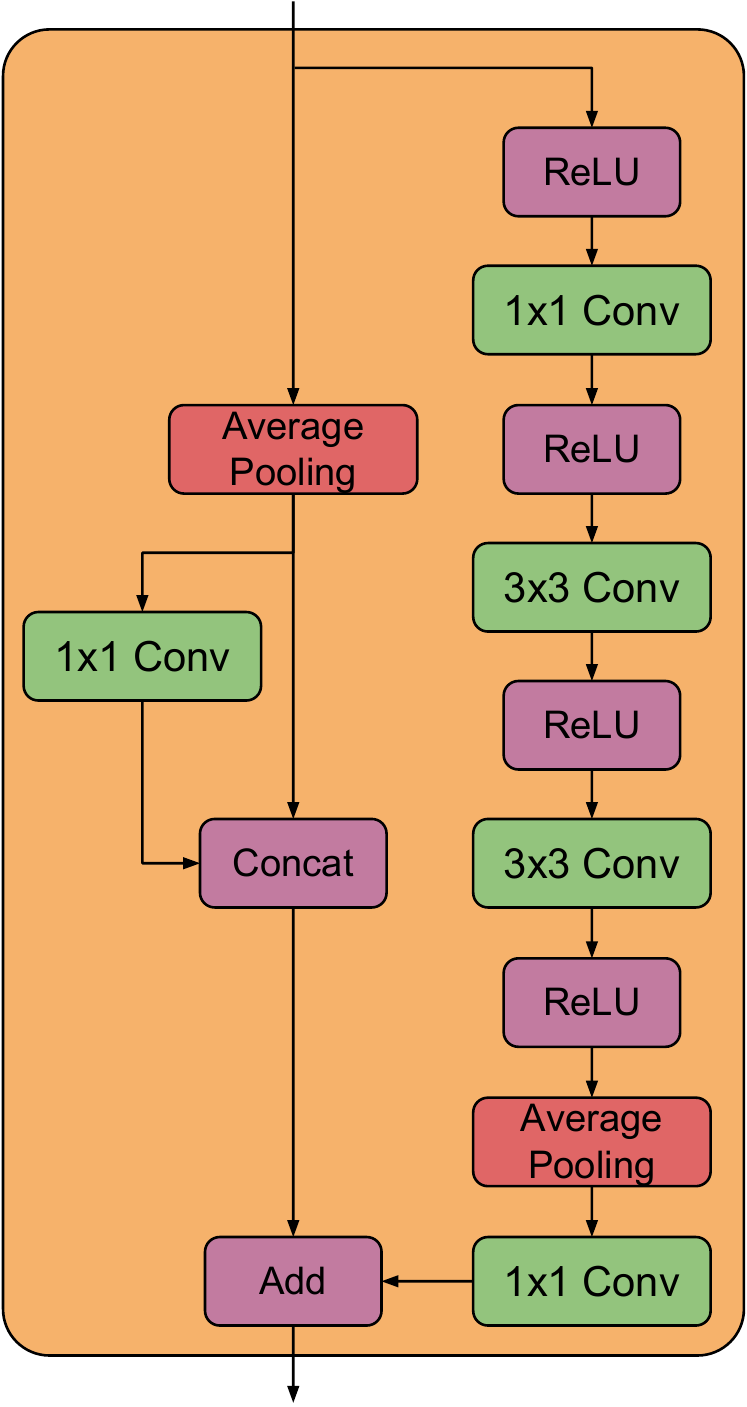}
        \caption{Residual blocks in the discriminator}
    \end{subfigure}
    \hfill
    \begin{subfigure}[b]{0.3\textwidth}
        \includegraphics[width=\textwidth, height=8cm, keepaspectratio]{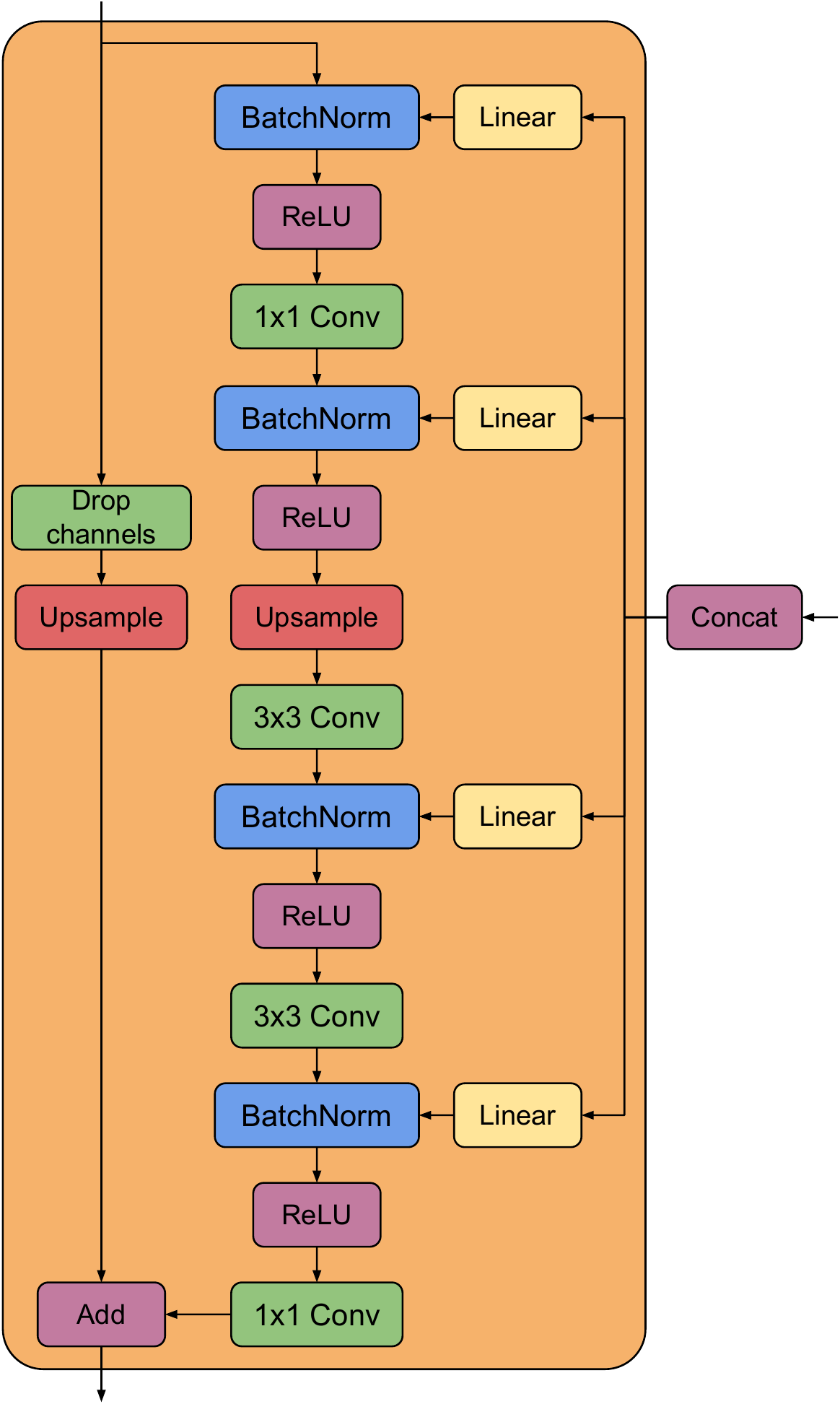}
        \caption{Residual blocks in the generator}
    \end{subfigure}
    \caption{BigGAN-deep's main building blocks, taken from~\cite{brock_large_2019}}
\end{figure}

With the right blend of these technologies and appropriate hyperparameter optimization, BigGAN and BigGAN-deep have achieved state-of-the-art results for high-resolution conditional image generation across numerous natural image datasets. However, these models have their shortcomings.
They tend to exhibit low stability during the training and lack compatibility with fine-grained datasets. These datasets require the ability to distinguish between visually similar objects from subordinate categories, where there are high inter-class similarities and low intra-class similarities.

\subsubsection{ContraGAN}
BigGANs with a projection discriminator suffer from overfitting issues, which can lead to the collapse of adversarial training, especially when the number of classes increases~\cite{kang_contragan_2021}. 
ContraGAN~\cite{kang_contragan_2021} alleviates this overfitting problem with its conditional contrastive loss (2C loss). Incorporated from deep metric learning in order to seize ``data-to-data'' relations or ``intra-class'' relations, they introduce the 2C~loss, derived from NT-Xent loss~\cite{chen_simple_2020},

\begin{equation}
\resizebox{.9\hsize}{!}{$\ell_{2C}(x_i,y_i) = -~\log(\frac{\exp{(S_c(h(x_i)^{\top}e(y_i)))}+\sum^m_{k=1}\mathbf{1}_{k=i}.\exp{(S_c(h(x_i)^{\top}h(x_k)))}}{\exp{(S_c(h(x_i)^{\top}e(y_i)))}+\sum^m_{k=1}\mathbf{1}_{k\neq i}.\exp{(S_c(h(x_i)^{\top}h(x_k)))}}).$}
\label{eq:2C_loss}
\end{equation}

Here, $x_i$ are the images, $y_i$ are the corresponding labels, $S_c(.,.)$ is a similarity metric, and $h(.)$ is the output of the image embeddings.
Although ContraGAN benefits from this loss by capturing the intra-class characteristics among the images that belong to the same class, it is prone to class-confusion~\cite{kang_rebooting_2021,rangwani_class_2021} as different classes could also show similarity among themselves since their vector representation in the embedding space might not be orthogonal to each other, which is precisely what we are dealing with in a fine-grained dataset.

In this study, I try to address this issue by introducing an \emph{Intra-Event Aware mechanism} that not only captures ``intra-class'' relations but ``inter-class'' correspondence. 
This is, in particular, very important in Particle Physics and Detector Simulation, where the signatures of the detector can be grouped together for each collision event as they originate from the same readout window.

\subsection{Point Cloud Generation}
We define a point cloud by a triplet $P = (V, X, Y)$, where $V$ is the set of points, $X \in \mathbb{R}^{N\times D}$ is the feature matrix that stores the features~(usually the coordinates) associated with each point, and $Y \in \mathbb{R}^{N\times F}$ is an optional attribute matrix that stores additional attributes associated with each point. $D$ is the dimensionality of the point coordinates~(usually 2 or 3), and $F$ is the dimensionality of the optional attributes. Given a set of $M$ observed point clouds $P = \{P_i\}_{i=1}^M$, a point cloud generative model learns the distribution of these point clouds $p(P)$, from which new point clouds can be sampled $P_{\text{new}} \sim p(P)$.

For point clouds as inherently unordered collections of points akin to set objects, exchangeability is often viewed as a pivotal characteristic of independent sampling. Let's dive into this deeper. 

Point clouds and sets~(multi-sets), being unordered, exhibit a fundamental symmetry to the group of permutations, denoted as $S = \bigcup_{n \in \mathbb{N}^{*}} S_n$, encapsulating all possible permutations. 
A specific permutation $\pi \in S_n$ can operate on an $n \times n$ matrix $A$ by reordering its rows and columns, which we denote as $\pi.A = \pi A \pi^{T}$. 
When dealing with a $n \times c$ matrix, the permutation alters the order of rows ($\pi.X = \pi X$), and leaves a vector $z \in \mathbb{R}^h$ untouched ($\pi.z = z$). 
This inherent symmetry serves as an irrelevant factor of variation in the data, which should be accounted for and removed in the latent space representation (that is, $\pi.z = z$).

Discriminative models account for these symmetries when a neural network $f$ exhibits equivariance to the action of a group, $\pi.f(X) = f(\pi.X)$. 
However, when the input to $f$ is a vector, imposing $\pi.f(z) = f(\pi.z) = f(z)$ only allows solutions where all rows are identical, which is overly limiting. 
To circumvent this constraint, Vignac et al.~\cite{vignac_top-n_2022} introduce a definition called \emph{$(F, l)$-equivariance}, which extends the standard notion of equivariance and affords a more flexible condition in generative settings. In discriminative models, this is typically achieved by using an equivariant model with an invariant loss function. 
Formally, if $F_{\Theta} = \{f_{\theta}: X \rightarrow Y; \theta \in \Theta\}$ represents a class of hypothesis comprising $G$-equivariant functions mapping from $X$ to $Y$ (a neural network architecture parameterized by $\theta$ serves as an instance), then the loss function $l$ must meet the following requirement:

For all $f$ in $F$, for all $g$ in $G$, and for all $(X, Y)$ in $X \times Y$,

\[ l(g.f(X), g.Y) = l(f(X), Y) \]

while the gradients with respect to the parameters satisfy, 

\[\nabla_\theta l(f(g.X), g.Y) = \nabla_\theta l(f(X), Y).\]
This highlights that every adjustment to the parameters is independent of the group elements utilized in representing $X$ and $Y$. 
As a consequence, the overall training dynamics become indifferent to the group elements used in data representation, thus eliminating the necessity for data augmentation. As a result, Vignac et al.~\cite{vignac_top-n_2022}, proposed the following definition for equivariance:

\textbf{Definition 1: ($(F, l)$- equivariance)}. Consider a hypothesis class $F_{\Theta} \subset Y^X$, a group $G$ that acts on $X$ and $Y$, and a loss function $l$ defined on $Y$. The pair $(F_{\Theta}, l)$ is equivariant to the action of $G$ if the dynamics of $\theta \in \Theta$ trained with gradient descent on $l$ do not depend on the group elements that are used to represent the training data,

\[
\forall \theta \in \Theta, \forall g \in G, \forall (X, Y) \in X \times Y,
\nabla_{\theta} l(f_{\theta}(g.X), g.Y) = \nabla_{\theta} l(f_{\theta}(X), Y).
\]

Inherently, using an equivariant architecture in conjunction with an invariant loss is sufficient to achieve $(F, l)$-equivariance in discriminative scenarios. For typical generative architectures for sets and point clouds, they propose the following sufficient conditions:

\textbf{Lemma 1.} Sufficient conditions for $(F, l)$ - equivariance:
\begin{itemize}
    \item \textbf{GANs:} Given $F$ as a GAN architecture featuring a permutation-invariant discriminator, and $l$ as the standard GAN loss, the pair $(F, l)$ is permutation equivariant. There are no restrictions imposed on the generator.
    \newline
    \textbf{Proof:} given a set generator $f$, a discriminator function $d$, and $X_1, \dots, X_m$ as a training dataset, the standard loss function for GANs is formulated as

    \[
    \frac{1}{m} l(f, d, X_1, \dots, X_m) = \sum_{i=1}^{m} \log(d(X_i)) + \mathbb{E}_Z[\log(1 - d(f(z)))].
    \]

    In order to obtain $l(f, d, X_1, \dots, X_m) = l(f, d, \pi_1.X_1, \dots, \pi_m.X_m)$ for every choice of $\pi_i$, it is therefore sufficient to choose a permutation invariant discriminator.
    
    \item \textbf{VAEs:} Given $F$ as an encoder-decoder architecture with a permutation-invariant encoder, and the reconstruction loss $\hat{l}$ satisfies $\forall \pi \in S, \hat{l}(\pi.X, X) = \hat{l}(X, X)$, then the pair $(F, l)$ is permutation equivariant. The decoder function is exempted from any constraints.
    \newline
    \textbf{Proof:} Assuming that the Autoencoder is comprised of a permutation invariant encoder $\text{enc}$ and an arbitrary decoder $f$. For any set size $n$, set $X \in \mathbb{R}^{n \times d}$ and permutation $\pi \in S_n$, we have

    \begin{align*}
        l(\pi.X, \hat{X}) & = l(X, \hat{X}) \\
        \Rightarrow l(\pi.X, f(\text{enc}(X)) & = l(X, f(\text{enc}(X)) \\
        \Rightarrow l(\pi.X, f(\text{enc}(\pi.X)) & = l(X, f(\text{enc}(X)) \quad \text{(\text{enc} is invariant)}\\
        \Rightarrow \nabla_{\theta} l(\pi.X, f(\text{enc}(\pi.X))) & = \nabla_{\theta} l(X, f(\text{enc}(X)))
    \end{align*}

\end{itemize}

Within the context of generative models for point cloud generation, the aspired $(F, l)$ - equivariance doesn't demand $\pi.f(z) = f(\pi.z)$.
The stipulations offered by Lemma \num{1} are in accord with all point cloud generative models, and the common permutation invariant loss functions like the Chamfer loss, Wasserstein-2 distance, and Hungarian loss meet the requirements for the VAE loss. 
In the context of GANs and VAEs, the exchangeability condition is absent. A subtle point is that the model output in GANs and VAEs involves an implied mat-to-set conversion. Hence, a model creating uniformly permuted matrices equates to one producing exchangeable matrices. Thus, the output, being a set, not a matrix, is a presumption, not a proven fact. This insight explains why independent sampling doesn't surpass non-exchangeable set creation techniques like MLPs and First-n~(see \Cref{chap:6}), leading us to incorporate a new creation mechanism, \emph{Adaptive Top-q Sampling}, uncoupled from the constraint of model exchangeability, in \Cref{chap:6}.

\subsubsection{Permutation Invariant loss}
Thus far, we discussed the concept of $(F, l)$ - equivariance and how it affects the inherent inductive bias in the generator~(decoder) and the discriminator~(encoder) model. Now, we discuss some commonly used permutation invariant assignment-based loss functions, such as the Chamfer loss~\cite{fan_point_2016} and the Hungarian loss~\cite{noauthor_hungarian_nodate}. Given the true set $Y = \{y_1, ..., y_l\}$ and the predicted set, $\hat{Y} = \{\hat{y}_1,..., \hat{y}_k\}$, the assignment-based loss is as follows

\[
L_A(\hat{Y}, Y) = \sum_{i}d(\hat{y}_i, y_{\pi(i)}) + \sum_{j}d(\hat{y}_{\sigma(j)}, y_j)
\]

where $d$ is a distance metric such as a Frobenius norm, and $\pi : \{k\} \rightarrow \{l\}$, $\sigma : \{k\} \rightarrow \{l\}$ are assignment functions that map indices from one set to another. 
The inherent lack of order in sets and point clouds brings about the challenge of deciding which elements should be compared as a combinatorial optimization problem. 
In the most generalized version, the problem can be depicted as follows: The instance of the problem involves a number of generated points in a set and a number of truth-level points. Any generated set can be matched with any truth level set, with the cost of such a match possibly varying based on the specific generated-truth level set pairings. The aim is to align as many generated sets with truth level sets as possible, adhering to the condition that no more than one generated set is matched to a single truth level set and vice versa, with the intention of minimizing the overall cost of these matching. 
In graph theory language~\cite{noauthor_assignment_2023}, the assignment problem consists of finding, in a weighted bipartite graph, a matching of a given size, in which the sum of weights of the edges is minimum.
The variation in assignment-based set losses primarily lies in the strategies for assignment, which are represented by the choices for $\pi$ and $\sigma$.

The Chamfer loss is one such example that solves an unbalanced assignment where each element in $\hat{Y}$ is mapped to the nearest element in $Y$ and vice versa. 
It is defined by $\pi(i) = \mathop{arg min}_j d(\hat{y}_i, y_j)$ and $\sigma(j) = \mathop{arg min}_i d(\hat{y}_i, y_j)$, resulting in:

\[
L_C(\hat{Y}, Y) = \sum_i \mathop{min}_j d(\hat{y}_i, y_j) + \sum_j \mathop{min}_i d(\hat{y}_i, y_j)
\]

For the Hungarian loss, $\pi$ and $\sigma$ create a bijection, thus necessitating identical set sizes $n = m$ to solve a balanced assignment problem,

\begin{align}
L_H(\hat{Y}, Y) = \frac{1}{2}\left[\mathop{min}_{\pi \in S_k} \sum_i d(\hat{y}_i, y_{\pi(i)}) + \mathop{min}_{\sigma \in S_k} \sum_i d(\hat{y}_{\sigma(j)}, y_j)\right]
= \mathop{min}_{\pi \in S_k} \sum_i d(\hat{y}_i, y_{\pi(i)}) 
\label{eq:hungarian}
\end{align}

where $S_k$ is the collection of all permutations on sets of size $k$.
These varying strategies of assignment bring about different metric spaces on sets. Both Chamfer and Hungarian losses present unique strengths and weaknesses. The Chamfer loss scales with the set sizes $k$ and $l$ at a computational cost of $O(kl)$, while the Hungarian loss is substantially more costly with a complexity of $O(k^3)$~\cite{tomizawa_techniques_1971,noauthor_theoretical_nodate}. 
The Chamfer loss's lack of one-to-one assignments can be a disadvantage when comparing multi-sets or sets with numerous similar elements up to numerical precision. 
On the contrary, the stringent requirement for bijective assignments in the Hungarian loss becomes a disadvantage when comparing sets of different sizes in unbalanced assignment problems. Later, in the development of YonedaVAE, we will use the Hungarian loss as the reconstruction loss.


\section{Self-Supervised Representation Learning}
Self-supervised learning~(SSL), dubbed ``the dark matter of intelligence''~\cite{noauthor_self-supervised_nodate}, allows for the utilization of naturally occurring labels within data, harnessing the the abundant availability of unlabeled data. 
The process involves carefully crafting learning objectives to enable data self-supervision. 
The tasks within self-supervised learning, often referred to as pretext tasks, lead to a supervised loss function. The focus, however, isn't on the performance of these generated tasks but on the intermediate representation they develop. 
This representation is expected to hold semantic or structural significance, beneficial for multiple real-world applications. 
For instance, randomly rotating images and training a model to predict these rotations serve as a pretext task. While the precision of this task isn't crucial, it assists the model in deriving high-quality latent variables for practical tasks, like developing an object recognition classifier with minimal labeled samples. 
From a representation learning perspective, it is possible to classify all generative models as falling under the umbrella of self-supervised learning, albeit with varied objectives. A common thread linking them is the concept of \textbf{contrastive learning}

\begin{figure}[!htb]
    \centering
    \includegraphics[width=0.85\textwidth]{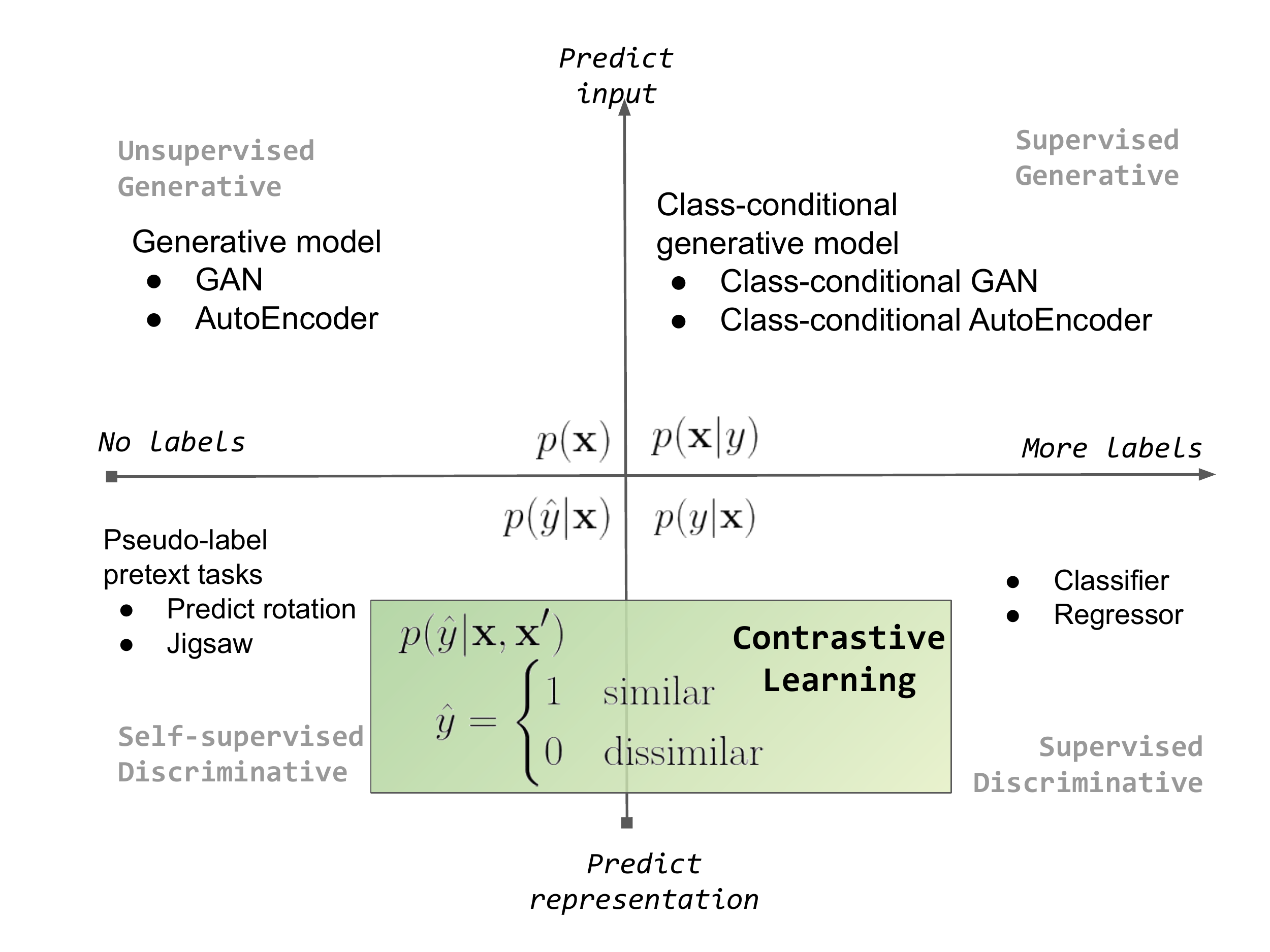}
    \caption{Contrastive learning at the interface of supervised and self-supervised learning, adopted from~\cite{le-khac_contrastive_2020}}
    \label{fig:contrastive}
\end{figure}

While distinct in their methods, self-supervised learning and contrastive representation learning are closely intertwined~(shown in~\cref{fig:contrastive})., with both aiming to leverage the inherent structures or patterns in data for learning representations. 
Contrastive learning, often employed as a method within the self-supervised learning framework, operates on the principle of learning by comparison, distinguishing between similar and dissimilar data pairs. It complements self-supervised learning, which focuses on leveraging the inherent labels in the data, by further enhancing the data representation through comparison-based learning. 
The powerful combination of these two learning paradigms allows for the development of rich and meaningful representations without relying heavily on manually annotated data, thus maximizing the utility of vast amounts of unlabeled data.

Now, let's explore different three main families within the realm of SSL, each possessing unique features and techniques for handling data and deriving valuable insights.

\subsection{The Deep Metric Learning~(Contrastive) Family}
\begin{figure}[!htb]
    \centering
    \includegraphics[width=0.95\textwidth]{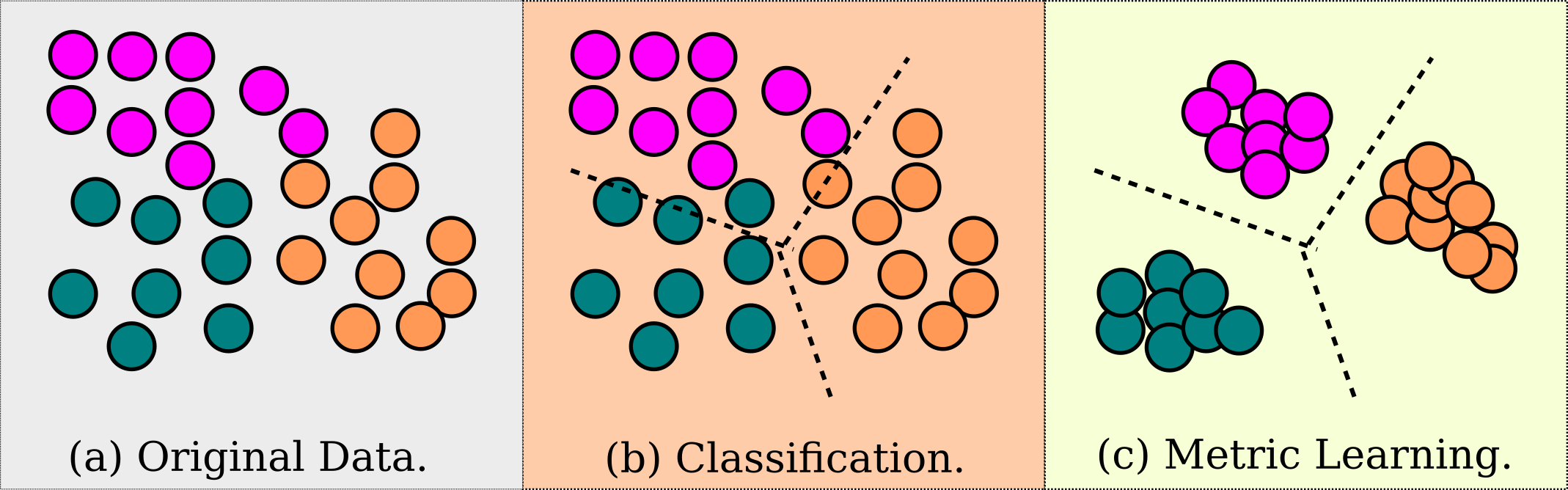}
    \caption{Deep Metric Learning vs. vanilla Classification}
    \label{fig:dml}
\end{figure}

The Deep Metric Learning~(DML) family of methods is based on the principle of encouraging similarity between semantically transformed versions of an input.  
DML originated with the idea of a contrastive loss, which allows the model to learn a discriminative embedding space~(metric) that both maximizes inter-class distance and minimizes intra-class distance~(see ~\Cref{fig:dml}).

Classification in machine learning involves assigning predefined labels to instances based on learned patterns, focusing on differentiating classes. DML, however, is concerned with learning a distance function to measure the similarity or dissimilarity between instances, aiming to bring similar instances~(positive pairs) closer and push dissimilar~(negative pairs) ones apart in the representation space. The Contrastive loss~\cite{bromley_signature_nodate,noauthor_learning_nodate} was first introduced as the earliest training objective used for DML,

\[
L_{\text{cont}}(\mathbf{Z}) = \sum_{(i,j) \in \mathbf{P}} \left\| z_j - z_i \right\|^2 + \sum_{(i,j) \notin \mathbf{P}} \text{ReLU}\left(m - \left\| z_i - z_j \right\|^2\right)^2, \quad m > 0,
\]

where $z_i \in \mathbf{Z}$ denotes the model representation of sample $i$, $\mathbf{P}$ denotes the set of positive samples, and $m$ is the margin hyperparameter, defining the lower bound distance between samples of different classes. The contrastive loss takes a pair of inputs and minimizes the embedding distance when they are from the same class but maximizes the distance otherwise. 
Improving the contrastive loss by focusing on relative distances rather than absolute distances, Triplet loss~\cite{weinberger_distance_2005,chechik_large_2010} considered the relationships between three samples, an anchor, a positive sample~(similar to the anchor), and a negative sample~(dissimilar to the anchor), at a time instead of just two, leading to better generalization and improved performance.

\[
L_{\text{triplet}}(Z) = \sum_{(i,j) \in \mathbf{P}} \sum_{(k,l) \notin \mathbf{P}|k=i} \text{ReLU}\left( \left\| z_i - z_j \right\|^2 - \left\| z_i - z_k \right\|^2 + m \right), \quad m > 0.
\]

After the Triplet loss, many DML methods were introduced, but the transition from DML to the modern SSL can be traced back to~\cite{sohn_improved_2016, oord_representation_2019} presented the Contrastive Predictive Coding~(CPC) loss. 
The innovative approach of using alternate positive sample views as the negative view for other pairs along with the incorporation of the Information Noise Contrastive Estimation~(InfoNCE) loss were pivotal elements of this transition.
Formally, the InfoNCE loss can be expressed as:

\[
L_{\text{InfoNCE}} = - \sum_{(i,j) \in \mathbf{P}} \log \left( \frac{e^{\text{Sim}(z_i, z_j) / \tau}}{\sum_{k=1}^{N} e^{\text{Sim}(z_i, z_k) / \tau}} \right),
\]

The innovation of the InfoNCE loss lies in its design. Instead of calculating the similarity with one negative sample at a time, InfoNCE loss considers multiple negative samples simultaneously, transforming the learning objective into a softmax-based, multi-class classification problem. 
This method introduces strong competition among the negative samples, which leads to a more balanced and effective gradient during the learning process. 
Additionally, by normalizing the similarities between all the negative samples and the anchor, InfoNCE loss ensures that the model's focus is not only on the hardest negatives but also on easy and moderate ones, thereby providing a more comprehensive understanding of the representation space. 
Also, InfoNCE loss is closely related to the mutual information between the representations of positive examples, which provides a solid theoretical foundation for the approach.

\begin{figure}[!htb]
\small
    \centering
\begin{tikzpicture}
    \node at (0,1.8) (h) {$\longleftarrow\,$Representation$\,\longrightarrow$};
    \node[draw, circle] at (0,-1) (x) {$\,~\bm{x}~\,$};
    \node[draw, circle] at (-2.5,0) (x1) {$\tilde{\bm{x}}_i$};
    \node[draw, circle] at (2.5,0) (x2) {$\tilde{\bm{x}}_j$};
    \node at (-2.5,1.8) (h) {$\bm h_i$};
    \node at (2.5,1.8) (c) {$\bm h_j$};
    \node at (-2.5,3) (hh) {$\bm z_i$};
    \node at (2.5,3) (cc) {$\bm z_j$};
    \path[->] 
        (x)  edge [>=latex] node[below,rotate=-25] {$t\sim\mathcal{T}$} (x1)
        (x)  edge [>=latex] node[below,rotate=25] {$t'\sim \mathcal{T}$} (x2)
        (x1)  edge [>=latex] node[left,rotate=0] {$f(\cdot)$} (h)
        (x2)  edge [>=latex] node[right,rotate=0] {$f(\cdot)$} (c)
        (h)  edge [>=latex] node[left,rotate=0] {$g(\cdot)$} (hh)
        (c)  edge [>=latex] node[right,rotate=0] {$g(\cdot)$} (cc);
    \path[<->]
        (hh)  edge [>=latex] node[above,rotate=0] {Maximize agreement} (cc);
    \end{tikzpicture}
    \caption{The framework for SimCLR. 
    Two separate data augmentation operators are sampled from the same family of augmentations ($t\sim \mathcal{T}$ and $t'\sim \mathcal{T}$) and applied to each data example to obtain two related views.
    A base encoder network $f(\cdot)$ and a projection head $g(\cdot)$ are trained to maximize agreement using a variant of InfoNCE loss, NT-Xent loss.}
    \label{fig:simclr}
\end{figure}
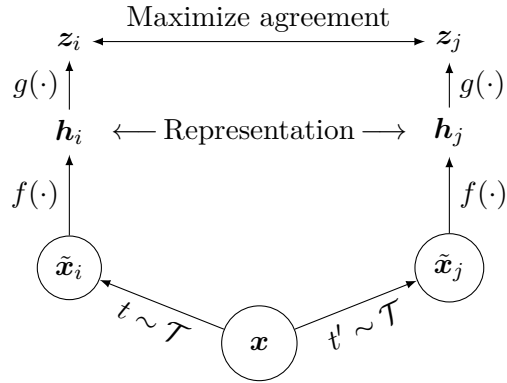

Among the most notable methodologies arising from this shift towards SSL within the domain of deep learning is SimCLR~\cite{chen_simple_2020}. SimCLR, depicted in~\cref{fig:simclr}, learns visual representations by formulating a correspondence between two altered views of the same image with contrastive learning. These views are derived by a series of transformations, which consist of random resizing, cropping, color distortion, and arbitrary blurring. Upon encoding every view, SimCLR utilizes a projector, typically an MLP, to relocate the initial embeddings into a compact manifold, a hypersphere, to prevent dimensional collapse in the representation space~\cite{jing_understanding_2022}. 
In this space, the contrastive loss is applied to promote congruity between the views.  
To maximize agreement between the representations, SimCLR modifies the InfoNCE loss by removing the i'th from the denominator to introduce the NT-Xent loss,

\[
L_{\text{NT-Xent}}(\mathbf{Z}) = -\sum_{(i, j) \in \mathbf{P}} \log \left( \frac{\exp\left(\text{Sim}(z_i, z_j)\right)}{\sum_{k=1, k \neq i}^{N} \exp\left(\text{Sim}(z_i, z_k)\right)} \right).
\]

The negative sign is because it wants to maximize the similarities of positive pairs, which is equivalent to minimizing the negative log of this ratio. As a training challenge, SimCLR requires a large batch size and heavy augmentations to incorporate enough negative samples to achieve good performance.

The significant shift from DML to Contrastive SSL is rooted in several pivotal modifications. These include the adoption of data augmentation to procure positive/negative pairs instead of sampling, the utilization of deeper network structures with possible sources of noise, and the employment of a projector network. Next, we describe an alternative to deep metric learning based on self-distillation.

\subsection{The Self-Distillation Family}
Knowledge distillation~\cite{hinton_distilling_2015} is the procedure of transferring knowledge from a large model~(the target or teacher) to a more compact one~(the online or student), often being used in the context of model compression. 
When both models have the same architecture, this procedure is called \textbf{self-distillation}~\cite{furlanello_born_2018} or the dark-knowledge~\cite{furlanello_born_2018}. 
Various forms of knowledge transfer can be broadly categorized into three types: Response-based knowledge, Feature-based knowledge, and Relation-based knowledge.

\begin{itemize}
    \item \textbf{Response-Based Knowledge:} This type of knowledge transfer depends heavily on the output of the teacher model. Specifically, the teacher's predictions~(also called 'soft targets') are used to guide the learning of the student. This is often done by using the teacher's class probabilities as targets in training the student, which can provide more information than the original hard labels, especially when the teacher model is confident about certain classes. 
    It is the most direct form of knowledge transfer in distillation, and it is at the core of the original knowledge distillation method proposed by Hinton et al.~\cite{hinton_distilling_2015}.
    
    \item \textbf{Feature-Based Knowledge:} Feature-based knowledge refers to the intermediate activations~(features) produced by the target model. 
    In this case, the student model is trained to mimic the internal representations of the teacher model, with the aim of capturing the teacher's way of processing and representing the data. 
    This approach~\cite{romero_fitnets_2014} can be beneficial in cases where the soft targets do not provide enough guidance for the student, and the student can learn more by mimicking the teacher's feature maps.
    
    \item \textbf{Relation-Based Knowledge:} Relation-based knowledge takes into account the correspondence among different data points or features within the teacher model. 
    It may include the pairwise similarities or relative rankings among different data points in the feature space or the correlations between different feature channels. 
    By learning such relational knowledge~\cite{yim_gift_2017}, the student model may learn to capture more abstract and high-level information, which can be crucial for complex tasks.
\end{itemize}

Self-distillation techniques, such as BYOL (Bootstrap Your Own Latent)~\cite{grill_bootstrap_2020}, SimSIAM~\cite{chen_exploring_2020}, and DINO (DIstillation of knowledge using NO labels)\cite{caron_emerging_2021}, function based on a straightforward yet effective mechanism. 
This mechanism involves presenting two distinct views of the same data point to a pair of encoder networks and subsequently using a predictor network to map the output of one encoder to that of the other. This process facilitates the transfer of learning from one view to the other, thus achieving self-distillation. 

The self-distillation mechanism employed by these models is a significant contribution towards addressing one of the central challenges in SSL: \textbf{model collapse}. 
Model collapse, a situation where a model generates trivial or meaningless representations of data, often results in poor model performance. This issue is especially prevalent in contrastive representation learning methods, where the model's task is to distinguish between different instances of data. A model can fall into the trap of model collapse if there is a shortage of negative samples or if the learning dynamics do not have enough constraints to prevent the model from collapsing toward these trivial solutions. However, self-distillation methods, like BYOL, SimSIAM, and DINO, have been developed to circumvent this problem by introducing an asymmetry in the learning process.
As such, they effectively counter model collapse by preserving the diversity in the representation space and encouraging meaningful learning from the input data. We will briefly review them in the following.

\begin{figure}[!htb]
    \centering
    \includegraphics[width=0.95\textwidth]{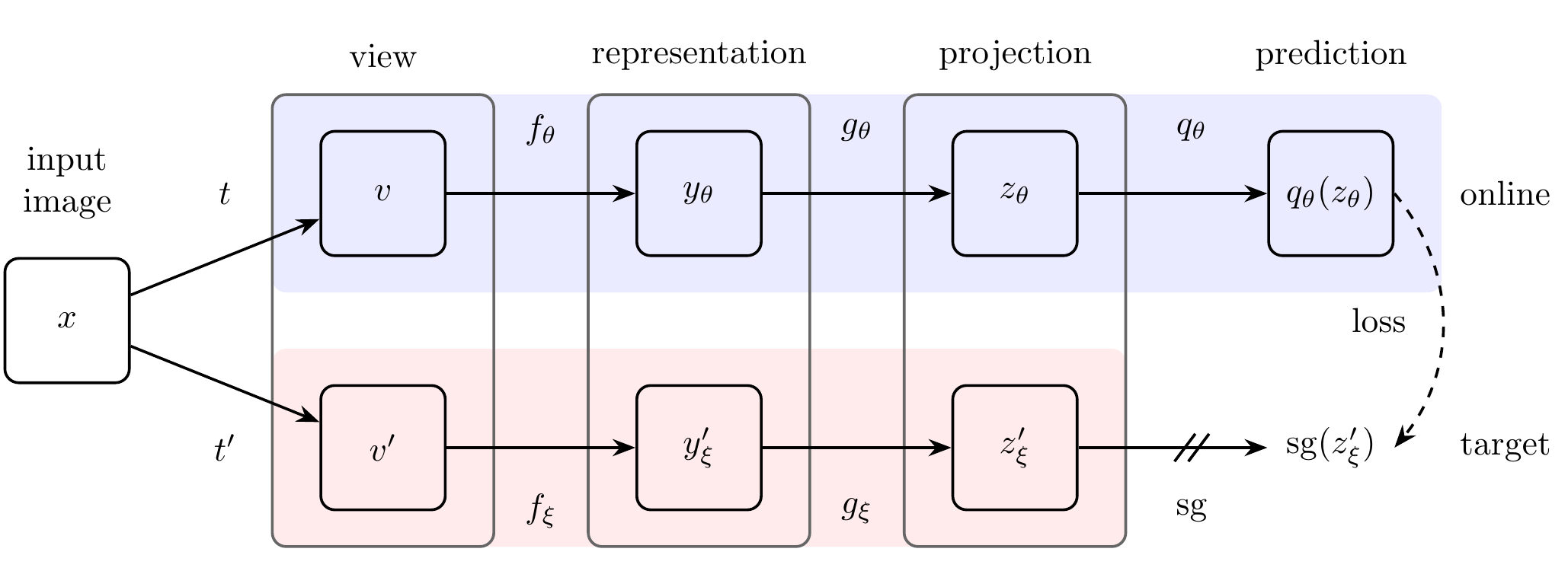}
    \caption{BYOL's overall architecture, adopted from~\cite{grill_bootstrap_2020}}
    \label{fig:byol}
\end{figure}

\textbf{BYOL}, or Bootstrap Your Own Latent~\cite{grill_bootstrap_2020}, depicted in~\cref{fig:byol}, is a novel approach that initially deployed the concept of self-distillation with a mechanism to prevent model collapse. This method employs a pair of neural networks along with a predictor tasked with the duty of aligning the outputs from one network onto the other. The model operates by feeding each network with a uniquely transformed view of the same image, using transformations that include operations like random resizing, cropping, and alterations to color and brightness. The student network is dynamically updated during the training process through gradient descent. On the other hand, the teacher network is updated using the Exponential Moving Average~(EMA) of the weights of the online network. The slow update pace, which is a consequence of using EMA, introduces a crucial asymmetry that significantly contributes to the successful performance of BYOL. The loss function for BYOL can be expressed as

\[
L_{\text{BYOL}}(\theta) = \mathbb{E}_{(x,t,t') \sim (X,T,T')} \left\| \text{renorm}\left(q_\theta\left(g_{\theta}(t(x))\right)\right) - \text{renorm}\left(g_{\xi}(t'(x))\right) \right\|_2^2,
\]

where the two vectors in the representation space are L2-normalized automatically. This normalization can be represented as:

\[
\text{renorm}(v) = \frac{v}{\max\left(\|v\|_2 + \epsilon, \right)}.
\]

$g_{\theta}$ is the online~(student) encoder network, parameterized by $\theta$, while $q_{\theta}$ is the predictor network parameterized by $\theta$. Here, $x \sim X$ is the input drawn from the data distribution $X$, and $t(x)$, $t'(x)$ are two augmented views of $x$ where $t \sim T$, $t' \sim T'$ are two data augmentations. 
The target network $g_{\xi}$ shares the same architecture as the student and is only updated via EMA with $\tau$ controlling the extent to which the target network retains its history, as shown:

\[
\xi \leftarrow \tau \theta + (1 - \tau) \theta.
\]

\begin{figure}[!htb]
    \centering
    \includegraphics[width=0.5\textwidth]{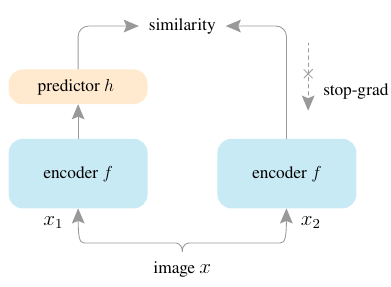}
    \caption{SimSiam's overall architecture, adopted from~\cite{chen_exploring_2020}}
    \label{fig:simsiam}
\end{figure}

\textbf{SimSiam:} SimSiam~\cite{chen_exploring_2020} can be thought of as a simplified version of  the BYOL framework, as depicted in~\cref{fig:simsiam}. 
Through their exploration, they found that the use of EMA was not strictly necessary, even though its inclusion slightly improved performance. This discovery allowed for the formulation of a more streamlined loss function. 
The loss function for SimSiam is defined as follows:

\[
L_{\text{SimSIAM}}(\theta) = \mathbb{E}_{(x,t,t') \sim (X,T,T')}\left[\left\| \text{renorm}(q_\theta(f_{\theta}(t(x)))) - \text{sg}(\text{renorm}(f_{\theta}(t'(x)))) \right\|^2_2\right],
\]
where $\text{sg}(.)$ is the stop-gradient operation. The reason why BYOL and SimSiam evade model collapse and are vital to the success of these methods is the asymmetry introduced between the two network branches, along with certain dynamics of the training process, which implicitly regularizes the variance of the embeddings. It is this dynamic process and the careful management of variance that ensure the maintenance of a diverse and meaningful representation space. It has been observed~\cite{noauthor_understanding_nodate,moon_embedding-dynamic_2022,zhang_how_2022} that the performance of these self-supervised learning methods tends to degenerate to the level of random guessing when the component of batch normalization is eliminated. Batch normalization indirectly institutes a form of contrastive learning, thereby playing a crucial role in these methods. The dependency on negative samples that arises through batch normalization is a significant element in preventing model collapse. For instance, if we were to represent every data point as an all-zero vector, it would trigger a model collapse. However, the process of batch normalization inherently redistributes values, even if the batch of inputs is remarkably similar. By ensuring a diverse distribution of values, it keeps the model from falling into the trap of collapse. 

The transformative journey from the Contrastive SSL to the Self-Distillation paradigm revolves around a number of crucial evolutions that try to address the central challenges in SSL, namely, model collapse and large batch size requirements.
This includes the deployment of dual views of the same data instance, rather than generating positive/negative pairs via data augmentation. 
Furthermore, the strategic asymmetric updating of the networks using various mechanisms contributes significantly to this progression. 
Self-distillation methods offer several other advantages over contrastive methods for SSL, such as needing lower batch size due to not using negative pairs thus lower computation power and simpler Learning Dynamics.
These innovations have led to a promising alternative to contrastive SSL. 

In this thesis, we heavily incorporate self-distillation ideas, and introduce novel methods for both IEA-GAN~(see \cref{chap:5}) and YonedaVAE~(see \cref{chap:6}) models.


  \chapter{Deep Generative Models for Detector Simulation: A Taxonomy}
\label{chap:4}

\section{Introduction}
The core concept of generative models is derived from the training of a density estimator that produces samples approximating the distribution of the training data. The initial generation of neural network-based generative models, also known as energy-based models~\cite{hinton_optimal_1983,hinton_training_2002}, attempted to accomplish this by establishing an energy function proportional to the likelihood for data points. However, these models faced challenges in scaling up to high-dimensional, complex data, like natural images. They also necessitated the use of Markov Chain Monte Carlo~(MCMC) sampling~\cite{hinton_fast_2006}, a method required during both training and inference stages. This method, characterized by its iterative nature, often resulted in a slow and inefficient process. 
The resurgence of interest in generative models in the past few years can be attributed to the availability of larger datasets and significant advancements in deep generative models~(DGMs) for natural images. These DGMs have pushed the boundaries in terms of visual quality, sample diversity, and speed of sampling. In Particle Physics, the application of DGMs was first studied in the ``Fast and Efficient Simulation'' campaign~\cite{de_oliveira_learning_2017} that sparked the search for faster and more storage-efficient simulation methods of collider physics experiments. 

The remainder of this chapter is structured as follows. Firstly, I formulate the problem of detector simulation using deep generative models. Then, I give an algorithm taxonomy of existing methods in the category of latent variable approaches. This section presents a general framework, discusses common generation strategies and tasks in detail, and introduces a thorough review of works of each type.

\section{Problem Deﬁnition}
The detector response can be defined by a triplet $D_{e} = (\mathbf{L}_e, \mathbf{C}_e, \mathbf{H}_e)$ for each event $e$, where $\mathbf{L} \in \mathbb{Z}$ is the set of detector layers/sensors indicators, $\mathbf{C} \in \mathbb{R}^{n}$ is the global attribute of the corresponding event, $\mathbf{H} \in \mathbb{R}^{d}$ is the hit manifold per sensor/layer like tracks or clusters which can be represented for instance by a grid-like an image and sequence, or a set like point clouds and graphs. 
Given a set of $M$ observed detector responses $\mathbf{D} = \{D_i\}_{i=1}^M$, DGMs learn the distribution of these signatures $p(\mathbf{D})$, from which new responses can be sampled $D_{\text{new}} \sim p(\mathbf{D})$.

This chapter presents both the latent variable and non-latent variable approaches as the mainstream detector response generation models. latent variable approaches follow an encoder–sampler–decoder pipeline. It firstly maps the data into a hidden space through an encoding function, manipulates the hidden variables to reﬂect the desired properties of the detector response to be generated, and then generates new samples based on latent codes through a decoding function.

\section{Algorithm Taxonomy}
\subsection{Latent Variable Approaches}
Within the scope of latent variable techniques, the given data is mapped into a stochastic latent space. An \textit{i.i.d} sample from this distribution is then fed into a decoder that reconstructs the original data structures. 
Prior to delving into an in-depth exploration of distinct models, I initially established a bird-eye-view pipeline that encompasses encoding, sampling, and decoding stages. This pipeline allows us to encapsulate the majority of the existing generative models used for detector simulation within a single unified framework. 
In accordance with this framework, I classify various methodologies based on their interaction with three pivotal components:

\textbf{The Encoder}. The encoding function $f_{\theta}(z|D)$ maps the detector response triplets to a dense, continuous~(or quantized) topological space. To ensure the learned latent space is meaningful for generation, depending on the data type and the inductive bias, one incorporates various morphisms~(e.g., Convolutional Neural Networks~\cite{fukushima_neocognitron_1980}, Graph Neural Networks~\cite{velickovic_everything_2023}, DeepSets~\cite{zaheer_deep_2018}, etc) as the encoder. In this way, the encoder function $f_{\theta}$ generates the parameters for a stochastic distribution that adheres to a prior distribution denoted as $p(z)$.

\textbf{The Sampler}. Following the encoding process, the model that generates the detector response draws samples from the marginal distribution $z \sim p(z)$. There are generally two prevalent strategies for this sampling process: \textbf{random sampling} and \textbf{controlled sampling}~\cite{goodman_controlled_1950}. Random sampling refers to randomly selecting latent codes from the learned or prior distribution. On the other hand, guided or controlled sampling is designed to sample the stratified latent code with the specific goal of generating samples that exhibit certain preferred characteristics. In most tasks, controlled sampling is model-dependent and necessitates an additional optimization component that goes beyond the scope of random generation.

\textbf{The Decoder}. Upon obtaining the latent codes drawn from the learned distribution, the decoding mechanism is responsible for stochastically 
reconstituting them into a data manifold. Due to the multi-objective and fine-grained learning nature of detector simulation, the decoding phase is inherently more complex than the encoding stage. 
Typically, the decoders could be grouped into three categories: \textbf{sequential generation}, \textbf{one-shot generation}, and \textbf{zero-shot generation}. 
Sequential generation refers to generating sensor/layer information in a set of consecutive and autoregressive steps, usually done sensor-by-sensor or hit-by-hit. One-shot generation, instead, refers to generating the whole detector signature in one single step. Zero-shot generation is when the model generates an unseen set of detector information where the control parameter~(e.g. incident energy, luminosity, or hit multiplicity) goes beyond the training data, thus the model has to generalize well to the Out-Of-Distribution~(OOD) domain.

\subsubsection{Sampling Strategies}
After learning a latent space for representing the input data, the generative model samples from the learned distribution during the generation inference. 
The sampling strategies could be divided into two categories, \emph{random sampling} and \emph{controllable sampling}. Random sampling simply draws latent samples from the prior distribution, in which the model learns to approximate the distribution of the observed detector responses. The latter, on the contrary, samples detector responses with controls over the desired properties~(i.e. detector geometry, angles, energy, luminosity, and beam parameters). Therefore, for latent variable approaches, random sampling is relatively trivial, while controllable sampling usually requires extra effort in algorithm design.

Controllable generation usually manipulates the randomly sampled $z \sim p(z)$ or the encoded vector $z \sim p(z|D)$ in the latent space to obtain a ﬁnal representation vector $z$, which is later decoded to the detector response representation with expected properties. There are three types of commonly used approaches:

\begin{itemize}
    \item \textbf{Disentangled sampling} factorizes the latent vector $z$ with each dimension $z_n$ focusing on one property $p_n$, following the disentanglement regularization that encourages the learned latent variables to be disentangled from each other. Therefore, varying one latent dimension $z_n$ of the latent vector $z$ will lead to property change in the generated detector responses.
    
    \item \textbf{Conditional sampling} incorporates a conditional code $c$ that explicitly controls the property of generated detector information. In this case, the ﬁnal representation $\hat{z}$ is usually a modulation of $z$ and $c$.
    
    \item \textbf{Traverse-based sampling} searches over the latent space by directly optimizing the continuous latent vector $z$ to obtain $\hat{z}$ with speciﬁc properties or uses heuristic-based approach~(e.g., linear ~ or nonlinear interpolation from $z$ to obtain $\hat{z}$), to control the property of the generated detector hits.
    
\end{itemize}

In this thesis, all the developed models, PE-GAN, IEA-GAN, and YonedaVAE highly depend on Conditional sampling. 
IEA-GAN uses adversarial conditional sampling~((see~\cref{chap:5})), and YonedaVAE introduces a self-distilled Adaptive Top-q sampling~(see~\cref{chap:6}) as a specialized version of conditional sampling. 
In my exploration of PE-GAN, I experimented with traverse-based sampling through latent optimization~\cite{wu_logan_2020}. However, both the training and testing phases proved to be computationally expensive.

\subsubsection{Generation Strategies}

The decoder restores the latent code back to the data manifold. In this study, due to the sparse, high dimensional, and unordered nature of detector data, the resulting outcome of the decoder often faced challenges in accurately reconstructing the data manifold or generating new ones. This led to artifacts or inaccuracies in the generated detector responses.
To address this issue, existing works take three types of generation strategies for detector response generation, one-shot generation, sequential generation, and zero-shot generation.

\textbf{One-shot generation}. One-shot generation usually generates an event in one single step. It is achieved by feeding the latent representations to neural networks to obtain the desired representation. In practice, various neural networks could be utilized, including 2D Locally Connected Networks~\cite{de_oliveira_learning_2017}, 2D and 3D Convolutional Neural Networks~(CNN), Graph Neural Networks~(GNN), Multi-layer Perceptron~(MLP), according to different types of detectors and representations to be generated. The advantage of one-shot generation is that it generates the whole event in a single step without sequential dependency on layer ordering or hit ordering. If the detector topology has a non-sequential ordering of sensors~(non-cylindrical topology), it is not feasible to treat the sensor information, $\mathbf{L}$ the same as the hit manifold $\mathbf{H}$. Such types of non-trivial topologies are also present in the Mesh representation of 3D objects which consists of a collection of vertices~(like detector hits) and polygon faces~(like sensor information). 

The PXD detector carries such a non-trivial topology. Thus, in order to capture such topology-related intra-event relationship, I will introduce the \emph{Relational Reasoning Module}~(see~\cref{chap:5}) which is a modified Transformer Encoder model that acts over the event information with the advantage of bi-directional information aggregation.

\textbf{Sequential generation}. In contrast to one-shot generation, sequential generation generates the detector responses consecutively in a few steps. As there is ordering naturally deﬁned for detector layers such as a Calorimeter detector, sequential generation has to follow a certain sequential inductive bias for the generation. This is usually done by generating probabilistic sensor features while sampling and feeding step-by-step the reconstructed detector response following a predeﬁned node ordering. Despite slow sampling, sequential generative models, enjoy the beneﬁt of auto-regressive reasoning which prompts a precise correlation modeling of the data. Therefore, it could be easily combined with constraint checking in each of the generation steps, when the responses to be generated should obey certain restrictions. Using sequential sampling, when generating a large detector with either high-granularity or long detector layers, the error will accumulate at each step, possibly resulting in discrepancies in the ﬁnal generated and observed detector signatures. Also, in the case of non-cylindrical topologies like PXD, it could introduce a non-existing sequential bias into the generated data.

To alleviate these issues while benefiting from the auto-regressive reasoning, I will incorporate a proper relative positional embedding to resolve the bias and introduce a gating mechanism in YonedaVAE to mitigate the error-aggregation~(see chapter~\cref{chap:6}).

\textbf{Zero-shot generation}. Zero-shot detector response generation refers to the process of generating detector responses without any prior exposure to the specific representation or structure of the ``new'' target detector signatures. The new target detector signatures could belong to new beam parameters, higher incident energies, higher luminosity, or detector responses for uninstalled sensors.
The term \textit{Zero-Shot} comes from the concept of zero-shot learning, which is when a model can recognize or generate outputs for new, unseen categories or tasks without any training examples. This is achieved by leveraging the model's pre-existing knowledge and generalization abilities, typically acquired during the pre-training phase on large datasets by incorporating symmetries and constraints directly into the generative model.
This strategy leverages the latent space's ability to capture the essential characteristics of the data manifold, enabling the generation of plausible detector responses for regions and conditions beyond the current data at hand. The core idea that is introduced in this study, is to extrapolate to the Out-Of-Distribution~(OOD) data by designing a model to be more flexible and adaptive, allowing it to accommodate detector geometries, representations, and conditions beyond the training data. 

In particular, to robustly model such capabilities, in~\cref{chap:6} I introduce YonedaVAE, trained on only low luminosity data to integrate domain-specific knowledge into the generative framework, to extend its ability to reason about unobserved and generate plausible high luminosity PXD background.



\subsubsection{Representative Work}

In this subsection, I briefly discuss representative works for deep  generative models in detector simulation with an emphasis on how they handle generation and sampling.

\textbf{Variational Autoencoders}. VAE is a simple yet ﬂexible framework and could be adopted for controllable sampling by either modifying the loss function to enforce latent variables to be correlated with properties of interest or to feed the conditional information to different parts of the model. In~\cite{ghosh_deep_2020}, the ATLAS collaboration conditioned the encoder and decoder directly on the energy of the incident particle to generate showers corresponding to a specific energy.
In~\cite{buhmann_decoding_2021,buhmann_getting_2021}, they utilize the BIB-AE~\cite{voloshynovskiy_information_2019} model, conditioned on the incident photon energy. Thus the latent manifold is conditioned on the Energy. They introduce a post-processing module that relaxes the trade-off between the accuracy of the emulated hit energy spectrum and the reproduced shower shape. This module is an MLP-based network that fixes the hit energy spectrum resolution between the input and generated images. In the later efforts~\cite{buhmann_hadrons_2021}, along with improving their model, inspired by~\cite{otten_event_2021}, the authors use a Kernel Density Estimator~(KDE)~\cite{parzen_estimation_1962} to fit the learned latent manifold for the inference time and use rejection sampling for the correct density estimation.
In~\cite{deja_end--end_2020}, they leverage the Sinkhorn Autoencoder~\cite{patrini_sinkhorn_2019} to have a trainable prior approximation, namely a noise generator, to encode and generate embeddings with the same distribution in the latent space. In order to overcome the mode collapse issue and promote diversity, they include additional regularisation on the autoencoder's latent space. Following~\cite{ayinde_regularizing_2019}, they compute a similarity matrix for the neural network’s weights according to the cosine similarity between its different layers to assess the diversity.
DeepRICH~\cite{fanelli_deeprich_2020}, designs a conditional latent space as a combination of CVAE~\cite{sohn_learning_2015} and infoVAE~\cite{zhao_infovae_2018}, where the latent variable $\sigma$ is determined using a Bayesian Optimization~\cite{snoek_practical_2012}. The control variables in this conditional VAE are the kinematic parameters of each particle learned by an auxiliary classifier over the encoded latent manifold as a regularization. They only consider the reconstruction~(emulation) of their dataset.
GVAE~\cite{hariri_graph_2021}, introduces a graph-based VAE architecture for learning the representation of collision events without any controllable sampling for emulation. 
Orzari et.al~\cite{orzari_sparse_2021,touranakou_particle-based_2022}, develop a VAE for generating constituents of hadronic jets as sets. Although they incorporate a permutation invariant loss, Chamfer distance~\cite{fan_point_2016}, instead of the typical mean squared error~(MSE) as the reconstruction loss, their model does break the permutation equivariance using 2d convolution layers in the encoder. They regularize the model by constraining the $p_T$ and the invariant mass to follow the desired jet characteristics. The authors later in~\cite{tsan_particle_2021}, fix this issue by using DGCNN~\cite{wang_dynamic_2019} permutation equivariant layers.
Abhishek at.al~\cite{abhishek_calodvae_2022}, incorporated a Discrete Variational Autoencoder~(DVAE) based model~\cite{rolfe_discrete_2017,vahdat_dvae_2018,khoshaman_gumbolt_2019} with hierarchical dependencies of latent variables and a Restricted Boltzmann Machine~(RBM)~\cite{montufar_restricted_2018} latent prior using block Gibbs sampling for generation of Calorimeter showers. They also tackle the sparsity of the showers with a learnable masking like~\cite{musella_fast_2018}. Cresswell et.al~\cite{cresswell_caloman_2022}, develops a manifold hypothesis-inspired model~(density estimation)~\cite{noauthor_representation_nodate,brehmer_flows_2020,brown_verifying_2023} to do a dimensional reduction to the Calorimeter data to speedup the inference-level sampling process. 

Otten et.al~\cite{otten_event_2021}, introduce a method called buffering density information given the encoded events. They construct a prior by aggregating a subset of the encoded training data by saving all the parameters of the Gaussian distributions for all events in the training data to a file, which constitutes the buffer. At inference time, to increase the variance and avoid overfitting to the training data, they sample also from the buffered Gaussian distributions with a variance control factor. 
Collins et.al~\cite{collins_exploration_2022,collins_machine-learning_2022}, using a ParticleFlow~\cite{komiske_energy_2019} beta-VAE~\cite{higgins_beta-vae_2022} finds an interpretable and meaningful representation of the jets and their information complexity by analyzing the VAE's latent information. Moreover, they leverage $\beta$ from a fixed hyperparameter to an input of both the encoder and decoder networks.
Ilten et.al~\cite{ilten_modeling_2022}, for the first time, incorporates a conditional sliced-Wasserstein VAE~\cite{kolouri_sliced-wasserstein_2018} for full hadronization simulation.

\textbf{Generative Adversarial Networks}. GAN-based models, by design, allow easy implementation of controllable sampling due to introducing a property discriminator for desired properties. Although, taming its convergence is very difficult. The application of GAN-based generative models as implicit density estimators was embarked on by~\cite{de_oliveira_learning_2017} where they simulated 2D jet images for high energy W bosons and QCD jets as their conditional classes while introducing 2d locally connected layers~(LAGAN). 
CALOGAN~\cite{paganini_calogan_2018,de_oliveira_controlling_2018,paganini_accelerating_2018}employs the LAGAN layers to generate layer-wise two-dimensional images that were conditioned on the primary particle energy~($E_p$) ranging uniformly from 1–100 GeV. Vallecorsa et.al~\cite{khattak_three_2018,vallecorsa_3d_2019,khattak_fast_2021} uses 3D ACGAN~\cite{odena_conditional_2017}, to generate the calorimeter showers. In~\cite{belayneh_calorimetry_2020}, they add incident angle conditioning as well.
Erdmann et al~\cite{erdmann_generating_2018}, uses WGAN~\cite{arjovsky_wasserstein_2017} with continuous air shower energy conditioning using a constrainer network~(like ACGAN).
Musella et al~\cite{musella_fast_2018}, for the generation of sparse hadronic jets using a U-net~\cite{ronneberger_u-net_2015} module for the generator, introduce a decision-making method by adding an additional channel to their output as a mask probability to decide if a pixel should be zero or not. 
Srebre et al.~\cite{srebre_generation_2020,}, used the WGAN-gp~\cite{gulrajani_improved_2017} to simulate the PXD background hitmaps with random sampling. This model is the baseline for the IEA-GAN study, as they also used the same dataset.
In~\cite{chekalina_generative_2019}, they do controllable sampling by conditioning the WGAN-gp model for the LHCb Calorimeter images. 
In~\cite{alonso-monsalve_image-based_2020}, they introduce an emulator-simulator setup that benefits from the Siamese Network~\cite{bromley_signature_1993,chopra_learning_2005,koch_siamese_nodate}. They show that using their parameter-to-image grid 2-stage training pipeline, and they can model complex functions. In the pre-training stage, the goal is to learn an emulator distribution that matches the Monte-Carlo simulator distribution using the Siamese network to learn the similarity of the simulated and emulated images. Then, at the next stage, a generator will be trained to learn to map the random noise to the parameter space. All these stages follow an adversarial training regime.
Diefenbacher et al.~\cite{diefenbacher_dctrgan_2020} introduces a method for refining the precision of GANs using a post-processing re-weighting and tuning function based on~\cite{andreassen_neural_2020,badiali_efficiency_2020}. Kansal et al.~\cite{kansal_graph_2021,kansal_particle_2022} for the first time choose a more sparse representation of the detector data and introduce a graph GAN based on~\cite{gilmer_neural_2017} layers.
In~\cite{winterhalder_latent_2021}, they introduce the Latent Space Refinement~(LaSeR) protocol to enhance the precision and the topological obstruction of sampling by refining the predictions of a generator. In LaSeR, each generated sample is assigned a weight, which is then mapped to the corresponding latent space point. Rather than directly sampling from this weighted latent space, which could lead to biased results, the authors propose training a second generative model, the refiner, to transform the weighted latent space into an unweighted one. 
Shirobokov et.al~\cite{shirobokov_black-box_2020}, introduce a new approach that synergizes deep generative models and non-differentiable simulators. They show that one can both approximate the stochastic behavior of the simulator and enable direct gradient-based optimization of an objective by parameterizing the latent variable model with the relevant parameters of the simulator.
Jaruskova et.al~\cite{jaruskova_ensemble_2023}, improves the calorimeter simulations over the lower energy depositions with AdaGAN-based~\cite{tolstikhin_adagan_2017} ensemble of GANs.

Hashemi et.al~\cite{hashemi_lhc_2019}, for the first time uses GAN for directly emulate high-level features computed
from the reconstructed $Z\rightarrow \mu \mu$ events. DijetGAN~\cite{di_sipio_dijetgan_2019} a simple GAN with random sampling for the simulation of QCD dijet events. Butter et.al~\cite{butter_how_2019}, do top pair generation with a modified MMD-GAN~\cite{li_mmd_2017} where the MMD kernel helps to describe on-shell resonances as well as tails of distributions, an improvement to~\cite{hashemi_lhc_2019}. They also study the statistical uncertainties and do ablation studies of the GAN approach to the event simulation using GANs.
Carrazza et al. ~\cite{carrazza_lund_2019} employ CycleGAN~\cite{zhu_unpaired_2020} with the cycle-consistency loss to create mappings between two domains of Lund images~\cite{dreyer_lund_2018}, different categories of jets.
Farrell et al.~\cite{farrell_next_2019} apply GANs to generate full particle physics events conditioned on  physics theory parameters. 
Li et.al~\cite{li_polarization_2022}, introduces a style-based~\cite{karras_style-based_2019} conditional GAN~\cite{mirza_conditional_2014} to predict lepton decay angles in the rest frames of W bosons in Vector Boson Scattering~(VBS) processes. Their approach addresses the challenge of missing neutrino information in the final state, which traditionally hampers the full determination of lepton angles. 
Alanazi et.al~\cite{alanazi_simulation_2021,velasco_cfat-gan_2022} develop a GAN-based model that does an importance sampling over the generated features that improve the sensitivity
of the discriminator. 
Prieto et.al~\cite{bravo-prieto_style-based_2022} proposes a style-based quantum GAN to generate events with a 3-qubit model. 
Howard et.al~\cite{howard_learning_2022} incorporates the Sliced-Wasserstein VAE~\cite{kolouri_sliced-wasserstein_2018} and the theory-based physics constraints in an unsupervised setting for event generation.
Anderlini et.al~\cite{anderlini_generative_2023} introduce a distilled GAN from an ensemble of models~\cite{malinin_ensemble_2019} to reduce the variance for a maximally diverse set of models. In the inference time, they test the performance of their model in the Out-Of-Distribution~(OOD) regions of the phase space.

\subsection{Non-Latent Variable Approaches}

Apart from the latent variable models, some generation methods do not map the observation into the latent space. In other words, they perform the generation in the raw space and directly generate the data representation.

\subsubsection{Representative Work}

\textbf{Flow-based Models}.
Flow-based Models~\cite{rezende_variational_2016,berg_sylvester_2019} learn an explicit bijection between the data distribution and a known distribution~(usually a simple one, like a multivariate normal). This allows for exact likelihood computation and exact sampling but does not involve a latent space in the same way as models like VAEs or GANs. The primary limitation of this category of models is their inadequacy in generating detector data with high resolution or granularity. Specifically, these models often struggle to capture the fine details and subtle variations that are crucial for applications requiring a high level of precision and fidelity, such as high-granularity detector simulation. 
Therefore, while normalizing flows offers advantages such as exact likelihood optimization and tractable inference, they might not be the most suitable choice for tasks that require the generation of high-resolution or granular detector data.

For the first time, the CaloFlow~\cite{krause_caloflow_2023} applies the normalizing flow to the simplified calorimeter geometry with \num{504} cells. Their model is a combination of MADE blocks ~\cite{germain_made_2015} and RQS transformations~\cite{durkan_neural_2019}. CaloFlow provides the additional benefit of tractable likelihoods with application to parameter inference for particle reconstruction. 
In CaloFlow~II~\cite{krause_caloflow_2023}, they incorporate Knowledge distillation to transfer the probability
density of the stronger model~(teacher) to the student, an Inverse Autoregressive Flow~(IAF)~\cite{kingma_improving_2017} model which is much faster in sampling.
Xu at al.~\cite{xu_generative_2023}, develop a conditional Normalizing Flow based on~\cite{papamakarios_masked_2018} with the emphasis on modeling the correlation between the kinematic variables.

\textbf{Diffusion Models}.
Diffusion models or deep score-based generative models~\cite{song_generative_2020,song_how_2021,ho_denoising_2020,kingma_variational_2023,ramesh_hierarchical_2022} employ the concept of a stochastic diffusion process to gradually transform a known noise distribution into a target data distribution. This is very much like simulating a random walk over a series of time steps, each guided by a set of transition probabilities. One of the notable benefits of using diffusion models for high-resolution detector simulation data is their ability to capture intricate data structures. By taking numerous small steps to evolve from a simple distribution to a complex, high-dimensional target, diffusion models can effectively model the subtle nuances and details often required in high-fidelity simulations. 
However, there are downsides to consider when applying diffusion models to high-resolution detector data such as PXD. One such drawback is computational cost. The diffusion process involves multiple steps, each of which typically requires its own round of computation, making them somewhat slow samplers and computationally intensive. 

CaloScore~\cite{mikuni_score-based_2022} marks the first application of a score-based generative model for detector simulation. They construct the score function using Conv3D U-nets models~\cite{ronneberger_u-net_2015}, conditioned over the normalized incident energy.
Mikuni et al.~\cite{mikuni_fast_2023} apply a Transformer-based diffusion model to particle jets~\cite{kansal_particle_2022} conditioned fully on the initial jet type, kinematics, and multiplicity. In order to increase the sampling process, they use the progressive distillation mechanism~\cite{salimans_progressive_2022}, to transfer the knowledge of the Transformer-based teacher to an MLP-based student. 

\textbf{Autoregressive Models~(ARM)}.
Autoregressive Models~\cite{oord_conditional_2016,chen_pixelsnail_2018,you_graphrnn_nodate}, or ARMs, are designed to generate data in a sequential manner. Each new data point or feature generated by these models is conditioned on the preceding elements. Although these models might utilize internal hidden states during the generation process, it's worth noting that these states are not considered latent variables in the traditional probabilistic sense.
In the context of detector simulation, ARMs offer the advantage of generating highly correlated data, thanks to their sequential conditioning on previously generated elements. This enables intricate probabilistic relationships to be captured, offering the potential for more realistic simulations. However, this strength comes with notable weaknesses, including computational inefficiency and potential error accumulation. The inherently sequential nature of ARMs makes them computationally intensive and difficult to parallelize, leading to slower data generation processes. Moreover, errors at the early stages of the sequential generation can propagate, potentially causing significant inaccuracies in the final high-resolution output.

Lu et al.~\cite{lu_sarm_2021} model the joint distribution of the data directly by breaking it down into a product of conditional distributions. In other words, each data point as pixels is modeled autoregressively as being conditionally dependent on the previous data points. This model takes sparseness into account during learning and generation, which is a key feature of this work. 
Liu et al.~\cite{liu_geometry-aware_2022,liu_generalizing_2023}, also generate Calorimeter responses autoregressively while taking into account variable detector sizes as a geometrical (detector size) conditioning for OOD detector geometries.
Di Bello et al.~\cite{di_bello_conditional_2022}, introduce a a modified Transformer model that is a combination of GNN as an encoding module and Slot-Attention~\cite{locatello_object-centric_2020} layers as a context-injecting layer to the probabilistic decoder which is a GRU-based model~\cite{cho_learning_2014}. 
Finke et al.~\cite{finke_learning_2023}, develops a masked Transformer Encoder-based model~\cite{vaswani_attention_2017}, based on TraDE (Transformers for Density Estimation)~\cite{fakoor_trade_2020} for discretized and ordered jet constituents.

\section{Applications}
In this section, I give an overview of the applications of DGMs in detector simulation. Speciﬁcally, I categorize them in three concrete branches, \emph{statistics amplification}, \emph{Amortised generation}, and \emph{OOD simulation}. Then, I illustrate their formulations in detector response generation and how different approaches and incorporation of inductive biases lead to success in various applications. 
``Inductive biases'' in DGMs refer to the assumptions the model makes to predict outputs for both in- and out-of-distribution inputs. These biases are inherent in the model's architecture and learning algorithm such as symmetries and adversarial robustness that guide the model's learning process~\cite{zhao_bias_2018,battaglia_relational_2018,zietlow_demystifying_2021,krippendorf_detecting_2020,dillon_symmetries_2021,barenboim_symmetry_2021,tombs_method_2022,desai_symmetrygan_2022,hashemi_ultra-high-resolution_2023,haochen_theoretical_2023}. In general, for DGMs in particle physics, the inductive biases could include, distributional assumptions such as smoothness inductive bias, structural assumptions such as relational inductive bias, and Physics-informed Assumptions such as Energy-Momentum conservation bias.

\subsection{Statistics Amplification}
As surrogate models are being used for detector signature emulation to do sample amplification~\cite{axelrod_sample_2019}, it is very important to quantify the statistical power of the generated dataset. Sample amplification is a procedure where there exists a map that takes a finite, initial subset of data and generates an extended one.
In this regime, DGMs work as classical parametric fits~\cite{carrazza_compressing_2021,chahrour_comparing_2022}to the training that also works as data augmentation methods. The question then would be to which extent they can interpolate. In other words, \emph{how can one quantify the diversity and uncertainty of DGMs?} This involves understanding the limitations and potential biases of the DGMs and possibly introducing  methods to quantify the diversity or amplification factors~\cite{hao_data_2019,axelrod_statistical_2022,butter_ganplifying_2021}. 
For amplification of statistics, the smoothness assumption where the physics probability densities are smooth is very natural. In DGMs, the smoothness inductive bias is an assumption that similar inputs will produce similar outputs up to model interpolation. Regularization techniques in machine learning, which are used to encourage the model to learn smoother functions, are a common manifestation of the smoothness inductive bias.

The smoothness inductive bias is studied in~\cite{butter_amplifying_nodate,butter_ganplifying_2021,bieringer_calomplification_2022} for a simplified VAE-GAN~\cite{buhmann_getting_2021}. Their study revealed that individual-generated shower samples contain less information compared to a single real data point. However, when the number of generated shower samples increases significantly, the information contained within the generated sample set eventually levels off. As a result, they demonstrate the ability of DGMs to not only sample from implicitly defined distributions but also to leverage enhanced interpolation or fitting capabilities.

In \cref{chap:6}, similar results are being achieved using an Information Theory approach incorporating a robust diversity measure, Vendi Score~\cite{friedman_vendi_2023}. 

To improve interpolation, it's crucial to consider how to diversify samples. 
In this study, by~\cref{chap:5} I enhance the diversity and complexity~\cite{haochen_theoretical_2023,shwartz-ziv_compress_2023} of generated samples by introducing Self-Supervised Learning~\cite{balestriero_cookbook_2023} techniques that can be incorporated in any DGMs. In particular, I introduce a Uniformity Loss that helps the model to avoid mode collapse and to generate more diverse samples by maintaining the ``uniformity of information'' inductive bias for the discriminator.

\subsection{Amortised Generation}
For detector simulation, amortized generation generally refers to a strategy where a computationally expensive simulation process is replaced or approximated by a faster and more efficient surrogate model. Detector effect simulations and parametrization can be computationally intensive due to a high space- and time-computation complexity. An ``amortized'' approach would involve training DGMs on a large set of data. Once trained, this model can generate new simulations much faster than the original simulation process. The term amortized in this context refers to the fact that the upfront cost of training the model is spread out, or ``amortized,'' over the many simulations that the model generates. 

In an amortized generation, the sub-tasks are either ``Fast Simulation'' to overcome the high time complexity of detector simulation or ``Data Compression'' to overcome the high space complexity of detector simulation.
Based on the different types of data manifolds, various inductive biases are being incorporated. One can categorize different representations into two main classes, ``geometry-dependent'' and ``geometry-independent'' representations. 
Geometry-dependent approaches look at the detector signatures as grid-like structures. For example, 1d fixed representations come with sequential local and translation invariance~\cite{mu_photon_2021,regadio_synthesis_2022}. 
2d representation with local and translation-invariant inductive bias, which was studied for showers, jets, and detector hits in depth for different detectors.
Then, in order to capture the detector's layer-by-layer association and correspondence, 3d grid-based models~\cite{khattak_three_2018,vallecorsa_3d_2019,belayneh_calorimetry_2020,khattak_fast_2021,mikuni_score-based_2022} are studied. However, as a consequence of translation invariance, it has the drawback of stationary assumption over the temporal/spatial features. 

Geometry-independent approaches, on the other hand, are more suitable for detector signature simulation due to the unordered and variable length of sensor data, and their heterogeneity and sparsity. 
For example, graph-based models~\cite{hariri_graph_2021,kansal_graph_2021,tsan_particle_2021,kansal_particle_2022,di_bello_conditional_2022}, besides the variable-length assumption and relational inductive bias, it assumes graph isomorphism inductive bias as well. This bias ensures that the model focuses on the structural information contained in the graph, rather than the specific labeling of nodes.
Another geometry-independent approach is considering the set representation of detector signatures. The most important property with set-based models are the permutation equivariant encoders for jet and shower constituents~\cite{finke_learning_2023,buhmann_epic-gan_2023,leigh_pc-jedi_2023,mikuni_fast_2023,buhmann_caloclouds_2023} 
, and possibly permutation invariant loss functions~\cite{tsan_particle_2021,di_bello_conditional_2022,kansal_evaluating_2023}.
For normalizing flows equivariance under permutation group action is more non-trivial. Proved by Köhler et al.~\cite{kohler_equivariant_2020}, given a Flow-based model $F$ such that the set creation yields an exchangeable distribution, the update is permutation equivariant and invertible, and $p_{\theta}$ denotes the model likelihood, then $(F, -\log p_{\theta})$ is permutation equivariant.
While maintaining the variable-length assumption, dropping the permutation equivariance of the encoders would correspond to a sequential inductive bias of detector responses either layer-by-layer~\cite{diefenbacher_l2lflows_2023,butter_jet_2023}, or hit-by-hit~\cite{lu_sarm_2021,liu_geometry-aware_2022,liu_generalizing_2023}.
While autoregressive models show a better predictive capability~\cite{butter_jet_2023}, the sequential nature of these models can be a disadvantage when it comes to computation as the sampling is rather slow. Moreover, it is noteworthy that ARM-only models lack a latent representation, therefore, it is not straightforward to manipulate their internal data representation which makes it less appealing for tasks like conditional compression and metric learning.


In an amortized simulation, in all sub-tasks, there is a huge computational and engineering bottleneck, which is the size of the detector to be simulated. Given the previous works, I classify the detector simulation datasets into four categories, namely \emph{the low granularity} such as the jets data with $\mathcal{O}(100)$ number of hits per event, \emph{mid-granularity} such as simplified detector~(calorimeter) data with $\mathcal{O}(1000)$ number of hits per event, \emph{high-granularity} data such as ILD Calorimeter prototype with $\mathcal{O}(1e4)$ number of hits per event, and finally the \emph{ultra-high-granularity} data such as PXD or High Granularity Calorimeter~(HGCAL) with $+\mathcal{O}(1e6)$ number of hits per event.
Previous work on mid-granularity calorimeter simulation is recent approaches~\cite{buhmann_getting_2021,krause_caloflow_2023}, incorporating a combination of several GAN-like or VAE-like and Flow-based architectures with less than 30k simulated channels, and 3DGAN~\cite{khattak_fast_2021} for high-granularity calorimeter simulation with only 65k pixel channels. 
Nonetheless, these studies barely scratch the surface of the profound challenges posed by the PXD detector simulation or the future detectors. 
Take, for instance, the impending High Granularity Calorimeter~(HGCAL) - a component of the High Luminosity Large Hadron Collider~(HL-LHC)~\cite{bruning_chapter_2020} upgrade program at the Compact Muon Solenoid~(CMS) experiment~\cite{strobbe_status_2021}. With an estimated \num{6.5} million detector channels distributed across \num{50} layers, the HGCAL's complexity far surpasses the capacity of existing methods, pointing to the urgency of developing more advanced simulation approaches.

\subsection{Out-Of-Distribution~(OOD) Generation}
DGMs for OOD and zero-shot learning is an exciting area of research that holds significant potential across various fields including, but not limited to, Drug Discovery, material design, and weather forecasting. Traditional methods for the synthetic generation of objects with enhanced or specific properties are often iterative and costly, requiring extensive manual work or heavy computational resources. 
In contrast, DGMs with zero-shot capability can deal with new scenarios that are not explicitly present in the training data, making them highly desirable in a wide range of applications~\cite{anishchenko_novo_2021,rajak_autonomous_2021,ravuri_skilful_2021,madani_large_2023,li_zero-knowledge_2023}. 
OOD generation of detector signatures is an emerging field in High Energy Physics~(HEP) as well, especially through Simulation-Based Inference methods~\cite{brehmer_simulation-based_2021,boelts_flexible_2022}.
The current main challenge remains to be the optimization of DGMs, based on available information while avoiding overfitting, and the generalization to cases in which information is scarce or altogether absent, such as extrapolation to beam parameters, energies, luminosities, and geometries where there are no data. 

The first example of such OOD detector response generation is the CaloGAN paper~\cite{paganini_calogan_2018}, where they very briefly show they can generate showers beyond the training incident energy conditions. DijetGAN~\cite{di_sipio_dijetgan_2019}, also discusses extrapolation to new BSM-dependent OOD regions of dijet invariant mass.
Anderlini et al.~\cite{anderlini_generative_2023}, evaluates the uncertainty of GANs in new momentum regions through Background efficiency comparison. Their analysis shows that for Kaons where the background efficiency does not decrease linearly in the OOD, high-momentum phase-space regions, their ensemble model fails to capture this behavior. Whereas, for Muons, due to the monotonic behavior of the background efficiency, the extrapolation shows better results.

A very important inductive bias for OOD generation is the variable-length assumption, especially in the context of detector simulation, due to the inherently variable nature of the data, characterized by a fluctuating number of detector responses in individual events. DGMs with a variable-length inductive bias have the capability to generate sets/sequences of differing lengths, contingent on the inherent complexity or dynamism of the event being generated. This equips the model with a heightened adaptability when faced with novel and unseen scenarios.
Liu et al.~\cite{liu_geometry-aware_2022,liu_generalizing_2023}, by using an ARM approach~(length-independent) tries to extrapolate to unseen detector geometries where the extrapolation domain is calorimeter cell sizes. However, they show that the high granularity is a bottleneck to their approach.


\section{Conclusion}
The aim of this chapter has been to provide a structured overview of the landscape of deep generative models applied to the challenging domain of detector simulation in experimental particle physics. This survey is particularly important for the community as it not only categorizes various approaches but also highlights key areas that are currently under-explored or inadequately addressed. These include: 

\begin{itemize}
    \item An absence of effective methods and studies for capturing inter-sensor correlations within detector setups.
    \item Inadequate metrics for assessing the diversity of generated samples.
    \item A lack of focused research on length and attribute extrapolation in detector simulation.
    \item A complete dearth of studies for handling ultra-high granularity detector signatures.
\end{itemize}

Within this framework, my thesis offers two distinct contributions to the above issues. 
First, in \cref{chap:5}, I present a fresh perspective for capturing inter-sensor correspondence that introduces a Bert-like transformer as a bidirectional encoder. This is aimed at encapsulating the complex correlations that exist between different sensors within a single event over the Geant4 simulated PXD data. 
Next, in \cref{chap:6}, I introduce a unified approach that fuses the benefits of bidirectional encoding with auto-regressive reasoning. This results in a "geometry-aware" generative model that offers a more nuanced balance between efficient simulation and the ability to operate in out-of-distribution settings. Incorporating such a geometry-sensitive methodology promises to be a pivotal factor in enhancing the model's extrapolative capabilities.
Moreover, I will introduce a data-driven and robust diversity measure for the generated detector responses.  
  \chapter{PXD Background Generation: Simulation Data}
\label{chap:5}

\section{Introduction}
As discussed in~\cref{chap:2}, classical PXD background simulation is inefficient both from the time-complexity~(CPU demand) and space-complexity~(storage demand) perspective. That is, this thesis incorporates Deep Generative Models~(DGM) as the surrogate model to replace the old Geant4 detector simulation at Belle2.

This chapter elucidates the complexities inherent in emulating PXD data. First, it introduces the concept of Fine-Grained Image Analysis and argues that the PXD background generation task is a Fine-Grained generative one.
Then, it presents a comprehensive elaboration of the surrogate model candidates, PE-GAN and IEA-GAN. These generative models, painstakingly crafted and continuously refined, aim to reproduce the multifaceted and intricate nature of ultra-high granularity data of the PXD, a task that presents unique challenges due to their inherent structure and characteristics. 
We delve into the methodological approaches employed, the obstacles encountered and overcome, and the improvements in both the data-level and downstream physics-level analysis. 
As we navigate the terrain of simulated PXD data emulation, I will also address the pressing questions surrounding the applicability, robustness, and limitations of deep generative models. 

\section{PXD Images: Fine-Grained Image Analysis}
The innate abilities of the human visual system enable us to discern visual differences, allowing us to distinguish not just between broad categories like dogs and birds but between closely related ones such as a Siberian Husky and an Alaskan Malamute. 
This prowess led to the conceptualization of Fine-Grained Image Analysis~(FGIA), intending to impart AI with a refined visual understanding. FGIA's central objective in computer vision involves recognizing and generating images from a variety of subcategories~(objects) within a larger category, such as different animal species or various car models. 
The core challenge lies in discerning the underlying distinctions that set apart objects that may otherwise appear highly similar.

\begin{figure}[!htb]
\centering
{\includegraphics[width=0.8\columnwidth]{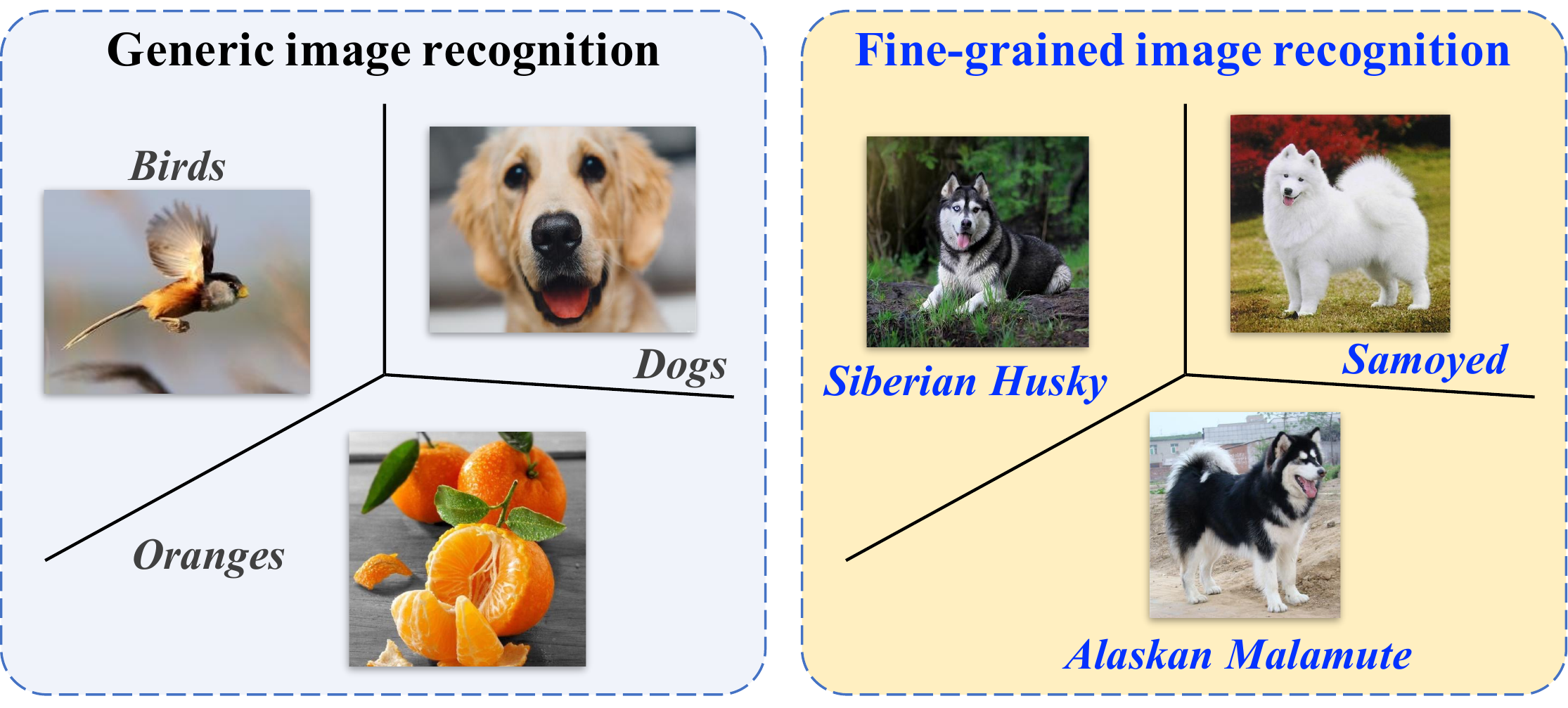}}
\caption{Fine-grained image analysis vs. generic image analysis (using visual classification as an example), adopted from~\cite{noauthor_211106119_nodate}.
}
\label{fig:fgvsgeneric}
\end{figure}

\begin{figure}[!htb]
\centering
{\includegraphics[width=0.8\columnwidth]{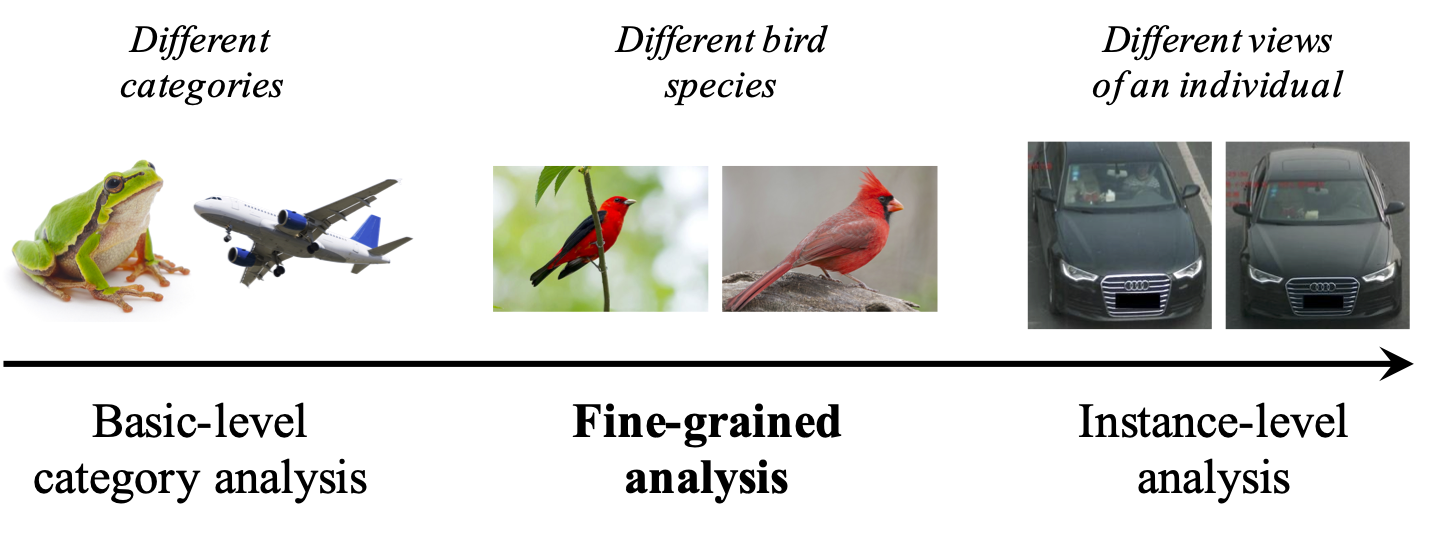}}
\caption{An illustration of fine-grained image analysis which lies in the continuum between the basic-level category analysis (i.e., generic image analysis) and the instance-level analysis (e.g. car identification), adopted from~\cite{noauthor_211106119_nodate}}
\label{fig:multigrained}
\end{figure}

Positioned between basic-level image analysis~(generic image categorization) and instance-level analysis~(identifying individual entities), FGIA, as shown in~\cref{fig:fgvsgeneric}, seeks to distinguish objects from multiple subordinate categories within a broader category~(see Figure~\ref{fig:multigrained}). 

\begin{figure}[!htb]
\centering
{\includegraphics[width=0.85\columnwidth]{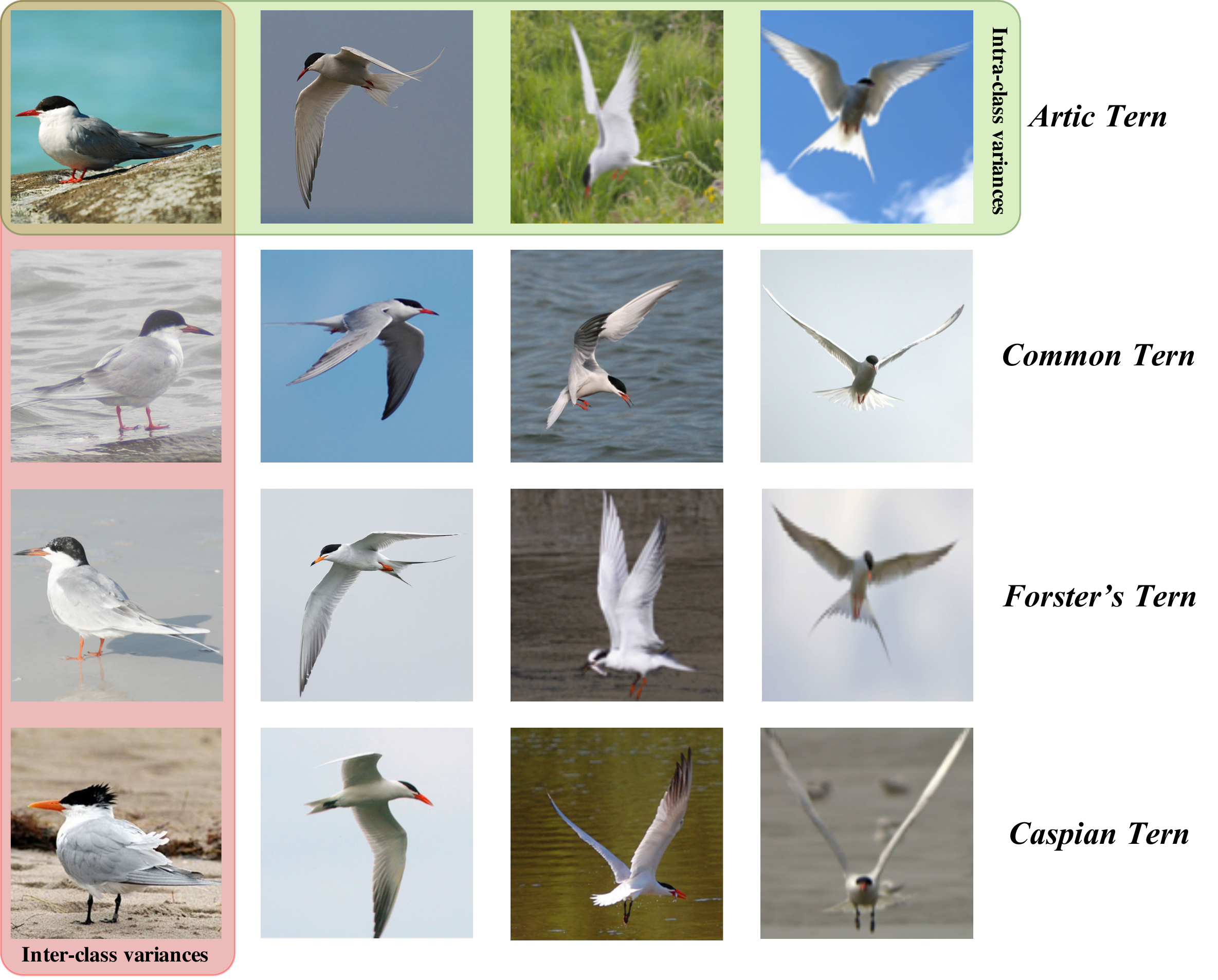}}
\caption{Key challenges of fine-grained image analysis, small inter-class variations, and large intra-class variations. Different Tern species, one species per row, with different instances in the columns, adopted from~\cite{noauthor_211106119_nodate}}
\label{fig:cub_example}
\end{figure}

For instance, while generic image analysis might categorize images into broad classes like bird, fruit, or dog, FGIA delves deeper into categorizing breeds of the same species. 
To achieve this, it becomes imperative to pinpoint subtle visual nuances, such as distinct ear shapes or tail lengths. 
Recognizing these intricate features is also essential for other FGIA tasks, like generation. 
The challenge in FGIA stems from the minimal visual differences between sub-categories and the large differences that can exist within a sub-category due to varying poses, scales, and rotations~(refer to Figure~\ref{fig:cub_example}). This complexity distinguishes FGIA from generic image analysis and marks its uniqueness.

On the other hand, instance-level analysis targets specific entities, not merely categories or sub-categories. 
Descending the granularity spectrum, identifying individuals, such as through face recognition, emerges as a unique form of fine-grained recognition where individual identity becomes the focal point. Consider person/vehicle re-identification tasks. Here, the aim is to discern if two images depict the same person or vehicle. These tasks deploy FGIA-like techniques, capturing distinguishing object features, harnessing coarse-to-fine structural information, or utilizing attribute-centric models.

\begin{figure}[!htb]
\centering
{\includegraphics[width=0.95\columnwidth]{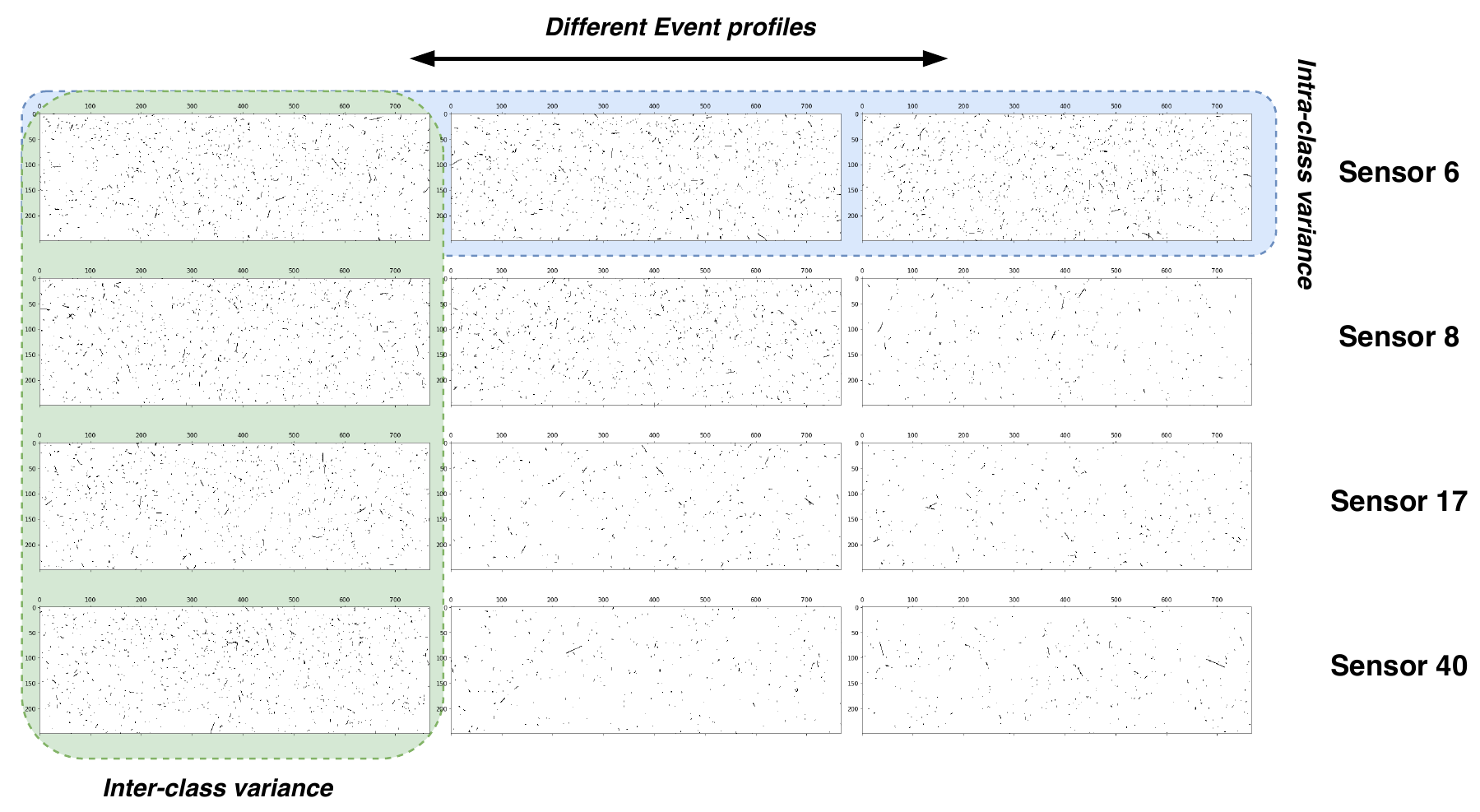}}
\caption{PXD background signatures across events for different sensors. The horizontal cases demonstrate various profiles~(different clustering, occupancy, etc.) across different events. The vertical cases exhibit similar profiles across sensors.}
\label{fig:pxd_fine}
\end{figure}

The PXD data as pixelated images are created in high resolution and contain rich information that can be similar to the nuanced details considered in FGIA.
The data captured by the PXD is extremely granular, allowing for very detailed information on the particle tracks and the different types of background manifested in various image-level artifacts.
Inter-class variations in PXD background data refer to different types of background tracks being recorded. The features such as energy levels, trajectories~(cluster information), and occupancy slightly vary from one class of sensors to another~(see \cref{fig:occupancy-intensity}).
On the other hand, within each class~(PXD sensor), intra-class variations occur due to differences in beam parameters, amount of background, and incident energies across various events.

\subsection{PXD Generation Formulation} 
\label{sec:finepxd}
In PXD image generation, one is provided with a training dataset denoted as 

\begin{equation}
\mathcal{D}=\left\{ \left( \bm{x}^{(n)}, y^{(n)}\right) | i=1, ..., N  \right\},
\end{equation}

encompassing numerous images and their corresponding class labels, represented by $\bm{x}$ and $y$, respectively. Here, $y$ is drawn from the set $[1, ..., 40]$, which corresponds to each PXD sensor. Each data pair $\left(\bm{x}, y\right)$ is a member of the combined space comprising image and label spaces, represented as $\mathcal{X}$ and $\mathcal{Y}$. This pair is drawn according to the distribution $p_r(\bm{x}, y)$.

\begin{equation}
\left(\bm{x}, y\right) \subset \mathcal{X}\times \mathcal{Y}\,.
\end{equation}

Notably, the label space $\mathcal{Y}$ is a combined space of all the $C$ subspaces linked to the $C$ categories~(number of sensors), that is, $\mathcal{Y} = \bigcup_{c=1}^{C} \mathcal{Y}_c$. A generative deep network, symbolized by $g_{\phi}(\bm{z};y)$, parameterized by $\phi$, can then be trained for this generic conditional image generation. 
The training process aims to minimize the generative divergence:

\begin{equation}
\min_{\phi} \mathbb{E}_{(\bm{z},y)\sim p_g(\bm{z}, y)} \left[ \mathcal{D}(x, g_{\phi}(\bm{z};y))\right],,
\end{equation}

where $\mathcal{D}(\cdot,\cdot)$ is a divergence measure assessing the difference between the desired and generated distributions. 
In the framework of fine-grained generation, the goal is to precisely generate PXD instances of different subordinate granular categories, such as various background profiles, clustering, and occupancy information per sensor, from a certain meta-category~(sensor labels), represented as:

\begin{equation}
\left(\bm{x}, y'\right) \subset \mathcal{X}\times \mathcal{Y}c,,
\end{equation}

with $y'$ standing for the fine-grained label, and $\mathcal{Y}c$ being the label space of class $c$ regarded as the meta-category. A meta category is a high-level category that encompasses multiple subordinate categories. It's the umbrella term under which various more specific categories fall. A subordinate category is a specific, low-level category within a broader meta category in a hierarchical way. It is more specialized and is usually distinguished by more nuanced attributes or features.
The optimization aim for fine-grained generation is:

\begin{equation}
\min_{\phi} \mathbb{E}_{(\bm{z},y')\sim p_g(\bm{z}, y')} \left[ \mathcal{D}(x, g_{\phi}(\bm{z};y'))\right],,
\end{equation}

In this chapter, I only introduce a control parameter over the meta-category~(the sensor numbers) and allow the model, IEA-GAN, to explore and learn the subordinate categories~(various intra-class variations) in a self-supervised way. In the next chapter~(\cref{chap:6}), YonedaVAE utilizes a control parameter over the occupancy~(amount of background) of each sensor and event to generate samples with subordinate category attributes.

\subsection{Challenges Of DGMs for PXD simulation}

I have discussed in \cref{chap:3} and \cref{chap:4}, the limitations of DGMs. However, the task of learning to generate ultra-high resolution detector responses such as the PXD background signatures poses several more intricate challenges. Here, I discuss them in detail:

\begin{itemize}
    \item \textbf{Data dimensionality:} 
    For PXD data, one is dealing with spatially asymmetric high-resolution hitmaps. Each event comes with $40\times250\times768$ pixel signatures, even more than the notorious high resolution, $1024\times1024\times3$ ImageNet dataset. The sheer volume and dimensionality of such data pose significant difficulties. Firstly, it necessitates the handling of vast quantities of information, demanding substantial computational resources and sophisticated techniques for efficient data processing and model training. Secondly, it complicates the task of effectively learning the latent variables and dependencies in the data due to the increased complexity associated with such high-dimensional spaces. Thirdly, the risk of mode-collapse escalates with the model potentially memorizing the training data rather than generalizing and learning underlying patterns. Lastly, the spatial asymmetry of the hitmaps adds another layer of complexity, requiring a nuanced understanding and careful modeling of these asymmetric features within the DGM framework. 
    \item \textbf{Data frequency and sparsity:}
    This is underscored by the stochastic nature of background processes. The PXD captures a wide range of charged tracks, that prompts to a dramatic variation of occupancies from order of $\mathcal{O}(10)$ to as high as $\mathcal{O}(1M)$. This variability introduces a significant degree of sparsity in the data - a challenge that amplifies the complexity of generating realistic data using deep generative models as will be discussed in the next chapter.
    This variation in frequencies means that the model must be robust enough to generate events that span across these vast ranges even with imbalance.
    The sparsity of the data poses its own unique set of problems. It makes it harder for the model to learn meaningful representations, as the presence of many zero or near-zero elements can make it difficult for the model to discern the underlying structure or patterns in the data. This issue is further compounded when one considers the high dimensionality of the PXD data, as discussed previously. This sparsity can also lead to computational inefficiency. Traditional representations may lead to wastage of computational resources when dealing with sparse data. Therefore, appropriate data structures and algorithms are required to handle this sparsity effectively~(see~\cref{chap:6}).
    \item \textbf{Fine-graininess and correlation:} 
    The detector responses in an event, a single readout window after the collision of particles, share both statistical and semantic similarities with each other~(see \cref{fig:pxd_fine}). For example, the sparsity~(occupancy) of each image within a class, defined as the fraction of pixels with a non-zero value, shows statistical similarities between detector components as shown in~\Cref{fig:occupancy-intensity}. 
    As the detector response images show extreme resemblance at the semantic and visual levels~\cite{noauthor_visual_nodate}, they can be classified as fine-grained images. The small inter-class and considerable intra-class variation inherent to fine-grained image analysis make it a challenging problem~\cite{noauthor_211106119_nodate}. The current state-of-the-art conditional models focus on class and intra-class level image similarity, in which intra-image~\cite{zhang_self-attention_2019}, data-to-class~\cite{miyato_spectral_2018}, and data-to-data~\cite{kang_contragan_2021} relations are considered. 
    However, in the case of detector simulation, classes become hierarchical and fine-grained, and the discrimination between generated classes that are semantically and visually similar becomes harder. 
    Therefore, a wide range of SOTA models on popular natural datasets show extensive class confusion~\cite{kang_rebooting_2021,rangwani_class_2021} at the inter-class level. 
    In addition, since the information in an event comes from a single readout window of the detector, the processes happening in this window affect all sensors simultaneously, leading to a correlation among them, as shown in~\Cref{fig:spearman_corr}. This study will show how this fine-grain intra-event correlation plays an important role in the downstream physics analysis.
\end{itemize}

\section{Prior Embedding GAN~(PE-GAN)}

In the realm of high-resolution image generation, especially in the intricate domain of PXD images, the choice of architecture, modeling techniques, and training strategies can significantly impact the efficacy of the generated output. With such complexity of data and increasing demands for computational efficiency, there's a pressing need to identify and adapt the right methodologies that strike a balance between performance and computational feasibility. 
Over the years, a plethora of generative models have been proposed, each with its unique strengths and weaknesses. The challenge lies not just in selecting the right model, but in tuning it to cater to specific domain constraints and data peculiarities.

This section delves into my journey of identifying, customizing, and refining a generative model tailored for the PXD image generation task. After rigorous experimentation with various model structures and hyperparameters, informed by the evaluation metrics, I converged on a modified and deeper version of the SAGAN model, supplemented with contrastive conditioning, consistency regularization, and integration of prior knowledge on PXD sensors. 
The ensuing sections provide a detailed account of the model selection process, the challenges encountered, and the strategies employed to overcome these hurdles, culminating in the proposal of the PE-GAN model. This section is based on the paper~\cite{hashemi_pixel_2021}.

\subsection{Transition to BigGAN}
\begin{figure}[!htb]
  \centering
  \begin{subfigure}{0.48\textwidth}
    \centering
    \includegraphics[width=\linewidth]{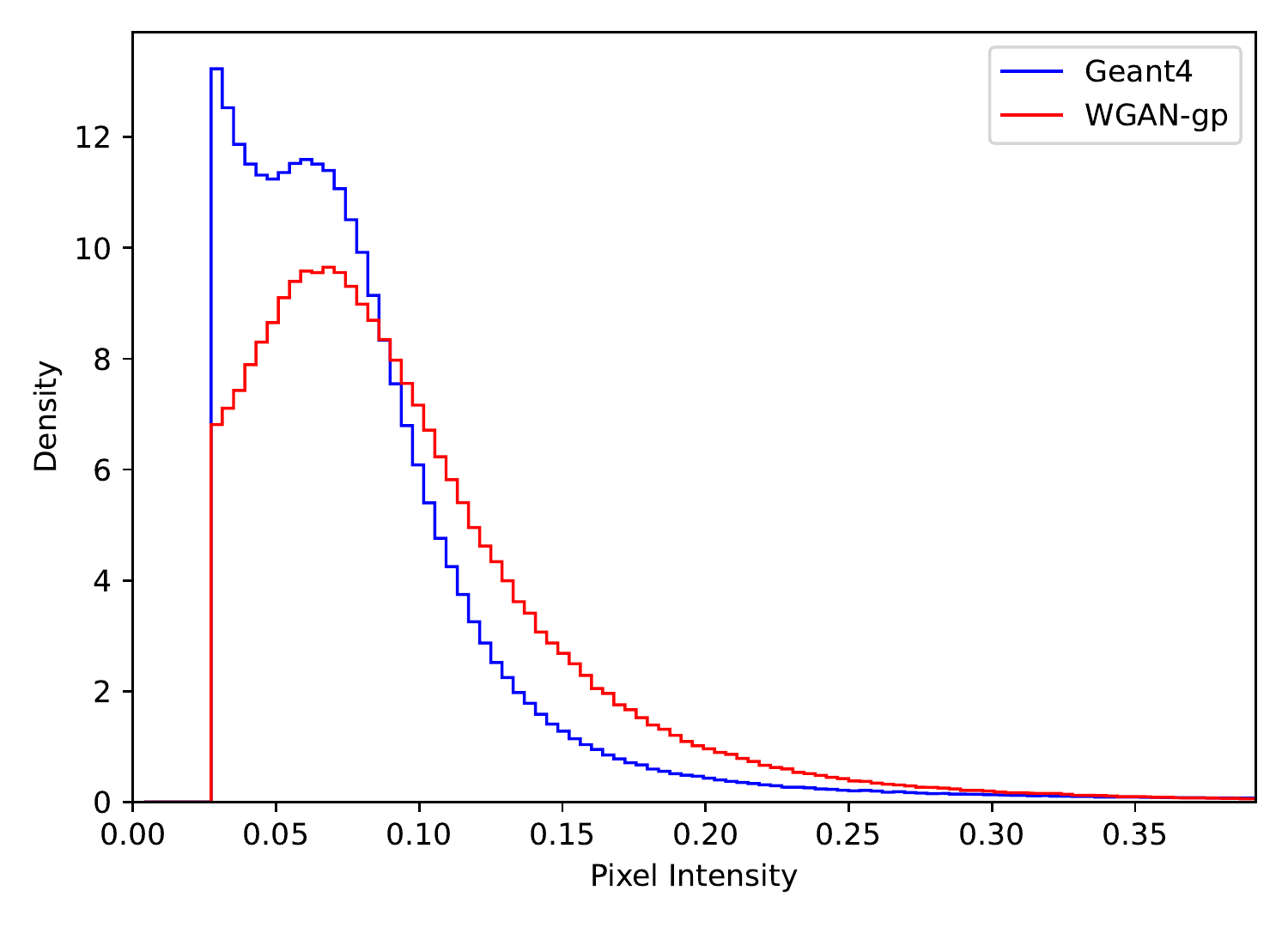}
  \end{subfigure}
  \hfill 
  \begin{subfigure}{0.48\textwidth}
    \centering
    \includegraphics[width=\linewidth]{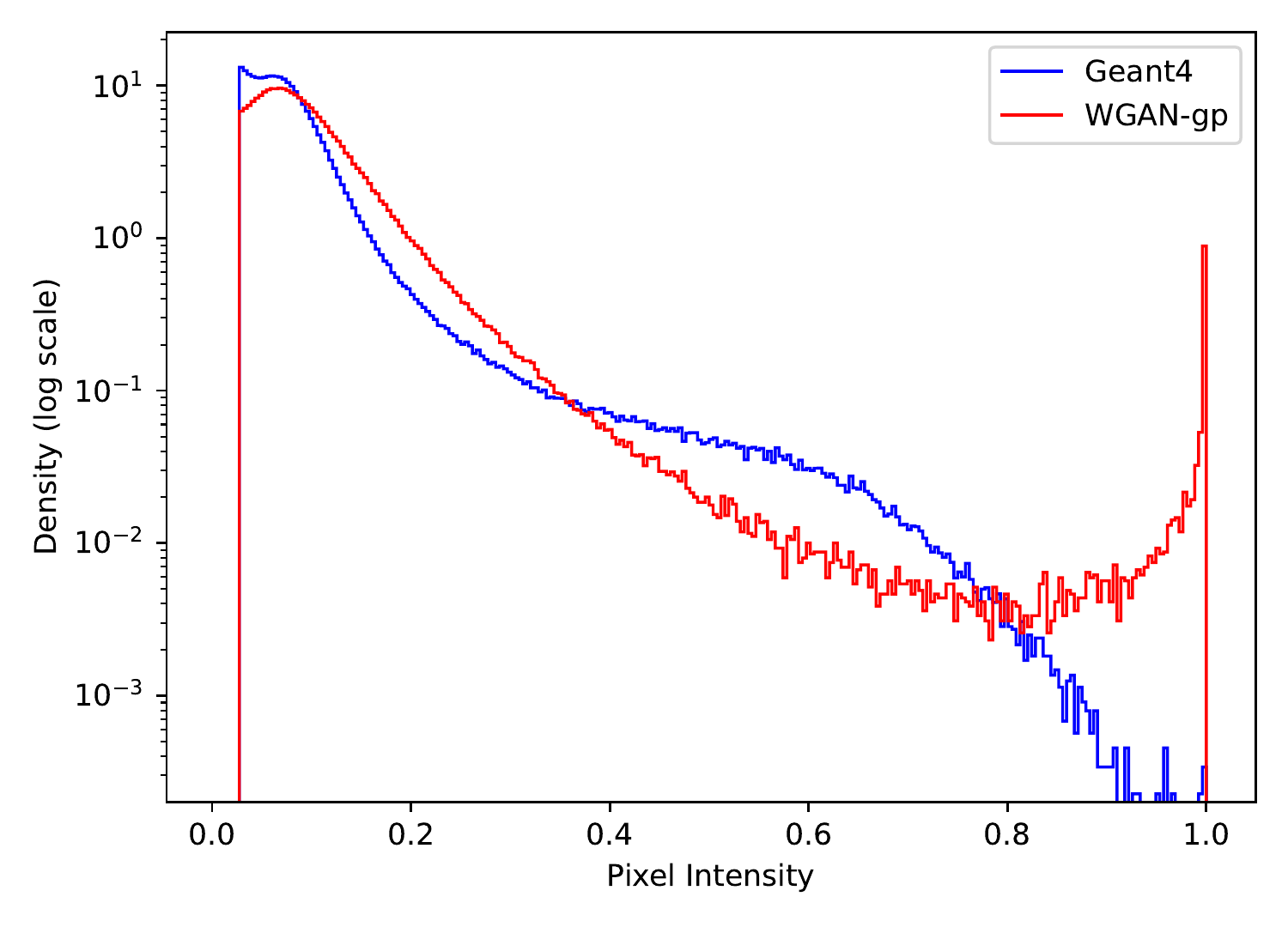}
  \end{subfigure}
  \caption{Charge intensity distribution for the WGAN-gp without conditioning~\cite{srebre_generation_2020} on PXD as the baseline.}
  \label{fig:wgan_int}
\end{figure}

My journey commenced with the integration of class labels, representing the 40 sensor numbers, into the previous model for PXD background generation, WGAN-gp~\cite{srebre_generation_2020}. 
While WGAN-gp served as the initial baseline, the primary challenges encountered were its slow event-based image generation due to its large kernel filter size and its inability to wholly capture the pixel intensity distribution, as depicted in~\cref{fig:wgan_int}.

Subsequently, my attention was directed towards more intricate but rapid models like BigGAN\cite{brock_large_2019}, which utilizes a reduced kernel size and a specialized residual block tailor-made for high-resolution image generation. During its training, however, I observed training collapse, emphasizing the need for early stopping. As outlined in BigGAN's paper \cite{brock_large_2019}, the first singular values of the weight matrices for both the generator and discriminator generally increased throughout training. However, any indications of mode collapse or discriminator overfitting were accompanied by a drastic surge in these singular values for the generator.

Numerous countermeasures were attempted, including decreasing the discriminator's learning rate and experimenting with various regularization techniques like Orthogonal Regularization, Dropout, and L2 regularization. Other approaches, such as Latent Optimization~\cite{wu_logan_2020} and Feature Quantisation~\cite{zhao_feature_2020}, also proved futile. This cascade of unsuccessful attempts guided my decision to adopt a more straightforward model.

\subsection{Adopting Self-Attention GAN}

Drawing from the intricacies of our dataset, memory constraints, and the feedback accrued during training, I transitioned to the Self-Attention GAN~(SAGAN)~\cite{zhang_self-attention_2019} model. SAGAN equips both the generator and discriminator to model extended dependencies over the image manifold, enabling the generator to assimilate both local and global image details. 
Despite the inherent advantages of convolutional blocks like parameter sharing and translation invariance, they sometimes fall short in replicating complex structures necessitating long-range information. 
This deficiency is aptly addressed by the attention mechanism intrinsic to SAGAN. 
This architectural pivot offered invaluable insights regarding the influence of model variations on the results. 
The culmination of my efforts was a model deeper than SAGAN but relatively simpler than BigGAN.

\subsubsection{Model Refinements}
In the Prior Embedding GAN~(PE-GAN) model, the depth of each block was increased for both the generator and discriminator, with each layer incorporating \num{32} multiplication channels. 
An in-depth evaluation of SAGAN prompted me to eliminate Projection Discrimination, a concept originally introduced in SNGAN~\cite{miyato_cgans_2018}. 
The primary objective of Projection Discrimination is to enhance the projection of the real image embeddings onto the corresponding target embeddings while minimizing the inner-product values for fake images.
Its limitation of solely harnessing the data-to-class relationship often led to discriminator overfitting and training collapse~\cite{kang_contragan_2021}. 
To address this, I incorporated the contrastive conditioning discrimination mechanism~\cite{kang_contragan_2021}. 
This technique is rooted in a conditional contrastive loss derived from metric learning that evaluates data-to-data relationships within each batch. 
Consequently, the discriminator can discern not just the data-to-class but also data-to-data interrelations among samples.

Given that our task is a fine-conditional image generation, I further appended embeddings of prior global sensor occupancy data along with the class labels to both the discriminator and generator.
Including prior data provides more context, enabling the model to understand and internalize the nuances of each class more effectively. 
Such enriched representations can be crucial in tasks where class boundaries are intricate, and minor variations can lead to significant differences in outcomes.
Let $\mathbf{X}=\{x_1,...,x_m\}$, where $x_i \in \mathbb{R}^{W\times H}$ be sampled training images and $\mathbf{y}=\{y_1,...,y_m\}$, where $y_i \in \mathbf{Z}$ be the corresponding labels from 1 to 40, and $\mathbf{s}=\{s_1,...,s_m\}$, where $s_i \in \mathbf{R}$ be the mean occupancies for each class of sensors. Also, a linear projection function $h^{(d)}:\mathbb{R}^k \rightarrow \mathbb{S}^d$ and an embedding function $\mathbf{e}:y_i \rightarrow \mathbb{R}^{d}$ is defined.
Then, using the linear projection of the output of the discriminator, one can define the conditional contrastive loss as, 

\[L(x_i,y_i,s_i;t)=-\log\left(\frac{\exp{(h_i^Ta_i/t)}+\sum^m_{k=1}\mathbf{1}_{k=i}.\exp{(h_i^Th_k/t)}}{\exp{(h_i^Ta_i/t)}+\sum^m_{k=1}\mathbf{1}_{k\neq i}.\exp{(h_i^Th_k/t)}}\right)
\] 

By minimizing the above-mentioned contrastive loss, the distances between the embeddings of images with matching labels are reduced while distances between differing labels are maximized. This mechanism considers both the data-to-data relations \( h^T_i h_k \) and data-to-class relations \( h^T_i a_i \) without any augmentations. As a result, the discriminator is guided not solely by the image but is also informed by fine-grained prior information, functioning as an injection of inductive bias. This incorporation of prior semantic information, rooted in the distinct mean occupancies extracted from the simulated images, substantially enhanced the model's performance concerning the mean-occupancy distribution as shown in~\Cref{fig:occupancy-intensity}. Additionally, it augmented the stability of the training process for nearly 20,000 iterations.

Lastly, the introduction of consistency regularization~\cite{zhang_consistency_2020} significantly extended our training's stability and delayed potential collapses. The fundamental idea behind this technique is to ensure that the classifier remains invariant to specific semantics-preserving transformations. 

Given an image \(\bm{x}\) and its perturbed counterpart \(\bm{x'} = T(\bm{x}\), which is generated through semantics-preserving noise such as flipping or random shifts, the discriminator's outputs \(D(\bm{x};\theta)\) and \(D(\bm{x'};\theta)\) should ideally be consistent. Formally, the consistency regularization loss \(\mathcal{L}_{\text{consistency}}\) can be expressed as:

\begin{equation*}
\mathcal{L}_{\text{consistency}} = \mathbb{E}_{\bm{x} \sim p_g} \left[ \left\| D(\bm{x};\theta) - D(T(\bm{x});\theta) \right\|_2^2 \right],
\end{equation*}

where \(T(\cdot)\) represents the semantics-preserving transformation (e.g., image flip, random shift) applied to the image, and \(p_g\) is the distribution of the generated images. In our scenario, due to the generator's tendency to produce images with artifacts such as flips and random shifts, introducing such a regularization ensured that the discriminator did not penalize these perturbations excessively. By optimizing the model with this regularization, the robustness of the training process was enhanced, leading to improved stability and longevity in training cycles.

Eventually, PE-GAN, as depicted in Table~\ref{tab:1}, using the above technologies along with orthogonal initialized~\cite{brock_neural_2017} residual blocks~\cite{brock_large_2019} for both the generator and the discriminator and Hinge loss~\cite{zhang_self-attention_2019}, succeeded as the proof of concept sensor-dependent PXD hitmap generation. 
However, as I elaborate in the next section, PE-GAN had some serious problems including incomplete marginal distributions, low diversity, and very bad performance over the downstream Physics analysis.

\begin{table}[!htb]
\centering
\begin{minipage}{.5\textwidth}
\centering
\caption*{Generator}
\begin{tabular}{c} 
 \toprule
 \toprule
  $z\in \mathbb{R}^{128} \sim \mathcal{N}(0,I)$\\ Embed(y)~$\in\mathbb{R}^{128}$\\$ch=32$\\
 \midrule
  Linear(128+128)~$\rightarrow4\times12\times16ch$\\
 ResBlock~$16ch\rightarrow16ch$\\
 ResBlock up~$16ch\rightarrow16ch$\\
 ResBlock~$16ch\rightarrow16ch$\\
 ResBlock up~$16ch\rightarrow8ch$\\
 ResBlock~$8ch\rightarrow8ch$\\
 ResBlock up~$8ch\rightarrow8ch$\\
 ResBlock~$8ch\rightarrow8ch$\\
 ResBlock up~$8ch\rightarrow4ch$\\
 Non-Local Attention Block\\
 ResBlock~$4ch\rightarrow4ch$\\
 ResBlock up~$4ch\rightarrow2ch$\\
 ResBlock~$2ch\rightarrow2ch$\\
 ResBlock up~$2ch\rightarrow ch$\\
 BN, ReLU, $3\times3$~Conv~$ch\rightarrow 1$\\
 Tanh\\
 \bottomrule
 \bottomrule
\end{tabular}
\end{minipage}%
\begin{minipage}{.5\textwidth}
\centering
\caption*{Discriminator}
\begin{tabular}{c} 
 \toprule
 \toprule
  zero-padded gray-scale image\\$x\in\mathbb{R}^{256\times768\times1}$\\
 \midrule
 ConvBlock~$1\rightarrow ch$\\
 ResBlock down~$ch\rightarrow 2ch$\\
 ResBlock~$2ch\rightarrow 2ch$\\
 ResBlock down~$2ch\rightarrow4ch$\\
 ResBlock~$4ch\rightarrow 4ch$\\
 ResBlock down~$4ch\rightarrow8ch$\\
 ResBlock~$8ch\rightarrow 8ch$\\
 ResBlock down~$8ch\rightarrow8ch$\\
 ResBlock~$8ch\rightarrow 8ch$\\
 ResBlock down~$8ch\rightarrow16ch$\\
 ResBlock~$16ch\rightarrow 16ch$\\
 ResBlock down~$16ch\rightarrow16ch$\\
 ResBlock~$16ch\rightarrow 16ch$\\
 ReLU, Global sum pooling\\
 Self-Attention~[Proj(Embed(y)~||~Linear(s))],\\ Norm(Linear512$\rightarrow1024$)\\
 Conditional Contrastive loss\\
 \bottomrule
 \bottomrule
\end{tabular}
\end{minipage}
\caption{\label{tab:1} The Architecture of the Generator and the Discriminator. ``$ch$'' denotes the number of multiplicative channels, ``ResBlock'', the residual blocks introduced in BigGAN-deep, ``BN'', the batch normalization, ``ReLU'', the rectified linear unit, ``Norm'', a normalization operator, ``Proj'' and ``Linear'', a linear projection or an MLP.}
\end{table}

\section{Intra-Event Aware GAN~(IEA-GAN)}
Although PE-GAN introduced a conditional setup for PXD background generation and provided a significant speedup to the WGAN-gp~\cite{srebre_generation_2020}, it lacks an intra-event grounding. Moreover, the use of PE-GAN's generated samples in the downstream physics analysis is still not satisfactory~(see \cref{fig:helix_param_res}). 
Thus, I embarked on another journey to introduce Intra-Event Aware GAN. This section is based on the pre-print~\cite{hashemi_ultra-high-resolution_2023}.

\subsection{Introduction and Overview}
\begin{figure}[!htb]
    \centering
    \begin{subfigure}[t]{0.85\textwidth}
    \centering
    \includegraphics[width=\textwidth]{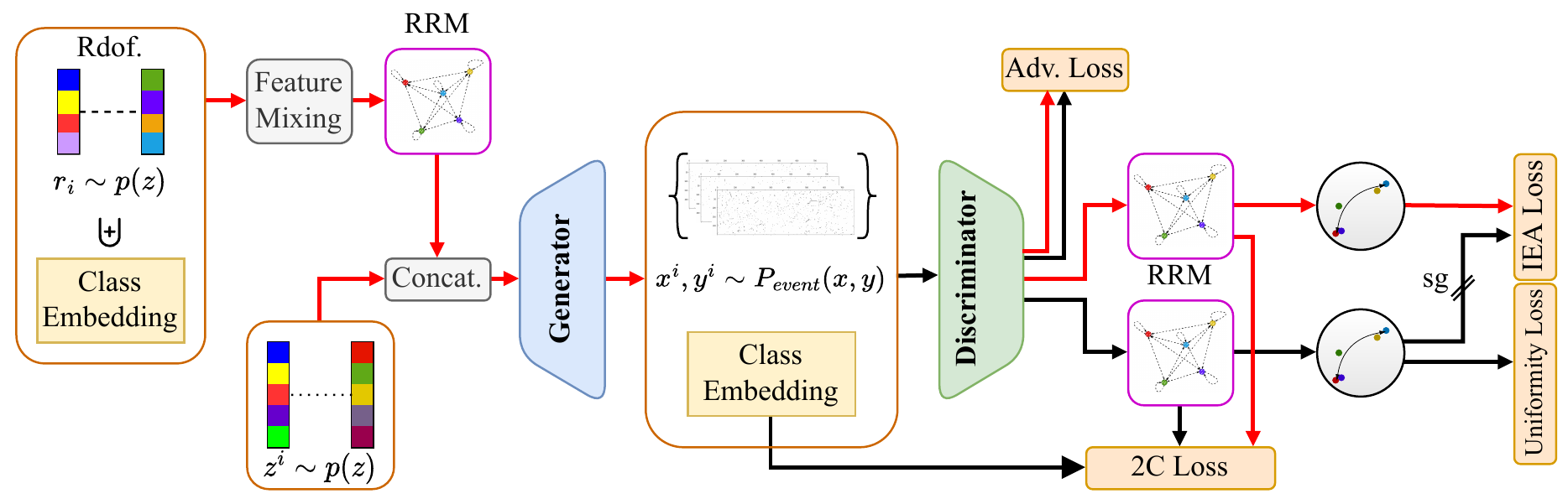}
    \caption{Rdof stands for Random degrees of freedom, which decorates the generator's sensor/layer embedding with an event-level learnable embedding responsible for the generator's intra-event correlation. The Relational Reasoning Modules~(RRM) in the generator and the discriminator do the intra-event reasoning by clustering class/image embeddings based on their contextual similarity, respectively.
    The red lines correspond to the forward and backward passes of the generator. The black lines correspond to the forward and backward passes of the discriminator. The discriminator is trained with the Adversarial Hinge loss, see~\cref{chap:3}, 2C loss, see~\cref{eq:2C_loss} and the Uniformity loss, see~\cref{eq:uniformity_loss}. On the other hand, the generator uses the Adversarial loss, 2C loss, and the IEA loss, see~\cref{eq:iea_loss}. Sg means stop-gradient for the discriminator from the IEA loss, a self-supervised dyadic-aware loss for the generator.}
    \label{fig:IEA_GAN}
    \end{subfigure}
    \hfill%
    \begin{subfigure}[t]{0.85\textwidth}
    \centering
    \includegraphics[width=0.7\textwidth]{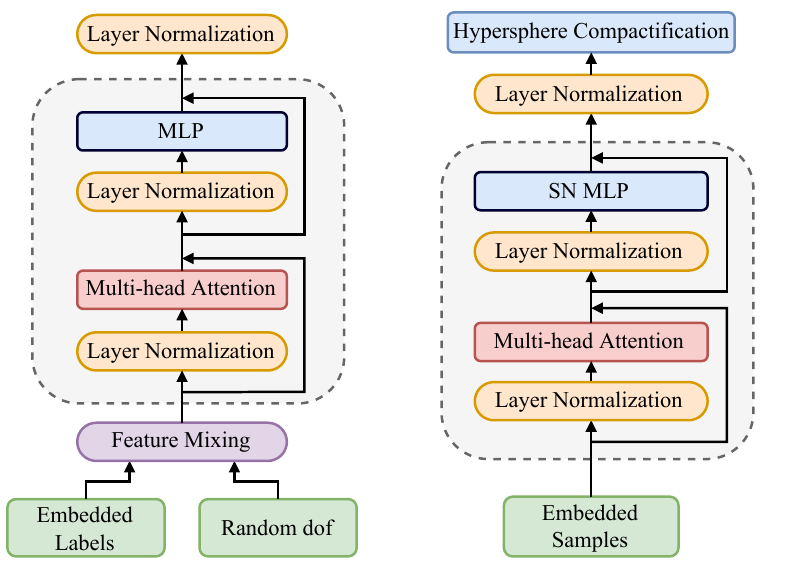}
    \caption{The Relational Reasoning Module for the generator~(left) and the discriminator~(right)}
    \label{fig:RRM_components}
    \end{subfigure}
    \caption{IEA-GAN architecture~(\textbf{a}) and Relational Reasoning Module components~(\textbf{b}).}
    \label{fig:IEAGAN_arch}
\end{figure}

IEA-GAN is a GAN-based deep generative model based on self-supervised relational reasoning. 
IEA-GAN's discriminator, $D$, takes the set of detector response images $x_i \in \mathbf{R}^d$ coming from one event and embeds them as input nodes within a fully connected event graph in a self-supervised way. An Event graph is a weighted graph where the nodes are the embedded detector images in an event and the edges are weighted by the degree of similarity between the detector images in each event~(detailed description follows in~\cref{sub:event_approx}).
It approximates the concept of an event by contextual reasoning using a permutation equivariant Relational Reasoning Module~(RRM). RRM is a novel, GAN-compatible, fully connected, multi-head Graph Transformer Network~\cite{battaglia_relational_2018,sharifzadeh_classification_2020} that groups the image tokens in an event based on their contextual similarity. For multi-modal contrastive reasoning, the discriminator also takes the sensor embedding of the detector as class tokens. In the end, it compactifies both image and class modalities information by projecting the normalized graph onto a hypersphere, as shown in~\Cref{fig:IEA_GAN}. 

To ensure that the Generator $G$ has a proper understanding of an event and captures the intra-event correlation, it first samples from a Normal distribution, $\mathcal{N}(0,1)$, at each event as random degrees of freedom~(Rdof), and decorates the sensor embeddings with this four-dimensional learnable Rdof. Then, for a self-supervised contextual embedding of each event, the RRM acts on top of this. Notably, Rdof differs from the original GAN~\cite{goodfellow_generative_2014} Gaussian latent vectors. Rdof can be considered as an event-level learnable segment embedding~\cite{devlin_bert_2019} or perturbation~\cite{zhang_word_2018} to the token embeddings, which can leverage the diversity of generated images. Combining these modules with the IEA Loss allows the Generator to gain insight and establish correlation among the samples in an event, thus improving its overall performance. 

Apart from the adversarial loss, IEA-GAN also benefits from a self-supervised and contrastive set of losses. The model understands the geometry of the PXD detector through a proxy-based contrastive 2C loss~\cite{kang_contragan_2021} where the learnable proxies are the sensor embeddings over the hypersphere. Moreover, to improve the diversity and stability of the training, I introduce a Uniformity loss for the discriminator. The Uniformity loss can encourage the discriminator to give equal weight to all regions of the hypersphere~\cite{wang_understanding_2022}, rather than just focusing on the areas where it can easily distinguish between real and fake data. Encouraging the discriminator to impose uniformity not only promotes more diverse and varied outputs but also mitigates issues such as mode collapse.

Another essential part of IEA-GAN is the IEA loss that addresses the class confusion~\cite{kang_rebooting_2021,rangwani_class_2021} problem of the conditional generative models for fine-grained datasets. In the IEA-loss the generator tries to imitate the discriminator's understanding of each event through a dyadic information transfer with a stop-gradient for the discriminator. This can improve the ability of the generator to generate more fine-grained samples in the simulation process by being aware of the variability of conditions at each event. 

In the next section, we delve deeper into each of these methods and discuss their importance.

\subsection{Relational Reasoning}
Transformers~\cite{vaswani_attention_2017} are widely used in different contexts. However, their application in Generative Adversarial Networks is either over the image manifold to learn long-range interactions between pixels~\cite{zhang_self-attention_2019,hudson_generative_2021} or via pure Vision-Transformer based GANs~\cite{jiang_transgan_2021} in which they utilize a fully Vision-Transformer~\cite{dosovitskiy_image_2021} based generator and discriminator.
Given the fact that training the Transformers is notoriously difficult~\cite{liu_understanding_2020} and task-agnostic when determining the best learning rate schedule, warm-up strategy, decay settings, and gradient clipping, fusing and adapting a Transformer encoder over a GAN learning regime is a highly non-trivial task. 
IEA-GAN successfully merges a Transformer-based module adapted to the GAN training schemes for the discriminator's image and the generator's class modalities without any of the aforementioned problems.

\subsubsection{Event Approximation}
\label{sub:event_approx}
An event, a single readout window after the collision of particles consisted of \num{40} of images each of which a sensor hitmap~(image) of size $256\times768$. Thus, each event represents a round of detector signature collection. 
In order to approximate the concept of an event, at each iteration, IEA-GAN should take an event with \num{40} sensor images. Therefore, I am conditioning the model with the sensor type $[[1,40]]$ which can be thought of as a mixture of angle and radius conditioning. These conditions have to enter the model as learnable \emph{tokens} as they are not absolute and are context-based. It is impossible to pre-define meaningful sparse connections among the sample nodes in an event. For instance, the relation between images from different sensors can vary from event to event, albeit cumulatively, they follow a particular distribution. 
Ergo, the model has to learn any dynamical inherited conditions from the data in context~(through the Relational Reasoning Module).

To model the context-based similarity between the different detector sensors in each event rather than their absolute properties, I have to use a permutation-equivariant~\cite{guttenberg_permutation-equivariant_2016,ravanbakhsh_equivariance_2017} relational block such that can encode pairwise correspondence among elements in the input set.
For instance, Max-Pooling~(e.g.~DeepSets~\cite{zaheer_deep_2018}) and Self-Attention~\cite{vaswani_attention_2017} are the common permutation equivariant modules for set-based problems.
Performing attention on all token pairs in a set to identify which pairs are the most interesting enables Transformers like Bert~\cite{devlin_bert_2019} to learn a context-specific syntax as the different heads in the multi-head attention might be looking at different syntactic properties~\cite{clark_what_2019,yun_are_2020}.

Hence, I incorporate a self-attention mechanism with weighted sum pooling as a form of information routing to process meaningful connections between elements in the input set and create an event graph. Each sample in an event is viewed as a node in a fully connected event graph, where the edges represent the learnable degree of similarity.
Samples in each event go into message propagation steps of the Relational Reasoning Module~(RRM), a GAN-compatible fully connected multi-head Graph Transformer Network~\cite{battaglia_relational_2018,sharifzadeh_classification_2020}.

\subsubsection{Relational Reasoning Module}

Specifically designed to be compatible with GAN training policies, the Relational Reasoning Module~(RRM) can capture contextualized embeddings and cluster the image or class tokens in an event based on their inherent similarity. 

Let $\mathbf{X}=\{x_1,...,x_m\}$ be the set of the sampled images in each event, where $x_i \in \mathbb{R}^{d}$, and $\mathbf{y}=\{y_1,...,y_m\}$ be the set of labels, with $y_i \in [\![1,40]\!]$ for \num{40} detector~(PXD) sensors. I also define two linear hypersphere projection diffeomorphisms, $\mathbf{h}_x:\mathbb{R}^k \rightarrow \mathbb{S}^n$ and $\mathbf{h}_y:\mathbb{Z} \rightarrow \mathbb{S}^n$, which map the image embedding manifold and the set of labels to a unit n-sphere, respectively.
The unit n-sphere is the set of points, $\mathbb{S}^n = \{s\in\mathbb{R}^{n+1}\mid \|s\|_2=1 \}$, that is always convex and connected.
The Relational Reasoning Module benefits from a variant of the Pre-Norm Transformer~\cite{vaswani_attention_2017} with a dot-product Multi-head Attention block such that

\begin{align}
\mathbf{p}'^{(l)}_i &= \mathbf{p}^{(l)}_i +\sum_{k=1}^h\sum_{j=1}^{m} a_{ij}^{(l,k)}\mathbf{W}_{\mathrm{SN}}^{(l)}\mathbf{LN}(\mathbf{p}_j^{(l)}) ~, 
\label{eq:dis_1}
\\
\mathbf{p}^{(L)}_i &= \mathbf{h}_x^{\mathrm{LN}}\left(\mathbf{LN}(\mathbf{\bigcirc}_{l=0}^{L}(\mathbf{p}'^{(l)}_i + \mathcal{F}_{\mathrm{SN}}[\mathbf{LN}(\mathbf{p}'^{(l)}_i)]))\right) ~,
\label{eq:dis_2}
\end{align}

where $\mathbf{p}'^{(l)}_i\in \mathbb{R}^{k}$ is the embedding of each image via the discriminator for layer $l$ of the RRM.
$\mathbf{LN}$ is the Layer Norm function~\cite{noauthor_160706450_nodate} and $h$ is the number of heads defined in \Cref{eq:attention_multi_head}.
$\mathcal{F}[~.~]$ is a two layer MLP functional defined as $\mathcal{F}_{\mathrm{SN}}[\mathbf{p}_i^{(l)}] = \textrm{ReLU}(p_i^{(l)}\mathbf{W}_{\mathrm{SN}}^{(l,1)})\mathbf{W}_{\mathrm{SN}}^{(l,2)}$ with Spectral Normalization~\cite{miyato_spectral_2018}.
The logits $a_{ij}^{(l,k)}$ are the normalized Attention weights of the bilinear function that monitor the dyadic interaction between image embeddings in layer $l$ and head $k$ defined in \Cref{eq:attention_map}. $\mathbf{W}_{\mathrm{SN}}^{(l)}$ in~\Cref{eq:dis_1} is the learnable multi-head projector at layer $l$ defined in \Cref{eq:attention_multi_head} with Spectral Normalization.
The output of the composition of all layers via the composition of $L$ functionals, $\bigcirc_{l=0}^{L}\Phi^l:= \phi_{w_L}\circ ... \circ \phi_{w_0}[\mathbf{p}_i^{(l=0)}]\in \mathbb{R}^{m\times k}$, goes into a Layer Normalization layer where $\Phi^l = \mathbf{p}'^{(l)}_i + \mathcal{F}[\mathbf{LN}(\mathbf{p}'^{(l)}_i)]$.
$\mathbf{h}_x^{\mathrm{LN}}(~.~)$ in~\Cref{eq:dis_2} is the hypersphere compactification while the vectors are being standardized over the unit n-sphere $\mathbb{S}^n$ by a Layer Normalization.

For the discriminator, this module takes the set of image embeddings as input nodes within a fully connected event graph applies a dot-product self-attention over them, and then updates each sample or node's embedding via the attentive message passing, as shown on the right of~\Cref{fig:RRM_components}.
In the end, it compactifies the information by projecting the normalized graph onto a hypersphere, as shown in~\Cref{fig:IEA_GAN}.
Embedding the samples in an event on the unit hypersphere provides several benefits.
In modern machine learning tasks such as face verification and face recognition~\cite{wang_normface_2017}, when dot products are omnipresent, fixed-norm vectors are known to increase training stability.
In our case, this avoids gradient explosion in the discriminator.
Furthermore, as $S^n$ is homeomorphic to the 1-point compactification of $\mathbb{R}^n$ when classes are densely grouped on the n-sphere as a compact convex manifold, they are linearly separable, which is not the case for the Euclidean space~\cite{noauthor_170301203_nodate}. 

For the generator's RRM, I use a simpler version of the above dot-product Multi-head Attention block without the last hypersphere compactification due to the stability issues., as shown on the left of \Cref{fig:RRM_components}.
It finds a learnable contextual embedding for each event that will be fused to each class token via the feature mixing layer, which is a matrix factorization linear layer $\mathbf{W}_{\mathrm{SN}}(.)$. Formally we have,

\begin{align}
\mathbf{q}^{(0)}_i &= \mathbf{W}_{\mathrm{SN}}(\mathbf{r}_i\uplus\mathbf{e}_i) ~,
\label{eq:feat_mix_1}
\\
\mathbf{q}'^{(l)}_i &= \mathbf{q}^{(l)}_i+\sum_{k=1}^M\sum_{j=1}^{m} a_{ij}^{(l,k)}\mathbf{W}^{(l)}\mathbf{LN}(\mathbf{q}_j^{(l)}) ~,
\label{eq:feat_mix_2}
\\
\mathbf{q}^{L}_i &=\mathbf{LN}\left(\mathbf{\bigcirc}_{l=0}^{L}( \mathbf{q}'^{(l)}_i + \mathcal{F}[\mathbf{LN}(\mathbf{q}'^{(l)}_i)])\right) ~,
\label{eq:feat_mix_3}
\end{align}

where $\mathbf{e}_i:\mathbb{Z} \rightarrow \mathbb{R}^t$ is the embedding of each class token via the embedding layer of the generator. The logits $a_{ij}^{(l,k)}$ are the normalized Attention weights of the bilinear function that monitor the dyadic interaction between classes in the event embeddings in layer $l$ and head $k$ defined in~\Cref{eq:attention_map}. $\mathbf{W}^{(l)}$ in~\Cref{eq:feat_mix_2} is the learnable multi-head projector at layer $l$ defined in \Cref{eq:attention_multi_head}. The output of the composition of all layers via the composition of $L$ functionals, $\bigcirc_{l=0}^{L}\Phi^l:= \phi_{w_L}\circ ... \circ \phi_{w_0}[\mathbf{q}_i^{(l=0)}]\in \mathbb{R}^{m\times t}$, goes into a Layer Normalization layer where $\Phi^l = \mathbf{q}'^{(l)}_i + \mathcal{F}[\mathbf{LN}(\mathbf{q}'^{(l)}_i)]$ as shown in~\Cref{eq:feat_mix_3}.

One input to the generator is the embedded labels, which can be considered rigid token embeddings that will be learned as a global representation bias of each sensor. As sensor conditions change for each event as a set, having merely class embeddings, as used in conditional GANs~\cite{mirza_conditional_2014}, is insufficient because the context-based information will not be learned.
Thus, the generator samples from a per-event shared distribution at each event as random degrees of freedom~(Rdof).
Rdofs are random samples from a shared Normal distribution for each class, $\mathbf{r}_i \sim \mathcal{N}(0,1)$, that introduces four-dimensional learnable degrees of freedom for the generator, see~\cref{eq:feat_mix_1} 
This way, I ensure that the generator is aware of intra-event local changes, culminating in having an intra-event correlation among the generated images. Rdof can be interpreted as both perturbation~\cite{zhang_word_2018} to the token embeddings and an event-level segment embedding~\cite{devlin_bert_2019}, which can enhance the diversity of the generated images.

\subsection{Intra-Event Aware Loss}
 
\begin{figure}[!htb]
     \centering
     \includegraphics[width=\textwidth,clip]{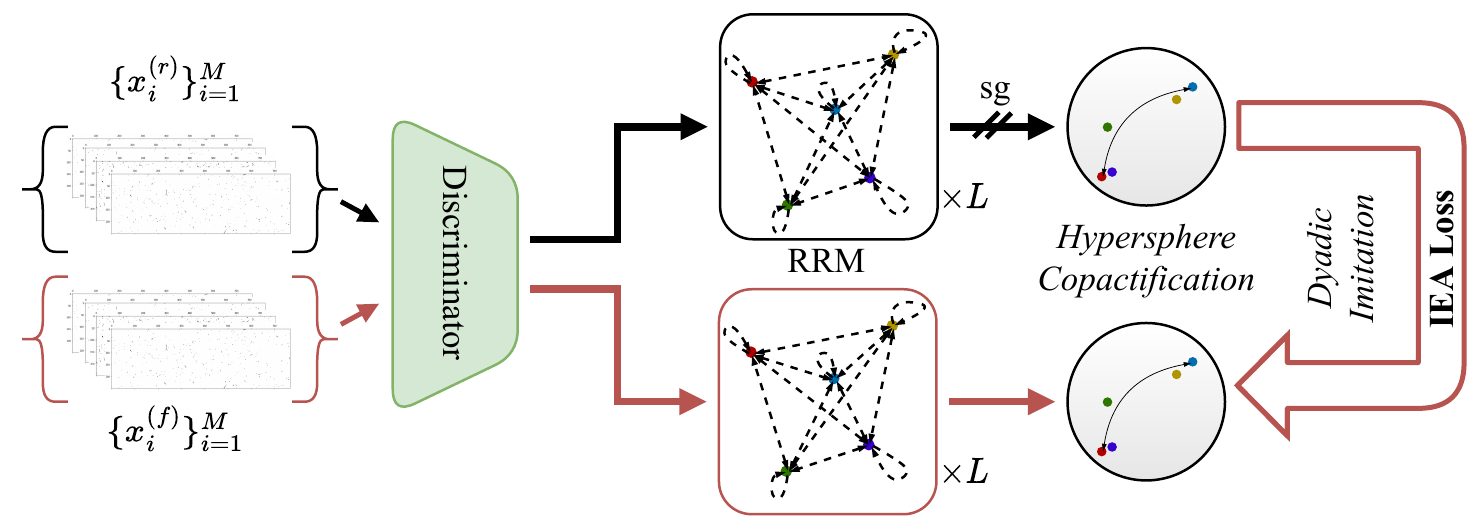}
     \caption{IEA-loss imposes a pair-wise fine-grained class-to-class imitation force for the generator. sg indicates that for the discriminator gradients are stopped and only the generator's gradients will be updated.}
     \label{fig:IEA_loss}
 \end{figure}

Motivated by Self-Supervised Learning~\cite{weng_self-supervised_2019,balestriero_cookbook_2023}, to transfer the intra-event contextualized knowledge of the discriminator to the generator in an explicit way, I introduce an Intra-Event Aware~(IEA) loss, depicted in~\cref{fig:IEA_loss}, for the generator that captures class-to-class relations,

\begin{equation}
\ell_\mathrm{IEA}(\mathbf{x}_r,\mathbf{x}_f) = \sum_{i,j} D_\mathrm{KL} \left( \sigma \left( \mathbf{h}(x^{(r)}_i)^{\top}\mathbf{h}(x^{(r)}_j) \right) \Big\Vert \sigma \left(\mathbf{h}(x^{(f)}_i)^{\top}\mathbf{h}(x^{(f)}_j)\right)\right),
\label{eq:iea_loss}
\end{equation}

where $\mathbf{x}_r=\{x^{(r)}_i\}_{i=1}^m$ is the set of real images, and $\mathbf{x}_f=\{G(z^i,y^i,r^i)=x^{(f)}_i\}_{i=1}^m$ the set of generated images. The softmax function, $\sigma:\mathbb{R}^m\rightarrow [0,1]^m$, normalizes the dot-product self-attention between the image embeddings.
The map $\mathbf{h}:\mathbb{R}^k \rightarrow \mathbb{S}^n$ is the unit hypersphere projection of the discriminator.
Therefore, the dot product is equivalent to the cosine distance.
$D_{\mathrm{KL}}(.\vert \vert.)$ is the Kullback-Leibler~(KL) divergence~\cite{noauthor_information_nodate} which takes two $m\times m$ matrices that have values in the closed unit interval (due to the softmax function).
Hence, having a KL divergence is natural here as one wants to compare one probability density with another in an event.
I also tested other distance functions reported in \Cref{sec:ablation_studies}.
By considering the linear interaction~\cite{cao_coupling_2015} between every sample in an event and assigning a weight to their similarity, the generator mimics the fine-grained class-to-class relations within each event and incorporates this information in its RRM module as shown in \Cref{fig:IEA_GAN}.

The KL divergence, a fundamental concept in information theory, has diverse interpretations that enrich our understanding of IEA loss. 
Primarily, the KL divergence is a measure of how much the generator's intra-event understanding \(Q\) deviates from the true distribution \(P\) of sensor-by-sensor relationships, or in simpler terms, how much \(P\) and \(Q\) differ in a context where \(P\) is the true distribution. 
This lack of symmetry is evident in several ways. For instance, it represents the expected ``surprise'' when observing data with distribution \(P\), assuming falsely that the intra-event distribution is \(Q\). 
In hypothesis testing, it corresponds to the expected evidence for \(P\) over \(Q\) when \(P\) is true. 
In relation to Maximum Likelihood Estimators~(MLEs), when \(P\) is an empirical distribution of data, \(D_{KL}(P||Q)\) is minimized when the generated intra-event distribution \(Q\) is the MLE for \(P\). 
Moreover, from the information theory perspective, it denotes the inefficiency or extra bits used when compressing a data source with distribution \(P\) using a code optimized for \(Q\). Lastly, in the domain of Game Theory, it can indicate potential winnings when one understands the true distribution while another party~(the generator) operates under false assumptions. 

Upon minimizing it for the generator~(having the stop-gradient for the discriminator), it is putting a discriminator-supervised penalizing system over the intra-event awareness of the generator by encouraging it to look for more detailed dyadic connections among the images and be sensitive to even slight differences.
Ultimately, it will maximize the consensus of data points on two unit hyperspheres of real images and generated image embeddings. 

\subsection{Uniformity Loss}
\label{sec:uniformity}

The other crucial loss function comes from contrastive representation learning.
With the task of learning fine-grained class-to-class relations among the images, I also want to ensure the feature vectors have as much hyperspherical diversity as possible.
Thus, by imposing a uniformity condition over the feature vectors on the unit hypersphere, they preserve as much information as possible since the uniform distribution carries a high entropy.
This idea stems from the Thomson problem~\cite{noauthor_xxiv_nodate}, where a static equilibrium with minimal potential energy is sought to distribute N electrons on the unit sphere in the evenest manner. It encourages hyperspherical diversity by pursuing the following learning objective for $n$ $d$-dimensional vectors $\bm{H}_n=\{\bm{h}_1,\cdots,\bm{h}_n\in\mathbb{R}^{d}\}$,

\begin{equation}
\label{energy}
\footnotesize
    \min_{\{\bm{h}_1,\cdots,\bm{h}_n\in\mathbb{S}^{d-1}\}}\big{\{} E_s(\bm{H}_n):=\sum_{i=1}^{n}\sum_{j=1,j\neq i}^{n}
    K_s(\bm{h}_i,\bm{h}_j)\big{\}}
\end{equation}

where $h_i \equiv h(x_i)$, is the $i$-th vector projected onto the unit hypersphere $\mathbb{S}^{d-1}=\{\bm{h}\in\mathbb{R}^{d}|\norm{\bm{h}}=1\}$. $K_s(\cdot,\cdot)$ models the interaction between two vectors, typically following Riesz $s$-kernel function

\begin{equation}
\label{eq:measure}
\footnotesize
   K_s(\bm{h}_i,\bm{h}_j) =\left\{
{\begin{array}{*{20}{l}}
{\rho(\bm{h}_i,\bm{h}}_j)^{-s},\ \ \ s>0\\
{\log(\rho(\bm{h}}_i,\hat{\bm{h}}_j)^{-1}),\ \ \ s=0\\
{-\rho(\bm{h}}_i,\bm{h}_j)^{-s},\ \ \ s<0
\end{array}} \right.
\end{equation}

where $\rho(\cdot,\cdot)$ is defined to measure the geodesic similarity on the unit hypersphere. In general, I can use either $\rho(\bm{h}_i,\bm{h}_j)=\|\bm{h}_i-\bm{h}_j\|_2$~(the standard Riesz $s$-kernel~\cite{liu_learning_2020,wang_understanding_2022}) or hyperspherical geodesic distance~(angular distance) $\rho(\bm{h}_i,\bm{h}_j)=\arccos(\bm{h}_i^\top\bm{h}_j)$. Minimizing this pairwise energy sum asymptotically corresponds to the uniform distribution on the hypersphere~\cite{noauthor_asymptotics_nodate}. To do this, I use the Riesz $s$-kernel as an auxiliary loss function,

\begin{equation}
\Lb_{\mathrm{uniform}}(x;s) = \log \mathbb{E}_{x_i,x_j\sim p_{\mathrm{event}}} [\exp(s\|\mathbf{h}(x_i)-\mathbf{h}(x_j)\|_2^2)].
\label{eq:uniformity_loss}
\end{equation}

Upon minimizing this loss for the discriminator, it tries to maintain a uniform distance among the samples that are not well-clustered and thus not similar.
In other words, eventually, we want to reach a maximum geodesic separation incorporating the Riesz s-kernel with $s=-2$ as a measure of geodesic similarity, to preserve maximal information over the Hypersphere. 

\begin{figure}[!htb]
     \centering
     \includegraphics[width=0.2\textwidth,clip]{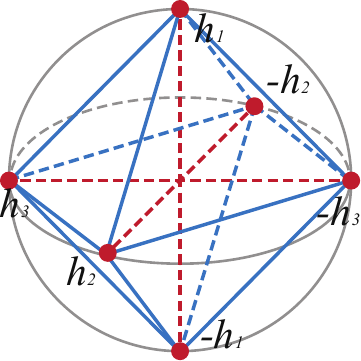}
     \caption{Uniformity over a 3-dimensional cross-polytope immersed in a unit hypersphere}
     \label{fig:crossp}
 \end{figure}
 
By ensuring uniform distribution on the hypersphere for \(2d+2\) vectors in \(\mathbb{S}^d\), one obtains a cross-polytope as described in~\cite{noauthor_minimum_nodate}. The illustration in Fig.~\ref{fig:crossp} showcases a 3-dimensional cross-polytope within \(\mathbb{S}^2\) (consisting of 6 vectors). Supposing one has a set of unit-vectors with \(d+1\) vectors: \(\{\bm{h}_1,\ldots,\bm{h}_{d+1}\in\mathbb{S}^{d}\}\). By incorporating vectors in the opposite direction to this set, I generate a new collection comprising \(2d+2\) vectors: \(\{\bm{h}_1,\ldots,\bm{h}_{d+1},-\bm{h}_1,\ldots,-\bm{h}_{d+1}\in\mathbb{S}^{d}\}\). Aiming for uniformity on the hypersphere for vectors in this expanded set translates to pursuing orthogonality between the initial \(d+1\) vectors. 
This observation establishes a profound link between hyperspherical uniformity and vector orthogonality.

Inspired by~\cite{liu_learning_2021}, in order to examine the hyperspherical uniformity from a statistical test perspective uniformity, one can use the Sobolev test~\cite{beran_testing_1968,noauthor_new_nodate,jammalamadaka_sobolev_2020}.
Using independent, identically distributed samples \(\bm{h}_1,...,\bm{h}_n\in\mathbb{S}^{d-1}\) of a unit random vector \(\bm{h}\), one can define the challenge of verifying uniformity on the hypersphere as the examination of the hypothesis \(\mathcal{H}_0:\bm{P}=\textnormal{Uniform}(\mathbb{S}^{d-1})\) against its alternative \(\mathcal{H}_1:\bm{P}\neq\textnormal{Uniform}(\mathbb{S}^{d-1})\), with \(\bm{P}\) signifying \(\bm{h}\)'s probability distribution. 
The underlying principle behind the Sobolev test is the transformation of the hypersphere \(\mathbb{S}^{d-1}\) into the Hilbert space \(L^2(\mathbb{S}^{d-1},\mu)\) of square-integrable functions with finite integral on \(\mathbb{S}^{d-1}\). 
This is achieved via a function \(t:\mathbb{S}^{d-1}\rightarrow L^2(\mathbb{S}^{d-1},\mu)\) ensuring that when \(\bm{h}\) follows a uniform distribution on the hypersphere, the expectation value of \(t(\bm{h})\) is zero. Denoting \(\bm{\epsilon}_k\)~($p_{d,k} = \text{dim}\bm{\epsilon}_k$), being associated with the \(k\)-th distinct eigenvalue of the Laplacian, the space of eigenfunctions from \(\mathbb{S}^{d-1}\) to \(\mathbb{R}\). 
A mapping \(t_k:\mathbb{S}^{d-1}\rightarrow\bm{\epsilon}_k\) can be expressed as \(t_k(\bm{h})=\sum_{i=1}^{p_{d,k}}g_{i,k}(\bm{h})g_{i,k}\) where \(\{g_{i,k}\}_{i=1}^{p_{d,k}}\) creates an orthogonal basis of \(\bm{\epsilon}_k\). With \(\{v_k\}_{k=1}^\infty\) being a series satisfying \(\sum_{k=1}^\infty v_k^2 p_{d,k}<\infty\), the function defined by \(\bm{h}\rightarrow t(\bm{h}):=\sum_{k=1}^\infty\bm{v}_k t_k(\bm{h})\) is a transformation from \(\mathbb{S}^{d-1}\) into the Hilbert space \(L^2(\mathbb{S}^{d-1},\mu)\) of square integrable real functions real functions over \(\mathbb{S}^{d-1}\) with respect to the uniform measure \(\mu\). 
The Sobolev test, refutes \(\mathcal{H}_0\) for large values of the following test statistic,
\begin{equation}
    S_n:=\frac{1}{n}\sum_{i,j}^n\sum_{k=1}^\infty v_k^2\langle t_k(\bm{h}_i),t_k(\bm{h}_j)\rangle
\end{equation}
where $\langle f,g \rangle:=\int_{\mathbb{S}^{d-1}}f(\bm{h})g(\bm{h})\textnormal{d}\mu(\bm{h})$ denotes the inner product on $L^2(\mathbb{S}^{d-1},\mu)$.
For even values of \(k\), let \(v_k=0\) and for odd \(k\), set \(v_k=(\pi k)^{-1}\). Then a variant of the Sobolev test, the Ajne test~\cite{ajne_simple_1968,garcia-portugues_overview_2018}, employs the following simplified test statistic.

\begin{equation}
\label{ajne}
\footnotesize
    A_n=\frac{n}{4}-\frac{1}{n\pi}\sum_{1\leq i <j\leq n}\arccos(\bm{h}_i^\top\bm{h}_j)
\end{equation}

This test rejects \(\mathcal{H}_0\) when the $A_n$ values are substantially large. Its relevance to Eq.\ref{eq:uniformity_loss} is evident as both of them derive from pairwise relationships. 
Specifically, minimizing the this equation in relation to \(\{\bm{h}_i\}_{i=1}^n\) corresponds directly to Eq.\ref{eq:measure} with a Riesz \(s\)-kernel, where \(s=-2\) which is our Uniformity loss~\cref{eq:uniformity_loss}.

This loss is beneficial for capturing the exact distribution of the mean occupancy distribution and balancing the inter-class pulling force of the Relational Reasoning module. As a result, not only does it help generate more diverse and varied outputs, but it also can prevent issues such as mode collapse or overfitting.

\subsection{Model Details and Hyperparameters}
\label{sec:dataset_and_model_details}

For training and evaluation, \num{40000} and \num{10000} Monte Carlo simulated~\cite{kuhr_computing_2011} events are used respectively. The data in each event consists of \num{40} grey-scale $256\times 768$ zero-padded images. They are zero-padded on both sides from their original size of $250 \times 768$ to be divisible by \num{16} for training purposes.

To capture the intra-event mutual information among the images using the RRM and approximate the concept of an event, I have to sample properly at each iteration.
Hence, a one-sample per class sampler is used in data loading.
Using this sampler, one can ensure that in each event, one has \num{40} unique classes of images from all \num{40} sensors that belong to the same event in the Monte-Carlo simulation.
All hyperparameters are chosen based on the model's stability and performance upon the evaluation set.
The learning rates for the Generator and Discriminator are $5\times10^{-5}$.
The Relational Reasoning Module of the Generator has two heads and one layer of non-spectrally normalized message propagation with an embedding dimension of \num{128} and ReLU non-linearity.
The input to the generator's RRM is embedded class tokens mixed with \num{4} random degrees of freedom by a spectrally normalized linear layer. 

For the Discriminator, the RRM has four heads with one layer of spectrally normalized message propagation with the embedding dimension \num{1024} as the hypersphere dimension and ReLU non-linearity.
All Generator and Discriminator modules use Orthogonal initialization~\cite{hu_provable_2020}.
For the IEA-loss in \Cref{eq:iea_loss}, the coefficient $\lambda_{\mathrm{IEA}} = 1.0$, defined in \Cref{alg:iea} gives the best result.
The most stable contribution of the Uniformity loss, defined in \Cref{eq:uniformity_loss} and \Cref{alg:iea}, is with $\lambda_{\mathrm{uniform}}=0.1$.
For the backbone of both the discriminator and the generator, BigGAN-deep~\cite{brock_large_2019} with a non-local block at channel \num{32} for the discriminator only is used.
Since in GAN training, there is no meaningful way to define a minimal loss, my stopping point is the divergence of the Frechet Inception Score (FID)~\cite{heusel_gans_2018}, which is significantly correlated with the quality of other metrics.

\begin{algorithm}[!htb]
\caption{Intra-Event Aware GAN}
\label{alg:iea}
\begin{algorithmic}[1]
\Require{generator and discriminator parameters $\theta_G$, $\theta_D$, Intra-Event-aware coefficient $\lambda_{\mathrm{IEA}}$, Uniformity coefficient $\lambda_{\mathrm{uniform}}$ and hyperparameter $s$, Adam~\cite{kingma_adam_2017} hyperparameters $\alpha$, $\beta_1$, $\beta_2$, event size $M$, number of discriminator iteration steps per generator iteration $N_D$}
\Statex
\For{number of training iterations}
    \For{$t=1,...,N_D$}
        \State sample $\{z^{i}\}^M_{i=1}\sim p(z)$ 
        \State $\{x^{i},y^i\}^M_{i=1}\sim p_{\mathrm{event}}(x, y)$, $\{r^i\}^M_{i=1} \sim p_{\mathrm{Rdof}}(z)$ \Comment{\scriptsize Event Sampling}
        \For{$i=1,...,M$}
            \State $\ell_{D_{\mathrm{hinge}}}^{(i)}\leftarrow \ell_{D_{\mathrm{hinge}}}(x^{(i)};G(z^{i}, y^{i}, r^i))$
        \EndFor
        \State $\mathcal{L}_{D_{\mathrm{hinge}}}\leftarrow \frac{1}{M}\sum_{i=1}^M\ell_{D_{\mathrm{hinge}}}^{(i)}$
        \State $\mathcal{L}_{\mathrm{uniform}}\leftarrow \mathcal{L}_{\mathrm{uniform}}(x;s)$ \Comment{\scriptsize The Uniformity Loss}
        \State $\mathcal{L}_{2C}^{\mathrm{real}}\leftarrow \frac{1}{M}\sum_{i=1}^M\ell_{2C}(x^i,y^i)$
        \State $\theta_D\leftarrow \text{Adam}(\mathcal{L}_{D_{\mathrm{hinge}}}+\lambda_{2C}\mathcal{L}_{2C}^{\mathrm{real}}+\lambda_{\mathrm{uniform}} \mathcal{L}_{\mathrm{uniform}}, \alpha, \beta_1, \beta_2)$
    \EndFor
    \State sample $\{z^{i}\}^M_{i=1}\sim p(z)$ 
    \State sample $\{r^i\}^M_{i=1} \sim p_{\mathrm{Rdof}}(z)$ \Comment{\scriptsize Event Sampling}
    \For{$i=1,...,M$}
        \State $\ell_{G_{\mathrm{hinge}}}^{(i)}\leftarrow \ell_{G_{\mathrm{hinge}}}(G(z^{i}, y^{i}, r^{i}))$
    \EndFor
    \State $\mathcal{L}_{G_{\mathrm{hinge}}}\leftarrow \frac{1}{M}\sum_{i=1}^M\ell_{G_{\mathrm{hinge}}}^{(i)}$
    \State $\mathcal{L}_{\mathrm{IEA}}\leftarrow \frac{1}{M}\sum_{i=1}^M\ell_{\mathrm{IEA}}(G(z^i,y^i, r^i),x^{i})$ \Comment{\scriptsize The Intra-Event Aware Loss}
    \State $\mathcal{L}_{2C}^{\mathrm{fake}}\leftarrow \frac{1}{M}\sum_{i=1}^M\ell_{2C}(G(z^i,y^i,r^i),y^i)$
    \State $\theta_G\leftarrow \text{Adam}(\mathcal{L}_{G_{\mathrm{hinge}}}+\lambda_{2C}\mathcal{L}_{2C}^{\mathrm{fake}}+\lambda_{\mathrm{IEA}}\mathcal{L}_{\mathrm{IEA}},\alpha, \beta_1, \beta_2)$
\EndFor
\end{algorithmic}
\end{algorithm}

\section{Evaluation Results: IEA-GAN vs SOTAs}
This study showcases a series of analyses and evaluations that demonstrate the performance of IEA-GAN in generating ultra-high-granularity detector responses of the Pixel Vertex Detector, consisting of over 7.5 million pixel channels. 
Furthermore, for the first time, this finding reveals that the FID~\cite{heusel_gans_2018} and KID~\cite{binkowski_demystifying_2021} metrics for detector simulation is a very versatile estimator in conjunction with the marginal distributions~(mentioned in~\cref{chap:2}), and is associated with the other image level metrics. 
I show that by using IEA-GAN, one is able to capture the underlying distributions such that one can generate and amplify detector response information with a very good agreement with the Geant4 distributions. I also found out that the SOTA models in high-resolution image generation even with an in-depth hyperparameter tuning analysis do not perform well in comparison.

For evaluation, I have two categories of metrics: image level and physics level. As I am interested in having the best pixel-level properties, diversity, and correlation of the generated images simultaneously while adhering to minimal generator complexity due to computational limitations, choosing the best iteration to compare results is challenging.
Hence, I chose the models' weights with the best FID for all comparisons. 

This study compares IEA-GAN with three other models and the reference, which is the Geant4-simulated~\cite{agostinelli_geant4simulation_2003} dataset. 
The baselines are the SOTA in conditional image generation: BigGAN-deep~\cite{brock_large_2019} and ContraGAN~\cite{kang_contragan_2021}.
I also compare IEA-GAN with the previous works on the PXD image generation task: PE-GAN~\cite{hashemi_pixel_2021} and WGAN-gp~\cite{srebre_generation_2020}\footnote{Only for FID.}. 

\subsection{Neural Network-based Metrics: FID and KID}
To compute the FID and KID scores, based on the recent Clean-FID project~\cite{parmar_aliased_2022}, I entirely fine-tuned the Inception-V3~\cite{szegedy_rethinking_2015} model on the PXD images, as the PXD images are very different from the natural images used in their initial training. The downstream task for the fine-tuning was multi-class classification, involving \num{40} different sensors with which it acquired the ability to discriminate sensors and their corresponding data manifold. The former FID implementations use the fixed-width bilinear interpolation which is independent of the resizing ratio. 
In contrast, the Clean-FID follows standard signal processing principles and adaptively stretches the filter to prevent aliasing~\cite{parmar_aliased_2022}. 
This process can be done for any other detector dataset. 
FID measures the similarity of the generated samples’ representations to those of samples from the real distribution. 
Given large sampling statistics, for each hidden activation of the Inception model, the FID evaluates the Fréchet distance, also known as Wasserstein-2 distance, between the first two moments of the activation distributions. As demonstrated to be useful and practical in the natural image analysis domain~\cite{noauthor_pdf_nodate}, FID performs  well in terms of discriminability, diversity, and robustness, despite only modeling the first two moments of the distributions in the feature space. The lower the FID score, the more similar the distributions of the real and generated samples are. Kernel Inception Distance~(KID) is another metric similar to FID, used for evaluating the quality of generative models. Unlike FID, KID uses a kernel two-sample test, which provides an unbiased estimate of the distance between distributions and is more robust to small sample sizes.

\Cref{tab:fid} demonstrates that generated images by IEA-GAN have the lowest FID and KID score compared to the other models and outperform them by \num{42}\%. 
This indicates that IEA-GAN is able to generate synthetic samples that are much closer to the target data than the samples generated by the other models.

\begin{table}[!htb]
\begin{minipage}{\textwidth}
    \begin{center}
    \caption{
    FID and KID comparison between models~(all models in the benchmark are highly tuned to the current problem and dataset), averaged across six random seeds.
    The lower the FID and KID the better the image quality and diversity.
    }
    \label{tab:fid}
    \setlength{\tabcolsep}{4pt} 
    \begin{tabular}{@{}l|lllll@{}}
        \toprule
        & WGAN-gp & BigGAN-deep & ContraGAN & PE-GAN & IEA-GAN \\ 
        \midrule
        \textbf{FID}  & $12.09$ & $4.40\pm 0.88$ & $3.14\pm 0.74$ & $2.61\pm 0.91$ & $\mathbf{1.50\pm 0.16}$ \\
        \midrule
        \textbf{KID} & $0.0096$ & $0.0031\pm 0.0001$ & $0.0015\pm 0.0002$& $0.0021\pm 0.0004$ & $\mathbf{0.0010\pm 0.0002}$\\
        \bottomrule
    \end{tabular}
    \end{center}
\end{minipage}
\end{table}

\begin{figure}[!htb]
    \centering
    \begin{subfigure}{0.7\textwidth}
        \centering
        \includegraphics[width=\textwidth]{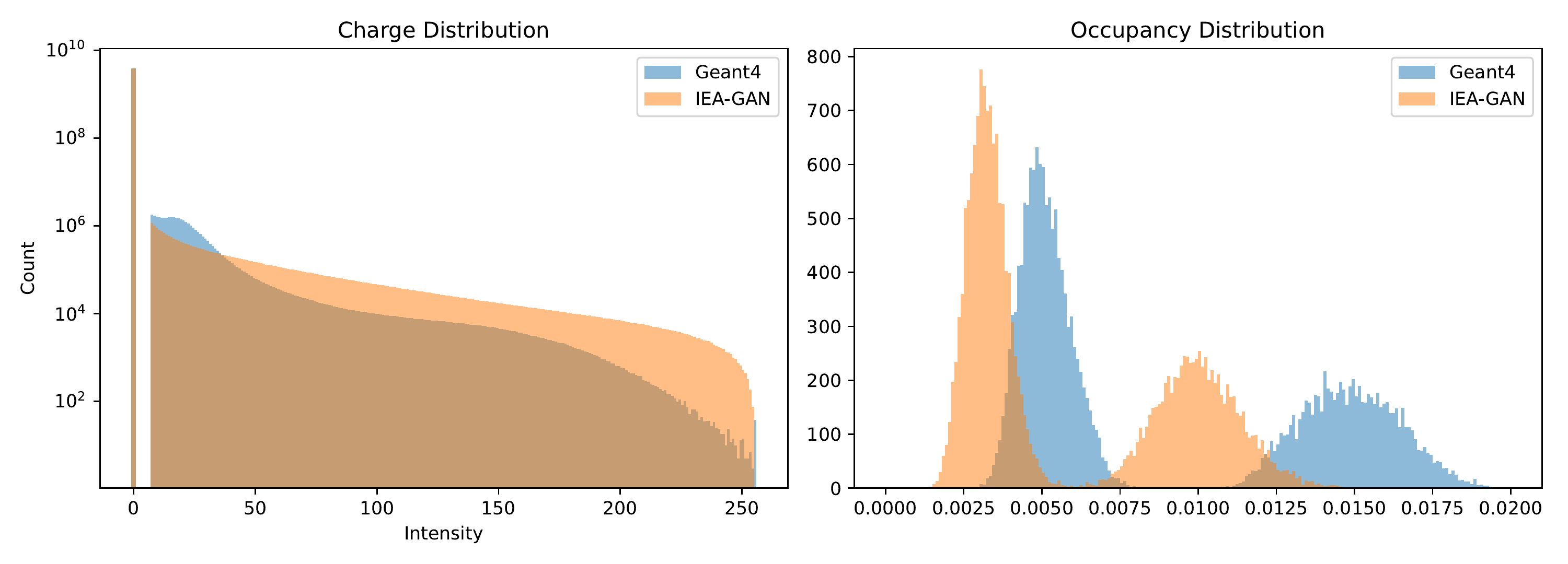}
        \caption{FID 54.43}
    \end{subfigure}
    
    \begin{subfigure}{0.7\textwidth}
        \centering
        \includegraphics[width=\textwidth]{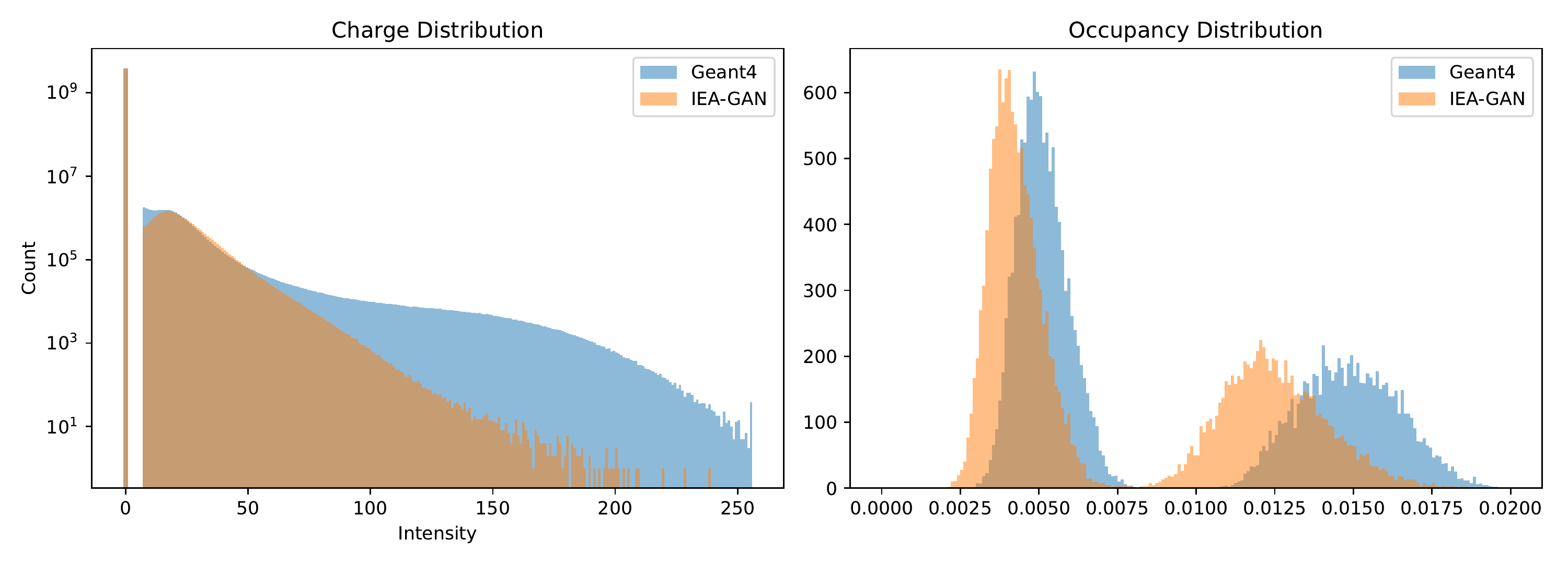}
        \caption{FID 7.81}
    \end{subfigure}
    
    \begin{subfigure}{0.7\textwidth}
        \centering
        \includegraphics[width=\textwidth]{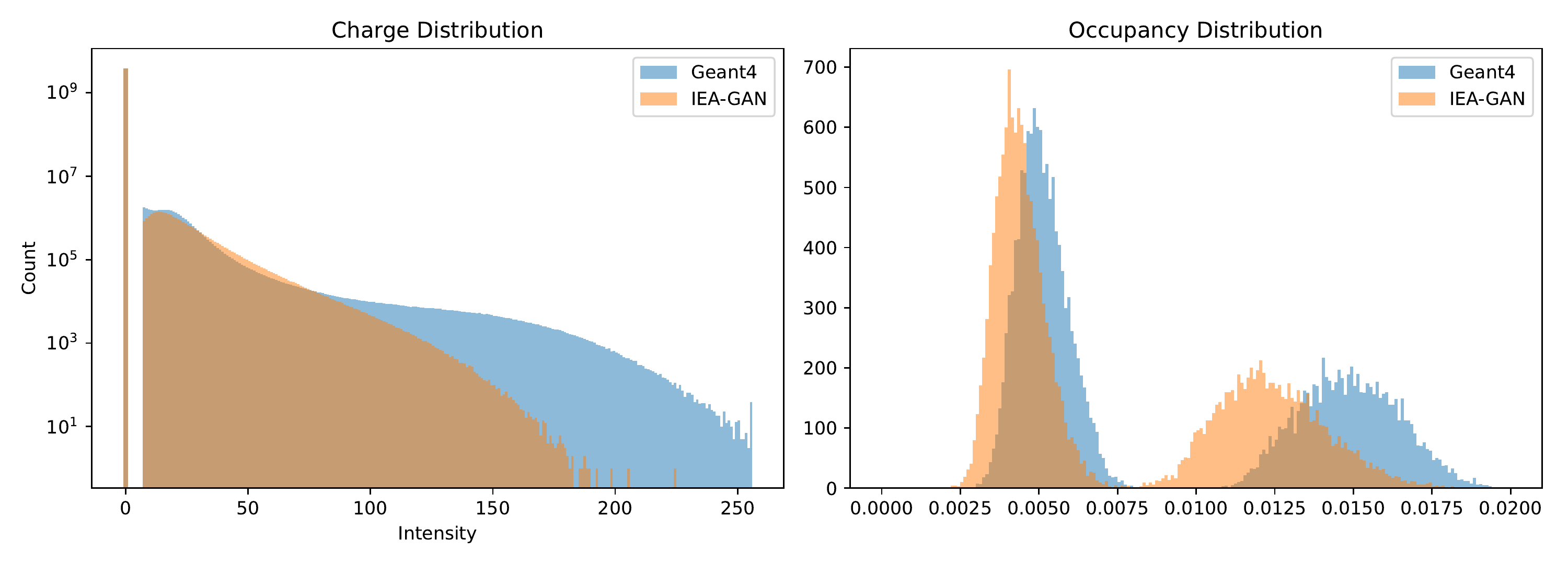}
        \caption{FID 7.51}
    \end{subfigure}
    
    \begin{subfigure}{0.7\textwidth}
        \centering
        \includegraphics[width=\textwidth]{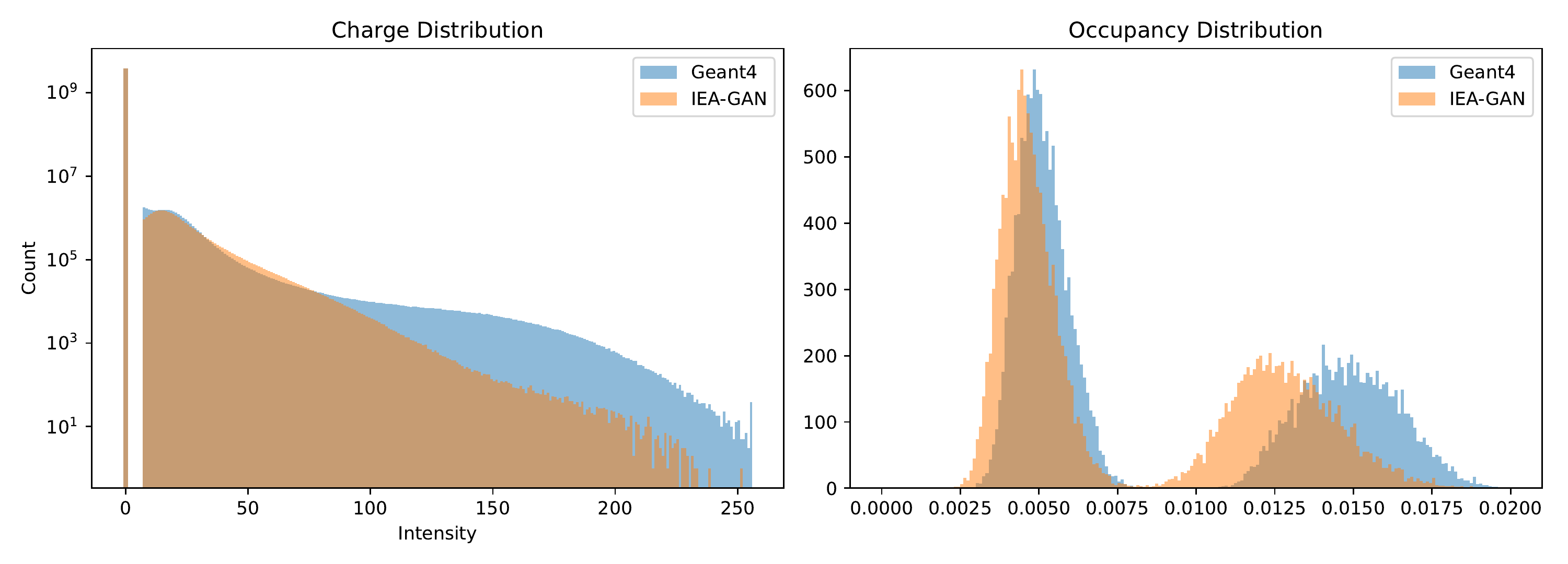}
        \caption{FID 4.12}
    \end{subfigure}
    
    \begin{subfigure}{0.7\textwidth}
        \centering
        \includegraphics[width=\textwidth]{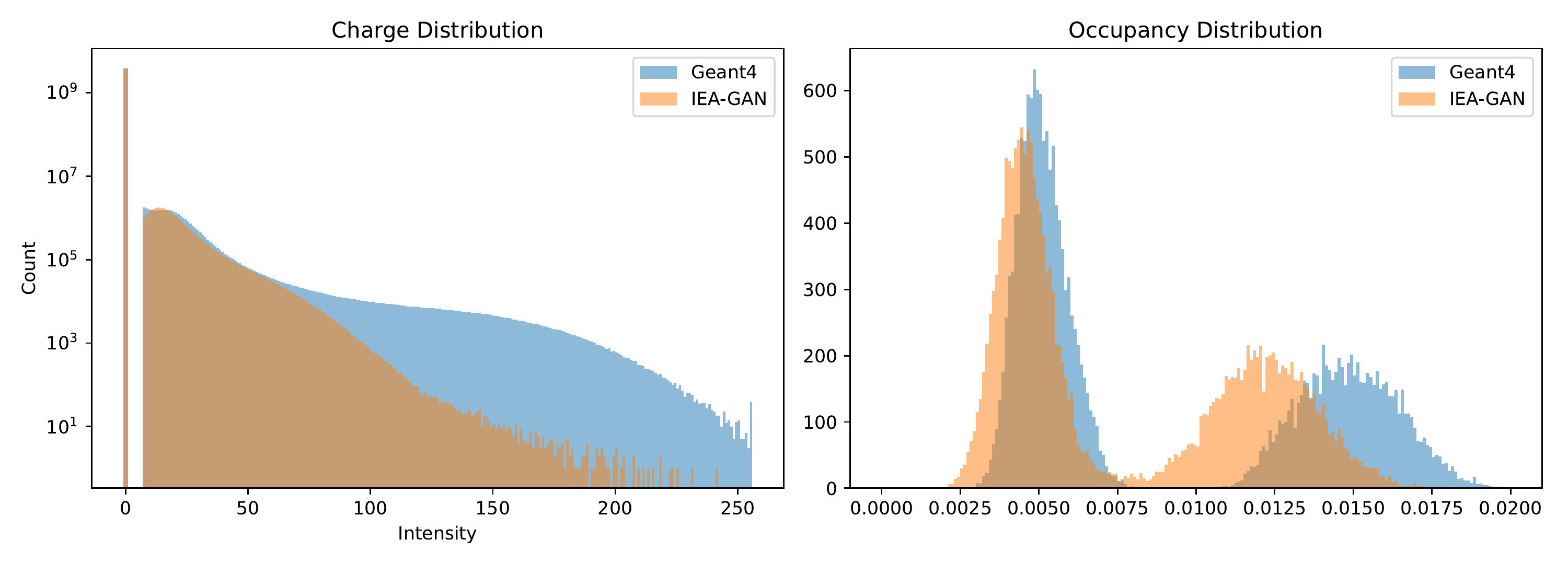}
        \caption{FID 14.39}
    \end{subfigure}

    \caption{Different FID values with respect to the Charge and Occupancy distributions}
    \label{fig:fid_stages}
\end{figure}

In order to qualitatively analyze and interpret the FID flow during training, first I showcase the change of FID value with respect to the occupancy and charge distribution at different stages of the training as depicted in~\cref{fig:fid_stages}. 

Additionally, in~\cref{tab:fid_jitter} I demonstrate the sensitivity of FID to various types of image distortions directly linked to the underlying physics recorded by the corresponding sensor. This is achieved by introducing controlled changes or 'jitters' to the images and tracking their impact on the FID score.

\begin{table}[!htb]
    \centering
    \begin{tabular}{@{}C{6cm}C{3cm}@{}}
    \toprule
    \textbf{Image Jitterings} & \textbf{FID}  \\ [0.5ex] 
    \midrule
    None &  0\\
    Random Masking~(dead zones) &  14.58\\ 
    Random Noise &  87.23\\ 
    Random Rotation~(30 degrees) &  23.69\\
    Random Rotation~(10 degrees) &  2.81\\
    Random Translation~(0.1, 0.1) &  1.99\\
    Random Shear~(10, 10) &  23.53\\
    Random Zoom &  9.06\\
    High Intensity smearing & 3.16\\
    Low Intensity smearing & 47.24\\
    \bottomrule
    \end{tabular}
    \caption{FID Score after Different Jittering Methods applied to the images}
    \label{tab:fid_jitter}
\end{table}

\subsection{Marginal Distributions}
\begin{figure}[!htb]
    \centering
    \includegraphics[width=0.49\textwidth]{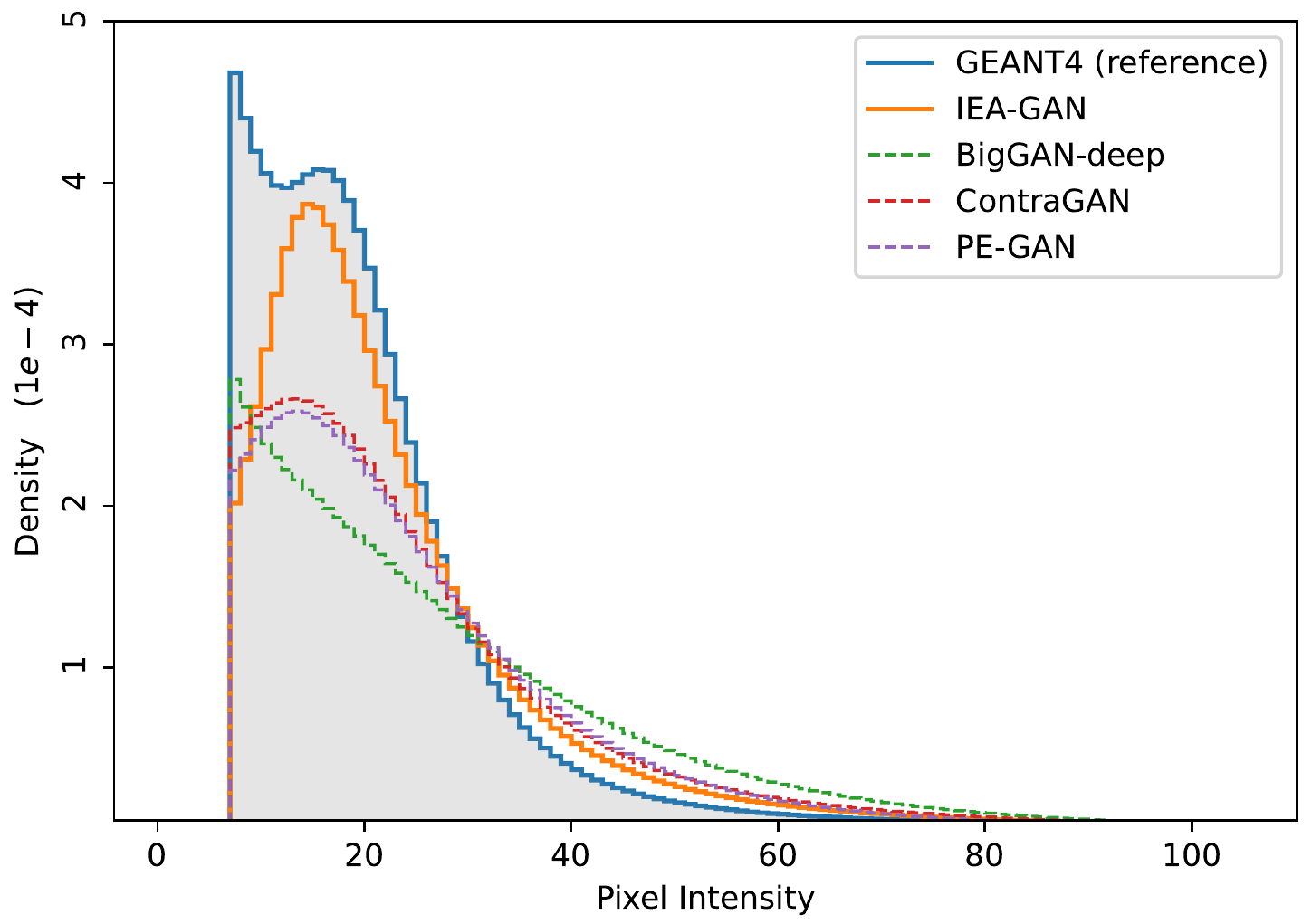}%
    \hfill%
    \includegraphics[width=0.49\textwidth]{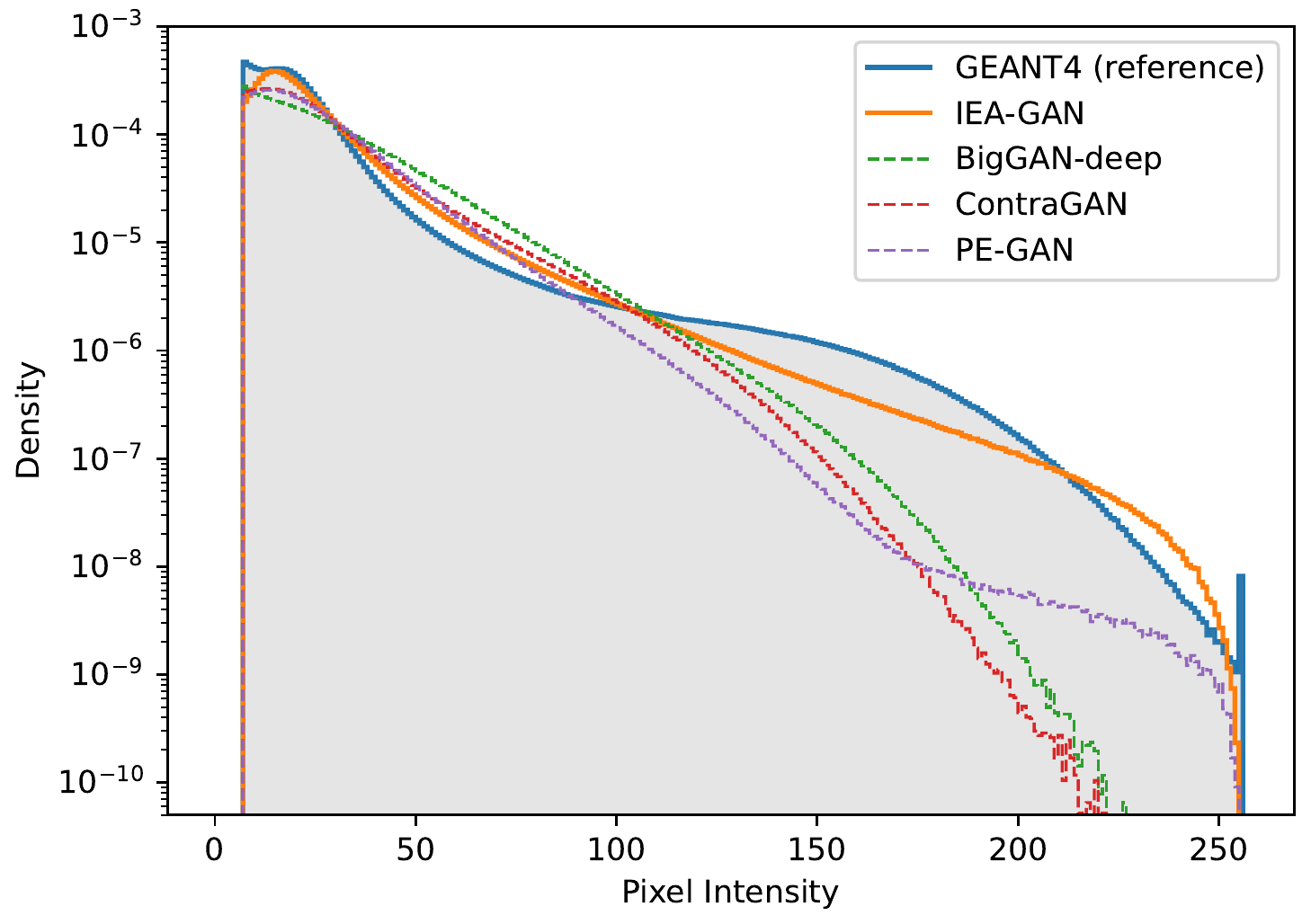}
    \includegraphics[width=0.49\textwidth]{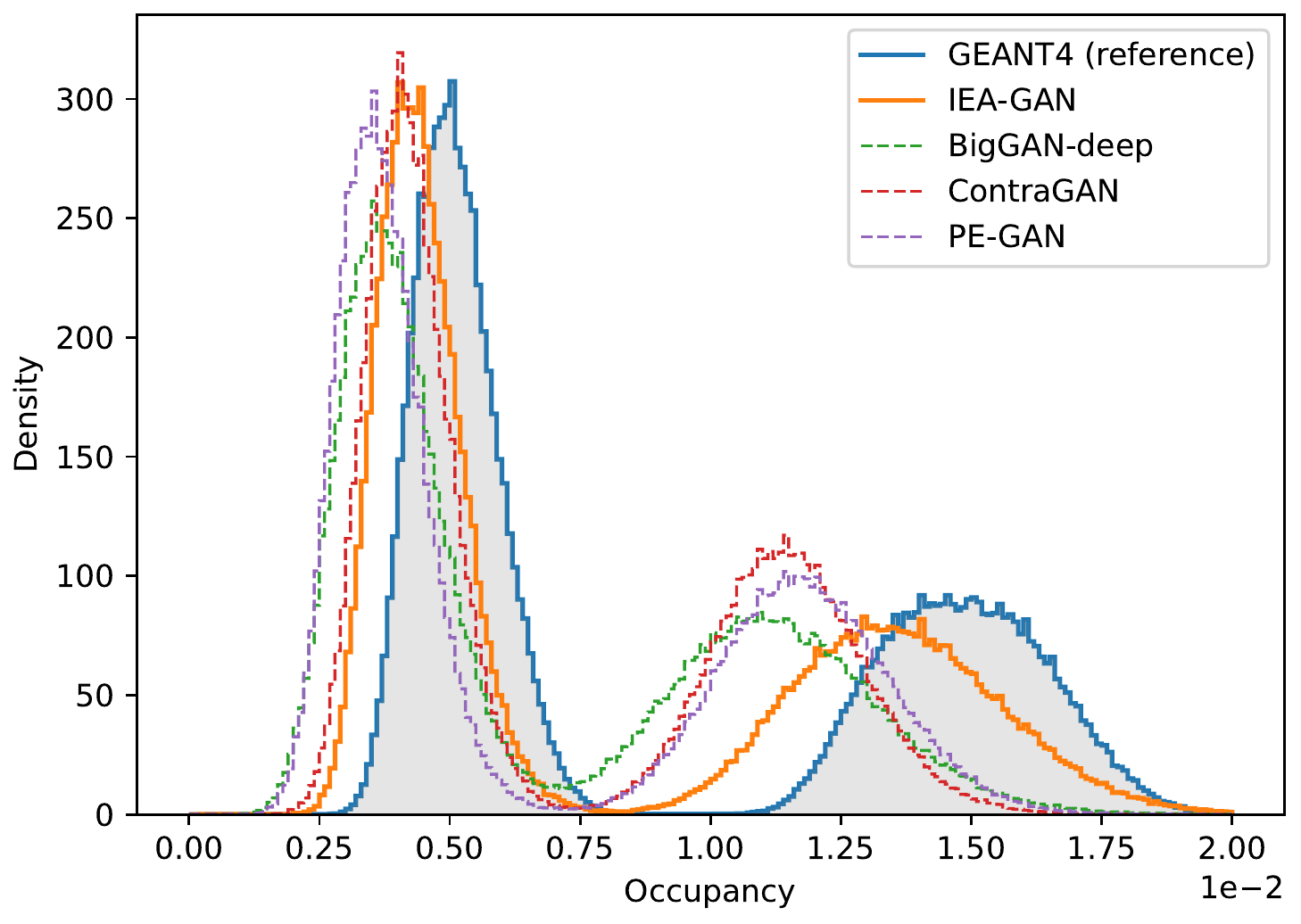}
    \hfill%
    \includegraphics[width=0.49\textwidth]{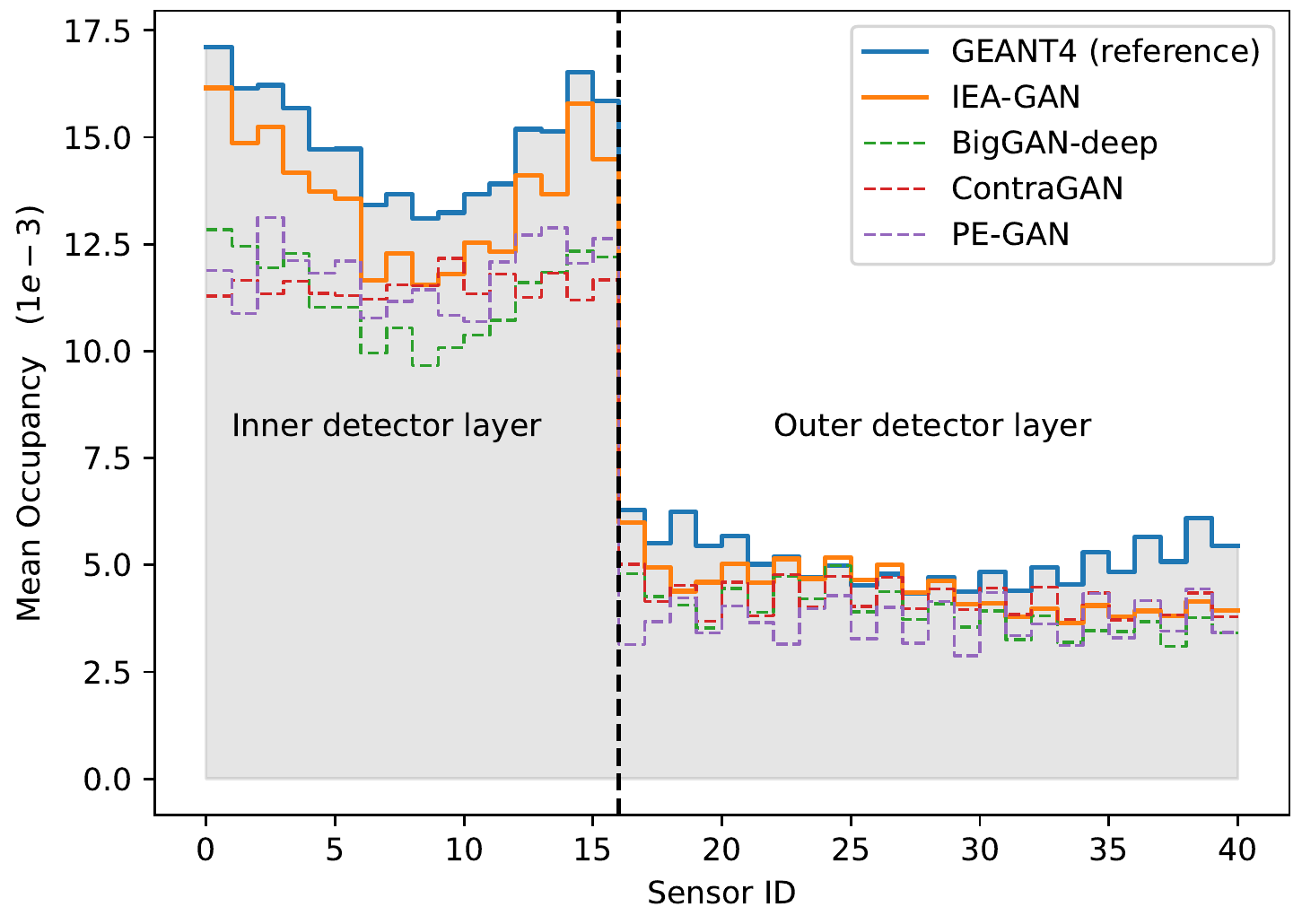}
    \caption{Pixel intensity distribution in linear~(top left) and logarithmic scale~(top right), the distribution of the occupancy~(bottom left) and the mean occupancy per sensor~(bottom right) for \num{10000} events.}
    \label{fig:occupancy-intensity}
\end{figure}

At the pixel level, there are the pixel intensity distribution, occupancy distribution, and mean occupancy.
The pixel intensity distribution defines the distribution of the deposited charge of the background hits. 
The occupancy distribution and the pixel intensity distribution are evaluated over all sensors of a given number of events, while the mean occupancy corresponds to the mean value of sparsity across events for each sensor.
This pixel-level information is essential since upon physics analysis via the basf2 software~\cite{kuhr_belle_2019}, when one wants to use the images and overlay the extracted information on the signal hits, the sparsity of the image defines the volume of the background hits on each sensor.
The pixel intensity distribution, the occupancy distribution, as well as the mean occupancy per sensor are shown in~\cref{fig:occupancy-intensity}. 
The distributions for the IEA-GAN model show the closest agreement with the reference. 
The bimodal distribution of the occupancy comes from the topology of the detector as the sensors are not in a cylindrical shape like a calorimeter but in an annulus shape in two layers. This indicates how challenging generating this detector signature is concerning both its geometry and resolution. In order to capture the correct bi-modality of the occupancy distribution, the RRM and the Uniformity loss play an important role. By using the Uniformity loss in the discriminator, the generator is incentivized to produce samples that are not biased towards a particular mode or class, leading to a wider bimodal distribution of generated samples. 

Moreover, by utilizing the RRM module that considers the inter-dependencies and correlations among the samples within an event, the IEA-GAN exhibits a superior consistency with both high-intensity hits, which enhances the diversity of generated samples in regions with lower occurrence rates.

\begin{figure}[!htb]
    \centering
    \includegraphics[width=\textwidth]{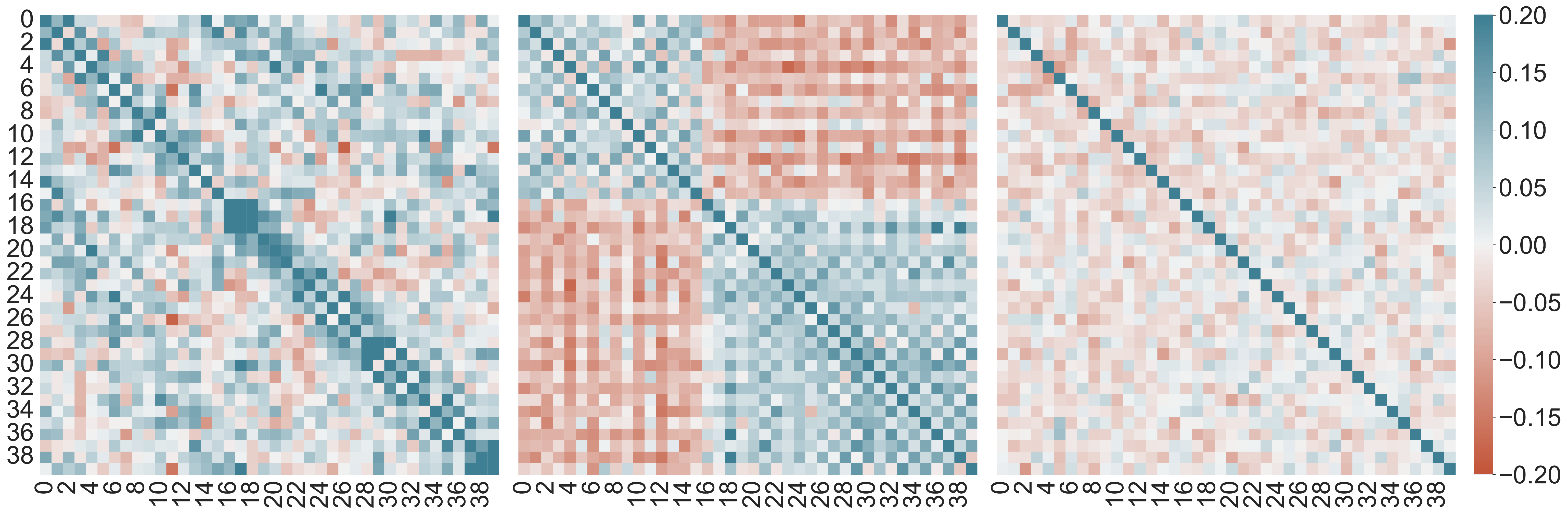}
    \caption{Spearman's correlation between the occupancy of Geant4 sensor images~(left), and sensor images from IEA-GAN~(center), sensor images from PE-GAN~(right).}
    \label{fig:spearman_corr}
\end{figure}

Along with all these image-level metrics, one also needs an intra-event sensitive metric.
All the above metrics are equivariant under permutation between the samples among events. In other words, if I randomly shuffle the samples between events while fixing the sensor number, all the discussed metrics are unchanged. Hence, a metric that looks at the context of each event individually in its event space and goes beyond the sample space is needed.
Ergo, I compute the Spearman's correlation between the occupancy of the sensors along the population of generated events,

\begin{equation*}
r_s = \mathbf{Corr_p}(R(\biguplus_{i=1}^{M=40}(\|x_i\|_0)),R(\biguplus_{i=1}^{M=40}(\|x_i\|_0))),
\label{eq:spearman_corr}
\end{equation*}

where $R(.)$ is the rank operator, a function that assigns a rank to each number in a list as in the definition of Spearman's correlation, and $\mathbf{Corr_p}(.)$ is the Pearson Correlation function. $\biguplus$ is the disjoint union operator that symbolizes the concatenation operation. The norm with subscript 0, denoted by $\|.\|_0$, is the L0 measure. It is a function that counts the number of non-zero elements in a vector. In this work, it is used to calculate the occupancy of the sensors, i.e., the number of non-zero elements in the sensor image $x_i$.
The coefficients by PE-GAN are random values in the range $[-0.2,0.2]$, whereas IEA-GAN images show a meaningful correlation among their generated images.
Even though the desired correlation is different from the reference, as shown in \Cref{fig:spearman_corr}, IEA-GAN understands a monotonically positive correlation for intra-layer sensors and a primarily negative correlation for inter-layer sensors. 

In order to demonstrate that the learned correlation is actually meaningful, I incorporate the Mantel test~\cite{mantel_detection_1967,noauthor_biometry_nodate}, which is a significance test of the correlation between two distance/correlation matrices excluding the diagonal part.
The Mantel test works by comparing each pair of corresponding elements in the two matrices. The null hypothesis is that there's no relationship between the two sets of correlations, and the test statistic is a correlation coefficient.
The significance of the observed correlation is evaluated using permutation testing. This involves randomly rearranging the elements of one matrix many times, recalculating the test statistic each time, and then seeing how extreme the observed test statistic is relative to this null distribution of test statistics. If the observed test statistic is very extreme, then the p-value is less than \num{0.05}, and the null hypothesis is rejected.
For IEA-GAN, the Mantel test results show a veridical correlation 0f $0.18 \pm 0.02$ with empirical p-value $0.0013$. 
As the p-value is less than 0.05, the null hypothesis can be rejected, as there is significant evidence for the correlation between the two sets of matrices.
This suggests that the sensor classes that are more correlated in the Geant4 samples tend to also be correlated in the generated ones by IEA-GAN.
Whereas for PE-GAN, the Mantel test results show a veridical correlation 0f $0.002$ with empirical p-value $0.96$ in support of the null hypothesis.

\subsection{Physics-Level Metric: Tracking}
While image level metrics indicate the low-level quality of simulations, one must also confirm that the resulting simulations are reasonable physics-wise when the entire detector is considered as a whole. For this, a tracking analysis will be done to examine the Helix Parameter Resolutions~(HPR). The quality of the tracking and HPRs directly impacts the precision and accuracy of the measurements.
At the Belle~II experiment, after each collision event, charged particle tracks propagating in vacuum in a uniform magnetic ﬁeld move roughly along a helix path described by the ﬁve helix parameters $\{d_0, z_0, \phi_0,\omega, \tan\lambda\}$ with respect to a pivot point~\cite{kuhr_belle_2019}~(introduced in\cref{chap:2}). These five helix parameters provide a comprehensive description of the trajectory of a charged particle moving in a uniform magnetic field, capturing its shape, orientation, and position in space relative to a reference or pivot point.
The difference between the true and reconstructed helix parameters defines the resolution for the corresponding helix parameter.
The track parameter resolution is affected by the number of hits, the hit intensity, and the underlying intra-event correlation. Understanding how the background effects impact the HPR can give us crucial insights into the overall performance of the detector and the quality of the data it produces. 
A better~(lower) tracking resolution means that the path of a particle can be determined more precisely, which is critical for making accurate measurements and distinguishing between different particles or decay modes. This study utilizes the same event generation and track reconstruction for the comparison, implying that the signal hits used in all simulations are essentially identical. Thus, the true track information is similar.
The primary point of difference lies in the origin of the background. This distinct differentiation allows any disparities identified in the tracking parameter resolutions to be attributed largely to the different backgrounds and their generation origin, enabling a direct evaluation of the quality and performance of the IEA-GAN model in comparison to Geant4. 
First, let's see how important is the Intra-Event sensor correlation from the underlying tacking analysis.
\FloatBarrier
\subsubsection{Physics Motivation for Intra-Event sensor correlation}
As a physics motivation, I highlight the impact of the intra-event correlation by shuffling Geant4 samples. In other words, I show that in a physics analysis, the intra-event sensor-by-sensor correlation influences the performance of the tracking parameters.
I examine the results by comparing the unbiased variance of the parameter resolutions and the 2-sample Kolmogorov–Smirnov test~(KS test) between the shuffled and unshuffled Geant4 PXD background. 
The 2-sample Kolmogorov–Smirnov~(KS) test is a non-parametric hypothesis test used to compare the distributions of two independent samples. Specifically, it tests the null hypothesis \(H_0\) that the two samples are drawn from the same continuous distribution. The test statistic, \(D\), is computed as the maximum absolute difference between the empirical cumulative distribution functions~(ECDF) of the two samples. Formally, if \(F_n(x)\) and \(G_m(x)\) are the ECDFs of two samples of sizes \(n\) and \(m\), respectively, the test statistic is defined as:
\[ D = \max_x |F_n(x) - G_m(x)| \]
Under the null hypothesis \(H_0\), the expected value of \(D\) would be close to 0, while a larger value of \(D\) would indicate a significant difference between the two distributions, leading us to reject \(H_0\) in favor of the alternative hypothesis \(H_1\) which states that the two samples come from different distributions. The critical value or p-value for the test statistic can be obtained from the Kolmogorov distribution, enabling a formal test of the hypothesis.

The results for the high momentum tracks, with more than \SI{0.4}{\giga\electronvolt} show that there is strong evidence that losing the intra-event sensor-by-sensor correlation would impact the resolution and thus the precision of the $\mathbf{d}_0$, $\phi_0$ and $\omega$ Helix parameters. 
For the $\mathbf{z}_0$ and $\tan\lambda$ parameter, there is no significant difference between the resolutions; however, the KS test for these parameters yields low p-values, indicating a high discrepancy between the shape of the two distributions.

\begin{figure}[!htb]
 \centering
     \centering
     \includegraphics[width=0.85\textwidth]{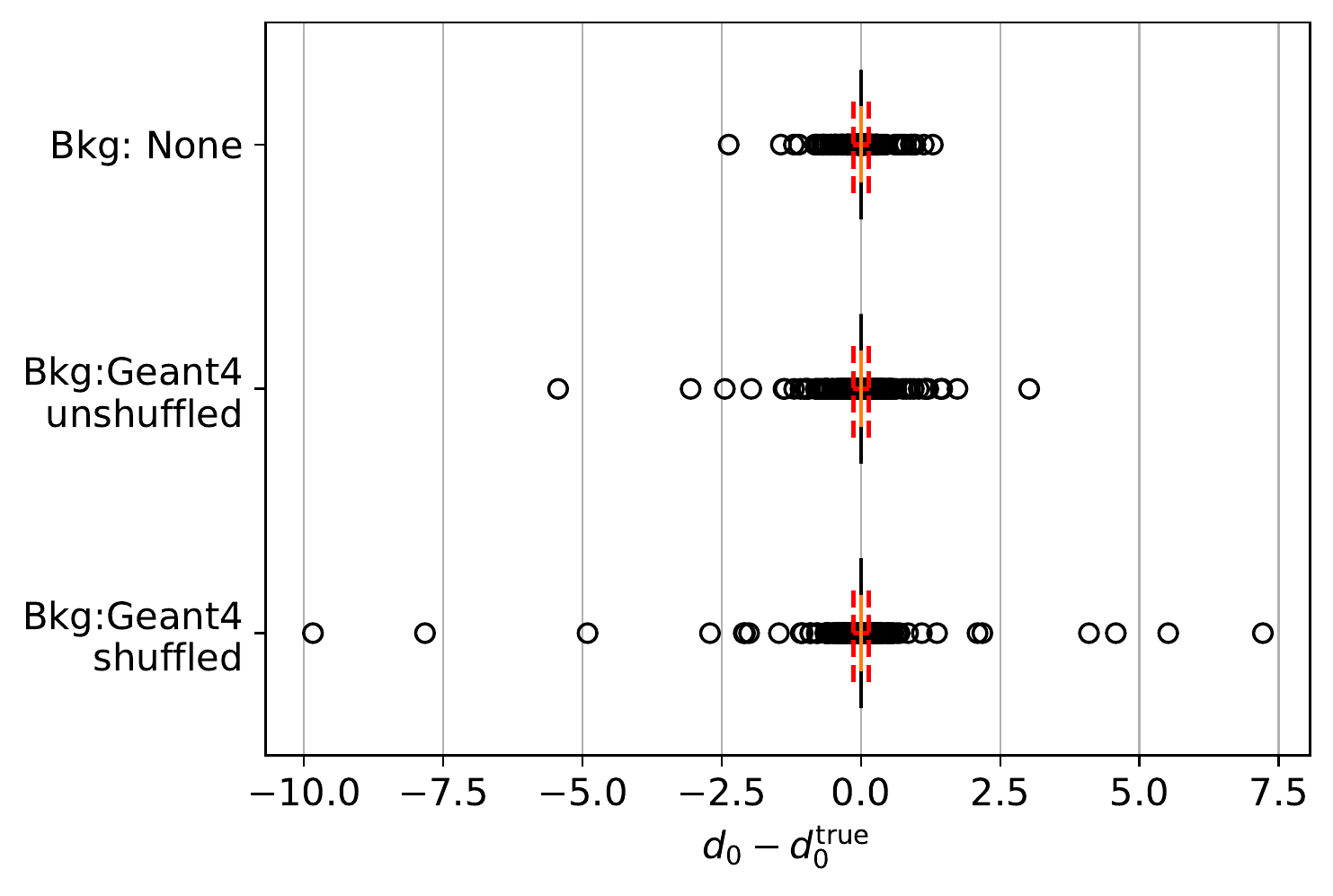}
     \caption{The boxplot for comparing the $d_0$ resolution in the presence of correlated background and uncorrelated background at high momentum regime. The $\pm 5$ standard deviations interval for the no-background case is shown in all 3 cases for reference as a red dashed line.}
     \label{fig:d0_boxplot}
\end{figure}

Let's interpret the results in the context of each Helix parameter. For $\mathbf{d}_0$ impact parameter, the significant variance in resolution shows that the loss of correlation directly impacts how well one can measure the particle's closest approach to the origin in the transverse plane. Given that sensor-by-sensor correlations help to correctly associate track hits, losing it leads to a more spread out distribution of reconstructed values as shown in~\cref{fig:d0_boxplot}. This could affect subsequent analyses, such as identifying primary and secondary vertices, especially in scenarios where particles have negligible deflection.

For the $\phi_0$ parameter, the insignificant resolution variance difference and KS test result suggest that the lack of layer correlation doesn't significantly impact the distribution and precision of measurements of the azimuthal for high momentum tracks.
The higher error in the variance of $\Delta \mathbf{z}_0$ in the shuffled data suggests that the lack of correlation between detector layers introduces more uncertainty in determining the longitudinal interaction point. High momentum tracks are less likely to deviate significantly in the z-direction. Combined with the significant KS test result, this indicates a fundamental difference in how particle trajectories are reconstructed in the z-direction without layer correlation.
$\omega$ is a measure of the curvature of the particle's track, and it's inversely proportional to the particle's momentum. For high-momentum particles, one would expect the curvature to be smaller since higher-momentum particles travel more linearly.
The variance for $\Delta\omega$ shows a slight discrepancy between the shuffled and unshuffled data, but the KS test doesn't show a significant difference. This indicates that while the overall distributions of the curvature resolution don't significantly differ, there's a minor difference in the precision with which the curvature is reconstructed, which could have implications for subsequent physics analyses that depend on accurate momentum information.
Despite the insignificant resolution difference, the significant KS test result for $\Delta \tan\lambda$ suggests differences in the inclination distributions of high momentum tracks between the shuffled and unshuffled data. This might indicate that the inclination of the track, which is also related to the momentum in the longitudinal direction, is affected by the loss of correlation between layers and sensors.

\begin{table}[!htb]
\centering
\begin{tabular}{@{}lcc|cc@{}}
\toprule
Parameter & \multicolumn{2}{c}{Unbiased Variance $\pm$ error} & KS statistic & p-value \\ 
\cmidrule(lr){2-3}
          & Shuffled Geant4 & Unshuffled Geant4 \\
\midrule
$\Delta d_0$        & $0.1343 \pm 0.0007$ & $0.0732 \pm 0.0004$ & $0.0067$ & $0.7655$ \\
$\Delta \phi_0$      & $0.2158 \pm 0.0011$ & $0.1859 \pm 0.0009$ & $0.0066$ & $0.7899$ \\
$\Delta z_0$        & $5.0076 \pm 0.0253$ & $4.9341 \pm 0.0249$ & $0.0152$ & $0.0211$ \\
$\Delta \omega$     & $0.0010 \pm 0.0001$ & $0.0008 \pm 0.0000$ & $0.0138$ & $0.0485$ \\
$\Delta \tan\lambda$ & $0.0388 \pm 0.0002$ & $0.0382 \pm 0.0002$ & $0.0167$ & $0.0086$ \\
\bottomrule
\end{tabular}
\caption{Unbiased variance and KS test results for the shuffled and unshuffled Geant4 data across the 5 Helix parameters.}
\label{tab:shuf_helix}
\end{table}

\subsubsection{IEA-GAN's Tracking Performance}
\begin{figure}[!htbp]
\centering
\begin{subfigure}{.5\textwidth}
\centering
\includegraphics[width=0.8\linewidth]{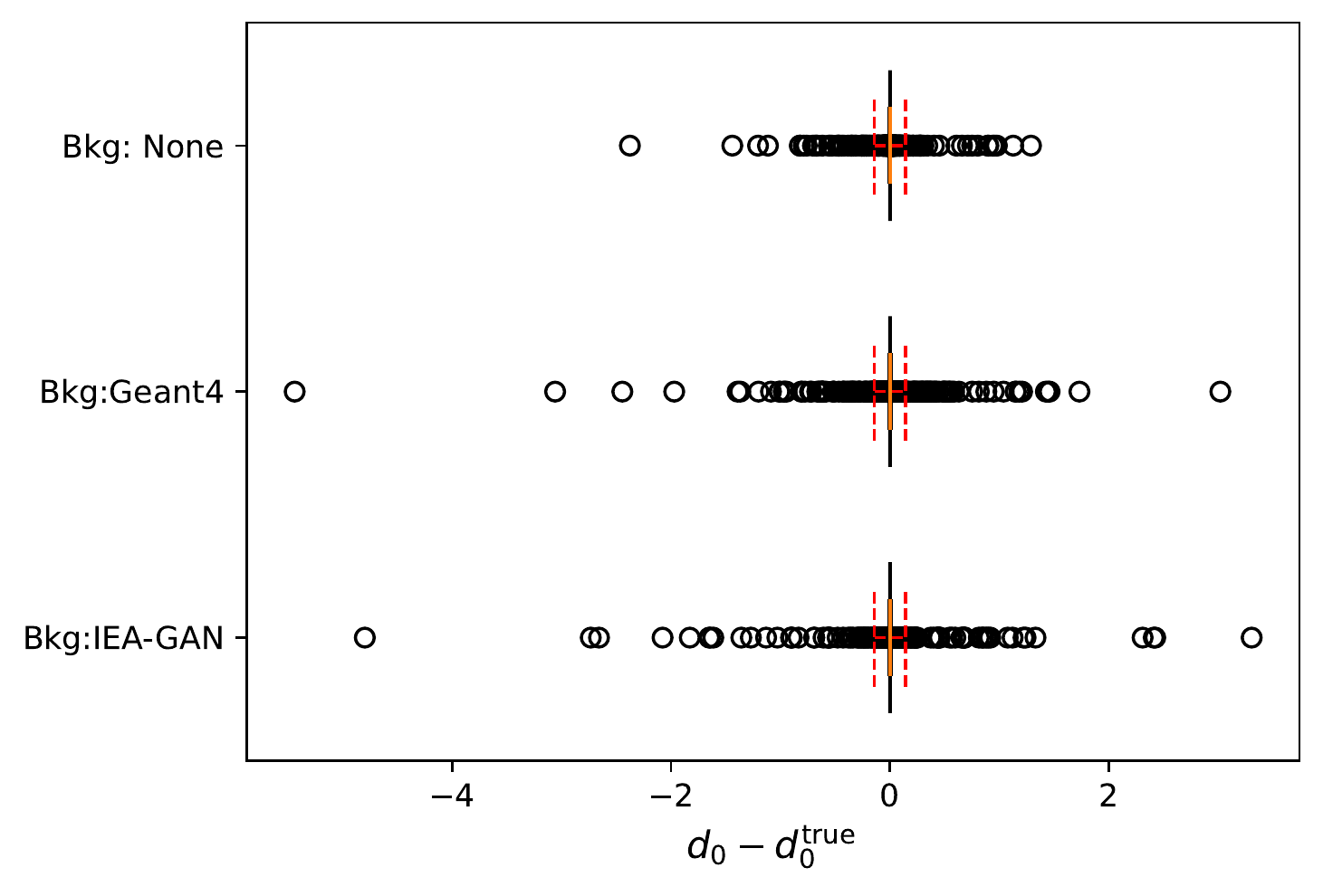}
\includegraphics[width=0.8\linewidth]{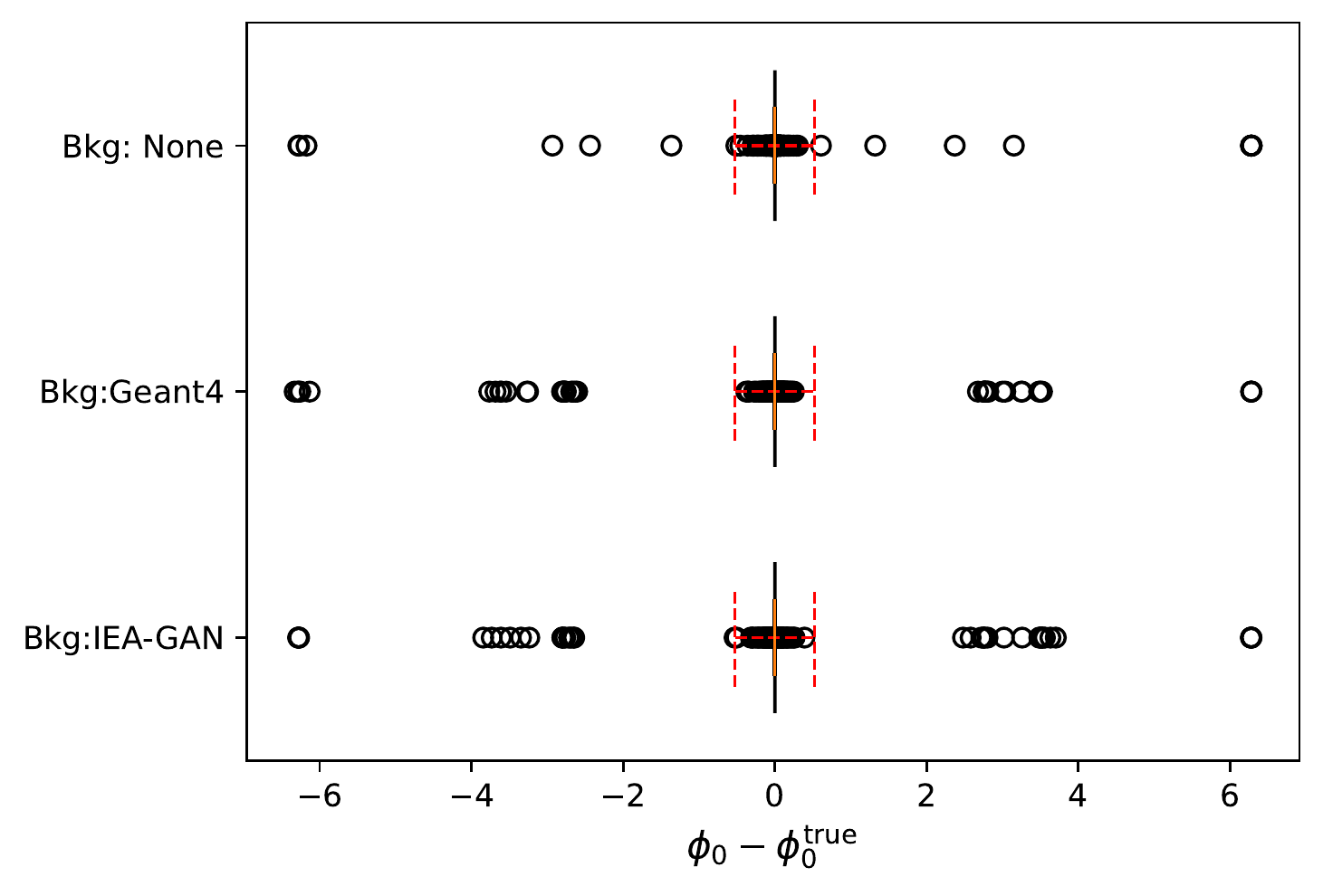}
\includegraphics[width=0.8\linewidth]{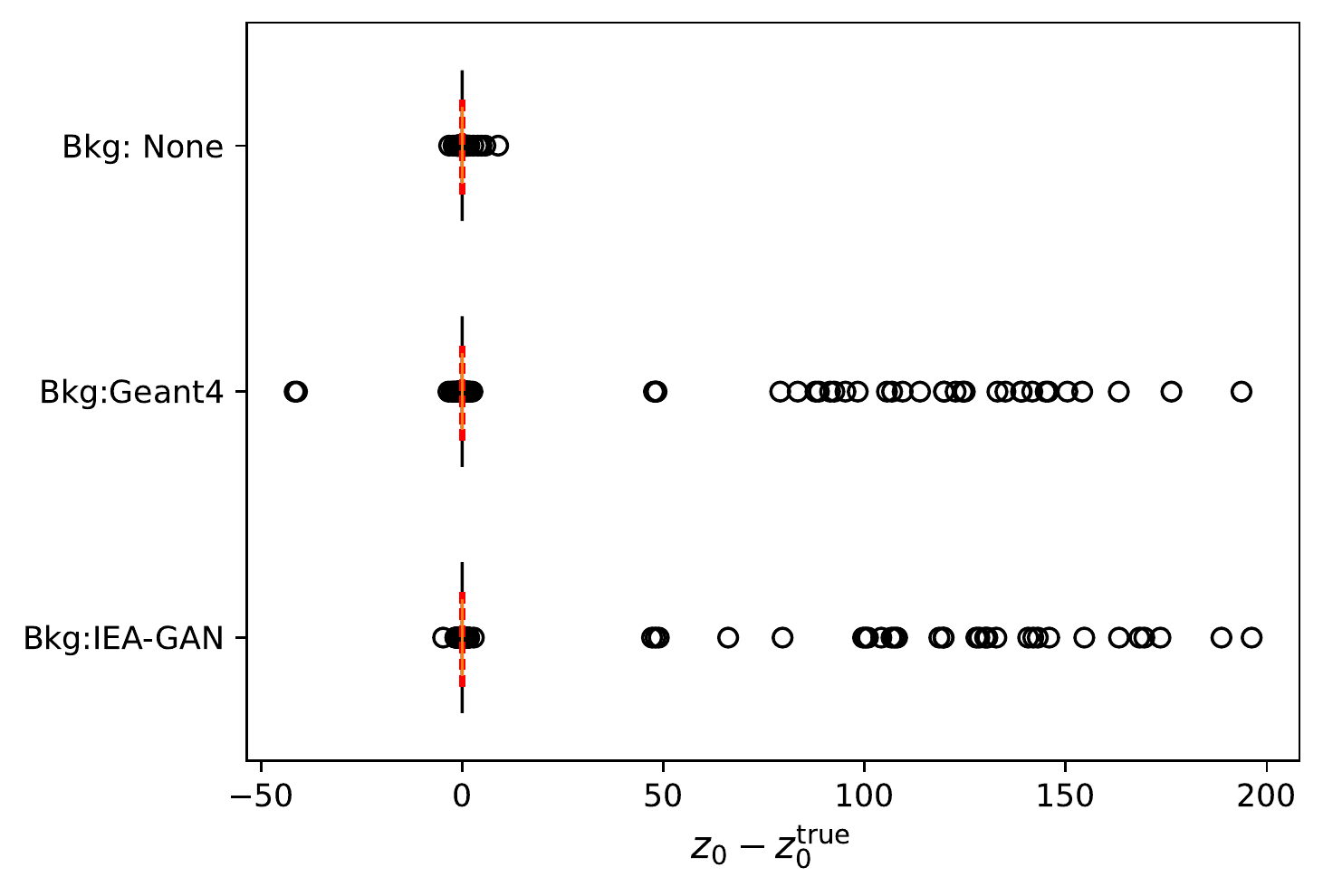}
\includegraphics[width=0.8\linewidth]{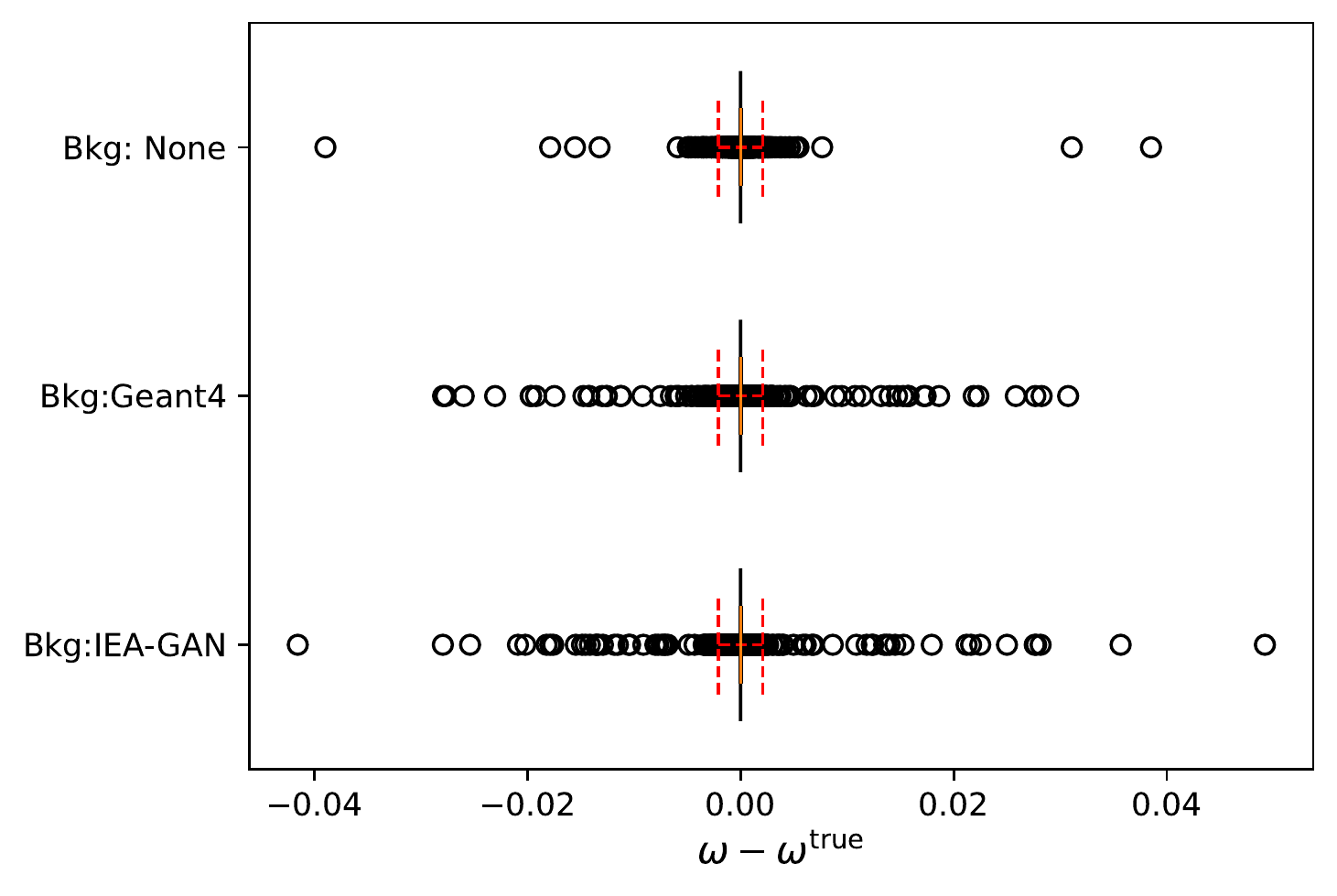}
\includegraphics[width=0.8\linewidth]{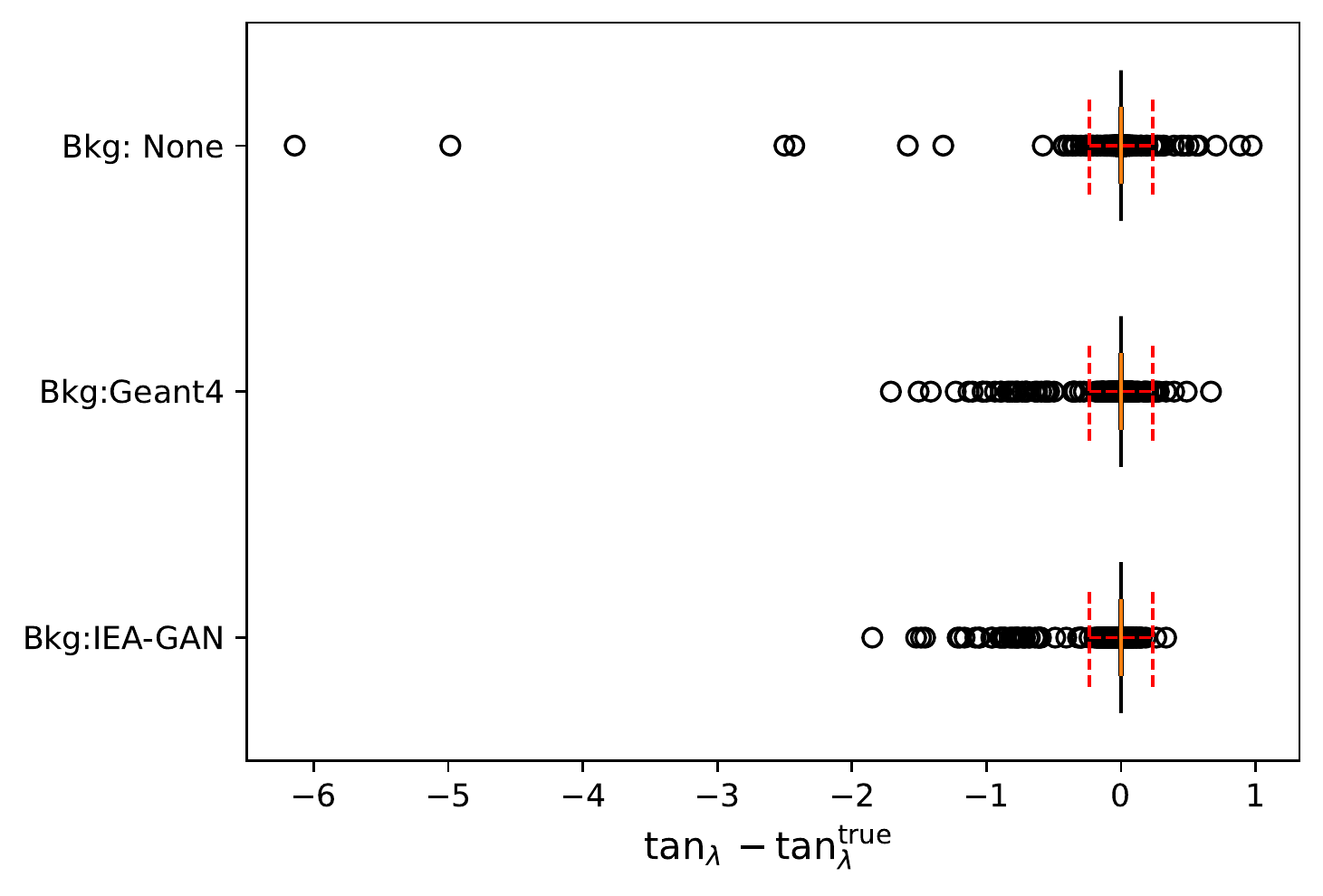}
\end{subfigure}%
\begin{subfigure}{.5\textwidth}
\centering
\includegraphics[width=0.8\linewidth]{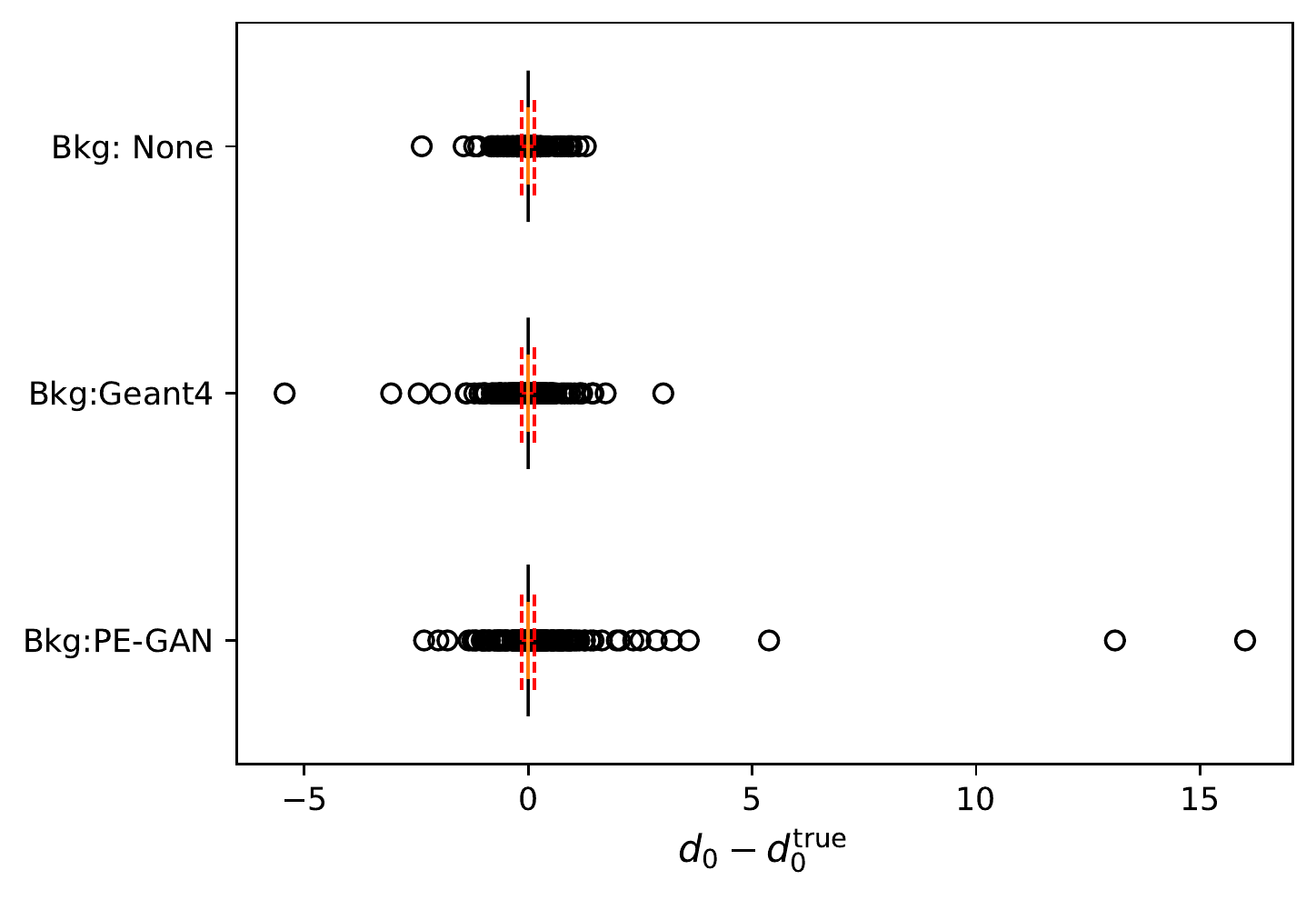}
\includegraphics[width=0.8\linewidth]{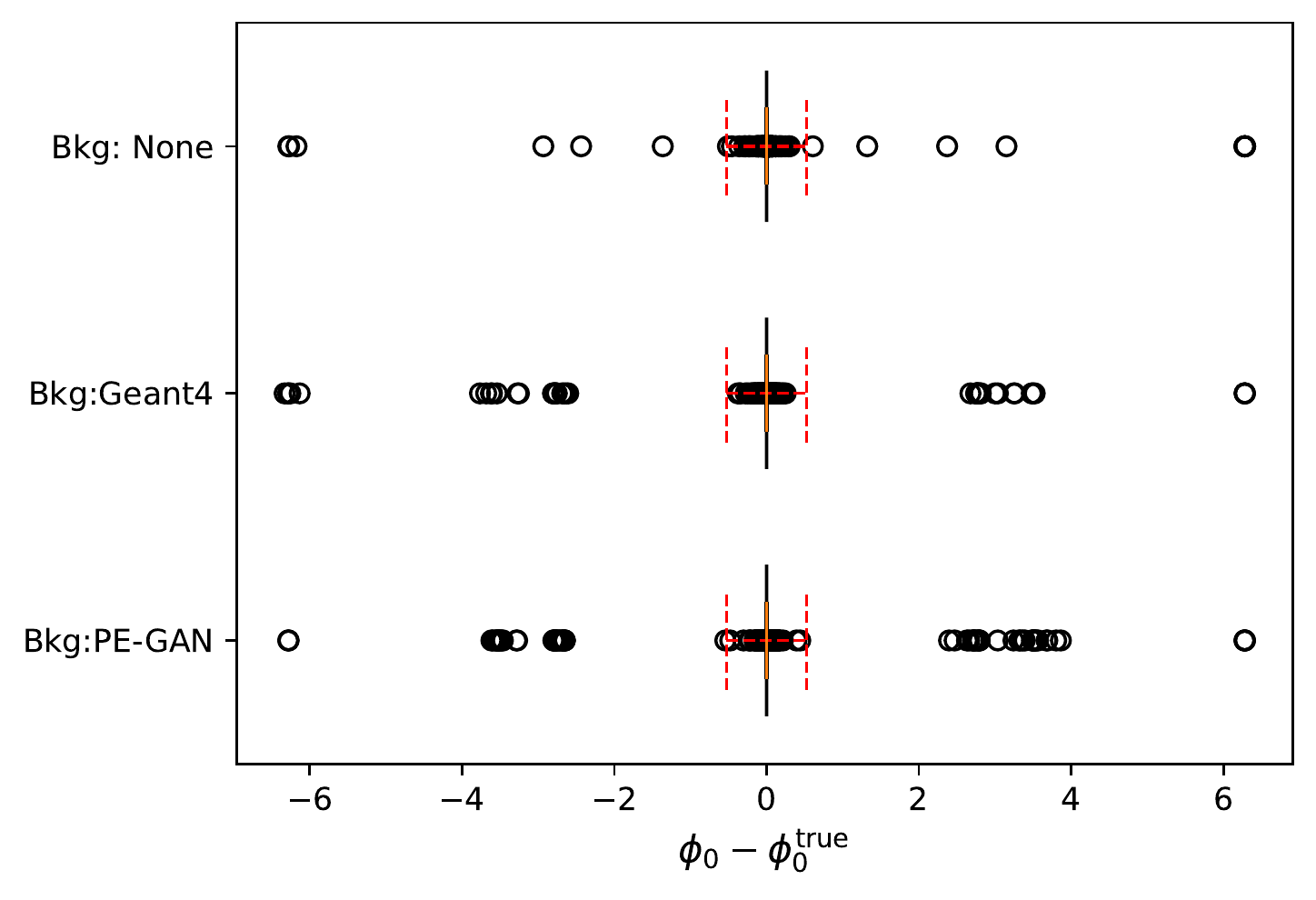}
\includegraphics[width=0.8\linewidth]{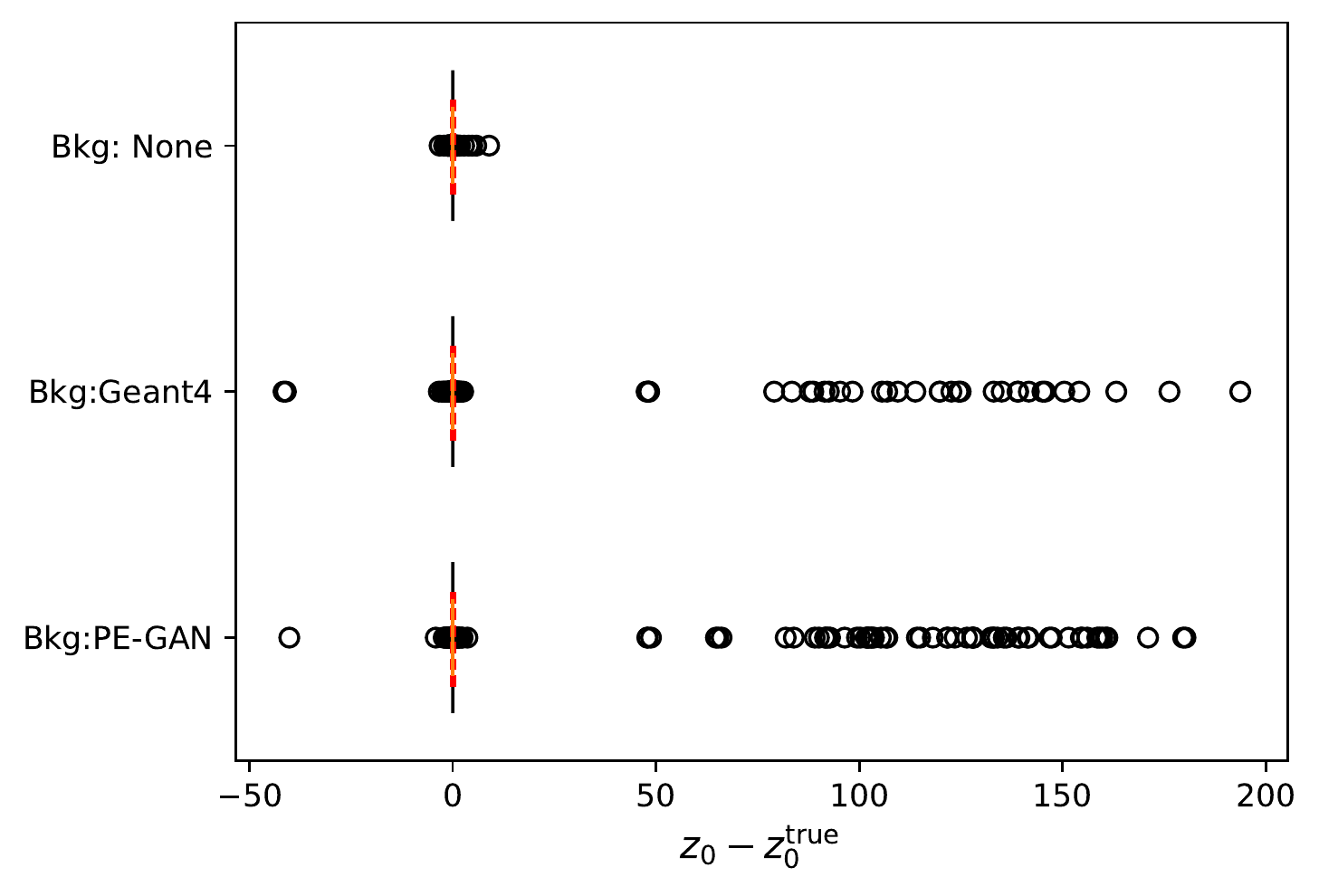}
\includegraphics[width=0.8\linewidth]{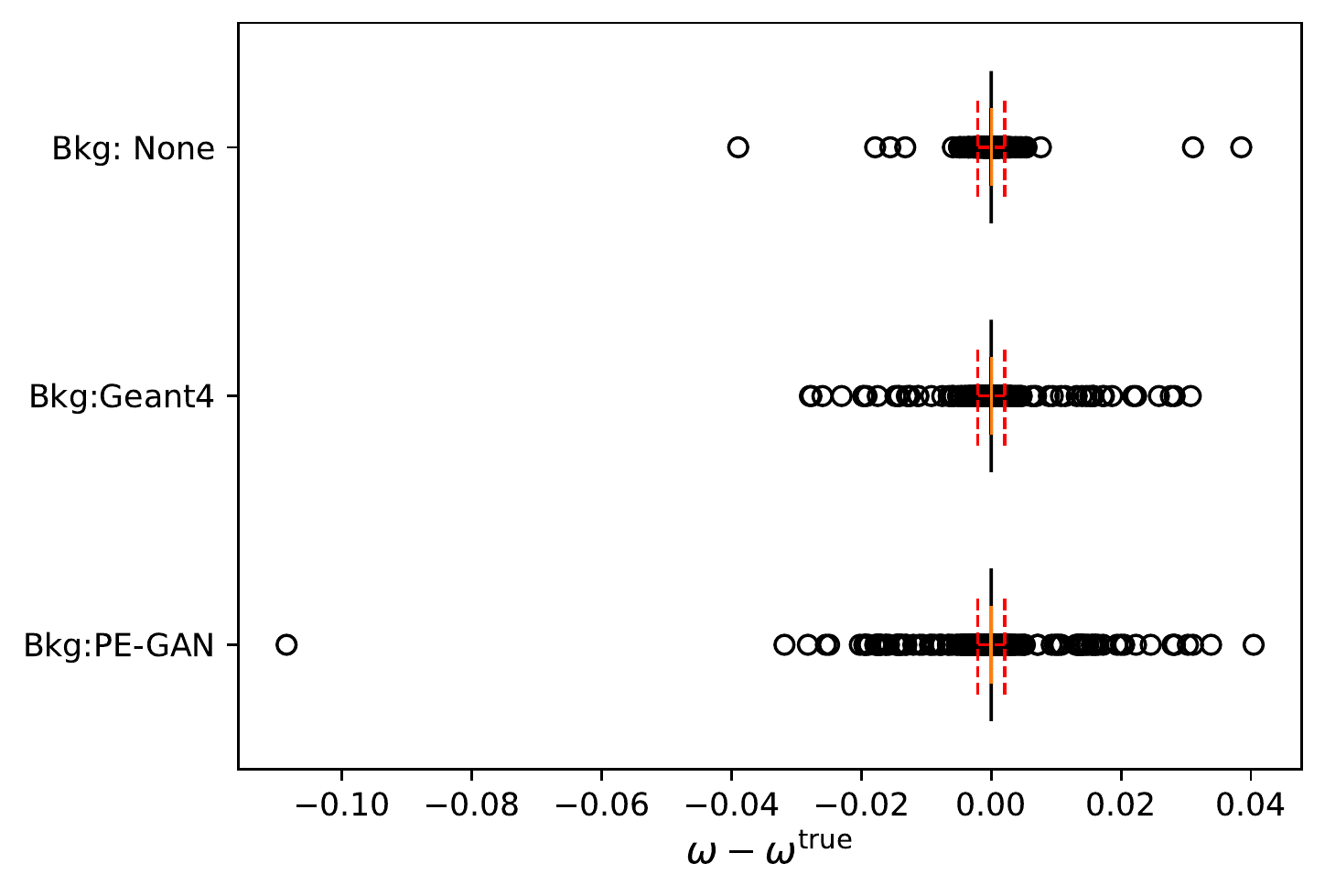}
\includegraphics[width=0.8\linewidth]{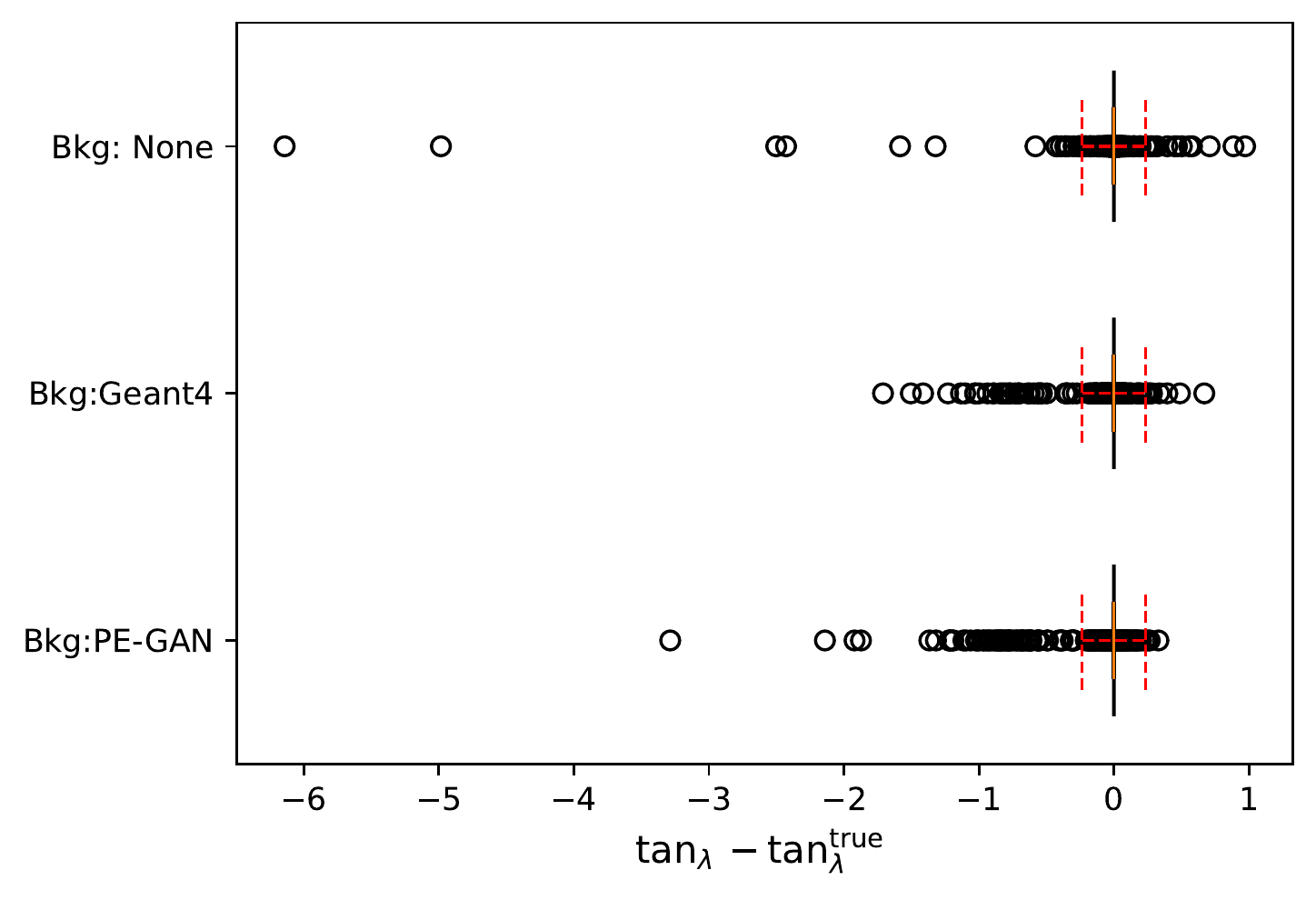}
\end{subfigure}
\caption{
Helix parameter resolutions for $d_0$~(row 1), $z_0$~(row 2), $\phi_0$~(row 3), $\omega_0$~(row 4), and $\tan \lambda$~(row 5).
For each parameter, the left figure corresponds to the IEA-GAN simulated background, and the right figure corresponds to the PE-GAN simulated background. The $\pm 5$ standard deviations interval for the no-background case is shown in all 3 cases for reference as a red dashed line.
}
\label{fig:helix_param_res}
\end{figure}

This study compares IEA-GAN with PE-GAN for the resolutions of all five helix parameters as shown in \Cref{fig:helix_param_res} and \Cref{tab:model_compare} for \num{5000} events for high momentum tracks $(P_T>0.4)$ \si{\giga\electronvolt}. 

In the low momentum region, the resolution performance of the models is on par.
My meticulous comparison revealed that the unbiased variance of these parameters, produced by the IEA-GAN model, approximates more closely to the Geant4 reference, outperforming the PE-GAN model in each instance. Moreover, the Kolmogorov-Smirnov test results further consolidated my findings, showing higher p-values for the IEA-GAN model, thus adhering more accurately to the Geant4 reference. 
Another interesting observation is that, in comparison with the shuffled Geant4, IEA-GAN shows a more significant KS test p-value for $z_0$, $\omega$, and $\tan\lambda$ resolutions and a more precise $\mathbf{d}_0$ reconstruction.

As a result, I observe a good agreement between the IEA-GAN and Geant4, both in the tail segments~(variance) and precision of the resolutions where the largest difference between Geant4 and no background is found. 
Hence, not only does IEA-GAN demonstrate a close image level agreement with Geant4, but it maintains a proper reconstructed physical behavior during track reconstruction as well.

\begin{table}[!htb]
\centering
\begin{tabular}{@{}lcccccc@{}}
\toprule
Model & Parameter & \multicolumn{2}{c}{Unbiased Variance $\pm$ error} & KS statistic & p-value \\ 
\cmidrule(lr){3-4}
      &           & Model & Geant4 \\
\midrule
\multirow{5}{*}{PE-GAN} & $d_0$       & $0.1709 \pm 0.0009$ & $0.0732 \pm 0.0004$ & $0.0156$ & $0.0164$ \\
                           & $\phi_0$     & $0.2207 \pm 0.0011$ & $0.1859 \pm 0.0009$ & $0.0120$ & $0.1193$ \\
                           & $z_0$       & $6.9073 \pm 0.0349$ & $4.9341 \pm 0.0249$ & $0.0183$ & $0.0029$ \\
                           & $\omega$    & $0.0014 \pm 0.0001$ & $0.0008 \pm 0.0001$ & $0.0116$ & $0.1425$ \\
                           & $\tan\lambda$     & $0.0579 \pm 0.0003$ & $0.0382 \pm 0.0002$ & $0.0179$ & $0.0037$ \\
\midrule
\multirow{5}{*}{IEA-GAN}   & $d_0$       & $0.0762 \pm 0.0004$ & $0.0732 \pm 0.0004$ & $0.0104$ & $0.2373$ \\
                           & $\phi_0$     & $0.1905 \pm 0.0010$ & $0.1859 \pm 0.0009$ & $0.0109$ & $0.1939$ \\
                           & $z_0$       & $5.1467 \pm 0.0261$ & $4.9341 \pm 0.0249$ & $0.0073$ & $0.6814$ \\
                           & $\omega$    & $0.0010 \pm 0.0001$ & $0.0008 \pm 0.0001$ & $0.0103$ & $0.2537$ \\
                           & $\tan\lambda$     & $0.0412 \pm 0.0002$ & $0.0382 \pm 0.0002$ & $0.0068$ & $0.7538$ \\
\bottomrule
\end{tabular}
\caption{Comparison of Unbiased Variance and KS test results for the PE-GAN and IEA-GAN models with the Geant4 reference across 5 Helix parameters for high momentum tracks.}
\label{tab:model_compare}
\end{table}

\FloatBarrier
\section{Ablation Studies and Things I Tried But Did Not Work}
\label{sec:ablation_studies}

\begin{figure}[H]
    \centering
    \includegraphics[width=0.6\textwidth,clip]{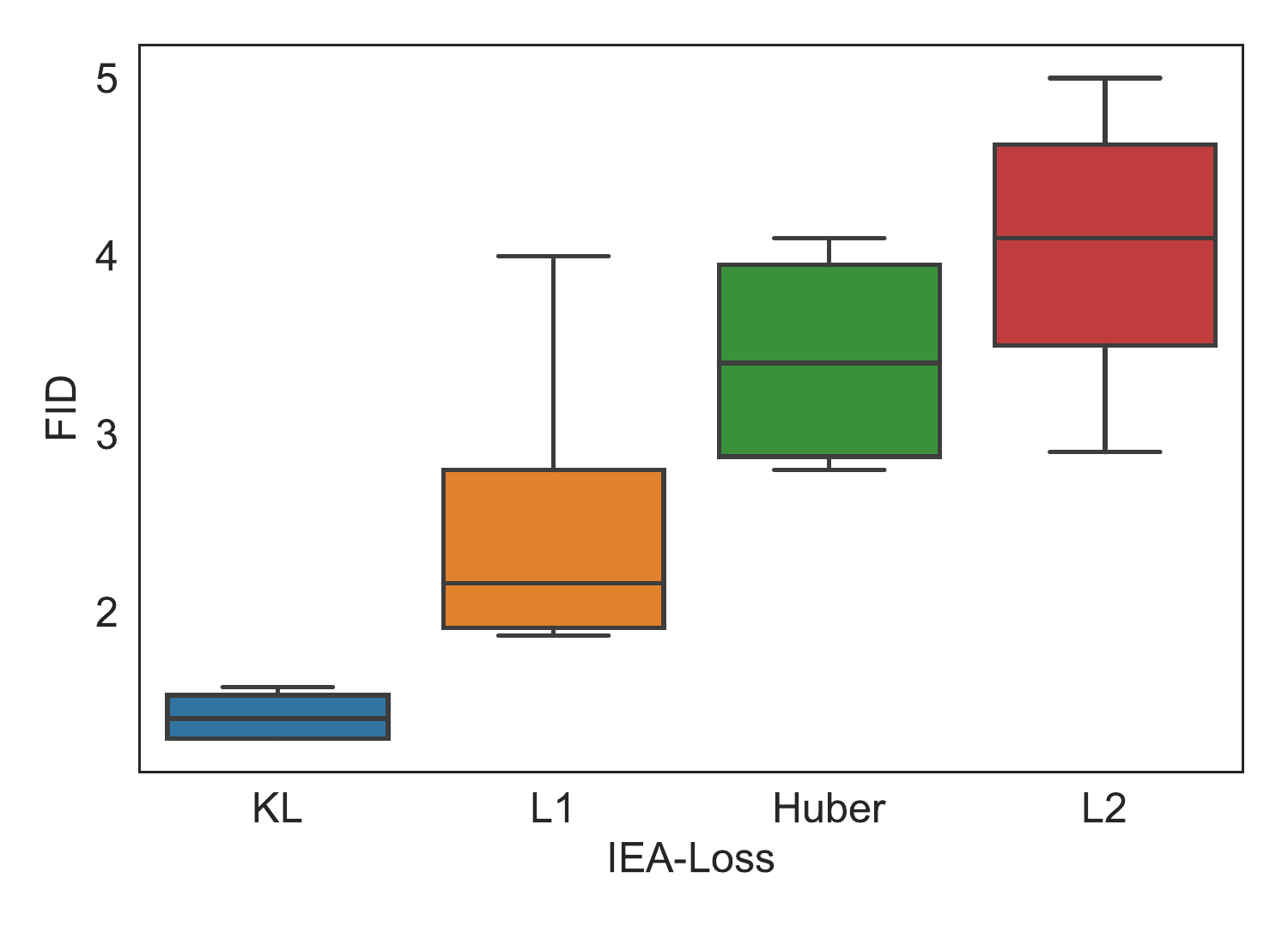}
    \caption{Comparison of the FID between different IEA-losses}
    \label{fig:FID_IEAloss}
\end{figure}

For the IEA-loss, I tested several losses in order to achieve the best stability, shown in \Cref{fig:FID_IEAloss}.
Some of them have their own merits and downsides. I explored the KL divergence, the L1 loss, the Huber loss~\cite{noauthor_robust_nodate}, and the L2 loss. KL divergence was more stable to capture differences between the real and fake self-similarities and more robust to outliers.

I probed a range of coefficients for the IEA-loss and the Uniformity loss.
For the KL divergence as the IEA-loss, I tried the values \{0.1, 1, 5, 10\} and selected $1$.
For the L1 loss, as well as the IEA-loss, the best $\lambda_{\mathrm{IEA}}$ value is $10$.
For the Uniformity loss, I probed the values \{0.01, 0.1, 0.5, 0.75, 1, 5, 10\} and selected $0.1$.
Moreover, IEA-GAN, without the IEA-loss and Uniformity loss suffers from the lack of agreement maximization penalty for the generator and information maximization for the discriminator.
This study shows that having either of these losses without the other causes training instability, divergence, and lower fidelity as shown in~\Cref{tab:fid_ablation_losses}.

\begin{table}[!htb]
\begin{minipage}{\textwidth}
 \begin{center}
    \caption{FID comparison between IEA-GAN,  IEA-GAN with RRM only, IEA-GAN with Uniformity loss only, and IEA-GAN with both IEA-loss, averaged across six random seeds.}
    \label{tab:fid_ablation_losses}
    \begin{tabular}{@{}lllll@{}}
    \toprule
    & IEA-GAN & Only RRM & RRM with Uniformity & RRM with IEA-loss \\ 
    \midrule
    \textbf{FID}  & $\mathbf{1.50\pm 0.16}$ & $2.74\pm 0.62$& $2.29\pm 0.14$& $3.42\pm 0.52$ \\
    \bottomrule
    \end{tabular}
  \end{center}
\end{minipage}
\end{table}

For the hypersphere dimension, I probed the values \{512, 768, 1024, 2048\} and selected \num{1024}.
For dimensions smaller than \num{512} the discriminator fails to converge.
I also changed the position of the hypersphere projection layer and put it before and after the Multi-head attention.
The best position for the hypersphere projection is after the Multi-head attention and two layers of MLP.
Moreover, for the hypersphere projection, I also tried an inverse Stereographic Projection $h:\mathbb{R}^N \rightarrow \mathbb{S}^N/\{p\}$ with p as a north pole on the n-sphere~\cite{eybpoosh_applying_2022} instead of L2 compactification.
This map is conformal thus it locally preserves angles between the data points.
The results were more stable but the average FID was better with L2 compactification as shown in \Cref{tab:fid_hyper_comparison}.

\begin{table}[!htb]
\begin{minipage}{\textwidth}
  \begin{center}
     \caption{FID comparison between two different Hypersphere projections for IEA-GAN's discriminator, averaged across six random seeds.}
     \label{tab:fid_hyper_comparison}
    \begin{tabular}{@{}lll@{}}
    \toprule
    & L2 compactification & Inverse Stereographic projection \\ 
    \midrule
    \textbf{FID} & $\mathbf{1.50\pm 0.16}$ & $2.01\pm 0.07$ \\
    \bottomrule
    \end{tabular}
  \end{center}
\end{minipage}
\end{table}

Inside the RRM I tried a GeLU~\cite{hendrycks_gaussian_2023} non-linearity instead of ReLU and the result was in favor of the latter.
I also put the layer normalization before and after the Multi-head Attention.
The pre-norm version seems to be much more stable and adaptable to GAN training intricacies.
Another observation related to RRM is the weight normalization of the linear layers.
I observed that for the discriminator spectrally normalized MLPs show the best results.
For the Generator, applying Spectral Normalization to the linear layers destabilizes the training.
My observation regarding the effect of RRM over the generator's label embedding shows that without it, the RRM in the discriminator becomes also unstable and the training diverges very early.

For the random degrees of freedom~(Rdof), first, I utilized the random vectors that are fed to the generator and applied the RRM on top of it. However, the FID did not reach values below \num{20}, and there was no correlation. Hence, I introduced separate random sampling for event generation for which I probed dimensions \{2, 4, 8, 16, 32\}. \num{4} degrees of freedom was the most optimal choice. I observed that as the dimension of Rdof increases, the intra-event correlation fades away between the generated images. I also checked the Uniform distribution for event Rdof sampling, which did not lead to any stable result. 
Several ways to fuse the Rdof to the class embeddings were tested such as learnable neural network layer~(matrix factorization), concatenating, summing, and having an MLP with non-linear part, but eventually chose a learnable neural network layer~(matrix factorization) for the feature mixing layer.

I also looked at different combinations of learning rates for G and D.
Using TTUR~\cite{heusel_gans_2018} regime results in a severe mode collapse.
Thus, I used the same learning rates for both G and D.
I swept through $\{1\times10^{-5}, 2.5\times10^{-5}, 5\times10^{-5}, 7.5\times10^{-5}, 1\times10^{-4}\}$ and selected $5\times10^{-5}$.
For the backbone model, the shallow version of BigGAN-deep, BigGAN, leads to mode-collapse, therefore, I chose BigGAN-deep.

\FloatBarrier
\section{Conclusion and Discussion}
This study has proposed a series of new methods for ultra-high-resolution, fine-grained, correlated PXD detector background response generation and conditional sampling. 
First, I argued that PXD background generation can be categorized as a fine-grained data analysis task with meta-categories being the sensor labels, and subordinate categories being the fine features such as various amounts of occupancy or different background types.
Then PE-GAN was introduced as a proof-of-concept of having sensor-number conditioning. However, despite its performance in conditional sampling, it did not have any intra-event understanding.

As a result, IEA-GAN was proposed to capture the dyadic class-to-class relations and exhibit an explainable~(weak) intra-event correlation among the generated detector images while all other models fail to capture any correlation. 
To achieve this, novel components, the Relational Reasoning Module~(RRM) and the IEA-loss, with the Uniformity loss were introduced. 
The RRM introduced a self-supervised relational contextual embedding for the samples in an event, which is compatible with GAN training policies, a task that is very challenging. 
RRM dynamically clusters the images in a collider~(PXD) event based on their inherent correlation culminating in approximating a collision event.
The IEA-loss, a discriminator-supervised loss, helps the generator reach a consensus over the discriminator's dyadic relations between samples in each event.
Finally, I have demonstrated that the Uniformity loss plays a crucial role for the discriminator in maximizing the homogeneity of the information entropy over the embedding space, thus helping the model to overcome mode-collapse and to capture a better bi-modality of generated occupancy. As a result, an improvement in all metrics compared to the previous SOTA occurs. Let's quickly review the results:

\begin{itemize}
    \item Introducing the application of the FID and KID metric, for the detector simulation as a powerful tool for evaluating the performance of deep generative models in detector simulation. Moreover, it provides detector-response level interpretations of the FID as a data-driven robust NN-based figure of merit for PXD. 
    \item Achieving an FID score of $1.50$, an over $42\%$ improvement, and a KID score of $0.0010$, as presented in~\Cref{tab:fid}.
    \item Proving an improvement to all marginal distributions in comparison to SOTA models for high-resolution PXD generation. In particular, IEA-GAN provides a weak~(showed by the Mantel test) intra-event correlations among the generated detector images while all other models fail to capture any correlation. 
    \item Illustrating the vital role of intra-event sensor-by-sensor correlation in the downstream Physics analysis.
    \item Doing a tracking analysis over the IEA-GAN's generated PXD background where it shows a very close performance to the Geant4 simulated background. Consequently, This study revealed that IEA-GAN, despite only capturing a weak correlation, surpasses PE-GAN and even outperforms the inter-event-shuffled~(uncorrelated) Geant4 in certain HPRs.
\end{itemize}

Using IEA-GAN comes also with a storage release of more than $2$ orders of magnitude. 
Furthermore, due to the dramatic CPU speed-up of $\times \num{147}$ as shown in~\Cref{tab:computational_perf}, It is now possible to employ the IEA-GAN as an online surrogate model for the ultra high-granularity PXD background simulation on the fly, a task that was unattainable before for such a high-resolution detector simulation. This work has a significant impact on high-granularity fast and efficient detector response and collider event simulations. Since they require fine-grained intra-event-correlated data generation, I believe that the Intra-Event Aware GAN~(IEA-GAN) offers a robust controllable sampling for high granular particle physics simulations.
For instance, the High-Luminosity Large Hadron Collider~(HL-LHC)~\cite{bruning_chapter_2020} is expected to surpass the LHC's design integrated luminosity by increasing it by a factor of \num{10}. The upcoming high-granularity Calorimeter~(HGCAL) with roughly \num{6.5}M channels, or the ITk 3D pixel detector at the HL-LHC~\cite{noauthor_frontiers_nodate} with around \num{1}M information channels, will massively increase the geometry and precision complexity, leading to a dramatic increase in the time and storage to simulate the detector~\cite{pedro_current_2019}.
As a result, much more effective and efficient high-resolution detector simulations are required.
IEA-GAN is the first potential candidate for simulating the corresponding high-resolution and high-granular detector signatures with the remarkable capability of generating more than 7.5M pixel channels.

\begin{table}[!htb]
\begin{minipage}{\textwidth}
    \centering
    \caption{
    Computational performance of IEA-GAN and PE-GAN generators on a
    single core of an Intel Xeon Silver~\num{4108}~\num{1.80}GHz~(CPU) and NVIDIA V100 with~\num{32} GB of memory~(GPU) compared to Geant4. For the generative models, the mean and standard deviation were obtained for sets of~\num{10000} events, meaning that the model generates these events one at a time, not in a batch of \num{10000}.
    The time for Geant4 refers to the theoretical time it would take to run the simulation of background processes on the fly, one event at a time. The storage consumption corresponds to storing~\num{10000} events of \num{1} times the PXD background simulated information.}
    
    \label{tab:computational_perf}
    \begin{tabular}{lllll}  
        \toprule
        Hardware & Simulator & time/event~[s] & Storage~[Mb] & Speed-up \\
        \midrule
        \multirow{3}{*}{CPU}  & Geant4 & $\approx 1500$ & $\approx 2000$ & $1$\\
                              & PE-GAN & $11.781\pm 0.357$ & $\approx 47$ & $\approx \times127$ \\
                              & IEA-GAN & $10.159\pm 0.208$ & $\approx 47$ & $\approx \times\textbf{147}$ \\
        \midrule
        \multirow{2}{*}{GPU}  & PE-GAN & $0.090\pm 0.010$ & $\approx 47$ & $\approx \times16667$ \\
                              & IEA-GAN & $0.070\pm 0.006$ & $\approx 47$ & $\approx \times21429$ \\
        \bottomrule
    \end{tabular}
\end{minipage}
\end{table}

The ability to capture the underlying correlation structure of the data in particle physics experiments where the physical interpretation of the results heavily relies on it is very important. The true correlation between the occupancy of the sensors is determined by the underlying physical processes within the simulation. 
Although the true correlation differs from the one captured by IEA-GAN, the model might be learning biases or artifacts introduced by the training data or the discriminator. Therefore, while the IEA-GAN can provide valuable insights into the correlations and patterns present in the data, it is important to interpret its results in conjunction with the domain knowledge. To alleviate the discrepancy, I expect that incorporating perturbations directly into the discriminator's RRM module would improve its contextual understanding and generalization. For example, using random masking~\cite{liu_roberta_2019} or inter-event permutation~\cite{yang_xlnet_2020} over the samples and asking the RRM module to predict the representation of perturbed sample could improve the robustness of the model, a concept that will be studied in the next chapter. 

In the next chapter, I will be dealing with the real~(random trigger) PXD background data. I will show that the sparsity~(occupancy) profile of the real PXD background data is much more diverse than the Geant4 simulated data. 
Moreover, the intra-event correlation of the real PXD data is dynamic across Belle~II runs and has different monotonic and linear behavior. 
Hence, the main question will be whether IEA-GAN can handle the real PXD sparsity and complex intra-event correlation or whether I need a fresh perspective if I want to deal with these new challenges. 

  \chapter[PXD Background Generation:
Real Data]{PXD Background Generation: Deus Ex Machina the Real Detector Data}
\label{chap:6}

\section{Introduction}
During the last chapter, with the story of IEA-GAN conditioned on sensor numbers~(as meta categories), many new concepts were introduced. The whole study was based on the Geant4 simulated PXD background data. 
The current chapter approaches the final challenge of the PXD detector simulation, studying the real~(random trigger) PXD background conditioned on the rate of background or occupancy~(as subordinate categories). 
During this path, I ask a more fundamental question. Is it possible to develop a surrogate model not only for amortized generation but also for Out-Of-Distribution~(OOD) simulation?
One of the most intriguing open problems in the realm of Deep Generative Models for Particle Physics is the possibility of simulating particle detector signatures that extend beyond the known experimental conditions. 
Can one extend these simulations into the OOD territories in a truly daring venture?

The essence of this query is embedded in our quest to transcend the simulation boundaries. In~\cref{chap:4}, I elucidated the foundation of OOD generation in detector simulation, reviewing the current works being done in this domain. 
From the real PXD data perspective, the idea of generating PXD hits conditioned over the amount of background~(rate of background) and luminosity even beyond the current experimental data is still an open problem. 
In this chapter, I dive deeper into this regime, elaborate on the challenges, and propose possible solutions.

This is where \emph{YonedaVAE} enters, striving to attack this problem with Self-Supervised context extrapolation. YonedaVAE not only emulates real PXD signatures~(an In-Distribution simulation regime) but also extrapolates to the amount and profile of background beyond the training data~(Out-Of-Distribution simulation regime).
Along the way, I set the stage for a unified perspective of relational reasoning, the Yoneda perspective. The Yoneda perspective, rooted in the foundations of Category Theory, is an abstract mathematical lens that deals with objects and their relationships in an algebraic context. This perspective emphasizes a shift in focus from studying objects in isolation to studying them in relation to others. 
Moreover, I also introduce a new set creation mechanism, \emph{Adaptive Top-q sampling}, to adaptively create sets with variable inter-category~(inter-event) and intra-category~(intra-event) cardinalities. 
As a result, YonedaVAE, a point cloud generative model trained on low luminosity data, not only possesses the power to generate sparse ultra-high granularity PXD point cloud data but to do so for double the training luminosity region with a finesse and adaptability that holds the promise to redefine OOD simulations.

Toward the end, I also provide novel evaluation methods for detector simulations using Topological Data Analysis~(TDA). I demonstrate the fresh perspective of TDA to analyze how and to what order of pattern complexity the PXD background hits are oriented and clustered. 

\section{Real PXD data and OOD Simulation: Challenges and Motivations}

\begin{figure}[!htb]
    \centering
    \begin{subfigure}[b]{0.45\textwidth}
        \includegraphics[width=\textwidth]{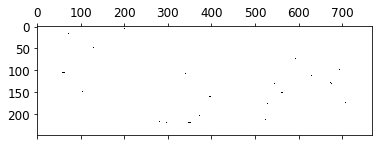}
    \end{subfigure}
    \begin{subfigure}[b]{0.45\textwidth}
        \includegraphics[width=\textwidth]{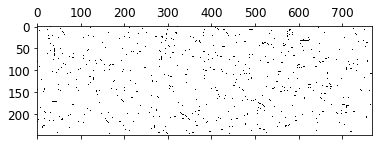}
    \end{subfigure}
    
    \begin{subfigure}[b]{0.45\textwidth}
        \includegraphics[width=\textwidth]{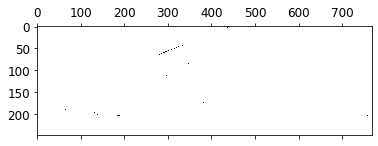}
    \end{subfigure}
    \begin{subfigure}[b]{0.45\textwidth}
        \includegraphics[width=\textwidth]{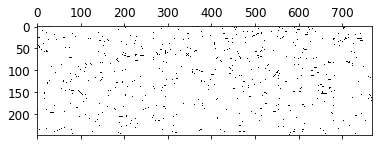}
    \end{subfigure}
    
    \begin{subfigure}[b]{0.45\textwidth}
        \includegraphics[width=\textwidth]{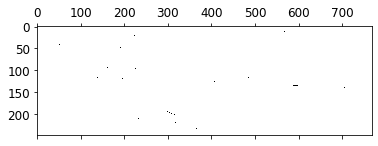}
    \end{subfigure}
    \begin{subfigure}[b]{0.45\textwidth}
        \includegraphics[width=\textwidth]{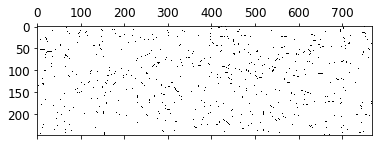}
    \end{subfigure}
    
    \caption{Random-trigger~(real) PXD background~(left) and Geant4 generated PXD background~(right) in a hitmap format. As stated before, the sparsity of the real PXD background is much less than the Geant4 version.}
    \label{fig:pxd_realvsgeant}
\end{figure}

In the realm of detector physics data, achieving OOD~(Out-of-Distribution) simulation demands transcending conventional homogeneous data representations, such as images and sequences typically used in experimental particle physics. The intent is to amplify the adaptability of generative models, thus enabling them to cater to unordered and variable-sized sets~\cite{kansal_particle_2022,di_bello_conditional_2022,buhmann_epic-gan_2023}. 
The first motivation for adopting unordered and variable-sized representations such as point clouds for the real~(random trigger) PXD background is its high sparsity and vast variance in number of hits compared to the Geant4 version~(see~\cref{fig:pxd_realvsgeant}). For example, in the recent PXD data taking, Experiment 26, the number of hits of sensors varies in the range $[10,5300]$ with a standard deviation $\sigma = 400$ and the mean around $500$ as shown in~\cref{fig:card_extr}. That is why models based on homogeneous data representations~(image-based) such as IEA-GAN perform poorly~(as will be shown in~\cref{tab:fid}). This is because Convolutional Neural Network~(CNN)-based methods rely on dense tensors, which makes them suboptimal for spatially sparse data. 

Thus, data representation versatile multi-sets becomes crucial in detector simulation, given the data irregularities. Examples of these irregularities include high variances in number of hits, gaps in data due to developmental factors~(handling scarce or absent detector information), irregular detector geometries, and parameter extrapolation. 

Historically, the primary objectives behind the current multi-set~(point cloud) generative methods were two-fold: either to replicate detector responses based on specific sensor attributes and overarching conditions or to generate incidents within a target kinematic zone. 
However, these methodologies often stumbled when it came to generalizing to situations where the data is sparse or entirely absent~\cite{di_sipio_dijetgan_2019,butter_jet_2023,alanazi_survey_2021,noauthor_learning_nodate} like going beyond the kinematic regions of the training data or the situations where the upper bound of set cardinalities is well-beyond the training data like a geometry-independent detector simulation. 
For the PXD background case, an issue with using real random trigger data is its dependence on the experimental runs. Thus, real luminosity and beam-parameter-dependent PXD background data beyond the current experiments will be absent. This signals a need for a surrogate model to generalize well to OOD luminosity regions.

Further complicating matters is the intrinsic challenge deep neural networks face in generating outputs that deviate from their training distribution~\cite{arora_generalization_2017,xu_how_2021,feng_principled_2022,kocaoglu_causalgan_2017}. Looking beyond the domain of Experimental Particle Physics, the concept of OOD generation encompasses intriguing applications. One notable example is the \emph{de Novo design}, prevalent in producing innovative protein structures or molecular compounds with desired attributes~\cite{lotfollahi_conditional_2020,lee_exploring_2022}. Rather than merely reflecting existing data distributions, the de Novo approach leverages generative models to conjure entirely novel samples. The focus is tailoring these samples to exhibit preferred traits, be it stability, binding affinity, or protein foldability~\cite{noauthor_novo_nodate,meyers_novo_2021,chen_rise_2018}.

As one delves deeper into the OOD simulation for non-uniform data structures~(multi-sets), three pivotal elements emerge:

\begin{enumerate}
    \item The mechanism the model employs for generating sets that mirror the detector hit representation, whether represented as graphs or point clouds~(multi-sets).
    \item The model's capacity to do length extrapolation beyond the training data.
    \item The model's adeptness at handling control parameters and performing context extrapolation beyond the training data.
\end{enumerate}

Inspired by this question, this chapter develops a set of novel methods that enable a point cloud deep generative model to be able to extrapolate beyond the training data in a zero-shot manner. The proposed model trained on the detector response data with a specific intra-event correlation and limited cardinality profile can be generalized to a new set of intra-event correlation and cardinality profiles. 
In the subsequent sections, we'll embark on a more in-depth exploration of these elements.

\subsection{Set Generation:}
\label{section:problem}
Graph or point cloud generation commonly engages in a two-fold process. Initially, point sampling occurs in a stochastic manner, primarily from a standard Gaussian distribution. Subsequently, the latent vector will be fused with each sampled point. Notably, this results in the manifestation of exchangeable distributions, whereby all potential permutations of a given set are equally probable. Such a property is often hailed as a generative model's counterpart to the concept of equivariance.

The main objective revolves around understanding the intricacies of designing a probabilistic decoder, denoted as $f$, whose goal is to transform latent vectors $\mathbf{z} \in \mathbb{R}^{d\_lat}$ to sets~(multi-sets). These sets, denoted as $\mathcal{X} = \{\mathbf{x}_1, \ldots, \mathbf{x}_n\}$, encompass a variable number $n$ of points, cardinality, $\mathbf{x}_i \in \mathbb{R}^d$. 
The challenge is to shape $f$ in a manner that when $\mathbf{z}$ is sourced from a prior distribution $p_Z(\mathbf{z})$, the resulting push-forward measure, represented as $f_\#(p_Z)$, aligns closely with an unknown distribution $\mathcal{D}$.

\begin{figure}[!htb]
\centering
    \includegraphics[width=0.7\textwidth]{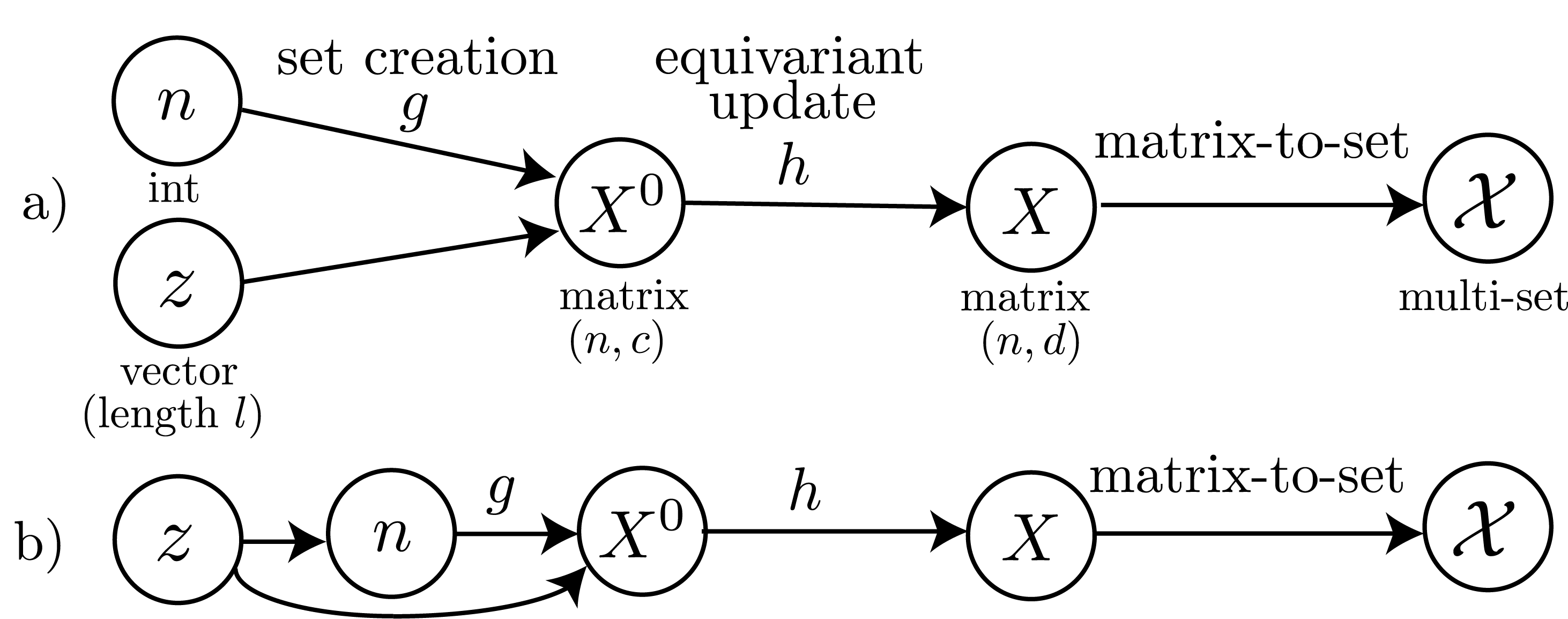} 
    \caption{Illustrative models for set generation borrowed from~\cite{vignac_top-n_2022}. The set's point count can either be derived from the dataset's distribution (a) or inferred from the latent vector (b). Although any equivariant function can serve for the update $h$, the set creation $g$ encapsulates the intricacies of set generation. When generating graphs, edge weights complement the node features matrices $\mathbf{M}^0$ and $\mathbf{M}$.}
    \label{fig:graphical-model}
\end{figure}

Focusing on one-shot generation architectures, they predominantly align with the models depicted in Figure \ref{fig:graphical-model}. The process commences by sampling a predetermined number of points for the set. A majority of methodologies presuppose the availability of set cardinalities during the training phase. During the generative phase, $n$ is usually sampled from the distribution corresponding to set cardinalities from the training samples. Under this strategy, the latent vector $\mathbf{z}$ remains independent of the point count $n$, leading to the generative task of $p(\mathcal{X} | n, \mathbf{z}) \cdot p(n) \cdot p(\mathbf{z})$.

In contrast,~\cite{kosiorek_conditional_2020} advocates for deducing the value of $n$ directly from the latent vector via an MLP. Despite training this layer through an auxiliary loss, its predicted outcome only occurs during the generation stage. When relying on ground truth during training, the generative model transforms to $p(\mathcal{X} \| \mathbf{z}, n) \cdot p(n | \mathbf{z}) \cdot p(\mathbf{z})$, indicating the interdependence of $\mathbf{z}$ and $n$.

Upon determining $n$, the one-shot generation blueprints can be dissected into distinct components. The primary function, $g$ (the \textit{creation function}), maps the latent vector to an initial set $\mathbf{M}^0 \in \mathbb{R}^{n \times d\_hid}$. Given the inherent simplicity of this function, it often falls short in modeling intricate inter-dependencies within each set. To overcome this, the initial set $\mathbf{M}^0$ may undergo further refinement via a secondary \textit{update} functions, $h$ that can be any equivariant modules. 
The succeeding segments delve into the proposed formulations for the set creation modules and the challenges they face with OOD set generation. 

\begin{figure}
    \centering
    \includegraphics[width=\textwidth]{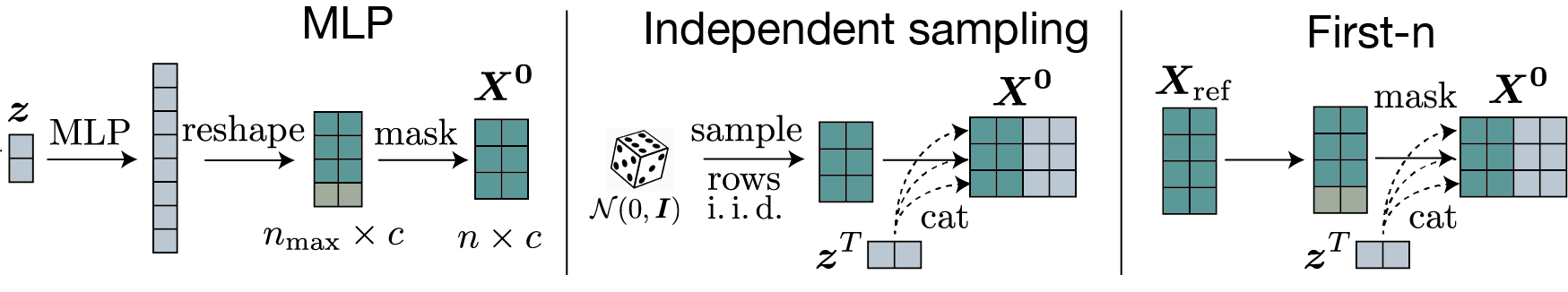}
    \caption{Existing creation methods, borrowed from~\cite{vignac_top-n_2022}, for mapping a latent vector $ \mathbf{z} $ to a set of points $ \mathbf{X}^0 $. First-n creation empirically gives the best performance. It learns a reference set represented by a matrix $ \mathbf{X}_{\text{ref}} $, and concatenates the latent vector to each point of this set.}
    \label{fig:creation-methods}
    \vspace{-0.3cm}
\end{figure}

\textbf{MLP Approach}
The adoption of MLPs is a recurring theme across several existing methodologies. These methods conventionally train an MLP to transform from $\mathbb{R}^{d_{\text{lat}}}$ to $\mathbb{R}^{\text{n}_{max} \times d_{\text{ch}}}$, where $\text{n}_{max}$ symbolizes the maximum set cardinality encountered in the training dataset. The output, post-training, is a vector that is subsequently reshaped into a $\text{n}_{max} \times d_{\text{ch}}$ matrix. A masking procedure is applied to cater to the problem at hand, retaining only the initial $n$ rows, as visualized in~\cref{fig:creation-methods}. 

MLPs, however, come with inherent constraints. Foremost among them is their inability to recognize and account for the symmetries intrinsic to the problem, which will be a huge problem with OOD set creation. Additionally, their training restricts them from generating a maximum of $\text{n}_{max}$ points, effectively limiting their capacity to venture into extrapolating larger set sizes. 

\textbf{i.i.d. Method}
A widely embraced approach is the strategy of independent and identically distributed~(i.i.d.) sampling. In this method, $n$ points are drawn i.i.d. from a standard or low-dimensional Gaussian distribution. Following this, the latent vector is appended to each of these samples \cite{kohler_equivariant_2020,luo_diffusion_2021, kosiorek_conditional_2020,kansal_particle_2022,di_bello_conditional_2022,buhmann_epic-gan_2023}. 
The primary advantage of this methodology is its inherent preservation of permutation invariance, which is essential for many set and graph-related tasks. 

However, a notable shortcoming is that DGMs built with an i.i.d. set generation mechanism fail to fit the train data correctly~\cite{wang_graphgan_2017,krawczuk_gg-gan_2020, zhang_understanding_2021}, which reflects in the poor quality/variability of the sampled sets due to excessive sources of randomness.

\textbf{The First-n Approach}
Some researchers have adopted a distinct technique rather than the traditional practice of point sampling. Proposed by~\cite{noauthor_deepfunc_nodate} and~\cite{krawczuk_gg-gan_2020}, this approach entails the use of a constant, learnable set, denoted as $\mathbf{X}_{\text{ref}} \in \mathbb{R}^{\text{n}_{max} \times d_{\text{ch}}}$. From this set, only the initial $n$ rows are retained, leading to the terming of this technique as the \emph{First-n creation}. Once established, this set has the latent vector attached to each of its individual points. 

From empirical analyses, it has been discerned that the First-n strategy tends to achieve convergence at a more rapid pace when used with different sampling methods~\cite{krawczuk_gg-gan_2020}. However, due to its deterministic nature, this method inherently restricts generation to a ceiling of $\text{n}_{max}$ points, making it incapable of extrapolating towards larger sets. 
An implicit bias is also introduced due to the recurrent selection of the first $n$ rows from the reference set $\mathbf{X}_{\text{ref}}$. This results in a disproportionate frequency of selection for the leading rows compared to those towards the end.

\textbf{The Top-N Mechanism}
Moving away from the conventional i.i.d. sampling for set generation, Vignac et al. introduced a new approach~\cite{vignac_top-n_2022}. Their method revolves around a trainable reference set, $\mathbf{R}_{\text{ref}}$. Instead of indiscriminately sampling from this set, the latent vector is employed to select the top $N$ most relevant points, giving rise to the name \emph{Top-N creation}. Once these points are identified, they are processed in conjunction with the prior vector. Empirical studies have shown that the Top-N mechanism surpasses traditional set generation in various benchmarks. 

However, despite solving many issues in set generation for ID~(In-Distribution) generation regime, they still lack the possibility to generate sets with variable sizes that belong to the same category. They do the sampling by grouping an equivalent class of sets with the same cardinality, which can easily bias the model toward more frequent cardinalities at inference time, indicating a lack of diversity in the generated set.
Another limitation of this approach in OOD generation is its dependency on the predefined reference set $\mathbf{R}_{\text{ref}}$, which might not always capture the full diversity of the data if the inference set cardinality is much higher than the training set cardinality. 

In this chapter, I go to the next level for OOD set generation by introducing \emph{Adaptive Top-q Sampling}. With Adaptive Top-q Sampling, instead of relying on a fixed top-k or using a temperature parameter to control the shape of the distribution without sufficiently suppressing the unreliable tail, I propose sampling from the top-$q_i$ portion of the probability mass where $q_i$ is adaptively learned in context for each sample $i$, expanding and contracting the candidate pool dynamically for each sample.

\subsection{Length Extrapolation}
\label{sec:pos_emb}
While sampling sets with correct cardinality stochastically and maintaining the symmetries of the problem during training is of paramount importance, it does not guarantee proper extrapolation during decoding inference. In other words, given a generated set, it has to be mapped to the final representation~(point cloud or graph) in such a way that the model can extrapolatively reason~(update) to a longer set size.
The length~(cardinality) of the sets in the ID training, $n_{tr}$, is an important factor.  
Due to the above-mentioned motivations, one will use the model for substantially longer evaluation set cardinalities $n_{ex}\gg n_{tr}$ where only control parameters/attributes exist for the OOD. 
The discrepancy between $n_{tr}$ and $n_{ex}$ motivates the task of~\textbf{length extrapolation}~\cite{bevilacqua_size-invariant_2021,press_train_2022}:~\emph{Can a set decoder model maintain equally good, if not better, perplexities when longer sequences are used in the testing stage?}

It is a well know fact that DeepSet-based models~\cite{zaheer_deep_2018, zhang_deep_2023, buhmann_epic-gan_2023} have difficulty extrapolating to set sizes~\cite{kosiorek_conditional_2020,jurewicz_set--sequence_2021,kim_transformers_2022}. As an alternative, Attention-based models are the other equivariant candidates for set updates. 
However, It has been recently~\cite{press_train_2022,ontanon_making_2022} demonstrated that the vanilla Transformer Encoder fails to extrapolate to longer sequences, which is caused by the position embedding method. On the other hand, any positional information injection to the attention-based models will break the exchangeability of the set by order variance.
From the DGM perspective, as I have shown in \cref{chap:3}, sufficient conditions for $(F, l)$ - equivariance, the VAE's decoder, and GAN's generator does not need to carry exchangeability condition for their corresponding set update. 
As a result, their length extrapolation can be enabled by changing the positional embedding policies. 

Transformers, unlike RNNs, are designed to process data in parallel and require positional encoding to capture point order. 
Positional encodings can be either \emph{absolute}, which signifies each exact position (like 1, 2, 3, ...), or \emph{relative}, denoting the distance between tokens. Let's dive into various ways of incorporating positional information and their ability in length extrapolation.

\textbf{Absolute Position Embedding}~(APE) converts each position $i$ into a unique vector $p_i$, which is added to word embeddings before introducing them to the model as:
\[ e_i = w_i + p_i \]
Where \(e_i\) represents the combined embedding for the token at position \(i\), incorporating both its semantic information (from \(w_i\)) and its positional information (from \(p_i\)). For a given position \(i\), the non-parametric approach to APE, as introduced in the original Transformer model~\cite{vaswani_attention_2017}, calculates the positional encoding using sine and cosine functions:
\[ p_i^{(2k)} = \sin\left(\frac{i}{10000^{2k/d}}\right) \]
\[ p_i^{(2k+1)} = \cos\left(\frac{i}{10000^{2k/d}}\right) \]
Where \(d\) is the dimensionality of the embeddings and \(k\) ranges from \(0\) to \(d/2-1\). These functions ensure the model can generate positional embeddings for any position, even if it hasn't seen it during training. 

In contrast, some models, including GPT3~\cite{brown_language_2020} and OPT~\cite{zhang_opt_2022} instead of relying on fixed functions like sine and cosine, models, introduce trainable position embeddings. In this approach:
\[ p_i = \text{LearnableParameter}(i) \]
Here, \(p_i\) is initialized randomly and updated during the training process along with other model parameters. The drawback, however, is that since these embeddings are learned from the training data, the model cannot naturally extrapolate to positions beyond the range it has seen during training. However, it's restricted to known positions and cannot extrapolate. 

\textbf{T5's Relative bias}~\cite{raffel_exploring_2020} computes the bias between token positions $i$ and $j$ using a lookup function, adding this bias to the self-attention mechanism's dot product. Distances beyond a specific range map to a common parameter, ensuring adaptability to unknown distances.
Specifically, for any two token positions \(i\) and \(j\), the relative distance \(d\) is given by:
\[ d = i - j \]

The lookup function \( f \) is used to determine the bias associated with this relative distance:
\[ \text{bias}(d) = f(d) \]

In the self-attention mechanism, the attention score between two positions is typically computed using the dot product of the query and key vectors. Let's denote the query for position \(i\) as \( Q_i \) and the key for position \(j\) as \( K_j \). Without considering any bias, the attention score \( S \) would be:
\[ S(i,j) = Q_i \cdot K_j^T \]

However, with T5's relative bias, the attention score is modified as follows:
\[ S(i,j) = Q_i \cdot K_j^T + \text{bias}(d) \]

For values of \(d\) that exceed a certain threshold, say \( \Delta \), the lookup function maps them to a consistent bias value:
\[ 
f(d) = 
\begin{cases} 
f(d) & \text{if } |d| \leq \Delta \\
f_{\text{max}} & \text{otherwise}
\end{cases}
\]

Where \( f_{\text{max}} \) is a common parameter value representing the maximum bias. This feature ensures the model remains resilient and adaptable even when encountering unfamiliar relative distances between tokens.

\textbf{Rotary},(ROPE) introduces a novel technique for encoding relative positions in the context of the self-attention mechanism of Transformers.
In essence, the ROPE method revolves around rotation operations applied to the query and key representations based on their absolute positions, which in turn ensures that the attention mechanism effectively focuses on the relative distances between tokens.
For any given token position $i$, the Rotary mechanism calculates a rotational positional encoding:
\[ R(i) = \begin{bmatrix} \cos(\theta(i)) & -\sin(\theta(i)) \\ \sin(\theta(i)) & \cos(\theta(i)) \end{bmatrix} \]
Where \(\theta(i)\) is a function of the absolute position \(i\). Then, before feeding the query and key representations into the attention mechanism, they are rotated using their corresponding positional encodings. Let's denote the query for position \(i\) as \( Q_i \) and the key for position \(j\) as \( K_j \). Their adjusted representations are:
\[ Q'_i = Q_i \times R(i) \]
\[ K'_j = K_j \times R(j) \]
Afterward, the dot product attention score between the adjusted query and key representations will be $S(i,j) = Q'_i \cdot {K'_j}^T$. Because both \( Q'_i \) and \( K'_j \) have been rotated based on their absolute positions, the attention score \( S(i,j) \) is inherently based on the relative distance \( d = i - j \). In other words, despite the Rotary mechanism using absolute positional information to adjust the representations, the attention mechanism ultimately focuses on the relative distances between tokens, ensuring a more context-aware attention distribution.

By this method, the Rotary mechanism in models such as PaLM~\cite{chowdhery_palm_2022} and LLaMA~\cite{touvron_llama_2023} effectively leverages both absolute and relative positional information to enhance the self-attention mechanism's contextual understanding~\cite{su_roformer_2022}.

\textbf{ALiBi},~\cite{press_train_2022} in BLOOM~\cite{workshop_bloom_2023}, acts similarly to T5's Relative Bias but reduces a scalar bias from the attention score. The bias increases with the distance between tokens, leading to a tendency towards recent tokens. When using ALiBi, position embeddings are not added at any point in the network. The only modification applied is after the query-key dot product, where a static, non-learned bias is added:
\[ \text{softmax}(q_i K^{\top} + m \cdot [- (i - 1), \ldots, -2, -1, 0]) \]
The scalar \(m\) is a head-specific slope that is fixed before training. This bias penalizes the attention scores between distant query-key pairs, with the penalty increasing as the distance between a key and a query grows. The different heads increase their penalties at different rates, depending on the slope magnitude. For models with 8 heads, the slopes used are the geometric sequence:
\[ \frac{1}{2^1}, \frac{1}{2^2}, \ldots, \frac{1}{2^8} \]
This method introduces an inductive bias towards \emph{recency}, penalizing attention scores between distant query-key pairs. The penalty increases as the distance between a key and a query grows. Depending on the slope magnitude, different heads increase their penalties at different rates.

\textbf{XPOS}, introduced in the Length-Extrapolatable Transformer~\cite{sun_length-extrapolatable_2022}, offers a novel approach to relative position modeling in Transformers. 
The core idea behind XPOS is to add the same decay as in ALiBi that leverages the extrapolation to all distances. 

For a given token position $j$, XPOS calculates a position encoding based on exponential and rotational transformations:
\[ fq(q, j) = qe^{\xi j + i\theta j} \]
\[ fk(k, j) = ke^{-\xi j - i\theta j} \]
Where \( \xi \) and \( \theta \) are parameters that determine the rate of exponential decay and rotation, respectively. If \( \xi = 0 \), the encoding is similar to the ROPE method. The transformation provides a rotation on vectors, ensuring that the relative angle between vectors $q$ and $k$ determines the inner product. The attention scores between the transformed query and key representations inherently focus on the relative distance between tokens. This is achieved by ensuring that the inner product between vectors with a larger relative angle is smaller, thereby capturing the relative distances effectively.

It has been shown~\cite{press_train_2022,sun_length-extrapolatable_2022} showed that the APE method in practice has very limited extrapolation capabilities. On the other hand, the rotary position method improves over the sinusoidal one, but it still does not achieve satisfying results. Interestingly, they observe that the T5 bias method leads to better extrapolation than either of these, concluding that extrapolation ability depends heavily on the position
embedding.

In this study, I incorporate Rotary positional encoding for detector geometry and introduce modifications to ALiBi positional embeddings to be more compatible with bidirectional models~(they were introduced for autoregressive models) and, in general, with OOD generalization.

\subsection{Context Extrapolation:}

Although there are very few works on Context Extrapolation for generative models, the common theme amongst the remaining is ``controlled generation''.
When one is dealing with a controlled generation, it involves generating samples characterized by a particular attribute, feature, or context. 
In Natural Language Processing~(NLP), the context can be sentiment style as a prompt. In computational biology and chemistry, the context can represent attributes like stability, fluorescence, and binding affinity of a protein sequence.
In event generation and detector simulation, the context can be the kinematic or luminosity profile, the number of detector hits~(amount of background hits), and detector geometry. 

OOD generation focuses on generating samples with context values that extrapolate beyond the training distribution. 
For example, in protein design, the problem of \emph{de novo} design, simulation of novel protein sequences with respect to some context features such as binding afﬁnity to a speciﬁc target, is of vital importance to drug discovery~\cite{freschlin_machine_2022,yeh_novo_2023,gainza_novo_2023}. 
In event generation, going beyond the manifold of training events can bring physical insight into regions without any experimental data. 

While controlled generation is prevalent in literature, its capability for extrapolation remains largely unexplored. The current approaches can be categorized into the following,

\begin{itemize}
    \item \textbf{Control Code Methods:} These employ control codes encoding attributes either as discrete values~(similar to style transfer) or scalar values~\cite{keskar_ctrl_2019,noauthor_progen_nodate,madani_large_2023}. However, discrete codes often face challenges when exposed to unseen attribute values.
    
    \item \textbf{Iterative Editing Methods:} Edit-based methods focus on iterative refinements~\cite{guu_generating_2018,mallinson_edit5_2022,novak_iterative_2018}. Some recent works involved few-shot model learning by receiving feedback on generated samples and learning to edit itself~\cite{welleck_generating_2022,padmakumar_extrapolative_2023}.
    
    \item \textbf{Latent Variable Models:} With latent variable models, control features are modeled inside latent variables. For instance, Genhance introduces a method where the latent vector comprises both attribute-relevant and irrelevant components~\cite{chan_deep_2021}. However, stability issues arise when these models operate on set-based data.
    
    \item \textbf{Scorer Model for Attribute Control:} Some models incorporate attribute information through a scorer model during inference, guiding an unconditional generative model~\cite{dathathri_plug_2020,yang_fudge_2021,noauthor_denoising_nodate}. While effective, this method is not always suitable for context extrapolation, especially when it goes beyond the training data distribution. The classifier-type model, when used as a reward model for reinforcement learning, also presents challenges~\cite{gong_reinforcement_2019,angermueller_model-based_2020,amodei_concrete_2016,michaud_understanding_2020,pang_reward_2023} as the generator can exploit and amplify imperfections in the reward.
\end{itemize}

With YonedaVAE, I introduce methods as a fusion of Control Code Methods and Latent Variable Models and show that it can reach context extrapolation with Transformer-based models by synthesizing Self-Distillation and VAEs in conjunction with introducing a novel control-gating for the attention mechanism. However, before going deeper into the YonedaVAE, I first have to introduce a building block of this model: Category Theory. Using the concepts of category theory, I try to show how it relates mathematically to \emph{Yoneda pooling} module, a relational pooling mechanism introduced in the YonedaVAE's encoder. 

\section{Prologue: Category Theory and Yoneda Perspective}
\label{sec:category}
In this section, I briefly introduce the concept of Category theory~(CT). Consequently, I will demonstrate how CT acts as a universal language to formulate the Yoneda pooling mechanism that involves relational intra-event dependencies. 

\subsection{What is Category Theory?}
\label{sec:cats4ai}
Category theory, initiated by Eilenberg and Mac Lane~\cite{eilenberg_general_nodate}, provides a unifying and abstract framework in which mathematical structures and their interrelationships can be studied. CT can be seen as a language of structure, abstraction, and relationships. It allows us to describe mathematical~(and non-mathematical) concepts in a highly general and structural way, focusing on the relationships between these concepts rather than their internal details. This powerful bird's-eye view of mathematics makes CT an invaluable tool with diverse applications in areas such as logic, Physics, Chemistry, Computer Science, and AI~\cite{de_haan_natural_2020,dudzik_graph_2022}. 
A Category is any collection of objects that can relate to each other via morphisms like functions with specific features, like composition and associativity. So, the collection of sets with functions forms a category, as does the collection of groups with group homomorphisms and topological spaces with continuous functions.

\subsection{Basic Definitions}
A \textit{category} $\mathcal{C}$ consists of the following data:
\begin{itemize}
\item A class $\text{ob}(\mathcal{C})$ of \textit{objects}.
\item For every pair of objects $A, B \in \text{ob}(\mathcal{C})$, a set $\text{Hom}_{\mathcal{C}}(A, B)$ of \textit{morphisms} or \textit{arrows} from $A$ to $B$.
\item For every object $A \in \text{ob}(\mathcal{C})$, a specified morphism $1_A \in \text{Hom}_{\mathcal{C}}(A, A)$ called the \textit{identity} on $A$.
\item For every triple of objects $A, B, C \in \text{ob}(\mathcal{C})$, a binary operation $\text{Hom}_{\mathcal{C}}(B, C) \times \text{Hom}_{\mathcal{C}}(A, B) \rightarrow \text{Hom}_{\mathcal{C}}(A, C)$ called \textit{composition} of morphisms.
\end{itemize}

These data are subject to the following axioms:
\begin{itemize}
\item (Associativity) For every quadruple of objects $A, B, C, D \in \text{ob}(\mathcal{C})$ and every triple of morphisms $f \in \text{Hom}_{\mathcal{C}}(A, B)$, $g \in \text{Hom}_{\mathcal{C}}(B, C)$, $h \in \text{Hom}_{\mathcal{C}}(C, D)$, one has $(h \circ g) \circ f = h \circ (g \circ f)$.
\item (Identity) For every pair of objects $A, B \in \text{ob}(\mathcal{C})$ and every morphism $f \in \text{Hom}_{\mathcal{C}}(A, B)$, one has $f \circ 1_A = f = 1_B \circ f$.
\end{itemize}

\begin{figure}[ht]
\centering
\begin{tikzcd}
A \arrow{r}{f} \arrow[swap]{dr}{g \circ f} & B \arrow{d}{g} \\
 & C
\end{tikzcd}
\caption{Commutative diagram in a category}
\label{fig:commutative_diagram}
\end{figure}

\subsection{Functors and Natural Transformations}
A \textit{functor} $F: \mathcal{C} \rightarrow \mathcal{D}$ between two categories $\mathcal{C}$ and $\mathcal{D}$ is a mapping that assigns to each object $A \in \text{ob}(\mathcal{C})$ an object $F(A) \in \text{ob}
(\mathcal{D})$ and to each morphism $f \in \text{Hom}_{\mathcal{C}}(A, B)$ a morphism $F(f) \in \text{Hom}_{\mathcal{D}}(F(A), F(B))$, such that the following conditions hold:
\begin{itemize}
\item $F(g \circ f) = F(g) \circ F(f)$ for all morphisms $f, g \in \text{Hom}_{\mathcal{C}}$ for which the composition $g \circ f$ is defined.
\item $F(1_A) = 1_{F(A)}$ for all $A \in \text{ob}(\mathcal{C})$.
\end{itemize}

A \textit{natural transformation} $\eta: F \Rightarrow G$ between two functors $F, G: \mathcal{C} \rightarrow \mathcal{D}$ is a family of morphisms $\eta_A: F(A) \rightarrow G(A)$ in $\mathcal{D}$, one for each object $A \in \text{ob}(\mathcal{C})$, such that for every morphism $f: A \rightarrow B$ in $\mathcal{C}$, the following square commutes in $\mathcal{D}$:

\begin{figure}[!hbt]
\centering
\begin{tikzcd}
F(A) \arrow{r}{\eta_A} \arrow[swap]{d}{F(f)} & G(A) \arrow{d}{G(f)} \\
F(B) \arrow{r}{\eta_B} & G(B)
\end{tikzcd}
\caption{Commutative square for a natural transformation}
\label{fig:natural_transformation}
\end{figure}

The composition of natural transformations is defined component-wise. This turns the class of all functors between two fixed categories into a category itself, with natural transformations as morphisms. This concept of a category of functors is essential for the definition of a Yoneda lemma, a central concept in category theory.

\textbf{An experimental Example:} In an optical laboratory, think of each experimental setup - whether involving simple lenses, prisms, or mirrors - as \emph{objects} in Category Theory. The transformations we apply to these setups, such as adding a filter to a light source or replacing a flat mirror with a concave one, act as \emph{morphisms}, reflecting how one setup can evolve into another. The act of consecutively adding a filter and then changing a mirror represents the \emph{composition} of morphisms. Importantly, the final experimental outcome remains consistent regardless of the order of these transformations, highlighting \emph{associativity}. 
Some transformations, like placing an undistorted glass sheet in front of a light, might leave the setup unchanged, analogous to the \emph{identity} morphism in Category Theory. Now, picture a neighboring acoustic lab where experiments with sound waves parallel our optical experiments. Transitioning our thought process from optics to acoustics while retaining the foundational concepts (like translating optical reflection to acoustic echo) serves as a \emph{functor}, bridging these two distinct ``categories'' of experiments. Within the acoustics realm, subtle changes, such as transitioning from air to water as the medium and preserving the experiment's essence, can be linked to \emph{natural transformations} between functors. Through this lens, the optical lab analogy offers a tangible depiction of Category Theory's essence.

In the context of AI, category theory offers a novel framework for understanding and unifying learning algorithms~\cite{de_haan_natural_2020,dudzik_graph_2022,ong_learnable_2022,nemecek_coinductive_2023}. 
As I will demonstrate in this thesis, category theory can also provide a powerful tool to have a fresh and unified understanding of intra-event relational reasoning, enabling the development of our proposed system,\textit{YonedaVAE}.

\subsection{The Yoneda Perspective}
The Yoneda Perspective states that mathematical objects are completely determined by their relationships to other objects.
If one wants to understand objects~(sets, groups, topological spaces, etc.), then one has to understand the network of relationships they enjoy with all the other objects of their species~\cite{mazur_when_nodate}. 

\subsubsection{Yoneda Embedding and Yoneda Lemma}
Yoneda Embedding offers a deep insight into the structure of a category by analyzing the set of morphisms (or functors) from a given object (or functor) to all other objects (or functors) in the category. Consider a category \(\mathcal{C}\) and its category of functors \([ \mathcal{C}^{op}, \text{Set} ]\), where \(\mathcal{C}^{op}\) is the opposite category of \(\mathcal{C}\) and \(\text{Set}\) is the category of sets. For each object \(A\) in \(\mathcal{C}\), there's a functor \(h_A: \mathcal{C}^{op} \rightarrow \text{Set}\) given by:

\[ h_A(B) = \text{Hom}_{\mathcal{C}}(B, A) \]

for every object \(B\) in \(\mathcal{C}\), and for every morphism \(f: B \rightarrow B'\) in \(\mathcal{C}\), \(h_A(f)\) maps a morphism \(g: B' \rightarrow A\) to its composition \(g \circ f\). 

The Yoneda Embedding states that the object \(A\) in \(\mathcal{C}\) can be embedded into the category \([ \mathcal{C}^{op}, \text{Set} ]\) via the functor \(h_A\). 

The Yoneda Lemma establishes a one-to-one correspondence between the natural transformations from \(h_A\) to any functor \(F: \mathcal{C}^{op} \rightarrow \text{Set}\) and the elements of \(F(A)\). Formally, the Yoneda Lemma asserts that for any functor \(F: \mathcal{C}^{op} \rightarrow \text{Set}\), there is a natural isomorphism:

\[ \text{Nat}(h_A, F) \cong F(A) \]

where \(\text{Nat}(h_A, F)\) denotes the set of all natural transformations from \(h_A\) to \(F\).

\textbf{Optical Lab Example} Continuing with our experimental analogy, weaving in the Yoneda Lemma and Yoneda Embedding.

Within our optical lab, imagine there's a universally renowned experimental setup called the \emph{Yoneda Experiment}. This setup is known for its unique property: for any other experiment in the lab, we learn about that experiment based on how it interacts with the Yoneda Experiment. For example, consider how an experimental setup with prisms interacts with the Yoneda Experiment: does it amplify the light? Refract it in a unique pattern? Or perhaps it even dims it? These interactions~(or transformations) can be thought of as \textbf{morphisms} from our prism experiment to the Yoneda Experiment. Now, the collection of all such interactions from every experiment in the lab to the Yoneda Experiment is similar to the \textbf{Yoneda Embedding}. Just by looking at these interactions, we get a deeper insight into the structure of the whole lab and how each setup inherently behaves.

Meanwhile, the \textbf{Yoneda Lemma} would state something profound about our lab: If you know all the ways every setup in the lab can interact with the Yoneda Experiment, then you can predict how any given setup would behave in any conceivable scenario or with any conceivable tool or modification. For instance, imagine the lab gets a new, mysterious tool. If we understand how this tool interacts with the Yoneda Experiment and how every other setup interacts with the Yoneda Experiment, then we can predict how this tool will modify or affect any other setup in the lab. The Yoneda Lemma, in this context, is a powerful prediction tool. It asserts that if we understand these interactions deeply enough, we can foresee how any setup in our lab~(object in our category) will behave under any new circumstances~(with any functor).

\subsubsection{Presheaf}
A presheaf on a category \(\mathcal{C}\) is simply a functor from the opposite category \(\mathcal{C}^{op}\) to the category of sets \(\text{Set}\). Formally, a presheaf \(F\) is defined by:

\[ F: \mathcal{C}^{op} \rightarrow \text{Set} \]

For each object \(A\) in \(\mathcal{C}\), \(F(A)\) is a set, and for each morphism \(f: A \rightarrow B\) in \(\mathcal{C}\), \(F(f)\) is a function from \(F(B)\) to \(F(A)\).

Expanding on the notion of presheaves, one can consider cases where, instead of sets, one assigns vector spaces to each object of \(\mathcal{C}\). Such presheaves are referred to as vector-valued presheaves.

Formally, a vector-valued presheaf \(F\) on a category \(\mathcal{C}\) with values in the category \(\text{Vect}\) of vector spaces over a fixed field \(k\) is a functor:

\[ F: \mathcal{C}^{op} \rightarrow \text{Vect} \]

For each object \(A\) in \(\mathcal{C}\), \(F\) assigns a vector space \(F(A)\). For each morphism in \(\mathcal{C}\), \(F\) provides a corresponding linear transformation between these vector spaces.

For example, consider the category of open subsets of a topological space \(X\), where the morphisms are inclusions. A vector-valued presheaf on this category might assign to each open subset \(U\) a vector space of functions from \(U\) to \(k\) that satisfy certain properties (e.g., continuous functions, smooth functions, etc.). 

\textbf{Optical Lab Example} A \textbf{presheaf} in category theory is essentially a contravariant functor from a category to the category of sets. In the optical lab, it can be a systematic way of assigning, to every experimental setup, a set of \emph{observations} based on the properties and results of that setup. For instance, for every experimental setup involving lenses, prisms, or mirrors, our presheaf could assign a set of observed light patterns, intensities, or colors produced by that setup. If you make a modification to a setup (e.g., adding a filter or changing a lens), this could correspond to a morphism in our category, and our presheaf would tell us how the set of observations changes in response.

The Yoneda Embedding, in the context of presheaves, involves understanding how each setup~(or object) in the optical lab interacts with the Yoneda Experiment. In the context of our presheaf, this means looking at how the observations~(or sets assigned by the presheaf) for each setup relate to each other and to those of the Yoneda Experiment.

Specifically, for every setup in the lab, we would consider the set of all possible interactions that the setup can have with the Yoneda Experiment. This ``set of interactions'' forms a set for each object, and this assignment from setups to sets, based on interactions with the Yoneda Experiment, is a presheaf! This special presheaf is called the \textbf{representable presheaf} associated with the Yoneda Experiment. The Yoneda Lemma then tells us that, for any other presheaf~(any other way of assigning sets of observations to setups), the set of natural transformations from this representable presheaf to that presheaf is isomorphic to the set of observations assigned to the Yoneda Experiment by that other presheaf.

The most important message of the above analogy and discussion is as follows:
\newline
In our optical lab, presheaves are like systematic observation logs detailing the various outcomes of each experimental setup. The Yoneda Embedding highlights the importance of understanding interactions with the Yoneda Experiment, and the Yoneda Lemma provides a bridge between the special observation log of the Yoneda Experiment and any other possible observation log in the lab. Through this analogy, we see the power of the Yoneda perspective: \emph{by deeply understanding an experiment and its interactions with the other experiments, we can gain insights into the entire laboratory's structure and behavior.}

The following section introduces YonedaVAE and demonstrates an analogy between Yoneda perspective and intra-event reasoning. As a result, it introduces \emph{Yoneda-pooling}, a potent relational pooling layer that will be of paramount importance for the YonedaVAE to do OOD point cloud generation.

\section{YonedaVAE}
YonedaVAE is a VAE-based deep generative model for multiset~(point cloud and graph) generation with a self-distillation mechanism. It consists of an Encoder, Set Generator, and Decoder. In the following, I go through each module and the technologies YonedaVAE introduces to each to reach OOD simulation with context extrapolation.

\subsection{Problem setup}
Working with the real~(random trigger) PXD data, and the fact that the installed PXD is incomplete~(until experiment 26) and only \num{19} sensors are installed. 
For real PXD in the point cloud format, we denote an input event multiset with $N_{evt}$ samples from an event~(in our case 19 samples) where each sample comes with variable $\{N_{l}^i\}_{i=1}^{19}$ permutation invariant hit points as $\mathbf{x} = \{x_1,...,x_{l}\}\in \mathbb{R}^3$ and a sensor-level feature as $\mathbf{c} \in \mathbb{R}^{evt}$ such as cardinality, or energy deposits, and an event-level feature $\mathbf{e} \in \mathbb{R}$. 
$\mathbf{e}$ denotes any event-level attribute such as luminosity, maximum occupancy, or mean occupancy that defines the controllable context for sampling. Samples in each event would have different cardinalities as each sensor could have various numbers of hits.
During training, we are typically given a large supervised corpus $\mathcal{D}_{train} = \{ (\mathbf{x}^{(m)}, \mathbf{e}^{(m)}, \mathbf{c}^{(m)}) \}_{m=1}^{M_{train}}$ with maximum cardinality $N_{l_{max}}=400$. I refer to this region as the \emph{training range}.
On the other hand, during inference we have only a set of Event-level attribute values, $\mathcal{D}_{test} = \{ ( \mathbf{e}^{(m)}, \mathbf{c}^{(m)}) \}_{m=1}^{M_{test}}$ with maximum cardinality $N_{l_{max}}=5300$. I refer to this region as the \emph{extrapolation range}. In other words, a model trained on a max event cardinality of $19\times400 = 7600$ should extrapolate to a max event cardinality of $19\times5300 = 100700$.
Our goal is to generate detector responses that have context profile and cardinality beyond the training region. 

There are two main applications that I am targeting with this setup:
\begin{enumerate}
\label{par:le_ce}
    \item \textbf{PXD background simulation with length extrapolation:} Given that one has access to the fine sensor-level cardinalities of an event, for example, in the case of PXD background amplification, one can sample from the cardinality distribution~(even beyond the training region) to generate a background for any amount of multiplicity~(e.g. $20\times$ the nominal background). In this case, we are dealing with a few-shot conditional generation task where length extrapolation becomes very important~(see \cref{fig:card_extr}), but context extrapolation is not as important as we have access to the fine-level context~(sensor-level attributes). Therefore, during the training, the model has access to cardinalities within the $[1,400]$ range. During inference, the model must generate point clouds with cardinalities in the $[400,5300]$ range.
    
    \begin{figure}[!htb]
  \centering
  \begin{subfigure}{0.48\textwidth}
    \centering
    \includegraphics[width=\linewidth]{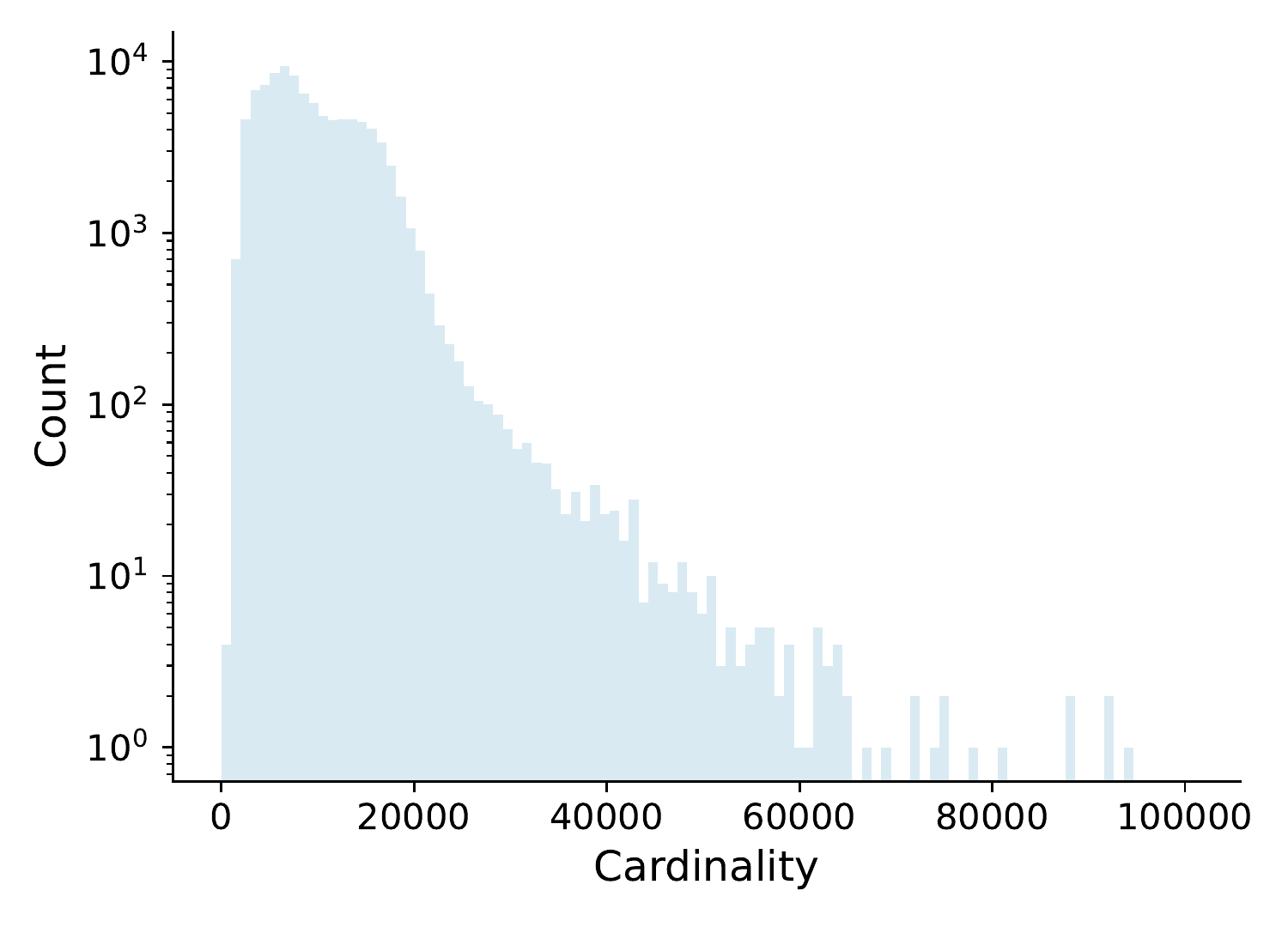}
  \end{subfigure}
  \begin{subfigure}{0.48\textwidth}
    \centering
    \includegraphics[width=\linewidth]{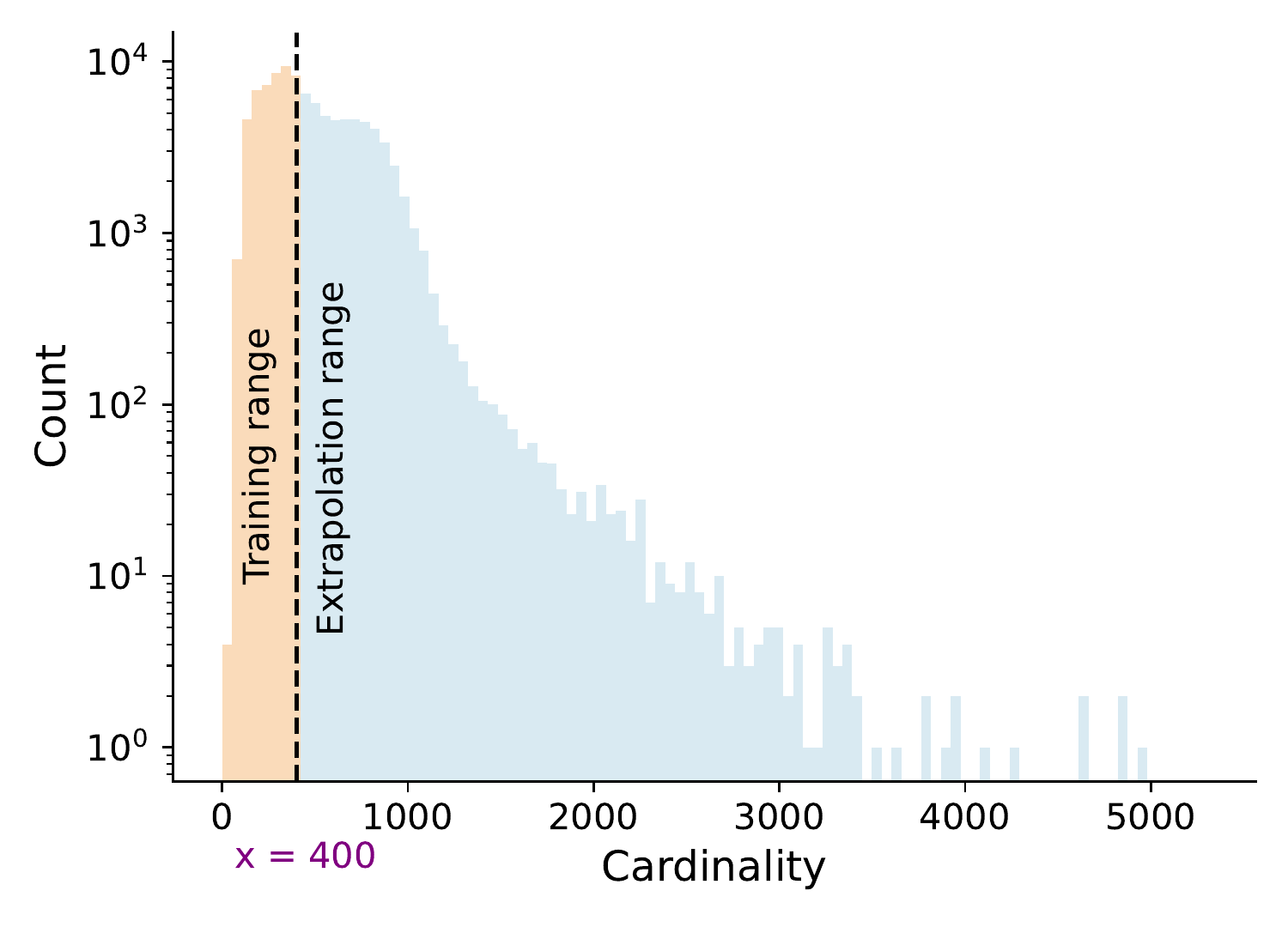}
  \end{subfigure}
\caption{(right)~The cardinality distribution in log scale for the training that belongs to Experiment 12 data and extrapolation~(inference) region that belongs to Experiment 26 data.~(left)~The overall cardinality distribution of each event can go up to 100k points per event.}
  \label{fig:card_extr}
\end{figure}

    \item \textbf{OOD PXD background simulation with context extrapolation:} When one does not have access to the fine sensor-level cardinality profile, and only knows the coarse event-level attributes such as luminosity or the max cardinality of the whole event in the extrapolation region. In this case, we are dealing with a zero-shot generative task, for which context extrapolation is extremely important~(see~\cref{fig:cont_extr}) as we do not have access to fine details and only have access to coarser information~(OOD event-level attributes). In particular, during the training, the model has access to the max cardinality of each event as the context within the range $[100,400]$. During inference, the model gets max cardinalities as the context in the range $[400,5300]$.
    
\begin{figure}[!htb]
      \centering
      \begin{subfigure}{0.48\textwidth}
        \centering
        \includegraphics[width=\linewidth]{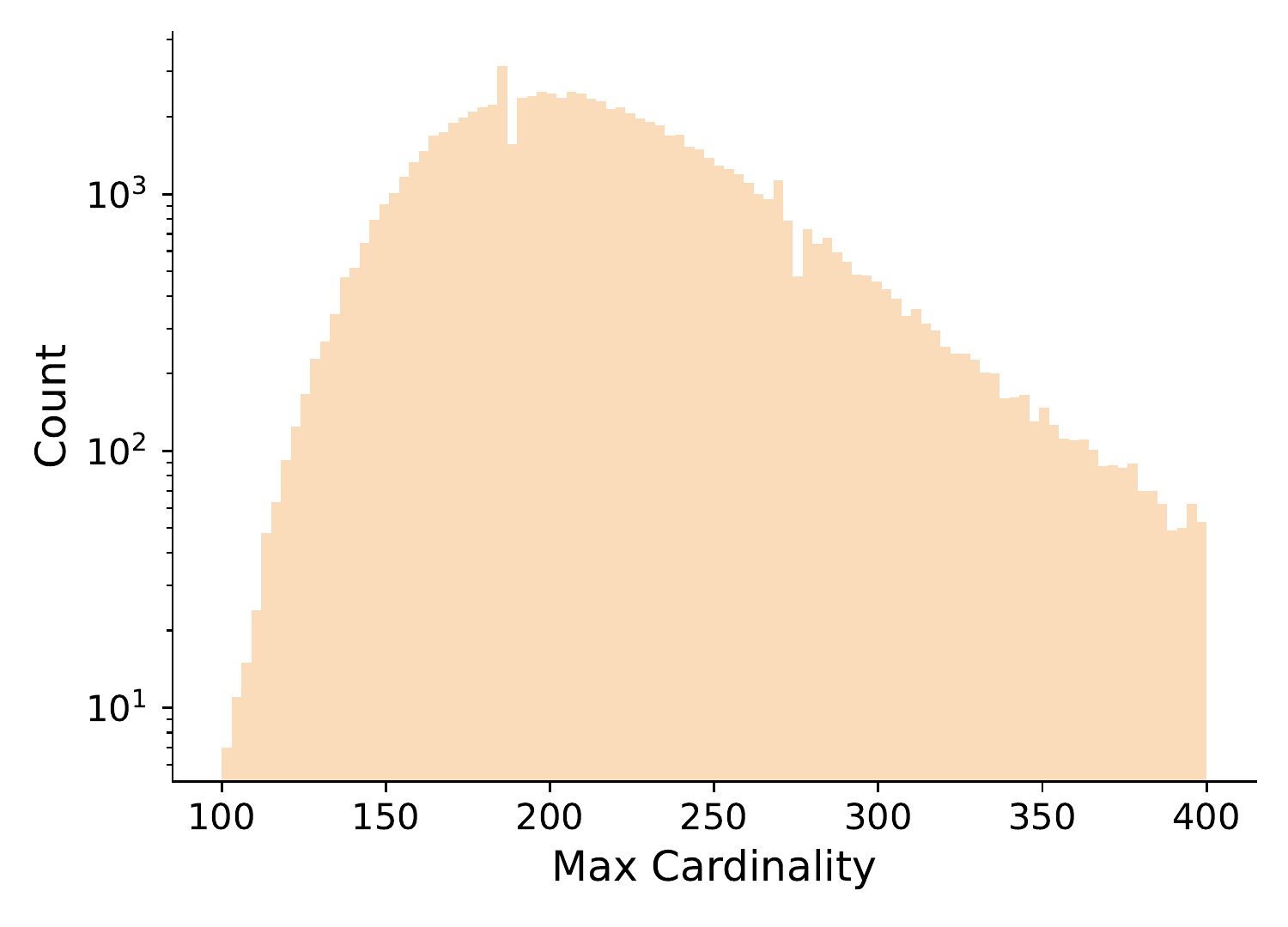}
      \end{subfigure}
      \begin{subfigure}{0.48\textwidth}
        \centering
        \includegraphics[width=\linewidth]{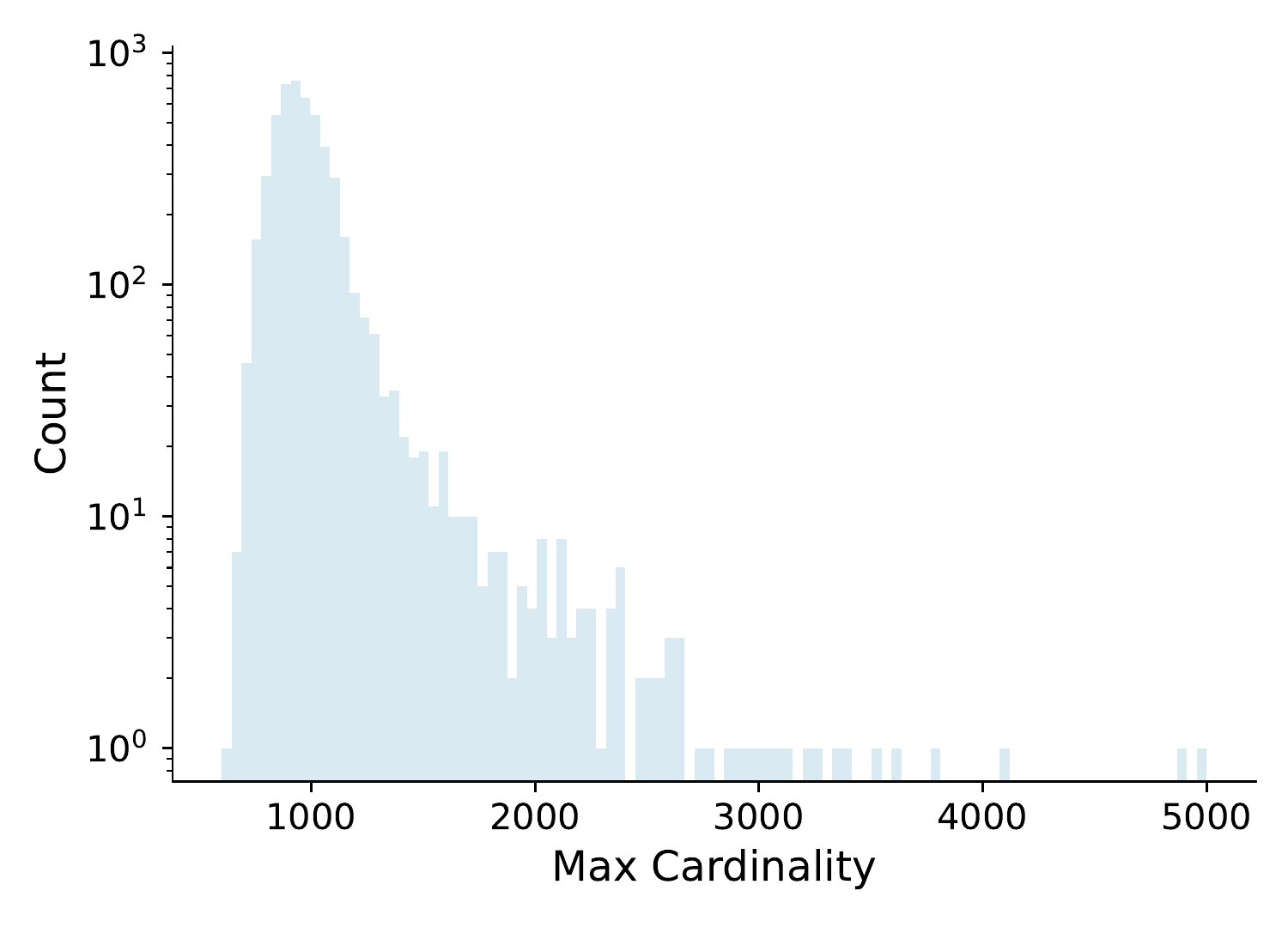}
      \end{subfigure}
    \caption{(left)~The max cardinality distribution in log scale for the training that belongs to Experiment 12 data as event-level attributes.~(left)~The max cardinality distribution in the log scale for inference that belongs to Experiment 26 data as event-level attributes.}
      \label{fig:cont_extr}
\end{figure}
\end{enumerate}

\subsection{YonedaVAE: Yoneda Encoder}
\begin{figure}[!htb]
    \centering
    \includegraphics[width=0.95\textwidth]{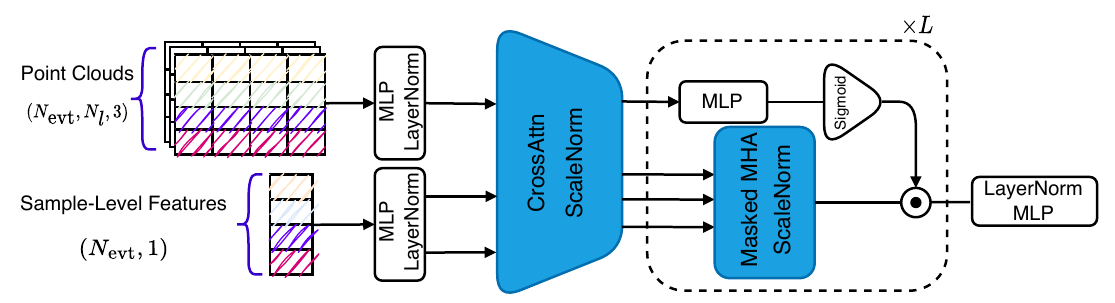}
    \caption{Eventformer}
    \label{fig:eventformer}
\end{figure}

YonedaVAE’s encoder, during training, takes the set of detector signatures event-by-event as a point cloud of shape $N_{evt}\times N_{l} \times 3$, along with each sensor's conditional parameter as inputs. The three feature channels in the input point cloud correspond to the (x, y, charge) data of the PXD hits.
In our setup, the conditional parameters are the normalized occupancies for each sensor. The encoder comprises two main stages. First, the trunk of the network processes the inputs through multiple layers of an equivariant neural network block that I term \emph{Eventformer}, as depicted in \cref{fig:eventformer}, to output $N_{evt}\times N_{l} \times ch$ array that represents, $N_{evt}$ number of samples in an event which is $19$ for real PXD data, $N_{l}$ the number of points representing the number of hits per sample, and $ch$ the embedding dimension. 
The Eventformer blocks contain a number of attention-based components with a gating mechanism and cross-attention as I discuss below. The key point in the Eventformer block is the mechanism to inject intra-event context within each point cloud representation which enables direct reasoning over both the sensor and event relationships. 

\subsubsection{Eventformer and Attention Gating}
After a two-layer MLP layer with Layer normalization, the point cloud information enters the Cross Attention module as the queries. From the other side, the embedded sample-level features~(attributes for each sample in the event) also enter the Cross Attention module as Keys and Values. This is a mechanism to assign the sensor attribute to its corresponding sensor hit information.

\textbf{Scale Normalization:} Within both Cross Attention and Multi-Head Attention~(MHA) I incorporate the residual connections~\cite{he_deep_2015} and Scale Normalization~\cite{nguyen_transformers_2019} instead of the Layer Normalization to achieve a faster convergence. Layer Normalization inspired by batch normalization~\cite{ioffe_batch_2015}, aims to reduce internal covariate shift by fixing the mean and variance of activation distributions. Both have been applied to self-attention~\cite{vaswani_attention_2017,kool_attention_2019}. However,~\cite{santurkar_how_2019} shows that batch normalization's success has little to do with covariate shift but comes instead from smoothing the loss landscape. As a result,~\cite{nguyen_transformers_2019} proposed replacing the Layer Normalization with Scale Normalization:

\begin{equation*}
    \mathbf{ScaleNorm}(\vx; g) = g\frac{\vx}{\norm{\vx}}.
\end{equation*}

With Scale Normalization can be viewed as compactifying $d$-dimensional vectors onto a $(d-1)$-dimensional hypersphere with learned radius $g$. This expresses the inductive bias that each sublayer's activations have an ideal ``global scale,'' by replacing the $2d$ scale and shift parameters of Layer Normalization with a single learned scalar, improving computational and parameter efficiency.

\textbf{Attention Gating:} After the fusion of point clouds and their corresponding attributes, I introduce a gating mechanism over the attention weights to overcome the \emph{over-smoothing} phenomenon that happens with multiple layers of MHA modules. 
Over-smoothing~\cite{rusch_survey_2023} refers to the exponential convergence of all point features towards the same constant value as the number of Attention layers increases. 
For Transformer-based models, this means that the excessive blending of representations, causes the model to assign very similar representation to each mode over successive layers and potentially leads to a loss in the granularity of information. 
In fine-grained cases, over-smoothed representations are less informative because distinct points become indistinguishable.

To address the issue of over-smoothing, this study introduces a \emph{gating mechanism} on every MHA layer, regulating the layer's output. Gating was initially devised to counter the Short-Term memory challenges seen in RNNs. 
RNNs, when handling extensive input sequences, sometimes struggle to retain information throughout the sequence's entirety. Additionally, they are vulnerable to the vanishing gradient dilemma in initial layers, negatively impacting the learning process. The LSTM~\cite{hochreiter_long_1997} architecture was designed to overcome these challenges. It employs gates within the standard RNN cells, which learn and decide which data to retain and which to dismiss. This decision-making process is facilitated by the sigmoid functions that adjust values between 0 and 1. In this setting, values closer to $0$ indicate that specific cell information should be discarded, while values nearing $1$ denote information worth retaining.

In our setup, the inclusion of the output gate signifies that certain information might need to be discarded after the self-attention by Hadamard multiplication of the sigmoid-transformed values to the attention matrix's linear transformation. This provides a strong regulation over what information should be passed on to the subsequent layers for further refinement. There are certain points with a higher likelihood to form larger hit clusters than others, and so they are more important in determining the consistency of PXD hits. Thus, the gating mechanism provides an effective mechanism to prevent over-smoothing over multiple Attention layers which in the end enables the model to not overfit and be able to extrapolate. Without this gating mechanism, context extrapolation is not possible.

\subsubsection{Yoneda-Pooling}
Inspired by the Yoneda embedding, the trunk of the network is followed by a novel aggregator layer, \emph{Yoneda-Pooling}, that introduces a learnable PNA-pooling~\cite{corso_principal_2020} module that is also aware of samples relationship with each other. This module maps the input point clouds to the VAE's latent manifold.
The Key innovations in the Yoneda-pooling mechanism are the simultaneous global refinement over all samples within an event with respect to the event-level attribute, using the attention mechanism, and adding learnable weights to the PNA-aggregator.

\textbf{PNA Aggregator:} The Principal Neighbourhood Aggregation~(PNA)~\cite{corso_principal_2020} mechanism offers an advanced approach to aggregate information in multisets. Instead of relying on traditional mean, sum, or max aggregations, PNA combines multiple aggregators and scalers in a systematic manner. 
The aggregation functions can include the mean, sum, max, min, and standard deviation, capturing diverse statistical properties of node features in the neighborhood. 
On the other hand, scalers are responsible for adjusting the aggregated values involving functions like identity, amplification, and attenuation. They propose the logarithmic scaler $S(d, \alpha)$:

\begin{equation}
\label{eq:S_amp}
S(d, \alpha) = \left(\frac{\log(d + 1)}{\delta} \right)^\alpha, \quad \delta = \frac{1}{|\text{train}|}\sum_{i \, \in \,  \text{train}}\log(d_i + 1), \quad d>0, \quad -1 \leq \alpha \leq 1
\end{equation}

where $\delta$ is a normalization parameter computed over the training set, $d$ is the degree of the point/node receiving the message, $\alpha$ is a variable parameter that is negative for attenuation, positive for amplification, or zero for no scaling. The integration of these aggregators can harness a broader spectrum of neighborhood information, leading to improved performance in multiset-related tasks. 

However, the preset scalers and aggregators are fixed in their operations, meaning they lack adaptability to datasets or tasks where certain types of aggregations might be more pertinent than others. 
This data-agnostic strategy leads to sub-optimal performance when dealing with a high number of nodes~(points) setup, the encoder does not show a generalizable representation learning.
Hence, I introduce the \emph{weighted learnable PNA} aggregation as a part of Yoneda Pooling. This data-centric method introduces learnable weights to the aggregation process, allowing the model to automatically prioritize certain types of aggregations over others based on the data and the task at hand: 

\begin{equation}
\label{eq:PNA_weighted}
{\bigoplus}_{\mathcal{W}} = 
\underbrace{
\begin{bmatrix}
I \\ S(D, \alpha=1) \\ 
S(D, \alpha=-1)
\end{bmatrix} }_{\text{scalers}}
\otimes
\underbrace{\begin{bmatrix}
w_1 \\ w_2 \\ w_3 \\ w_4 
\end{bmatrix}}_{\text{Learnable Weights}}
\odot
\underbrace{\begin{bmatrix}
\mu \\ \sigma \\ \max \\ \min
\end{bmatrix}}_{\text{aggregators}}
\end{equation}

\begin{figure}[!htb]
    \centering
    \includegraphics[width=0.95\textwidth]{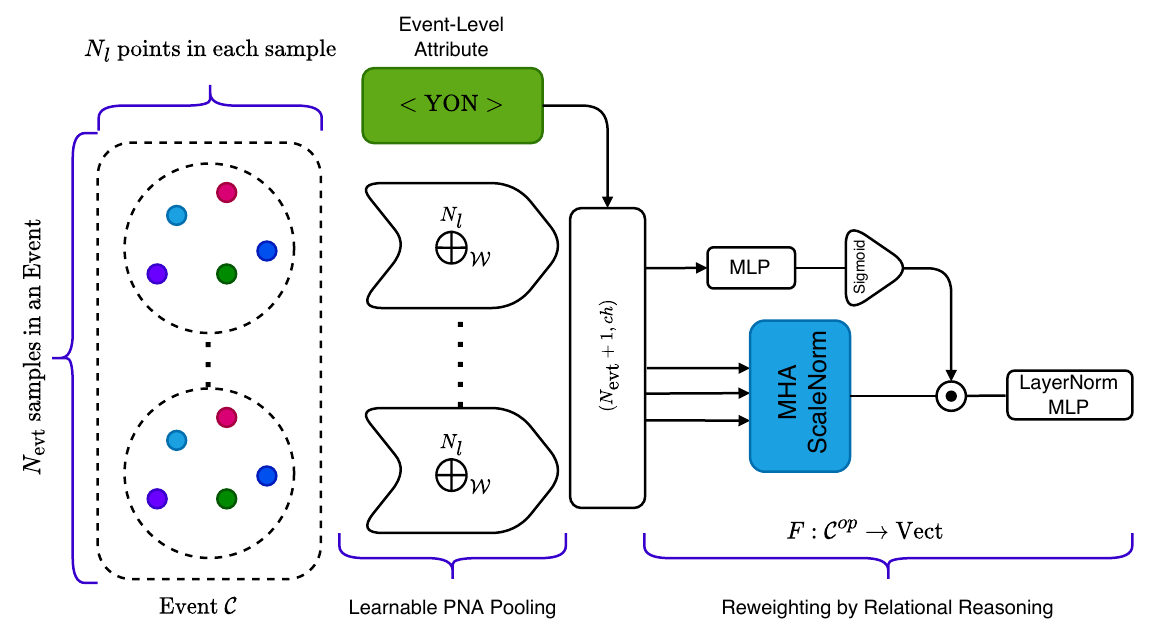}
    \caption{Yoneda-Pooling}
    \label{fig:yonedapooling}
\end{figure}

Where ${\bigoplus}_{\mathcal{W}}$ is representing the weighted~(learnable) PNA aggregator, $\otimes$ represents the tensor product, and $\odot$ represents the Hadamard product. This important modification extends the PNA formalism to learn the weights for each aggregation function resulting in dynamic adjustment of the fixed algebraic operations to prioritize certain aggregations over others based on their relevance to the data.\\
The Yoneda-pooling, depicted in~\cref{fig:yonedapooling}, along with incorporating the learnable PNA-aggregator, to provide dynamical summary statistics of points for each sample in an event, also introduces an Event-level attribute, $<Yon>$. 
$<Yon>$ can be thought of as a global summary statistics learnable token of the whole event that one has access to such as Luminosity, the maximum amount of background, or the type of background hits. In our case, it is the maximum cardinality of each event.
After concatenating this token to the other samples in the event~(we get now a mini-batch of size $[N_{evt}+1,ch]$), everything goes into a Gated-MHA module that acts over the samples in the event. 
By doing this relational reasoning, the representation of samples in the event~(mini-batch) will be reweighted with respect to the global~(event-level) $<Yon>$ token and also their relationship with each other.

\subsubsection{Yoneda-Pooling and Category Theory}
In order to create a universal mathematical language for relational reasoning, let's take a step back, and look at what Yoneda pooing is doing. 
Yoneda pooling is utilizing a learnable PNA aggregation of points, then mapping them to the VAE's latent manifold such each sample in the mini-batch is refined to log the information of the other samples in the mini-batch~(very much like the relational reasoning introduced in the last chapter).
This general process is similar to the Yoneda lemma, demonstrating how objects in a category can be fully understood in terms of the morphisms~(arrows) that go into them.
Here the category $\mathcal{C}$ is an event~(self-dual as the morphisms are bidirectional), where the objects are the PXD sensor information represented as a $\mathcal{G}$-invariant multi-sets where G is the permutation group $\mathcal{S}_n$~(permutation invariant point clouds). For two \( G_{n_i} \)-invariant multisets \( A \) and \( B \) with multiplicities \( n_a \) and \( n_b \), respectively, a morphism \( \rho: A \rightarrow B \) is a function such that

\begin{itemize}
    \item \( \rho(g \cdot a) = g \cdot f(a) \) where $g \in G$
    \item \( k \times \text{Multiplicity}(a) = \text{Multiplicity}(f(a)) \), where \( k = n_b / n_a \)
\end{itemize}

This category is like ``a bag of words'' in which it sets up the basic framework for objects and morphisms without adding additional structure or interpretation. It specifies the kinds of objects~(in this case, \( G_{n_i} \)-invariant multi-sets with elements in \( [0,1] \times [0,1] \times [0,1] \)) and the rules for how they can be transformed. Much like syntax in natural language, which outlines how words can be combined but doesn't specify their meaning, this category outlines how these specific multi-sets~(PXD sensor information) exist individually without detailing what those PXD sensors might represent in a given event. On the other hand, The ``contextual representation'' or ``semantics,'' so to speak, would come from how one interprets these objects~(PXD sensor information) and morphisms~(their relation) in a specific context~(event). 

The Yoneda embedding $P$ then assigns a vector-valued representation ~(presheaves) to each object in the category while taking the set of morphisms between the objects, by means of the Attention mechanism. 
Formally, one can define a presheaf $P: \mathcal{C}^{op} \to \text{Vect}_k$, where the category $\text{Vect}_k$ refers to vector spaces over a field $k$ and Linear transformations as the morphisms. For each object $A \in \mathcal{C}$, $P$ assigns a vector space $V_A$ over the field $k$. And for each morphism $\rho: A \rightarrow B$ in $\mathcal{C}$, $P$ assigns a linear transformation $T_{\rho}: V_B \to V_A$. Given that \( P \) maps each object \( A \in \mathcal{C} \) to a vector space \( V_A \in \text{Vect}_k \), the presheaf \( P \) also associates to each morphism \( \rho \in \mathcal{C} \) a k-linear map~(denoted as \( P(\rho) \)) between the corresponding vector spaces $P(\rho): V_B \rightarrow V_A$. Here \( P(\rho) \) satisfies the following functorial properties for \( P \)

\begin{itemize}
    \item Preservation of Identity: \( P(\text{id}_A) = \text{id}_{V_A} \) for every object \( A \) in \( \mathcal{C} \).
    \item Preservation of Composition: Given morphisms \( \rho_1: A \rightarrow B \) and \( \rho_2: B \rightarrow C \), one must have \( P(\rho_2 \circ \rho_1) = P(\rho_1) \circ P(\rho_2) \).
\end{itemize}

Hence, given an object $A\in \mathcal{C}$, the Yoneda embedding maps $A$ to the hom-functor $\text{Hom}_\mathcal{C}(-,A)$. This functor assigns to each object $B\in \mathcal{C}$ the vector space spanned by affine transformations from $B$ to $A$, and to each morphism $\rho: B \rightarrow C$ in $\mathcal{C}$, the corresponding linear transformation induced by composition with $g$, 

\[ \text{Hom}_\mathcal{C}(-,A): C^{op} \to \text{Vect}_k \]
\[ B \mapsto \text{Hom}_\mathcal{C}(B,A) \]
\[ g: B \to C \mapsto \text{Hom}_\mathcal{C}(g,A): \text{Hom}_\mathcal{C}(C,A) \to \text{Hom}_\mathcal{C}(B,A) \]

where $\text{Hom}_\mathcal{C}(g,A)$ is given by function composition with $g$. As a result, Yoneda Lemma for Category $\mathcal{C}$ states that for any presheaf $P: \mathcal{C}^{op} \to \text{Vect}_k$ and any object $A\in \mathcal{C}$, there is a natural isomorphism:

\[ \text{Nat}(\text{Hom}_\mathcal{C}(-,A), P) \cong P(A).\]

This states that the natural transformations from the hom-functor represented by $A$ to $P$ correspond bijectively to the elements of the vector space $P(A)$. In essence, the Yoneda Lemma captures the idea that a set of points in $\mathbb{R}^3$ is fully described by how other sets relate to it, reflected in the context of vector spaces and linear transformations in $\text{Vect}_k $.
In other words, an object $A\in \mathcal{C}$ is uniquely determined by its relationships to all other objects in $\mathcal{C}$.
Going back to Yoneda-Pooling, in order to encode a sample $A$ in an event, the Yoneda-Pooling mechanism considers all affine transformations from other samples~(samples) to $A$. Thus, learning the information about the transformations of $A$ could serve as a form of \emph{relational compression}. This way, one makes sure that the latent space of the VAE also includes not only local information but also relational information.

From the earlier analogy, syntax tells us ``which sensor is near which sensor,'' while semantics tells you ``what they mean in an event.''
In a similar manner, the Yoneda Embedding doesn't just tell us about the structure~(syntax) of the event; it also tells us how to understand each sensor's information in terms of its relationships/interactions with all other sensors~(semantics). Given this, one can say that the Yoneda embedding and the resulting presheaf representation capture some form of ``contextual representation'' or ``semantics'' for each event.

\subsection{YonedaVAE: Set Generation}
\begin{figure}[!htb]
    \centering
    \includegraphics[width=0.65\textwidth]{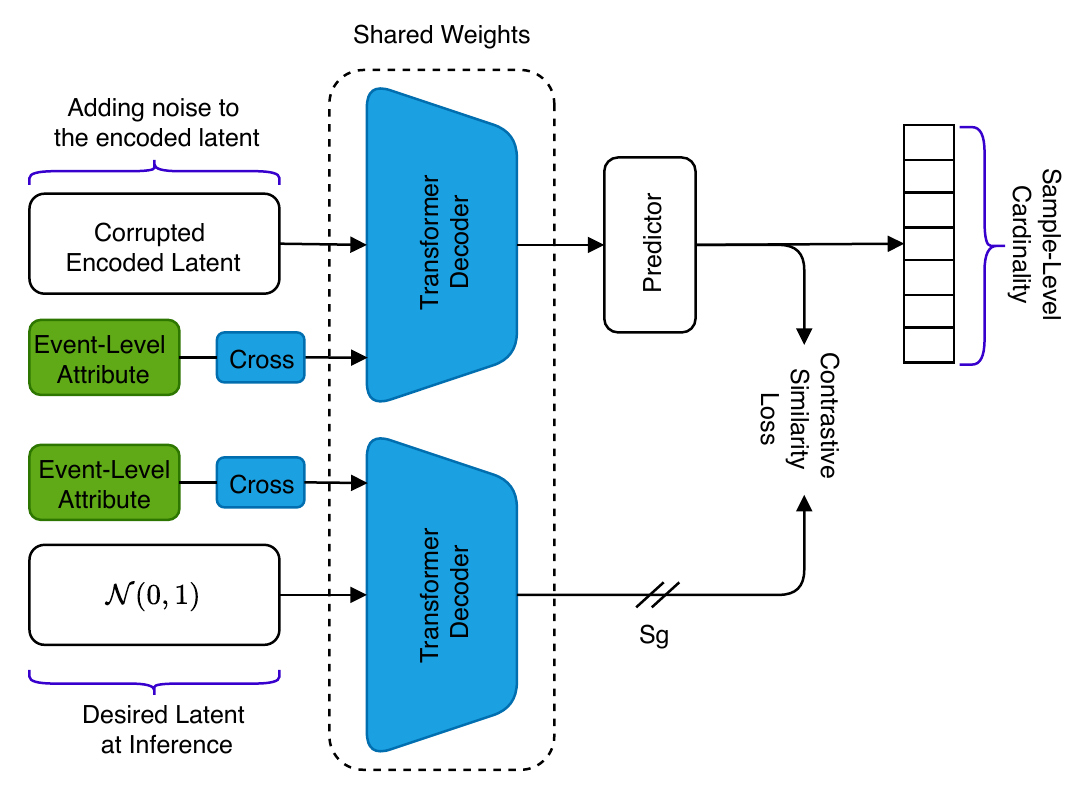}
    \caption{Self-Distillation of the learned latent space via the VAE encoder to the desired prior. It is termed ``self-distillation'' since the backbones models, the Transformer Decoders, share weights.}
    \label{fig:selfdistil}
\end{figure}

YonedaVAE's set generator has two components. The first part is to train a Self-Distilled Transformer decoder to learn to generate the cardinality of each sample given the event-level attribute~(such as event luminosity or Max cardinality) as the condition, depicted in~\cref{fig:selfdistil}. 
The second part is to use the Transformer decoder's latent embedding and the generated~(inference) cardinalities or true level~(training) cardinalities to do the Top-q sampling~(will be discussed shortly in~\cref{sec:top_p}).

\textbf{Training} During training, via self-distillation~(discussed in \cref{chap:3},\cref{sec:self_dis}), the set generator is trained on generating the cardinalities given the event-level attribute and the Normal distribution prior and the VAE encoded latent. Then, it uses the Transformer decoder's latent embedding and the true cardinalities to do the Top-q sampling. Afterward, the created set, along with the encoded latent~(by the Yoneda Encoder) will be fed to the VAE decoder where the output is the generated point cloud.

\textbf{Inference} At inference time, the set generator generates the cardinalities given only the event-level attribute and the Normal distribution prior. Then, after Top-q sampling, based on the generated cardinalities, the generated set, along with the prior latent~(the Normal distribution) will be fed to the VAE decoder where the output is the generated point cloud.

Now, let's dive into the details step-by-step.

\subsubsection{Self-Distillation}
Since I am targeting OOD detector response simulation, the model needs to be able to generalize and predict each sensor's number of hits and their correlation with each other accurately. 
In order to do so, YonedaVAE trains a Transformer decoder to generate them in a causal manner, given the latent vector~(prior) and event-level attribute. By causal, I mean that during training the Transformer decoder utilizes a causal masking mechanism where the self-attention layer is only allowed to attend to earlier positions in the output sequence. This is done by masking future positions. This way the model successfully generates the cardinalities of each sensor in an auto-regressive manner. In order to take into account also the topology of the PXD detector~(as they are not sequential but distributed in an annulus shape), and learn PXD sensor positions~(angles and radius), learnable Rotary positional encodings~(discussed in~\cref{sec:pos_emb}) of the PXD sensors are being added to the self-attention.

When one trains the model only given the encoded latent information, the model overfits the data~(encoded latent), and cannot generalize at inference where I don't have access to the true encoded data distribution during the inference. That is why I conjure Self-Supervised Learning, in particular, the \emph{self-distillation} method to train the set generator's transformer decoder. 

Training with self-distillation normally is when one has two different views~(positive pairs) of the same data and wants to train an encoder to subsequently use a predictor network to map the output of one encoder to that of the other. The upshot then would be to avoid mode collapse~(discussed in \cref{chap:3},\cref{sec:self_dis}). 
However, in our setup, the trivial representation is the output of the Encoder, encoded latent space. The dual representation over which we want the model to generalize is the normal prior, to which we have access at the inference time. This introduces a new way for the VAE to learn a latent representation to be as similar as possible to a normal prior.
In other words, the architecture takes as input two latent distributions that I want the model to consider being isomorphic, the corrupted encoded latent $z_1$ and the prior Normal distribution $z_2$.
The two latent spaces are processed by the Transformer decoder network $g$ and a projection MLP head. The Transformer decoder, $g$, shares weights between the two latents. A prediction MLP head, denoted as $h$, transforms the output of one latent space and matches it to the other latent space.
Denoting the two output vectors as $p_1 \equiv h(g(z_1)) $ and $x_2 \equiv g(z_2)$ , one minimizes their negative cosine similarity:

\begin{equation}
\mathcal{D}(\mathbf{p}_1, \mathbf{z}_2) = -\frac{\mathbf{p}_1}{\left\lVert{\mathbf{p}_1}\right\rVert _2} \cdot \frac{\mathbf{z}_2}{\left\lVert{\mathbf{z}_2}\right\rVert _2},
\label{eq:dist_cosine}
\end{equation}

where ${\left\lVert{\cdot}\right\rVert _2}$ is $\ell_2$-norm. Similar to Simsiam loss~\cite{chen_exploring_2020}, one can define the self-distillation loss as,

\begin{equation}
\mathcal{L}_{SD} = \frac{1}{2}\left( \mathcal{D}(\mathbf{p}_1, \textbf{sg}(\mathbf{z}_2)) + \mathcal{D}(\mathbf{p}_2, \textbf{sg}(\mathbf{z}_1)) \right),
\label{eq:sd}
\end{equation}

where \textbf{sg} represents the stop-gradient operation on the respective variables. Here, the encoder on $z_2$ receives no gradient from $x_2$ in the first term, but it receives gradients from $p_2$ in the second term~(and vice versa for $z_1$).

Incorporating self-distillation then has a vital role for the set generator to be able to extrapolate well beyond the training context by detaching itself from the learned latent space of the Encoder.
Since the model has access to the trivial encoded representation of the data through the learned latent vector, it still experiences a mode collapse~(since it is easier for the model to map the learned latent space to cardinalities). That is why, as a solution, I propose to corrupt the encoded latent vector by random noise injection to make the training more balanced. From the perspective of the mainstream definition of self-distillation~(like in BYOL), this can be viewed as using augmented views of the original input as a positive sample.
Since the downstream task for the Transformer decoder is to predict the cardinalities correctly, along with the self-distillation loss, the model also uses the supervised loss as well over the predictor's output,

\begin{equation}
\mathcal{L}_{\ell_2} = \left\lVert h(g(z_1)) - \textbf{e} \right\rVert_2^2 = \sum_i^{N_{\textrm{evt}}=19} \left( h(g(z_1))_i - e_i \right)^2,
\label{eq:sup}
\end{equation}

where $e_i$ is the true occupancy per sensor. It is important to note that since this is a predictive/generative setting, the set generator uses a Transformer decoder instead of an encoder. The transformer decoder has a causal inductive bias that, along with the Rotary positional encoding of each sensor, helps to generate cardinalities with the correct correlation. For the Transformer decoder to generalize well to the extrapolation region, it benefits from the attention gating mechanism that I introduced earlier, Rotary positional encoding~(to encode PXD sensor positions), and dropouts~(embedding dropout, attention dropout, and layer dropout). 
The contextual event-level attribute, in our case, the max cardinality of the event, also gets injected into the Transformer decoder through the cross-attention mechanism. 
An interesting observation was that in order to encourage context extrapolation, one has to use Layer Normalization instead of Batch Normalization, which is frequently used in self-distillation setups~\cite{chen_simple_2020}. 
I think the reason behind this phenomenon is that batch normalization uses the training batch statistics to learn its parameters~(as they do not care about OOD generalization), while Layer Normalization does not rely explicitly on the statistics of the training data and only learns from the individual features through normalization.

\subsubsection{Adaptive Top-q sampling}
\label{sec:top_p}
After encoding the point clouds and the corresponding attributes, the set generator creates the multiset from the latent vectors during training. Of course, the most important thing at this stage is to generate the points with the correct cardinality. However, the first challenge I face is that within an event, the cardinality of each sample differs from one another. 
Moreover, since the cardinality of an event describes the occupancy of the detector responses in that event, in the OOD regime, I do not have access to each sensor's occupancy during inference. I only have the maximum cardinality or the luminosity of the event and nothing more. 
That is why one cannot use any of the described set creation approaches in~\cref{section:problem} as either they limit the extrapolation power or they lack the versatility to generate variable intra-event length sets.

Thus, it is of paramount importance for the set generator to be able to not only generate points with inter-event and intra-event variable cardinality but also to learn this in a zero-shot manner to extrapolate beyond the training region, as depicted in~\cref{fig:card_extr}. 
By inter-event variable cardinality, I mean when one is dealing with variable cardinality in general in the training/test set, and by intra-event variable cardinality, I mean when an event or mini-batch has variable cardinality. The former can be achieved without extrapolation with the Top-k method; however, the latter, handling intra-event variable cardinality, has not been done before and is vital when one is dealing with irregular detector geometries and hit patterns through the detector. 

Hence, this study introduces a novel set creation mechanism, \emph{Adaptive Top-q Sampling}. The key idea is to use the shape of the probability distribution to adaptively determine the multiplicity of points to be sampled from. Thus, instead of relying on a fixed and predefined Top-k or using a temperature parameter to control the shape of the distribution without sufficiently suppressing the unreliable tail, I propose sampling from the top-q portion of the reference probability distribution for each event where $p$ is adaptively learned in context, expanding and contracting the candidate pool dynamically. 

One might wonder why one needs to sample at all. As I described in~\cref{section:problem} and ~\cref{chap:4}, in a multi-set~(point cloud) generation task, during training or inference, one has two stages of decoding, a sampling stage where the best ``candidate'' points are being proposed by the set generator, then the update stage where the points become aware of each other and get updated to be mapped to the final output. For creating sets of point clouds stochastically, sampling is essential for several reasons, similar to why it's used in natural language generation models. Sampling allows the model to capture the inherent variability in possible valid outputs for an event. If we were to deterministically choose the most probable points, we could end up with suboptimal and unrealistic point clouds that don't capture the complexities of the whole dataset. By sampling from a distribution, the model inherently captures its own uncertainty about the ``best'' point or set of points, adding a layer of robustness and confidence assessment to the generated set. This is particularly useful in the challenging OOD scenarios where the model has to generalize beyond the training data.

\begin{figure}[!htb]
    \centering
    \includegraphics[width=0.95\textwidth]{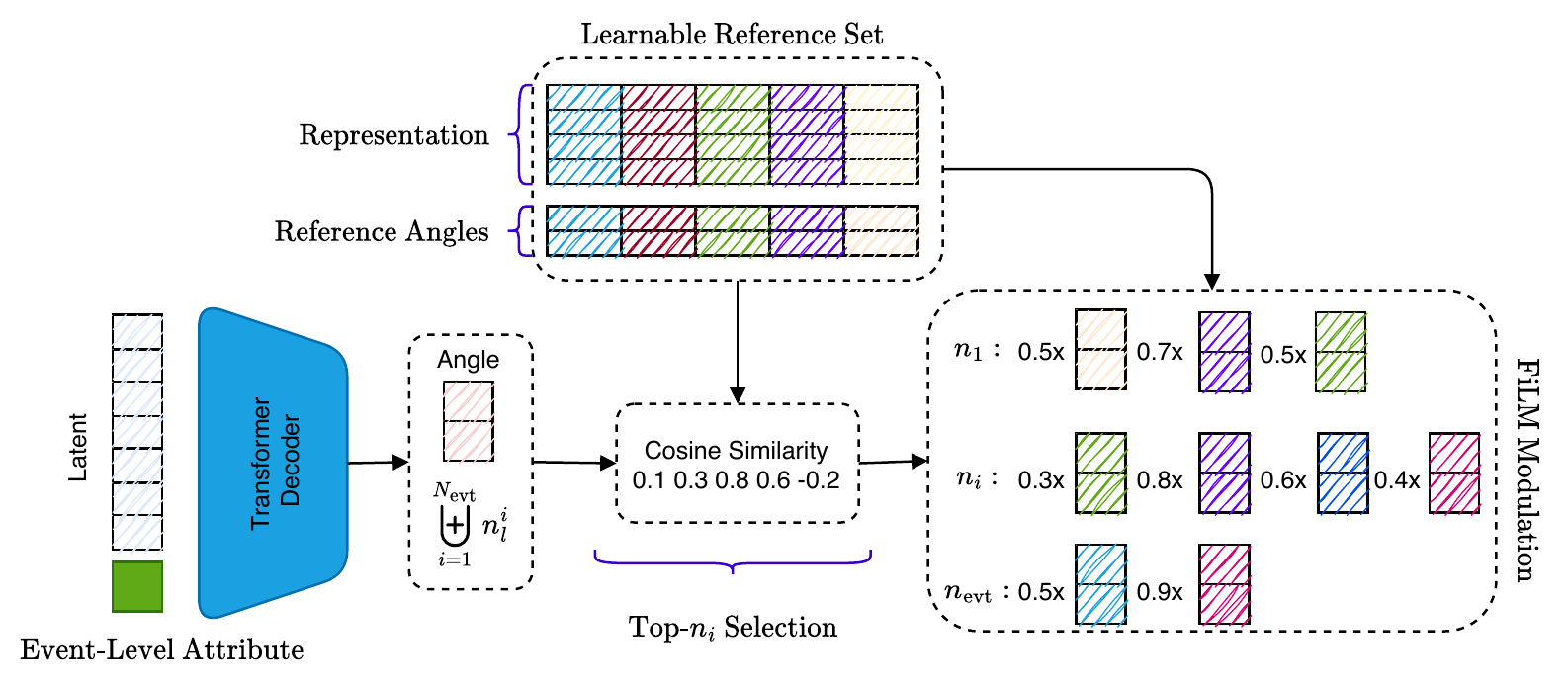}
    \caption{Adaptive Top-q sampling proposes a point sampling from the top $n_i$ portion of the probability mass where $n_i$ is adaptively learned in context, dynamically expanding and contracting the candidate pool in each event.}
    \label{fig:top-generator}
\end{figure}

Similarly to Top-k, Adaptive Top-q also uses a reference set with an arbitrary size $n_0$, as depicted in~\cref{fig:top-generator}. Each point in this set is a pair $(\bm\phi, \vr)$ : the \emph{reference angle} $\bm \phi \in \R^\dang$
is used to decide when to select the point, and $\vr \in \R^\dhid$ contains the \emph{reference representation} of the point. 
Given a latent vector $\vz \in \R^{\dlat \times 1}$, a reference set made of reference angles $\bm{\Phi} \in \R^{\nref \times \dang}$ and reference representations $\mR \in \R^{\nref \times \dhid}$, as well as learnable matrices $\mW_1$ and $\mW_2$ (respectively of sizes $1\times c, ~1\times c$), Adaptive Top-q creation computes

\begin{align}
    \va &= \operatorname{MLP}(\operatorname{Transformer Decoder}(\vz))  && \in \R^{\dang} \\
    \vn_{i} &= \operatorname{Transformer Decoder}(\vz)  && \in \R^{n_{\textrm{evt}}\times 1} \\
    \cosines &= \bm{\Phi}~ \va ~/~ \operatorname{vec}((||\bm\phi_i||_2)_{1\leq i \leq \nref}) && \in \R^{\nref}\\
    \tilde\vc &= \operatorname{softmax}(\cosines) && \in \R^{n_{i} \times 1} \\
    \idx_{i} &= \operatorname{argsort_\downarrow}(\cosines)[:n_{i}]  && \in \mathbb N^n_{i} \\
    \mX_{n_{i}} &= \mR[\idx_{i}] \odot \tilde\vc ~\mW_1 + \tilde\vc ~\mW_2 && \in \R^{n_{i} \times \dhid} \\
    \mX^0 &= \operatorname{ZeroPad}(\mX_{n_{i}})  && \in \R^{n \times \dhid}.
\end{align} 

In each event, the Transformer decoder adaptively proposes a set of cardinality for each sample in the event and a set of vectors~(Transformer decoder's intermediate layer), termed Angles, given the event-level attribute~(max cardinality) and the latent space~(or prior during inference). Then, the model computes the cosine similarity between these candidate's angles and the reference angles. Then, based on their normalized similarity, the top $n_i$ important points will be chosen for each sample. Note that, in Adaptive Top-q sampling, for each sample $i$, $n_i$ differs and represents the top $n_i$ portion of the reference probability mass. In the end, all samples get zero-padded to the maximum cardinality of the event for vectorization purposes. 

The crux of the Adaptive Top-q module is to select the points adaptively that will be used to generate a dynamic set with variable multiplicity. This is the intra-event variability I was talking about. In practice, this means selecting the highest probability points whose multiplicity does not exceed the learned intra-event $n_{i}$ for sample $i$. The size of the sampling set will adjust dynamically based on this criteria at each time step. For instance, for low values of $n_{i}$, this is a small subset of points that takes up a majority of the reference probability mass. 
 
\subsection{YonedaVAE: Decoder}
The decoder of YonedaVAE comprises three main parts: the sample-level feature injection, set-latent space fusion, and the point cloud final decoder.

\begin{figure}[!htb]
    \centering
    \includegraphics[width=0.95\textwidth]{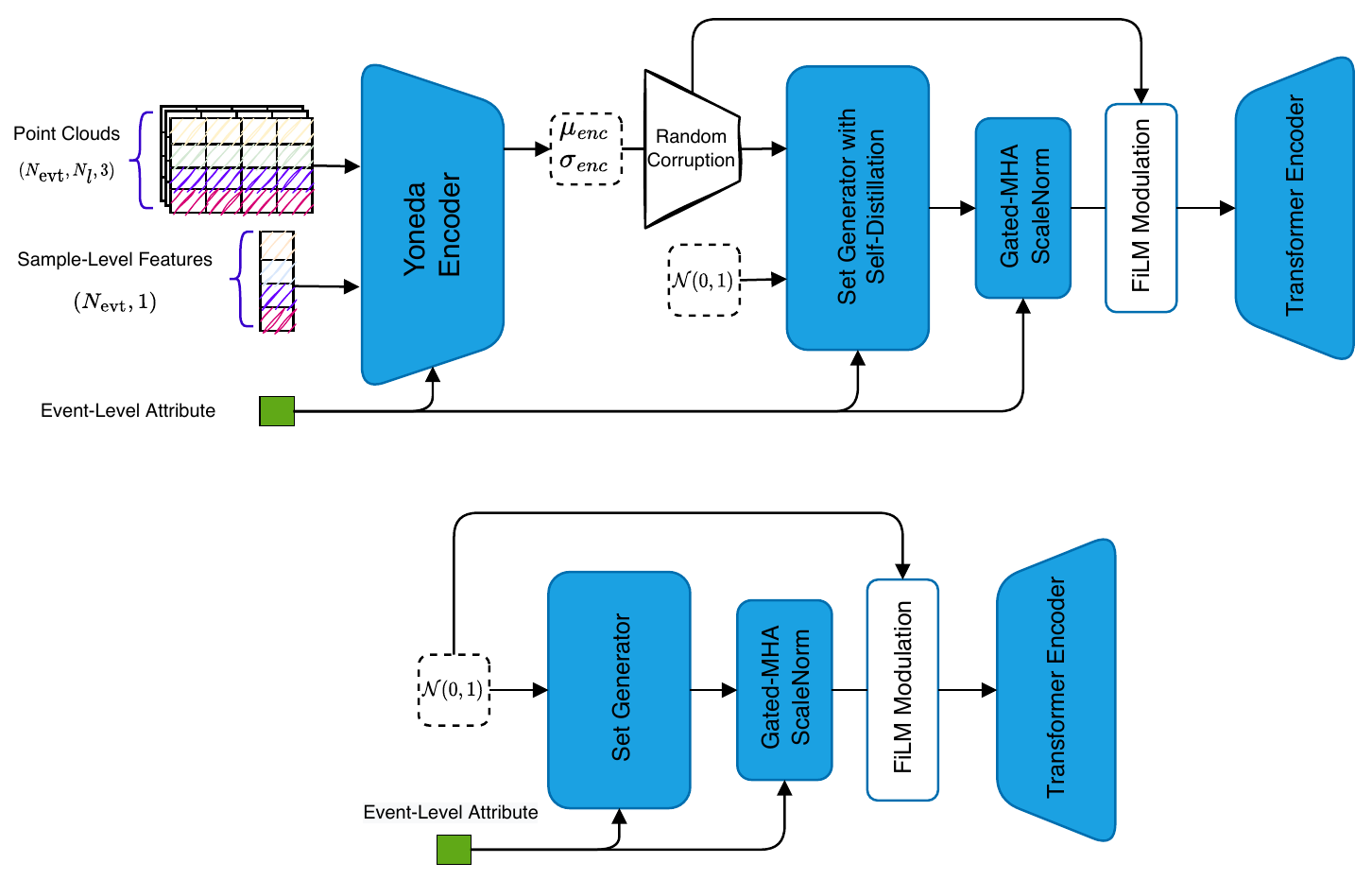}
    \caption{YonedaVAE in training~(top) and inference~(bottom)}
    \label{fig:yonedavae}
\end{figure}

\textbf{Training} At training time, the decoder can access the true sample-level features. They go through a Transformer encoder to encode their relationship. Afterward, they will be injected into the created set by the set generator via the cross-attention mechanism to create an enriched set. While previous works usually append the latent vector to the enriched set, this study exploits the equivalence between summation and concatenation when applying a linear layer. Hence, the enriched set gets fused to the latent vector~(still corrupted due to preventing overfitting) by a feature-wise linear modulation~(FiLM) layer~\cite{perez_film_2017}, which writes

\begin{equation}
\operatorname{cat}(\mX^0, ~ \bm 1_n \vz^T) ~ \mW = \mX^0 \mW_1 + \bm 1_n (\vz^T \mW_2),
\end{equation}

for $\mW = \operatorname{cat}(\mW_1, \mW_2)$. Contrary to the concatenation~(left-hand side), the sum~(right-hand side) does not compute $\vz^T \mW_2$ several times, which reduces the complexity of this layer from $O(n (\dhid + \dlat) \dhid)$ to $O(n \dhid^2 + \dhid \dlat + n \dlat)$. As a result, it combines the sum and multiplicative modulation in a FiLM layer to build a more expressive model. This feature-wise affine transformation based on conditional information generalizes well to challenging, new data of zero-shot regime.
Then, the latent-injected enriched set goes into a Transformer Encoder network with Scale normalization to learn to produce output points in $\mathbb{R}^3$ with the proper inter-dependencies, as depicted in~\cref{fig:yonedavae}. 
The Transformer Encoder learns this point-by-point association by benefiting from the gated self-attention mechanism. In order to reach length extrapolation, as I discussed at the beginning of the section, one needs to incorporate a positional encoding mechanism. 
After several trials and in-depth exploration of different methods, I incorporated a modified version of ALiBi~\cite{press_train_2022} embedding to be used in the bidirectional setup of the Transformer Encoder. Since it was originally proposed for autoregressive setups with causal masking, the modified version adopts it for the bidirectional setup of the Transformer Encoder without a causal mask. 
Another vital factor for extrapolation in this module is the dropout operation. The Transformer Encoder module benefits from $0.1$ point embedding dropout, $0.1$ layer dropout, and $0.1$ Attention dropout. 

For loss functions, the model needs to incorporate a permutation invariant reconstruction loss, as discussed in~\cref{chap:3}. YonedaVAE benefits from the Hungarian loss~\cref{eq:hungarian}, as the main reconstruction loss and the differentiable geometric loss based on Optimal Transport~\cite{feydy_geometric_nodate} as a loss over only the third channel of the point cloud, the charge. The version of geometric loss that this study uses computes the Sinkhorn divergence between two empirical measures. 
Given two sets of points \( \mathcal{X} = \{ x_1, x_2, \ldots, x_m \} \) and \( \mathcal{Y} = \{ y_1, y_2, \ldots, y_n \} \) in \( \mathbb{R}^d \), the empirical measures are \( \mu = \frac{1}{m} \sum_{i=1}^{m} \delta_{x_i} \) and \( \nu = \frac{1}{n} \sum_{j=1}^{n} \delta_{y_j} \). The Sinkhorn divergence \( S_{\epsilon}(\mu, \nu) \) is defined as:

\begin{equation}
S_{\epsilon}(\mu, \nu) = W_{\epsilon}^p(\mu, \nu) - \frac{1}{2} W_{\epsilon}^p(\mu, \mu) - \frac{1}{2} W_{\epsilon}^p(\nu, \nu),
\label{eq:geo}
\end{equation}

Where \( W_{\epsilon}^p(\mu, \nu) \) is the entropically-regularized Wasserstein distance defined by:

\[
W_{\epsilon}^p(\mu, \nu) = \min_{\gamma \in \Pi(\mu, \nu)} \left\langle \gamma, C \right\rangle + \epsilon H(\gamma)
\]

Where, \( \langle \gamma, C \rangle = \sum_{i=1}^{m} \sum_{j=1}^{n} \gamma_{ij} c(x_i, y_j) \) is the dot product of \( \gamma \) with the cost matrix \( C \), where \( c(x_i, y_j) = || x_i - y_j ||^p \), \( \Pi(\mu, \nu) \) is the set of joint measures with marginals \( \mu \) and \( \nu \), \( H(\gamma) = -\sum_{i,j} \gamma_{ij} \log(\gamma_{ij}) \) is the entropy of \( \gamma \), and \( \epsilon \) is the regularization parameter. 
The optimal transport problem for \( \gamma \) is typically solved using the Sinkhorn-Knopp algorithm~\cite{knight_sinkhornknopp_2008} to approximate the Wasserstein distance in high-dimensional spaces effectively with time complexity $O(n \log n)$. 
The parameter \( p = 2 \) specifies the power of the distance \( || x_i - y_j ||^p \) in the cost function \( c(x_i, y_j) \).
I apply this geometric loss only to the charge loss since, upon experiments, having this loss over all channels of the point cloud confuses the model and makes the training very long. However, since the essence of the third channel~(charge) is somewhat different than the first two channel~(x and y dimension of the PXD), using the geometric loss over the third channel actually acts as a learning disentanglement force for the model to learn the difference between charge and coordinate channels. Without this geometric loss over Charge, the model could not capture the full distribution of the PXD charge, especially the long-tail~(high intensity) of the distribution.

For the VAE's prior regularization term~(discussed in~\cref{chap:3}), I incorporate the Maximum mean discrepancy Variational Autoencoder~(MMD-VAE)~\cite{gretton_kernel_2006,zhao_infovae_2018} regularization approach. Maximum mean discrepancy~(MMD) \cite{gretton_kernel_2006} is based on the idea that two distributions are identical if and only if all their moments are the same. Therefore, one can define a divergence by measuring how different the moments of two distributions \( p(z) \) and \( q(z) \) are. MMD can accomplish this efficiently via the kernel embedding trick:

\begin{equation}
\text{MMD}(p(z) \| q(z)) = \mathbb{E}_{p(z), p(z')}[k(z, z')] + \mathbb{E}_{q(z), q(z')}[k(z, z')] - 2\mathbb{E}_{p(z), q(z')}[k(z, z')]
\label{eq:mmd}
\end{equation}

where $k(z, z')$ is any universal kernel, such as the Gaussian kernel $k(z, z') = e^{-\frac{\| z - z' \|^2}{2\sigma^2}}$. 
A rough intuition of MMD is that if two distributions are identical, then the average ``similarity'' between samples from each distribution should be identical to the average ``similarity'' between mixed samples from both distributions. MMD regularization loss always prefers to maximize the mutual information between the data and the prior and avoid overfitting the data during training. Notice that the self-distillation mechanism that was discussed earlier has the same objective. 

Hence, in the end, YonedaVAE benefits from the Hungarian loss~(\cref{eq:hungarian}) and geometric loss~(\cref{eq:geo}) as reconstruction losses, MMD loss~(\cref{eq:mmd}) as regularization loss, and Self-Distillation loss~(\cref{eq:sd}) in conjunction with the supervised auxiliary  loss to learn cardinalities~(\cref{eq:sup}), all together as follows:


\begin{align}
\mathcal{L}_{\text{YonedaVAE}} &= \mathcal{L}_{\text{Rec}} + \mathcal{L}_{\text{Reg}} + \mathcal{L}_{\text{SelfD}} \\
\mathcal{L}_{\text{Rec}} &= \sum_{i=1}^{m}\sum_{j=1}^{n} \min_{\pi} d(x_i, y_j) + S_{\epsilon}^{\text{geom}}(\mu, \nu) \\
S_{\epsilon}^{\text{geom}}(\mu, \nu) &= W_{\epsilon}^p(\mu, \nu) - \frac{1}{2} W_{\epsilon}^p(\mu, \mu) - \frac{1}{2} W_{\epsilon}^p(\nu, \nu) \\
\mathcal{L}_{\text{Reg}} &= \text{MMD}(p(z), q(z)) + D_{\text{KL}}(p(z) \| q(z)) \\
\text{MMD}(p(z), q(z)) &= \mathbb{E}_{p(z), p(z')}[k(z, z')] + \mathbb{E}_{q(z), q(z')}[k(z, z')] - 2\mathbb{E}_{p(z), q(z')}[k(z, z')] \\
\mathcal{L}_{\text{SelfD}} &= \frac{1}{2} \left( \mathcal{D}(\text{p}_1, \text{sg}(z_2)) + \mathcal{D}(\text{p}_2, \text{sg}(z_1)) \right) + \left\lVert h(g(z_1)) - \mathbf{e} \right\rVert_2^2.
\end{align}

\textbf{Inference} The only differences for the decoder at the inference time are the decoder's lack of access to the true sample-level features and the learned latent space. In the inference time, the decoder has only access to the generated sample-level features by the set generator and the Normal latent distribution, that is, the prior distribution of the VAE. Everything else is exactly the same as the training time setup.

With YonedaVAE, one is now able to not only emulate PXD background signatures given the desired event profile but also simulate OOD PXD responses. The following section demonstrates the extent of success of YonedaVAE on OOD simulation~(Length and context extrapolation), on low-level hit topology and geometry, and on the downstream physics analysis.

\section{Evaluation Results}
\label{sec:res_6}
\begin{figure}[!htb]
    \centering
    \begin{subfigure}[b]{0.45\textwidth}
        \includegraphics[width=\textwidth]{figures/chap_6/sens/rsen0_1.png}
    \end{subfigure}
    \begin{subfigure}[b]{0.45\textwidth}
        \includegraphics[width=\textwidth]{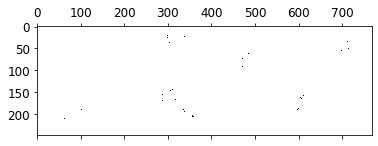}
    \end{subfigure}
    
    \begin{subfigure}[b]{0.45\textwidth}
        \includegraphics[width=\textwidth]{figures/chap_6/sens/rsen0_0.png}
    \end{subfigure}
    \begin{subfigure}[b]{0.45\textwidth}
        \includegraphics[width=\textwidth]{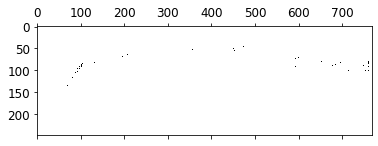}
    \end{subfigure}
    
    \begin{subfigure}[b]{0.45\textwidth}
        \includegraphics[width=\textwidth]{figures/chap_6/sens/rsen0_3.png}
    \end{subfigure}
    \begin{subfigure}[b]{0.45\textwidth}
        \includegraphics[width=\textwidth]{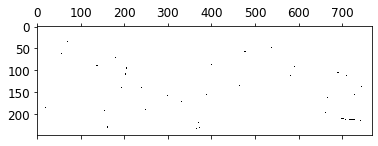}
    \end{subfigure}
    
    \caption{Random-trigger~(real) PXD background~(left) and YonedaVAE generated PXD background~(right) in a hitmap format. As stated before, the sparsity of the real PXD background is much less than the Geant4 version~(last chapter).}
    \label{fig:pxd_tirg}
\end{figure}

In order to evaluate the performance of YonedaVAE in generating ultra-high-granularity detector responses of the Pixel Vertex Detector, shown in~\cref{fig:pxd_tirg}, this section provides results divided into two parts, ``length extrapolation'' and ``context extrapolation''. As discussed in~\cref{par:le_ce}, with the length extrapolation, YonedaVAE has access to the sample-level occupancy/cardinality during inference and has to extrapolate to point cloud cardinalities beyond the training data. So, the extrapolation here is that the model has to infer how an event with a higher~(than training) cardinality looks like.
With the context extrapolation, YonedaVAE has only access to the event-level attribute~(amount of background per event as max cardinality of the event) during inference and has to learn to generate sample-level occupancy/cardinality by itself in a zero-shot manner and extrapolate to point cloud cardinalities and intra-event correlation beyond the training data. So, the extrapolation here is that the model has to produce the correct distribution of cardinalities over all sensors, given only the maximum cardinality per event. 
In general, from the last chapter's fine-grained image analysis perspective~(\cref{chap:5},\cref{sec:finepxd}), one can say that YonedaVAE is now conditioned on both the sensor's location as meta-conditions and the amount of PXD background as subordinate conditions. 

This thesis does the evaluation by first going through the same analysis that I did in the last chapter, namely
the low-level metrics at the hit level such as occupancy and charge distribution, mean occupancy and charge per sensor, sensor-by sensor intra-event correlation, FID~\cite{heusel_gans_2018} and KID~\cite{binkowski_demystifying_2021}. In this chapter, I also introduce two new metrics for detector simulation, namely the topological and geometrical profiling of the PXD background signatures and Vendi Score~\cite{friedman_vendi_2023}, a diversity measure to quantify the diversity of the generated samples. 

At high-level, this thesis goes through the downstream physics analysis of track reconstruction and studies the performance of YonedaVAE's generated samples, trained on experiment 12 random trigger data- recorded in experiment number 12 of Belle~II data taking, with peak recorded luminosity of $1.42\times10^{34} \text{cm}^{-2}\text{s}^{-1}$, and evaluated on experiment 26 random trigger data- recorded in experiment number 26 of Belle~II data taking, with almost double peak luminosity $2.68\times10^{34} \text{cm}^{-2}\text{s}^{-1}$ in order to demonstrate the extrapolation abilities of YonedaVAE.

\subsection{Marginal Distributions}
\label{sec:mar_dist}
\begin{figure}[!htb]
    \centering
    \begin{subfigure}{0.45\linewidth}
        \includegraphics[width=\linewidth]{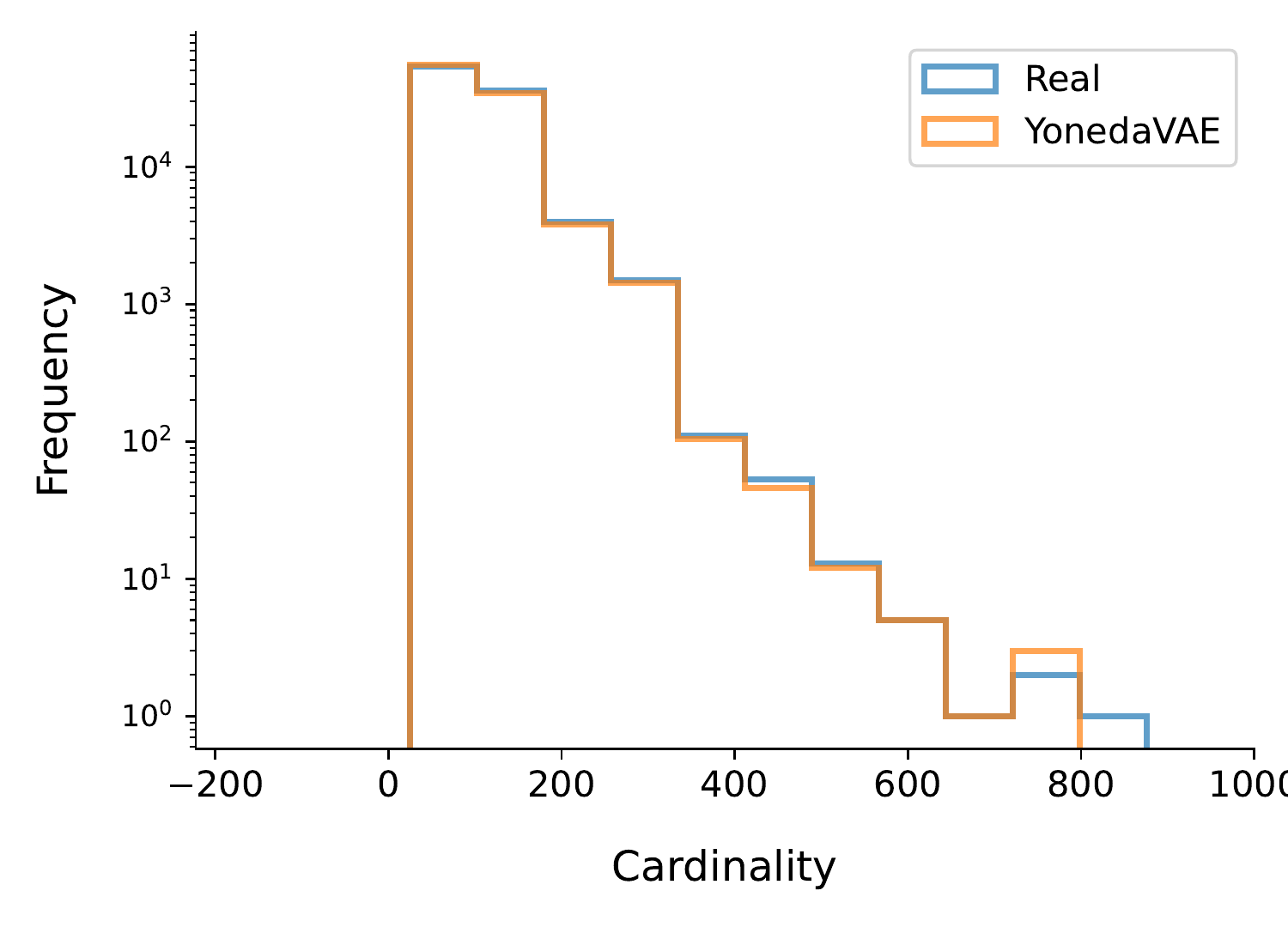}
        \caption{}
    \end{subfigure}
    \begin{subfigure}{0.45\linewidth}
        \includegraphics[width=\linewidth]{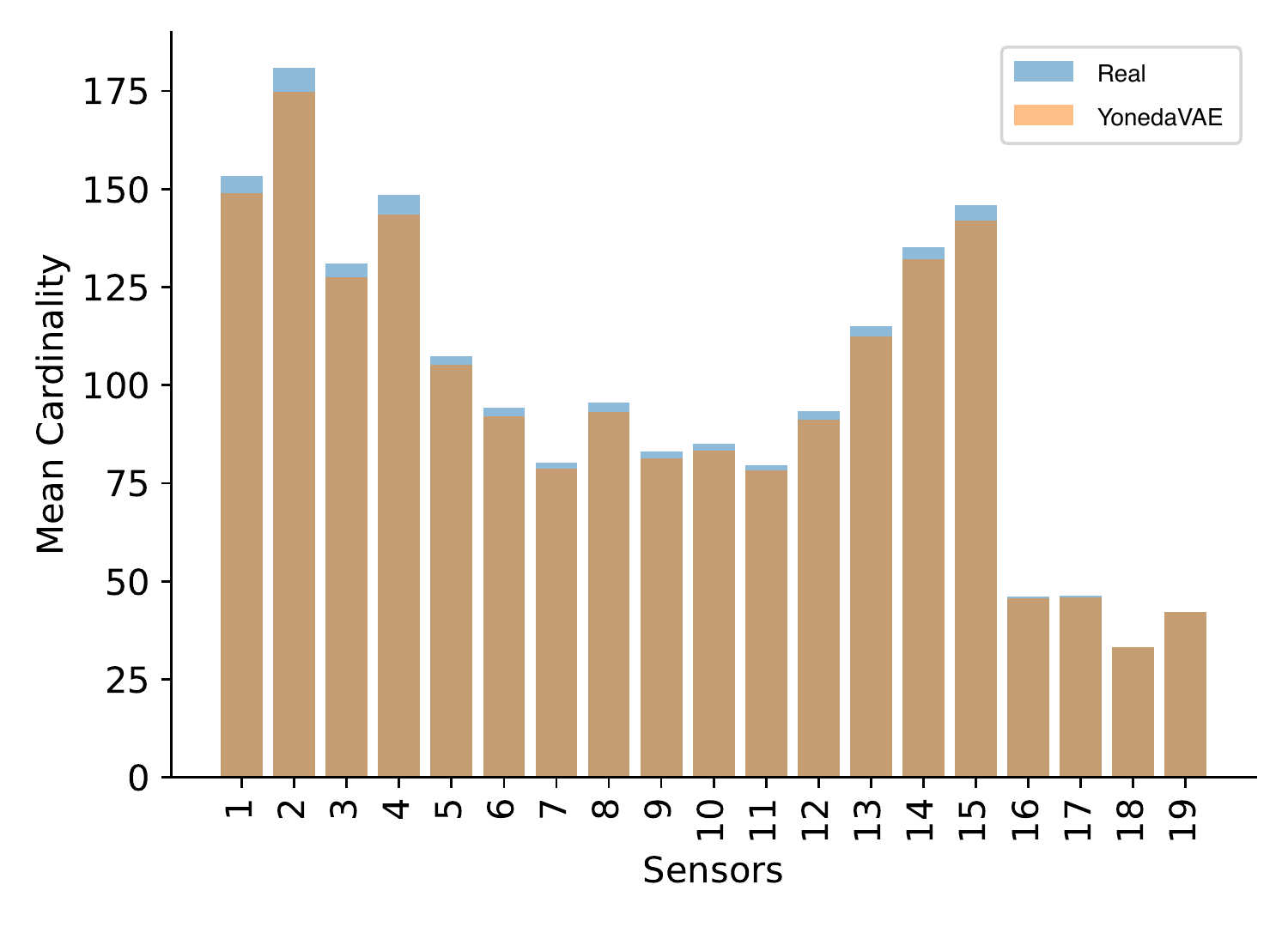}
        \caption{}
    \end{subfigure}\\
    
    \begin{subfigure}{0.45\linewidth}
        \includegraphics[width=\linewidth]{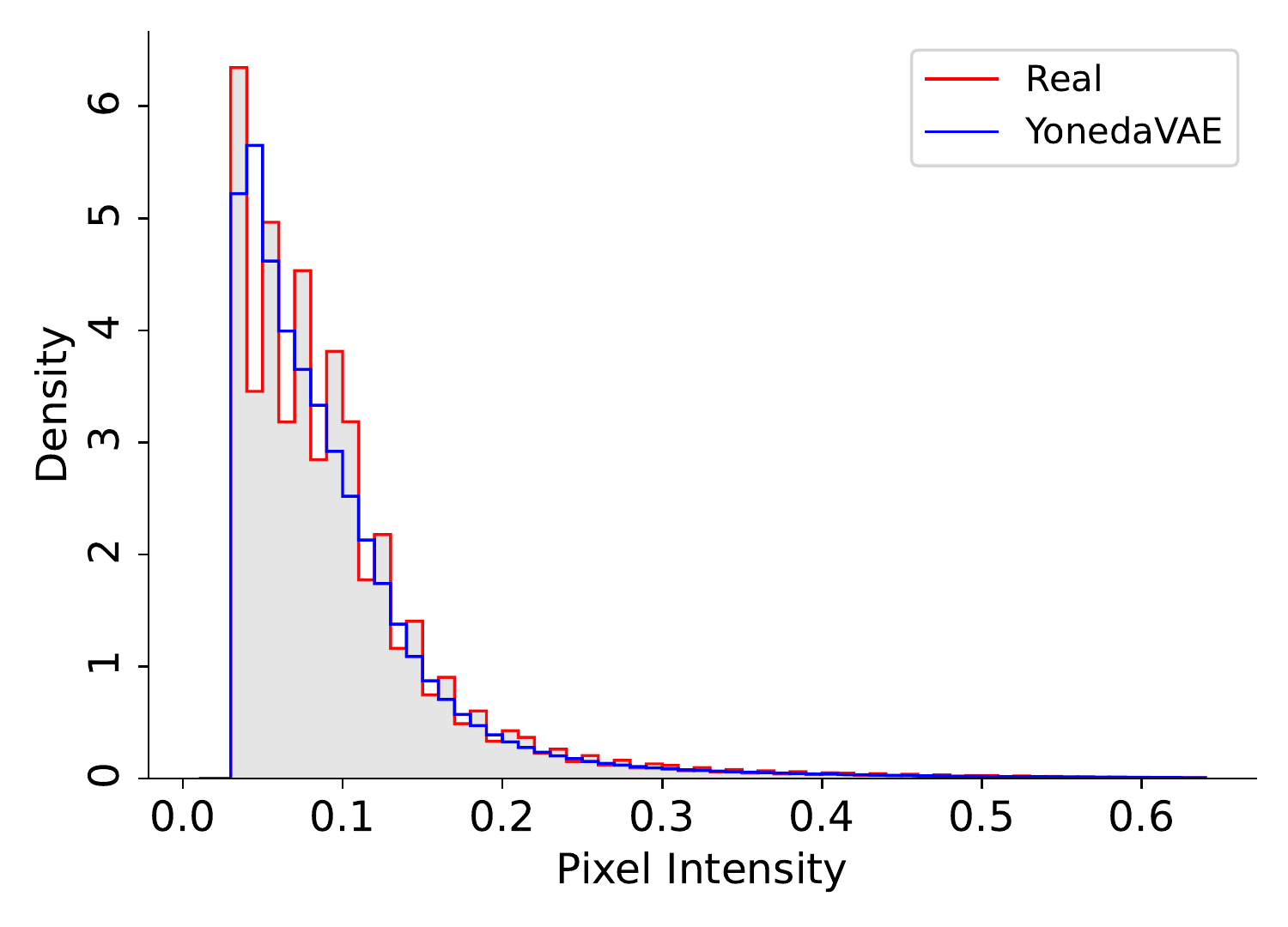}
        \caption{}
    \end{subfigure}
    \begin{subfigure}{0.45\linewidth}
        \includegraphics[width=\linewidth]{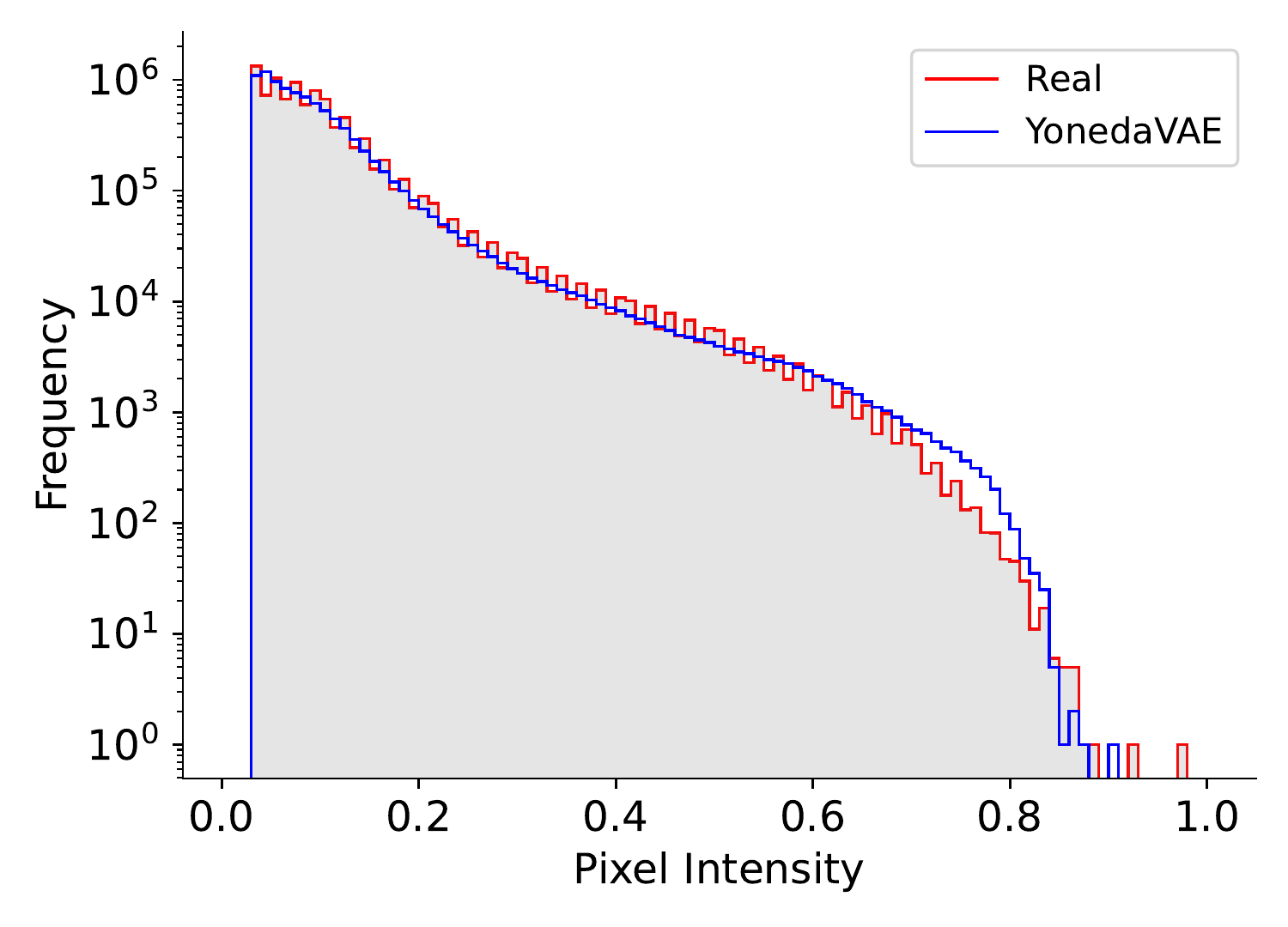}
        \caption{}
    \end{subfigure}\\
    
    \begin{subfigure}{0.45\linewidth}
        \includegraphics[width=\linewidth]{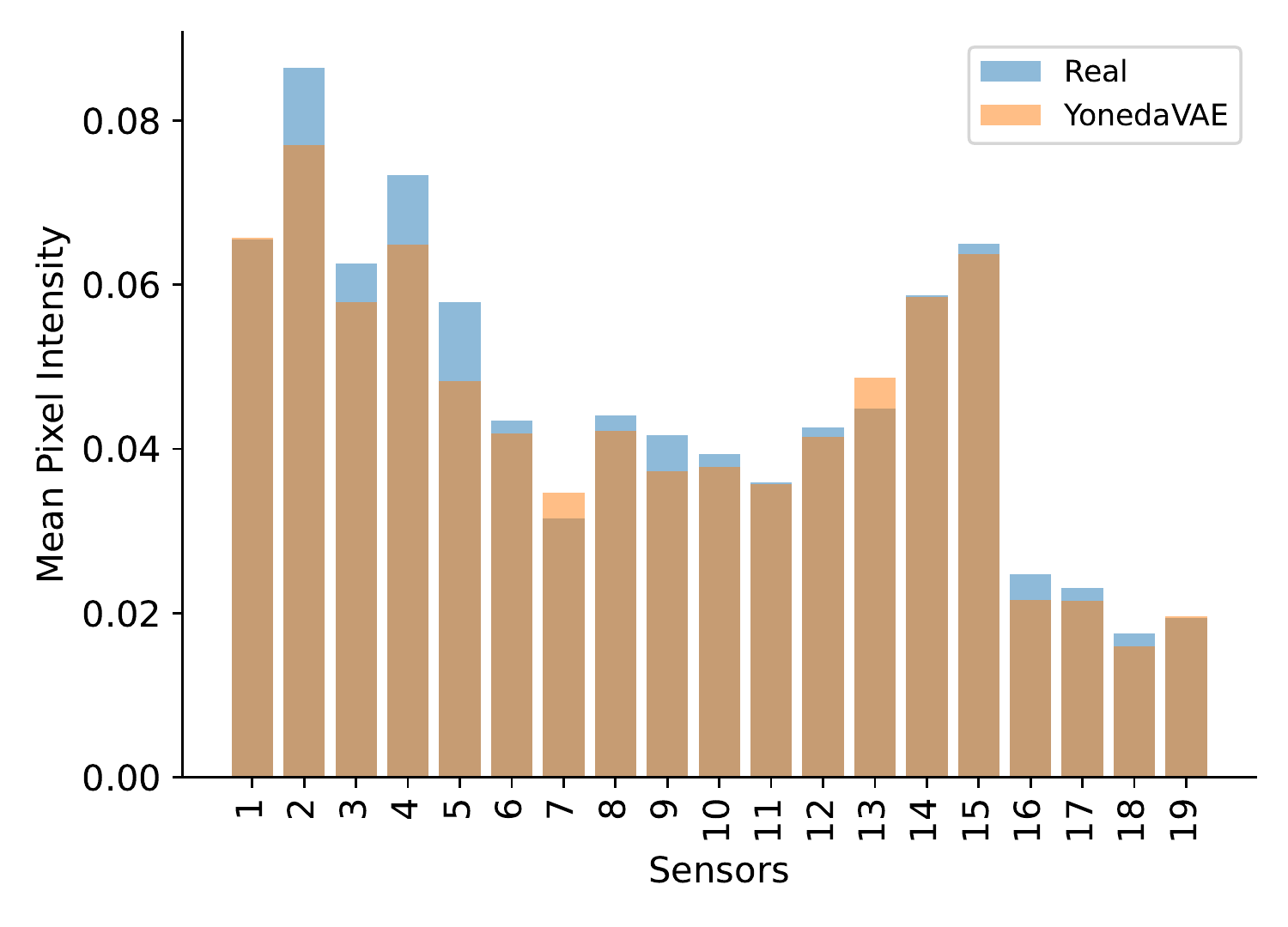}
        \caption{}
    \end{subfigure}
    
    \caption{Marginal Distributions for YonedaVAE on the validation data~(Experiment 12 with arbitrary cardinalities) with access to sensor occupancies as sensor-level features.}
    \label{fig:trmarg_le}
\end{figure}

\begin{figure}[!htb]
\centering
\includegraphics[width=0.9\textwidth]{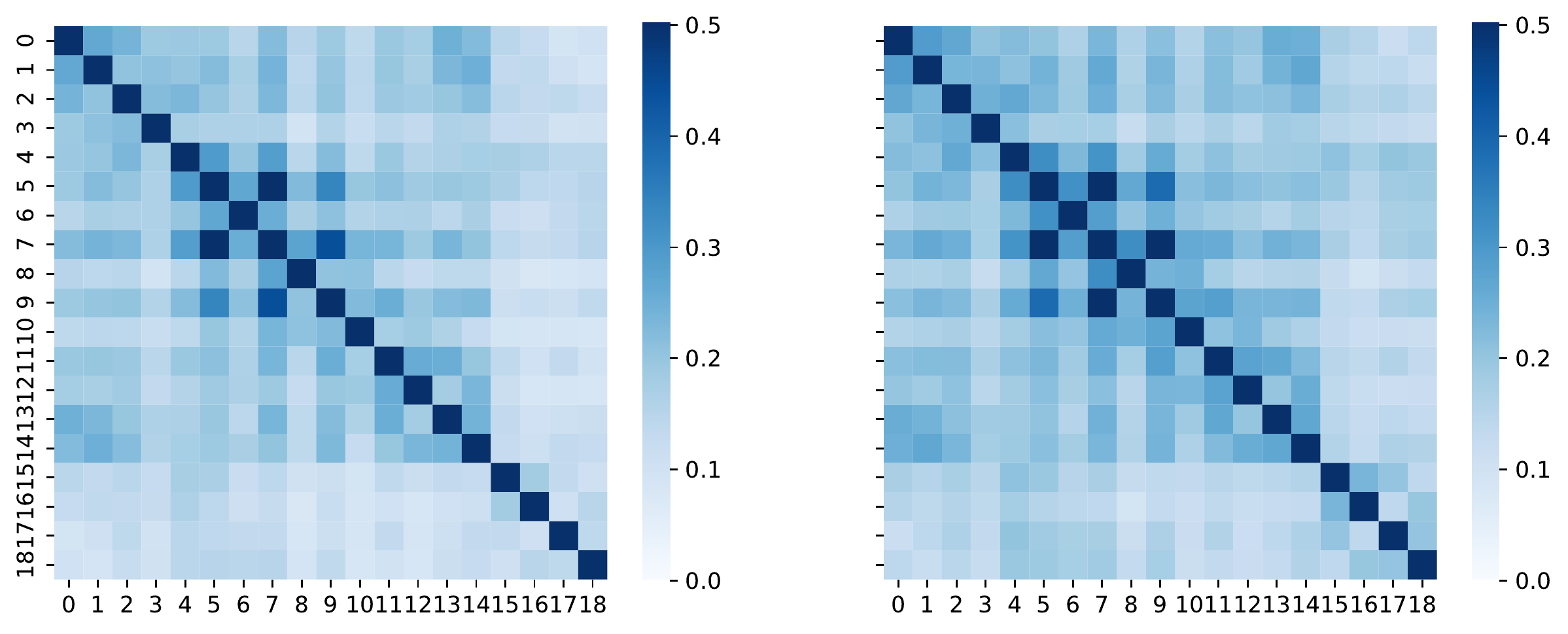} 
\caption{\label{fig:corr_trle}
Pearson Correlation between sensor's mean occupancy for YonedaVAE on the validation data~(Experiment 12 with arbitrary cardinalities) with access to sensor occupancies as sensor-level features.}
\end{figure}

\begin{figure}[!htb]
    \centering
    \begin{subfigure}{0.45\linewidth}
        \includegraphics[width=\linewidth]{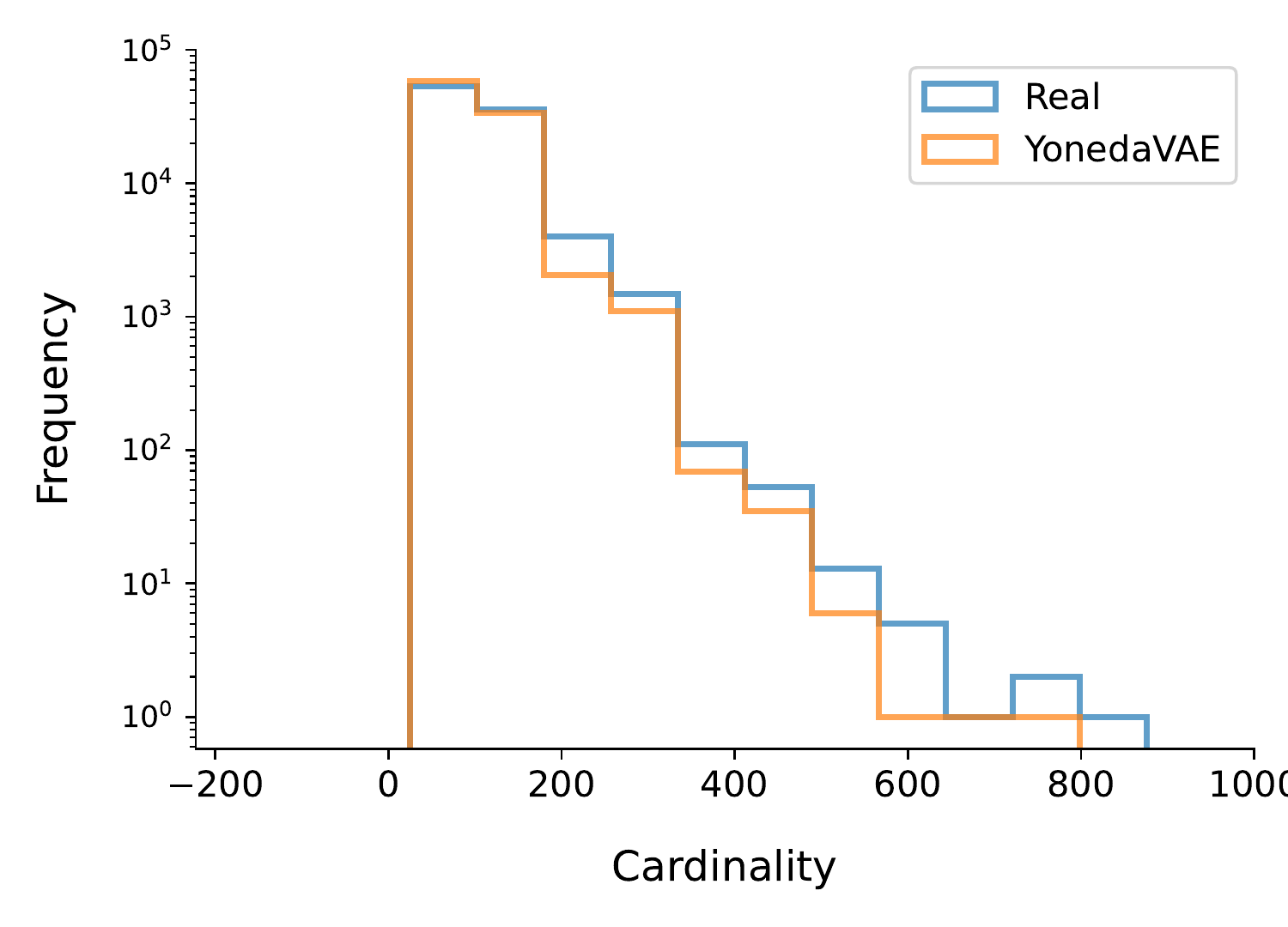}
        \caption{}
    \end{subfigure}
    \begin{subfigure}{0.45\linewidth}
        \includegraphics[width=\linewidth]{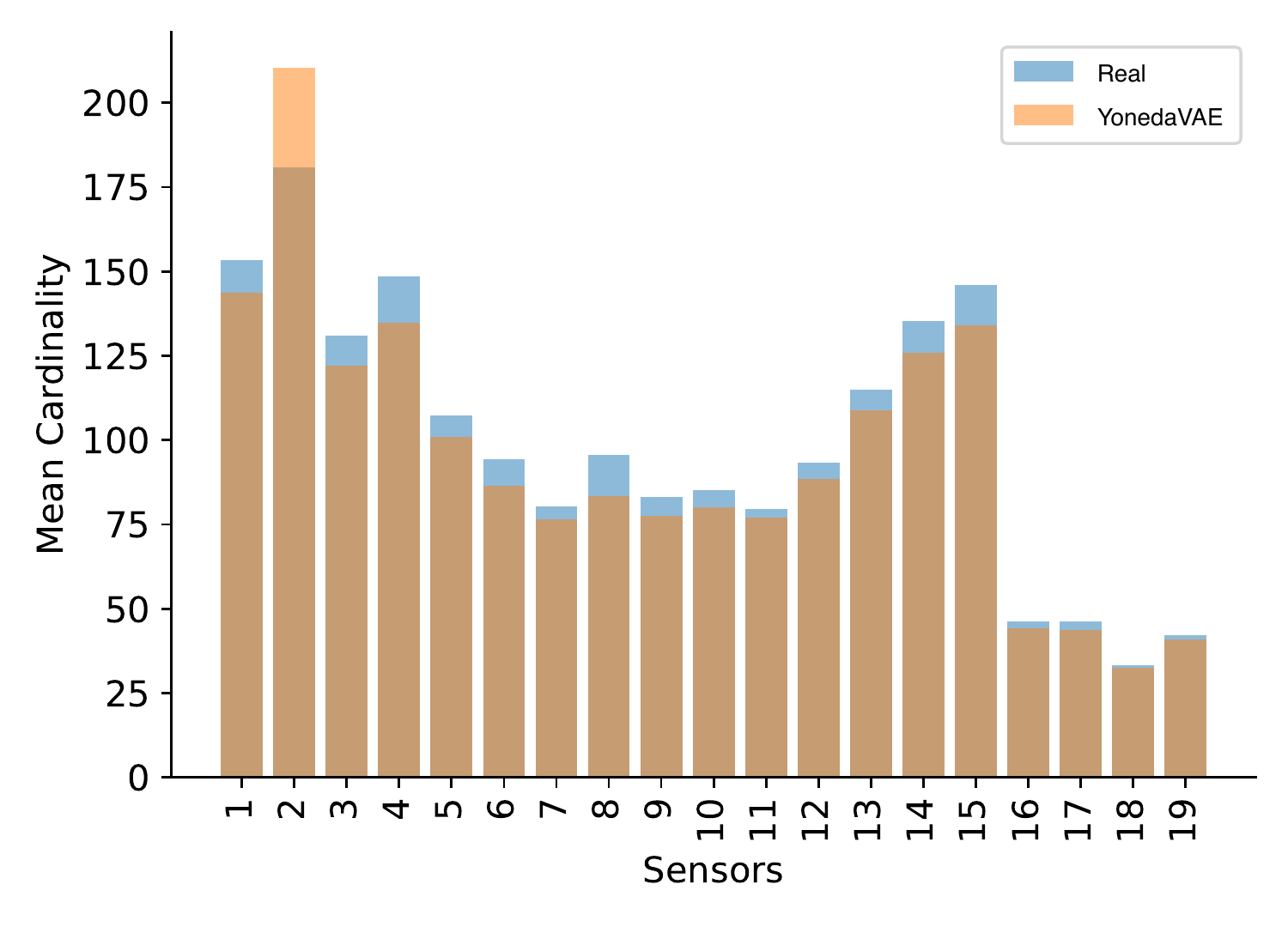}
        \caption{}
    \end{subfigure}\\
    
    \begin{subfigure}{0.45\linewidth}
        \includegraphics[width=\linewidth]{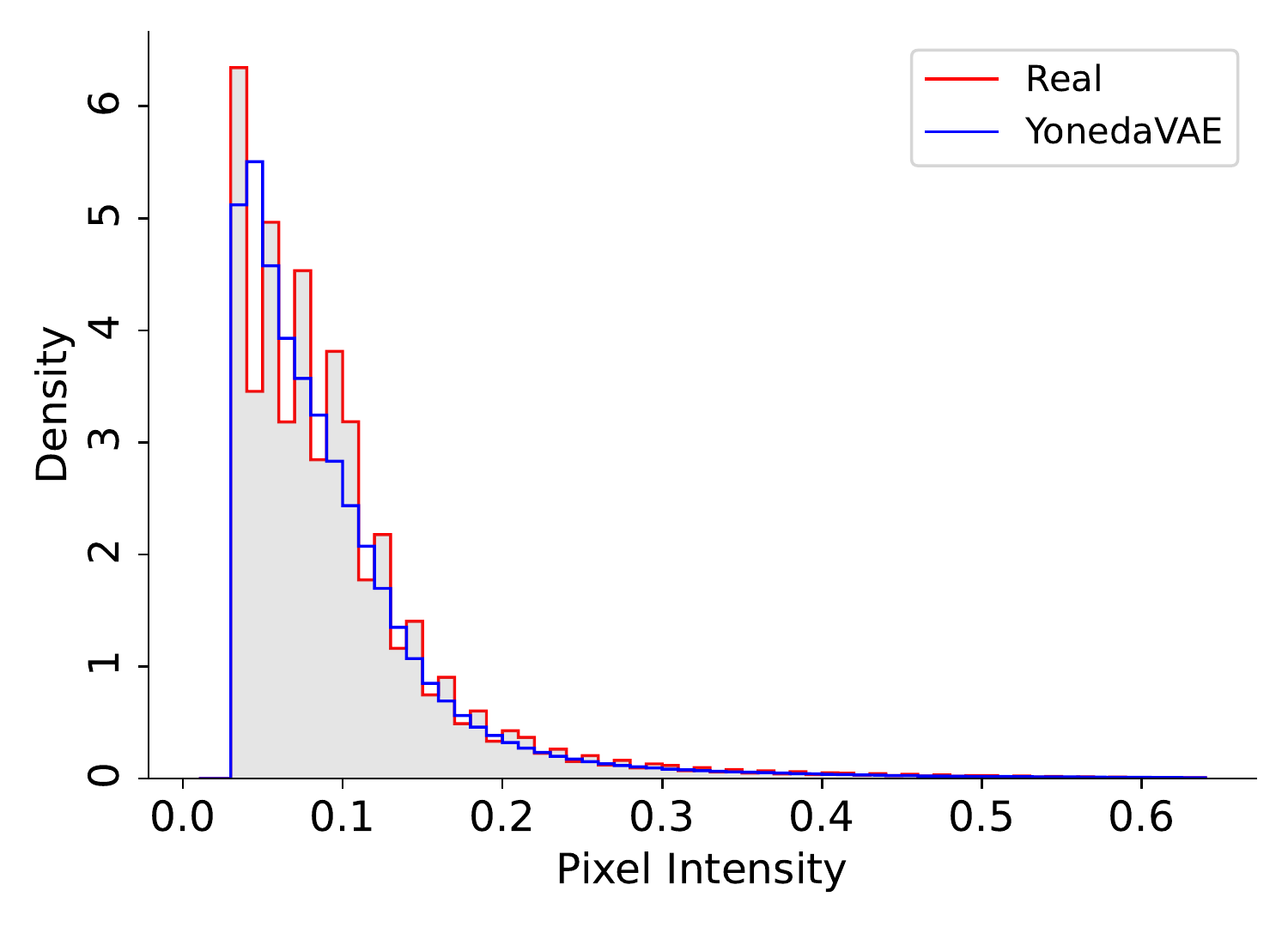}
        \caption{}
    \end{subfigure}
    \begin{subfigure}{0.45\linewidth}
        \includegraphics[width=\linewidth]{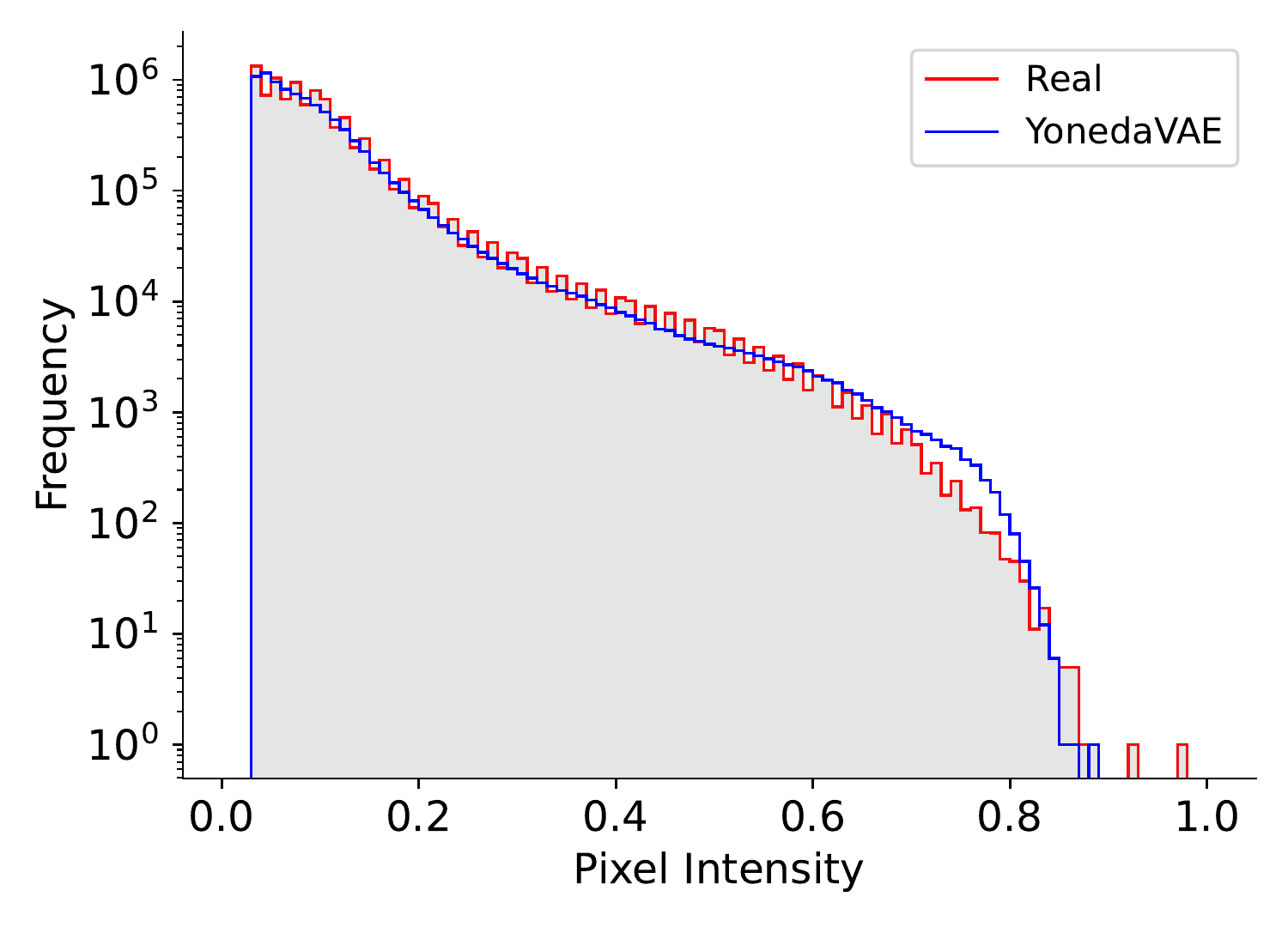}
        \caption{}
    \end{subfigure}\\
    
    \begin{subfigure}{0.45\linewidth}
        \includegraphics[width=\linewidth]{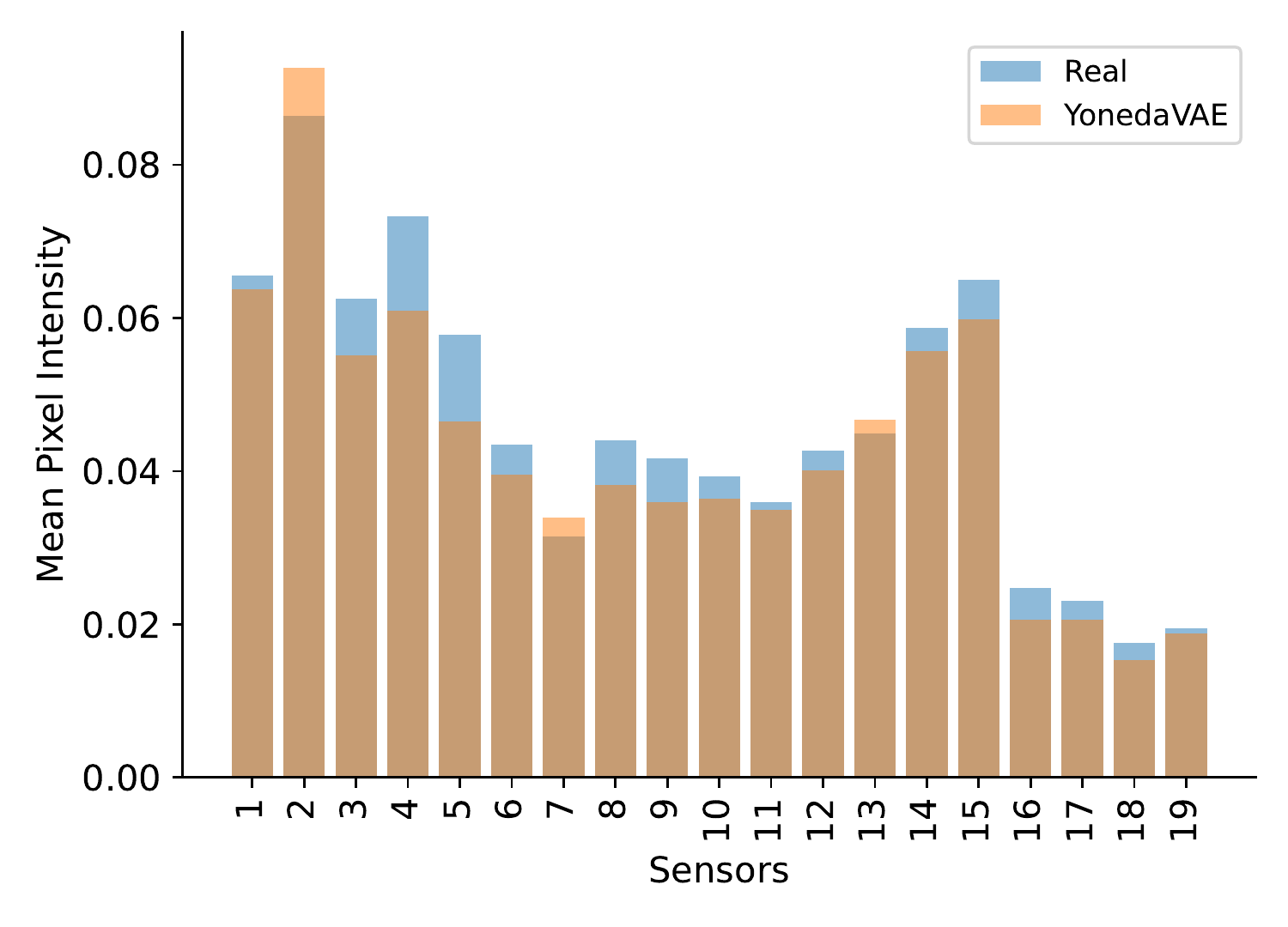}
        \caption{}
    \end{subfigure}
    
    \caption{Marginal Distributions for YonedaVAE on the validation data~(Experiment 12 with arbitrary cardinalities) with access to max cardinality per event as event-level features.}
    \label{fig:trmarg_ce}
\end{figure}

\begin{figure}[!htb]
\centering
\includegraphics[width=0.9\textwidth]{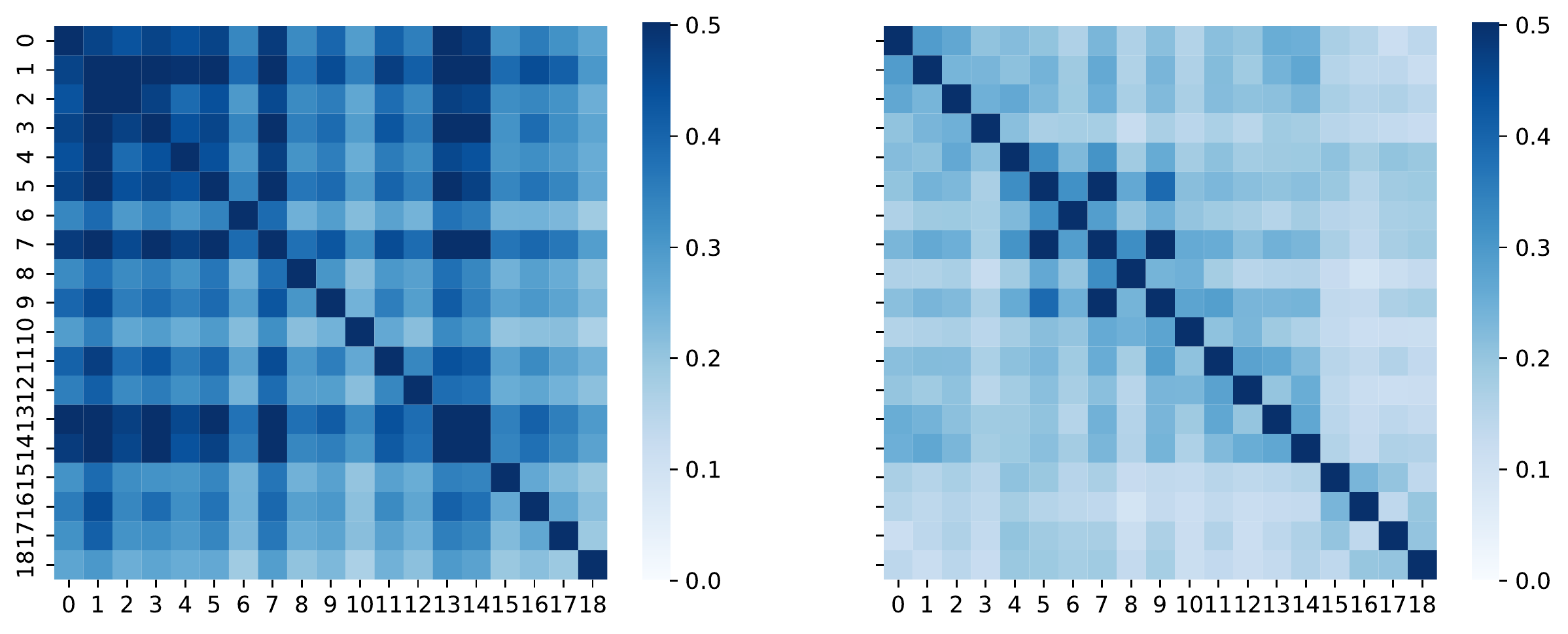} 
\caption{\label{fig:corr_trce}
Pearson Correlation between sensor's mean occupancy for YonedaVAE on the validation data~(Experiment 12 with arbitrary cardinalities) with access to max cardinality per event as event-level features.}
\end{figure}

As the first layer of evaluation, the hit-level marginal distributions are compared. The low-level metrics at the hit level are the occupancy distribution~(a) and charge distribution in linear~(c) and log~(d) scale, mean cardinality~(b), and charge~(c) per sensor and sensor-by-sensor intra-event correlation. I analyze both Pearson~(linear) and Spearman~(monotonic) sensor-by-sensor mean occupancy correlation. As a reminder, I have to emphasize that during training, the model only has access to events with up to $400$ sensor cardinality~(or $7600$ event cardinality). This is done to make sure that the model does not \emph{see} any data with cardinality beyond $400$.
During validation, the model is asked to generate point clouds with arbitrary cardinality over the Experiment 12 dataset. In~\cref{fig:trmarg_le} and~\cref{fig:corr_trle} model has access to the cardinality distribution and is sampling from it. In~\cref{fig:trmarg_ce} and~\cref{fig:corr_trce} model does not have access to the cardinality distribution and only has access to the max occupancy of each event~(event-level attribute). As depicted, YonedaVAE captures all the marginal distributions with good precision for the training and validation data.

\begin{figure}[!htb]
    \centering
    \begin{subfigure}{0.45\linewidth}
        \includegraphics[width=\linewidth]{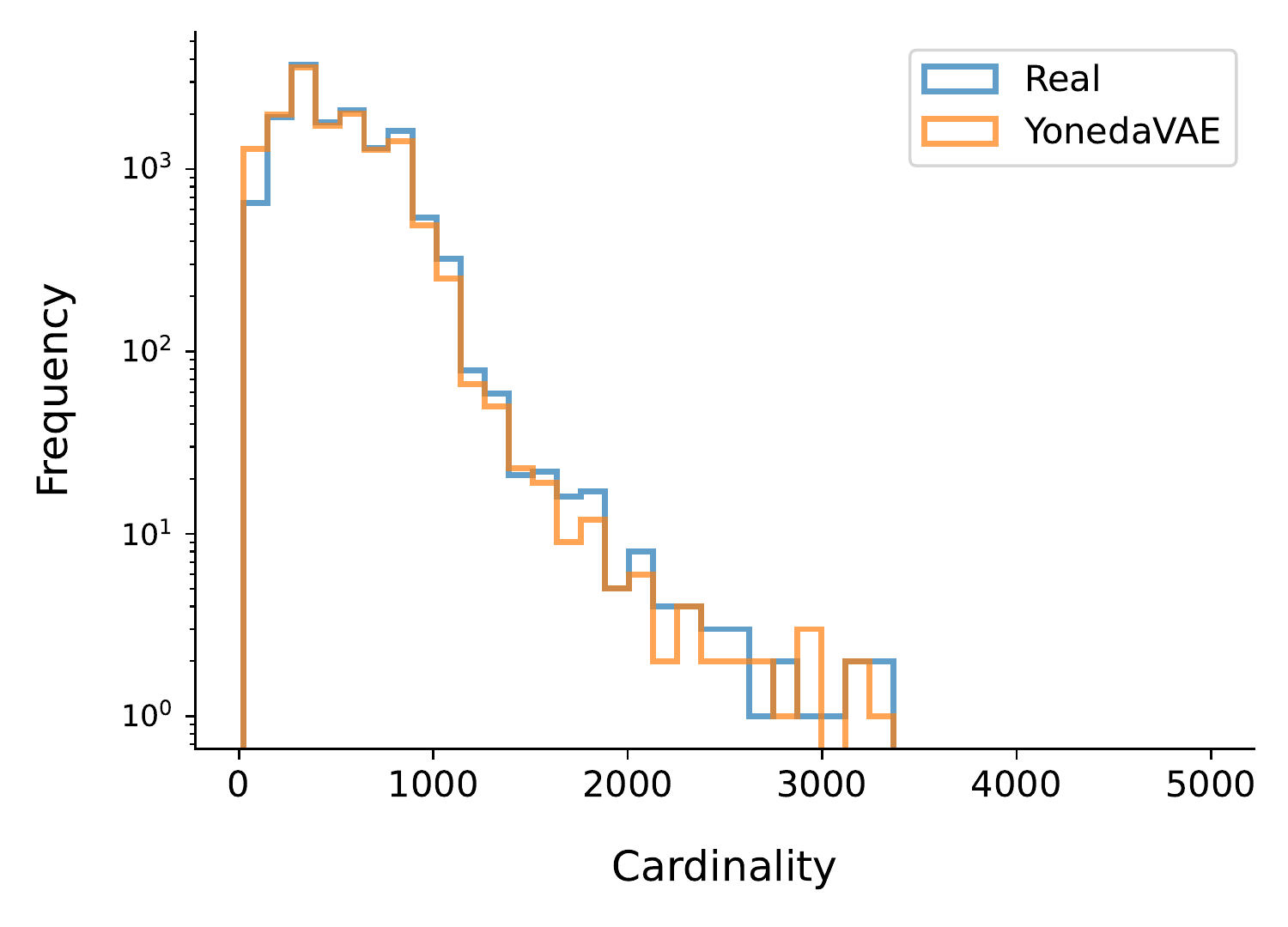}
        \caption{}
    \end{subfigure}
    \begin{subfigure}{0.45\linewidth}
        \includegraphics[width=\linewidth]{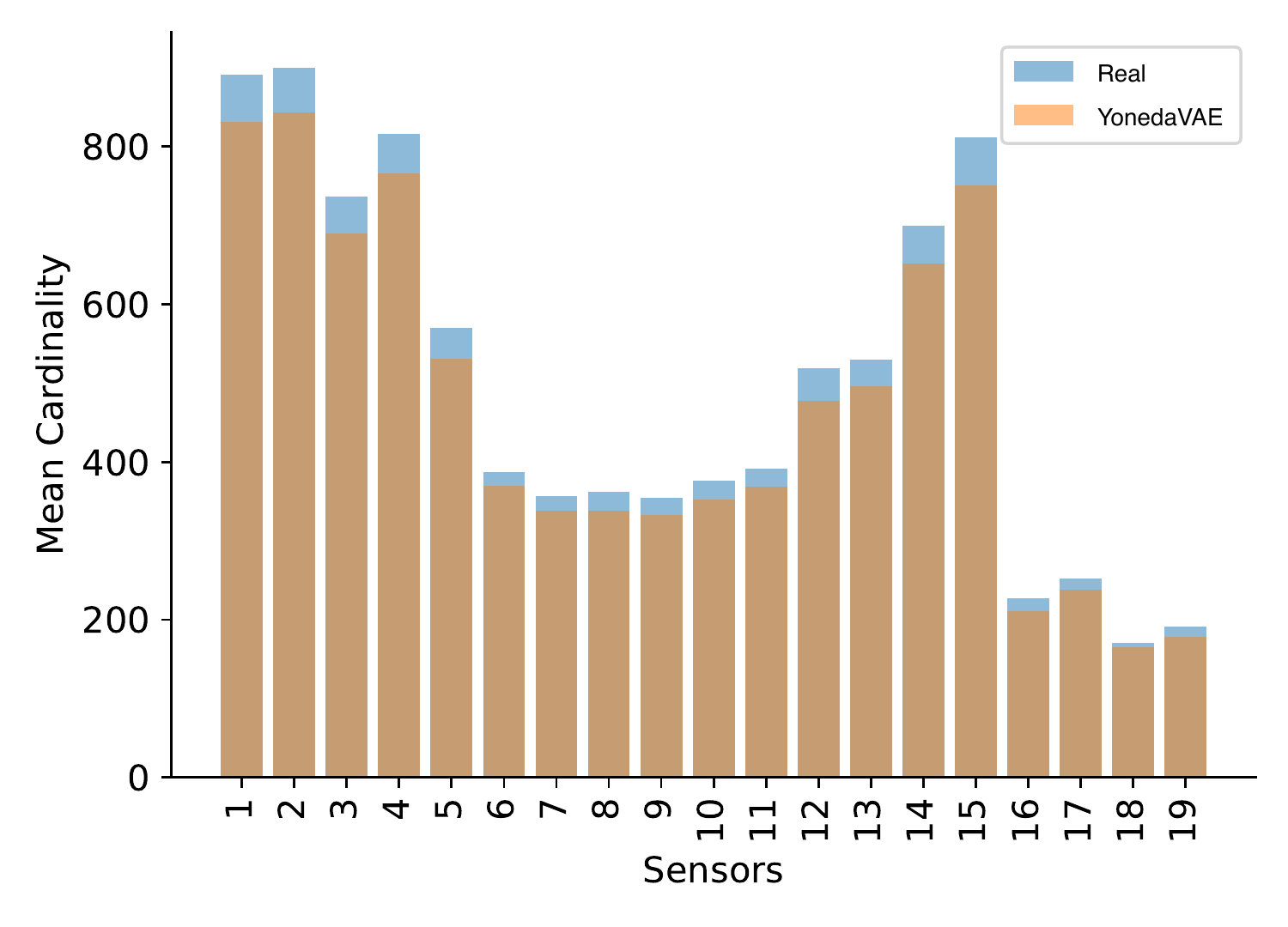}
        \caption{}
    \end{subfigure}\\
    
    \begin{subfigure}{0.45\linewidth}
        \includegraphics[width=\linewidth]{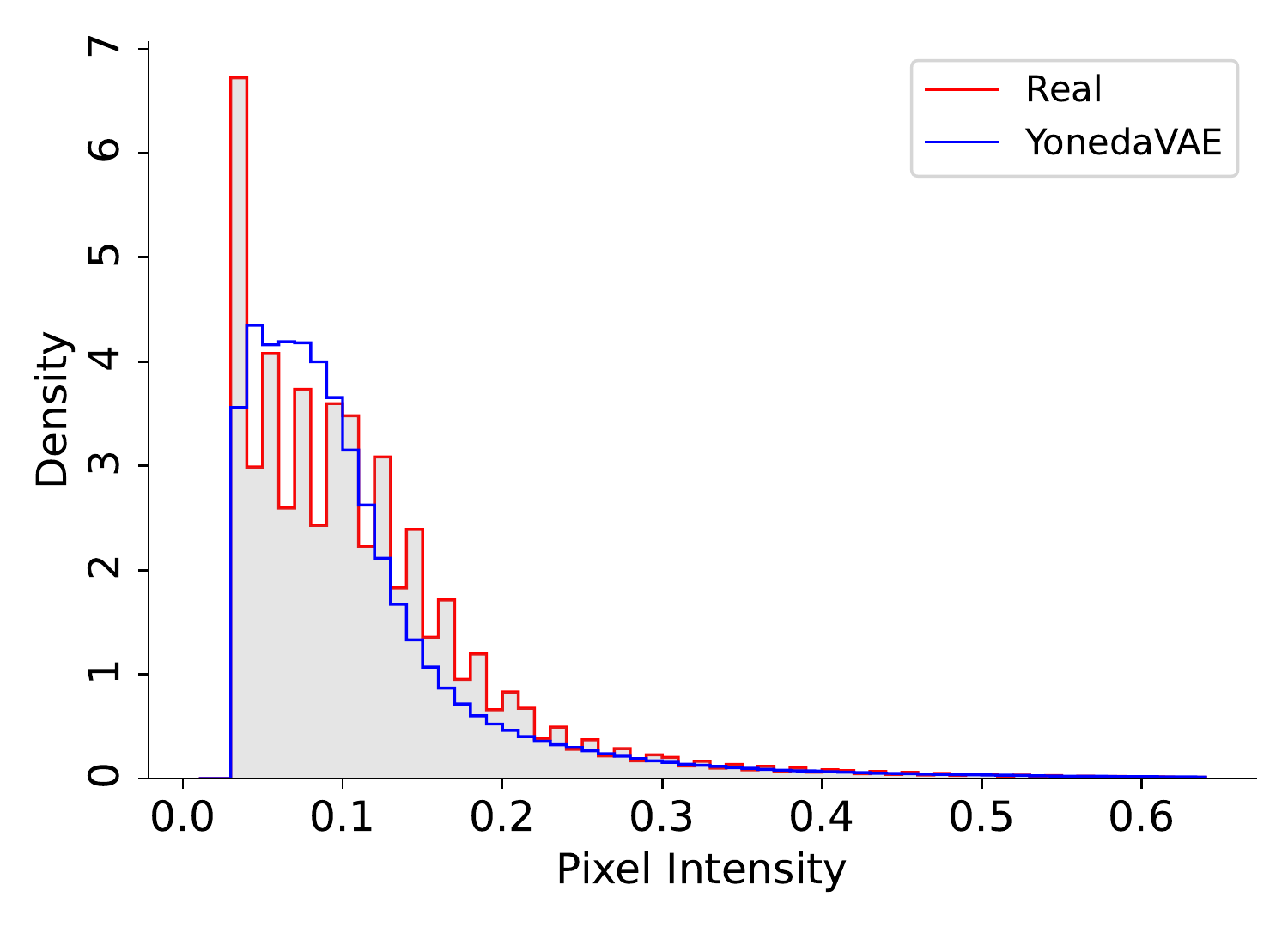}
        \caption{}
    \end{subfigure}
    \begin{subfigure}{0.45\linewidth}
        \includegraphics[width=\linewidth]{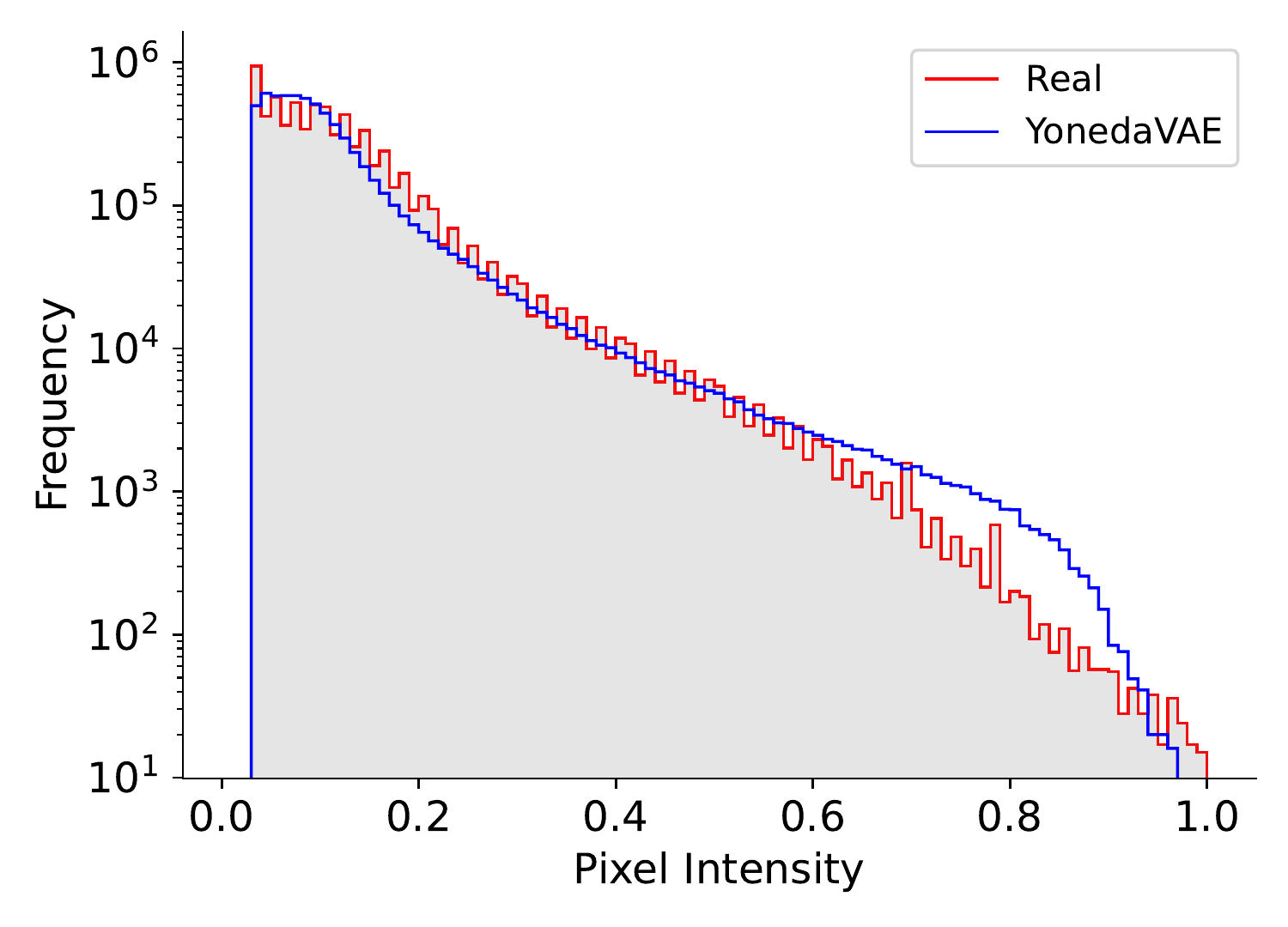}
        \caption{}
    \end{subfigure}\\
    
    \begin{subfigure}{0.45\linewidth}
        \includegraphics[width=\linewidth]{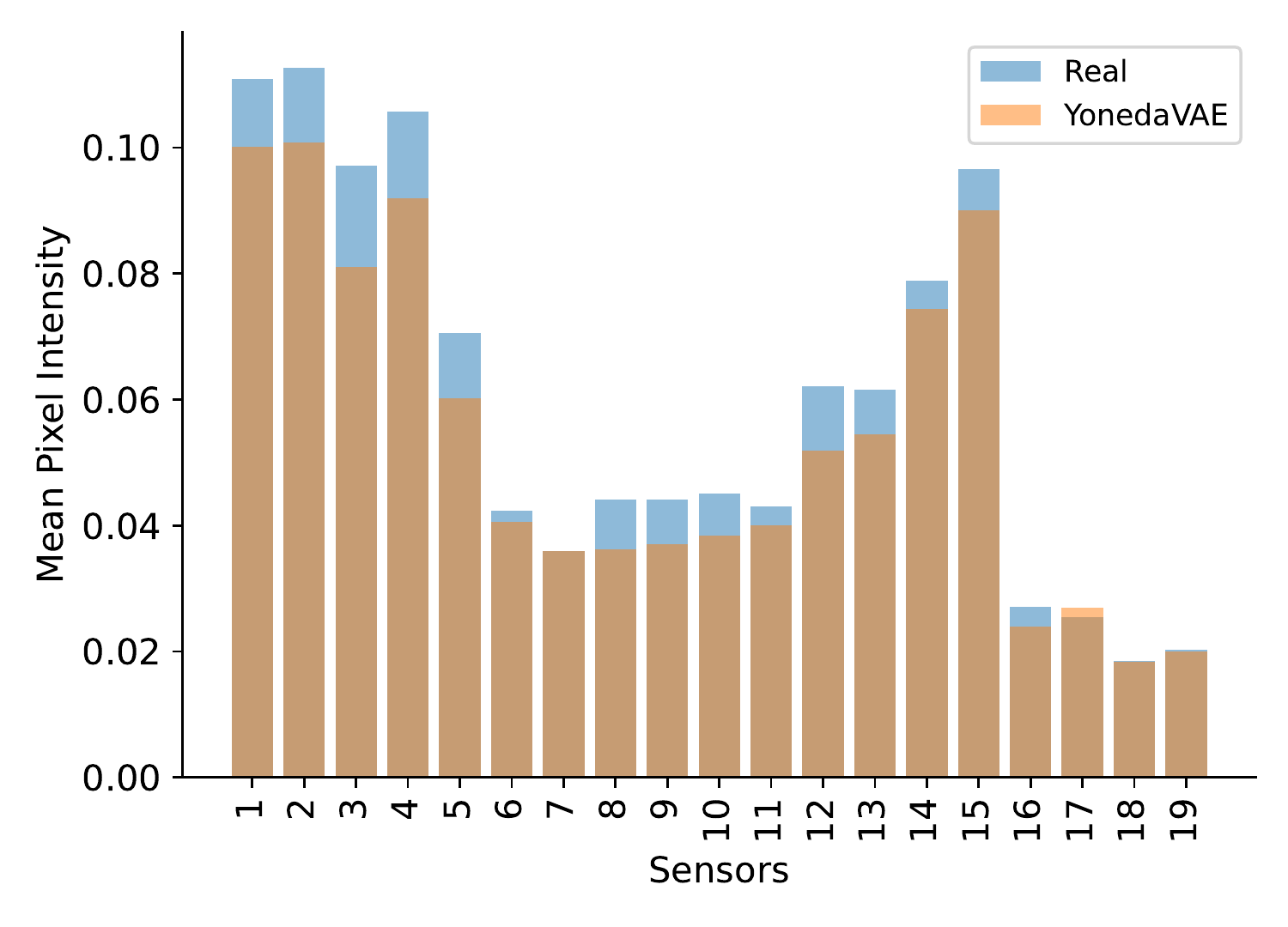}
        \caption{}
    \end{subfigure}
    
    \caption{Marginal Distributions for YonedaVAE with length extrapolation on the OOD data~(Experiment 26) with access to sensor occupancies, as sensor-level features.}
    \label{fig:marg_le}
\end{figure}

\begin{figure}[!htb]
\centering
\includegraphics[width=0.9\textwidth]{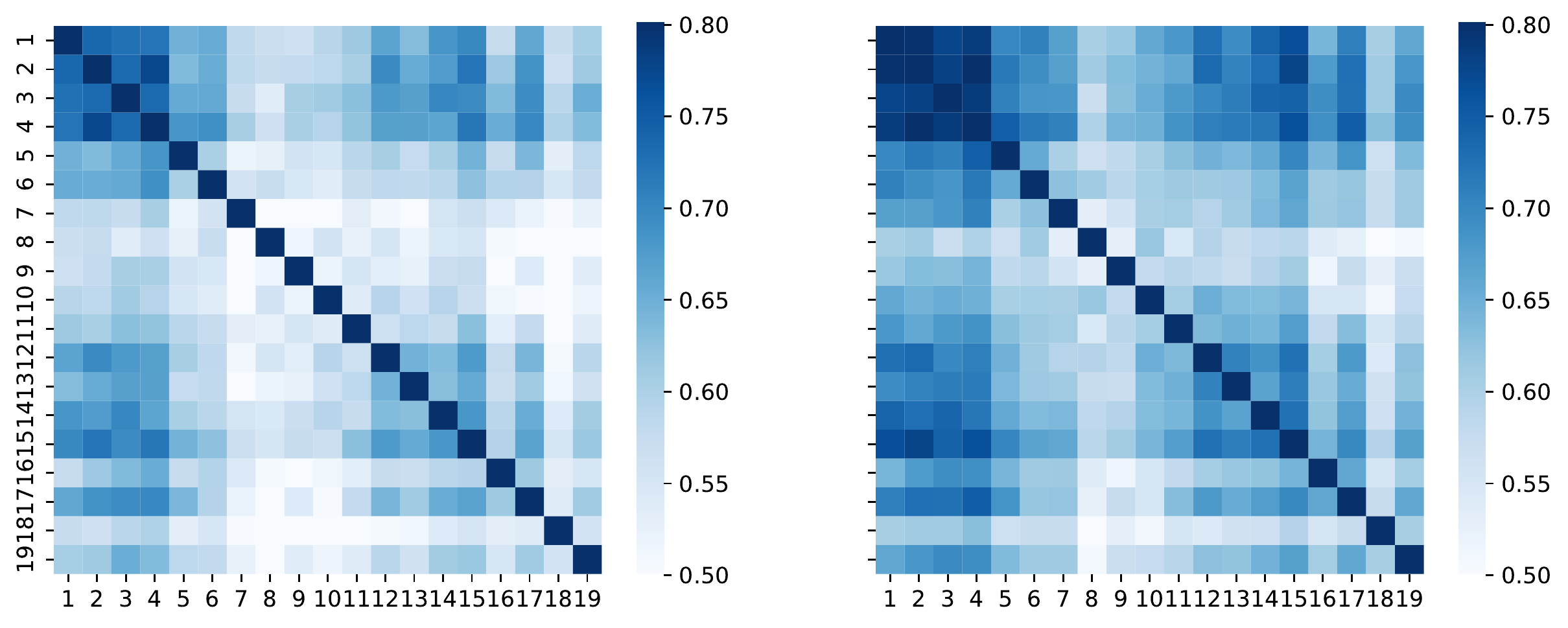} 
\caption{\label{fig:corr_le_pears}
Pearson Correlation between sensor's mean occupancy in the length extrapolation regime for YonedaVAE on the OOD data~(Experiment 26)}
\end{figure}

\begin{figure}[!htb]
\centering
\includegraphics[width=0.9\textwidth]{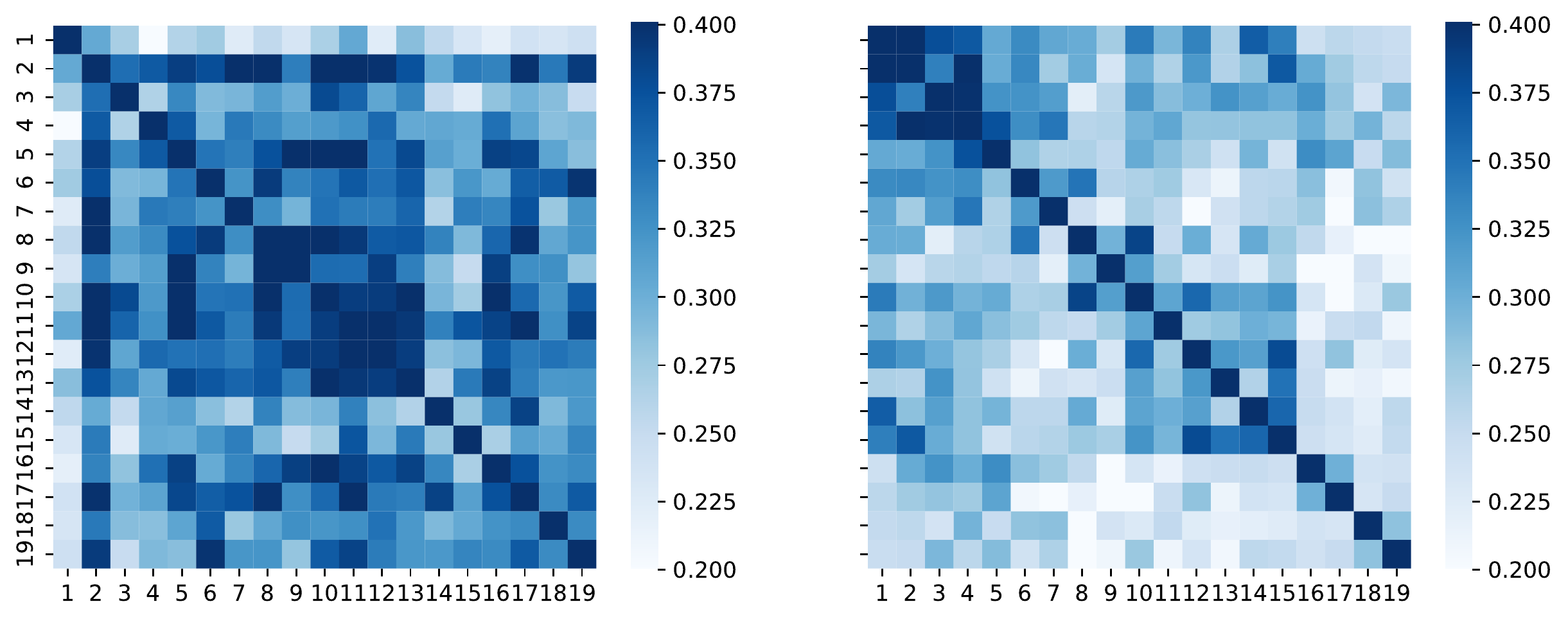} 
\caption{\label{fig:corr_ce_spear}
Spearman Correlation between sensor's mean occupancy in the context extrapolation regime for YonedaVAE on the OOD data~(Experiment 26)}
\end{figure}

As mentioned before, to demonstrate extrapolation beyond the training data, YonedaVAE generates previously unseen PXD background hits from experiment 26~(with almost double luminosity) in two regimes. First with length extrapolation regime where the model samples sensor-level features~(sensor occupancy) depicted in~\cref{fig:marg_le},~\cref{fig:corr_le_pears}, and~\cref{fig:corr_ce_spear}. As shown in~\cref{fig:corr_le_pears} and~\cref{fig:corr_ce_spear}, Experiment 26 interestingly has a different linear and monotonic intra-event correlation profile with higher linear correlation than the monotonic one. These results demonstrate that while the model is sampling sensor-level occupancies, it can \emph{fill} the previously unseen spots and infer how a background event with a higher cardinality looks like. For generic physics analysis where the analyst wants to do PXD background generation on the fly for analysis, conditioned on the amount of background and the geometry~(sensor's position) of the PXD, this result shows one does not need to retrain the generative model for each experiment as far as there will be no strange beam-background issues. I have to note that this is from the perspective of marginal and low-level distributions. In the following, I also discuss other low-level and physics-level metrics to further study the results. 

\begin{figure}[!htb]
\centering
\includegraphics[width=0.9\textwidth]{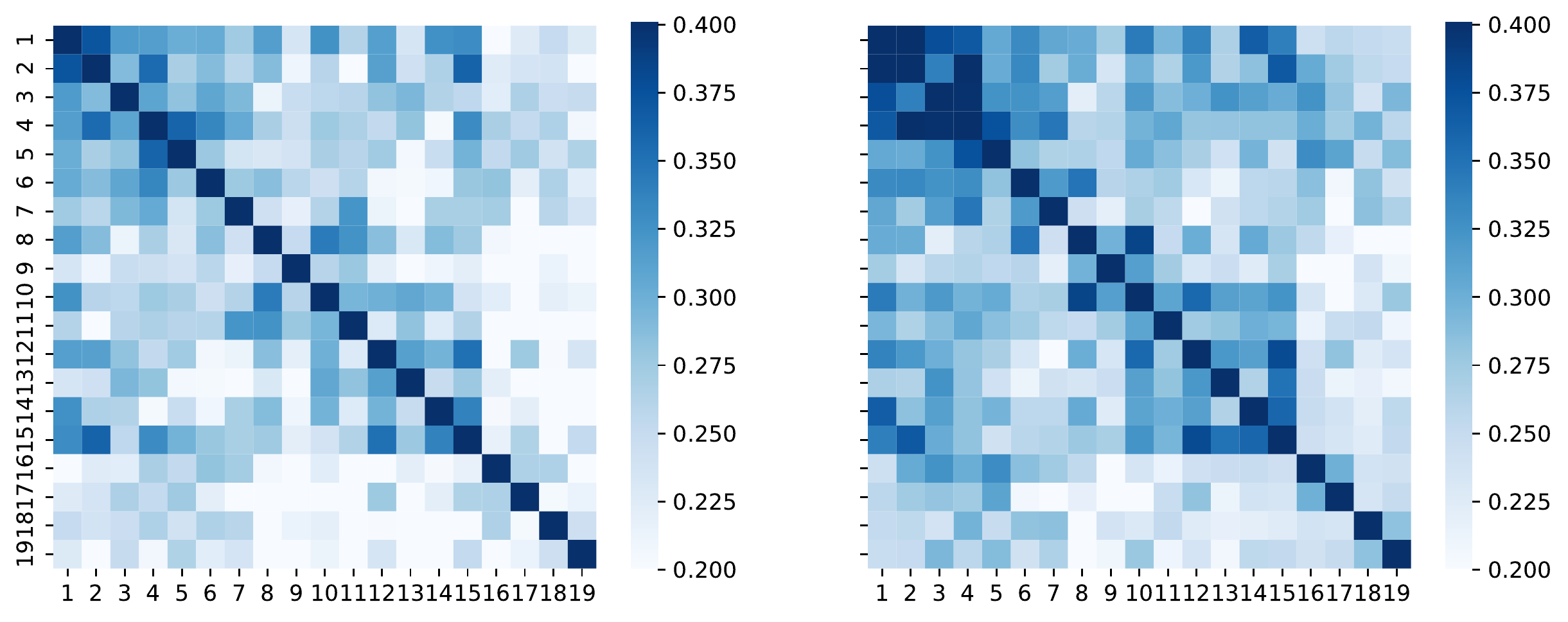} 
\caption{\label{fig:corr_le_spear}
Spearman Correlation between sensor's mean occupancy in the length extrapolation regime for YonedaVAE on the OOD data~(Experiment 26)}
\end{figure}

\begin{figure}[!htb]
    \centering
    \begin{subfigure}{0.45\linewidth}
        \includegraphics[width=\linewidth]{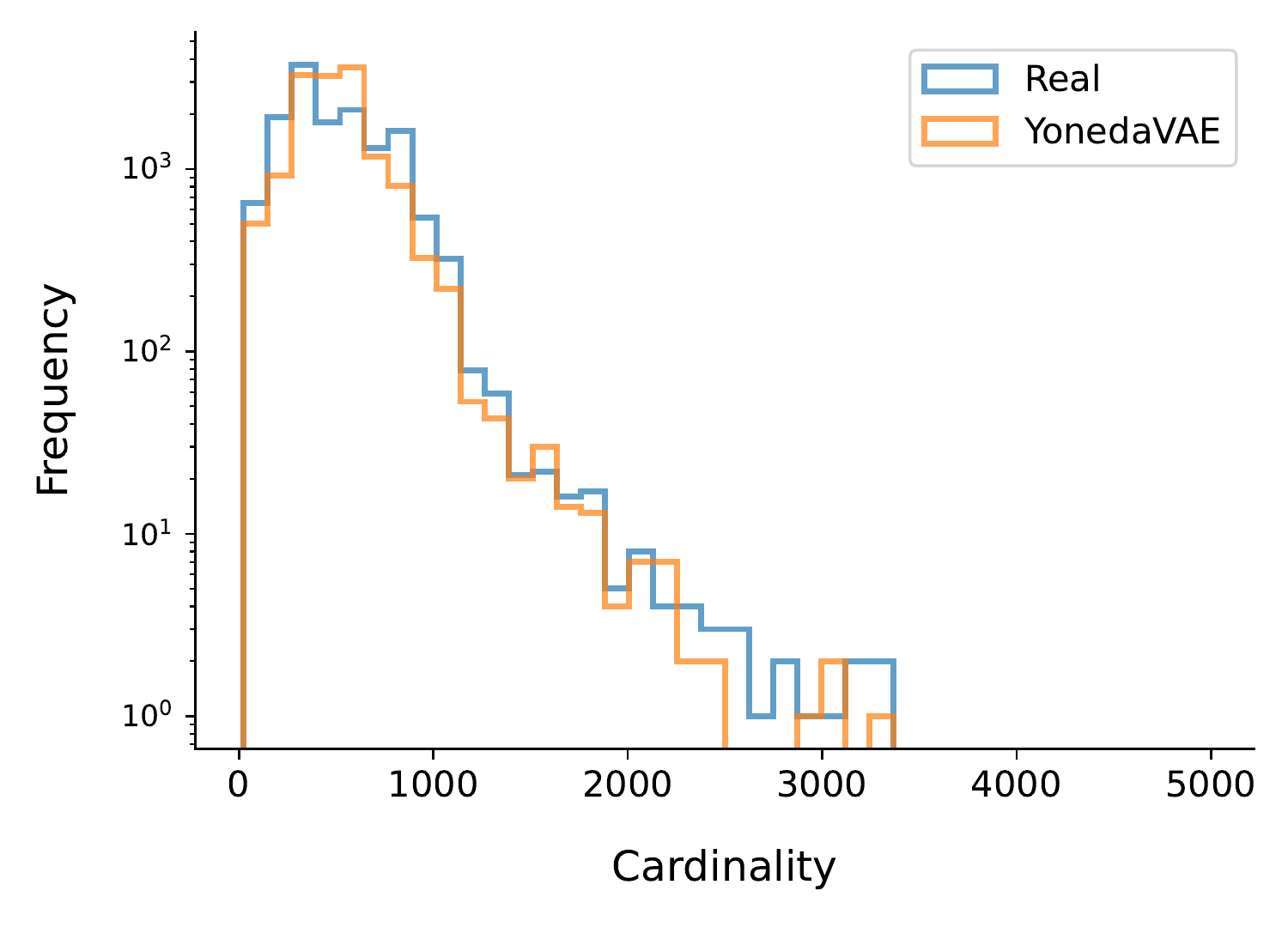}
        \caption{}
    \end{subfigure}
    \begin{subfigure}{0.45\linewidth}
        \includegraphics[width=\linewidth]{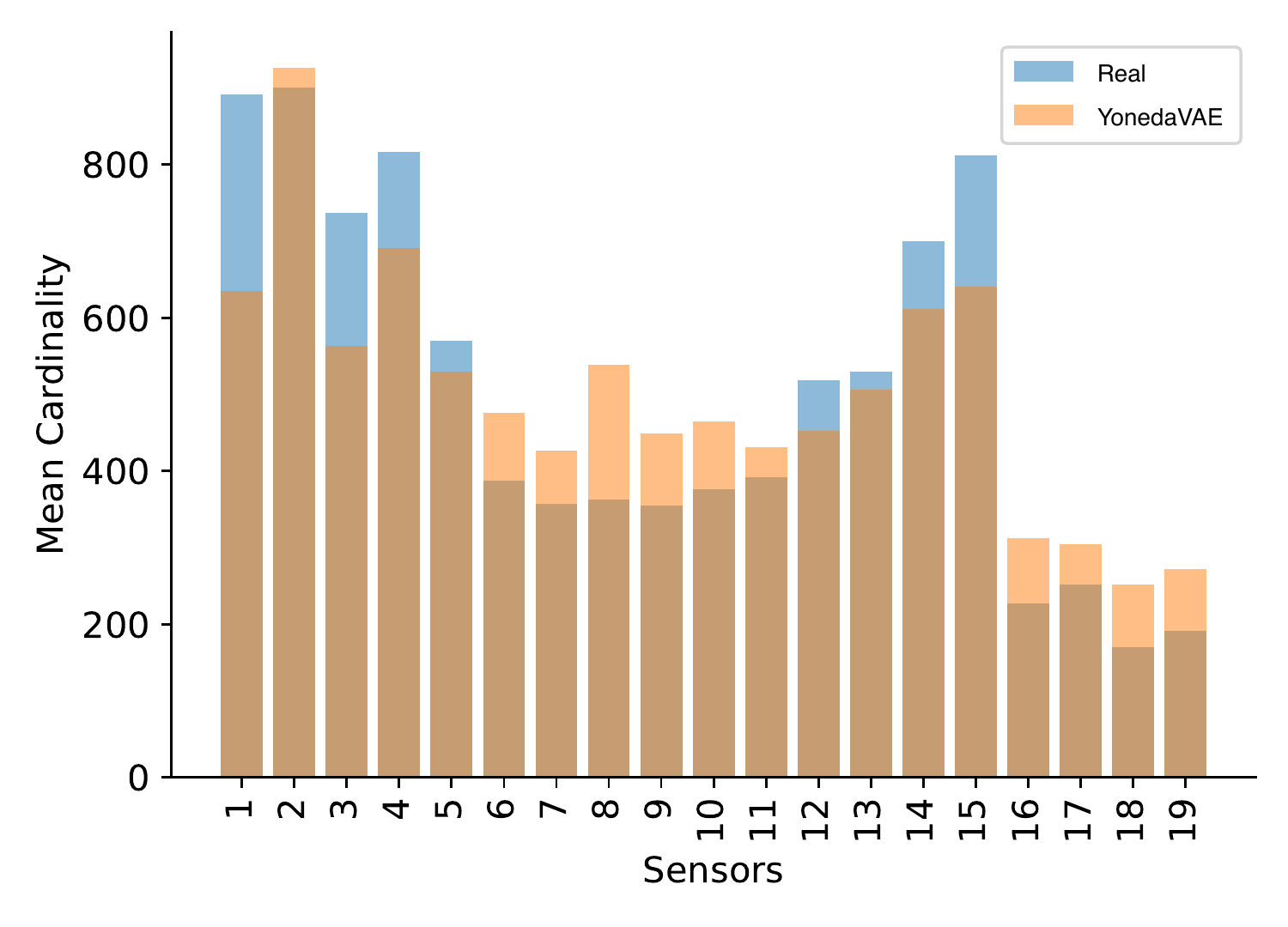}
        \caption{}
    \end{subfigure}\\
    
    \begin{subfigure}{0.45\linewidth}
        \includegraphics[width=\linewidth]{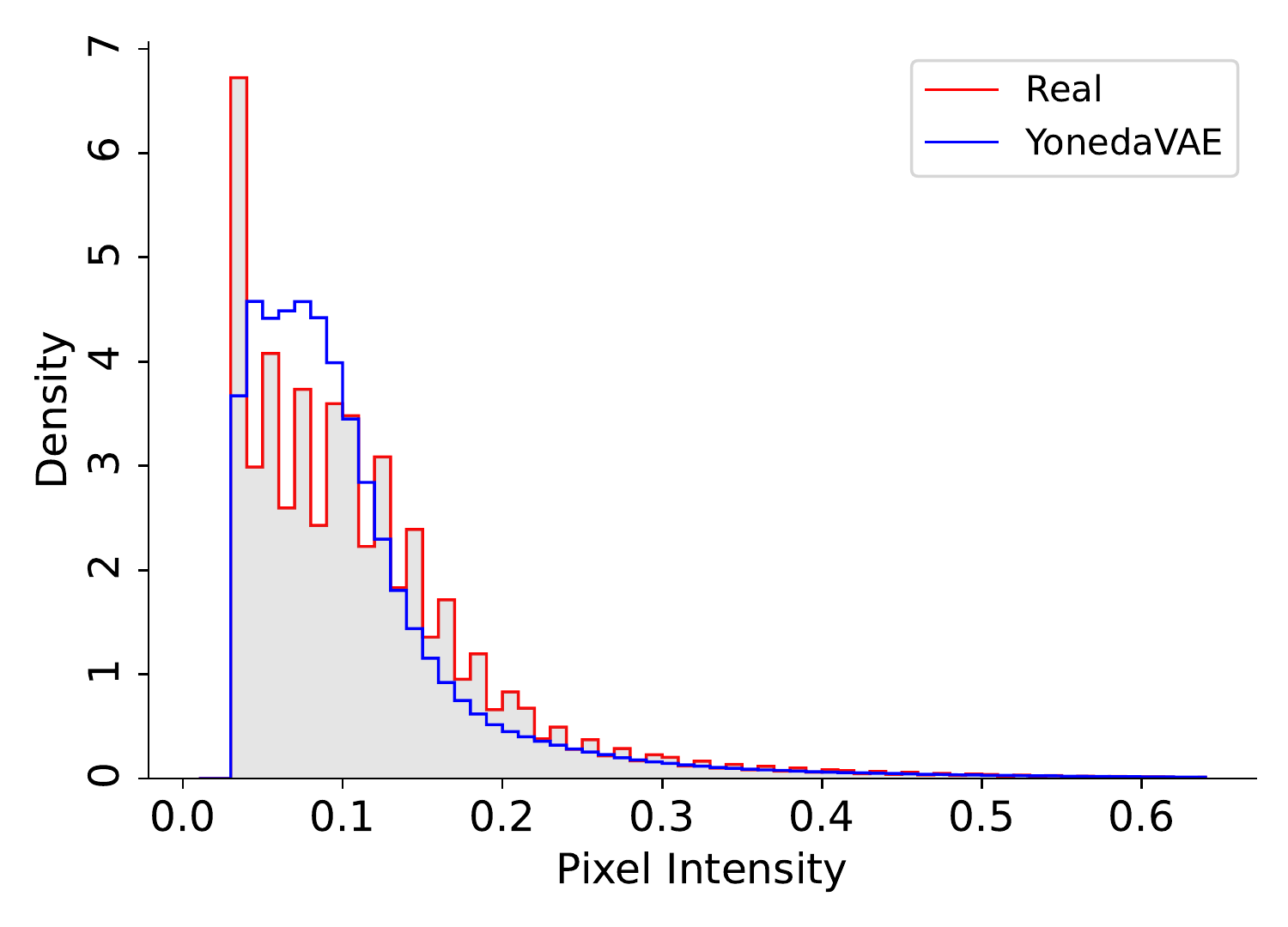}
        \caption{}
    \end{subfigure}
    \begin{subfigure}{0.45\linewidth}
        \includegraphics[width=\linewidth]{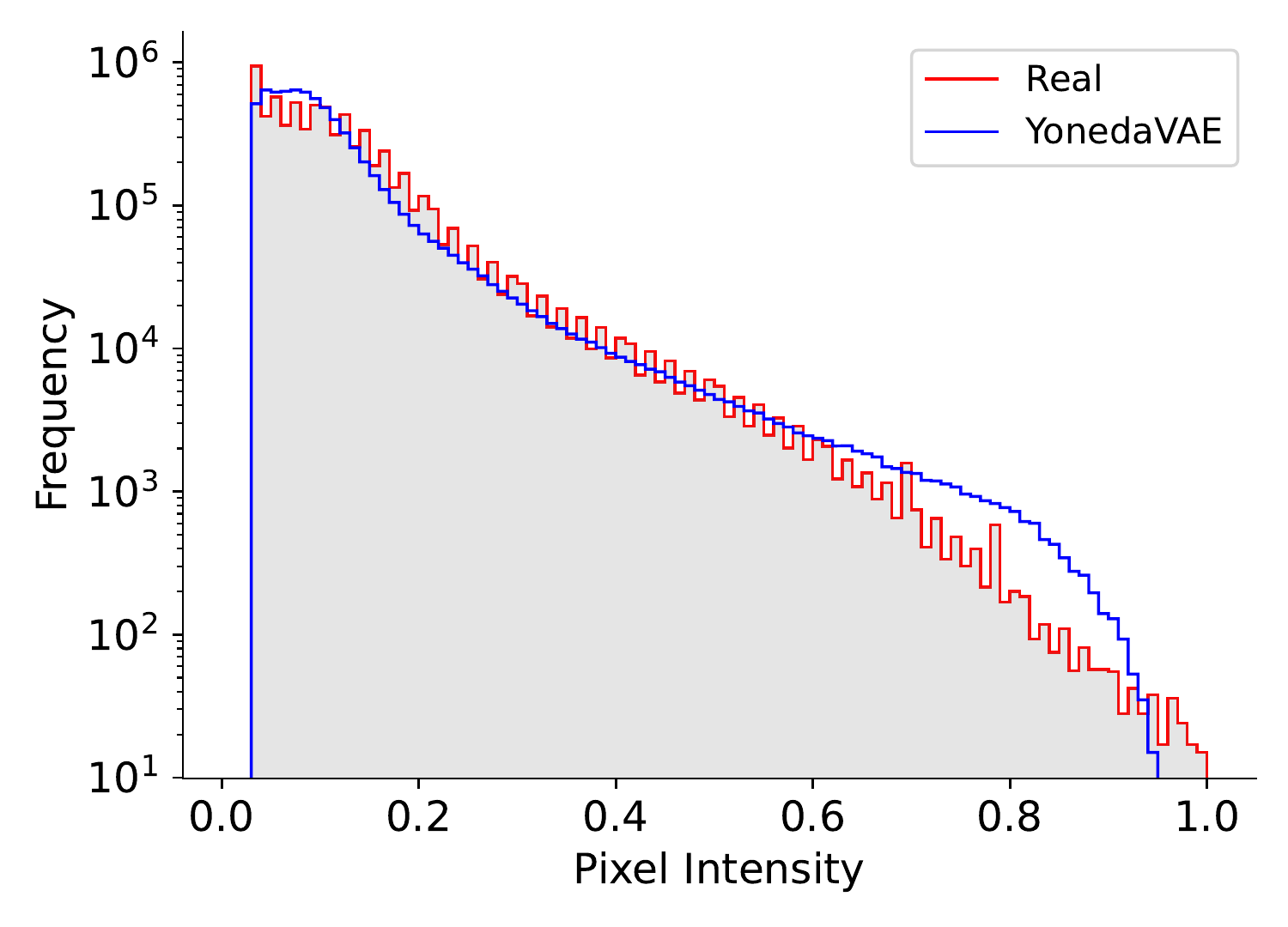}
        \caption{}
    \end{subfigure}\\
    
    \begin{subfigure}{0.45\linewidth}
        \includegraphics[width=\linewidth]{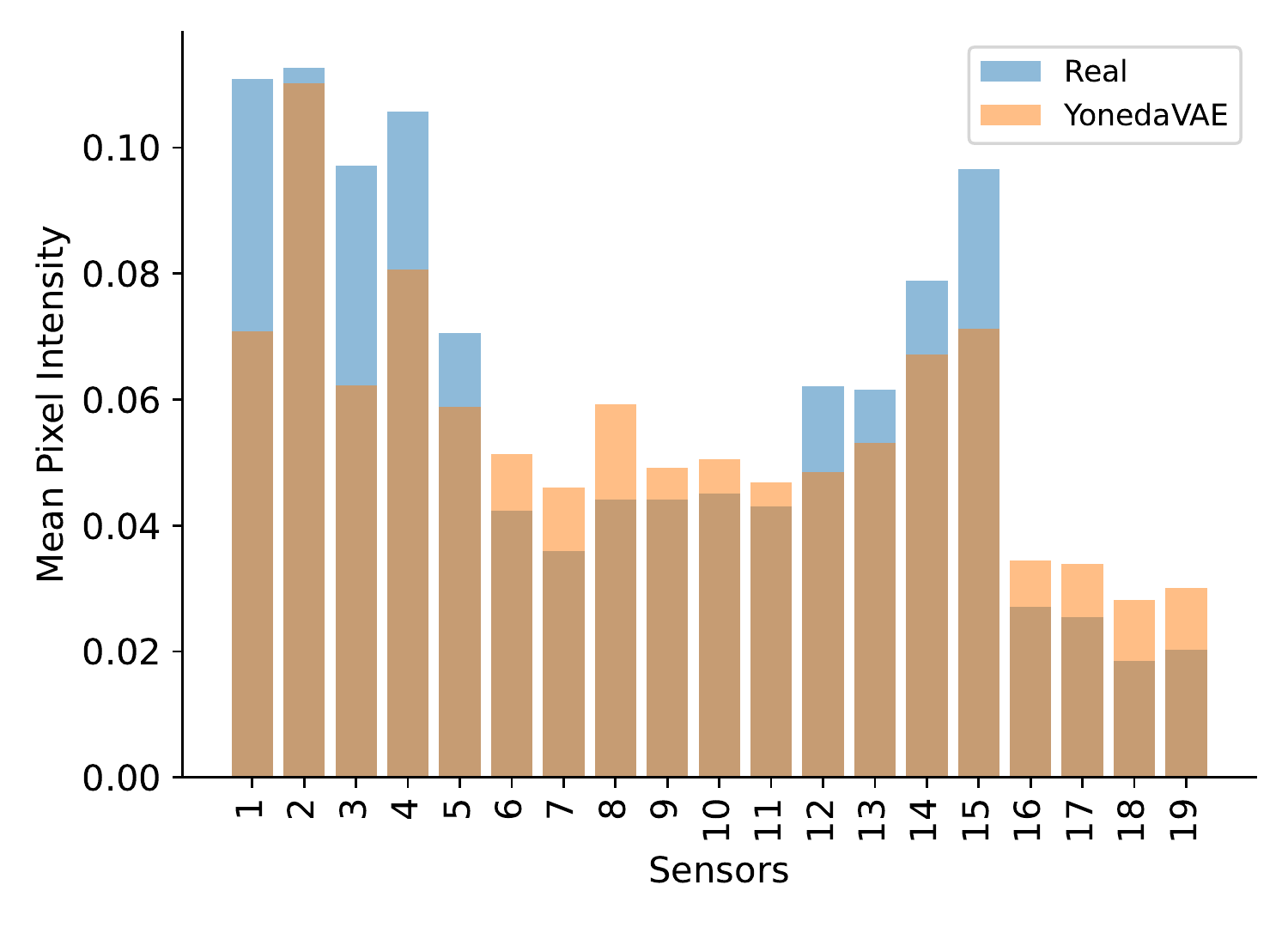}
        \caption{}
    \end{subfigure}
    
    \caption{Marginal Distributions for YonedaVAE with context extrapolation on the OOD data~(Experiment 26).}
    \label{fig:marg_ce}
\end{figure}

\begin{figure}[!htb]
\centering
\includegraphics[width=0.9\textwidth]{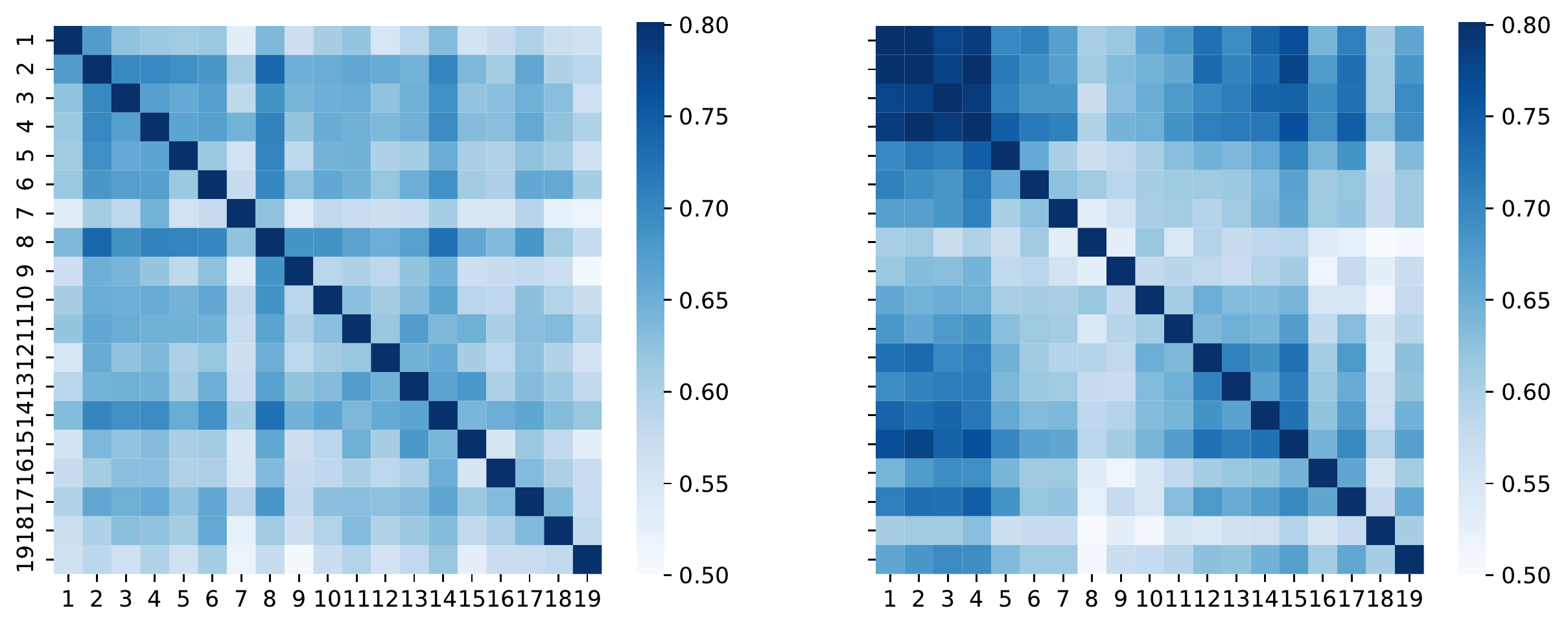} 
\caption{\label{fig:corr_ce_pears}
Pearson Correlation between sensor's mean occupancy in the context extrapolation regime for YonedaVAE on the OOD data~(Experiment 26)}
\end{figure}

Eventually, with context extrapolation model has access only to event-level attribute~(maximum cardinality of the event) depicted in~\cref{fig:marg_ce},~\cref{fig:corr_ce_pears}, and~\cref{fig:corr_ce_spear}. Thus, YonedaVAE not only needs to generate events with the desired amount of background but also has to learn to generalize to attribute profiles beyond the training data and generate point clouds with the correct cardinality and intra-event correlation. I have to again emphasize that during inference, the model is asked to generate events with up to $5300$ sensor cardinality~(or $100700$ event cardinality) coming from Experiment 26. This is almost the 3\% background occupancy limit discussed in~\cref{chap:2}. 
Context extrapolation results show that YonedaVAE can recover~(generate) sensor-level information given only the event-level attribute very much like Inverse Problems~\cite{bai_deep_2020,bingham_inverse_2024}. The inverse problem refers to using coarse observations to infer the values or the parameters that characterize the finer levels of the system and to estimate data that are not directly observed. 
For a physics analysis where the analyst wants to do PXD background generation on the fly of analysis, conditioned on the amount of background and the geometry~(sensor's position) of the PXD, this result provides evidence that one can also generate PXD background for around 100K event cardinality for luminosities beyond the current threshold of the experiment. There is still a minor discrepancy between the intra-event correlation in the context extrapolation regime. 
However, it's important to note that the model is not without its limitations. There are still minor inconsistencies in capturing the intra-event correlation in the context extrapolation domain. Despite these shortcomings, achieving the current level of accuracy in context extrapolation is an extraordinary feat in itself and represents a significant advance in the field.
Again, I have to note that this is from the perspective of marginal and low-level distributions. I also discuss other low-level and physics-level metrics to further consolidate these results.

\subsection{FID, KID, and Vendi Score}
To compute the FID and KID, again, I incorporate the Clean-FID project~\cite{parmar_aliased_2022}, fine-tuned on the training and test PXD data with more than 100,000 events. Thus, the FID model in this setup is aware of both OOD and ID distribution of data.
Since the YonedaVAE generates PXD responses in point cloud~(continuous) representation in $\mathbb{R}^3$, I quantize each point cloud into a grid-like representation of size $1\times250\times768$. The downstream task for this fine-tuning is multi-class classification between $19$ PXD images. 

Despite the fact that FID and KID implicitly measure the diversity and explicitly the quality of the generated samples, having a strong and reference-free metric is very absent in the Detector Simulation and event generation domain. Diversity is a criterion that has long been sought after in generative models, especially in surrogate modeling.  A lack of diversity in generated samples can hinder the usefulness of surrogate models in large sampling regimes.
It is, therefore, important to be able to measure diversity.
Thus, to quantify the diversity of generated samples, I incorporate the \emph{Vendi Score} for detector simulation, originating from ecology and statistical quantum mechanics. Although diversity is a vital property in detector simulation, limited efforts have been devoted to its rigorous understanding and quantification. 
The Vendi Score was first articulated in Friedman et al.~\cite{friedman_vendi_2023}. The metric is formalized as \( e^{H} \), where \( H \) is the Shannon entropy of the eigenvalues derived from a similarity matrix, which itself is generated using a data-centric similarity function. This makes the Vendi Score intrinsically connected to quantum statistical mechanics by being analogous to \( e^{S} \), where \( S \) is the von Neumann entropy of a quantum system.
In ecology, diversity is commonly described as \( e^{\log(n)} \), where \( n \) is the entropy of the species distribution under study. This serves as an adequate measure of ecological diversity. For a population evenly distributed across \( n \) species, the maximal diversity would be \( n \), equivalent to a population comprising \( n \) distinct species. The metric decreases as the species distribution deviates from uniformity, reaching a minimum value of one when all members belong to a single species. Ideally, a diversity metric should rely solely on the samples under evaluation and should attain its maximum value when the samples are entirely diverse and its minimum value when all samples are the same. Hence, the Vendi Score is introduced as a solution to this problem.

\begin{figure}[!htb]
     \centering
     \begin{subfigure}[c]{0.05\textwidth}
         \centering
         \caption{}
         \label{fig:effective_number}
     \end{subfigure}\hfill
     \begin{subfigure}[c]{0.95\textwidth}
         \centering
         \includegraphics[height=0.14\paperheight]{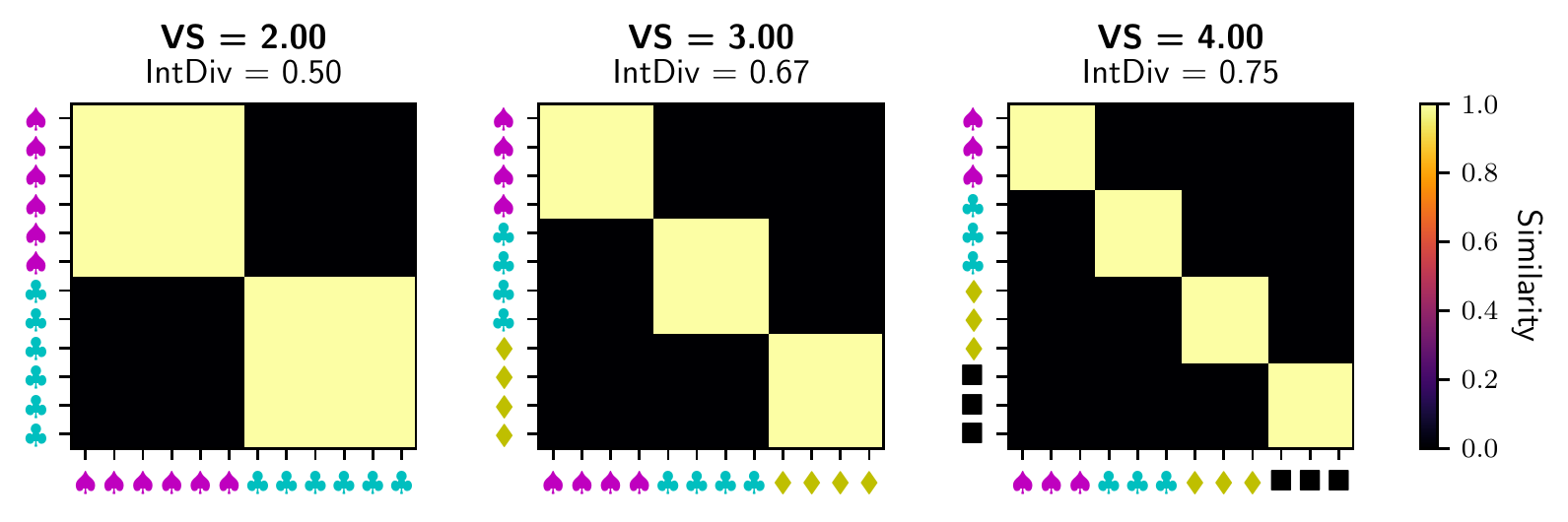}
     \end{subfigure}
     \begin{subfigure}[c]{0.05\textwidth}
         \centering
         \caption{}
         \label{fig:nonlinear}
     \end{subfigure}\hfill
     \begin{subfigure}[c]{0.95\textwidth}
         \centering
         \includegraphics[height=0.14\paperheight]{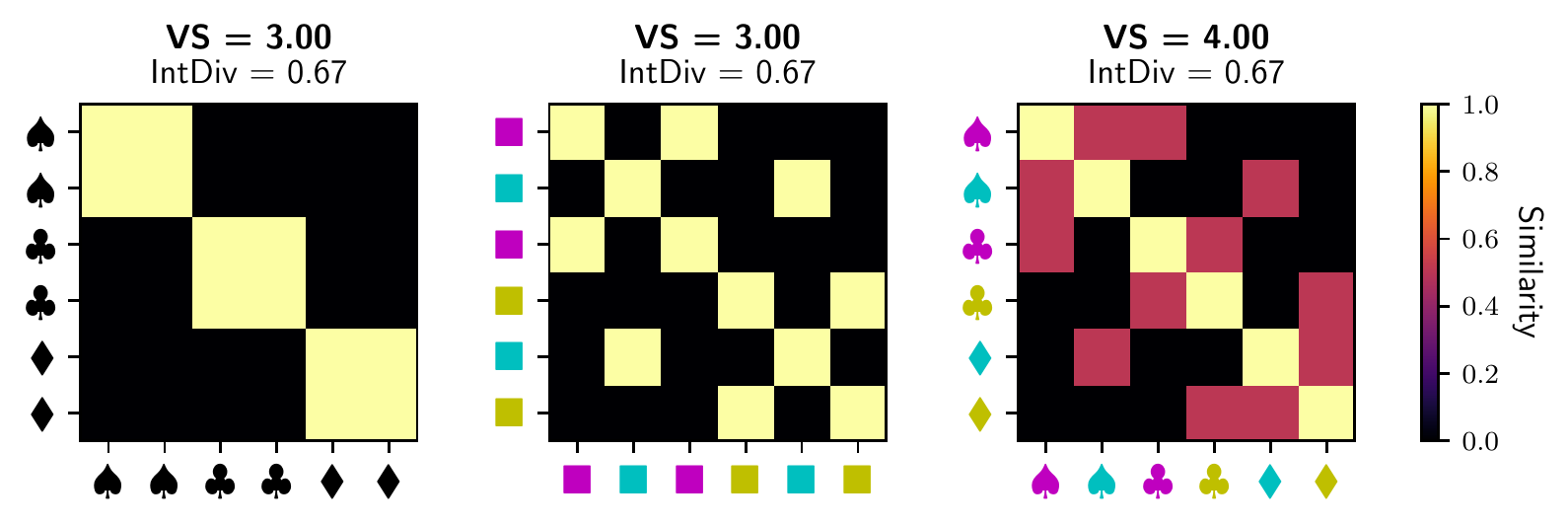}
     \end{subfigure}
     \begin{subfigure}[c]{0.05\textwidth}
         \centering
         \caption{}
         \label{fig:correlations}
     \end{subfigure}\hfill
     \begin{subfigure}[c]{0.95\textwidth}
         \centering
         \includegraphics[height=0.14\paperheight]{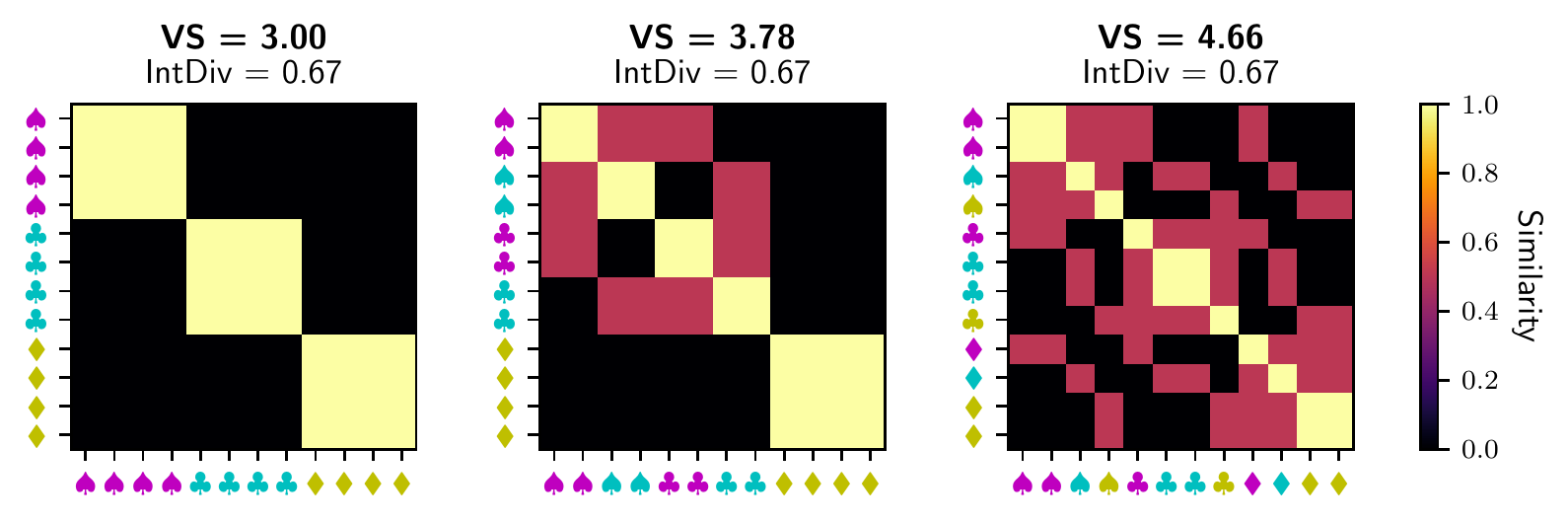}
     \end{subfigure}
     \caption{
       (a) The Vendi Score, \textrm{VS} in the figure, adopted from~\cite{friedman_vendi_2023}, can be interpreted as the effective number of unique elements in a set of events. It increases linearly with the number of modes in the dataset. IntDiv, the expected dissimilarity~(inverse correlation), becomes less sensitive as the number of modes increases, converging to 1~(or becoming uncorrelated).
      (b) Combining distinct similarity measures can increase the Vendi Score, as should be expected of a diversity metric while leaving IntDiv~(~(inverse correlation)) unchanged.
      (c) Vendi Score takes into account correlations between samples. The Vendi Score is highest when the samples differ in many attributes and the attributes are not correlated with each other.
     }
        \label{fig:properties}
\end{figure}

\begin{definition}[Vendi Score]
\label{def:vendi_score}
  Let $x_1, \ldots, x_n \in \mathcal{X}$ denote a collection of samples, let $k: \mathcal{X} \times \mathcal{X} \to \mathbb{R}$ be a positive semi-definite similarity function, with $k(x, x) = 1$ for all $x$, and let $\mathbf{K} \in \mathbb{R}^{n \times n}$ denote the kernel matrix with entry $K_{i, j} = k(x_i, x_j)$. Denote by $\lambda_1, \ldots, \lambda_n$ the eigenvalues of $\mathbf{K}/n$. The Vendi Score~(\textrm{VS}) is defined as the exponential of the Shannon entropy of the eigenvalues of $\mathbf{K}/n$:
  
  \begin{align}
  \label{eq:vendi-score}
    {\textrm{VS}}_k(x_1, \ldots, x_n) =  \exp\left( - \sum_{i=1}^n \lambda_i \log \lambda_i \right), 
  \end{align}
  
where I use the convention $0 \log 0 = 0$.
\end{definition}

To understand the validity of the Vendi Score as a mathematical object, note that the eigenvalues of $\mathbf{K}/n$ are non-negative~(because $k$ is positive semi-definite) and sum to one~(because the diagonal entries of $\mathbf{K}/n$ are equal to $1/n$). The Shannon entropy is, therefore, the Vendi Score is well-defined.
In this form, the Vendi Score can also be seen as the effective rank of the kernel matrix $\mathbf{K}$. The effective rank of a matrix is defined as the exponential of the entropy of the normalized singular values. To interpret the Vendi Score it represents the effective count of non-identical elements within the evaluated sample. For instance, a sample with ten unique elements would yield a Vendi Score of 10, a figure that remains constant even when elements are duplicated. 

In the case of PXD detector simulation, I incorporate the Vendi score and interpret it as the effective number of dissimilar elements in the PXD dataset. For example, if all PXD samples show that they belong to 3 clusters~(from high-level to low-level clustering), the VS also shows 3 modes as exemplified in~\cref{fig:properties}. 
One can understand VS over the FID embedding of the samples to quantify the inter-event diversity. The goal here is to provide a measurable metric to quantify the diversity of generated samples using the covariance matrix from the point of view of a model that had access to both the training and test data. 
In this case, the kernel matrix is the covariance matrix, and the sample feature is the FID embedding. This way, I get a Vendi score for each sensor based on a large number of samples. In~\cref{tab:fid}, I report this number. 

I compare YonedaVAE with three other models with the previously unseen real PXD detector data coming from experiment 26. 
The baselines are the SOTA in the conditional point cloud generation, Transformer Set Prediction Network~(TSPN) with i.i.d sampling~\cite{kosiorek_conditional_2020}, TSPN with Top-K set creation~\cite{vignac_top-n_2022}, and the previous PXD hitmap generation model~(trained now on the real PXD data), IEA-GAN. 
All these models are being compared in the length extrapolation~(le) setup, where they are conditioned on the sample-level feature~(occupancy per sensor) during inference. Other than YonedaVAE, none of the other models can handle the context-extrapolation~(ce) setup where only the event-level attribute in the OOD region is present during inference.

\begin{table}[!htb]
\begin{minipage}{\textwidth}
    \begin{center}
    \caption{
    FID, KID, and VS score comparison between SOTA models for PXD generation averaged across four random seeds. They are computed over 5000 generated events.
    The lower the FID and KID, the better the sample quality and diversity. For the VS score, higher values indicate better sample diversity.
    }
    \label{tab:fid}
    \setlength{\tabcolsep}{3pt} 
    \small 
    \begin{tabular}{@{}l|llllll@{}}
        \toprule
        & TSPN~(i.i.d) & TSPN~(Top-k) & IEA-GAN & YonedaVAE$^{\textrm{le}}$& YonedaVAE$^{\textrm{ce}}$& Test Data\\ 
        \midrule
        \textbf{FID}  & $49.46\pm 0.29$ & $41.40\pm 0.48$ & $37.84\pm 0.98$ & $\mathbf{20.19}\pm 0.31$ & $\mathbf{22.49\pm 0.11}$& 0\\
        \midrule
        \textbf{KID}$^{(\times 10^{-4})}$ & $339 \pm 7$ & $312\pm 1$ & $283\pm 8$& $\mathbf{130\pm 2}$ & $\mathbf{132\pm 4}$& 0\\
        \midrule
        \textbf{VS} & $2.81$ & $3.25$ & $1.71$& $\mathbf{4.91}$ & $\mathbf{4.73}$& $3.86$\\
        \bottomrule
    \end{tabular}
    \normalsize
    \end{center}
\end{minipage}
\end{table}

The~\cref{tab:fid} demonstrates that generated samples by YonedaVAE have the lowest FID and KID score compared to the other models both in context extrapolative and length extrapolative setups. Moreover, YonedaVAE also generates samples that with higher dissimilar modes. VS here means that across 5000 generated events, the number of main clusters of modes~(diversity of modes) in the test data is $3.86$.
A very interesting observation was that VS was increasing as YonedaVAE was generating more samples. For instance, the VS was $3.42$ in 100 samples, then it increased to $4.01$ in 1000 samples and $4.91$ in $5000$ samples. This result could be in accord with~\cite{bieringer_calomplification_2022} where generated samples by YonedaVAE outperform samples from a limited amount of test data from the diversity perspective.
IEA-GAN, despite being conditioned on sensor occupancy trained on the real data~(with very high instability), as expected, performs poorly. The main reason for this is that the real PXD background is much more sparse and diverse than the Geant4 simulated one, which makes it much harder for the IEA-GAN setup to adapt.

One can also utilize VS in order to quantify the diversity over the inter-sensor mean occupancy correlation, discussed in \cref{sec:mar_dist}. 
The idea here is to measure the number of clusters or unique sensors in an event that contribute to the mean occupancy correlation and compare the true value with the generated value. In this case, the kernel matrix is the correlation matrix, and the features for each sensor are the distribution of mean occupancy per sensor over a large number of samples. 
This way, one can measure the diversity of sensor information in an event based on the intra-event occupancy correlation.
In this scenario, the maximum VS diversity would be 19~(where no commonality is there between 19 sensors in an event), and the minimum VS diversity would be 1~(all sensors share the same information).  
Note that the VS results reported in~\cref{tab:fid}, is different. There, the VS counts the number of modes across the whole dataset. It is like saying that over all the PXD background data, there are $3.86$ unique clusters of information using the covariance matrix over their FID~(inception-v3) embedding as the measure. While the results in~\cref{tab:vs_corr} report that in one event, on average, there are $13.26$ unique clusters of information using the Spearman kernel over the mean occupancies as the measure.
As depicted in~\cref{tab:vs_corr}, the inter-sensor mean occupancy correlation shows a Vendi score of $4.92$ and $13.26$, which means that the effective number of unique contributions or clusters of similar behavior is around $4.92$ when considering the mean occupancy rates using the Pearson correlation, and $13.26$ when considering the mean occupancy rates using the Spearman correlation.
Spearman shows higher diversity~(more unique sensors in an event) than Pearson for the same data, possibly indicating that Spearman captures more non-linear relationships between the sensors. Thus, considering monotonic relationships, less sensors are similar. Although YonedaVAE shows a higher diversity~(dissimilar sensors in an event), it also exhibits the same behavior.

\begin{table}[!htb]
\begin{minipage}{\textwidth}
    \begin{center}
    \caption{VS over the inter-sensor mean occupancy correlation for the real test data and YonedaVAE
    }
    \label{tab:vs_corr}
    \setlength{\tabcolsep}{6pt} 
    \small 
    \begin{tabular}{@{}l|lll@{}}
        \toprule
        & Test Data & YonedaVAE$^{\textrm{le}}$& YonedaVAE$^{\textrm{ce}}$\\ 
        \midrule
        \textbf{VS}(Pearson kernel) & $4.92$ & $5.65$ & $5.94$ \\
        \midrule
        \textbf{VS}(Spearman kernel) & $13.26$ & $13.91$ & $11.89$\\
        \bottomrule
    \end{tabular}
    \normalsize
    \end{center}
\end{minipage}
\end{table}

As a result, it is shown that YonedaVAE is able to generate samples that are diverse and of higher quality than all other SOTA models for the high-granularity PXD background generation task. 

\begin{table}[!htb]
\begin{minipage}{\textwidth}
    \begin{center}
    \caption{
    FID, KID, and VS score comparison between SOTA models for PXD generation averaged across four random seeds. They are computed over 5000 generated events.
    The lower the FID and KID, the better the sample quality and diversity. For the VS score, higher values indicate better sample diversity.
    }
    \label{tab:fid}
    \setlength{\tabcolsep}{3pt} 
    \small 
    \begin{tabular}{@{}l|llllll@{}}
        \toprule
        & $\text{TSPN~(i.i.d)}^{1}$ & $\text{TSPN~(Top-k)}^{2}$ & $\text{IEA-GAN}^{3}$ & $\text{Set-VAE}^{4}$ & YonedaVAE& Test Data\\ 
        \midrule
        \textbf{FID}  & $49.46\pm 0.29$ & $41.40\pm 0.48$ & $37.84\pm 0.98$ & $33.49\pm 0.11$ & $\mathbf{20.19}\pm 0.31$& 0\\
        \midrule
        \textbf{KID}$^{(\times 10^{-4})}$ & $339 \pm 7$ & $312\pm 1$ & $283\pm 8$& $181\pm 2$ & $\mathbf{130\pm 2}$& 0\\
        \bottomrule
    \end{tabular}
    \normalsize
    \end{center}
\end{minipage}
\end{table}

\FloatBarrier
\subsection{Topological and Geometrical Clustering for PXD data}
This section wants to answer the following questions:

\emph{How are the points of a PXD point cloud clustered? How many complex clusters do they create? How many ``holes'' does a PXD point cloud have? How many pieces is it constructed out of? As one increases an observation threshold window, at what scale in the PXD point cloud do we observe changes in some geometric representation of the data?}

\subsubsection{Topological Data Analysis}
Topological Data Analysis~(TDA)~\cite{chazal_introduction_2017} is a way to explore these questions and the shape of data without concern for things like which metric to use. The knowledge of topological features~(such as connected components, loops, and higher dimensional cycles) that are present in data sets provides a better understanding of their structural properties at multiple scales. It can be leveraged to improve statistical inference as well. TDA is the branch of data science that aims to detect and encode such topological features. TDA has the following pipeline:

\begin{enumerate}
    \item The input is assumed to be a finite set of points~(like PXD hits as a point cloud), coming with a notion of distance - or similarity - between them. This distance can be induced by the metric in the ambient space~(e.g., the Euclidean metric when the data are embedded in $\mathbb{R}^d$) or come as an intrinsic metric defined by a pairwise distance matrix.
    \item  A continuous shape is built on top of the data in order to highlight the underlying topology or geometry. This is often a simplicial complex or a nested family of simplicial complexes, called filtration, that reflects the structure of the data at different scales. Simplicial complexes can be seen as higher dimensional generalizations of neighboring graphs that are classically built on top of data in many standard data analyses or learning algorithms.
    \item Topological or geometric information is extracted from the structures built on top of the data. This may either result in a full reconstruction, typically a triangulation, of the shape underlying the data from which topological/geometric features can be easily extracted or in crude summaries or approximations from which the extraction of relevant information requires specific methods, such as persistent homology.
    \item The extracted topological and geometric information provides new families of features and descriptors of the data. They can be used to understand the data better.
\end{enumerate}

\subsubsection{Persistent Homology and filtration}
Persistent homology is a powerful tool to compute, study, and encode efficiently multiscale topological features of simplicial complexes and topological spaces. A finite set of data points~(point cloud) can be viewed as a~(noisy) sampling from an underlying topological space. 
The most obvious way to convert these points in a metric space into a global object is to think of the points as the vertices of a combinatorial graph whose edges are determined by proximity~(vertices within some specified distance $\epsilon$). 
Such a graph, while capturing connectivity data, ignores a wealth of higher-order features beyond clustering. These features can be accurately discerned by thinking of the graph as a scaffold for a higher-dimensional object. Specifically, one completes the graph to a simplicial complex — a space built from simple pieces~(simplicies) identified combinatorially along faces. The choice of how to fill in the higher dimensional simplices of the proximity graph allows for different global representations.

\begin{figure}[!htb]
\centering
\includegraphics[width=0.9\textwidth]{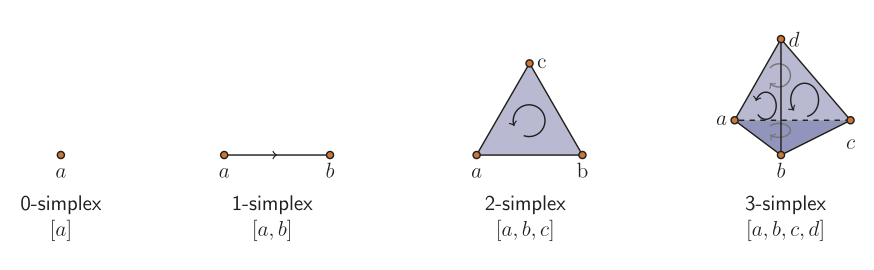} 
\caption{\label{fig:simplices}
{\bf Oriented $k$-simplices for $k = 0, 1, 2, 3$.} These $k$-simplices are the building blocks used to construct a simplicial complex from a point cloud of data.
}
\end{figure}

Simplicial complexes can be seen as higher dimensional generalizations of graphs. They are mathematical objects that are both topological and combinatorial, a property that makes them particularly useful for TDA. 
Given a set \( X = \{ x_0, \ldots, x_k \} \) which is a subset of \( \mathbb{R}^d \) and consists of \( k + 1 \) affinely independent points, I define a \( k \)-dimensional simplex \( \sigma \) as \([ x_0, \ldots, x_k ]\), which is essentially the convex hull of \( X \). The points in \( X \) are referred to as the vertices of \( \sigma \), and the simplices formed by the subsets of \( X \) are termed the faces of \( \sigma \).
For instance, a 0-simplex is a vertex, a 1-simplex is an edge, a 2-simplex is a triangle, a 3-simplex is a tetrahedron, and so on; see~\cref{fig:simplices}.

Constructing simplicial complexes can vary depending on the underlying data set. Multiple techniques are available for creating these structures depending on the inductive biases in the data—for instance, the Vietoris-Rips complex, Alpha Complexes, Čech Complexes, and Delaunay Triangulations. 

\begin{definition}[\textbf{Čech Complex}]
Let \( \{ x_\alpha \} \) be a collection of points in Euclidean space \( \mathbb{R}^d \). The \v{C}ech complex, \( \mathcal{C}_{\epsilon} \), is defined as an abstract simplicial complex. Its \( k \)-simplices are formed by unordered \( (k + 1) \)-tuples of points \( \{ x_{\alpha_0}, \ldots, x_{\alpha_k} \} \) such that their closed \( \frac{\epsilon}{2} \)-ball neighborhoods have a point of common intersection.
\end{definition}

\begin{definition}[\textbf{Rips Complex}]
Let \( \{ x_\alpha \} \) be a collection of points in Euclidean space \( \mathbb{R}^d \). The Rips complex, \( \mathcal{R}_{\epsilon} \), is an abstract simplicial complex. Its \( k \)-simplices are determined by unordered \( (k + 1) \)-tuples of points \( \{ x_{\alpha_0}, \ldots, x_{\alpha_k} \} \) such that each pair of points in the tuple are within distance \( \epsilon \).
\end{definition}

\begin{figure}[!htb]
	\centering
		\includegraphics[width = 0.8 \columnwidth]{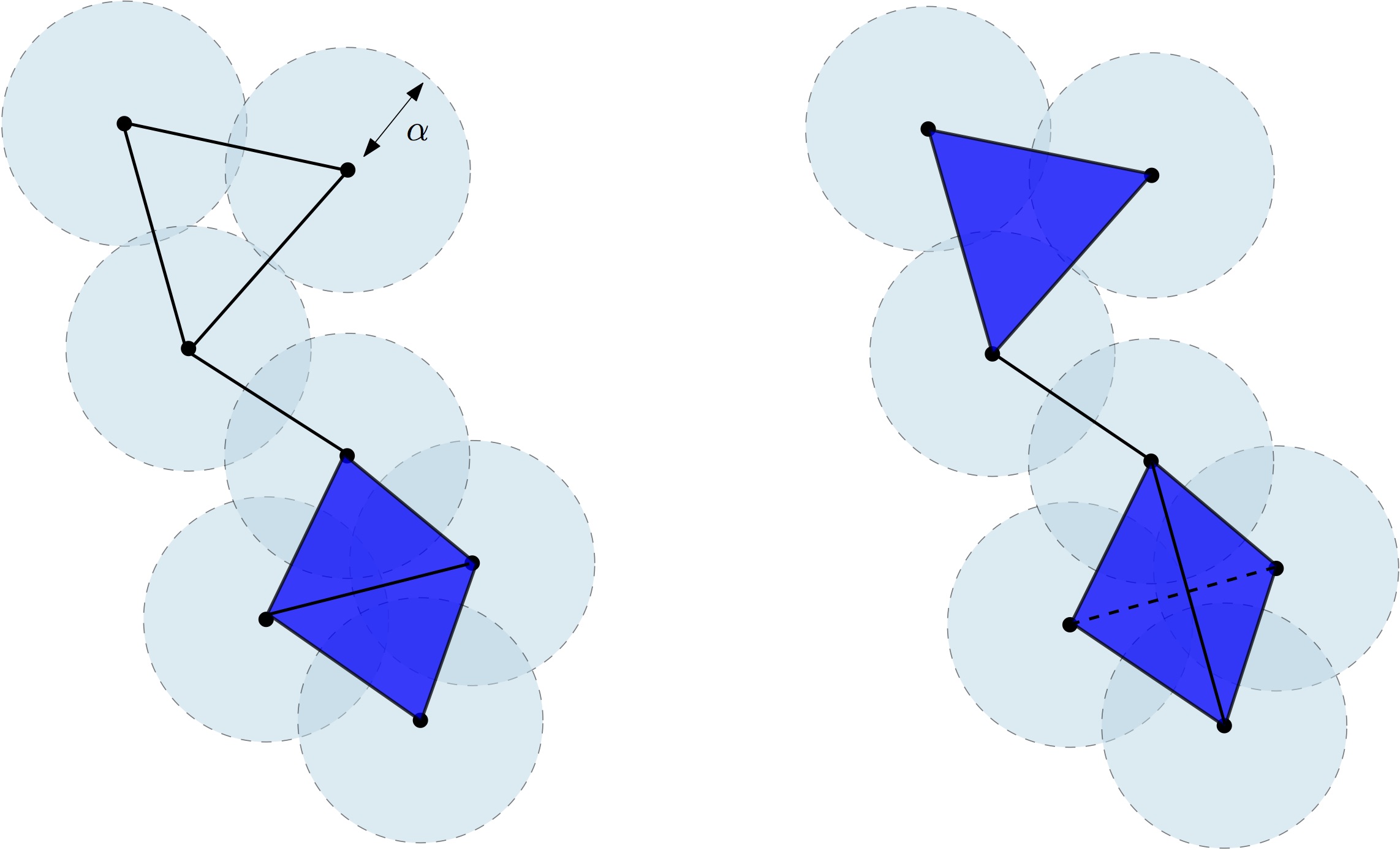} 
		\caption{The \v Cech complex~(left) and the and Vietoris-Rips~(right) of a finite point cloud in the plane $\R^2$. The bottom part of the Cech complex is the union of two adjacent triangles, while the bottom part of the Rips complex is the tetrahedron spanned by the four vertices and all its faces. The dimension of the \v Cech complex is $2$. The dimension of the Vietoris-Rips complex is $3$.} 
	\label{fig:RipsCech}
\end{figure}

Transforming a point cloud into an abstract simplicial complex, such as a \v{C}ech or Rips complex, for measuring the homology of data mandates the selection of a proximity scale parameter \( \epsilon \), the scale over which the simplicial connections are made. For small values of \( \epsilon \), the generated complex is essentially a disconnected set of points. Conversely, for large \( \epsilon \), it converges to a single, high-dimensional simplex. 

For example, to form $k$-simplices using the Vietoris-Rips complex, one first defines a distance metric, which can be realized as a symmetric $N \times N$ matrix of pairwise distances between points. For each $\epsilon>0$, I construct a Vietoris-Rips simplicial complex $\mathcal{R}_{\epsilon}$ in the following way. 
In $\mathcal{R}_{\epsilon}$, every collection of $k+1$ data points is a $k$-simplex if the pairwise distance between points is less than $\epsilon$. 
Thus, the 0-simplices are the data points themselves. A 1-simplex~(an edge) is formed whenever two points are within $\epsilon$ of one another. A 2-simplex~(a triangle) is formed whenever three points are pairwise within $\epsilon$ of one another; this occurs when there is a 3-cycle in the underlying graph formed by the vertices and edges in $S\mathcal{R}_{\epsilon}$. A 3-simplex~(a tetrahedron) is formed whenever four points are pairwise within $\epsilon$ of one another. This raises a question: \emph{Is there an optimal $\epsilon$ that accurately reflects the topology of the underlying data?}

Algebraic topology offers a mature set of tools for counting and collating holes and other topological features in spaces and maps between them. In the context of high-dimensional data, algebraic topology works like a telescope, revealing objects and features not visible to the naked eye. 
In order to handle topological measures, particularly numerical ones, that can be easily computed from the complex, one can consider the homology of the considered simplicial complexes. Homology is a classical concept in algebraic topology, providing a powerful tool to formalize and handle the notion of topological features of a simplicial complex in an algebraic way. 

For any dimension k, the k-dimensional hole is represented by a vector space $H_k$ whose dimension is intuitively the number of such independent features. 
For example, the 0-dimensional homology group $H_0$ represents the group of connected components of the complex, the 1-dimensional homology group $H_1$ represents the group of 1-dimensional loops, and the 2-dimensional homology group $H_2$ represents the group of 2-dimensional cavities. 
In terms of the topological measures~(invariants), the k'th \emph{betti number},$\mathsf{b}_k$ can be defined as the dimension of these groups.
$\mathsf{b}_k$ equals the number of independent holes of dimension $k$. For instance, $\mathsf{b}_0$ is the number of connected components, $\mathsf{b}_1$ is the number of topological circles, $\mathsf{b}_2$ is the number of trapped volumes, and so on.  The topology of a simplicial complex may be described by the sequence of Betti numbers, $\mathsf{b}=(\mathsf{b}_0, \mathsf{b}_1,\mathsf{b}_2,\ldots)$. For instance, a topological circle has $\mathsf{b}=(1,1,0,\ldots)$, a topological torus has $\mathsf{b}=(1,2,1,0,\ldots)$, and a topological sphere has $\mathsf{b}=(1,0,1,0,\ldots)$.  Betti numbers are homological invariants, meaning that homologically equivalent spaces have the same Betti number. 

\begin{figure}[!htb]
\begin{center}
\includegraphics[width=\textwidth]{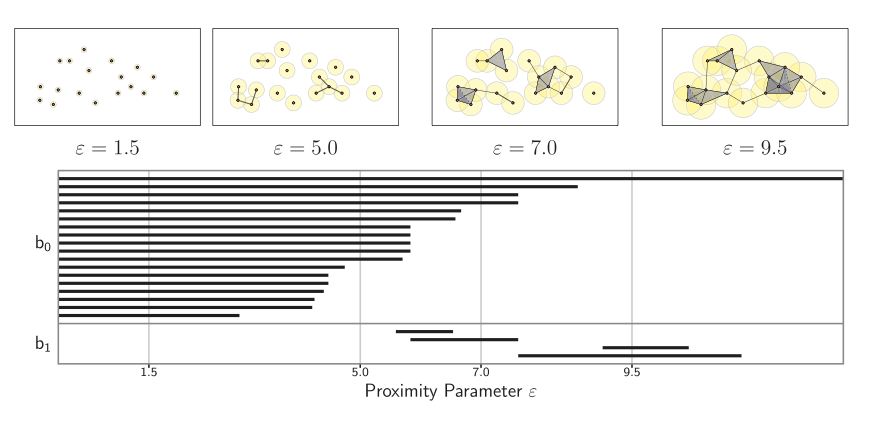}
\end{center}
\vspace{-0.3in}
\caption{
Example of the topological barcode of a Vietoris-Rips complex, adopted from~\cite{topaz_topological_2015}. The top four figures display the simplicial complex of 18 points for  different values of the proximity parameter $\epsilon$. The vertical lines in the barcode correspond to these four levels of $\epsilon$.  The number of horizontal bars intersecting each line gives the values of $\mathsf{b}(\epsilon) = (\mathsf{b}_0(\epsilon),\mathsf{b}_1(\epsilon))$. For the parameters selected, $\mathsf{b}(1.5)=(18,0)$, $\mathsf{b}(5.0)=(11,0),$ $\mathsf{b}(7.0)= (4,1),$ and $\mathsf{b}(9.5)=(1,2)$.}
\label{fig:pedExample}
\end{figure}

Despite being both computable and insightful, the homology of a complex associated with a point cloud at a particular $\epsilon$ is insufficient: it is a mistake to ask which
value of $\epsilon$ is optimal. Nor does it suffice to know a simple ``count'' of the number and
types of holes appearing at each parameter value $\epsilon$. Figure~\ref{fig:pedExample} presents an example. 
Homological invariants are not enough. One must declare which holes are essential and which can be safely ignored. The standard topological constructs of homology and homotopy offer no
such slack in their strident rigidity: a hole is a hole, no matter how fragile or fine. 

To address the above issue, the concept of persistence was initially formalized. In this framework, topological features that remain stable across a wide range of parameter values are considered as the ``signal,'' while features that exist for a short parameter range are treated as ``noise.'' One exploits the fact that as $\epsilon$ grows, so do the Rips complexes, giving an inclusion of complexes for small $\epsilon$ into those for larger values.
To illustrate, let \( \mathbf{R} = (R_i)_{i=1}^{N} \) be a sequence of Rips complexes generated from a fixed point cloud, using an increasing series of scale parameters \( (\epsilon_i)_{i=1}^{N} \). The relationship between these complexes can be understood through a sequence of natural inclusion maps,

\begin{equation}
R_1 \xhookrightarrow{\iota} R_2 \xhookrightarrow{\iota} \cdots \xhookrightarrow{\iota} R_{N-1} \xhookrightarrow{\iota} R_N.
\end{equation}

This sequence is called a \emph{filtration}. 
Persistent homology, then, tracks topological features which \emph{persist} across a range of values of $\epsilon$.  Those features that persist over a large range are considered signals of the underlying topology, while the short-lived features are taken to be noise inherent in approximating a topological space with a finite sample. Thus, rather than focusing solely on the homology groups \(H^*R_i\) of each individual Rips complex \(R_i\), one investigates the homomorphisms induced by the inclusion maps \(\iota : H^*R_i \rightarrow H^*R_j\) where \(i < j\). These induced maps disclose the persistent features within the sequence of complexes. The induced homomorphism \(\iota^*: H^*R_\epsilon \to H^*R_{\epsilon'}\) unveils details not immediately discernible when examining \(H^*R_\epsilon\) and \(H^*R_{\epsilon'}\) in isolation. This can be seen as a precursor to the more generalized notion of persistent homology within an arbitrary sequence of chain complexes.

Persistent homology provides efficient algorithms to compute the homological invariants of each complex in the considered families and encodes the evolution of the homology groups of the nested complexes across the scales.
If homology is to measure the shape of abstract spaces, then persistent homology is how the shape of a geometric data set can be quantified. 
Consequently, the topological features of a complex can be inferred through the persistent homology of the Vietoris-Rips filtration through which new connected components may emerge, existing ones may merge, and higher-order topological features like loops and voids may form and vanish. 
Persistent homology tracks these evolving features by associating each a ``lifetime.'' The output is typically a ``barcode,'' a collection of intervals, or a point cloud in \( \mathbb{R}^2 \), each point of which represents the birth and death times of a topological feature. 
Each homology space has a distinct barcode $H_k$ from which I infer the Betti number $\mathsf{b}_k$. 
As an example, see Figure~\ref{fig:pedExample}. 
The horizontal axis corresponds to the proximity parameter $\epsilon$, and the vertical axis is an~(arbitrary) ordering of the homology generators, \emph{i.e.}, the distinct homology classes of dimension $k$.  Each homology class is visualized by a bar that persists for a given range of $\epsilon$.  Its leftmost endpoint is at the $\epsilon$ value at which the homology class forms and its rightmost endpoint is the $\epsilon$ value at which it disappears. At any given $\epsilon$, the Betti number $\mathsf{b}_k(\epsilon)$ is the number of bars that intersect the vertical line through $\epsilon$. Those bars which persist over longer intervals generally correspond to real topological features, whereas short bars are considered noise. 

In general, the culmination of tracking the persistent features across multiple scales is referred to as a \emph{Persistence Diagram}. A persistence diagram is a multiset in \( \mathbb{R} \times \mathbb{R} \cup \{ \infty \} \), with each point \((b, d)\) signifying a topological feature that is born at the scale \(b\) and dies entering the scale \(d\). Formally, the \(i\)-th persistence diagram \( D_i \) is constituted by points \( (b_j, d_j) \) where the homological class \( \alpha_j \) is born at \( H^*(C^*_{b_j}) \) and dies entering \( H^*(C^*_{d_j}) \). Intuitively, a persistence diagram is a way to summarize the ``lifespan'' of topological features as one 'filters' through a point cloud. In a persistence diagram, points near the diagonal are inferred to live short, while points further from the diagonal are considered topologically persist longer.

\begin{figure}[!htb] 
	\centering
		\includegraphics[width = 1 \columnwidth]{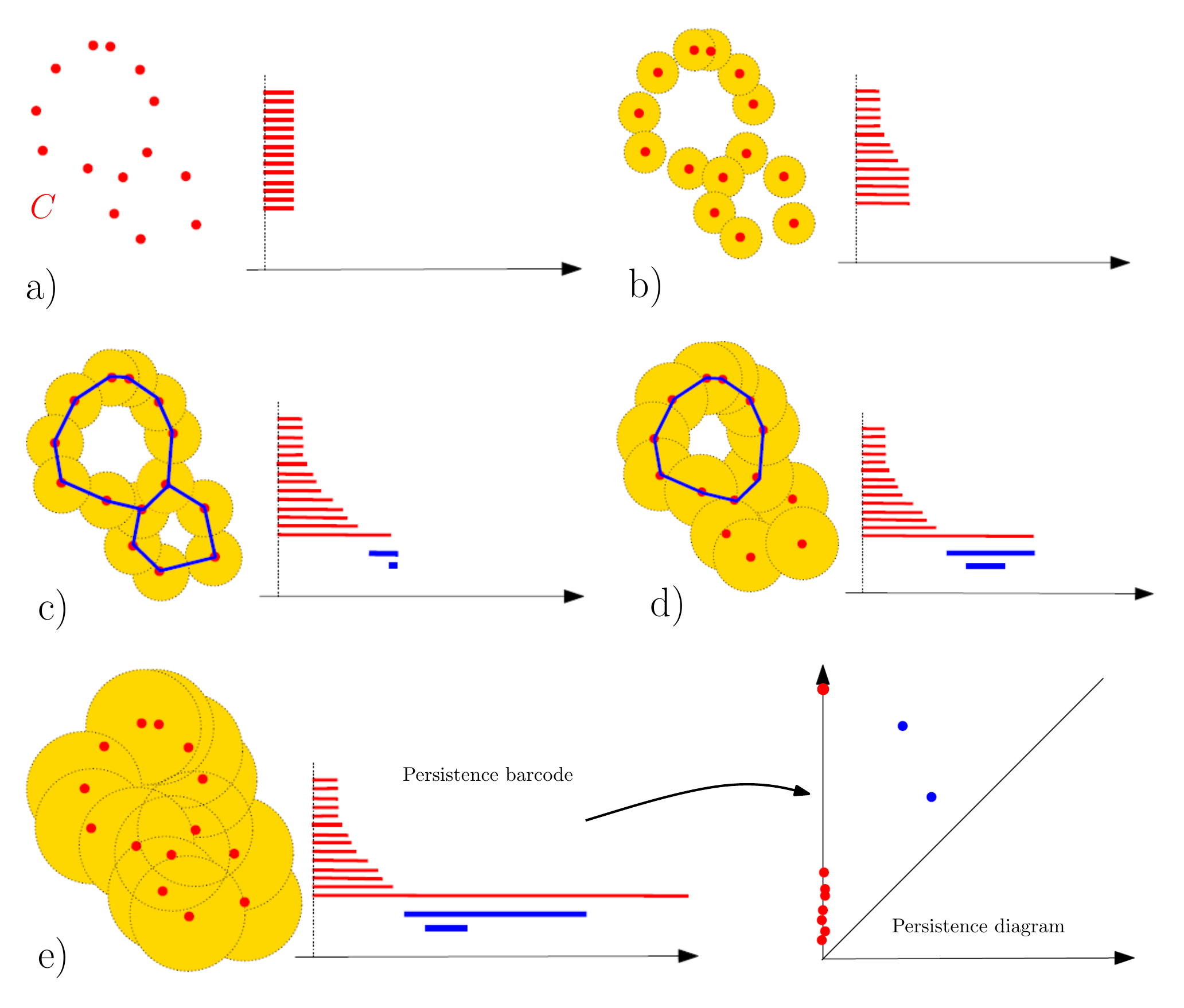}
		\caption{The set filtration of the distance function to a point cloud and the construction of its persistence barcode as the radius of balls increases, adopted from~\cite{chazal_introduction_2017}. The blue curves in the unions of balls represent $1$-cycles associated with the blue bars in the barcodes. 
}
	\label{fig:Cech-diag-example}
\end{figure}

Let's dive into an example in which I consider the filtration given by a union of growing balls centered on the finite set of points $C$, depicted in Figure~\cref{fig:Cech-diag-example}. This filtration is homotopy equivalent to the Čech filtration built on top of C. It shows several level sets of the filtration:

\begin{enumerate}
    \item For the radius $\epsilon = 0$, the union of balls is reduced to the initial finite set of points, each of them corresponding to a 0-dimensional feature, i.e., a connected component; an interval is created for the birth for each of these features at $\epsilon = 0$.
    \item Some of the balls started to overlap, resulting in the death of some connected components that get merged together; the persistence diagram keeps track of these deaths, putting an endpoint to the corresponding intervals as they disappear.
    \item New components have merged, giving rise to a single connected component, and so, all the intervals associated with a 0-dimensional feature have been ended, except the one corresponding to the remaining components; two new 1-dimensional features have appeared resulting in two new intervals~(in blue) starting at their birth scale.
    \item One of the two 1-dimensional cycles has been filled, resulting in its death in the filtration and the end of the corresponding blue interval.
    \item All the 1-dimensional features have died. It only remains the long~(and never dying) red interval. The final barcode can also be equivalently represented as a persistence diagram where every interval~(a,b) is represented by the point of coordinate~(a,b) in R2. Intuitively, the longer an interval in the barcode or, equivalently, the farther from the diagonal is the corresponding point in the diagram, the more persistent, and thus relevant, is the corresponding homological feature across the filtration. Notice also that for a given radius $\epsilon$, the k'th Betti~number of the corresponding union of balls is equal to the number of persistence intervals corresponding to k-dimensional homological features and containing $\epsilon$. So, the persistence diagram can be seen as a multiscale topological signature encoding the homology of the union of balls for all radii as well as its evolution across the values of $\epsilon$.
\end{enumerate} 
After this brief introduction to TDA, let's find out how these methods can be applied to the PXD background data to analyze their underlying clusters.

\subsubsection{Topological Data Analysis of PXD Background}
The clusters and distribution of PXD hits point clouds can also be studied through the lens of TDA. Traditional clustering algorithms may lack the capability to capture intricate spatial and topological features. 
Persistent homology, particularly in combination with Rips filtrations, offers a multi-scale, topologically rigorous approach to understanding this information.
Clusters of PXD hits can be intuitively thought of as connected components in a topological space. Each hit will be its own connected component at low values of \(\epsilon\). 
As \(\epsilon\) grows, these components will merge, capturing the notion of hits that are ``close'' to each other. 
The higher-dimensional features, such as loops and voids, could provide crucial insights into the organization and relations between hit clusters coming from background processes. 
For example, a line cluster of hits can be represented through a short-lived $H_0$ feature in a PXD point cloud as the filtration balls merge instantly for very close points, or a loop may indicate a spiral-like arrangement of hits, while a void could indicate an area devoid of hits surrounded by a high concentration of hits. 
The persistent diagram will indicate the scale at which these clusters form and disappear, providing a multiscale summary of clustering. 

\begin{figure}[!htb]
    \centering
    \caption{Persistent diagrams for each example of possible dense PXD hits}
    \label{fig:topo_ex}
    
    \begin{subfigure}{.4\textwidth}
        \centering
        \includegraphics[width=.8\linewidth]{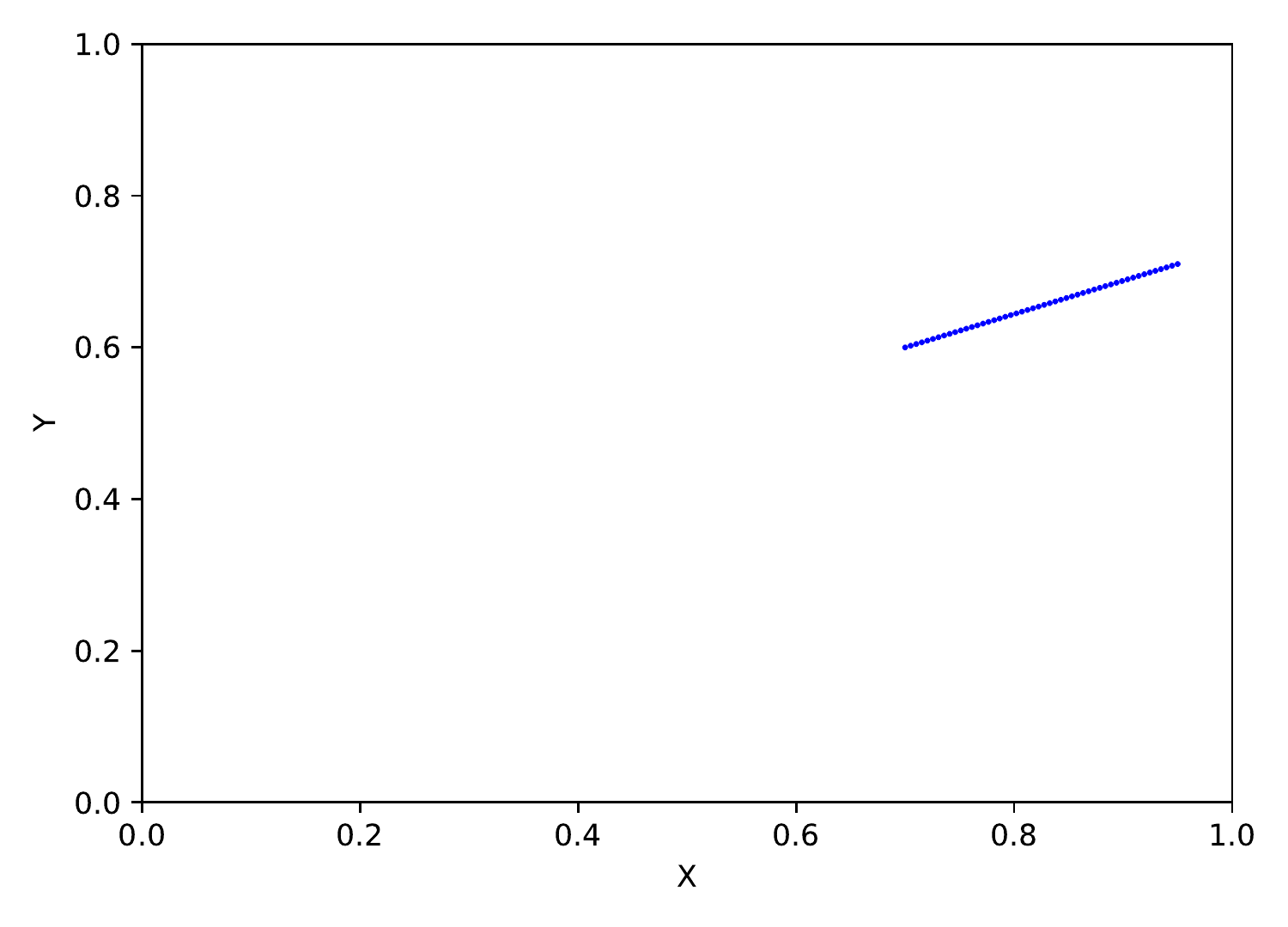}
        \caption{Having a dense linear artifact}
    \end{subfigure}%
    \begin{subfigure}{.4\textwidth}
        \centering
        \includegraphics[width=.8\linewidth]{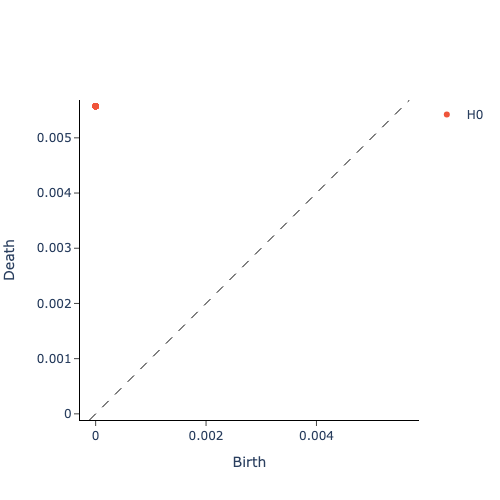}
    \end{subfigure}

    \begin{subfigure}{.4\textwidth}
        \centering
        \includegraphics[width=.8\linewidth]{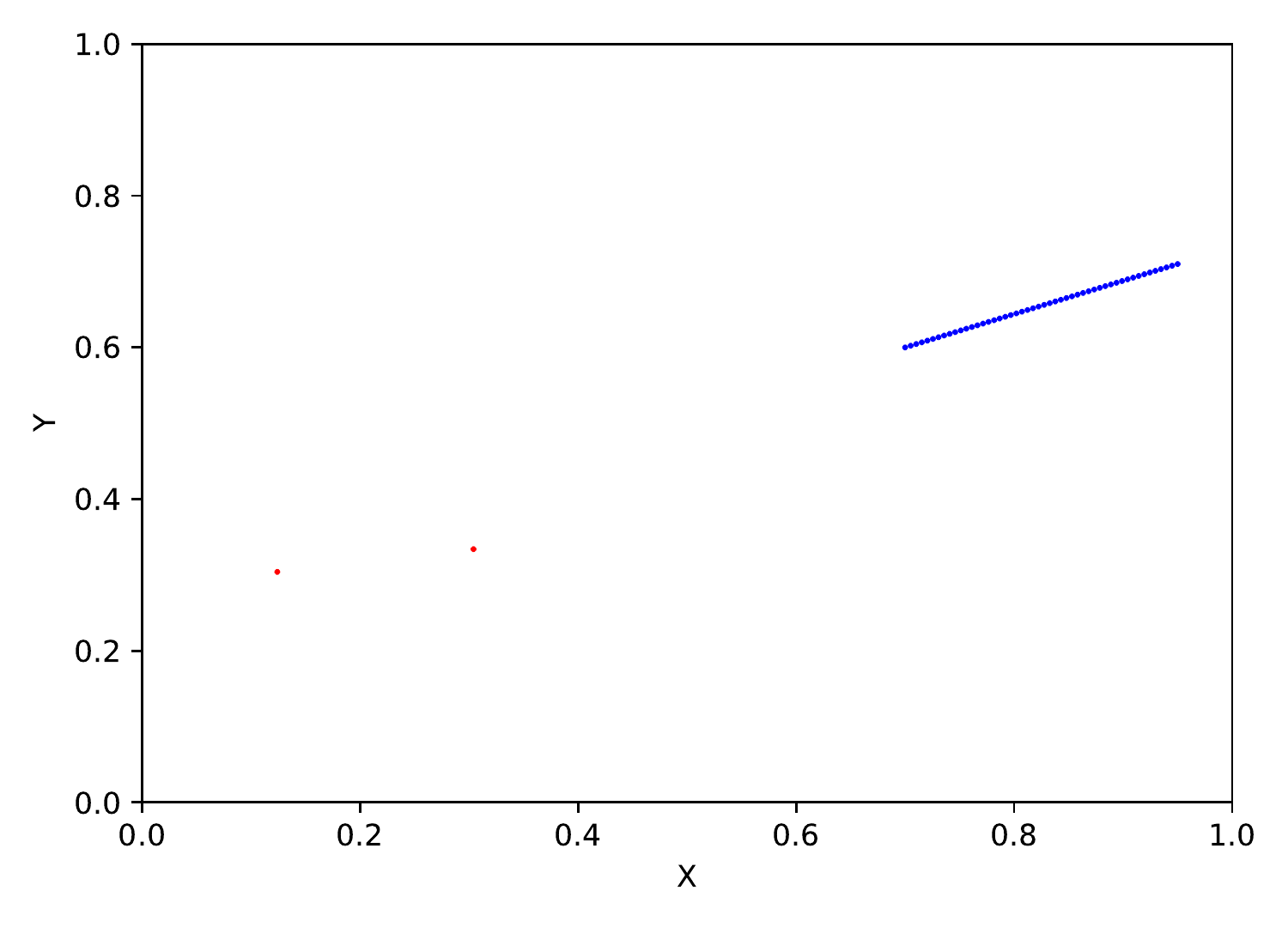}
        \caption{Having a dense linear artifact and two separate points}
    \end{subfigure}%
    \begin{subfigure}{.4\textwidth}
        \centering
        \includegraphics[width=.8\linewidth]{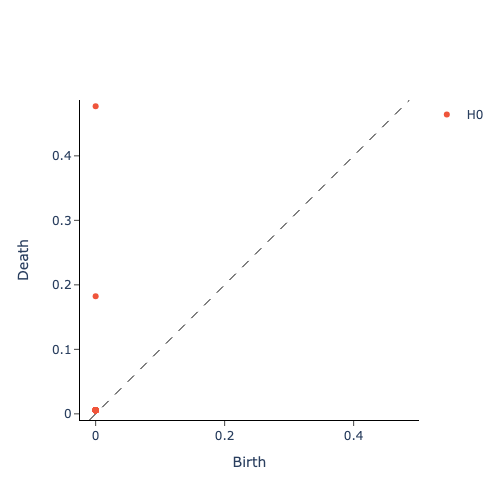}
    \end{subfigure}

    \begin{subfigure}{.4\textwidth}
        \centering
        \includegraphics[width=.8\linewidth]{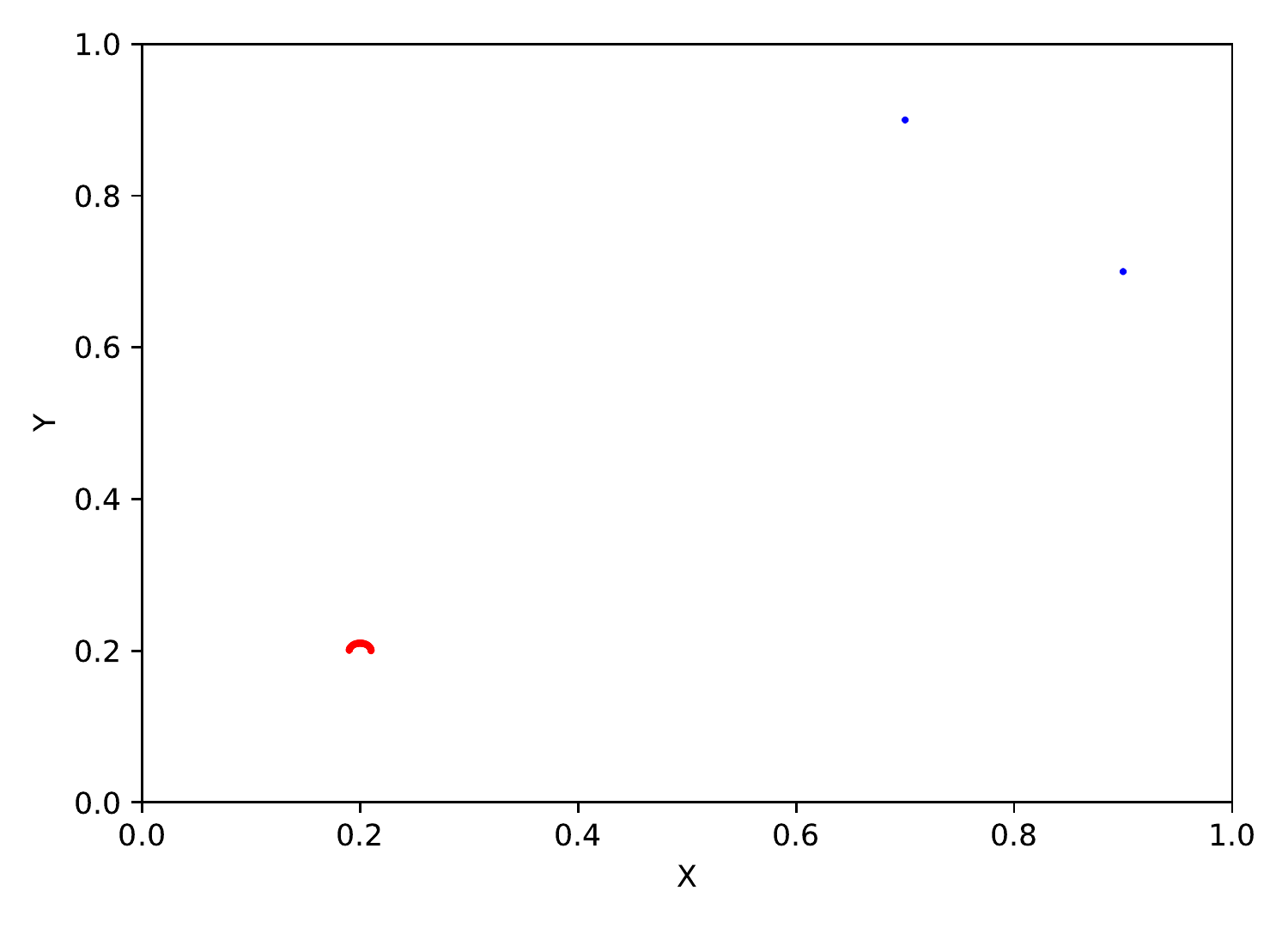}
        \caption{Having a dense curved-like artifact and two separate points}
    \end{subfigure}%
    \begin{subfigure}{.4\textwidth}
        \centering
        \includegraphics[width=.8\linewidth]{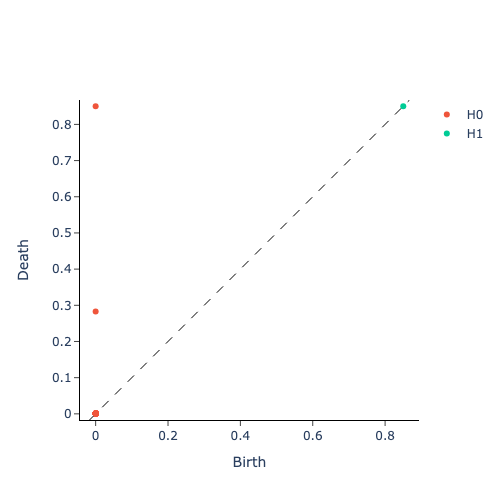}
    \end{subfigure}
    
    \begin{subfigure}{.4\textwidth}
        \centering
        \includegraphics[width=.8\linewidth]{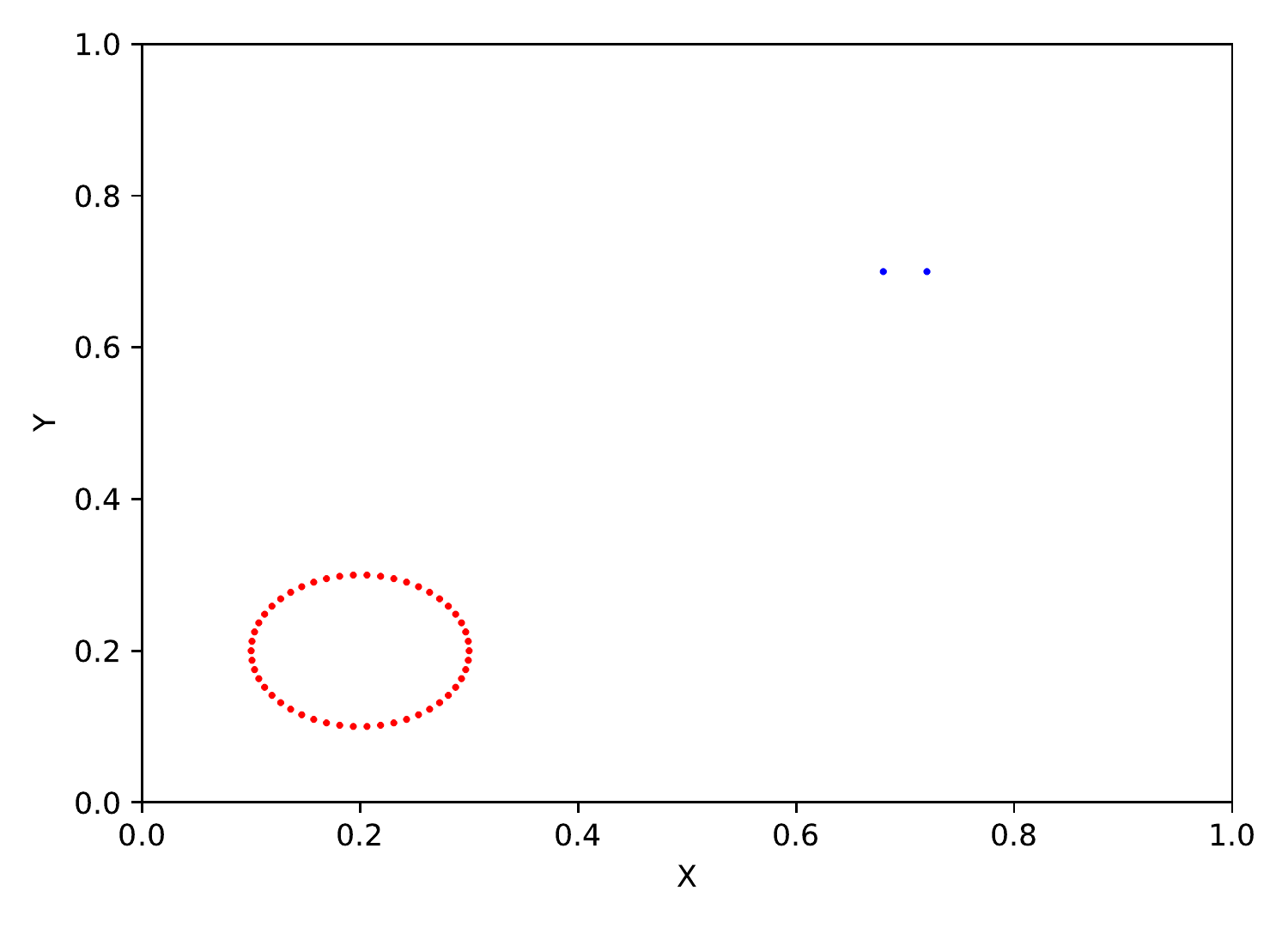}
        \caption{Having a dense circle-like artifact and two separate points}
    \end{subfigure}%
    \begin{subfigure}{.4\textwidth}
        \centering
        \includegraphics[width=.8\linewidth]{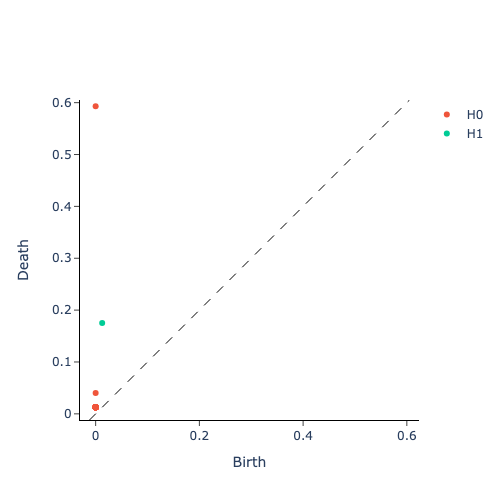}
    \end{subfigure}
    
\end{figure}

Let's dive into some controlled examples where we can observe the association between possible PXD artifacts and persistent homology, as depicted in~\cref{fig:topo_ex}. 
Figure~(a) represents a dense linear set of points that exemplifies a long continuous hit cluster and its corresponding persistent diagram. 
Due to the closeness of dense points in a linear cluster, the persistent diagram shows a very low lifetime of $0.0055$ for this line artifact. 
By adding two separate points, figure ~(b) illustrates how having a point-like~(alone) set of points creates a long-lived persistent homology. In the persistent diagram, the upper points at death coordinate $0.47$ and $0.18$ belong to the two separated added point and their merge with the linear cluster, while the lower point at $0.0055$ still belongs to the linear cluster.
Figure~(c) shows a curved cluster of points along with two separated points. Short-lived points in the persistent diagram again belong to the cluster. I also have a persistent $H_1$ feature that is due to the half-circular cluster revealing a one-dimensional or circular hole. Short-lived hits in the persistent diagram of Figure~(d) again belong to the clusters in the dense circle-like points.

\begin{figure}[!htb]
    \centering
    \begin{subfigure}{.32\textwidth}
        \centering
        \includegraphics[width=\linewidth]{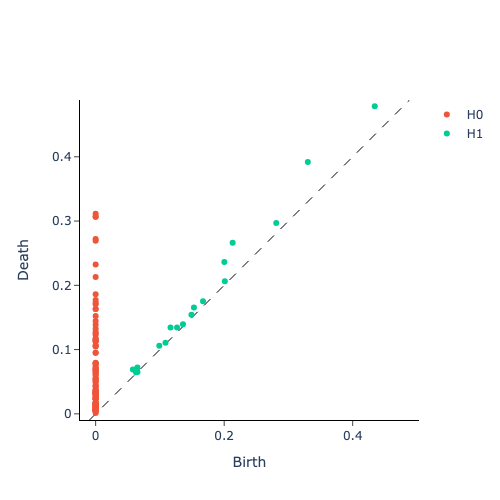}
        \caption{Real PXD, sensor 1}
    \end{subfigure}%
    \hspace{0.005\textwidth}%
    \begin{subfigure}{.32\textwidth}
        \centering
        \includegraphics[width=\linewidth]{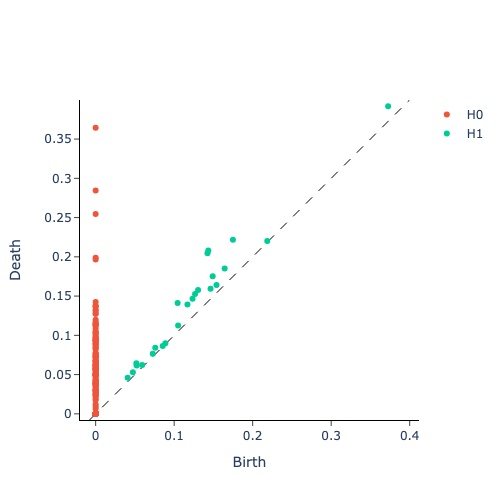}
        \caption{YonedaVAE, sensor 1}
    \end{subfigure}%
    \hspace{0.005\textwidth}%
    \begin{subfigure}{.32\textwidth}
        \centering
        \includegraphics[width=\linewidth]{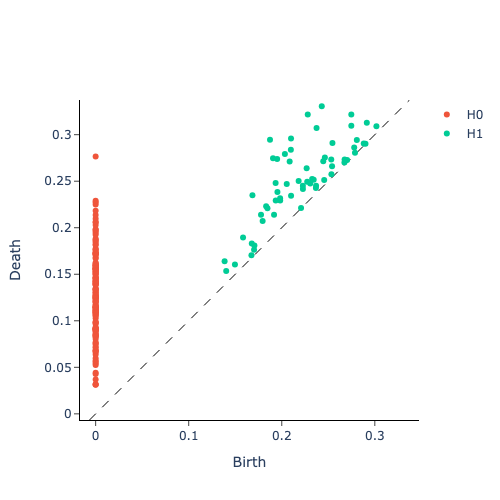}
        \caption{Random points}
    \end{subfigure}
    
    \caption{An example persistent diagram for a point cloud belonging to sensor 1 with 255 set cardinality. Notice the short-lived $H_0$ features from PXD and YonedaVAE.}
    \label{fig:pers_pxd}
\end{figure}

The most obvious conclusion from the examples in~\cref{fig:topo_ex}, is how dense and line-like clusters create $H_0$ persistent features and how curvy-like clusters reveal the existence of $H_1$ persistent features.
Persistence diagrams have been shown to be a powerful tool for quantifying shapes in geometric data. Moreover, one of their key properties is their stability with respect to perturbations in the input, which is crucial when dealing with noisy measurements. 
This study compares the real PXD hits with YonedaVAE's output and random set of points. 
In \cref{fig:pers_pxd}, as an example, compares the persistence diagrams of real PXD background hits with YonedaVAE's output and random set of points. 
As expected, the first noticeable difference is the lifetime of $H_0$ features. As opposed to random points where no meaningful geometric clusters exist, real PXD hits and YonedaVAE exhibit shorter-lifetime $H_0$ features that are caused by line-like clusters. 

\begin{figure}[!htbp]
    \centering
    \includegraphics[width=\linewidth]{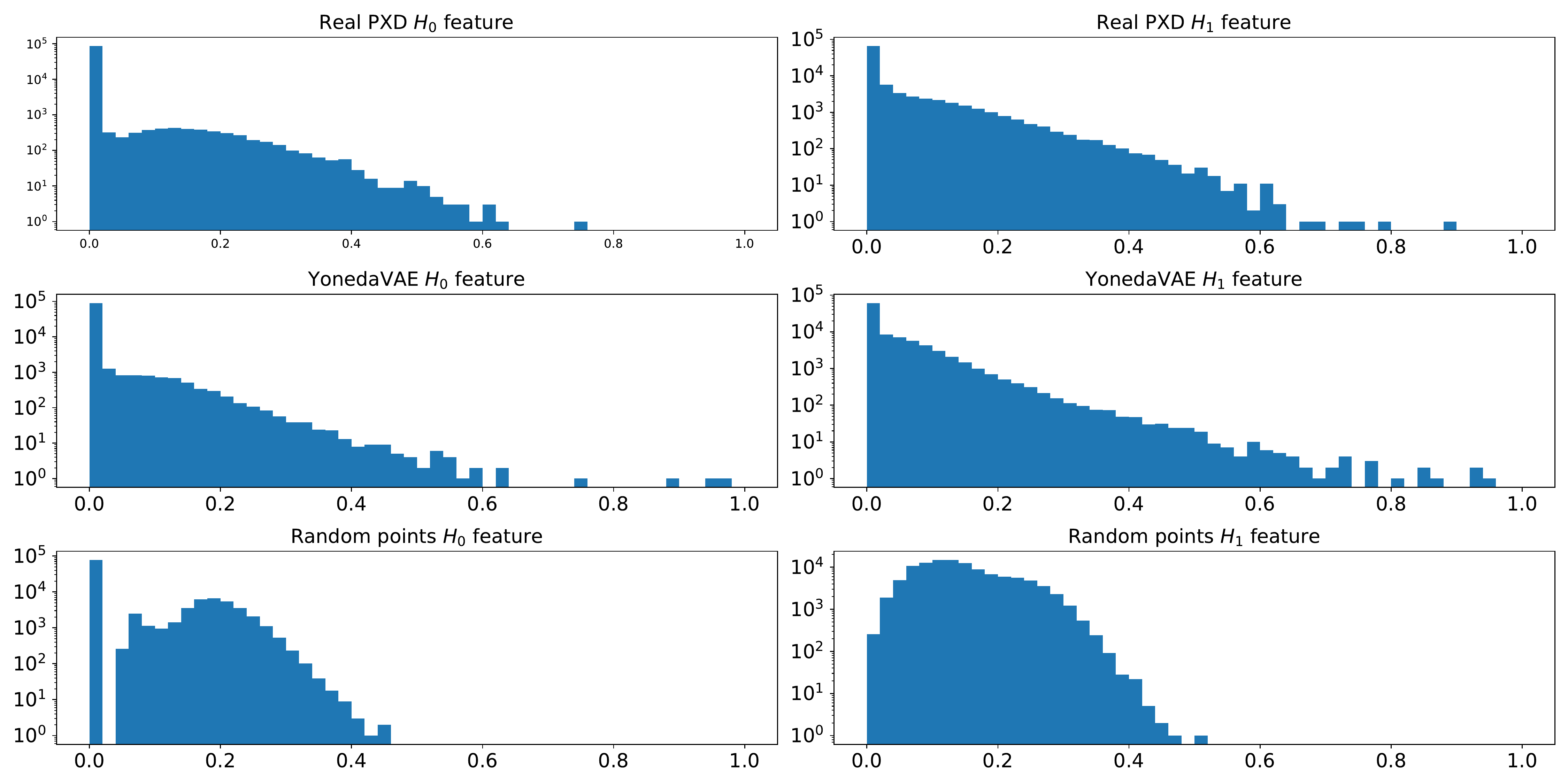}
    \caption{Histograms for $H_0$ and $H_1$ feature lifetimes for the real PXD hits, YonedaVAE generated samples, and samples with random hits.}
    \label{fig:pers_hist}
\end{figure}

In~\cref{fig:pers_hist}, the distribution of $H_0$ and $H_1$ lifetimes over \num{5000} samples are depicted for the three cases, real PXD background, YonedaVAE, and random set of points. The noticeable difference again is at the low lifetime $H_0$ feature that dies for clusters for both PXD and YonedaVAE samples. This signifies how well YonedaVAE captures the cluster of hits in the PXD data. 
On the other hand, both PXD and YonedaVAE exhibit more persistent $H_1$ features. In the persistent homology, a loop represents a 1-cycle that is not the boundary of a 2-chain in a simplicial complex and corresponds to a non-zero element in the first homology group $H_1$. 
A persistent $H_1$ feature in the persistent diagram signifies a topological loop structure in the data that is robust across a range of scales.
The emergence of a loop in the persistent diagram can be indicative of a spiral-like or curvy-like arrangement of hits. This could correspond to particles following a helical trajectory.

In order to numerically compare persistence diagrams of the real PXD data and from the YonedaVAE, one can incorporate different metrics such as Wasserstein distance, $L_p$ distance between betti curves~(the Betti numbers) as amplitudes, and the entropy of the persistence diagrams~\cite{rucco_characterisation_2016,atienza_persistent_2017}. 
Persistence entropy is a metric for characterizing the complexity of the persistence diagrams. It quantifies the uncertainty associated with the lifespan of topological features, thereby offering insights into the overall structural complexity of the data. 
The amplitude of a persistence diagram measures the ``spread'' or ``complexity'' of the points in the diagram. 
In other words, it quantifies how far the points in the diagram are from the diagonal, which in turn provides an indication of the prominence or importance of the corresponding topological features.
For each persistence diagram in a collection, an amplitude is measured according to the following steps. First, the diagram is partitioned into subdiagrams according to the homology dimension. Then, the amplitude of each subdiagram is calculated according to a metric that gives a vector of amplitudes over the available homology dimensions. I report the average of these amplitudes as summary statistics across all tested samples.

\begin{table}[!htbp]
\begin{minipage}{\textwidth}
    \begin{center}
    \caption{Quantitative comparison between persistence diagrams using amplitudes and entropy}
    \label{tab:amp}
    \setlength{\tabcolsep}{4pt} 
    \begin{tabular}{@{}l|lll@{}}
        \toprule
        & Real PXD & YonedaVAE & Random \\ 
        \midrule
        \textbf{Persistence Entropy} $H_0$  & $5.54$ & $5.64$ & $7.54$ \\
        \textbf{Persistence Entropy} $H_1$  & $2.34$ & $2.16$ & $5.78$ \\
        \midrule
        \textbf{Wasserstein Amplitude} $H_0$ & $0.45$ & $0.39$ & $0.88$\\
        \textbf{Wasserstein Amplitude} $H_1$ & $0.048$ & $0.048$ & $0.201$\\
        \midrule
        \textbf{betti Amplitude} $H_0$ & $16.45$ & $19.17$ & $62.36$\\
        \textbf{betti Amplitude} $H_1$& $0.56$ & $0.49$ & $7.52$\\
        \bottomrule
    \end{tabular}
    \end{center}
\end{minipage}
\end{table}

As depicted in~\cref{tab:amp}, across all metrics and both homology dimensions, the generated data by YonedaVAE closely mirrors the real PXD data. 
This indicates the model's effectiveness in capturing both continuous clustering features~($H_0$) and higher-order clustering structures~($H_1$). 
As expected, the random data set deviates significantly across all metrics, reinforcing the non-random nature of the real and generated PXD background data sets. Moreover, the persistence of $H_1$ features in the PXD data could be indicative of a specific type of interaction or decay process, which is an open question that can be studied in detail. 

\begin{figure}[!htbp]
    \centering
    \begin{subfigure}[b]{0.33\textwidth}
        \includegraphics[width=\textwidth]{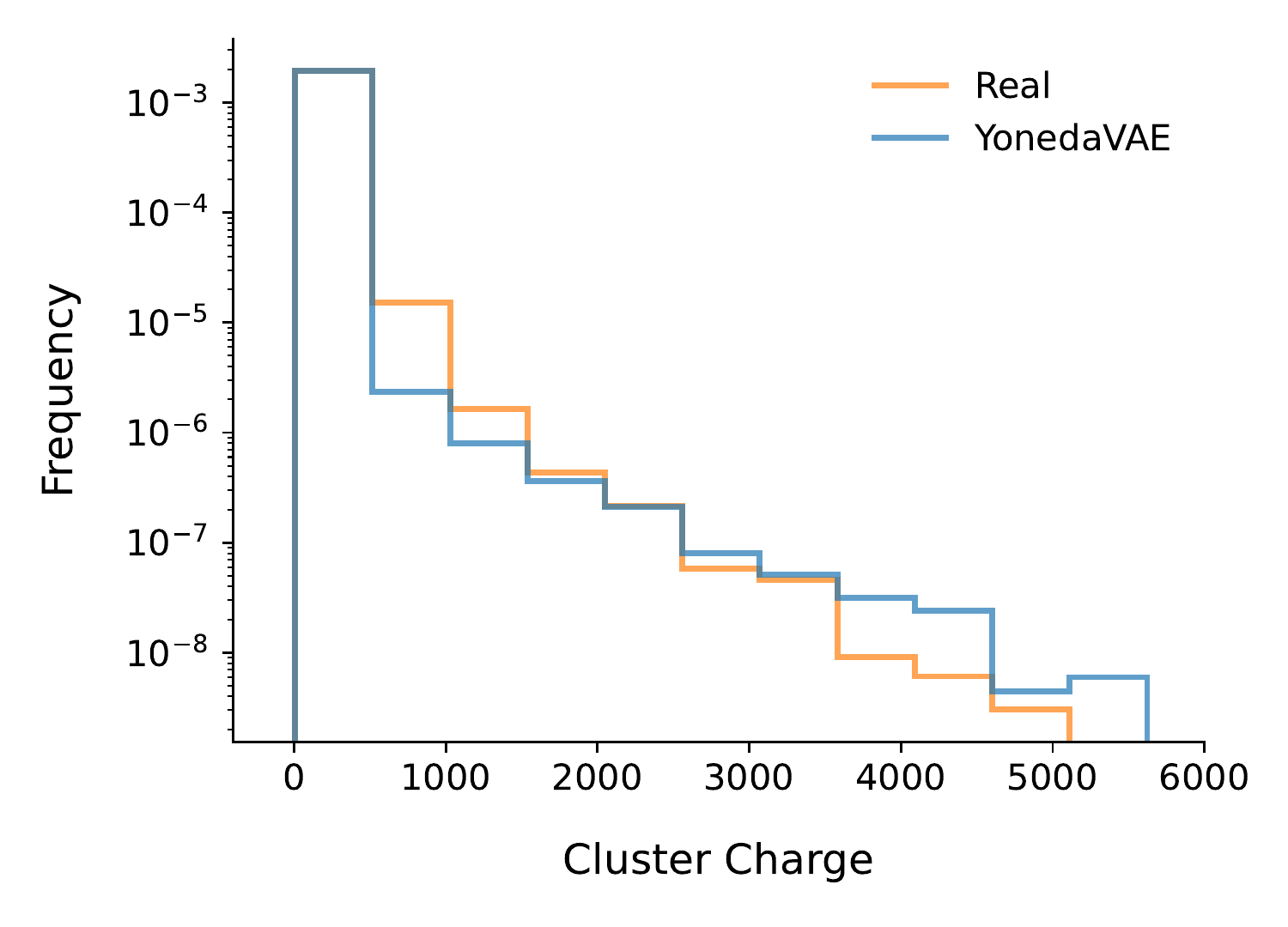}
        \caption{}
    \end{subfigure}%
    \begin{subfigure}[b]{0.33\textwidth}
        \includegraphics[width=\textwidth]{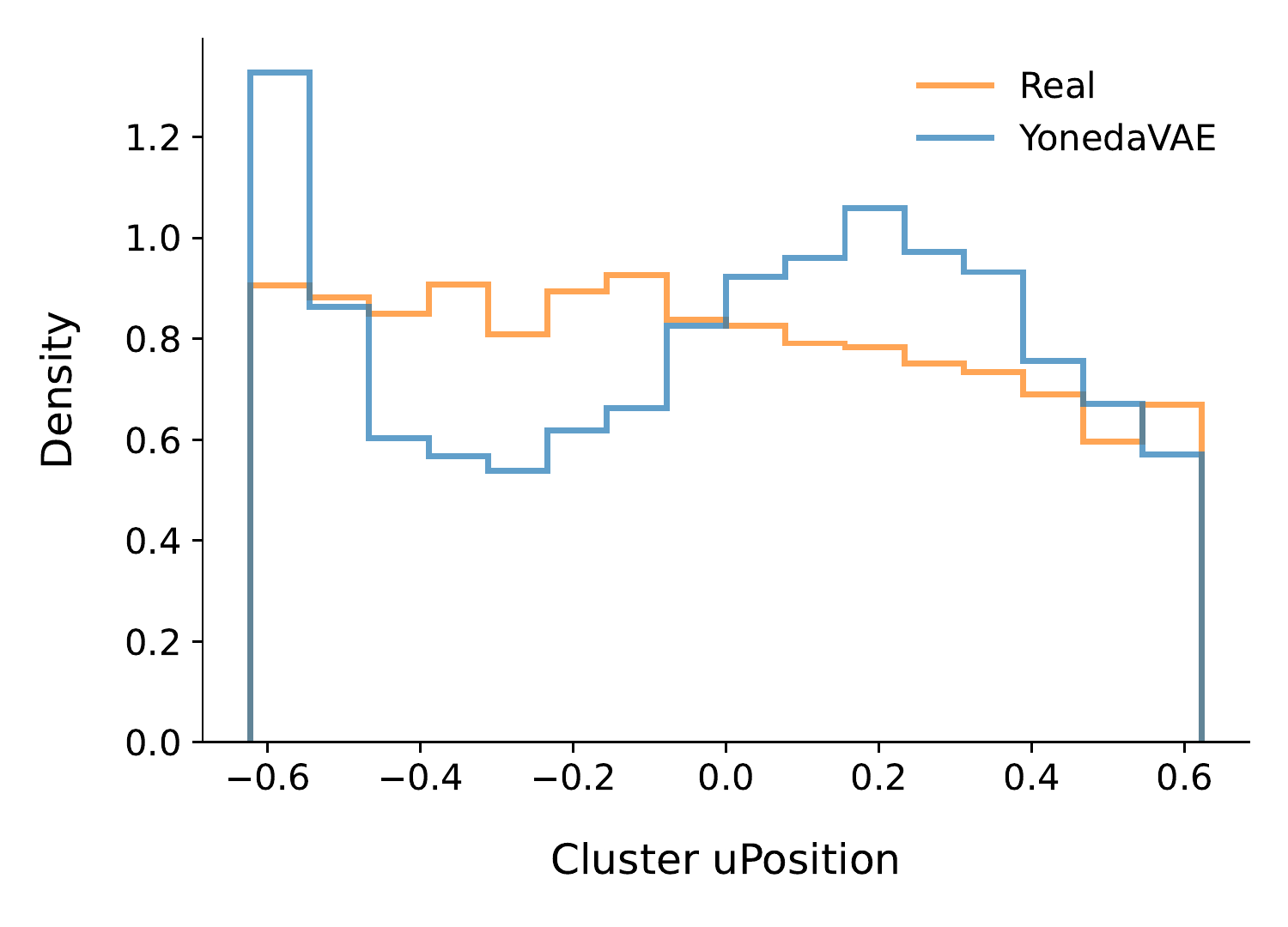}
        \caption{}
    \end{subfigure}%
    \begin{subfigure}[b]{0.33\textwidth}
        \includegraphics[width=\textwidth]{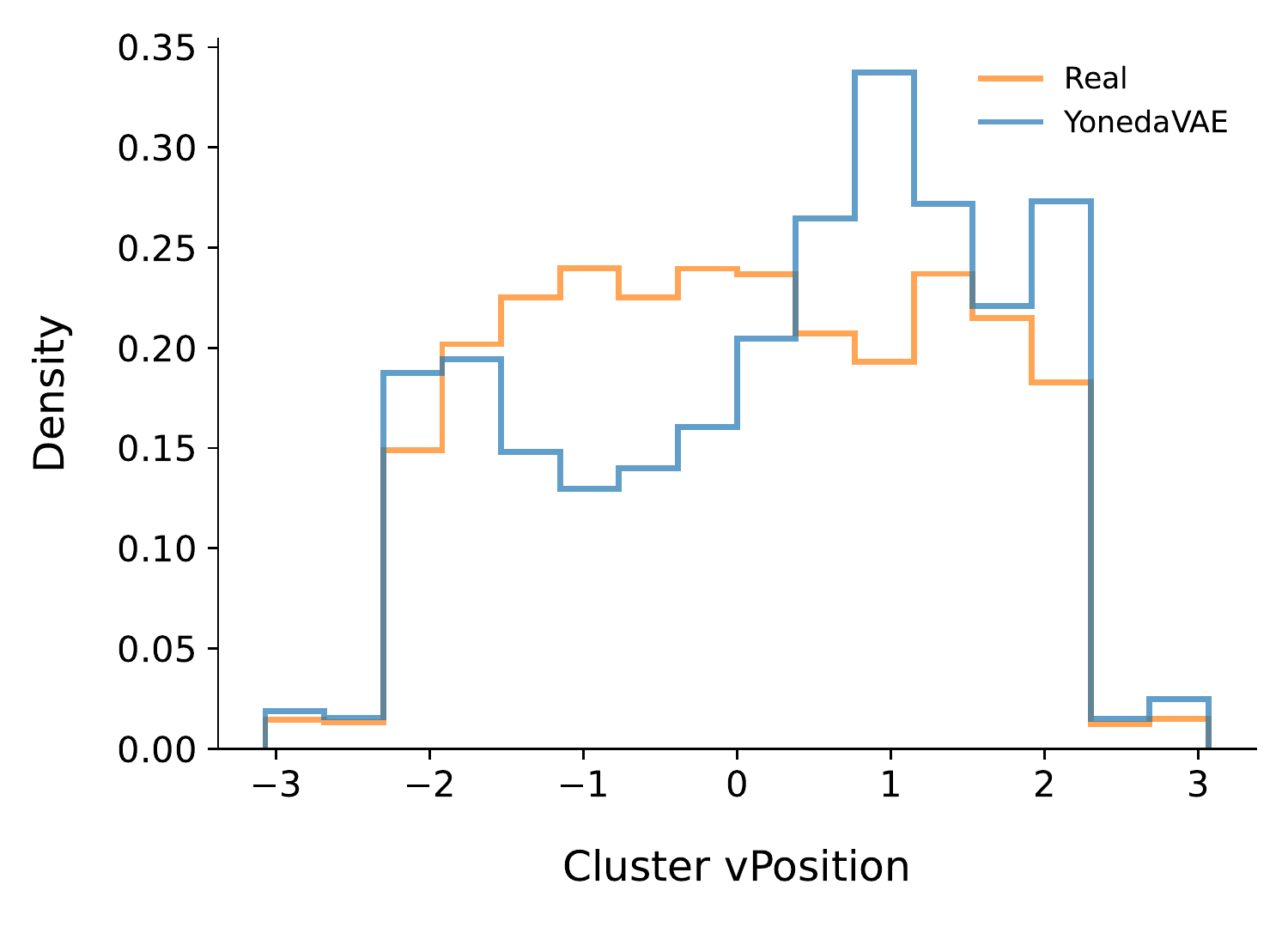}
        \caption{}
    \end{subfigure}\\
    \begin{subfigure}[b]{0.33\textwidth}
        \includegraphics[width=\textwidth]{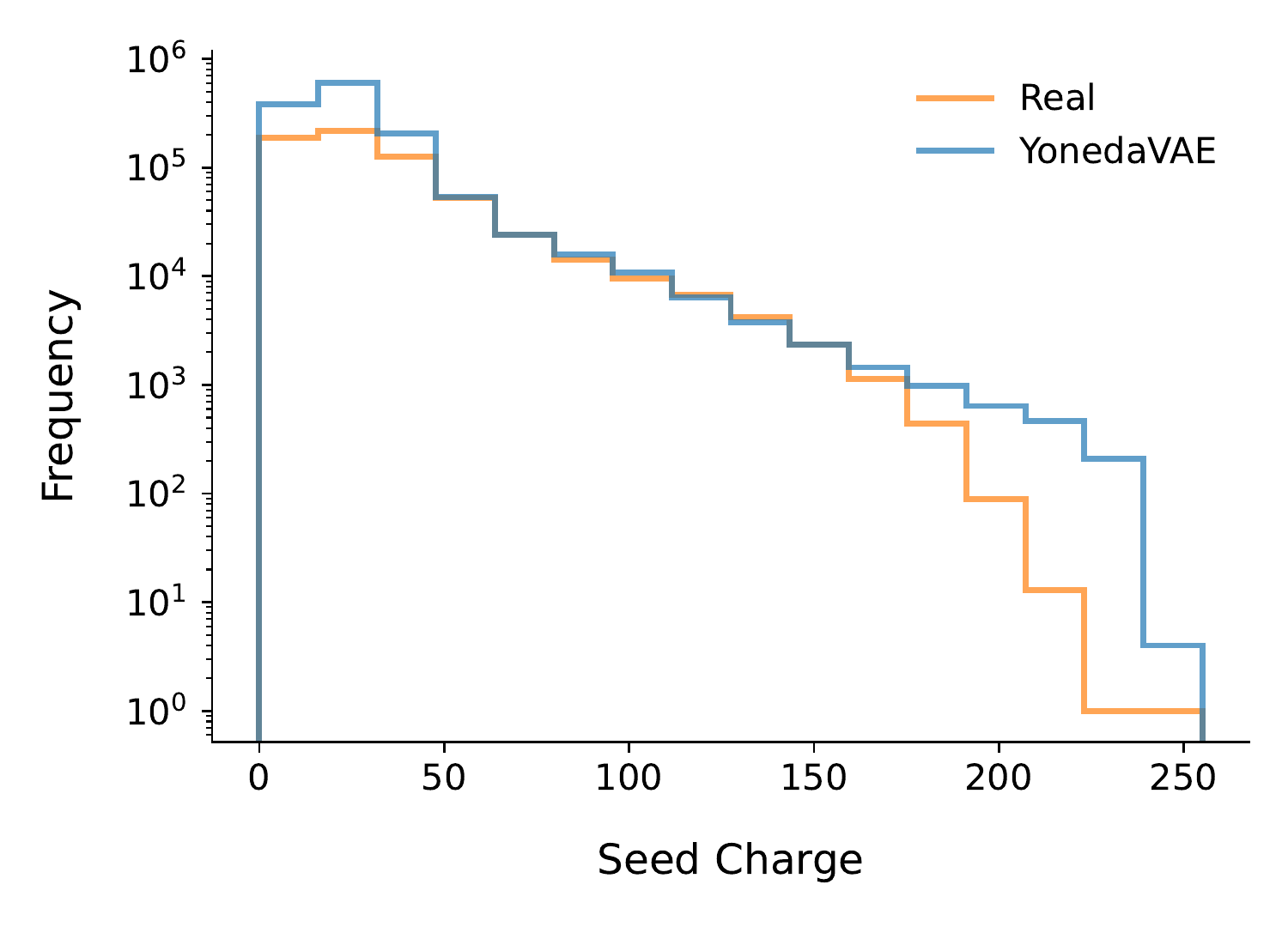}
        \caption{}
    \end{subfigure}%
    \begin{subfigure}[b]{0.33\textwidth}
        \includegraphics[width=\textwidth]{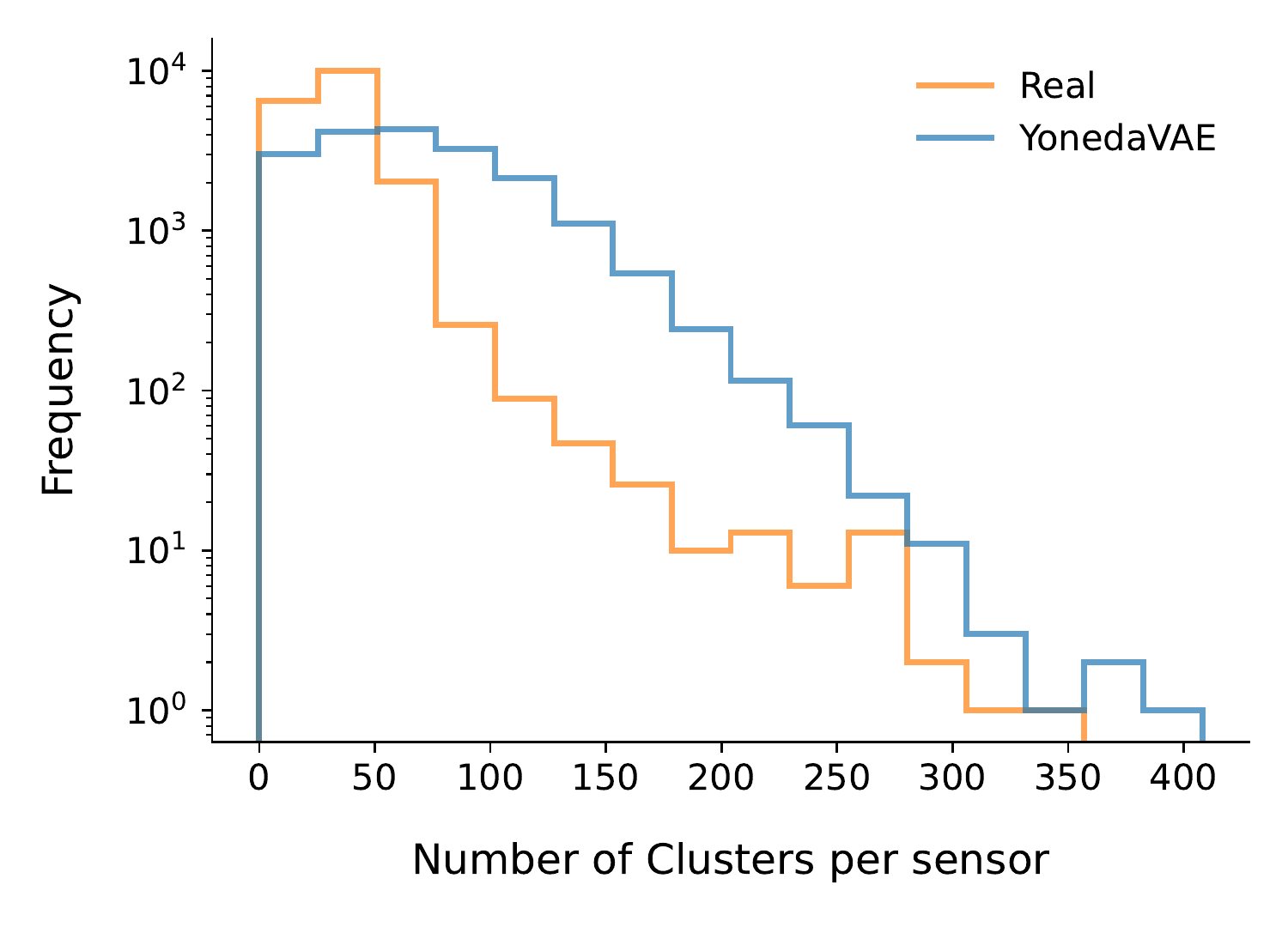}
        \caption{}
    \end{subfigure}%
    \begin{subfigure}[b]{0.33\textwidth}
        \includegraphics[width=\textwidth]{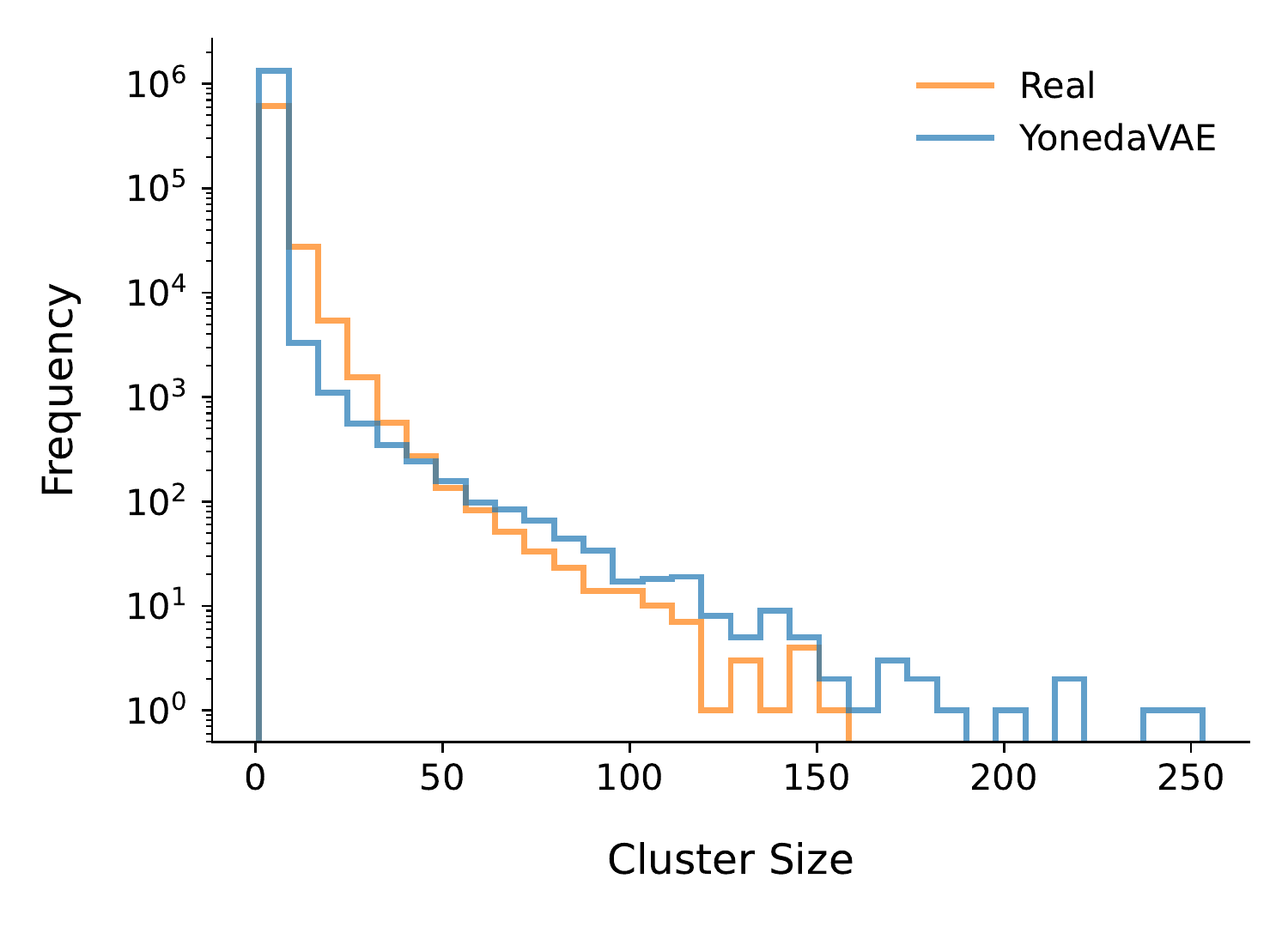}
        \caption{}
    \end{subfigure}
    \caption{Comparison between the real PXD clustering information and the YonedaVAE generated ones. }
    \label{fig:cluster_comp}
\end{figure}

\subsubsection{Geometrical Clustering}
In order to consolidate the clustering analysis, the PXD clustering algorithm, \emph{PXDClusterizer} module from the basf2~\cite{kuhr_computing_2011} is incorporated. 
As discussed in~\cref{chap:2}, the PXDClusterizer module is responsible for cluster formation and hit reconstruction. During its operation, the algorithm iteratively scans through all the digitized signals, or ``digits,'' from individual sensors. 
It then coalesces contiguous digits, centered around a seed pixel with a signal exceeding a predefined threshold, into consolidated cluster formations. 
After forming these clusters, they are then transformed into what are termed PXDHits. The transition to PXDHits involves deducing the hit location based on the cluster's charge distribution profile. Specifically, the center of gravity algorithm is employed for clusters comprising two pixels, while the analog head-to-tail algorithm is utilized for larger clusters. 
In scenarios where only a single digit in one dimension contributes to the cluster, the hit is assigned to the pixel's center. Subsequently, various attributes of the hit—such as its final position in the sensor, the cumulative charge, and its correlation to the originating cluster and particles—are recorded. 

Clusters serve as a proxy for the interaction of a single particle with the PXD. Their properties are subject to this analysis as depicted in~\cref{fig:cluster_comp}. These attributes can include the total number of pixels constituting the cluster~(Cluster size), the number of clusters for each sensor, the position along the u- and v-axes, the sum of charges across the cluster~(cluster charge), and the charge at the seed pixel~(seed charge). 
As depicted in \cref{fig:cluster_comp}, YonedaVAE shows a close agreement with the real PXD clustering attributes. 
On the other hand, although point cloud-based approaches can handle sparse data, they exhibit a drawback when utilized for PXD background generation that must be studied further to have a conclusive result. 
This drawback can be conceptualized as a trade-off incurred when transitioning from a discrete~(pixelated PXD hitmap) representation to a continuous~(point cloud) representation, which results in information loss. 
To leverage point cloud approaches, data must transition from a pixelated format, where the positions of hits are defined by their exact pixel locations in \( \mathbb{Z} \)~(e.g. [[1-768]]), to a continuous format, where the positions of hits are in \( \mathbb{R} \)~(e.g. [[0,1]]) that possess a precision uncertainty. This issue is particularly more significant in the PXD case due to its ultra-high-dimensional size; the error introduced when reverting back to a pixelated format from the continuous point cloud format is amplified by the number of pixels in the PXD.

Encountering challenges in accurately reconstructing large cluster patterns, let's explore in the following section how YonedaVAE performs in terms of downstream track reconstruction and its efficacy in generating hits that align with the tracks emerging from these clusters.

\FloatBarrier
\subsection{Physics Analysis}
This section dives into the tracking analysis by making a comparison between real PXD background and YonedaVAE-generated PXD background hits on the track parameter resolutions as introduced in~\cref{chap:5}. This study compares both the length extrapolative and context extrapolative YonedaVAE with the real PXD background of Experiment 26.

\begin{figure}[!htb]
    \centering
    \begin{subfigure}{0.45\linewidth}
        \includegraphics[width=\linewidth]{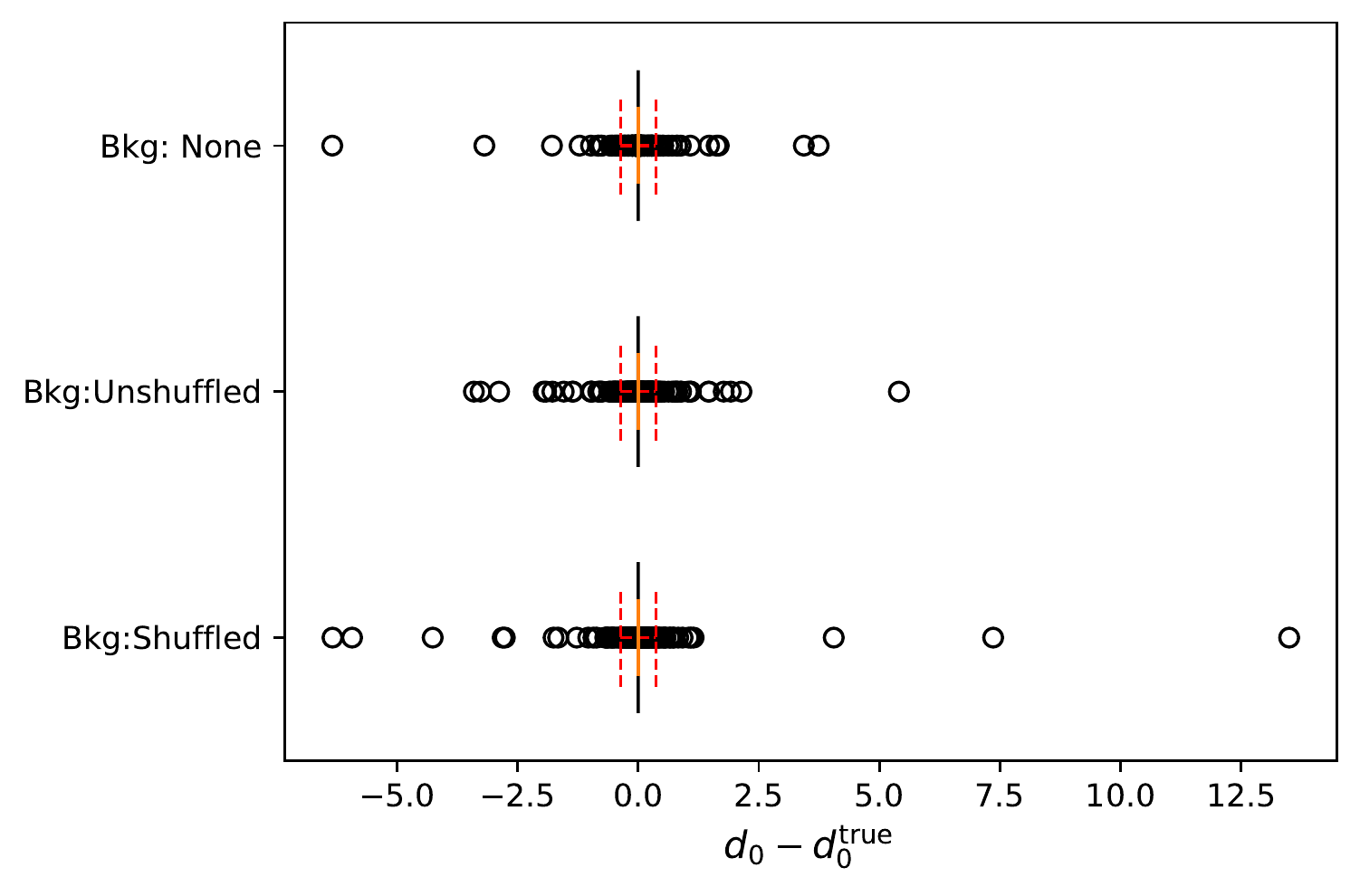}
    \end{subfigure}
    \begin{subfigure}{0.45\linewidth}
        \includegraphics[width=\linewidth]{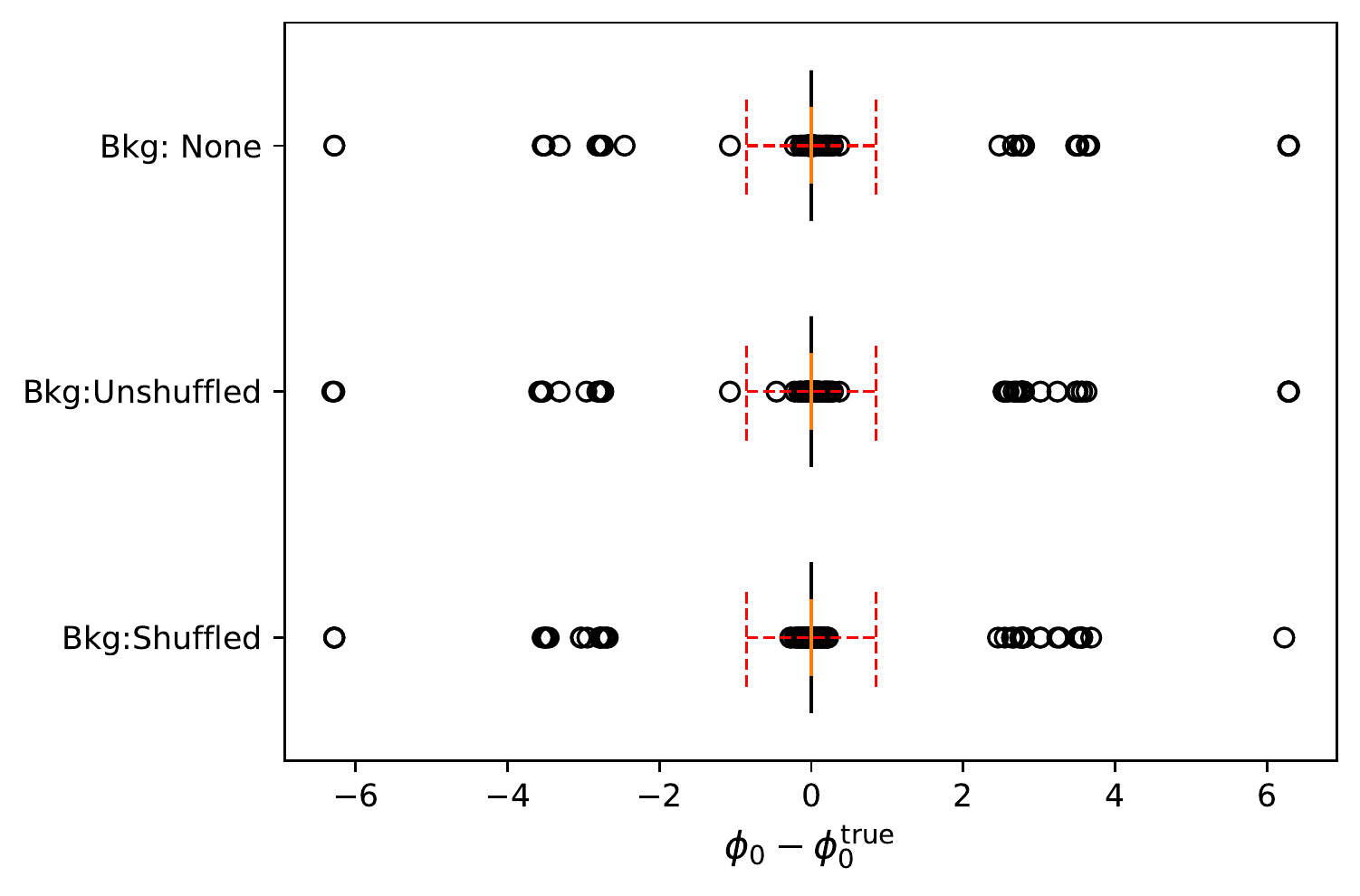}
    \end{subfigure}\\
    
    \begin{subfigure}{0.45\linewidth}
        \includegraphics[width=\linewidth]{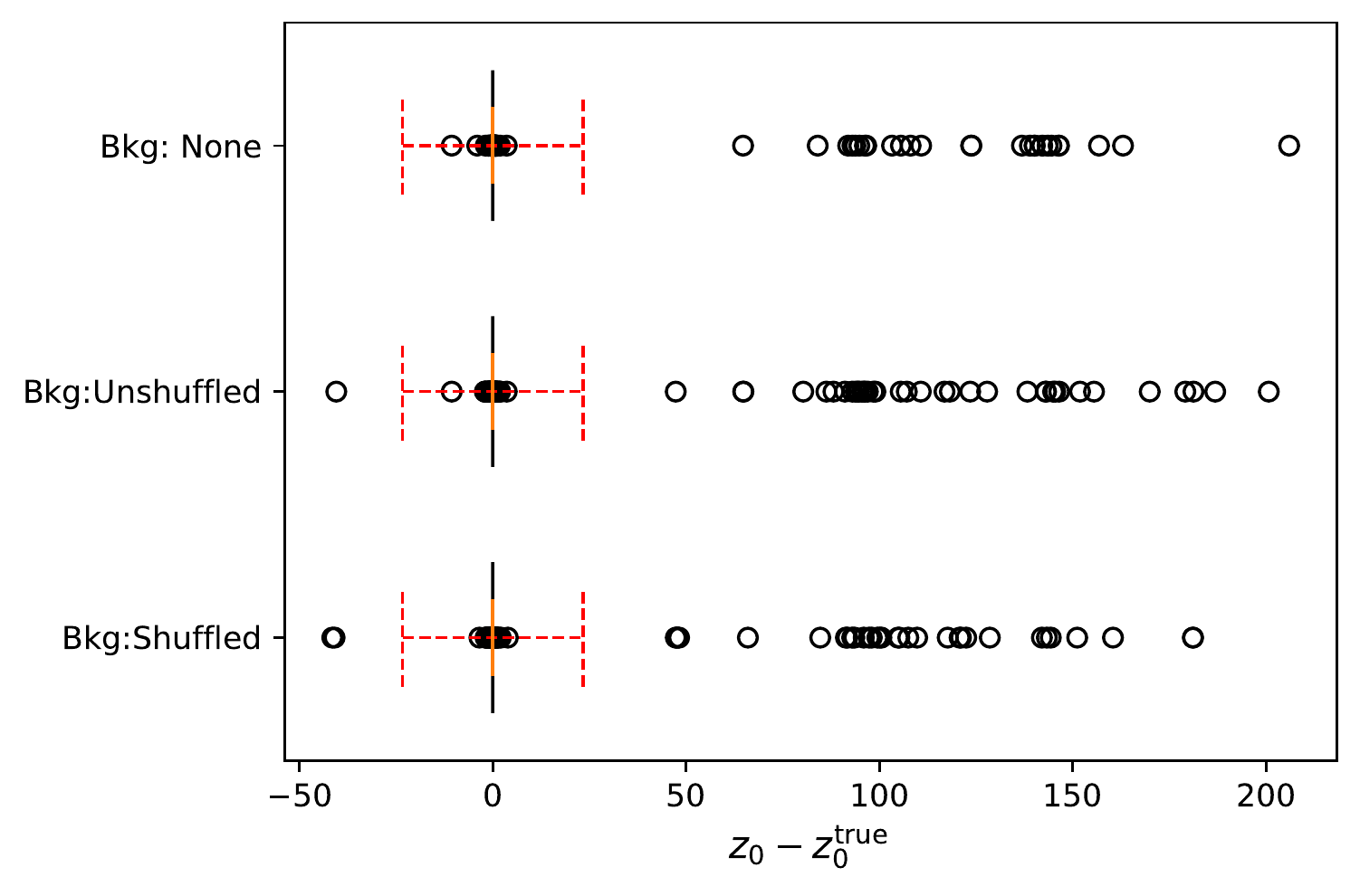}
    \end{subfigure}
    \begin{subfigure}{0.45\linewidth}
        \includegraphics[width=\linewidth]{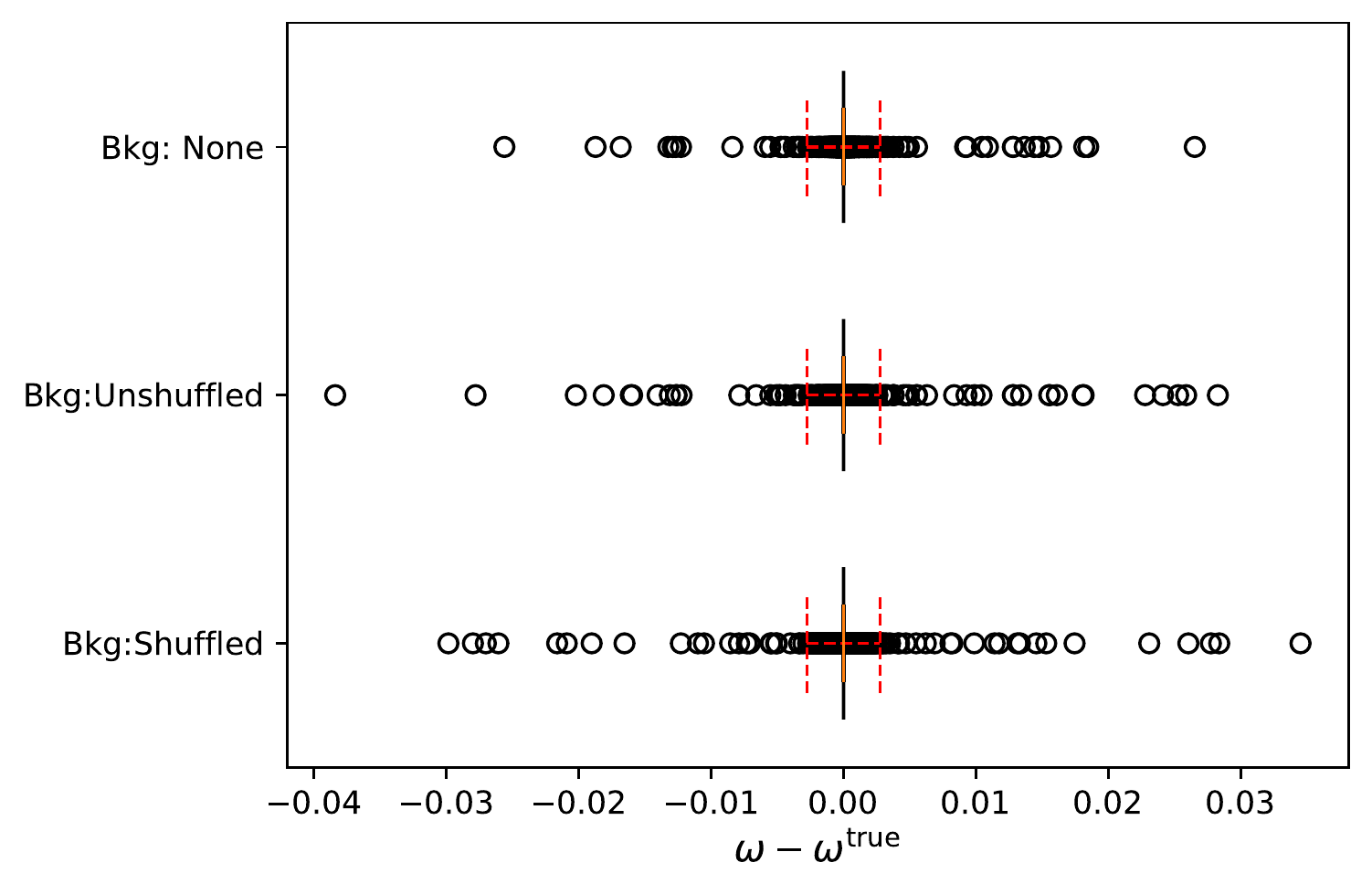}
    \end{subfigure}\\
    
    \begin{subfigure}{0.45\linewidth}
        \includegraphics[width=\linewidth]{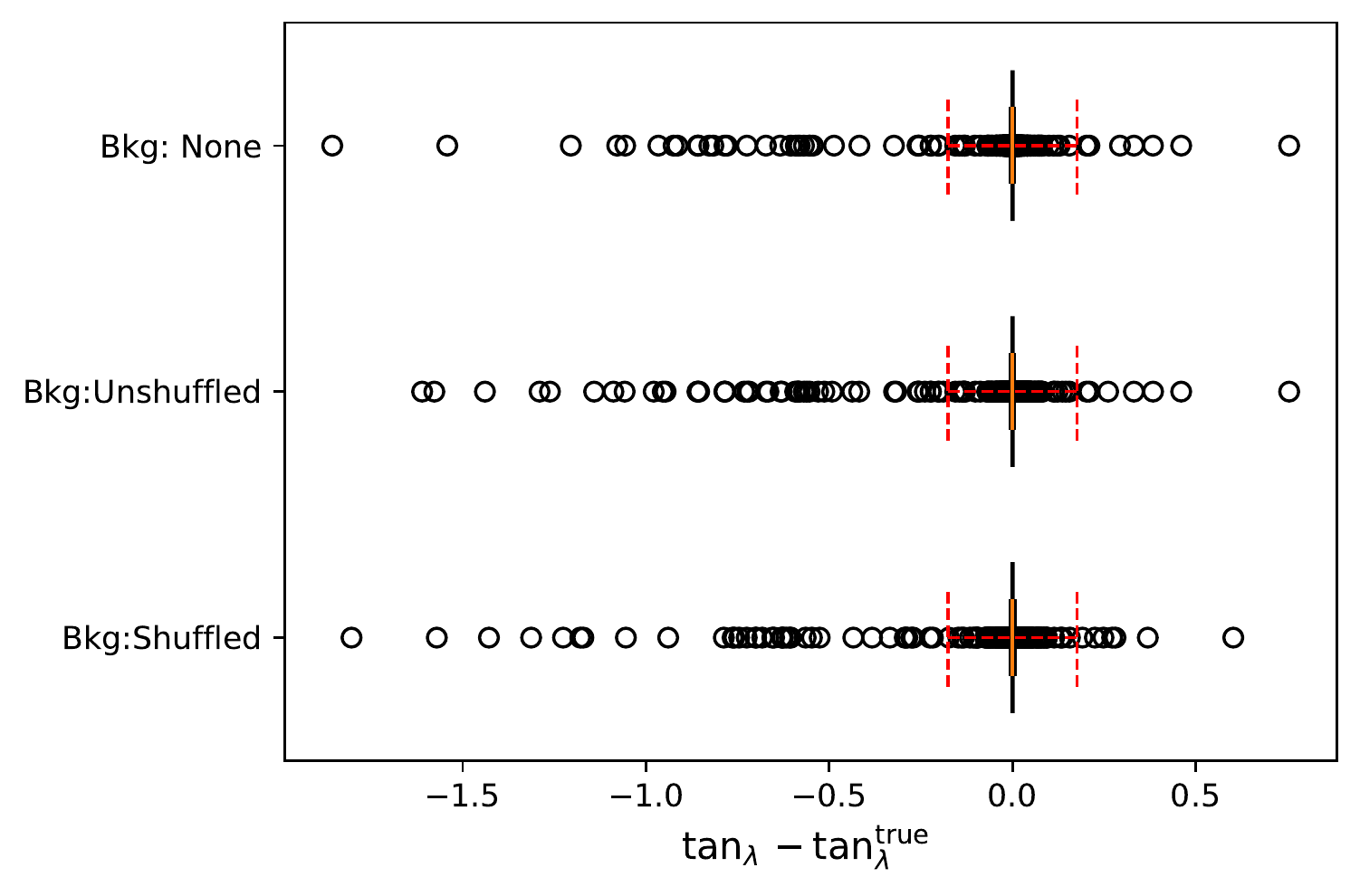}
    \end{subfigure}
    
    \caption{
    The boxplot for comparing the Helix parameter resolutions in the presence of correlated background and uncorrelated background at high momentum regime. The $\pm 5$ standard deviations interval for the no-background case is shown in all 3 cases for reference as a red dashed line.
    }
    \label{fig:helix_shu}
\end{figure}

First, let's study the effect of intra-event correlation on the track reconstruction, specifically the Helix parameters. In the last chapter, I studied this for the Geant4 simulated data and showed how the lack of intra-event correlation negatively influences the Helix parameter resolution. 
\cref{table:helix_shu}, and~\cref{fig:helix_shu} shows the effect of losing the intra-event correlation for high momentum tracks, $p_t>0.4$ GeV, is demonstrated through the huge discrepancy between the correlated~(unshuffled) and uncorrelated~(shuffled) events in the unbiased variance of the $\Delta d_0$ and $\Delta z_0$, thus in the precision of the $d_0$ and $z_0$ reconstruction.

\begin{table}[!htb]
    \centering
    \caption{Helix Parameter Resolution observable Comparison between correlated~(Unshuffled) and uncorrelated~(Shuffled) events with high momentum tracks, $p_t>0.4$ GeV.}
    \label{table:helix_shu}
    \begin{tabularx}{\textwidth}{|l|l|X|X|X|}
    \hline
    \multirow{2}{*}{Parameter} & \multirow{2}{*}{Observable} & \multicolumn{3}{c|}{Signal $+$ Background} \\
    \cline{3-5}
     & & Shuffled & Unshuffled & No Bkg. \\
    \hline
    \multirow{4}{*}{$\Delta d_0$} & Unbiased Variance & $0.1423 \pm 0.0007$ & $0.0770 \pm 0.0004$ & $0.0730 \pm 0.0004$ \\
    & KS Statistic & \multicolumn{2}{c|}{0.0074, p-value: 0.6638} & \\
    & Num. Tracks $<-5\sigma_{\text{sig.}}$ & 22 & 27 & 18 \\
    & Num. Tracks $>5\sigma_{\text{sig.}}$ & 19 & 24 & 17 \\
    \hline
    \multirow{4}{*}{$\Delta \phi_0$} & Unbiased Variance & $0.1650 \pm 0.0008$ & $0.1864 \pm 0.0010$ & $0.1706 \pm 0.0009$ \\
    & KS Statistic & \multicolumn{2}{c|}{0.0115, p-value: 0.1610} & \\
    & Num. Tracks $<-5\sigma_{\text{sig.}}$ & 19 & 18 & 12 \\
    & Num. Tracks $>5\sigma_{\text{sig.}}$ & 20 & 29 & 23 \\
    \hline
    \multirow{4}{*}{$\Delta z_0$} & Unbiased Variance & $4.6455 \pm 0.0238$ & $5.4540 \pm 0.0279$ & $4.6674 \pm 0.0239$ \\
    & KS Statistic & \multicolumn{2}{c|}{0.0189, p-value: 0.0022} & \\
    & Num. Tracks $<-5\sigma_{\text{sig.}}$ & 2 & 1 & 0 \\
    & Num. Tracks $>5\sigma_{\text{sig.}}$ & 32 & 37 & 26 \\
    \hline
    \multirow{4}{*}{$\Delta \omega$} & Unbiased Variance & $0.0008 \pm 0.0001$ & $0.0008 \pm 0.0001$ & $0.0006 \pm 0.0001$ \\
    & KS Statistic & \multicolumn{2}{c|}{0.0095, p-value: 0.3473} & \\
    & Num. Tracks $<-5\sigma_{\text{sig.}}$ & 23 & 25 & 20 \\
    & Num. Tracks $>5\sigma_{\text{sig.}}$ & 25 & 30 & 25 \\
    \hline
    \multirow{4}{*}{$\Delta \tan\lambda$} & Unbiased Variance & $0.0373 \pm 0.0002$ & $0.0399 \pm 0.0002$ & $0.0352 \pm 0.0002$ \\
    & KS Statistic & \multicolumn{2}{c|}{0.0237, p-value: 0.0000} & \\
    & Num. Tracks $<-5\sigma_{\text{sig.}}$ & 38 & 48 & 36 \\
    & Num. Tracks $>5\sigma_{\text{sig.}}$ & 7 & 7 & 7 \\
    \hline
    \end{tabularx}
\end{table}

\begin{figure}[!htb]
    \centering
    \includegraphics[width=0.9\linewidth]{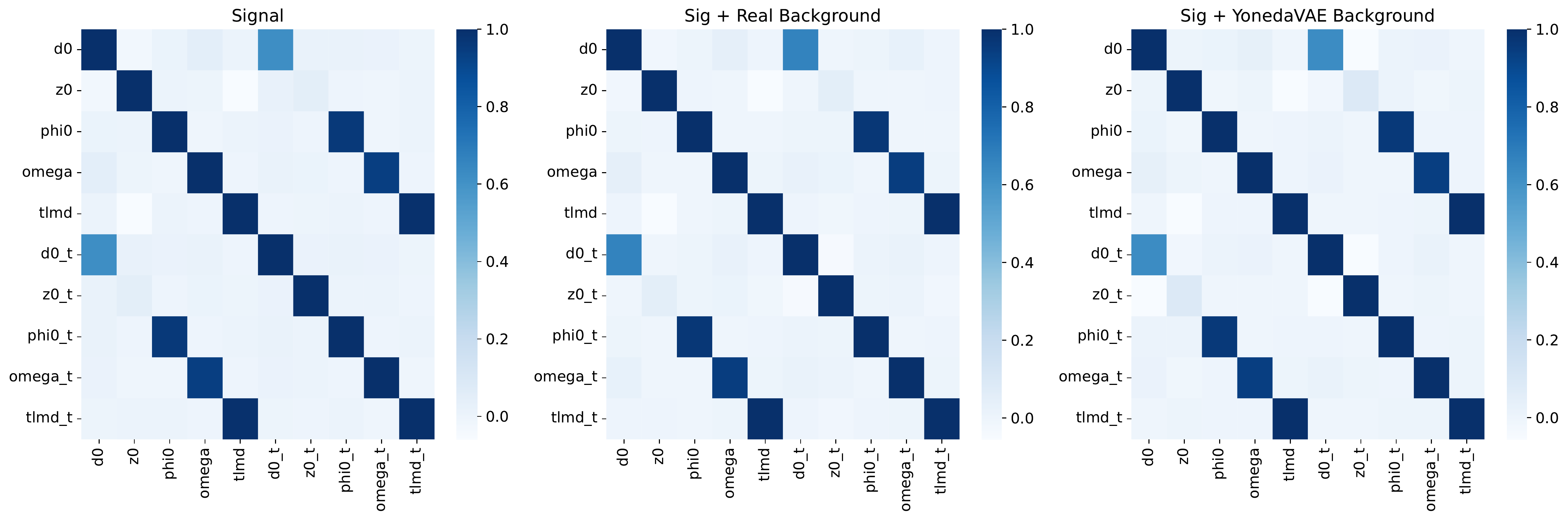}
    \includegraphics[width=0.9\linewidth]{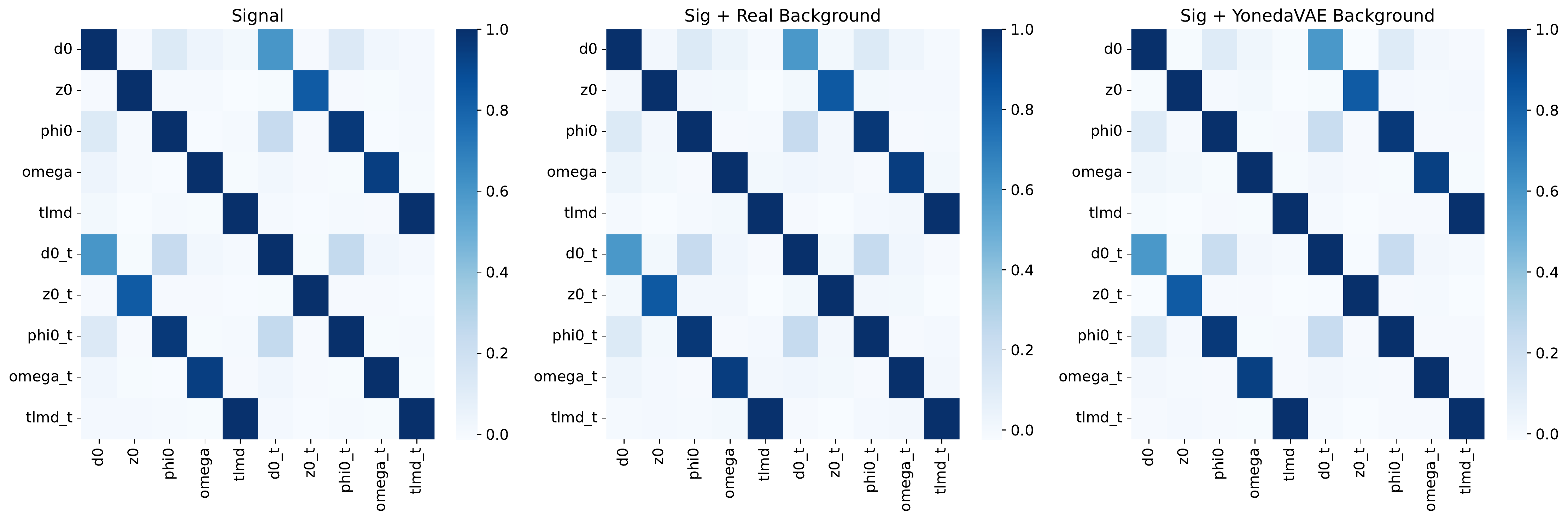}
    \caption{Pearson~(top) vs Spearman~(bottom) correlation between the true and reconstructed Helix parameters with and without background.}
    \label{fig:spear_pears}
\end{figure}

First, let's take a look at the correlation between the Helix parameters, both the true and the reconstructed ones. A very interesting observation in~\cref{fig:spear_pears} is the lack of linear correlation~(Pearson) but the monotonic correlation~(Spearman) between $z_0$ and $z_0^t$. YonedaVAE also mimics the same behavior.

\begin{figure}[!htb]
    \centering
    \fbox{\includegraphics[width=0.95\textwidth]{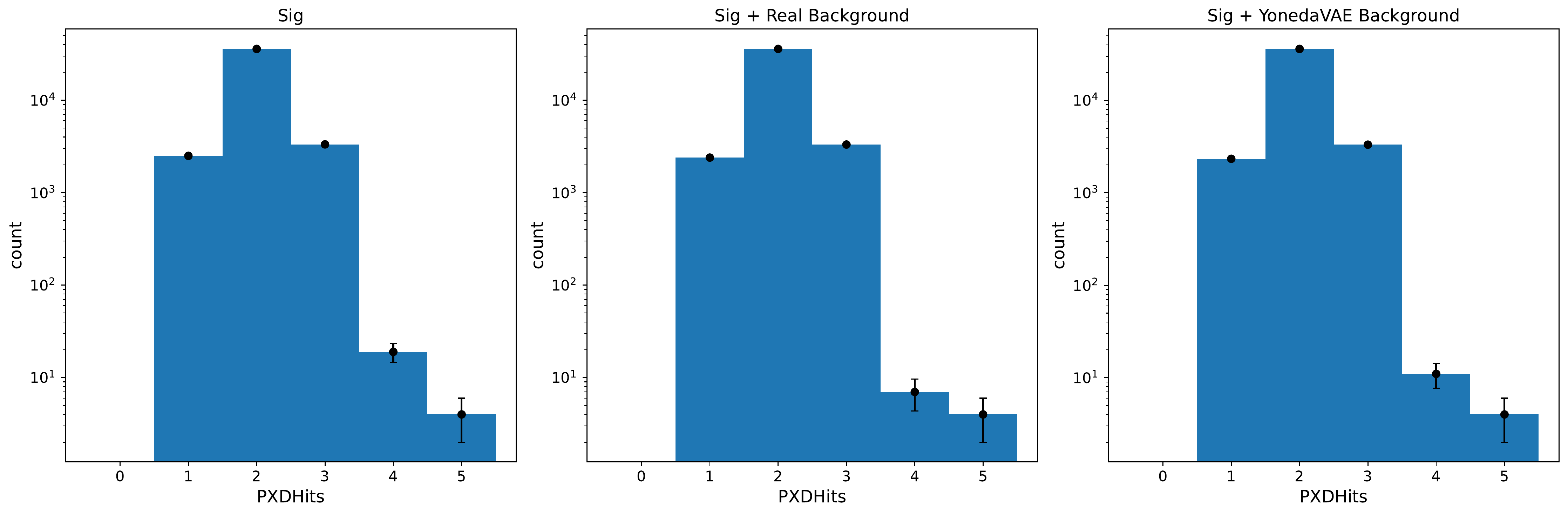}} 
    \caption[Short Caption]{The population plot of each number of PXD hits. Notice the count is in the log scale.}
    \label{fig:num_hits_count}
\end{figure}

One can also compare the population of the number of PXD hits as depicted in~\cref{fig:num_hits_count}. $4$ PXD hits show a slightly lower population as the background is overlaid. However, this could be coming from statistical fluctuations. 

\begin{figure}[!htb]
    \centering
    \includegraphics[width=0.9\linewidth]{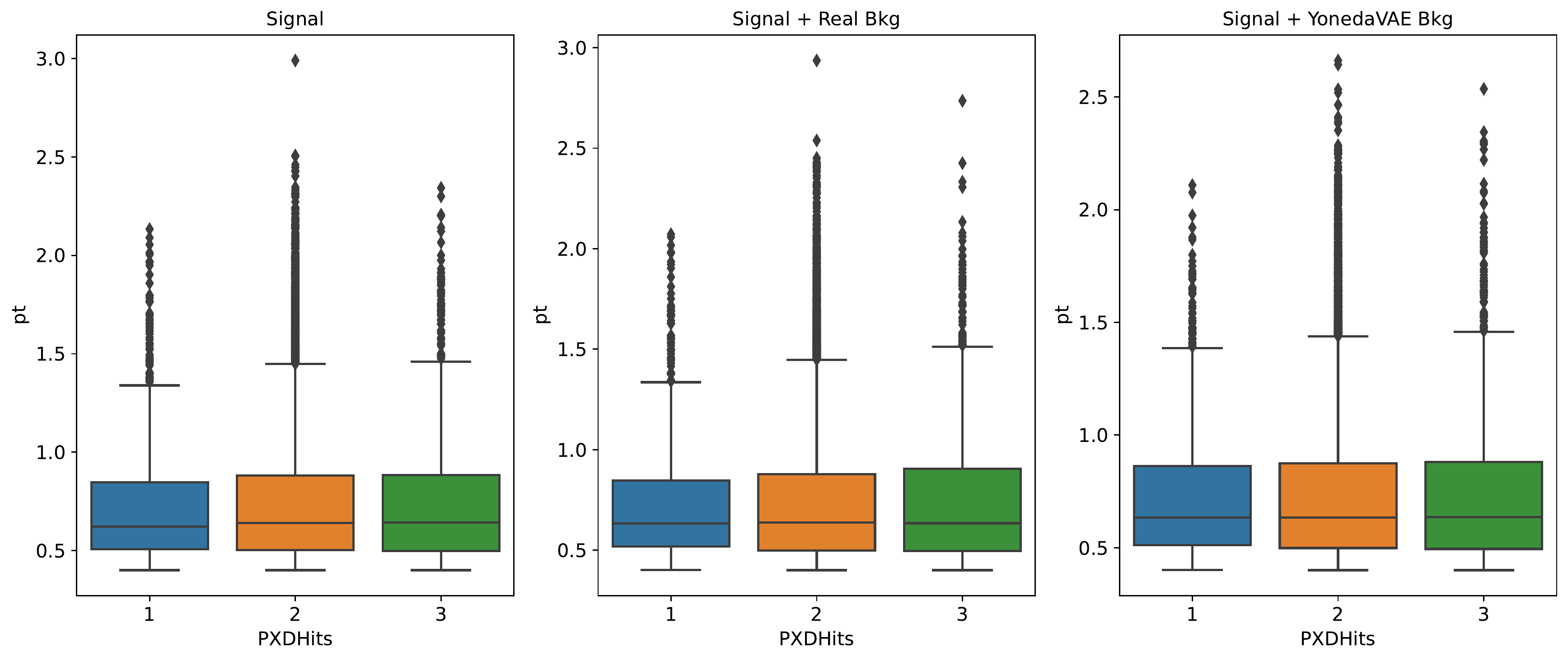}
    \includegraphics[width=0.9\linewidth]{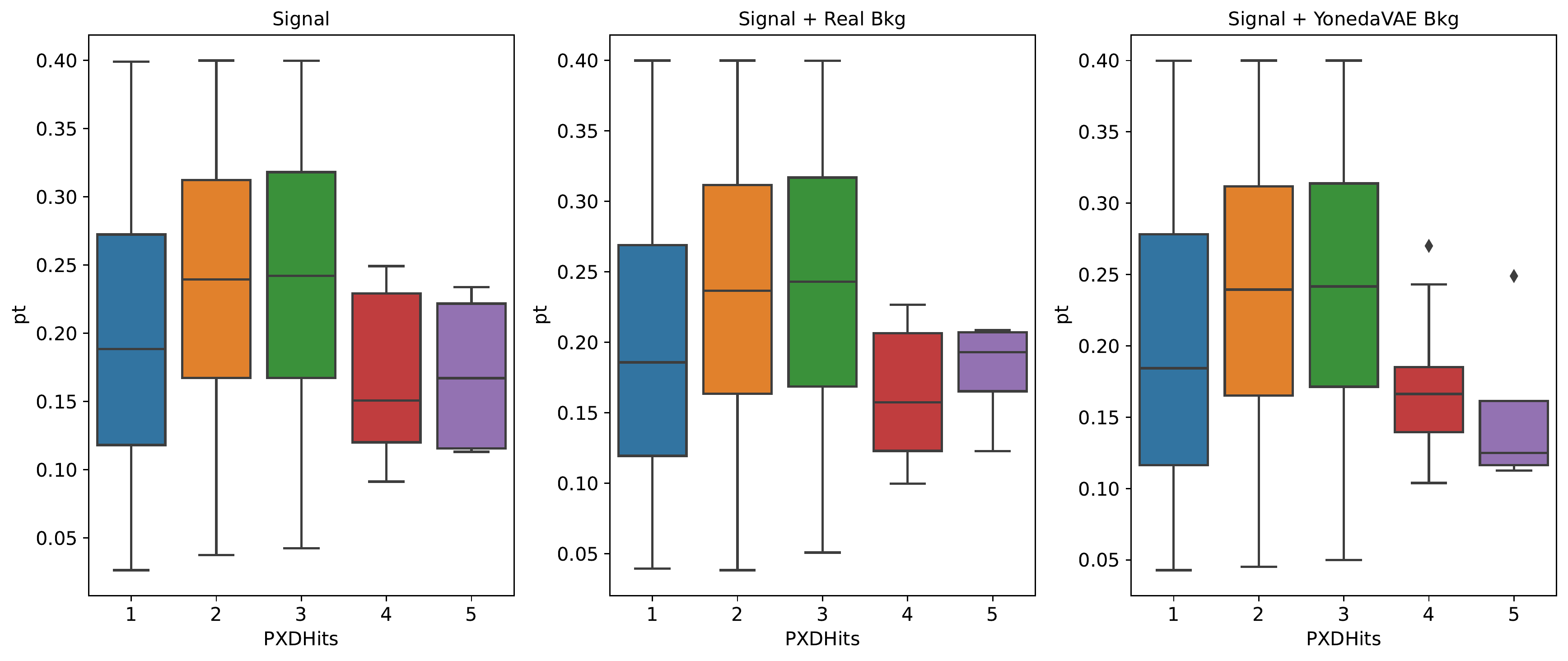}
    \caption{High momentum~(top), and low momentum~(bottom) momentum boxplot per number of PXD hits. The YonedaVAE here is with length extrapolation.}
    \label{fig:pt_pxdhits_le}
\end{figure}

\begin{figure}[!htb]
    \centering
    \includegraphics[width=0.9\linewidth]{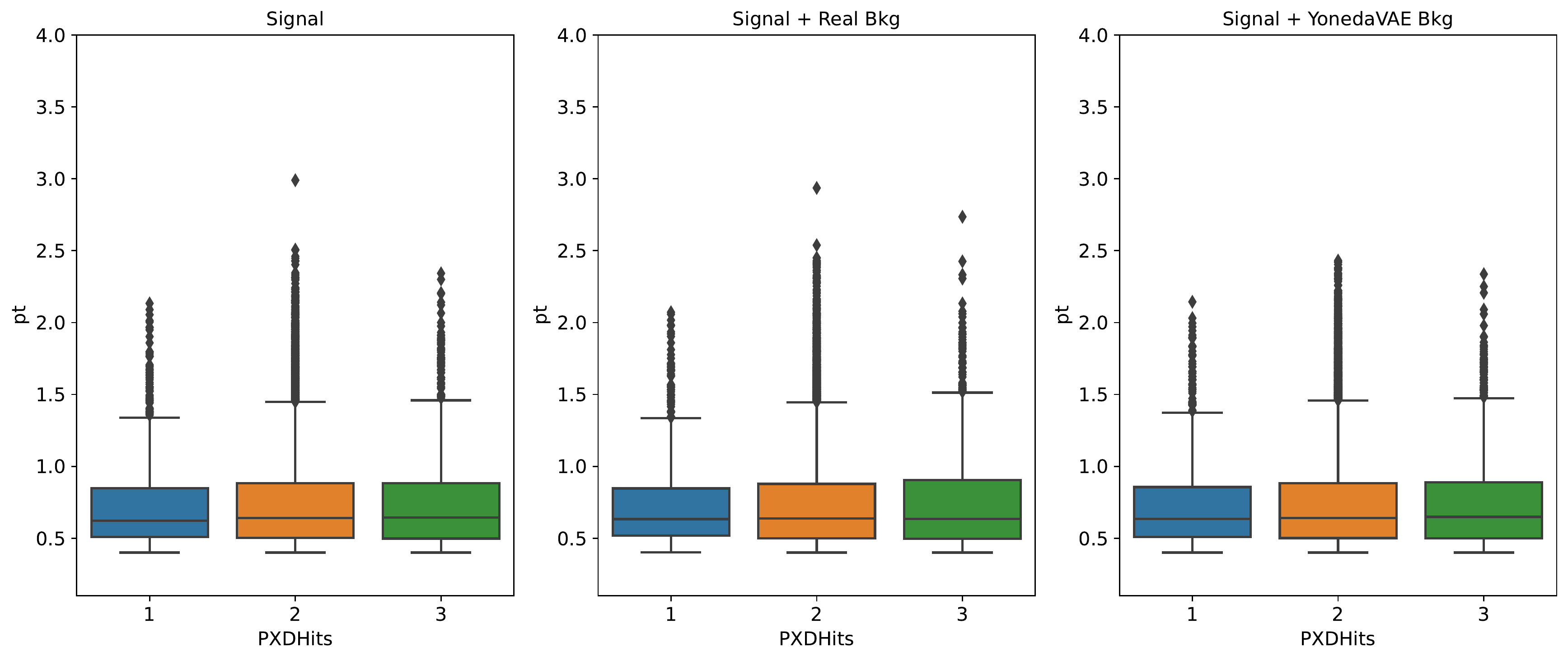}
    \includegraphics[width=0.9\linewidth]{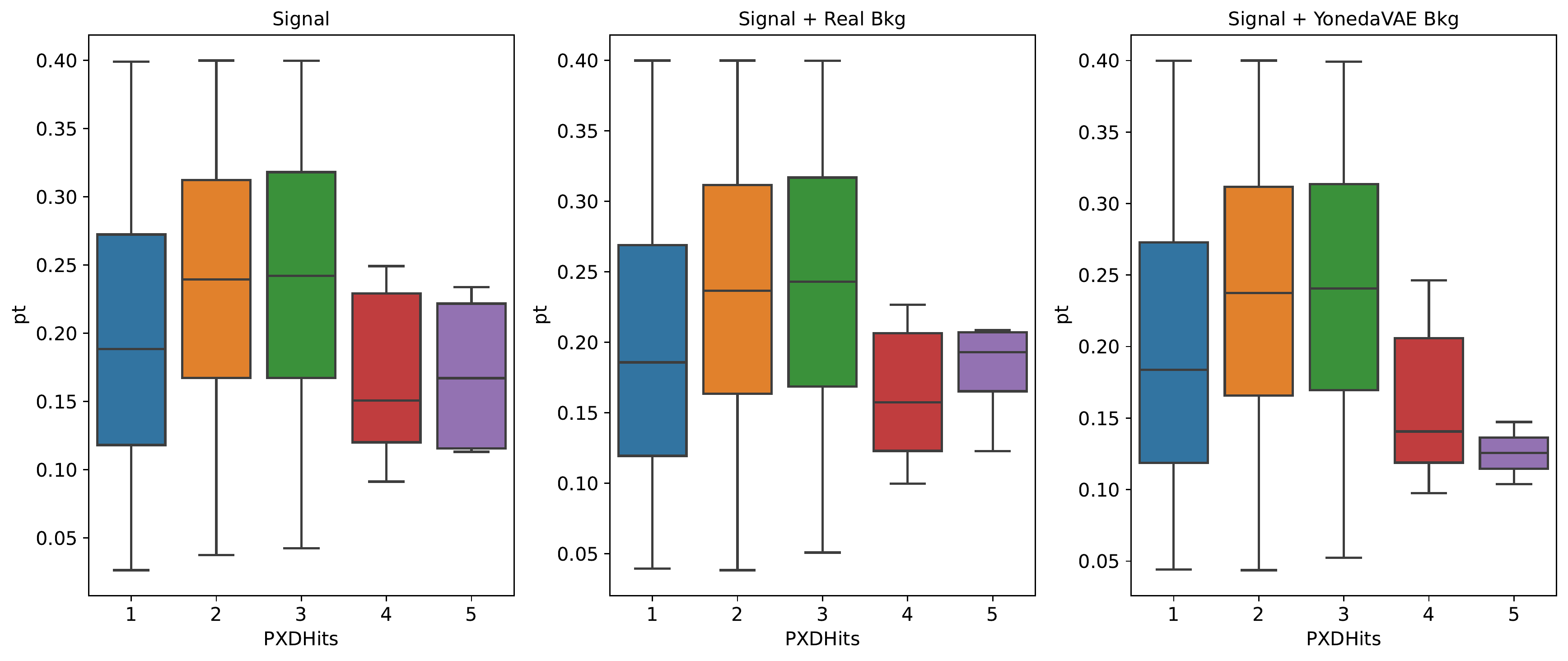}
    \caption{High momentum~(top), and low momentum~(bottom) momentum boxplot per number of PXD hits. The YonedaVAE here is with context extrapolation.}
    \label{fig:pt_pxdhits_ce}
\end{figure}

Dividing the hits to high momentum~$p_t>0.4$ GeV and low momentum~$p_t<0.4$ GeV regions, one can observe the effect of background in a more illuminating way in~\cref{fig:pt_pxdhits_le} and~\cref{fig:pt_pxdhits_ce}.

\begin{figure}[!htb]
    \centering
    \begin{subfigure}{0.45\linewidth}
        \includegraphics[width=\linewidth]{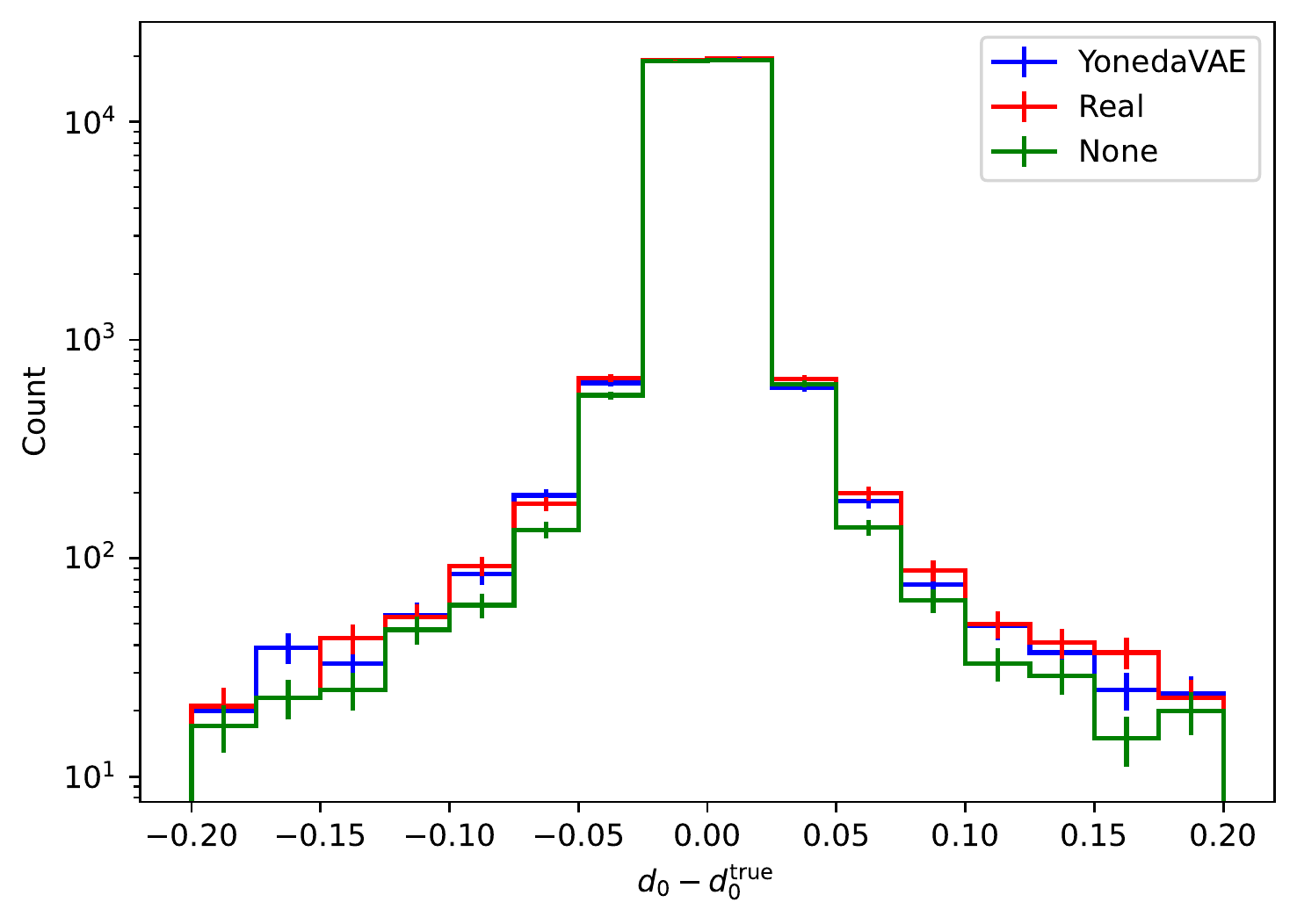}
    \end{subfigure}
    \begin{subfigure}{0.45\linewidth}
        \includegraphics[width=\linewidth]{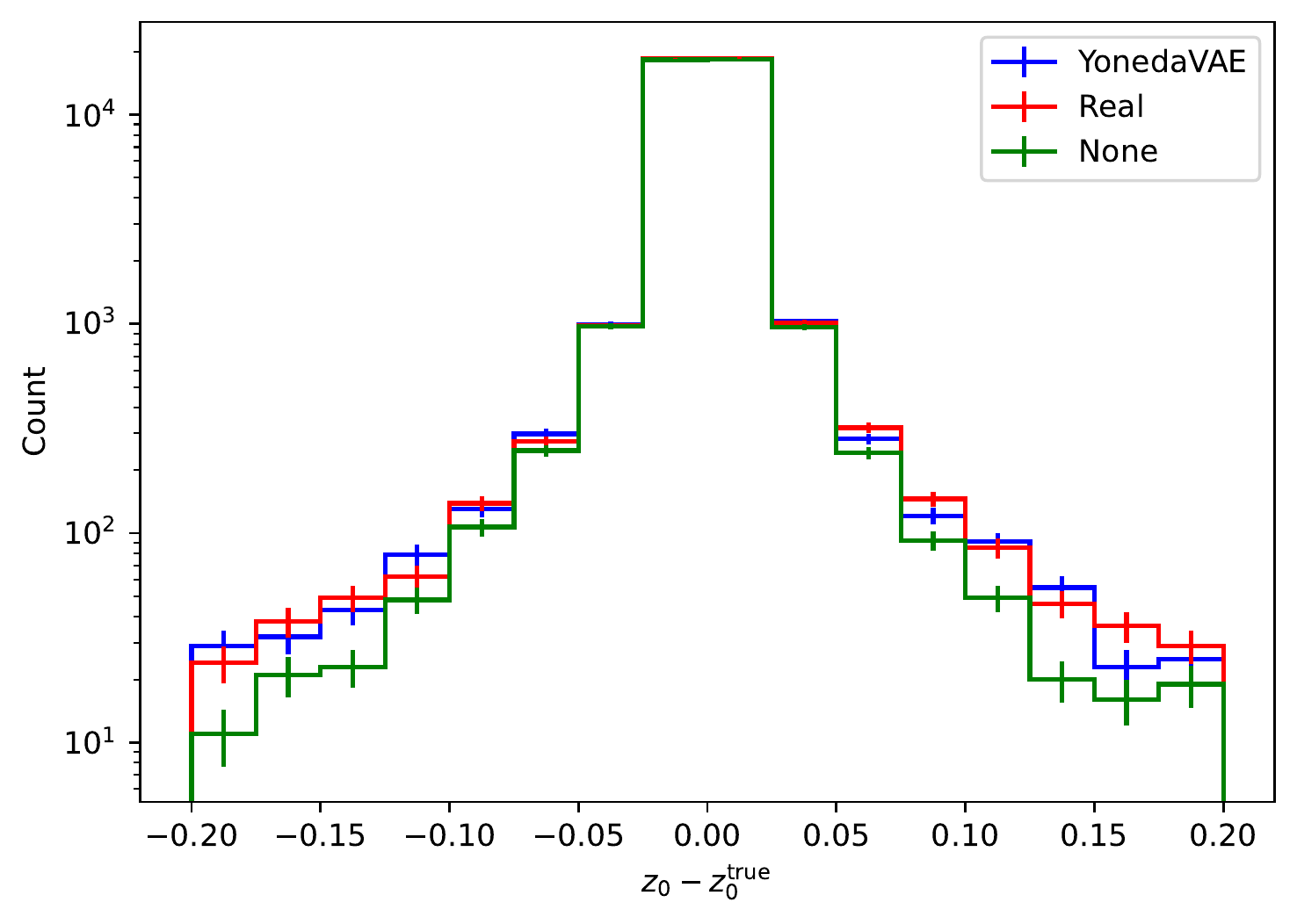}
    \end{subfigure}\\
    
    \begin{subfigure}{0.45\linewidth}
        \includegraphics[width=\linewidth]{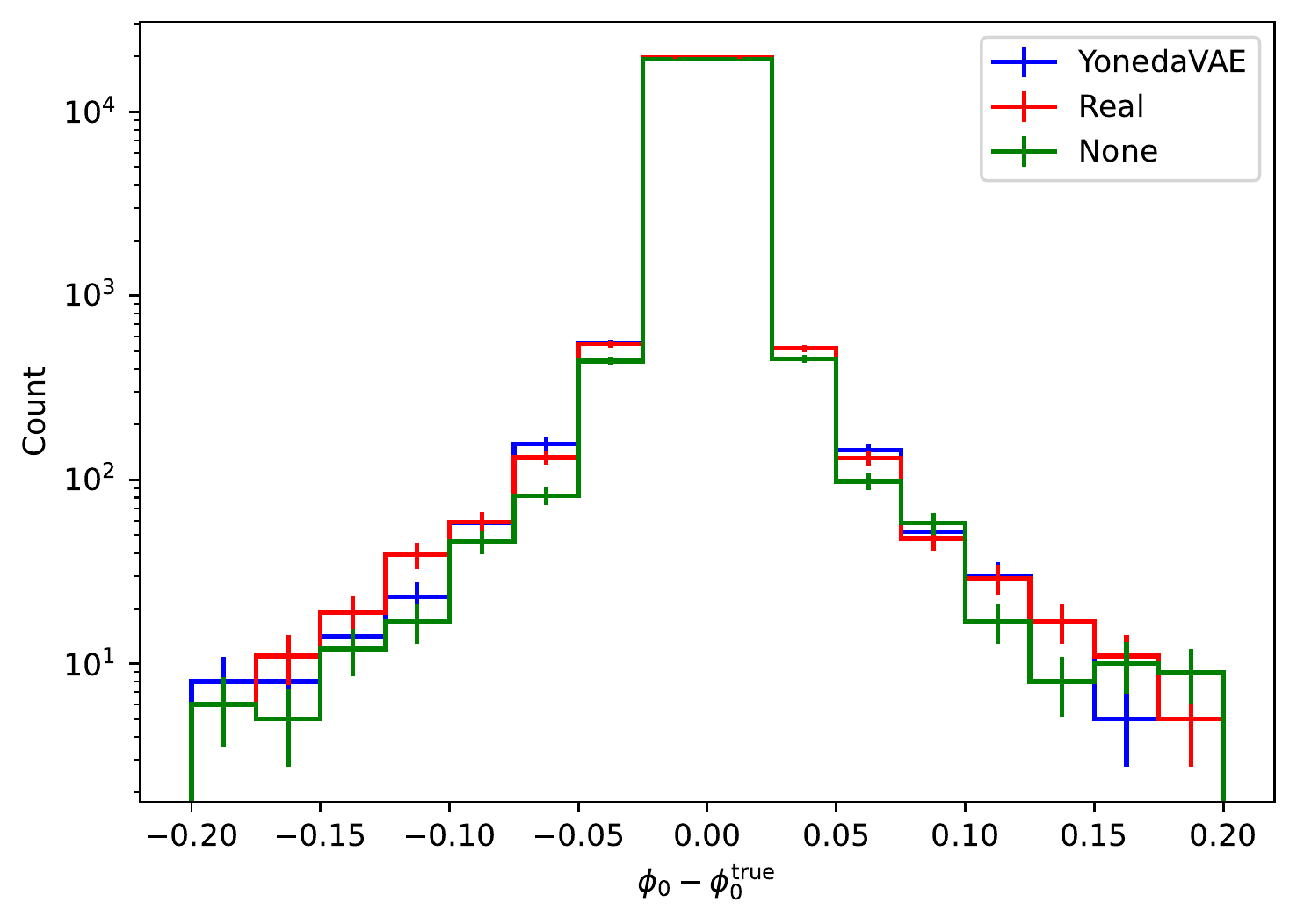}
    \end{subfigure}
    \begin{subfigure}{0.45\linewidth}
        \includegraphics[width=\linewidth]{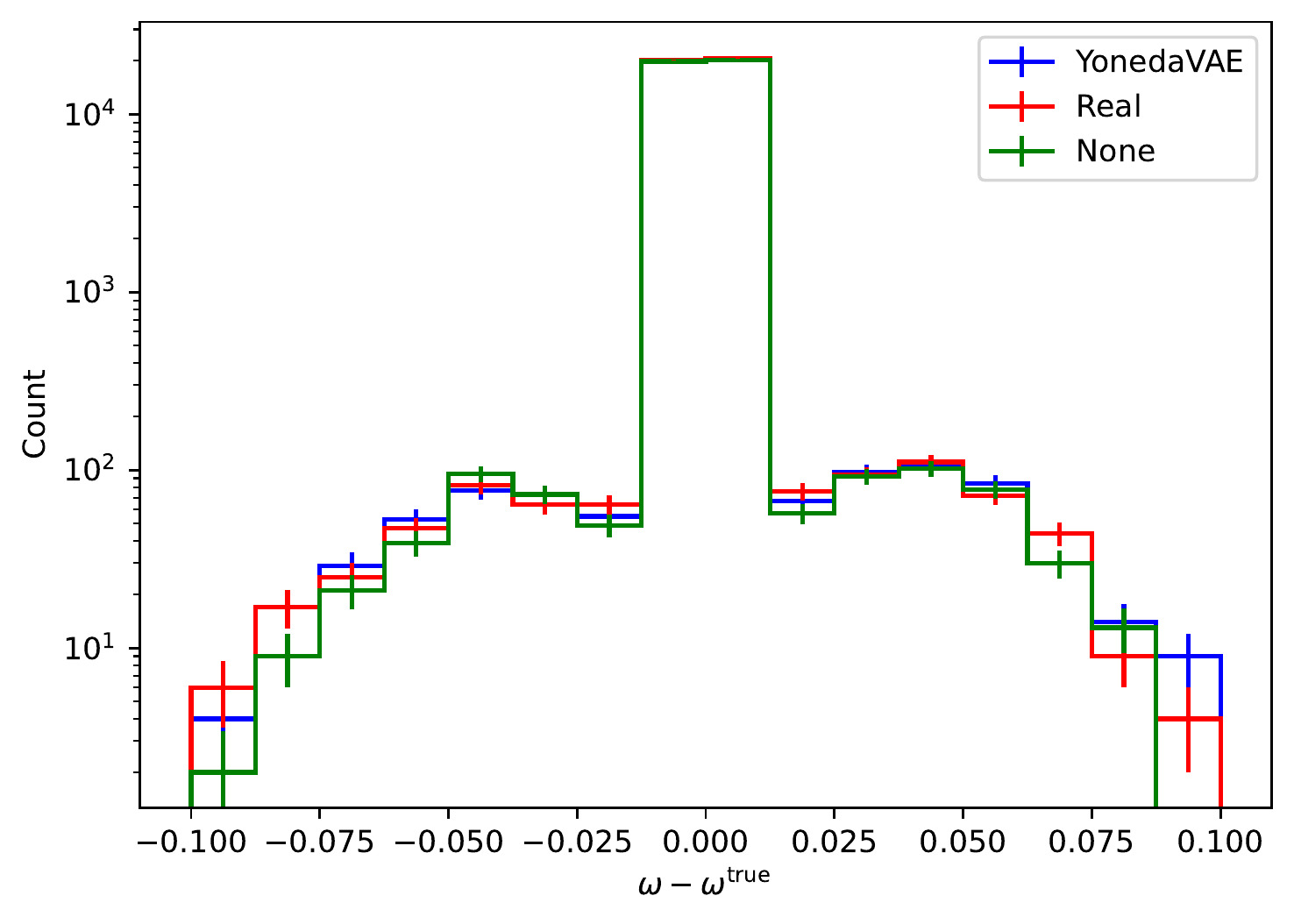}
    \end{subfigure}\\
    
    \begin{subfigure}{0.45\linewidth}
        \includegraphics[width=\linewidth]{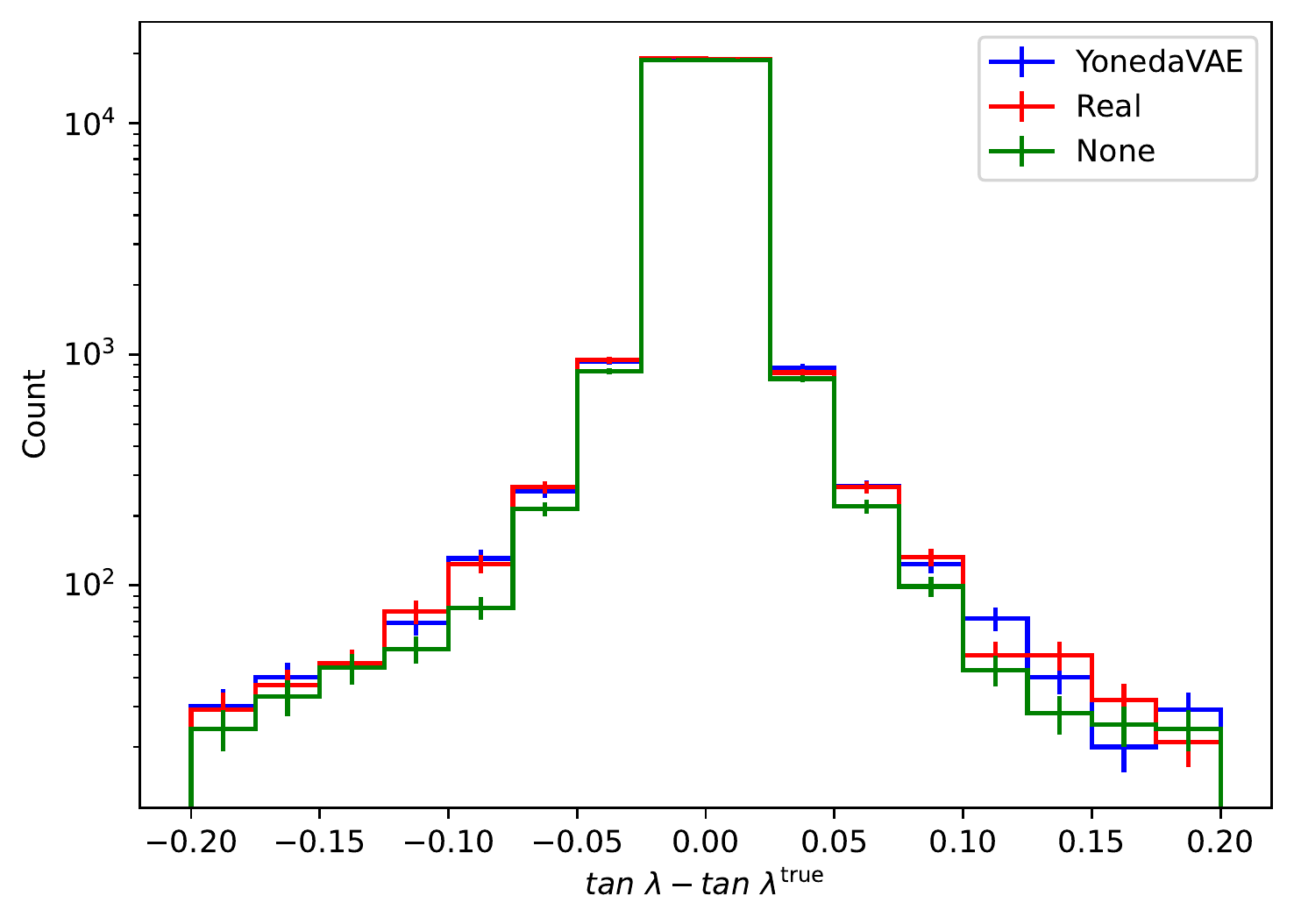}
    \end{subfigure}
    
    \caption{
    Helix parameter resolutions. For each parameter, blue represents YonedaVAE with length extrapolation, red represents Real PXD, and green represents resolution with no background overlay.
    }
    \label{fig:helix_le}
\end{figure}

\begin{figure}[!htb]
    \centering
    \begin{subfigure}{0.45\linewidth}
        \includegraphics[width=\linewidth]{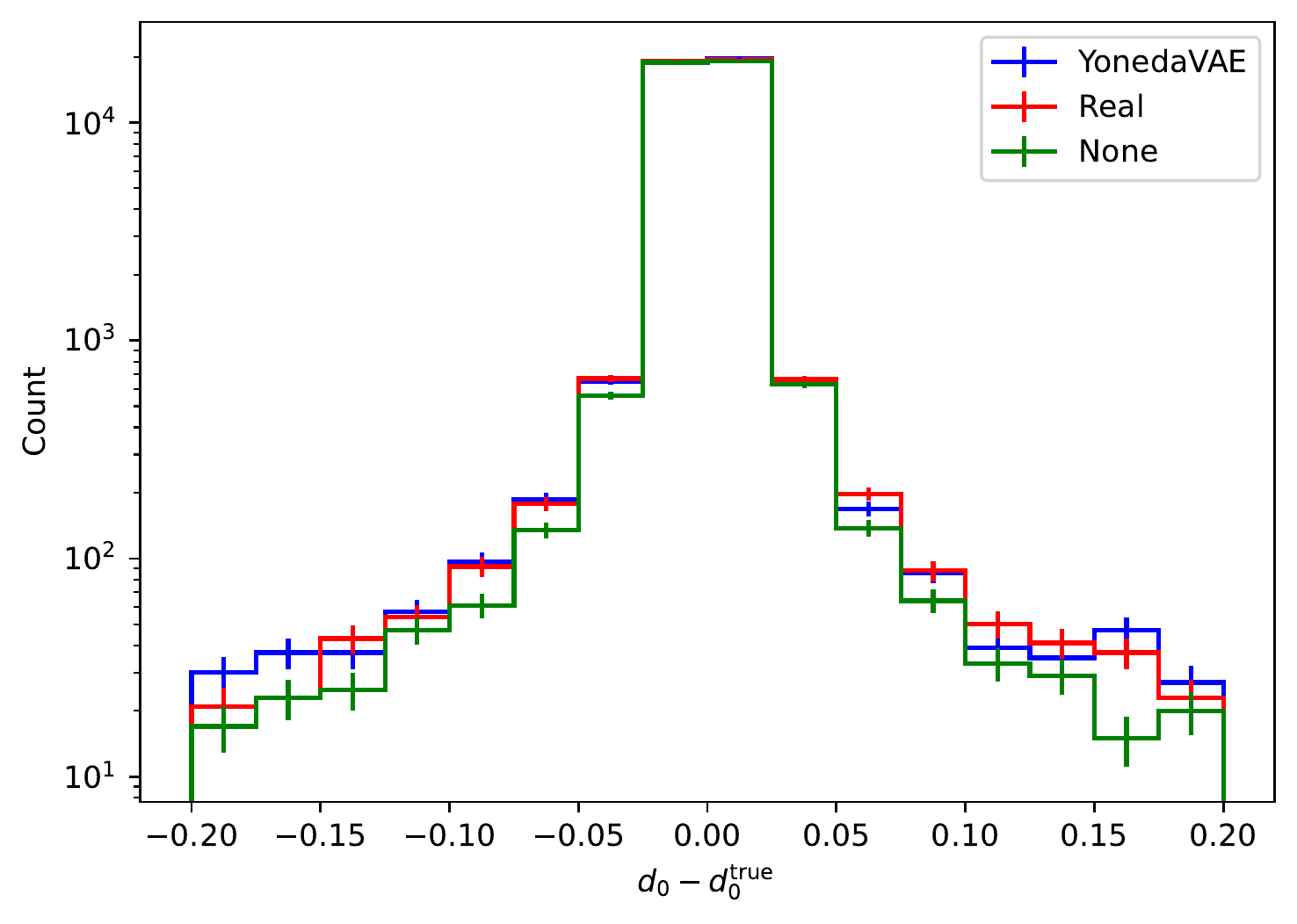}
    \end{subfigure}
    \begin{subfigure}{0.45\linewidth}
        \includegraphics[width=\linewidth]{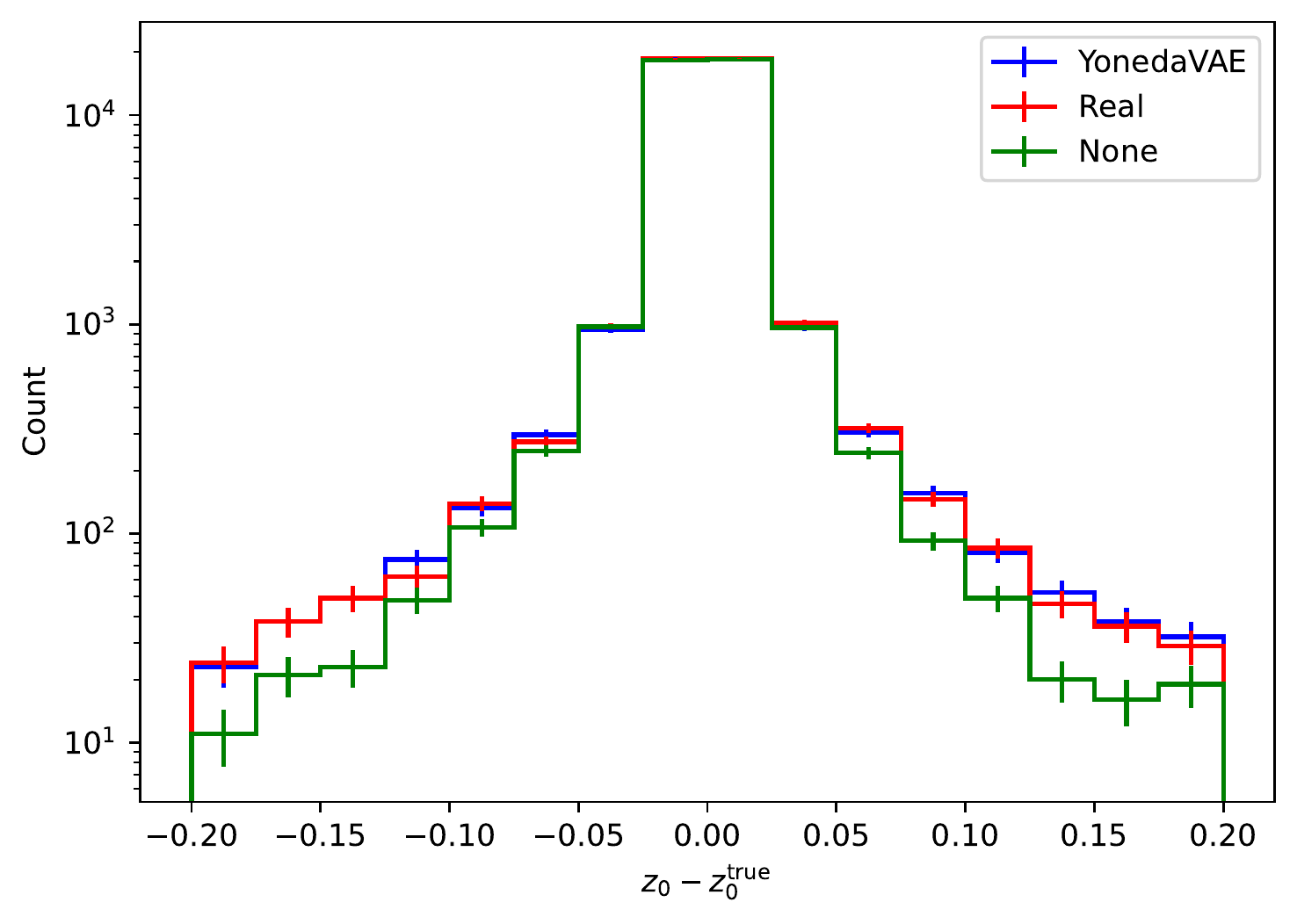}
    \end{subfigure}\\
    
    \begin{subfigure}{0.45\linewidth}
        \includegraphics[width=\linewidth]{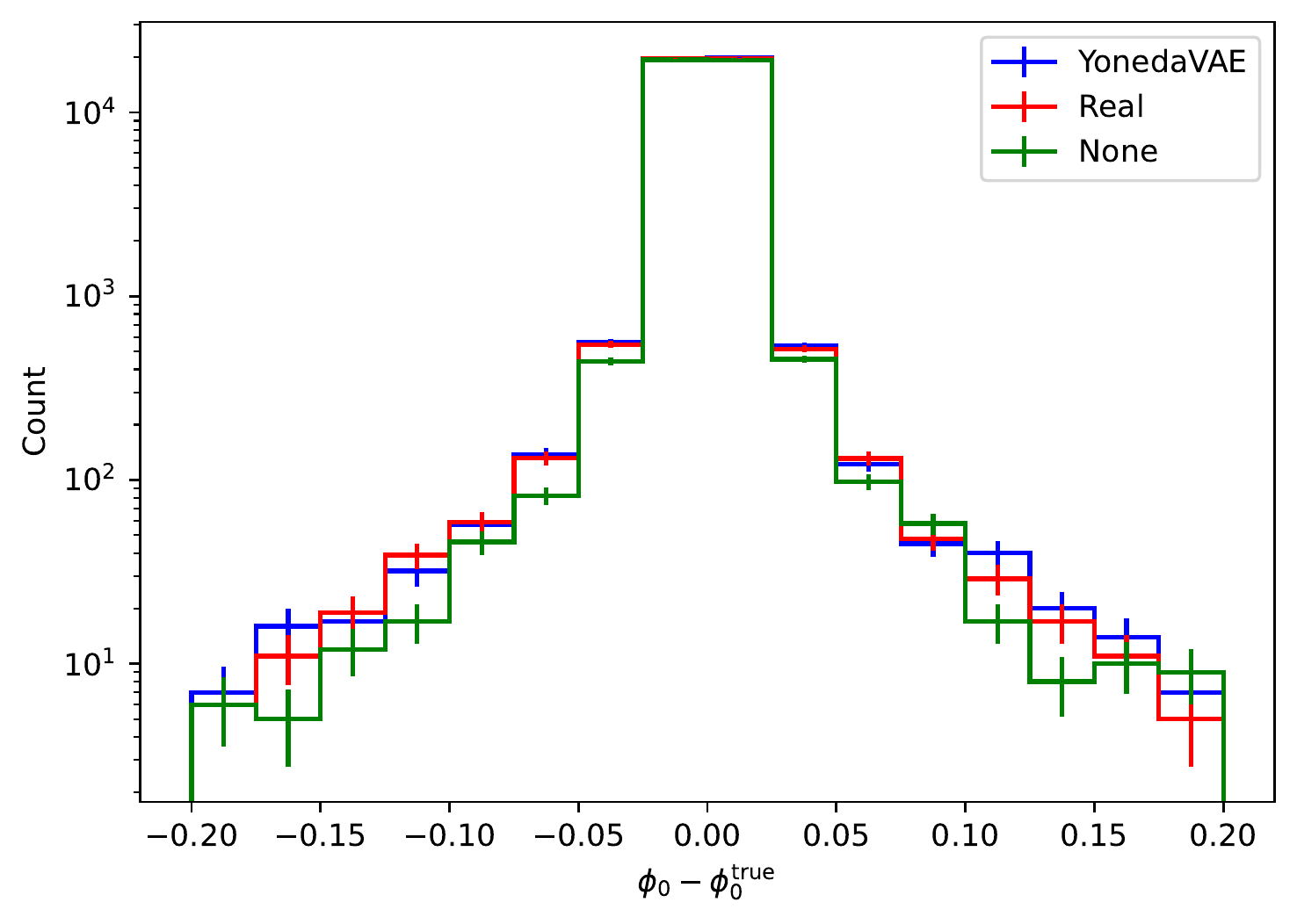}
    \end{subfigure}
    \begin{subfigure}{0.45\linewidth}
        \includegraphics[width=\linewidth]{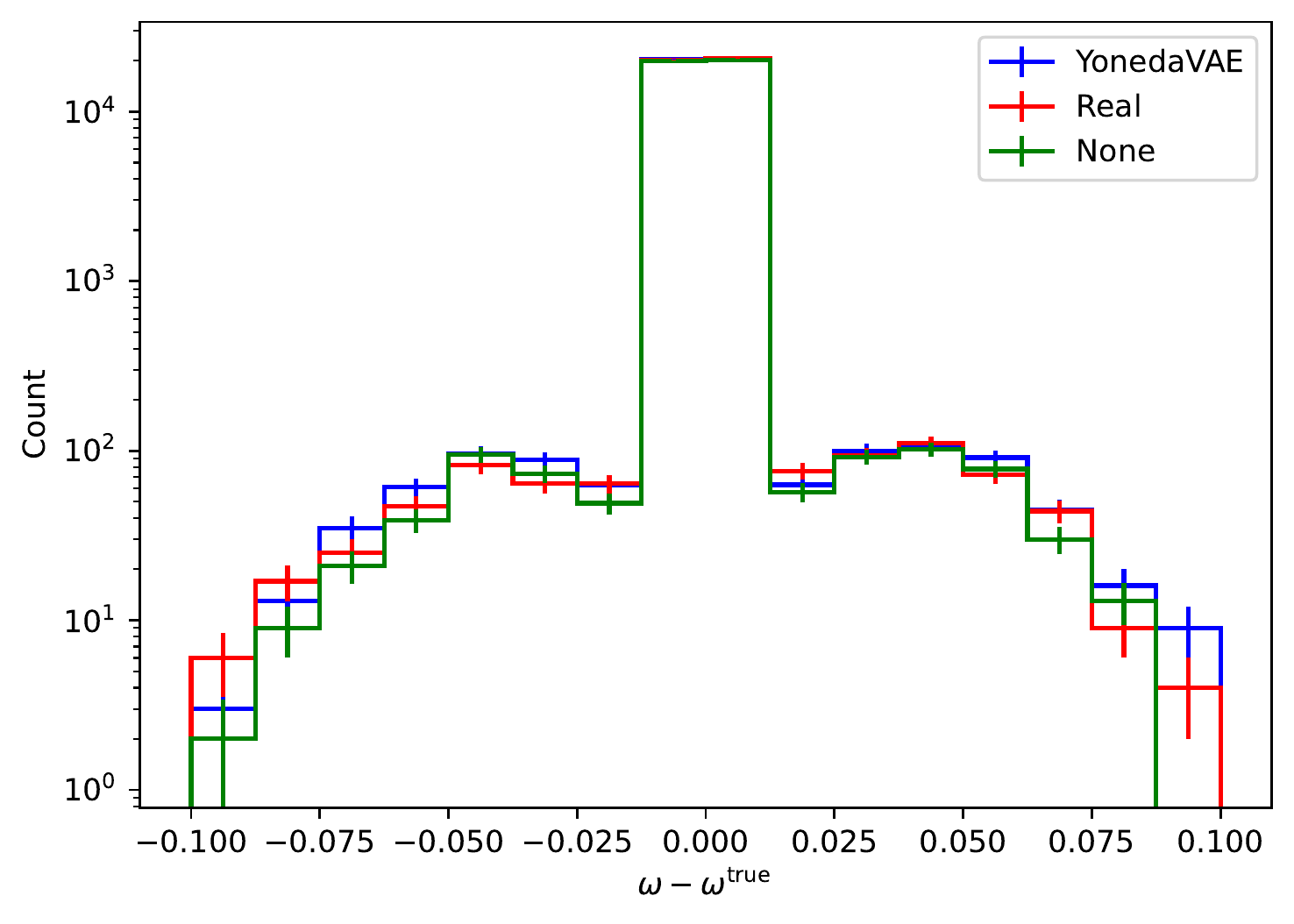}
    \end{subfigure}\\
    
    \begin{subfigure}{0.45\linewidth}
        \includegraphics[width=\linewidth]{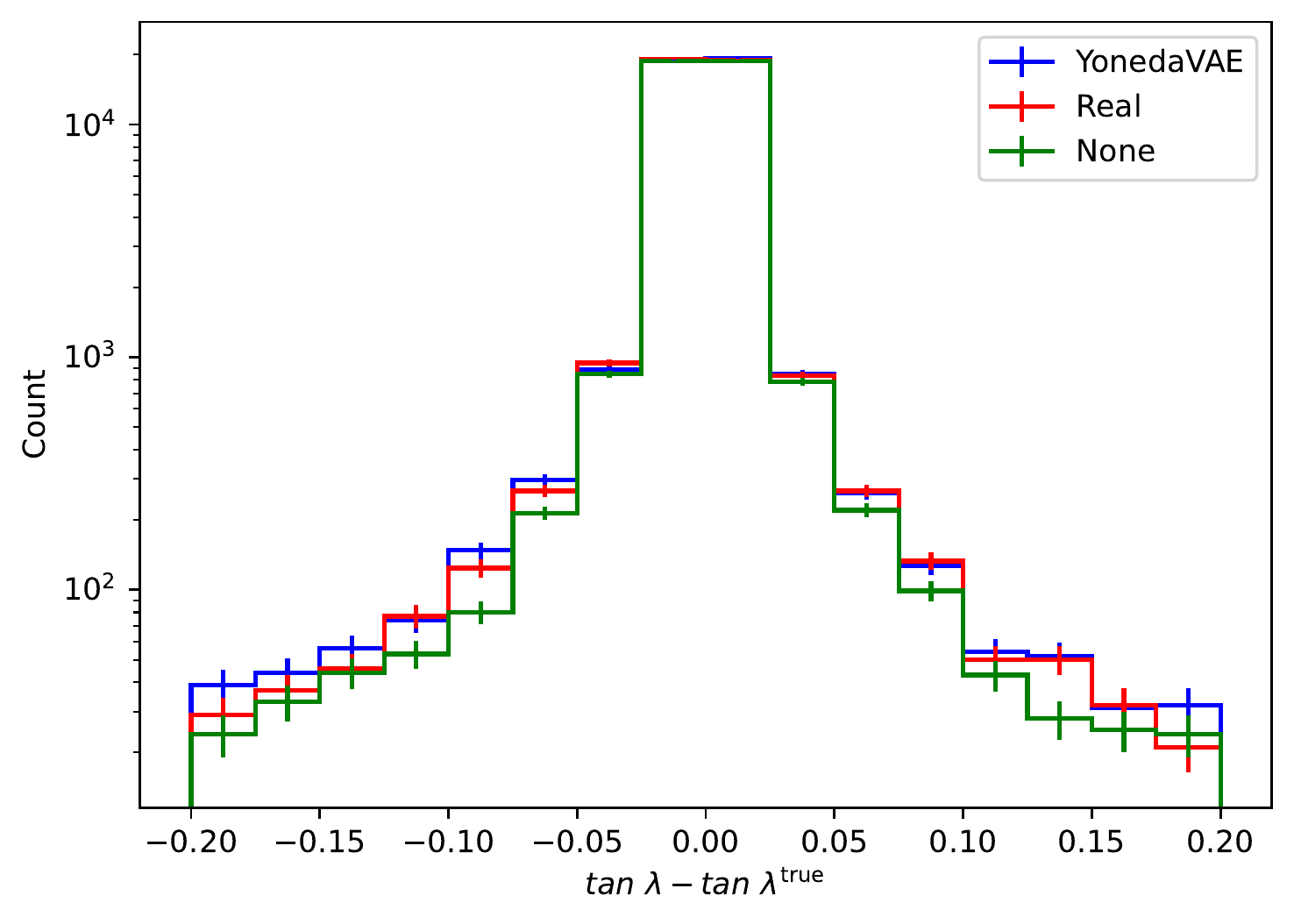}
    \end{subfigure}
    
    \caption{
    Helix parameter resolutions. For each parameter, blue represents YonedaVAE with context extrapolation, red represents Real PXD, and green represents resolution with no background overlay.
    }
    \label{fig:helix_ce}
\end{figure}

In the last chapter, analyzing the effect of the simulated background, the observation was that one of the effects of the background is the increased variance of the resolution. With a real PXD background overlay, this effect still remains true. 
The performance of YonedaVAE\(_{le}\) and YonedaVAE\(_{ce}\) in approximating helix parameter resolutions is particularly noteworthy when evaluated against the overlay of random-trigger real background conditions as depicted in~\cref{fig:helix_le} for YonedaVAE with length extrapolation and ~\cref{fig:helix_ce} for YonedaVAE with context extrapolation. 
As shown in~\cref{table:helix_le} and in~\cref{table:helix_ce}, quantitatively, the unbiased variances for helix parameters, \( \Delta d_0, \Delta \phi_0, \Delta z_0, \Delta \omega, \) and \( \Delta \tan\lambda \), are in close proximity to those obtained with real background data, often within the statistical uncertainties. Additionally, the Kolmogorov-Smirnov statistics yield p-values greater than 0.05, suggesting that the parameter distributions generated by YonedaVAE\(_{le}\) and YonedaVAE\(_{ce}\) are statistically indistinguishable from those generated under real background conditions. This congruence underscores YonedaVAE's robustness in mimicking the complex behaviors introduced by real background interference.

\begin{table}[!htb]
    \centering
    \caption{Helix Parameter Resolution observable Comparison with the length extrapolative YonedaVAE}
    \label{table:helix_le}
    \begin{tabularx}{\textwidth}{|l|l|X|X|X|}
    \hline
    \multirow{2}{*}{Parameter} & \multirow{2}{*}{Observable} & \multicolumn{3}{c|}{Signal $+$ Background} \\
    \cline{3-5}
     & & YonedaVAE$_{le}$ & Real & No Bkg. \\
    \hline
    \multirow{4}{*}{$\Delta d_0$} & Unbiased Variance & $0.1713 \pm 0.0006$ & $0.1643 \pm 0.0006$ & $0.1594 \pm 0.0006$ \\
    & KS Statistic & \multicolumn{2}{c|}{0.0058, p-value: 0.4910} & \\
    & Num. Tracks $<-5\sigma_{\text{sig.}}$ & 138 & 146 & 106 \\
    & Num. Tracks $>5\sigma_{\text{sig.}}$ & 148 & 136 & 112 \\
    \hline
    \multirow{4}{*}{$\Delta \phi_0$} & Unbiased Variance & $0.4444 \pm 0.0015$ & $0.4480 \pm 0.0016$ & $0.4429 \pm 0.0016$ \\
    & KS Statistic & \multicolumn{2}{c|}{0.0035, p-value: 0.9591} & \\
    & Num. Tracks $<-5\sigma_{\text{sig.}}$ & 339 & 317 & 311 \\
    & Num. Tracks $>5\sigma_{\text{sig.}}$ & 362 & 389 & 357 \\
    \hline
    \multirow{4}{*}{$\Delta z_0$} & Unbiased Variance & $5.9991 \pm 0.0209$ & $5.8735 \pm 0.0204$ & $5.7120 \pm 0.0201$ \\
    & KS Statistic & \multicolumn{2}{c|}{0.0040, p-value: 0.8993} & \\
    & Num. Tracks $<-5\sigma_{\text{sig.}}$ & 64 & 71 & 62 \\
    & Num. Tracks $>5\sigma_{\text{sig.}}$ & 273 & 261 & 238 \\
    \hline
    \multirow{4}{*}{$\Delta \omega$} & Unbiased Variance & $0.0066 \pm 0.0001$ & $0.0065 \pm 0.0001$ & $0.0061 \pm 0.0001$ \\
    & KS Statistic & \multicolumn{2}{c|}{0.0052, p-value: 0.6296} & \\
    & Num. Tracks $<-5\sigma_{\text{sig.}}$ & 226 & 226 & 225 \\
    & Num. Tracks $>5\sigma_{\text{sig.}}$ & 333 & 321 & 299 \\
    \hline
    \multirow{4}{*}{$\Delta \tan\lambda$} & Unbiased Variance & $0.0753 \pm 0.0003$ & $0.0726 \pm 0.0003$ & $0.0734 \pm 0.0003$ \\
    & KS Statistic & \multicolumn{2}{c|}{0.0044, p-value: 0.8238} & \\
    & Num. Tracks $<-5\sigma_{\text{sig.}}$ & 248 & 250 & 226 \\
    & Num. Tracks $>5\sigma_{\text{sig.}}$ & 66 & 69 & 59 \\
    \hline
    \end{tabularx}
\end{table}

\begin{table}[!htb]
    \centering
    \caption{Helix Parameter Resolution Observable Comparison with the context extrapolative YonedaVAE}
    \label{table:helix_ce}
    \begin{tabularx}{\textwidth}{|l|l|X|X|X|}
    \hline
    \multirow{2}{*}{Parameter} & \multirow{2}{*}{Observable} & \multicolumn{3}{c|}{Signal $+$ Background} \\
    \cline{3-5}
     & & YonedaVAE$_{ce}$ & Real & No Bkg. \\
    \hline
    \multirow{4}{*}{$\Delta d_0$} & Unbiased Variance & $0.1721 \pm 0.0006$ & $0.1643 \pm 0.0006$ & $0.1594 \pm 0.0006$ \\
    & KS Statistic & \multicolumn{2}{c|}{0.0066, p-value: 0.3221} & \\
    & Num. Tracks $<-5\sigma_{\text{sig.}}$ & 122 & 146 & 106 \\
    & Num. Tracks $>5\sigma_{\text{sig.}}$ & 143 & 136 & 112 \\
    \hline
    \multirow{4}{*}{$\Delta \phi_0$} & Unbiased Variance & $0.4825 \pm 0.0017$ & $0.4480 \pm 0.0016$ & $0.4429 \pm 0.0016$ \\
    & KS Statistic & \multicolumn{2}{c|}{0.0066, p-value: 0.3178} & \\
    & Num. Tracks $<-5\sigma_{\text{sig.}}$ & 400 & 317 & 311 \\
    & Num. Tracks $>5\sigma_{\text{sig.}}$ & 387 & 389 & 357 \\
    \hline
    \multirow{4}{*}{$\Delta z_0$} & Unbiased Variance & $5.9736 \pm 0.0207$ & $5.8735 \pm 0.0204$ & $5.7120 \pm 0.0201$ \\
    & KS Statistic & \multicolumn{2}{c|}{0.0066, p-value: 0.3309} & \\
    & Num. Tracks $<-5\sigma_{\text{sig.}}$ & 77 & 71 & 62 \\
    & Num. Tracks $>5\sigma_{\text{sig.}}$ & 275 & 261 & 238 \\
    \hline
    \multirow{4}{*}{$\Delta \omega$} & Unbiased Variance & $0.0068 \pm 0.0001$ & $0.0065 \pm 0.0001$ & $0.0061 \pm 0.0001$ \\
    & KS Statistic & \multicolumn{2}{c|}{0.0063, p-value: 0.3833} & \\
    & Num. Tracks $<-5\sigma_{\text{sig.}}$ & 281 & 226 & 225 \\
    & Num. Tracks $>5\sigma_{\text{sig.}}$ & 357 & 321 & 299 \\
    \hline
    \multirow{4}{*}{$\Delta \tan\lambda$} & Unbiased Variance & $0.0790 \pm 0.0003$ & $0.0726 \pm 0.0003$ & $0.0734 \pm 0.0003$ \\
    & KS Statistic & \multicolumn{2}{c|}{0.0069, p-value: 0.2803} & \\
    & Num. Tracks $<-5\sigma_{\text{sig.}}$ & 276 & 250 & 226 \\
    & Num. Tracks $>5\sigma_{\text{sig.}}$ & 77 & 69 & 59 \\
    \hline
    \end{tabularx}
\end{table}

\FloatBarrier
\section{Ablation Studies}
\begin{figure}[!htb]
  \begin{minipage}{0.48\textwidth}
    \includegraphics[width=\linewidth]{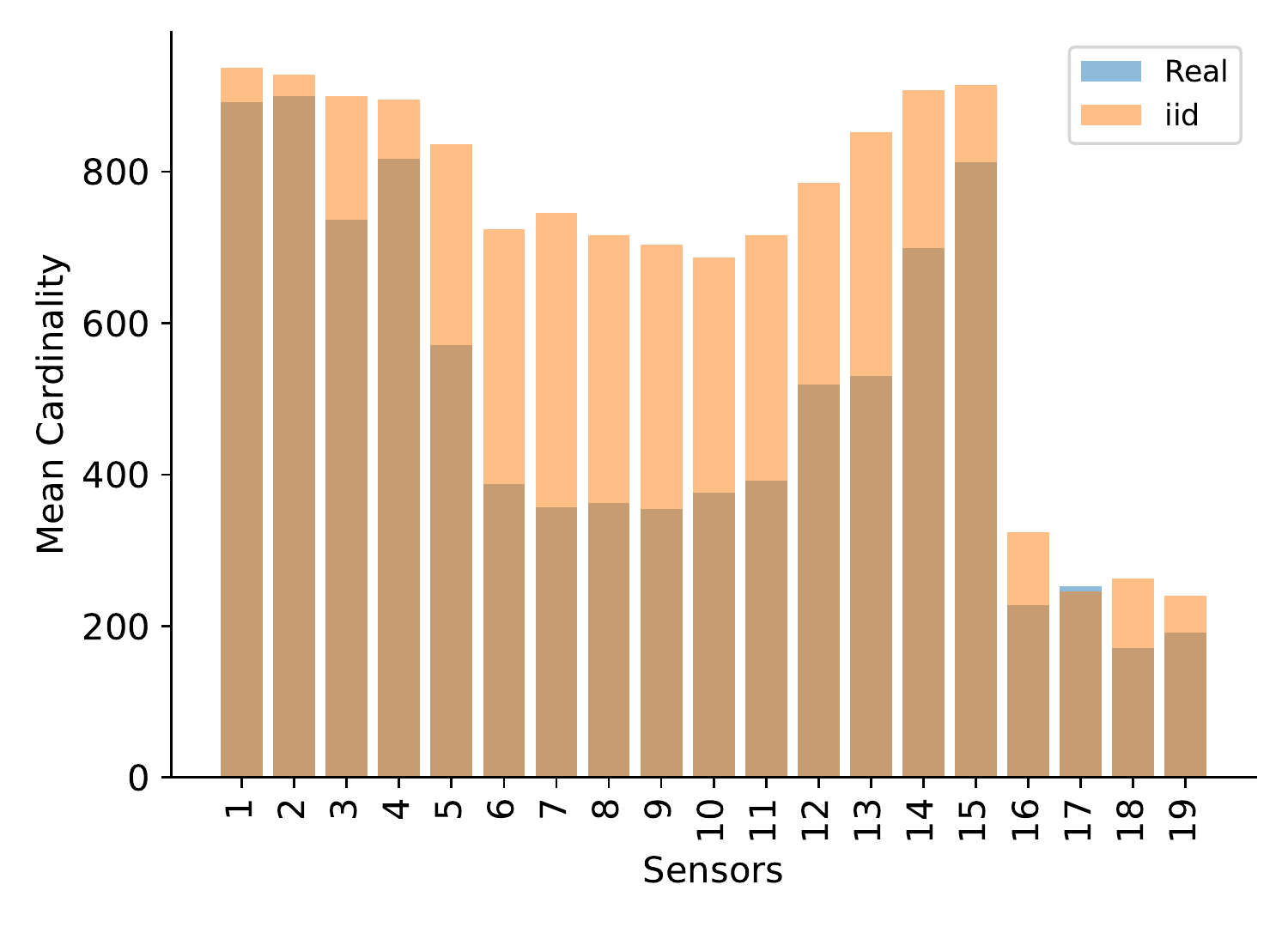}
  \end{minipage}\hfill 
  \begin{minipage}{0.48\textwidth}
    \includegraphics[width=\linewidth]{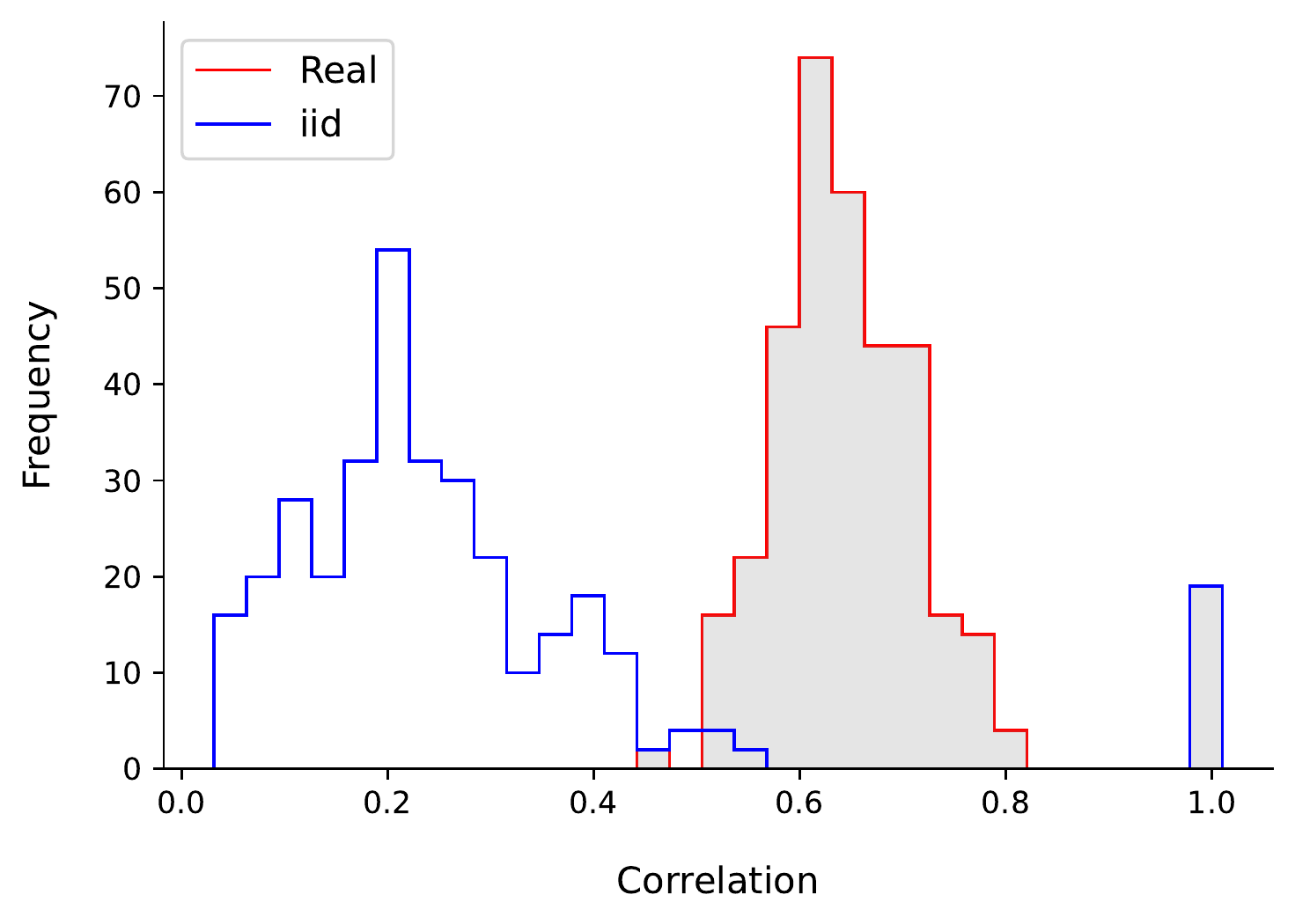}
  \end{minipage}
   \caption{Mean cardinality~(left) and mean cardinality correlation~(right), for YonedaVAE without the Adaptive Top-q sampling, and only with i.i.d sampling.}
   \label{fig:iid}
\end{figure}

\begin{figure}[!htb]
  \begin{minipage}{0.48\textwidth}
    \includegraphics[width=\linewidth]{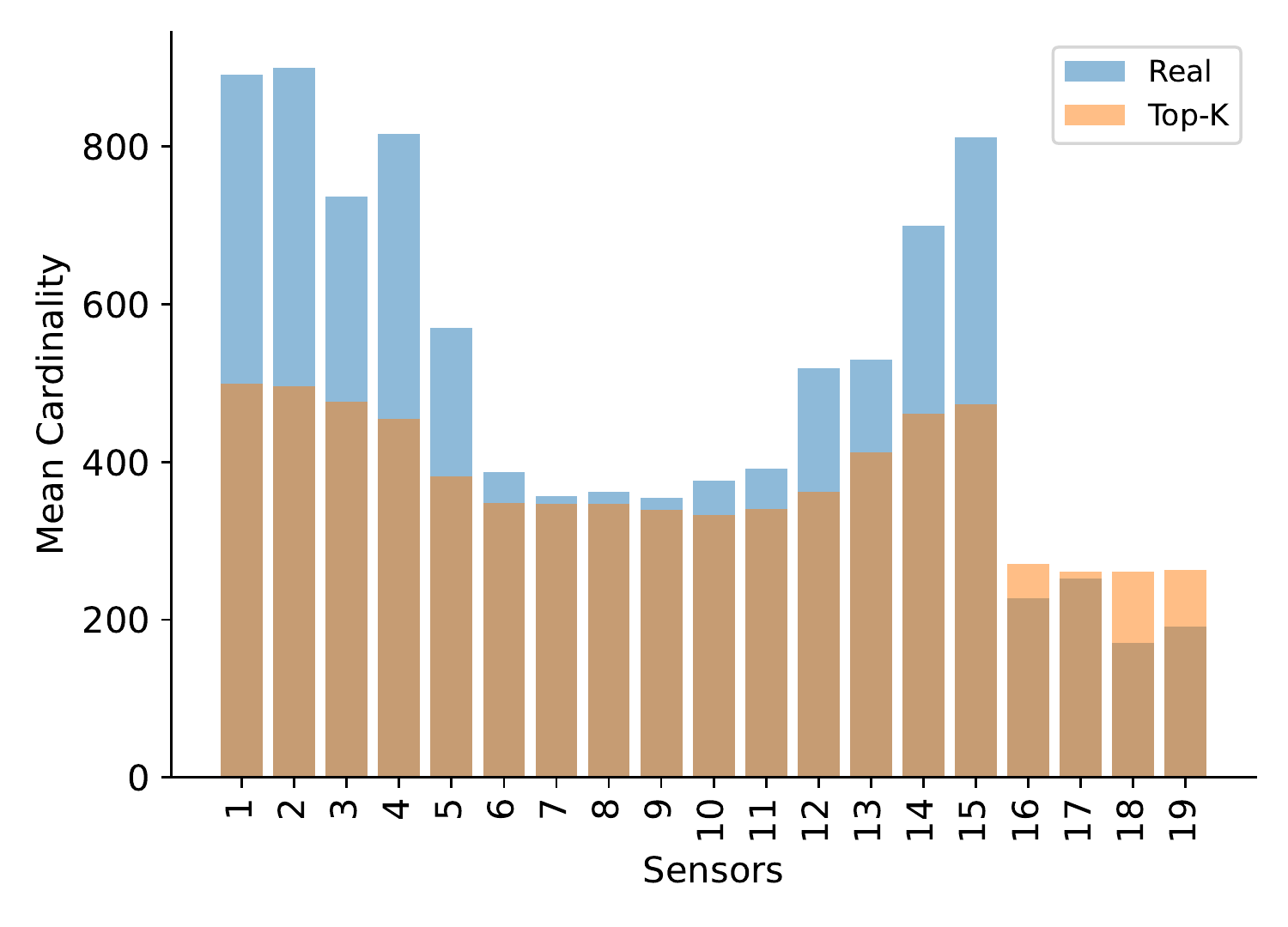}
  \end{minipage}\hfill 
  \begin{minipage}{0.48\textwidth}
    \includegraphics[width=\linewidth]{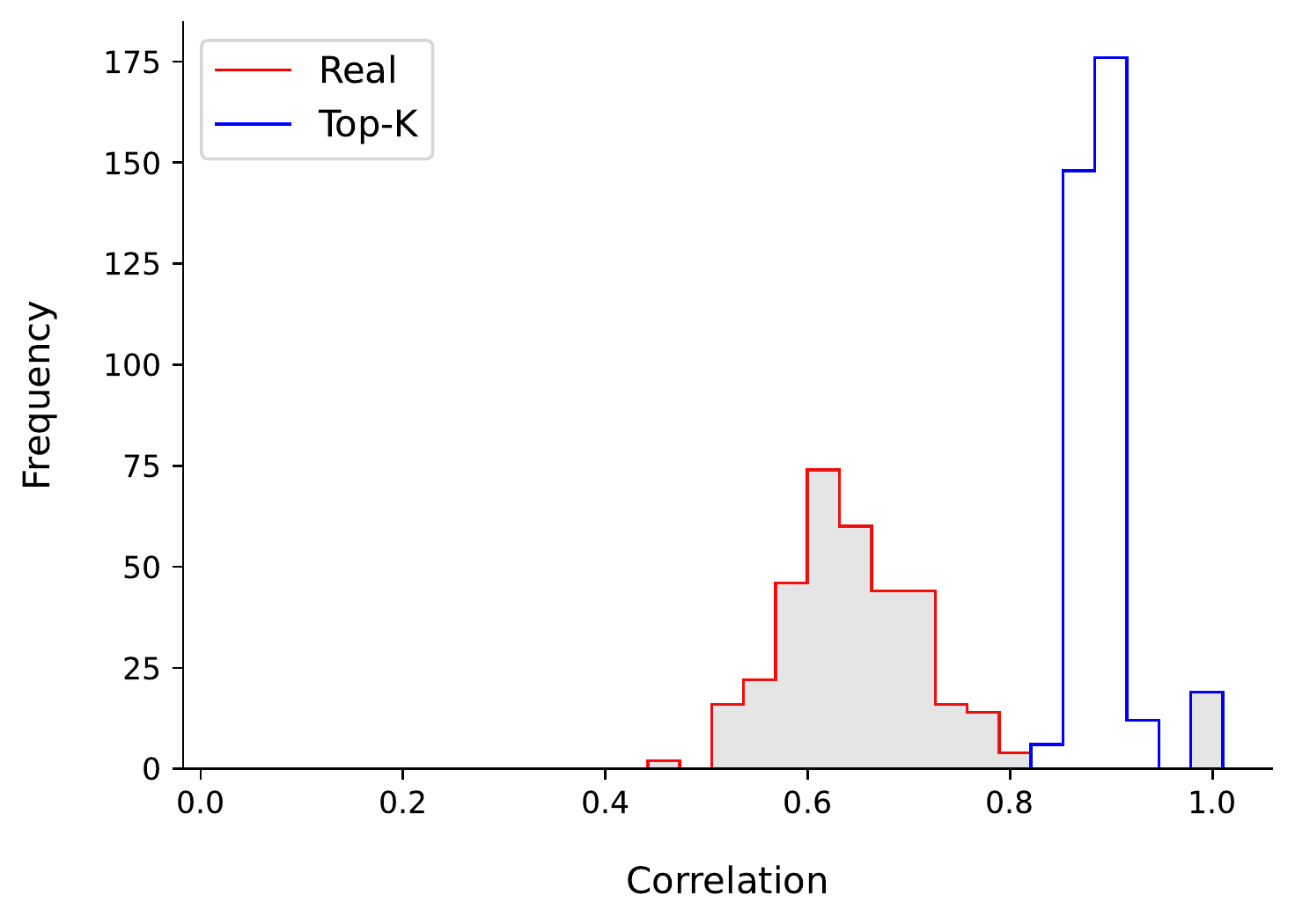}
  \end{minipage}
   \caption{Mean cardinality~(left) and mean cardinality correlation~(right), for YonedaVAE without the Adaptive Top-q sampling, and only with Top-k sampling.}
   \label{fig:topk}
\end{figure}

\begin{figure}[!htb]
  \begin{minipage}{0.48\textwidth}
    \includegraphics[width=\linewidth]{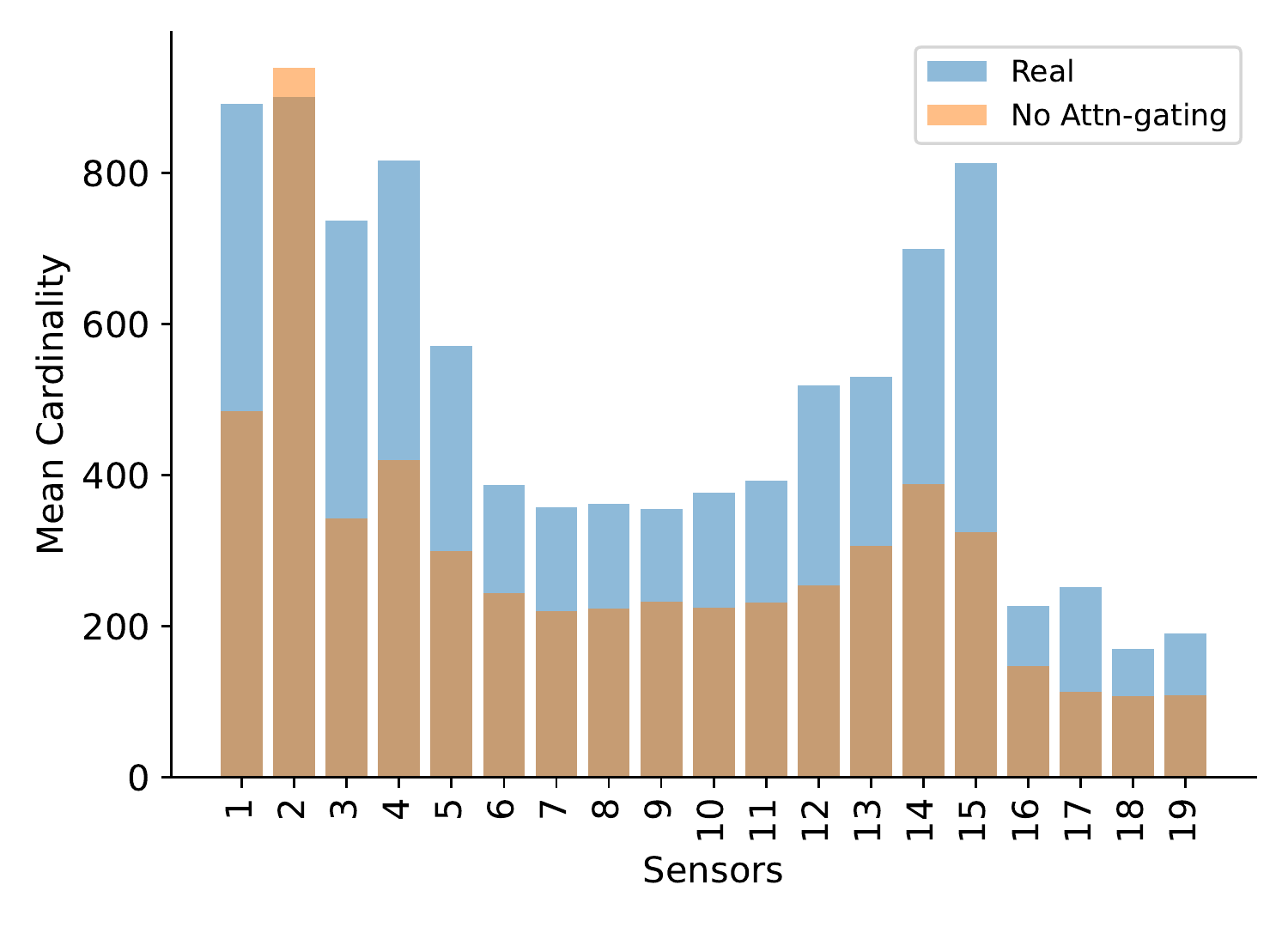}
  \end{minipage}\hfill 
  \begin{minipage}{0.48\textwidth}
    \includegraphics[width=\linewidth]{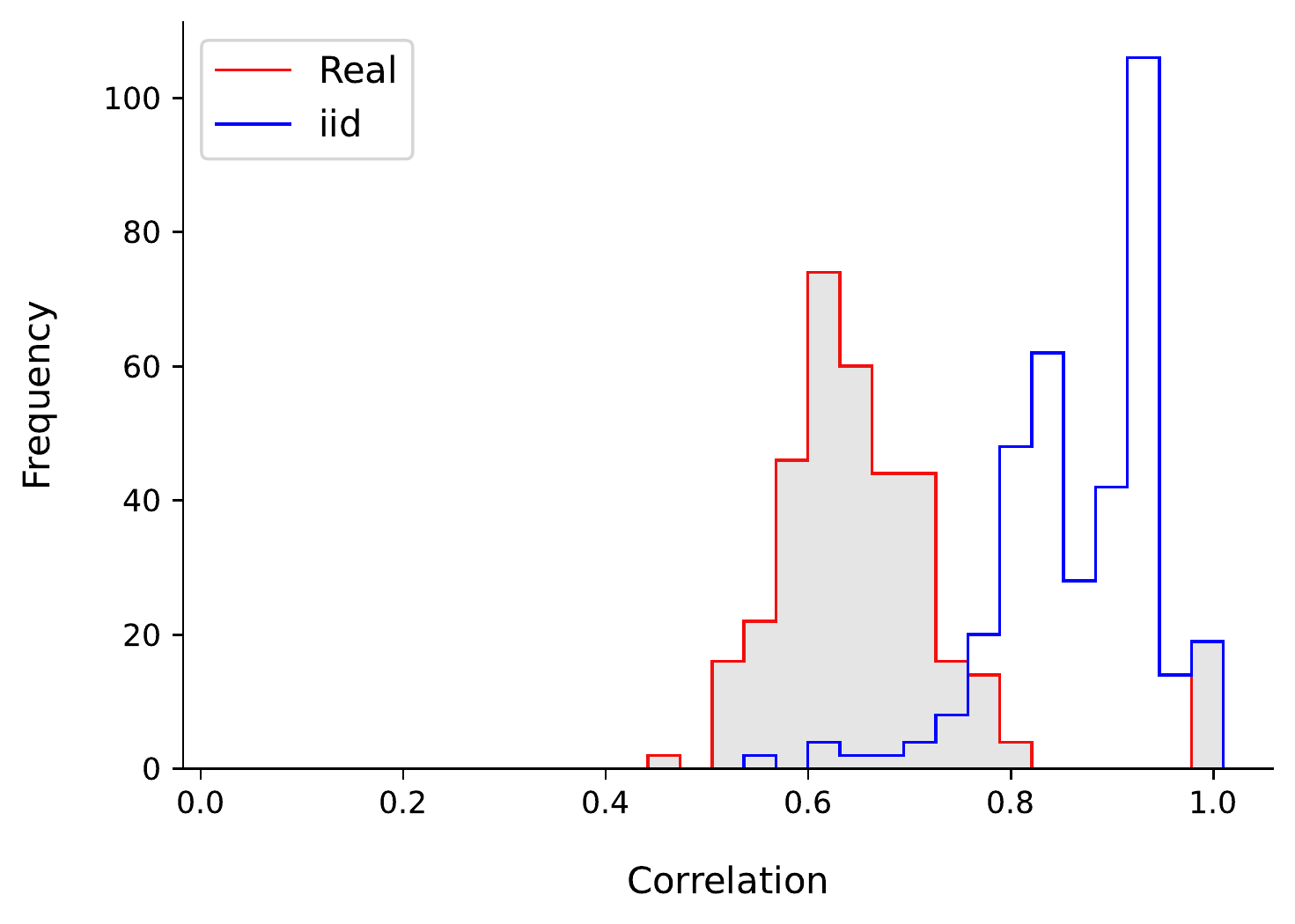}
  \end{minipage}
   \caption{Mean cardinality~(left) and mean cardinality correlation~(right), for YonedaVAE without the Attention gating mechanism.}
   \label{fig:attn_gating}
\end{figure}

\begin{figure}[!htb]
  \begin{minipage}{0.48\textwidth}
    \includegraphics[width=\linewidth]{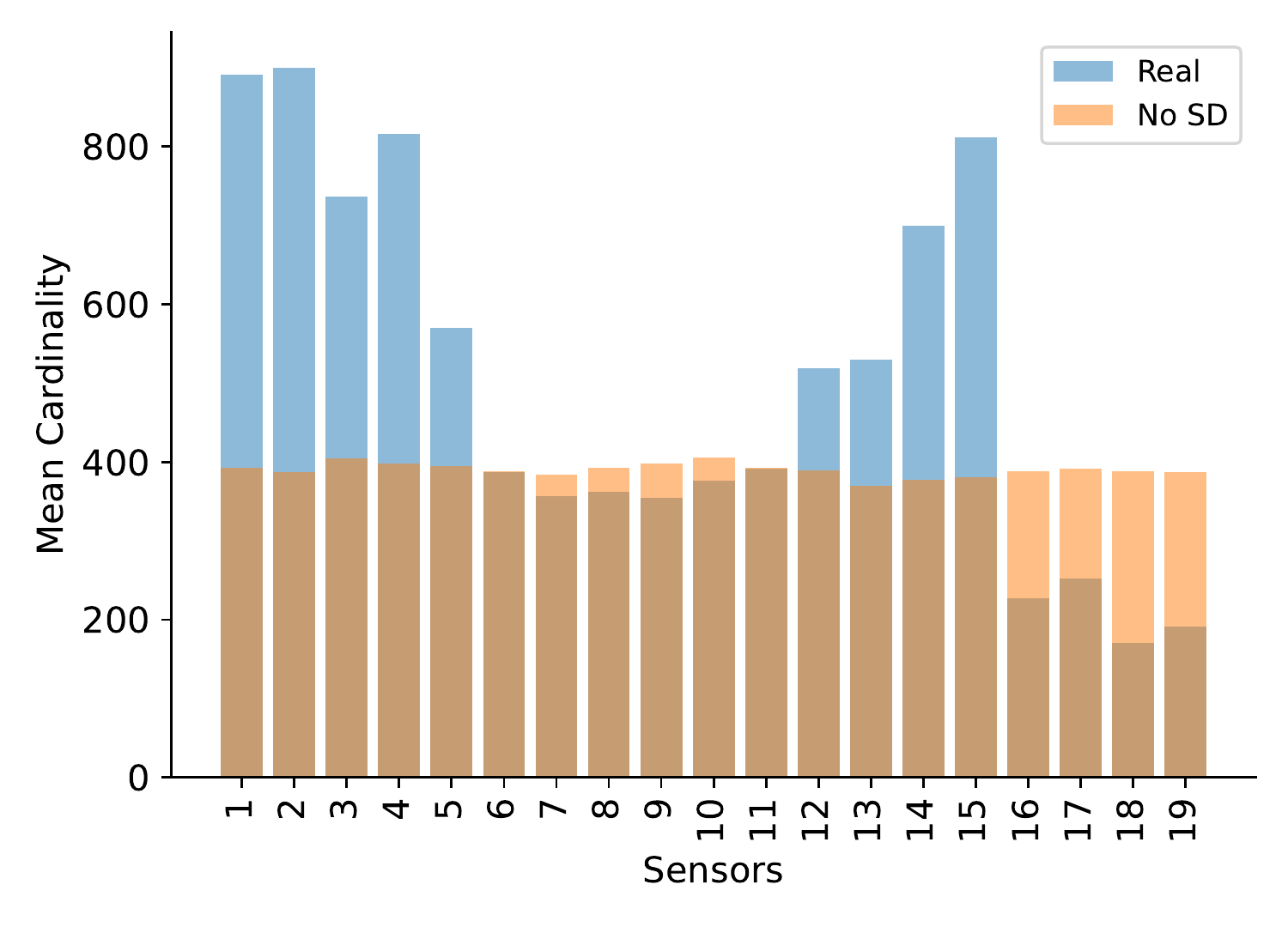}
  \end{minipage}\hfill 
  \begin{minipage}{0.48\textwidth}
    \includegraphics[width=\linewidth]{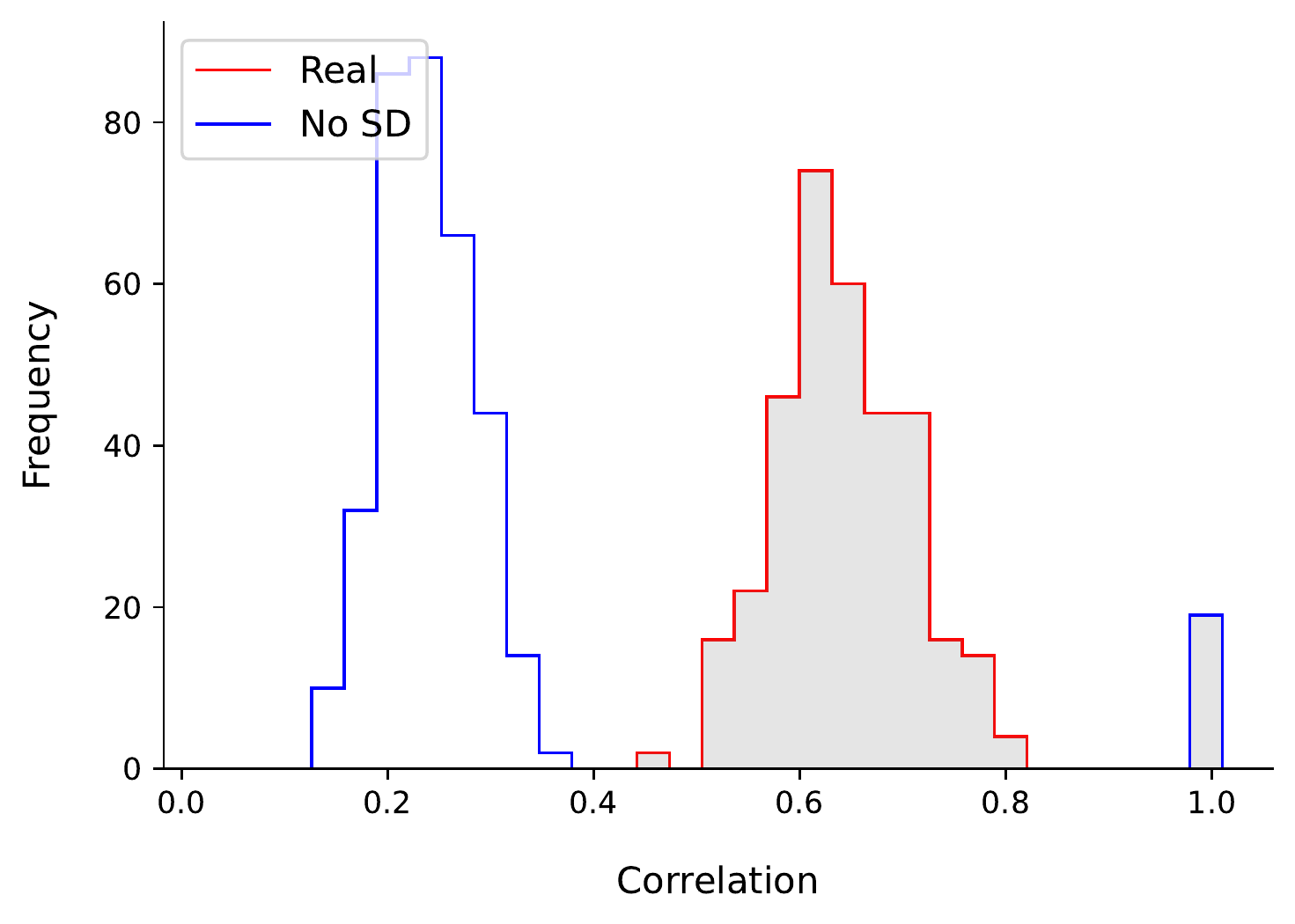}
  \end{minipage}
   \caption{Mean cardinality~(left) and mean cardinality correlation~(right), for YonedaVAE without the self-distillation mechanism during training.}
   \label{fig:sd}
\end{figure}

\begin{figure}[!htb]
  \begin{minipage}{0.48\textwidth}
    \includegraphics[width=\linewidth]{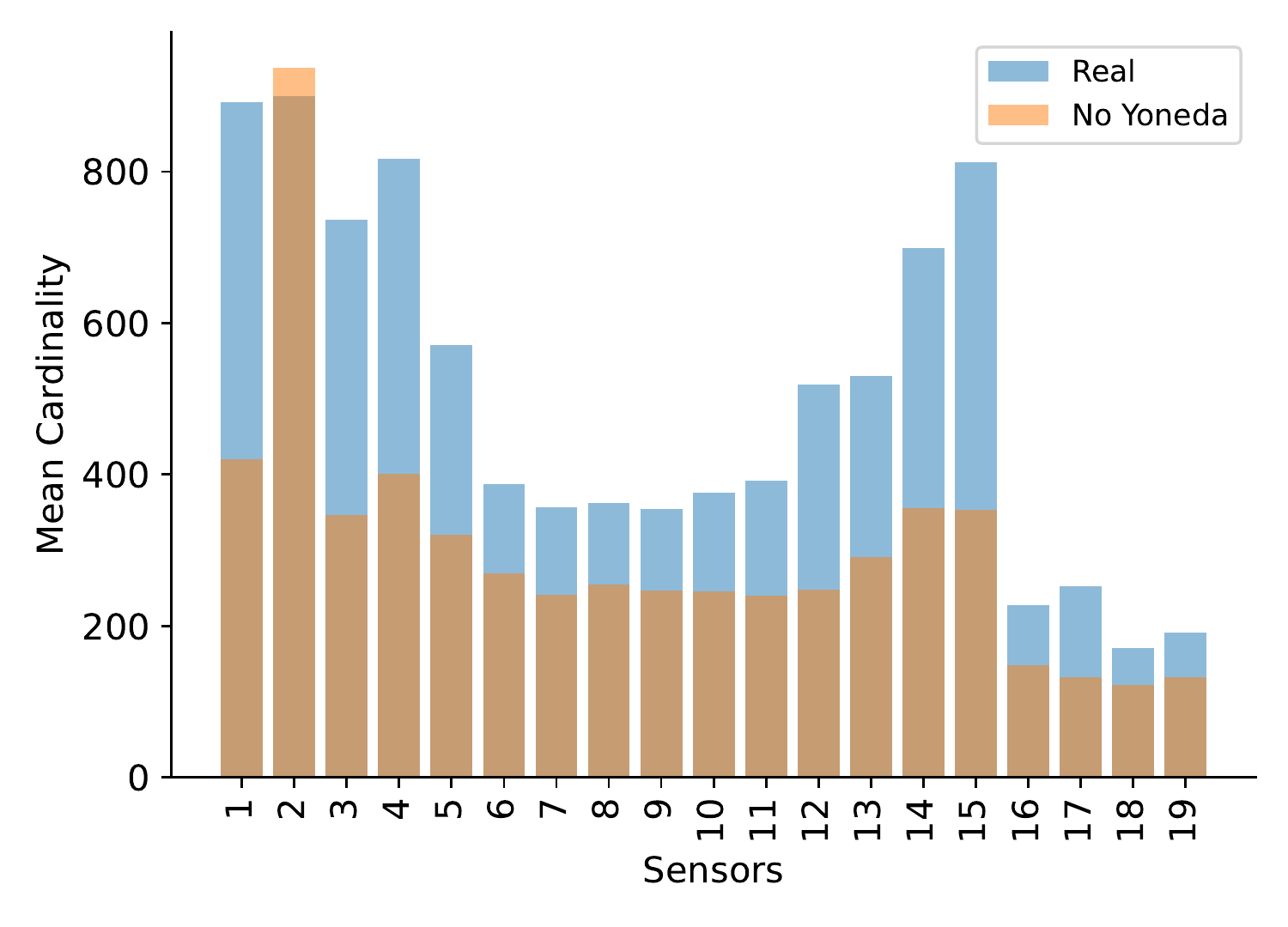}
  \end{minipage}\hfill 
  \begin{minipage}{0.48\textwidth}
    \includegraphics[width=\linewidth]{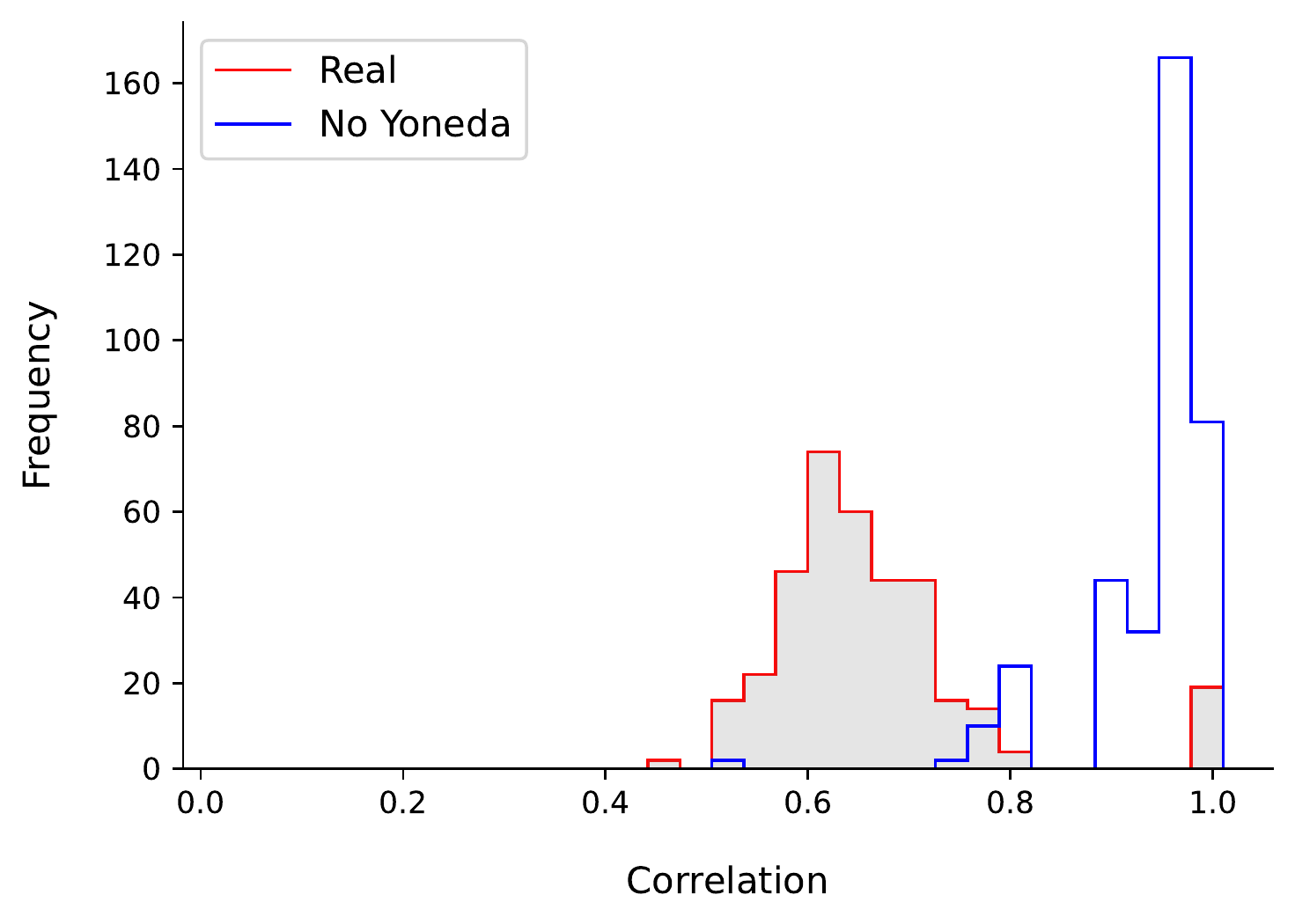}
  \end{minipage}
   \caption{Mean cardinality~(left) and mean cardinality correlation~(right), for YonedaVAE without the Yoneda pooling mechanism during training~(only using the PNA).}
   \label{fig:Nyoneda}
\end{figure}

This section tries to motivate and justify various components of YonedaVAE through an ablation study. I do this by fixing the YonedaVAE's architecture as far as possible and retraining it while a component of YonedaVAE is removed or changed as an intervention to observe the effect of the corresponding module. These results are for the context extrapolation setup where the model does not have access to the individual sensor feature~(sensor occupancy) during inference.
Due to the long training time, I could do this ablation only for Adaptive Top-q sampling~(see \cref{fig:iid} and \cref{fig:topk}), Attention gating~(see \cref{fig:attn_gating}), self-distillation mechanism~(see \cref{fig:sd}), and Yoneda Pooling~(see \cref{fig:Nyoneda}). 
In this study, Adaptive Top-q sampling is compared with top-k and i.i.d sampling~(described in~\cref{section:problem}). Note that I did not compare the original models that use these sampling methods~(doing so would have ended in a huge difference in comparison as they were not designed for context extrapolation). 
Attention-gating and self-distillation mechanisms are compared with the situation where these mechanisms are removed. 
Yoneda pooling is compared with the original PNA pooling~\cite{corso_principal_2020}.
I have to emphasize that only these modules were removed/changed, and the other components stayed intact.

\begin{figure}[!htb]
  \begin{minipage}{0.48\textwidth}
    \includegraphics[width=\linewidth]{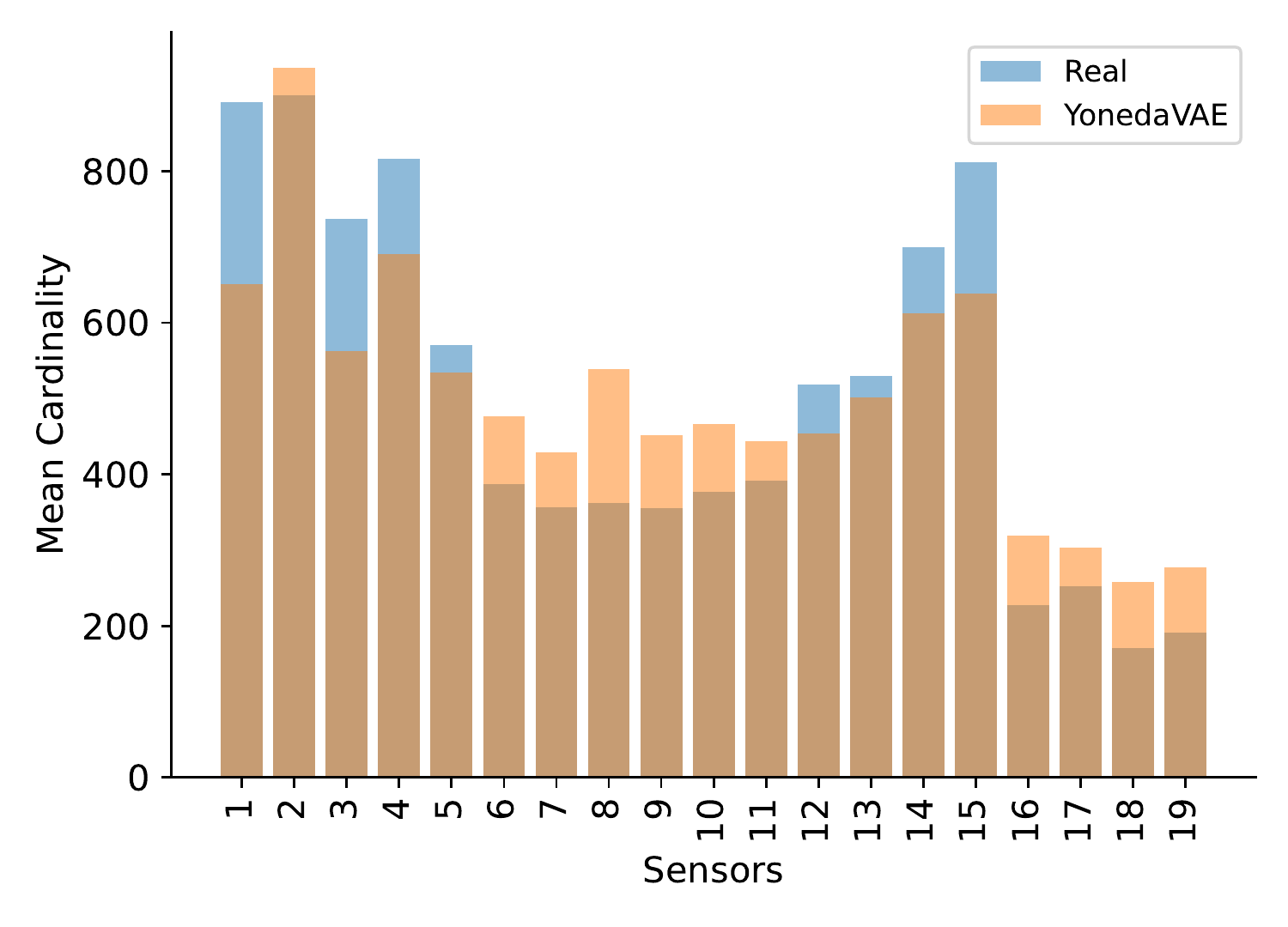}
  \end{minipage}\hfill 
  \begin{minipage}{0.48\textwidth}
    \includegraphics[width=\linewidth]{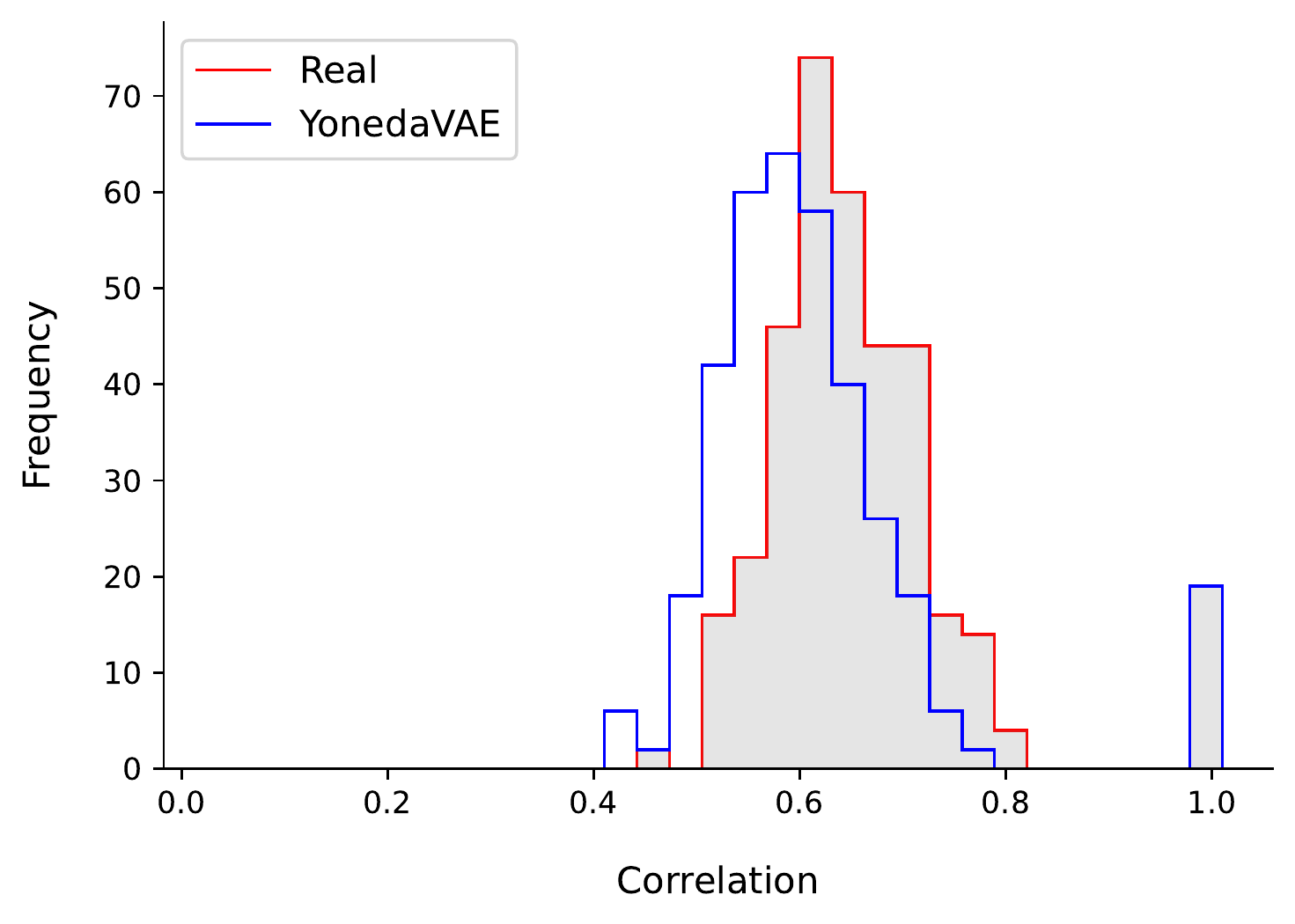}
  \end{minipage}
   \caption{Mean cardinality~(left) and mean cardinality correlation~(right), for YonedaVAE as a comparison baseline.}
\end{figure}

\FloatBarrier
\section{Conclusion}
This study provides an in-depth exploration of the complexities involved in simulating the real PXD background data with fine-grained features. The narrative started by discussing the motivation behind adopting  unordered and variable-sized representations of the data and going to the Out-Of-Distribution~(OOD) domain. The study then takes a closer look at the key elements crucial for OOD generation, specifically focusing on set generation, length extrapolation, and context extrapolation. It also addresses the associated challenges and limitations.

The issues discussed in this chapter are recognized as open questions within both Multi-Set generation and detector simulation communities. On the Multi-Set generation side, OOD simulation is particularly a notorious problem, for instance, in biological protein design, which is the problem of generating novel sequences/sets beyond the training data that exceeds the existing natural sequences in terms of certain attributes.

In the field of detector simulation, recent work has shown limitations. For example, the most current study~\cite{buhmann_caloclouds_2023} on detector point cloud simulation was only able to reconstruct in-distribution events with a cardinality of up to 6,000. In contrast, YonedaVAE has broken new ground by generating out-of-distribution events with a cardinality of up to 100,700.

This work begins with an introductory overview of Category Theory and the Yoneda perspective, leading to the introduction of YonedaVAE. This new VAE-based model is inspired by Category Theory for the generation of sparse ultra-high-granularity, variable-length, and correlated real PXD detector background responses. It does so while conditioning on both the amount of background and the geometry of the detector. The introduction of YonedaVAE is a paradigm shift by extending the capability of simulating PXD hits under conditions that have not been encountered during training.

During this path, first, I conjectured a unified and formal correspondence between relational reasoning and Yoneda lemma. Specifically, within the context of event-based reasoning, the idea is that Yoneda Embedding as an intra-event reasoning module offers more than just structural~(syntactic) insights into an event. It also elucidates how to interpret sensor data in the context of its relationships and interactions with other sensors, thus adding a semantic layer. Consequently, the Yoneda Embedding and its resulting presheaf representation serve as a pooling mechanism, capturing what can be termed ``contextual representation'' for each event.

Then, for YonedaVAE's set generator to be able to extrapolate well beyond the training data during inference, a self-distillation mechanism was introduced to distill the encoded~(relational) information from the YonedaVAE's encoder to a Normal prior distribution. 
Additionally, to handle detector hits with inter-event and intra-event variable cardinality, the novel \emph{Adaptive Top-q sampling} was introduced. This method dynamically determines the number of points to be sampled for each PXD sensor based on the shape of the probability distribution for each event. This approach avoids relying on fixed parameters, offering a more flexible way to sample data and making the model more adaptive to the complexity of individual events.

During this study, many tricks and modifications were introduced, such as attention gating that facilitates the extrapolative abilities of the model. 
As a result of introducing YonedaVAE, all In Distribution~level marginal distributions were accurately generated, surpassing the capabilities of IEA-GAN~(\cref{chap:5}). 
In particular, finally, the elusive intra-event correlation was captured. 

In the Out-Of-Distribution region, YonedaVAE excelled at both length extrapolation and context extrapolation, maintaining high accuracy over marginal distributions and other lower-level metrics. 
The length extrapolation was manifested in the case where the model has to populate the previously unseen PXD hit spots and infer what a background event with a higher cardinality looks like.
The context extrapolation was demonstrated through the model's ability to generate events with the desired amount of background and learn to generalize to attribute~(context) profile beyond the training data and generate point clouds with the correct cardinality and intra-event correlation. 

During the evaluation phase, this study also showcased the application Vendi Score, a metric to quantify the diversity of the generated samples, for PXD background diversity measurement.
Then, for the first time, Topological Data Analysis~(TDA) was studied for the PXD background data.
Along with the PXD clustering analysis, It offered valuable insights into the patterns and intricacies of how the PXD point cloud hits cluster.

Yet, some open challenges persist. While it's crucial to acknowledge that the model has some limitations, particularly in the intra-event correlation \emph{only} during context extrapolation, which is a luxury by itself, the current achievements shouldn't be understated. Attaining this level of precision in context extrapolation is marking a considerable step forward in the discipline.
Moreover, the YonedaVAE's ability to capture long cluster patterns needs further exploration and possibly enhancement, especially considering its influence over the downstream tracking analysis and the fake rate of reconstruction. 
The efficacy of YonedaVAE in approximating helix parameter resolutions, as indicated by closely matching unbiased variances and statistically consistent Kolmogorov-Smirnov p-values, paints a promising picture of its utility for simulating real background conditions. However, the under-representation of such long clusters in the YonedaVAE-generated data as a point cloud generative model could introduce systematic biases or limitations in the track reconstruction efficiency. 
This aspect may not immediately manifest in global performance metrics like unbiased variance or p-values but could have repercussions in specialized scenarios, including track reconstruction in high-density environments or the estimation of rare track configurations. Hence, a deeper investigation into the relationship between cluster size and helix parameter resolution could provide additional layers of validation for YonedaVAE's robustness and reliability.

  \chapter{Summary and Outlook}
\label{chap:7}

\section{Summery}

Rising storage cost and increasing computational time for detector simulations pose a considerable challenge for both existing and upcoming collider experiments.
For Belle~II with the full simulation of PXD background hits, the highest granularity sub-detector at Belle~II is the most intense bottleneck of the detector simulation chain. 
This leads to a desire for new, fast, efficient, amortized, and accurate surrogate methods. This thesis narrated the story of how the current state-of-the-art deep generative models fall short in this task, and highlights imperative to to develop new ideas while incorporating the inductive biases of this specific domain and our problem. \vspace{1ex}

A significant challenge I encountered during my research was the absence of a thorough, taxonomic, and experiment-focused review or survey of deep generative models in the context of detector simulation and particle physics. This gap led to confusion and resulted in the loss of valuable insights in this domain, as well as unnecessary and time-consuming failed experiments. To address this issue, in~\cref{chap:4}, I aimed to offer a comprehensive taxonomy of deep generative models for detector simulation, drawing on nearly all published papers and works available as of the writing date. This survey delves into detailed discussions and categorizations from both algorithmic and application standpoints. I explore the three main use-cases: statistics amplification, amortized generation, and out-of-distribution~(OOD) simulation, while also highlighting the strengths and weaknesses of current methods across various particle physics experiments. Additionally, I identify the limitations of state-of-the-art approaches in the specific context of ultra-high-granularity PXD background simulation.\vspace{1ex}

The critical question in employing generative models for efficient simulation of PXD background centers on designing a model capable of sampling PXD data while preserving accurate correlations among sensors within a single event. This issue is intimately connected to the model's ability to grasp the inherent intra-event correlations, a task that becomes especially challenging due to the ultra-high resolution of PXD data. 
If a generative model can accurately capture these correlations, facilitated by the correct inductive biases, it can be employed to simulate the entire PXD  detector. While most existing surrogate methods concentrate on generating samples that account for intra-sensor~(or layer) correlations, none have yet explored the significance of intra-event correlation or methods to accurately capture it.\vspace{1ex}

Therefore,~\cref{chap:5} tackles this issue by initially exploring the importance of intra-event correlation in subsequent track reconstruction analyses. The chapter demonstrates that a loss of intra-event correlation leads to less accurate Helix parameter resolution in the process of track reconstruction. 
Subsequently, I introduce the Intra-Event Aware GAN~(IEA-GAN), a GAN-based surrogate model enhanced with self-supervised relational reasoning. This model is capable of generating PXD events with an unprecedented granularity of \(7.68 \times 10^6\), a level of resolution that has never before been attempted. 
Developing IEA-GAN was a story of patience and an extreme amount of hyper-parameter tuning, where each run took between 3 to 5 days over about \num{300} runs. 
The chapter classifies the PXD generation task as a fine-grained image analysis problem, characterized by large intra-class and small inter-class variations. 
This adds an additional layer of complexity to the generative task, given the ultra-high resolution of PXD data. 
As a solution to these challenges, I propose various methods, namely the relational reasoning module, Uniformity Loss, and Intra-Event Aware Loss. 
Each of these methods are designed to address specific issues and introduce domain-specific inductive biases to the model. The relational reasoning module is responsible of capturing the correlation among the samples in an event. The Uniformity Loss ensures that the generated samples exhibit as much diversity as possible. The Intra-Event Aware Loss is a novel, non-adversarial, self-supervised loss formulated to enhance the generator's understanding of intra-event dyadic relationships.

For evaluation purposes, I not only compared image-level marginal distributions, where IEA-GAN outperformed the highly-tuned state-of-the-art models in the conditional image generation task, but also introduced the data-driven FID~\cite{heusel_gans_2018} and KID~\cite{binkowski_demystifying_2021} metrics for detector simulation. This study also provided interpretations for these metrics in the context of PXD's low-level image features, proving useful during both validation and hyper-parameter tuning as reliable, generic metrics. 
Ultimately, even though IEA-GAN captured only weak correlations, it proved to be a more potent candidate for replacing Geant4 PXD data compared to other state-of-the-art models. As a consequence, IEA-GAN has been implemented into the basf2 software as an amortized surrogate model for simulation. Geant4 PXD data has also been publicly released as the first ultra-high granularity dataset for future research. 
The integration of IEA-GAN as a surrogate model resulted in a speed-up factor of up to 147 and storage release by half. \vspace{1ex}

Two main concerns remained regarding the results from IEA-GAN. 
First, although IEA-GAN had some success in capturing intra-event correlations, these correlations were relatively weak, and the true underlying relationships between the mean occupancies of sensors had yet to be fully understood. 
Second, there was the issue of data sparsity; the real PXD background data is significantly more sparse than the Geant4 simulated data, and this sparsity also exhibited high variance~(or greater intra-class variation). 
As a result, it could be challenging for CNN-based models like IEA-GAN to meaningfully generate grid-based~(image) representations for PXD background data. \vspace{1ex}

An important challenge in working with real PXD background data is its reliance on actual experiments for gathering the necessary random triggers. Consequently, real luminosity and beam-parameter-dependent PXD background data that go beyond current experimental limits are unavailable, leaving us reliant on computationally demanding simulations. 
This underscores the critical need for a surrogate model that can effectively generalize to Out-Of-Distribution~(OOD) luminosity regions. \vspace{1ex}

To address the aforementioned issues and problems,~\cref{chap:7} introduces YonedaVAE, an equivariant point cloud VAE-based model featuring a self-distillation set generation mechanism and Adaptive Top-q sampling. 
YonedaVAE is conditioned on both the geometry of the PXD detector's sensors— using rotary positional embedding— and the amount of PXD background~(or occupancy) as control parameter. The model is trained on random trigger PXD background data from Experiment 12 of Belle~II, which had a peak recorded luminosity of \(1.42 \times 10^{34} \text{cm}^{-2}\text{s}^{-1}\) and a mean occupancy of 0.06\%. 
It was then tested on data from Experiment 26 of Belle II, which had nearly double the peak luminosity—\(2.68 \times 10^{34} \text{cm}^{-2}\text{s}^{-1}\)—and a mean occupancy of 0.32\%. 
This evaluation serves to demonstrate YonedaVAE's capabilities for extrapolation into OOD and new luminosity regions.

YonedaVAE introduces novel technologies for zero-shot irregular multi-set generation, where both variable intra-category~(intra-event) and variable inter-category~(inter-event) multiplicity are crucial. 
In the encoder of YonedaVAE, Yoneda pooling serves as a learnable, relational pooling mechanism that extends PNA pooling to account for the relationships between objects within an event. As such, I provide a Category Theoretical formulation for relational reasoning and the intra-event aware mechanism. It conjectures that Yoneda Pooling generates a relational observable embedding that incorporates both instance-level information and inter-object relations. Unlike PNA pooling, which only describes the structure~(syntax) of an event, Yoneda pooling delves into understanding each sensor's information in the context of its relationships with other sensors~(semantics). This allows one to assert that Yoneda pooling and the ensuing presheaf representation capture a form of ``contextual representation.''

Illustrated in~\cref{sec:res_6}, YonedaVAE's set generator introduces a novel training mechanism is of paramount importance for OOD generation: the self-distillation mechanism for the set generator. 
Drawing inspiration from self-supervised learning, this mechanism enables self-distillation in the Transformer decoder set generator to enforce an isomorphism between two distinct views of the latent space— the learned one by the encoder and the prior distribution. 
Through this approach, YonedaVAE attains enhanced generalization during inference where it only samples from from the prior Normal distribution.

Another crucial element of YonedaVAE is its Adaptive top-q sampling mechanism. This method accounts for the sparsity of the data and learns the potential skewness in the cardinality distribution within an event~(or mini-batch). Consequently, it selects the most relevant set of top points, having the highest probability, ranging from $\mathcal{O}(10)$ to $\mathcal{O}(100000)$ number of points for each sample within an event.\vspace{1ex}

As a result of deploying YonedaVAE, all In-Distribution~(ID) level marginal distributions were accurately replicated, surpassing the performance of IEA-GAN. Specifically, the elusive intra-event correlation was finally captured. In Out-Of-Distribution~(OOD) scenarios, YonedaVAE excelled in both length and context extrapolation, sustaining high accuracy in marginal distributions and other low-level metrics. 
Length extrapolation was evident when the model had to populate previously unseen PXD hit spots and infer the characteristics of a background event with higher cardinality. 
Context extrapolation was showcased by the model's capability to generate events with the specified higher amount of background, learn to generalize to attribute~(context) profiles beyond the training data, and produce point clouds with accurate cardinality and intra-event correlation. \vspace{1ex}

After demonstrating YonedaVAE's performance in marginal distributions and its advantages in FID and KID metrics, this thesis introduces additional metrics aimed at assessing the diversity of the generated samples. 
A key question in using surrogate models for detector simulation is the extent to which the model can sample diversely compared to real data. 
This question is intrinsically tied to the measurement and comprehension of a dataset's diversity, given a particular similarity measure. Rooted in Ecology, the Vendi Score~\cite{friedman_vendi_2023} provides a quantitative framework for evaluating data diversity. This thesis incorporated this metric in the context of PXD detector simulation and exhibited the amplification results that align with previous work~\cite{butter_amplifying_nodate}, but from an information-theoretical standpoint. \vspace{1ex}

Another important question that must be asked in detector signature simulation, especially for PXD, is how complex the hits are distributed from a geometrical and topological perspective. In particular, how are the points of a PXD point cloud clustered? And How many complex clusters do they create?
This study provides, for the first time in particle physics detector simulation, a Topological Data Analysis~(TDA) approach to answer these questions. TDA leverages the interpretable knowledge of topological features, such as connected points and loops in PXD data sets, to better understand their structural properties at multiple scales. 
As a result, this chapter provides insights into the clustering profile of PXD data. 
I complimented this study by analyzing the PXD background cluster information from the basf2 clusterizer's point of view. 
Given the ultra-high granularity of the PXD background data, YonedaVAE, as a point cloud generative model, could reach a satisfactory and unpredictable result using the clustering algorithm of basf2. This, in particular, is very important from the downstream track reconstruction analysis as the clusters serve as a proxy for the interaction of charged particles with the PXD.
Towards the end, as the final stage of the evaluation, I provide the track reconstruction analysis to compare the effect of experiment 26 real random trigger PXD background to the YonedaVAE generated PXD background.
Despite some deviation in the clustering profile of YonedeVAE's generated background from the real PXD ones, track reconstruction shows a good agreement between the two, even over the extrapolation region. 
\vspace{1ex}

Another crucial question in detector signature simulation, particularly for PXD, concerns the geometric and topological complexity of hit distributions. Specifically, how are the points in a PXD point cloud clustered, and how many complex clusters do they form? For the first time in particle physics detector simulation, this study introduces a Topological Data Analysis~(TDA) approach to address these questions. TDA offers insights into topological features like connected points and loops in PXD hits, to better understand the multi-scale understanding of their structural properties. To complement this, I analyzed the clustering of PXD background data from the perspective of the basf2 clusterizer. Given the ultra-high granularity of PXD background data, YonedaVAE, as a point cloud generative model, achieved promising and unexpected results using basf2's clustering algorithm. This is particularly significant for downstream track reconstruction analysis, where clusters act as indicators of charged particle interactions with the PXD. In the final stage of evaluation, I present a track reconstruction analysis that compares the effects of experiment 26's real random trigger PXD background to those generated by YonedaVAE. Although there is some deviation in the clustering profile between the two, track reconstruction reveals a good agreement, even in extrapolation regions.\vspace{1ex}

This outcome represents a significant milestone for both the fields of deep generative models and fast simulation in particle physics. It demonstrates the feasibility of context and length extrapolation beyond the training data. Specifically, YonedaVAE, trained on a maximum event cardinality of \(19 \times 400 = 7600\), successfully extrapolated to a maximum event cardinality of \(19 \times 5300 = 100700\) with different intra-event correlation profile. 
Consequently, YonedaVAE achieved an unparalleled hit~(point cloud) cardinality of \(100700\) per event in detector simulation, far exceeding the previous record of \(6000\) set by point cloud-based fast simulation models~\cite{buhmann_caloclouds_2023}.

\section{Future Directions and Opportunities}

The High-Luminosity Large Hadron Collider~(HL-LHC)~\cite{bruning_chapter_2020} is expected to surpass the LHC's design integrated luminosity by increasing it by a factor of \num{10}. For instance, the upcoming high-granularity Calorimeter~(HGCAL) with roughly \num{6.5}M channels, or the ITk 3D pixel detector at the HL-LHC~\cite{noauthor_frontiers_nodate} with around \num{1}M information channels, will massively increase the geometry and precision complexity. This leads to a dramatic increase in the time and storage to simulate the detector~\cite{pedro_current_2019}.
However, these granularities far surpass the capacity of existing methods, pointing to the urgency of developing more advanced simulation approaches, while the current studies in generative models barely scratch the surface of the profound challenges posed by future detector simulations. Thus, much more effective and efficient ultra-high-granularity detector simulations are required. 

This work has significant impacts on high-granularity fast and efficient detector simulations. YonedaVAE and IEA-GAN, with robust controllable sampling, are the first potential candidates for simulating the corresponding high-resolution and ultra-high-granular detector signatures with the remarkable capability of generating more than 7.5M information channels. 

In addition, YonedaVAE presents OOD simulation for control parameters beyond the current experimental limits, which has applications in both event generation and real detector simulation. YonedaVAE also presents potential applications in de novo protein design, which is a process that involves the generation of novel amino acid sequences to produce proteins with desired functions, such as enhanced stability and foldability, new binding specificity, or enzymatic activity~\cite{huang_coming_2016}. Proteins can be grouped into different categories based on the arrangement and connectivity of their secondary structure features, such as alpha helices and beta sheets. Then, YonedaVAE can be incorporated for the extrapolative generation of novel foldable protein structures where category-level reasoning is of paramount importance. \vspace{1ex}

The higher goal of this study was to demonstrate the importance of incorporating relational inductive bias and intra-event reasoning in the design of deep generative models for the emulation and extrapolative generation of ultra-high granular PXD backgrounds; however, several challenges will still have to be addressed. 

For first and foremost important challenge for YonedaVAE was the clustering profile of the PXD hits. In general, capturing ``continuity'' is a notorious problem in point cloud generation, especially when the point cloud gets re-scaled by large and imbalanced factors ~(by $[255,250,768]$).
Thus, in order to fully capture the clustering behavior of real PXD background, the future models have to incorporate clustering as an inductive bias in a differentiable way~\cite{luo_differentiable_2020,kim_pllay_2020,xie_differentiable_2020}.
A related open problem is the possible interpretation of higher-order TDA artifacts~(e.g. $H_1$). This thesis provides an interpretation of the lifetime of $H_0$ topological feature in the PXD background point cloud as the lifetime of connected components of large line-like clusters. However, a concrete explanation of the $H_1$ feature, especially in relation to the types of background hits, would be very interesting. 
In the case of a relationship, this can augment the surrogate model to generate PXD hits conditioned on the ``type of background'' by incorporating the desired $H_1$ profile during training. 

Moreover, there existed a deviation from the true intra-event correlation of YonedaVAE only in the context extrapolation regime. 
It's worth mentioning that generating out-of-distribution~(OOD) data that perfectly captures the inherent context is a notorious challenge and the subject of extensive research across various domains. These include not just particle physics but also fields like drug discovery~\cite{freschlin_machine_2022,ji_drugood_2022}, protein design~\cite{watson_novo_2023}, cell design~\cite{lotfollahi_conditional_2020}, natural language processing~\cite{chan_deep_2021}, and weather forecasting~\cite{ravuri_skilful_2021}.
The complexities inherent in this task indicate that the generative models presented in this thesis do not suffice for flawless context extrapolation. 
Additional refinements are essential for future models. Specifically, the incorporation of more refined inductive biases and regularization methods during the training phase could offer a pathway toward more reliable context extrapolation. \vspace{1ex}

I think a promising avenue for future research lies in the direction of differentiable programming~(DP)~\cite{adelmann_new_2022}. Through DP, software becomes differentiable through automatic differentiation (AD)~\cite{baydin_automatic_2018}, enabling efficient gradient computation to understand the influence of input variations on output predictions. Such gradient-based insights could be invaluable for various downstream tasks. Utilizing DP frameworks would allow particle physics simulation tools to be integrated into machine learning pipelines, thereby facilitating joint optimization for enhanced computational efficiency. This integration paves the way for the creation of physics-informed surrogate models capable of more accurately addressing the complexities inherent in high-granularity detector simulations.


  \cleardoublepage
  \addcontentsline{toc}{chapter}{List of Publications}

\chapter*{List of Publications}
The findings presented in this thesis are the result of research conducted at the Ludwig-Maximilians-Universität München. The research outcomes span various stages of my academic dissemination. Specifically, the results detailed in Chapters 4, 5, and 6 have been submitted for publication or are actively being prepared for submission. Some of these findings have already been published, while others are currently under review or in preparation for submission to peer-reviewed journals.

\begin{itemize}
    \item[\textbf{[1]}] M. Srebre, P. Schmolz, H. Hashemi, M. Ritter, and T. Kuhr 2020. \textbf{Generation of Belle II Pixel Detector Background Data with a GAN}, EPJ Web of Conferences, 245, 02010. 
    \href{	https://doi.org/10.1051/epjconf/202024502010}{\url{10.1051/epjconf/202024502010}}.
    
    \item[\textbf{[2]}] H. Hashemi, N. Hartmann, T. Kuhr, M. Ritter, and M. Srebre 2021. \textbf{PE-GAN: Prior Embedding GAN for PXD images at Belle II}, EPJ Web of Conferences, 251, 03031. \href{https://doi.org/10.1051/epjconf/202125103031}{\url{10.1051/epjconf/202125103031}}.

    \item[\textbf{[3]}] H. Hashemi, N. Hartmann, S. Sharifzadeh, J. Kahn, and T. Kuhr 2022.
    \textbf{Intra-Event Aware Imitation Game for Fast Detector Simulation}, NeurIPS 2022 workshop paper~(Machine learning and the physical sciences). \href{https://ml4physicalsciences.github.io/2022/files/NeurIPS_ML4PS_2022_97.pdf}{\url{NeurIPS_ML4PS_2022_97}}.

    \item[\textbf{[4]}] H. Hashemi, N. Hartmann, S. Sharifzadeh, J. Kahn, and T. Kuhr 2023.
    \textbf{Ultra-High-Resolution Detector Simulation with Intra-Event Aware GAN and Self-Supervised Relational Reasoning}, Under Peer Review by \textit{Nature Communications}. \href{https://doi.org/10.48550/arXiv.2303.08046}{\url{10.48550/arXiv.2303.08046}}
    
    \item[\textbf{[5]}] B. Hashemi, C. Krause, 2023. \textbf{Deep Generative Models for Detector Signature Simulation: A Taxonomic Review}, Under Peer Review by \textit{Reviews in Physics}.
    \href{https://doi.org/10.48550/arXiv.2312.09597}{\url{10.48550/arXiv.2312.09597}}
        
    \item[\textbf{[6]}] B. Hashemi, 2023. \textbf{YonedaVAE: Self-Supervised Out-of-Distribution Multi-Set Generation for Amortized Simulations}, Under Submission to \textit{Machine Learning: Science and Technology}.

    \item[\textbf{[7]}] B. Hashemi, B. Rieck, 2023. \textbf{Topological Data Analysis for Particle Physics Detector signatures}, Under Submission to \textit{Machine Learning: Science and Technology}.

\end{itemize}
  \markboth{List of Publications}{List of Publications}
  
  \cleardoublepage
  \listoffigures
  \addcontentsline{toc}{chapter}{List of Figures}
  \markboth{List of Figures}{List of Figures}
  
  \cleardoublepage
  \listoftables
  \addcontentsline{toc}{chapter}{List of Tables}
  \markboth{List of Tables}{List of Tables}

  \backmatter

  \cleardoublepage
  \printbibliography

@article{topaz_topological_2015,
	title = {Topological Data Analysis of Biological Aggregation Models},
	volume = {10},
	issn = {1932-6203},
	url = {https://dx.plos.org/10.1371/journal.pone.0126383},
	doi = {10.1371/journal.pone.0126383},
	pages = {e0126383},
	number = {5},
	journaltitle = {{PLOS} {ONE}},
	shortjournal = {{PLoS} {ONE}},
	author = {Topaz, Chad M. and Ziegelmeier, Lori and Halverson, Tom},
	editor = {Ermentrout, Bard},
	urldate = {2023-10-10},
	date = {2015-05-13},
	langid = {english},
}

@online{chazal_introduction_2017,
	title = {An introduction to Topological Data Analysis: fundamental and practical aspects for data scientists},
	url = {https://arxiv.org/abs/1710.04019v2},
	shorttitle = {An introduction to Topological Data Analysis},
	titleaddon = {{arXiv}.org},
	author = {Chazal, Frédéric and Michel, Bertrand},
	urldate = {2023-10-10},
	date = {2017-10-11},
	langid = {english},
}

@online{zhang_dive_2021,
	title = {Dive into Deep Learning},
	url = {https://arxiv.org/abs/2106.11342v5},
	titleaddon = {{arXiv}.org},
	author = {Zhang, Aston and Lipton, Zachary C. and Li, Mu and Smola, Alexander J.},
	urldate = {2023-10-10},
	date = {2021-06-21},
	langid = {english},
}

@article{sakharov_violation_1967,
	title = {Violation of {CP} Invariance, C asymmetry, and baryon asymmetry of the universe},
	volume = {5},
	doi = {10.1070/PU1991v034n05ABEH002497},
	pages = {32--35},
	journaltitle = {Pisma Zh. Eksp. Teor. Fiz.},
	author = {Sakharov, A. D.},
	date = {1967},
}

@article{peebles_cosmological_2003,
	title = {The Cosmological Constant and Dark Energy},
	volume = {75},
	issn = {0034-6861, 1539-0756},
	url = {http://arxiv.org/abs/astro-ph/0207347},
	doi = {10.1103/RevModPhys.75.559},
	pages = {559--606},
	number = {2},
	journaltitle = {Reviews of Modern Physics},
	shortjournal = {Rev. Mod. Phys.},
	author = {Peebles, P. J. E. and Ratra, Bharat},
	urldate = {2023-10-09},
	date = {2003-04-22},
	eprinttype = {arxiv},
	eprint = {astro-ph/0207347},
}

@article{bennett_nine-year_2013,
	title = {Nine-Year Wilkinson Microwave Anisotropy Probe ({WMAP}) Observations: Final Maps and Results},
	volume = {208},
	issn = {0067-0049, 1538-4365},
	url = {http://arxiv.org/abs/1212.5225},
	doi = {10.1088/0067-0049/208/2/20},
	shorttitle = {Nine-Year Wilkinson Microwave Anisotropy Probe ({WMAP}) Observations},
	pages = {20},
	number = {2},
	journaltitle = {The Astrophysical Journal Supplement Series},
	shortjournal = {{ApJS}},
	author = {Bennett, C. L. and Larson, D. and Weiland, J. L. and Jarosik, N. and Hinshaw, G. and Odegard, N. and Smith, K. M. and Hill, R. S. and Gold, B. and Halpern, M. and Komatsu, E. and Nolta, M. R. and Page, L. and Spergel, D. N. and Wollack, E. and Dunkley, J. and Kogut, A. and Limon, M. and Meyer, S. S. and Tucker, G. S. and Wright, E. L.},
	urldate = {2023-10-09},
	date = {2013-09-20},
	eprinttype = {arxiv},
	eprint = {1212.5225 [astro-ph]},
}

@article{planck_collaboration_planck_2014,
	title = {Planck 2013 results. I. Overview of products and scientific results},
	volume = {571},
	issn = {0004-6361, 1432-0746},
	url = {http://arxiv.org/abs/1303.5062},
	doi = {10.1051/0004-6361/201321529},
	pages = {A1},
	journaltitle = {Astronomy \& Astrophysics},
	shortjournal = {A\&A},
	author = {Planck Collaboration and Ade, P. A. R. and Aghanim, N. and Alves, M. I. R. and Armitage-Caplan, C. and Arnaud, M. and Ashdown, M. and Atrio-Barandela, F. and Aumont, J. and Aussel, H. and Baccigalupi, C. and Banday, A. J. and Barreiro, R. B. and Barrena, R. and Bartelmann, M. and Bartlett, J. G. and Bartolo, N. and Basak, S. and Battaner, E. and Battye, R. and Benabed, K. and Benoît, A. and Benoit-Lévy, A. and Bernard, J.-P. and Bersanelli, M. and Bertincourt, B. and Bethermin, M. and Bielewicz, P. and Bikmaev, I. and Blanchard, A. and Bobin, J. and Bock, J. J. and Böhringer, H. and Bonaldi, A. and Bonavera, L. and Bond, J. R. and Borrill, J. and Bouchet, F. R. and Boulanger, F. and Bourdin, H. and Bowyer, J. W. and Bridges, M. and Brown, M. L. and Bucher, M. and Burenin, R. and Burigana, C. and Butler, R. C. and Calabrese, E. and Cappellini, B. and Cardoso, J.-F. and Carr, R. and Carvalho, P. and Casale, M. and Castex, G. and Catalano, A. and Challinor, A. and Chamballu, A. and Chary, R.-R. and Chen, X. and Chiang, H. C. and Chiang, L.-Y. and Chon, G. and Christensen, P. R. and Churazov, E. and Church, S. and Clemens, M. and Clements, D. L. and Colombi, S. and Colombo, L. P. L. and Combet, C. and Comis, B. and Couchot, F. and Coulais, A. and Crill, B. P. and Cruz, M. and Curto, A. and Cuttaia, F. and Da Silva, A. and Dahle, H. and Danese, L. and Davies, R. D. and Davis, R. J. and de Bernardis, P. and de Rosa, A. and de Zotti, G. and Déchelette, T. and Delabrouille, J. and Delouis, J.-M. and Démoclès, J. and Désert, F.-X. and Dick, J. and Dickinson, C. and Diego, J. M. and Dolag, K. and Dole, H. and Donzelli, S. and Doré, O. and Douspis, M. and Ducout, A. and Dunkley, J. and Dupac, X. and Efstathiou, G. and Elsner, F. and Enßlin, T. A. and Eriksen, H. K. and Fabre, O. and Falgarone, E. and Falvella, M. C. and Fantaye, Y. and Fergusson, J. and Filliard, C. and Finelli, F. and Flores-Cacho, I. and Foley, S. and Forni, O. and Fosalba, P. and Frailis, M. and Fraisse, A. A. and Franceschi, E. and Freschi, M. and Fromenteau, S. and Frommert, M. and Gaier, T. C. and Galeotta, S. and Gallegos, J. and Galli, S. and Gandolfo, B. and Ganga, K. and Gauthier, C. and Génova-Santos, R. T. and Ghosh, T. and Giard, M. and Giardino, G. and Gilfanov, M. and Girard, D. and Giraud-Héraud, Y. and Gjerløw, E. and González-Nuevo, J. and Górski, K. M. and Gratton, S. and Gregorio, A. and Gruppuso, A. and Gudmundsson, J. E. and Haissinski, J. and Hamann, J. and Hansen, F. K. and Hansen, M. and Hanson, D. and Harrison, D. L. and Heavens, A. and Helou, G. and Hempel, A. and Henrot-Versillé, S. and Hernández-Monteagudo, C. and Herranz, D. and Hildebrandt, S. R. and Hivon, E. and Ho, S. and Hobson, M. and Holmes, W. A. and Hornstrup, A. and Hou, Z. and Hovest, W. and Huey, G. and Huffenberger, K. M. and Hurier, G. and Ilić, S. and Jaffe, A. H. and Jaffe, T. R. and Jasche, J. and Jewell, J. and Jones, W. C. and Juvela, M. and Kalberla, P. and Kangaslahti, P. and Keihänen, E. and Kerp, J. and Keskitalo, R. and Khamitov, I. and Kiiveri, K. and Kim, J. and Kisner, T. S. and Kneissl, R. and Knoche, J. and Knox, L. and Kunz, M. and Kurki-Suonio, H. and Lacasa, F. and Lagache, G. and Lähteenmäki, A. and Lamarre, J.-M. and Langer, M. and Lasenby, A. and Lattanzi, M. and Laureijs, R. J. and Lavabre, A. and Lawrence, C. R. and Jeune, M. Le and Leach, S. and Leahy, J. P. and Leonardi, R. and León-Tavares, J. and Leroy, C. and Lesgourgues, J. and Lewis, A. and Li, C. and Liddle, A. and Liguori, M. and Lilje, P. B. and Linden-Vørnle, M. and Lindholm, V. and López-Caniego, M. and Lowe, S. and Lubin, P. M. and Macías-Pérez, J. F. and {MacTavish}, C. J. and Maffei, B. and Maggio, G. and Maino, D. and Mandolesi, N. and Mangilli, A. and Marcos-Caballero, A. and Marinucci, D. and Maris, M. and Marleau, F. and Marshall, D. J. and Martin, P. G. and Martínez-González, E. and Masi, S. and Massardi, M. and Matarrese, S. and Matsumura, T. and Matthai, F. and Maurin, L. and Mazzotta, P. and {McDonald}, A. and {McEwen}, J. D. and {McGehee}, P. and Mei, S. and Meinhold, P. R. and Melchiorri, A. and Melin, J.-B. and Mendes, L. and Menegoni, E. and Mennella, A. and Migliaccio, M. and Mikkelsen, K. and Millea, M. and Miniscalco, R. and Mitra, S. and Miville-Deschênes, M.-A. and Molinari, D. and Moneti, A. and Montier, L. and Morgante, G. and Morisset, N. and Mortlock, D. and Moss, A. and Munshi, D. and Murphy, J. A. and Naselsky, P. and Nati, F. and Natoli, P. and Negrello, M. and Nesvadba, N. P. H. and Netterfield, C. B. and Nørgaard-Nielsen, H. U. and North, C. and Noviello, F. and Novikov, D. and Novikov, I. and O'Dwyer, I. J. and Orieux, F. and Osborne, S. and O'Sullivan, C. and Oxborrow, C. A. and Paci, F. and Pagano, L. and Pajot, F. and Paladini, R. and Pandolfi, S. and Paoletti, D. and Partridge, B. and Pasian, F. and Patanchon, G. and Paykari, P. and Pearson, D. and Pearson, T. J. and Peel, M. and Peiris, H. V. and Perdereau, O. and Perotto, L. and Perrotta, F. and Pettorino, V. and Piacentini, F. and Piat, M. and Pierpaoli, E. and Pietrobon, D. and Plaszczynski, S. and Platania, P. and Pogosyan, D. and Pointecouteau, E. and Polenta, G. and Ponthieu, N. and Popa, L. and Poutanen, T. and Pratt, G. W. and Prézeau, G. and Prunet, S. and Puget, J.-L. and Pullen, A. R. and Rachen, J. P. and Racine, B. and Rahlin, A. and Räth, C. and Reach, W. T. and Rebolo, R. and Reinecke, M. and Remazeilles, M. and Renault, C. and Renzi, A. and Riazuelo, A. and Ricciardi, S. and Riller, T. and Ringeval, C. and Ristorcelli, I. and Robbers, G. and Rocha, G. and Roman, M. and Rosset, C. and Rossetti, M. and Roudier, G. and Rowan-Robinson, M. and Rubiño-Martín, J. A. and Ruiz-Granados, B. and Rusholme, B. and Salerno, E. and Sandri, M. and Sanselme, L. and Santos, D. and Savelainen, M. and Savini, G. and Schaefer, B. M. and Schiavon, F. and Scott, D. and Seiffert, M. D. and Serra, P. and Shellard, E. P. S. and Smith, K. and Smoot, G. F. and Souradeep, T. and Spencer, L. D. and Starck, J.-L. and Stolyarov, V. and Stompor, R. and Sudiwala, R. and Sunyaev, R. and Sureau, F. and Sutter, P. and Sutton, D. and Suur-Uski, A.-S. and Sygnet, J.-F. and Tauber, J. A. and Tavagnacco, D. and Taylor, D. and Terenzi, L. and Texier, D. and Toffolatti, L. and Tomasi, M. and Torre, J.-P. and Tristram, M. and Tucci, M. and Tuovinen, J. and Türler, M. and Tuttlebee, M. and Umana, G. and Valenziano, L. and Valiviita, J. and Van Tent, B. and Varis, J. and Vibert, L. and Viel, M. and Vielva, P. and Villa, F. and Vittorio, N. and Wade, L. A. and Wandelt, B. D. and Watson, C. and Watson, R. and Wehus, I. K. and Welikala, N. and Weller, J. and White, M. and White, S. D. M. and Wilkinson, A. and Winkel, B. and Xia, J.-Q. and Yvon, D. and Zacchei, A. and Zibin, J. P. and Zonca, A.},
	urldate = {2023-10-09},
	date = {2014-11},
	eprinttype = {arxiv},
	eprint = {1303.5062 [astro-ph]},
}

@article{ravuri_skilful_2021,
	title = {Skilful precipitation nowcasting using deep generative models of radar},
	volume = {597},
	rights = {2021 The Author(s)},
	issn = {1476-4687},
	url = {https://www.nature.com/articles/s41586-021-03854-z},
	doi = {10.1038/s41586-021-03854-z},
	pages = {672--677},
	number = {7878},
	journaltitle = {Nature},
	author = {Ravuri, Suman and Lenc, Karel and Willson, Matthew and Kangin, Dmitry and Lam, Remi and Mirowski, Piotr and Fitzsimons, Megan and Athanassiadou, Maria and Kashem, Sheleem and Madge, Sam and Prudden, Rachel and Mandhane, Amol and Clark, Aidan and Brock, Andrew and Simonyan, Karen and Hadsell, Raia and Robinson, Niall and Clancy, Ellen and Arribas, Alberto and Mohamed, Shakir},
	urldate = {2023-10-08},
	date = {2021-09},
	langid = {english},
	note = {Number: 7878
Publisher: Nature Publishing Group},
}

@article{watson_novo_2023,
	title = {De novo design of protein structure and function with {RFdiffusion}},
	volume = {620},
	rights = {2023 The Author(s)},
	issn = {1476-4687},
	url = {https://www.nature.com/articles/s41586-023-06415-8},
	doi = {10.1038/s41586-023-06415-8},
	pages = {1089--1100},
	number = {7976},
	journaltitle = {Nature},
	author = {Watson, Joseph L. and Juergens, David and Bennett, Nathaniel R. and Trippe, Brian L. and Yim, Jason and Eisenach, Helen E. and Ahern, Woody and Borst, Andrew J. and Ragotte, Robert J. and Milles, Lukas F. and Wicky, Basile I. M. and Hanikel, Nikita and Pellock, Samuel J. and Courbet, Alexis and Sheffler, William and Wang, Jue and Venkatesh, Preetham and Sappington, Isaac and Torres, Susana Vázquez and Lauko, Anna and De Bortoli, Valentin and Mathieu, Emile and Ovchinnikov, Sergey and Barzilay, Regina and Jaakkola, Tommi S. and {DiMaio}, Frank and Baek, Minkyung and Baker, David},
	urldate = {2023-10-08},
	date = {2023-08},
	langid = {english},
	note = {Number: 7976
Publisher: Nature Publishing Group},
}

@online{ji_drugood_2022,
	title = {{DrugOOD}: Out-of-Distribution ({OOD}) Dataset Curator and Benchmark for {AI}-aided Drug Discovery -- A Focus on Affinity Prediction Problems with Noise Annotations},
	url = {https://arxiv.org/abs/2201.09637v1},
	shorttitle = {{DrugOOD}},
	titleaddon = {{arXiv}.org},
	author = {Ji, Yuanfeng and Zhang, Lu and Wu, Jiaxiang and Wu, Bingzhe and Huang, Long-Kai and Xu, Tingyang and Rong, Yu and Li, Lanqing and Ren, Jie and Xue, Ding and Lai, Houtim and Xu, Shaoyong and Feng, Jing and Liu, Wei and Luo, Ping and Zhou, Shuigeng and Huang, Junzhou and Zhao, Peilin and Bian, Yatao},
	urldate = {2023-10-08},
	date = {2022-01-24},
	langid = {english},
}

@misc{baydin_automatic_2018,
	title = {Automatic differentiation in machine learning: a survey},
	url = {http://arxiv.org/abs/1502.05767},
	doi = {10.48550/arXiv.1502.05767},
	shorttitle = {Automatic differentiation in machine learning},
	number = {{arXiv}:1502.05767},
	publisher = {{arXiv}},
	author = {Baydin, Atilim Gunes and Pearlmutter, Barak A. and Radul, Alexey Andreyevich and Siskind, Jeffrey Mark},
	urldate = {2023-10-08},
	date = {2018-02-05},
	eprinttype = {arxiv},
	eprint = {1502.05767 [cs, stat]},
}

@misc{adelmann_new_2022,
	title = {New directions for surrogate models and differentiable programming for High Energy Physics detector simulation},
	url = {http://arxiv.org/abs/2203.08806},
	number = {{arXiv}:2203.08806},
	publisher = {{arXiv}},
	author = {Adelmann, Andreas and Hopkins, Walter and Kourlitis, Evangelos and Kagan, Michael and Kasieczka, Gregor and Krause, Claudius and Shih, David and Mikuni, Vinicius and Nachman, Benjamin and Pedro, Kevin and Winklehner, Daniel},
	urldate = {2023-10-08},
	date = {2022-03-15},
	eprinttype = {arxiv},
	eprint = {2203.08806 [hep-ex, physics:hep-ph, physics:physics]},
}

@article{huang_coming_2016,
	title = {The coming of age of de novo protein design},
	volume = {537},
	rights = {2016 Springer Nature Limited},
	issn = {1476-4687},
	url = {https://www.nature.com/articles/nature19946},
	doi = {10.1038/nature19946},
	pages = {320--327},
	number = {7620},
	journaltitle = {Nature},
	author = {Huang, Po-Ssu and Boyken, Scott E. and Baker, David},
	urldate = {2023-10-08},
	date = {2016-09},
	langid = {english},
	note = {Number: 7620
Publisher: Nature Publishing Group},
}

@inproceedings{xie_differentiable_2020,
	title = {Differentiable Top-k with Optimal Transport},
	volume = {33},
	url = {https://papers.nips.cc/paper/2020/hash/ec24a54d62ce57ba93a531b460fa8d18-Abstract.html},
	pages = {20520--20531},
	booktitle = {Advances in Neural Information Processing Systems},
	publisher = {Curran Associates, Inc.},
	author = {Xie, Yujia and Dai, Hanjun and Chen, Minshuo and Dai, Bo and Zhao, Tuo and Zha, Hongyuan and Wei, Wei and Pfister, Tomas},
	urldate = {2023-10-08},
	date = {2020},
}

@online{kim_pllay_2020,
	title = {{PLLay}: Efficient Topological Layer based on Persistence Landscapes},
	url = {https://arxiv.org/abs/2002.02778v4},
	shorttitle = {{PLLay}},
	titleaddon = {{arXiv}.org},
	author = {Kim, Kwangho and Kim, Jisu and Zaheer, Manzil and Kim, Joon Sik and Chazal, Frederic and Wasserman, Larry},
	urldate = {2023-10-08},
	date = {2020-02-07},
	langid = {english},
}

@inproceedings{luo_differentiable_2020,
	title = {Differentiable Manifold Reconstruction for Point Cloud Denoising},
	url = {http://arxiv.org/abs/2007.13551},
	doi = {10.1145/3394171.3413727},
	pages = {1330--1338},
	booktitle = {Proceedings of the 28th {ACM} International Conference on Multimedia},
	author = {Luo, Shitong and Hu, Wei},
	urldate = {2023-10-08},
	date = {2020-10-12},
	langid = {english},
	eprinttype = {arxiv},
	eprint = {2007.13551 [cs]},
}

@online{noauthor_learning_nodate,
	title = {Learning to simulate high energy particle collisions from unlabeled data {\textbar} Scientific Reports},
	url = {https://www.nature.com/articles/s41598-022-10966-7},
	urldate = {2023-10-08},
}

@online{noauthor_sad_nodate,
	title = {{SAD} Home Page – Strategic Accelerator Design},
	url = {https://acc-physics.kek.jp/SAD/},
	urldate = {2023-10-04},
}

@online{noauthor_accelerator_nodate,
	title = {Accelerator design at {SuperKEKB} {\textbar} Progress of Theoretical and Experimental Physics {\textbar} Oxford Academic},
	url = {https://academic.oup.com/ptep/article/2013/3/03A011/1556583},
	urldate = {2023-10-04},
}

@article{natochii_measured_2023,
	title = {Measured and projected beam backgrounds in the Belle {II} experiment at the {SuperKEKB} collider},
	volume = {1055},
	issn = {0168-9002},
	url = {https://www.sciencedirect.com/science/article/pii/S0168900223005405},
	doi = {10.1016/j.nima.2023.168550},
	pages = {168550},
	journaltitle = {Nuclear Instruments and Methods in Physics Research Section A: Accelerators, Spectrometers, Detectors and Associated Equipment},
	shortjournal = {Nuclear Instruments and Methods in Physics Research Section A: Accelerators, Spectrometers, Detectors and Associated Equipment},
	author = {Natochii, A. and Browder, T. E. and Cao, L. and Cautero, G. and Dreyer, S. and Frey, A. and Gabrielli, A. and Giuressi, D. and Ishibashi, T. and Jin, Y. and Kojima, K. and Kraetzschmar, T. and Lanceri, L. and Liptak, Z. and Liventsev, D. and Marinas, C. and Massaccesi, L. and Matsuoka, K. and Meier, F. and Miller, C. and Nakayama, H. and Niebuhr, C. and Novosel, A. and Parham, K. and Popov, I. and Rizzo, G. and Roney, J. M. and Ryu, S. Y. and Santelj, L. and Schneider, S. and Schueler, J. and Schwenker, B. and Shi, X. D. and Simon, F. and Stefkova, S. and Takahashi, M. and Tanigawa, H. and Taniguchi, N. and Terui, S. and Vahsen, S. E. and Vitale, L. and Vossen, A. and Wang, Z. and Wiechczynski, J. and Windel, H. and Yoshihara, K.},
	urldate = {2023-10-04},
	date = {2023-10-01},
}

@thesis{moll_comprehensive_2015,
	title = {Comprehensive study of the background for the Pixel Vertex Detector at Belle {II}},
	url = {https://edoc.ub.uni-muenchen.de/19106/},
	institution = {Ludwig-Maximilians-Universität München},
	type = {phdthesis},
	author = {Moll, Andreas},
	urldate = {2023-10-04},
	date = {2015-07-31},
	langid = {german},
}

@article{kou_belle_2019,
	title = {The Belle {II} Physics Book},
	volume = {2019},
	issn = {2050-3911},
	url = {http://arxiv.org/abs/1808.10567},
	doi = {10.1093/ptep/ptz106},
	pages = {123C01},
	number = {12},
	journaltitle = {Progress of Theoretical and Experimental Physics},
	author = {Kou, E. and Urquijo, P. and Altmannshofer, W. and Beaujean, F. and Bell, G. and Beneke, M. and Bigi, I. I. and Blanke, F. Bishara M. and Bobeth, C. and Bona, M. and Brambilla, N. and Braun, V. M. and Brod, J. and Buras, A. J. and Cheng, H. Y. and Chiang, C. W. and Colangelo, G. and Czyz, H. and Datta, A. and De Fazio, F. and Deppisch, T. and Dolan, M. J. and Fajfer, S. and Feldmann, T. and Godfrey, S. and Gronau, M. and Grossman, Y. and Guo, F. K. and Haisch, U. and Hanhart, C. and Hashimoto, S. and Hirose, S. and Hisano, J. and Hofer, L. and Hoferichter, M. and Hou, W. S. and Huber, T. and Jahn, S. Jaeger S. and Jamin, M. and Jones, J. and Jung, M. and Kagan, A. L. and Kahlhoefer, F. and Kamenik, J. F. and Kaneko, T. and Kiyo, Y. and Kokulu, A. and Kosnik, N. and Kronfeld, A. S. and Ligeti, Z. and Logan, H. and Lu, C. D. and Lubicz, V. and Mahmoudi, F. and Maltman, K. and Misiak, M. and Mishima, S. and Moats, K. and Moussallam, B. and Nefediev, A. and Nierste, U. and Nomura, D. and Offen, N. and Olsen, S. L. and Passemar, E. and Paul, A. and Paz, G. and Petrov, A. A. and Pich, A. and Polosa, A. D. and Pradler, J. and Prelovsek, S. and Procura, M. and Ricciardi, G. and Robinson, D. J. and Roig, P. and Schacht, S. and Schmidt-Hoberg, K. and Schwichtenberg, J. and Sharpe, S. R. and Shigemitsu, J. and Shimizu, N. and Shimizu, Y. and Silvestrini, L. and Simula, S. and Smith, C. and Stoffer, P. and Straub, D. and Tackmann, F. J. and Tanaka, M. and Tayduganov, A. and Tetlalmatzi-Xolocotzi, G. and Teubner, T. and Vairo, A. and van Dyk, D. and Virto, J. and Was, Z. and Watanabe, R. and Watson, I. and Zupan, J. and Zwicky, R. and Abudinen, F. and Adachi, I. and Adamczyk, K. and Ahlburg, P. and Aihara, H. and Aloisio, A. and Andricek, L. and Ky, N. Anh and Arndt, M. and Asner, D. M. and Atmacan, H. and Aushev, T. and Aushev, V. and Ayad, R. and Aziz, T. and Baehr, S. and Bahinipati, S. and Bambade, P. and Ban, Y. and Barrett, M. and Baudot, J. and Behera, P. and Belous, K. and Bender, M. and Bennett, J. and Berger, M. and Bernieri, E. and Bernlochner, F. U. and Bessner, M. and Besson, D. and Bettarini, S. and Bhardwaj, V. and Bhuyan, B. and Bilka, T. and Bilmis, S. and Bilokin, S. and Bonvicini, G. and Bozek, A. and Bracko, M. and Branchini, P. and Braun, N. and Briere, R. A. and Browder, T. E. and Burmistrov, L. and Bussino, S. and Cao, L. and Caria, G. and Casarosa, G. and Cecchi, C. and Cervenkov, D. and Chang, M.-C. and Chang, P. and Cheaib, R. and Chekelian, V. and Chen, Y. and Cheon, B. G. and Chilikin, K. and Cho, K. and Choi, J. and Choi, S.-K. and Choudhury, S. and Cinabro, D. and Cremaldi, L. M. and Cuesta, D. and Cunliffe, S. and Dash, N. and Burelo, E. de la Cruz and De Lucia, E. and De Nardo, G. and De Nuccio, M. and De Pietro, G. and Hernandez, A. De Yta and Deschamps, B. and Destefanis, M. and Dey, S. and Di Capua, F. and Di Carlo, S. and Dingfelder, J. and Dolezal, Z. and Jimenez, I. Dominguez and Dong, T. V. and Dossett, D. and Duell, S. and Eidelman, S. and Epifanov, D. and Fast, J. E. and Ferber, T. and Fiore, S. and Fodor, A. and Forti, F. and Frey, A. and Frost, O. and Fulsom, B. G. and Gabriel, M. and Gabyshev, N. and Ganiev, E. and Gao, X. and Gao, B. and Garg, R. and Garmash, A. and Gaur, V. and Gaz, A. and Gessler, T. and Gebauer, U. and Gelb, M. and Gellrich, A. and Getzkow, D. and Giordano, R. and Giri, A. and Glazov, A. and Gobbo, B. and Godang, R. and Gogota, O. and Goldenzweig, P. and Golob, B. and Gradl, W. and Graziani, E. and Greco, M. and Greenwald, D. and Gribanov, S. and Guan, Y. and Guido, E. and Guo, A. and Halder, S. and Hara, K. and Hartbrich, O. and Hauth, T. and Hayasaka, K. and Hayashii, H. and Hearty, C. and De La Cruz, I. Heredia and Villanueva, M. Hernandez and Hershenhorn, A. and Higuchi, T. and Hoek, M. and Hollitt, S. and Van, N. T. Hong and Hsu, C.-L. and Hu, Y. and Huang, K. and Iijima, T. and Inami, K. and Inguglia, G. and Ishikawa, A. and Itoh, R. and Iwasaki, Y. and Iwasaki, M. and Jackson, P. and Jacobs, W. W. and Jaegle, I. and Jeon, H. B. and Ji, X. and Jia, S. and Jin, Y. and Joo, C. and Kuenzel, M. and Kadenko, I. and Kahn, J. and Kakuno, H. and Kaliyar, A. B. and Kandra, J. and Kang, K. H. and Kawasaki, T. and Ketter, C. and Khasmidatul, M. and Kichimi, H. and Kim, J. B. and Kim, K. T. and Kim, H. J. and Kim, D. Y. and Kim, K. and Kim, Y. and Kimmel, T. D. and Kindo, H. and Kinoshita, K. and Konno, T. and Korobov, A. and Korpar, S. and Kotchetkov, D. and Kowalewski, R. and Krizan, P. and Kroeger, R. and Krohn, J.-F. and Krokovny, P. and Kuehn, W. and Kuhr, T. and Kulasiri, R. and Kumar, M. and Kumar, R. and Kumita, T. and Kuzmin, A. and Kwon, Y.-J. and Lacaprara, S. and Lai, Y.-T. and Lalwani, K. and Lange, J. S. and Lee, S. C. and Lee, J. Y. and Leitl, P. and Levit, D. and Levonian, S. and Li, S. and Li, L. K. and Li, Y. and Li, Y. B. and Li, Q. and Gioi, L. Li and Libby, J. and Liptak, Z. and Liventsev, D. and Longo, S. and Loos, A. and Castro, G. Lopez and Lubej, M. and Lueck, T. and Luetticke, F. and Luo, T. and Mueller, F. and Mueller, Th and {MacQueen}, C. and Maeda, Y. and Maggiora, M. and Maity, S. and Manoni, E. and Marcello, S. and Marinas, C. and Hernandez, M. Martinez and Martini, A. and Matvienko, D. and {McKenna}, J. A. and Meier, F. and Merola, M. and Metzner, F. and Miller, C. and Miyabayashi, K. and Miyake, H. and Miyata, H. and Mizuk, R. and Mohanty, G. B. and Moon, H. K. and Moon, T. and Morda, A. and Morii, T. and Mrvar, M. and Muroyama, G. and Mussa, R. and Nakamura, I. and Nakano, T. and Nakao, M. and Nakayama, H. and Nakazawa, H. and Nanut, T. and Naruki, M. and Nath, K. J. and Nayak, M. and Nellikunnummel, N. and Neverov, D. and Niebuhr, C. and Ninkovic, J. and Nishida, S. and Nishimura, K. and Nouxman, M. and Nowak, G. and Ogawa, K. and Onishchuk, Y. and Ono, H. and Onuki, Y. and Pakhlov, P. and Pakhlova, G. and Pal, B. and Paoloni, E. and Park, H. and Park, C.-S. and Paschen, B. and Passeri, A. and Paul, S. and Pedlar, T. K. and Perello, M. and Peruzzi, I. M. and Pestotnik, R. and Piilonen, L. E. and Lerma, L. Podesta and Popov, V. and Prasanth, K. and Prencipe, E. and Prim, M. and Purohit, M. V. and Rabusov, A. and Rasheed, R. and Reiter, S. and Remnev, M. and Resmi, P. K. and Ripp-Baudot, I. and Ritter, M. and Ritzert, M. and Rizzo, G. and Rizzuto, L. and Robertson, S. H. and Perez, D. Rodriguez and Roney, J. M. and Rosenfeld, C. and Rostomyan, A. and Rout, N. and Rummel, S. and Russo, G. and Sahoo, D. and Sakai, Y. and Salehi, M. and Sanders, D. A. and Sandilya, S. and Sangal, A. and Santelj, L. and Sasaki, J. and Sato, Y. and Savinov, V. and Scavino, B. and Schram, M. and Schreeck, H. and Schueler, J. and Schwanda, C. and Schwartz, A. J. and Seddon, R. M. and Seino, Y. and Senyo, K. and Seon, O. and Seong, I. S. and Sevior, M. E. and Sfienti, C. and Shapkin, M. and Shen, C. P. and Shimomura, M. and Shiu, J.-G. and Shwartz, B. and Sibidanov, A. and Simon, F. and Singh, J. B. and Sinha, R. and Skambraks, S. and Smith, K. and Sobie, R. J. and Soffer, A. and Sokolov, A. and Solovieva, E. and Spruck, B. and Stanic, S. and Staric, M. and Starinsky, N. and Stolzenberg, U. and Stottler, Z. and Stroili, R. and Strube, J. F. and Stypula, J. and Sumihama, M. and Sumisawa, K. and Sumiyoshi, T. and Summers, D. and Sutcliffe, W. and Suzuki, S. Y. and Tabata, M. and Takahashi, M. and Takizawa, M. and Tamponi, U. and Tan, J. and Tanaka, S. and Tanida, K. and Taniguchi, N. and Tao, Y. and Taras, P. and Munoz, G. Tejeda and Tenchini, F. and Tippawan, U. and Torassa, E. and Trabelsi, K. and Tsuboyama, T. and Uchida, M. and Uehara, S. and Uglov, T. and Unno, Y. and Uno, S. and Ushiroda, Y. and Usov, Y. and Vahsen, S. E. and van Tonder, R. and Varner, G. and Varvell, K. E. and Vinokurova, A. and Vitale, L. and Vos, M. and Vossen, A. and Waheed, E. and Wakeling, H. and Wan, K. and Wang, M.-Z. and Wang, X. L. and Wang, B. and Warburton, A. and Webb, J. and Wehle, S. and Wessel, C. and Wiechczynski, J. and Wieduwilt, P. and Won, E. and Xu, Q. and Xu, X. and Yabsley, B. D. and Yamada, S. and Yamamoto, H. and Yan, W. and Yan, W. and Yang, S. B. and Ye, H. and Yeo, I. and Yin, J. H. and Yonenaga, M. and Yoshinobu, T. and Yuan, W. and Yuan, C. Z. and Yusa, Y. and Zakharov, S. and Zani, L. and Zeyrek, M. and Zhang, J. and Zhang, Y. and Zhang, Y. and Zhou, X. and Zhukova, V. and Zhulanov, V. and Zupanc, A.},
	urldate = {2023-10-04},
	date = {2019-12-01},
	eprinttype = {arxiv},
	eprint = {1808.10567 [hep-ex, physics:hep-lat, physics:hep-ph]},
}

@article{kobayashi_cp-violation_1973,
	title = {{CP}-Violation in the Renormalizable Theory of Weak Interaction},
	volume = {49},
	issn = {0033-068X},
	url = {https://doi.org/10.1143/PTP.49.652},
	doi = {10.1143/PTP.49.652},
	pages = {652--657},
	number = {2},
	journaltitle = {Progress of Theoretical Physics},
	shortjournal = {Progress of Theoretical Physics},
	author = {Kobayashi, Makoto and Maskawa, Toshihide},
	urldate = {2023-10-04},
	date = {1973-02-01},
}

@article{belle_collaboration_observation_2002,
	title = {Observation of Mixing-induced {CP} Violation in the Neutral B Meson System},
	volume = {66},
	issn = {0556-2821, 1089-4918},
	url = {http://arxiv.org/abs/hep-ex/0202027},
	doi = {10.1103/PhysRevD.66.032007},
	pages = {032007},
	number = {3},
	journaltitle = {Physical Review D},
	shortjournal = {Phys. Rev. D},
	author = {Belle Collaboration and Abe, K.},
	urldate = {2023-10-04},
	date = {2002-08-29},
	eprinttype = {arxiv},
	eprint = {hep-ex/0202027},
}

@article{agostinelli_geant4simulation_2003,
	title = {Geant4—a simulation toolkit},
	volume = {506},
	issn = {0168-9002},
	url = {https://www.sciencedirect.com/science/article/pii/S0168900203013688},
	doi = {10.1016/S0168-9002(03)01368-8},
	pages = {250--303},
	number = {3},
	journaltitle = {Nuclear Instruments and Methods in Physics Research Section A: Accelerators, Spectrometers, Detectors and Associated Equipment},
	shortjournal = {Nuclear Instruments and Methods in Physics Research Section A: Accelerators, Spectrometers, Detectors and Associated Equipment},
	author = {Agostinelli, S. and Allison, J. and Amako, K. and Apostolakis, J. and Araujo, H. and Arce, P. and Asai, M. and Axen, D. and Banerjee, S. and Barrand, G. and Behner, F. and Bellagamba, L. and Boudreau, J. and Broglia, L. and Brunengo, A. and Burkhardt, H. and Chauvie, S. and Chuma, J. and Chytracek, R. and Cooperman, G. and Cosmo, G. and Degtyarenko, P. and Dell'Acqua, A. and Depaola, G. and Dietrich, D. and Enami, R. and Feliciello, A. and Ferguson, C. and Fesefeldt, H. and Folger, G. and Foppiano, F. and Forti, A. and Garelli, S. and Giani, S. and Giannitrapani, R. and Gibin, D. and Gómez Cadenas, J. J. and González, I. and Gracia Abril, G. and Greeniaus, G. and Greiner, W. and Grichine, V. and Grossheim, A. and Guatelli, S. and Gumplinger, P. and Hamatsu, R. and Hashimoto, K. and Hasui, H. and Heikkinen, A. and Howard, A. and Ivanchenko, V. and Johnson, A. and Jones, F. W. and Kallenbach, J. and Kanaya, N. and Kawabata, M. and Kawabata, Y. and Kawaguti, M. and Kelner, S. and Kent, P. and Kimura, A. and Kodama, T. and Kokoulin, R. and Kossov, M. and Kurashige, H. and Lamanna, E. and Lampén, T. and Lara, V. and Lefebure, V. and Lei, F. and Liendl, M. and Lockman, W. and Longo, F. and Magni, S. and Maire, M. and Medernach, E. and Minamimoto, K. and Mora de Freitas, P. and Morita, Y. and Murakami, K. and Nagamatu, M. and Nartallo, R. and Nieminen, P. and Nishimura, T. and Ohtsubo, K. and Okamura, M. and O'Neale, S. and Oohata, Y. and Paech, K. and Perl, J. and Pfeiffer, A. and Pia, M. G. and Ranjard, F. and Rybin, A. and Sadilov, S. and Di Salvo, E. and Santin, G. and Sasaki, T. and Savvas, N. and Sawada, Y. and Scherer, S. and Sei, S. and Sirotenko, V. and Smith, D. and Starkov, N. and Stoecker, H. and Sulkimo, J. and Takahata, M. and Tanaka, S. and Tcherniaev, E. and Safai Tehrani, E. and Tropeano, M. and Truscott, P. and Uno, H. and Urban, L. and Urban, P. and Verderi, M. and Walkden, A. and Wander, W. and Weber, H. and Wellisch, J. P. and Wenaus, T. and Williams, D. C. and Wright, D. and Yamada, T. and Yoshida, H. and Zschiesche, D.},
	urldate = {2023-10-03},
	date = {2003-07-01},
}

@online{noauthor_belle_nodate,
	title = {Belle {II} Software Documentation — basf2 development documentation},
	url = {https://software.belle2.org/development/sphinx/index.html},
	urldate = {2023-10-03},
}

@online{noauthor_nobel_nodate,
	title = {The Nobel Prize in Physics 2008},
	url = {https://www.nobelprize.org/prizes/physics/2008/popular-information/},
	titleaddon = {{NobelPrize}.org},
	urldate = {2023-10-03},
	langid = {american},
}

@misc{abe_belle_2010,
	title = {Belle {II} Technical Design Report},
	url = {http://arxiv.org/abs/1011.0352},
	doi = {10.48550/arXiv.1011.0352},
	number = {{arXiv}:1011.0352},
	publisher = {{arXiv}},
	author = {Abe, T. and Adachi, I. and Adamczyk, K. and Ahn, S. and Aihara, H. and Akai, K. and Aloi, M. and Andricek, L. and Aoki, K. and Arai, Y. and Arefiev, A. and Arinstein, K. and Arita, Y. and Asner, D. M. and Aulchenko, V. and Aushev, T. and Aziz, T. and Bakich, A. M. and Balagura, V. and Ban, Y. and Barberio, E. and Barvich, T. and Belous, K. and Bergauer, T. and Bhardwaj, V. and Bhuyan, B. and Blyth, S. and Bondar, A. and Bonvicini, G. and Bozek, A. and Bracko, M. and Brodzicka, J. and Brovchenko, O. and Browder, T. E. and Cao, G. and Chang, M.-C. and Chang, P. and Chao, Y. and Chekelian, V. and Chen, A. and Chen, K.-F. and Chen, P. and Cheon, B. G. and Chiang, C.-C. and Chistov, R. and Cho, K. and Choi, S.-K. and Chung, K. and Comerma, A. and Cooney, M. and Cowley, D. E. and Critchlow, T. and Dalseno, J. and Danilov, M. and Dieguez, A. and Dierlamm, A. and Dillon, M. and Dingfelder, J. and Dolenec, R. and Dolezal, Z. and Drasal, Z. and Drutskoy, A. and Dungel, W. and Dutta, D. and Eidelman, S. and Enomoto, A. and Epifanov, D. and Esen, S. and Fast, J. E. and Feindt, M. and Garcia, M. Fernandez and Fifield, T. and Fischer, P. and Flanagan, J. and Fourletov, S. and Fourletova, J. and Freixas, L. and Frey, A. and Friedl, M. and Fruehwirth, R. and Fujii, H. and Fujikawa, M. and Fukuma, Y. and Funakoshi, Y. and Furukawa, K. and Fuster, J. and Gabyshev, N. and Cueto, A. Gaspar de Valenzuela and Garmash, A. and Garrido, L. and Geisler, Ch and Gfall, I. and Goh, Y. M. and Golob, B. and Gorton, I. and Grzymkowski, R. and Guo, H. and Ha, H. and Haba, J. and Hara, K. and Hara, T. and Haruyama, T. and Hayasaka, K. and Hayashi, K. and Hayashii, H. and Heck, M. and Heindl, S. and Heller, C. and Hemperek, T. and Higuchi, T. and Horii, Y. and Hou, W.-S. and Hsiung, Y. B. and Huang, C.-H. and Hwang, S. and Hyun, H. J. and Igarashi, Y. and Iglesias, C. and Iida, Y. and Iijima, T. and Imamura, M. and Inami, K. and Irmler, C. and Ishizuka, M. and Itagaki, K. and Itoh, R. and Iwabuchi, M. and Iwai, G. and Iwai, M. and Iwasaki, M. and Iwasaki, M. and Iwasaki, Y. and Iwashita, T. and Iwata, S. and Jang, H. and Ji, X. and Jinno, T. and Jones, M. and Julius, T. and Kageyama, T. and Kah, D. H. and Kakuno, H. and Kamitani, T. and Kanazawa, K. and Kapusta, P. and Kataoka, S. U. and Katayama, N. and Kawai, M. and Kawai, Y. and Kawasaki, T. and Kennedy, J. and Kichimi, H. and Kikuchi, M. and Kiesling, C. and Kim, B. K. and Kim, G. N. and Kim, H. J. and Kim, H. O. and Kim, J.-B. and Kim, J. H. and Kim, M. J. and Kim, S. K. and Kim, K. T. and Kim, T. Y. and Kinoshita, K. and Kishi, K. and Kisielewski, B. and van Dam, K. Kleese and Knopf, J. and Ko, B. R. and Koch, M. and Kodys, P. and Koffmane, C. and Koga, Y. and Kohriki, T. and Koike, S. and Koiso, H. and Kondo, Y. and Korpar, S. and Kouzes, R. T. and Kreidl, Ch and Kreps, M. and Krizan, P. and Krokovny, P. and Krueger, H. and Kruth, A. and Kuhn, W. and Kuhr, T. and Kumar, R. and Kumita, T. and Kupper, S. and Kuzmin, A. and Kvasnicka, P. and Kwon, Y.-J. and Lacasta, C. and Lange, J. S. and Lee, I.-S. and Lee, M. J. and Lee, M. W. and Lee, S.-H. and Lemarenko, M. and Li, J. and Li, W. D. and Li, Y. and Libby, J. and Limosani, A. and Liu, C. and Liu, H. and Liu, Y. and Liu, Z. and Liventsev, D. and Virto, A. Lopez and Makida, Y. and Mao, Z. P. and Marinas, C. and Masuzawa, M. and Matvienko, D. and Mitaroff, W. and Miyabayashi, K. and Miyata, H. and Miyazaki, Y. and Miyoshi, T. and Mizuk, R. and Mohanty, G. B. and Mohapatra, D. and Moll, A. and Mori, T. and Morita, A. and Morita, Y. and Moser, H.-G. and Martin, D. Moya and Mueller, T. and Muenchow, D. and Murakami, J. and Myung, S. S. and Nagamine, T. and Nakamura, I. and Nakamura, T. T. and Nakano, E. and Nakano, H. and Nakao, M. and Nakazawa, H. and Nam, S.-H. and Natkaniec, Z. and Nedelkovska, E. and Negishi, K. and Neubauer, S. and Ng, C. and Ninkovic, J. and Nishida, S. and Nishimura, K. and Novikov, E. and Nozaki, T. and Ogawa, S. and Ohmi, K. and Ohnishi, Y. and Ohshima, T. and Ohuchi, N. and Oide, K. and Olsen, S. L. and Ono, M. and Ono, Y. and Onuki, Y. and Ostrowicz, W. and Ozaki, H. and Pakhlov, P. and Pakhlova, G. and Palka, H. and Park, H. and Park, H. K. and Peak, L. S. and Peng, T. and Peric, I. and Pernicka, M. and Pestotnik, R. and Petric, M. and Piilonen, L. E. and Poluektov, A. and Prim, M. and Prothmann, K. and Regimbal, K. and Reisert, B. and Richter, R. H. and Riera-Babures, J. and Ritter, A. and Ritter, A. and Ritter, M. and Roehrken, M. and Rorie, J. and Rosen, M. and Rozanska, M. and Ruckman, L. and Rummel, S. and Rusinov, V. and Russell, R. M. and Ryu, S. and Sahoo, H. and Sakai, K. and Sakai, Y. and Santelj, L. and Sasaki, T. and Sato, N. and Sato, Y. and Scheirich, J. and Schieck, J. and Schwanda, C. and Schwartz, A. J. and Schwenker, B. and Seljak, A. and Senyo, K. and Seon, O.-S. and Sevior, M. E. and Shapkin, M. and Shebalin, V. and Shen, C. P. and Shibuya, H. and Shiizuka, S. and Shiu, J.-G. and Shwartz, B. and Simon, F. and Simonis, H. J. and Singh, J. B. and Sinha, R. and Sitarz, M. and Smerkol, P. and Sokolov, A. and Solovieva, E. and Stanic, S. and Staric, M. and Stypula, J. and Suetsugu, Y. and Sugihara, S. and Sugimura, T. and Sumisawa, K. and Sumiyoshi, T. and Suzuki, K. and Suzuki, S. Y. and Takagaki, H. and Takasaki, F. and Takeichi, H. and Takubo, Y. and Tanaka, M. and Tanaka, S. and Taniguchi, N. and Tarkovsky, E. and Tatishvili, G. and Tawada, M. and Taylor, G. N. and Teramoto, Y. and Tikhomirov, I. and Trabelsi, K. and Tsuboyama, T. and Tsunada, K. and Tu, Y.-C. and Uchida, T. and Uehara, S. and Ueno, K. and Uglov, T. and Unno, Y. and Uno, S. and Urquijo, P. and Ushiroda, Y. and Usov, Y. and Vahsen, S. and Valentan, M. and Vanhoefer, P. and Varner, G. and Varvell, K. E. and Vazquez, P. and Vila, I. and Vilella, E. and Vinokurova, A. and Visniakov, J. and Vos, M. and Wang, C. H. and Wang, J. and Wang, M.-Z. and Wang, P. and Wassatch, A. and Watanabe, M. and Watase, Y. and Weiler, T. and Wermes, N. and Wescott, R. E. and White, E. and Wicht, J. and Widhalm, L. and Williams, K. M. and Won, E. and Xu, H. and Yabsley, B. D. and Yamamoto, H. and Yamaoka, H. and Yamaoka, Y. and Yamauchi, M. and Yin, Y. and Yoon, H. and Yu, J. and Yuan, C. Z. and Yusa, Y. and Zander, D. and Zdybal, M. and Zhang, Z. P. and Zhao, J. and Zhao, L. and Zhao, Z. and Zhilich, V. and Zhou, P. and Zhulanov, V. and Zivko, T. and Zupanc, A. and Zyukova, O.},
	urldate = {2023-10-03},
	date = {2010-11-01},
	eprinttype = {arxiv},
	eprint = {1011.0352 [hep-ex, physics:physics]},
}

@online{noauthor_superkekb_2019,
	title = {{SuperKEKB} Phase 3 (Belle {II} Physics Run) Starts},
	url = {https://www.kek.jp/en/newsroom/2019/03/11/1600/},
	titleaddon = {News},
	urldate = {2023-10-03},
	date = {2019-03-11},
	langid = {english},
}

@article{bai_deep_2020,
	title = {Deep learning methods for solving linear inverse problems: Research directions and paradigms},
	volume = {177},
	issn = {0165-1684},
	url = {https://www.sciencedirect.com/science/article/pii/S0165168420302723},
	doi = {10.1016/j.sigpro.2020.107729},
	shorttitle = {Deep learning methods for solving linear inverse problems},
	pages = {107729},
	journaltitle = {Signal Processing},
	shortjournal = {Signal Processing},
	author = {Bai, Yanna and Chen, Wei and Chen, Jie and Guo, Weisi},
	urldate = {2023-09-29},
	date = {2020-12-01},
}

@article{bingham_inverse_2024,
	title = {Inverse Problems for Physics-Based Process Models},
	volume = {11},
	url = {https://doi.org/10.1146/annurev-statistics-031017-100108},
	doi = {10.1146/annurev-statistics-031017-100108},
	pages = {null},
	number = {1},
	journaltitle = {Annual Review of Statistics and Its Application},
	author = {Bingham, Derek and Butler, Troy and Estep, Don},
	urldate = {2023-09-29},
	date = {2024},
	note = {\_eprint: https://doi.org/10.1146/annurev-statistics-031017-100108},
}

@article{knight_sinkhornknopp_2008,
	title = {The Sinkhorn–Knopp Algorithm: Convergence and Applications},
	volume = {30},
	issn = {0895-4798},
	url = {https://epubs.siam.org/doi/10.1137/060659624},
	doi = {10.1137/060659624},
	shorttitle = {The Sinkhorn–Knopp Algorithm},
	pages = {261--275},
	number = {1},
	journaltitle = {{SIAM} Journal on Matrix Analysis and Applications},
	shortjournal = {{SIAM} J. Matrix Anal. Appl.},
	author = {Knight, Philip A.},
	urldate = {2023-09-27},
	date = {2008-01},
	note = {Publisher: Society for Industrial and Applied Mathematics},
}

@misc{dudzik_graph_2022,
	title = {Graph Neural Networks are Dynamic Programmers},
	url = {http://arxiv.org/abs/2203.15544},
	doi = {10.48550/arXiv.2203.15544},
	number = {{arXiv}:2203.15544},
	publisher = {{arXiv}},
	author = {Dudzik, Andrew and Veličković, Petar},
	urldate = {2023-09-26},
	date = {2022-10-10},
	eprinttype = {arxiv},
	eprint = {2203.15544 [cs, math, stat]},
}

@inproceedings{de_haan_natural_2020,
	title = {Natural Graph Networks},
	volume = {33},
	url = {https://proceedings.neurips.cc/paper/2020/hash/2517756c5a9be6ac007fe9bb7fb92611-Abstract.html},
	pages = {3636--3646},
	booktitle = {Advances in Neural Information Processing Systems},
	publisher = {Curran Associates, Inc.},
	author = {de Haan, Pim and Cohen, Taco S and Welling, Max},
	urldate = {2023-09-26},
	date = {2020},
}

@misc{nemecek_coinductive_2023,
	title = {Coinductive guide to inductive transformer heads},
	url = {http://arxiv.org/abs/2302.01834},
	number = {{arXiv}:2302.01834},
	publisher = {{arXiv}},
	author = {Nemecek, Adam},
	urldate = {2023-09-26},
	date = {2023-02-03},
	eprinttype = {arxiv},
	eprint = {2302.01834 [cs]},
}

@misc{ong_learnable_2022,
	title = {Learnable Commutative Monoids for Graph Neural Networks},
	url = {http://arxiv.org/abs/2212.08541},
	number = {{arXiv}:2212.08541},
	publisher = {{arXiv}},
	author = {Ong, Euan and Veličković, Petar},
	urldate = {2023-09-26},
	date = {2022-12-16},
	eprinttype = {arxiv},
	eprint = {2212.08541 [cs, stat]},
}

@article{chen_rise_2018,
	title = {The rise of deep learning in drug discovery},
	volume = {23},
	issn = {1359-6446},
	url = {https://www.sciencedirect.com/science/article/pii/S1359644617303598},
	doi = {10.1016/j.drudis.2018.01.039},
	pages = {1241--1250},
	number = {6},
	journaltitle = {Drug Discovery Today},
	shortjournal = {Drug Discovery Today},
	author = {Chen, Hongming and Engkvist, Ola and Wang, Yinhai and Olivecrona, Marcus and Blaschke, Thomas},
	urldate = {2023-09-26},
	date = {2018-06-01},
}

@article{meyers_novo_2021,
	title = {De novo molecular design and generative models},
	volume = {26},
	issn = {1878-5832},
	doi = {10.1016/j.drudis.2021.05.019},
	pages = {2707--2715},
	number = {11},
	journaltitle = {Drug Discovery Today},
	shortjournal = {Drug Discov Today},
	author = {Meyers, Joshua and Fabian, Benedek and Brown, Nathan},
	date = {2021-11},
	pmid = {34082136},
}

@online{noauthor_novo_nodate,
	title = {De novo Molecular Design {\textbar} Wiley Online Books},
	url = {https://onlinelibrary.wiley.com/doi/book/10.1002/9783527677016},
	urldate = {2023-09-26},
}

@article{lee_exploring_2022,
	title = {Exploring Chemical Space with Score-based Out-of-distribution Generation},
	url = {https://openreview.net/forum?id=45TeQUJw9tn},
	author = {Lee, Seul and Jo, Jaehyeong and Hwang, Sung Ju},
	urldate = {2023-09-26},
	date = {2022-09-29},
	langid = {english},
}

@article{lotfollahi_conditional_2020,
	title = {Conditional out-of-distribution generation for unpaired data using transfer {VAE}},
	volume = {36},
	issn = {1367-4803},
	url = {https://doi.org/10.1093/bioinformatics/btaa800},
	doi = {10.1093/bioinformatics/btaa800},
	pages = {i610--i617},
	issue = {Supplement\_2},
	journaltitle = {Bioinformatics},
	shortjournal = {Bioinformatics},
	author = {Lotfollahi, Mohammad and Naghipourfar, Mohsen and Theis, Fabian J and Wolf, F Alexander},
	urldate = {2023-09-26},
	date = {2020-12-30},
}

@misc{kocaoglu_causalgan_2017,
	title = {{CausalGAN}: Learning Causal Implicit Generative Models with Adversarial Training},
	url = {http://arxiv.org/abs/1709.02023},
	doi = {10.48550/arXiv.1709.02023},
	shorttitle = {{CausalGAN}},
	number = {{arXiv}:1709.02023},
	publisher = {{arXiv}},
	author = {Kocaoglu, Murat and Snyder, Christopher and Dimakis, Alexandros G. and Vishwanath, Sriram},
	urldate = {2023-09-26},
	date = {2017-09-14},
	eprinttype = {arxiv},
	eprint = {1709.02023 [cs, math, stat]},
}

@misc{feng_principled_2022,
	title = {Principled Knowledge Extrapolation with {GANs}},
	url = {http://arxiv.org/abs/2205.13444},
	doi = {10.48550/arXiv.2205.13444},
	number = {{arXiv}:2205.13444},
	publisher = {{arXiv}},
	author = {Feng, Ruili and Xiao, Jie and Zheng, Kecheng and Zhao, Deli and Zhou, Jingren and Sun, Qibin and Zha, Zheng-Jun},
	urldate = {2023-09-26},
	date = {2022-05-21},
	eprinttype = {arxiv},
	eprint = {2205.13444 [cs]},
}

@misc{xu_how_2021,
	title = {How Neural Networks Extrapolate: From Feedforward to Graph Neural Networks},
	url = {http://arxiv.org/abs/2009.11848},
	doi = {10.48550/arXiv.2009.11848},
	shorttitle = {How Neural Networks Extrapolate},
	number = {{arXiv}:2009.11848},
	publisher = {{arXiv}},
	author = {Xu, Keyulu and Zhang, Mozhi and Li, Jingling and Du, Simon S. and Kawarabayashi, Ken-ichi and Jegelka, Stefanie},
	urldate = {2023-09-26},
	date = {2021-03-02},
	eprinttype = {arxiv},
	eprint = {2009.11848 [cs, stat]},
}

@misc{arora_generalization_2017,
	title = {Generalization and Equilibrium in Generative Adversarial Nets ({GANs})},
	url = {http://arxiv.org/abs/1703.00573},
	doi = {10.48550/arXiv.1703.00573},
	number = {{arXiv}:1703.00573},
	publisher = {{arXiv}},
	author = {Arora, Sanjeev and Ge, Rong and Liang, Yingyu and Ma, Tengyu and Zhang, Yi},
	urldate = {2023-09-26},
	date = {2017-08-01},
	eprinttype = {arxiv},
	eprint = {1703.00573 [cs, stat]},
}

@article{pedro_current_2019,
	title = {Current and Future Performance of the {CMS} Simulation},
	volume = {214},
	rights = {© The Authors, published by {EDP} Sciences, 2019},
	issn = {2100-014X},
	url = {https://www.epj-conferences.org/articles/epjconf/abs/2019/19/epjconf_chep2018_02036/epjconf_chep2018_02036.html},
	doi = {10.1051/epjconf/201921402036},
	pages = {02036},
	journaltitle = {{EPJ} Web of Conferences},
	shortjournal = {{EPJ} Web Conf.},
	author = {Pedro, Kevin},
	urldate = {2023-09-26},
	date = {2019},
	langid = {english},
	note = {Publisher: {EDP} Sciences},
}

@online{noauthor_frontiers_nodate,
	title = {Frontiers {\textbar} Novel 3D Pixel Sensors for the Upgrade of the {ATLAS} Inner Tracker},
	url = {https://www.frontiersin.org/articles/10.3389/fphy.2021.624668/full},
	urldate = {2023-09-26},
}

@misc{yang_xlnet_2020,
	title = {{XLNet}: Generalized Autoregressive Pretraining for Language Understanding},
	url = {http://arxiv.org/abs/1906.08237},
	doi = {10.48550/arXiv.1906.08237},
	shorttitle = {{XLNet}},
	number = {{arXiv}:1906.08237},
	publisher = {{arXiv}},
	author = {Yang, Zhilin and Dai, Zihang and Yang, Yiming and Carbonell, Jaime and Salakhutdinov, Ruslan and Le, Quoc V.},
	urldate = {2023-09-26},
	date = {2020-01-02},
	eprinttype = {arxiv},
	eprint = {1906.08237 [cs]},
}

@misc{liu_roberta_2019,
	title = {{RoBERTa}: A Robustly Optimized {BERT} Pretraining Approach},
	url = {http://arxiv.org/abs/1907.11692},
	doi = {10.48550/arXiv.1907.11692},
	shorttitle = {{RoBERTa}},
	number = {{arXiv}:1907.11692},
	publisher = {{arXiv}},
	author = {Liu, Yinhan and Ott, Myle and Goyal, Naman and Du, Jingfei and Joshi, Mandar and Chen, Danqi and Levy, Omer and Lewis, Mike and Zettlemoyer, Luke and Stoyanov, Veselin},
	urldate = {2023-09-26},
	date = {2019-07-26},
	eprinttype = {arxiv},
	eprint = {1907.11692 [cs]},
}

@online{noauthor_pdf_nodate,
	title = {[{PDF}] An empirical study on evaluation metrics of generative adversarial networks {\textbar} Semantic Scholar},
	url = {https://www.semanticscholar.org/paper/An-empirical-study-on-evaluation-metrics-of-Xu-Huang/a476f45d867c9a2f953c44e14fec35a6e6af27a0},
	urldate = {2023-09-25},
}

@misc{kingma_adam_2017,
	title = {Adam: A Method for Stochastic Optimization},
	url = {http://arxiv.org/abs/1412.6980},
	doi = {10.48550/arXiv.1412.6980},
	shorttitle = {Adam},
	number = {{arXiv}:1412.6980},
	publisher = {{arXiv}},
	author = {Kingma, Diederik P. and Ba, Jimmy},
	urldate = {2023-09-25},
	date = {2017-01-29},
	eprinttype = {arxiv},
	eprint = {1412.6980 [cs]},
}

@misc{liu_learning_2021,
	title = {Learning with Hyperspherical Uniformity},
	url = {http://arxiv.org/abs/2103.01649},
	number = {{arXiv}:2103.01649},
	publisher = {{arXiv}},
	author = {Liu, Weiyang and Lin, Rongmei and Liu, Zhen and Xiong, Li and Schölkopf, Bernhard and Weller, Adrian},
	urldate = {2023-09-25},
	date = {2021-11-16},
	eprinttype = {arxiv},
	eprint = {2103.01649 [cs]},
}

@article{beran_testing_1968,
	title = {Testing for uniformity on a compact homogeneous space},
	volume = {5},
	issn = {0021-9002, 1475-6072},
	url = {https://www.cambridge.org/core/journals/journal-of-applied-probability/article/abs/testing-for-uniformity-on-a-compact-homogeneous-space/1F98A87CC8BC7DE51D2C55D6CD66AA86},
	doi = {10.2307/3212085},
	pages = {177--195},
	number = {1},
	journaltitle = {Journal of Applied Probability},
	author = {Beran, R. J.},
	urldate = {2023-09-25},
	date = {1968-04},
	langid = {english},
	note = {Publisher: Cambridge University Press},
}

@article{ajne_simple_1968,
	title = {A simple test for uniformity of a circular distribution},
	volume = {55},
	issn = {0006-3444},
	url = {https://doi.org/10.1093/biomet/55.2.343},
	doi = {10.1093/biomet/55.2.343},
	pages = {343--354},
	number = {2},
	journaltitle = {Biometrika},
	shortjournal = {Biometrika},
	author = {{AJNE}, B.},
	urldate = {2023-09-25},
	date = {1968-07-01},
}

@misc{garcia-portugues_overview_2018,
	title = {An overview of uniformity tests on the hypersphere},
	url = {http://arxiv.org/abs/1804.00286},
	number = {{arXiv}:1804.00286},
	publisher = {{arXiv}},
	author = {García-Portugués, Eduardo and Verdebout, Thomas},
	urldate = {2023-09-25},
	date = {2018-04-03},
	eprinttype = {arxiv},
	eprint = {1804.00286 [stat]},
}

@article{jammalamadaka_sobolev_2020,
	title = {On Sobolev tests of uniformity on the circle with an extension to the sphere},
	volume = {26},
	issn = {1350-7265},
	url = {https://projecteuclid.org/journals/bernoulli/volume-26/issue-3/On-Sobolev-tests-of-uniformity-on-the-circle-with-an/10.3150/19-BEJ1191.full},
	doi = {10.3150/19-BEJ1191},
	pages = {2226--2252},
	number = {3},
	journaltitle = {Bernoulli},
	author = {Jammalamadaka, Sreenivasa Rao and Meintanis, Simos and Verdebout, Thomas},
	urldate = {2023-09-25},
	date = {2020-08},
	note = {Publisher: Bernoulli Society for Mathematical Statistics and Probability},
}

@online{noauthor_new_nodate,
	title = {A New Sobolev Test for Uniformity on the Circle on {JSTOR}},
	url = {https://www.jstor.org/stable/2336748},
	urldate = {2023-09-25},
}

@misc{zhao_feature_2020,
	title = {Feature Quantization Improves {GAN} Training},
	url = {http://arxiv.org/abs/2004.02088},
	doi = {10.48550/arXiv.2004.02088},
	number = {{arXiv}:2004.02088},
	publisher = {{arXiv}},
	author = {Zhao, Yang and Li, Chunyuan and Yu, Ping and Gao, Jianfeng and Chen, Changyou},
	urldate = {2023-09-25},
	date = {2020-07-14},
	eprinttype = {arxiv},
	eprint = {2004.02088 [cs, stat]},
}

@article{bruning_chapter_2020,
	title = {Chapter 1: High-Luminosity Large Hadron Collider},
	volume = {10},
	rights = {Copyright (c) 2020 {CERN}},
	issn = {2519-8076},
	url = {https://e-publishing.cern.ch/index.php/CYRM/article/view/1153},
	doi = {10.23731/CYRM-2020-0010.1},
	shorttitle = {Chapter 1},
	pages = {1--1},
	journaltitle = {{CERN} Yellow Reports: Monographs},
	author = {Brüning, O. and Rossi, L.},
	urldate = {2023-09-25},
	date = {2020-12-17},
	langid = {english},
}

@inproceedings{strobbe_status_2021,
	location = {zoom},
	title = {Status and progress on the {CMS} Phase-{II} detector upgrades},
	url = {https://pos.sissa.it/397/318},
	doi = {10.22323/1.397.0318},
	eventtitle = {The Ninth Annual Conference on Large Hadron Collider Physics},
	pages = {318},
	booktitle = {Proceedings of The Ninth Annual Conference on Large Hadron Collider Physics — {PoS}({LHCP}2021)},
	publisher = {Sissa Medialab},
	author = {Strobbe, Nadja and {CMS Collaboration}},
	urldate = {2023-09-25},
	date = {2021-10-26},
	langid = {english},
}

@article{boelts_flexible_2022,
	title = {Flexible and efficient simulation-based inference for models of decision-making},
	volume = {11},
	issn = {2050-084X},
	url = {https://doi.org/10.7554/eLife.77220},
	doi = {10.7554/eLife.77220},
	pages = {e77220},
	journaltitle = {{eLife}},
	author = {Boelts, Jan and Lueckmann, Jan-Matthis and Gao, Richard and Macke, Jakob H},
	editor = {Wyart, Valentin and Behrens, Timothy E and Acerbi, Luigi and Daunizeau, Jean},
	urldate = {2023-09-24},
	date = {2022-07-27},
	note = {Publisher: {eLife} Sciences Publications, Ltd},
}

@article{you_graphrnn_nodate,
	title = {{GraphRNN}: Generating Realistic Graphs with Deep Auto-regressive Models},
	author = {You, Jiaxuan and Ying, Rex and Ren, Xiang and Hamilton, William L and Leskovec, Jure},
	langid = {english},
}

@inproceedings{chen_pixelsnail_2018,
	title = {{PixelSNAIL}: An Improved Autoregressive Generative Model},
	url = {https://proceedings.mlr.press/v80/chen18h.html},
	shorttitle = {{PixelSNAIL}},
	eventtitle = {International Conference on Machine Learning},
	pages = {864--872},
	booktitle = {Proceedings of the 35th International Conference on Machine Learning},
	publisher = {{PMLR}},
	author = {Chen, X. I. and Mishra, Nikhil and Rohaninejad, Mostafa and Abbeel, Pieter},
	urldate = {2023-09-24},
	date = {2018-07-03},
	langid = {english},
	note = {{ISSN}: 2640-3498},
}

@misc{oord_conditional_2016,
	title = {Conditional Image Generation with {PixelCNN} Decoders},
	url = {http://arxiv.org/abs/1606.05328},
	doi = {10.48550/arXiv.1606.05328},
	number = {{arXiv}:1606.05328},
	publisher = {{arXiv}},
	author = {Oord, Aaron van den and Kalchbrenner, Nal and Vinyals, Oriol and Espeholt, Lasse and Graves, Alex and Kavukcuoglu, Koray},
	urldate = {2023-09-24},
	date = {2016-06-18},
	eprinttype = {arxiv},
	eprint = {1606.05328 [cs]},
}

@misc{wu_logan_2020,
	title = {{LOGAN}: Latent Optimisation for Generative Adversarial Networks},
	url = {http://arxiv.org/abs/1912.00953},
	doi = {10.48550/arXiv.1912.00953},
	shorttitle = {{LOGAN}},
	number = {{arXiv}:1912.00953},
	publisher = {{arXiv}},
	author = {Wu, Yan and Donahue, Jeff and Balduzzi, David and Simonyan, Karen and Lillicrap, Timothy},
	urldate = {2023-09-22},
	date = {2020-07-01},
	eprinttype = {arxiv},
	eprint = {1912.00953 [cs, stat]},
}

@article{velickovic_everything_2023,
	title = {Everything is Connected: Graph Neural Networks},
	volume = {79},
	issn = {0959440X},
	url = {http://arxiv.org/abs/2301.08210},
	doi = {10.1016/j.sbi.2023.102538},
	shorttitle = {Everything is Connected},
	pages = {102538},
	journaltitle = {Current Opinion in Structural Biology},
	shortjournal = {Current Opinion in Structural Biology},
	author = {Veličković, Petar},
	urldate = {2023-09-22},
	date = {2023-04},
	eprinttype = {arxiv},
	eprint = {2301.08210 [cs, stat]},
}

@article{fukushima_neocognitron_1980,
	title = {Neocognitron: A self-organizing neural network model for a mechanism of pattern recognition unaffected by shift in position},
	volume = {36},
	issn = {0340-1200, 1432-0770},
	url = {http://link.springer.com/10.1007/BF00344251},
	doi = {10.1007/BF00344251},
	shorttitle = {Neocognitron},
	pages = {193--202},
	number = {4},
	journaltitle = {Biological Cybernetics},
	shortjournal = {Biol. Cybernetics},
	author = {Fukushima, Kunihiko},
	urldate = {2023-09-22},
	date = {1980-04},
	langid = {english},
}

@article{buades_non-local_2011,
	title = {Non-Local Means Denoising},
	volume = {1},
	issn = {2105-1232},
	url = {https://www.ipol.im/pub/art/2011/bcm_nlm/?utm_source=doi},
	doi = {10.5201/ipol.2011.bcm_nlm},
	pages = {208--212},
	journaltitle = {Image Processing On Line},
	author = {Buades, Antoni and Coll, Bartomeu and Morel, Jean-Michel},
	urldate = {2023-09-21},
	date = {2011-09-13},
	langid = {english},
}

@incollection{giakoustidis_status_2023,
	title = {Status of the {BELLE} {II} Pixel Detector},
	volume = {420},
	url = {https://pos.sissa.it/420/005},
	pages = {005},
	booktitle = {Proceedings of 10th International Workshop on Semiconductor Pixel Detectors for Particles and Imaging — {PoS}(Pixel2022)},
	publisher = {{SISSA} Medialab},
	author = {Giakoustidis, Georgios and Belle-{II} {DEPFET} \{and\} {PXD} Collaboration and Abudinen, Fernando and Ackermann, Karlheinz and Ahlburg, Patrick and Albalawi, Mohammed and Alonso, Oscar and Andricek, Laci and Ayad, Rachid and Babu, Varghese and Baur, Anselm and Bernlochner, Florian and Bilka, Tadeas and Bolz, Arthur and Bozek, Andrzej and Camien, Christian and Caldwell, Allen Christopher and Cao, Lu and Chekelian, Vladimir and Dieguez, Angel and Dingfelder, Jochen Christian and Dolezal, Zdenek and Fras, Markus and Frey, Ariane and Gabriel, Miroslav and Gadow, Karsten and Gessler, Thomas and Getzkow, Dennis and Gioi, Luigi Li and Greenwald, Daniel and Heck, Martin and Hensel, Martin and Hoek, Matthias and Huber, Stefan and Kandra, Jakub and Kapusta, Pjotr and Karl, Robert and Kehl, Jasper and Kiesling, Christian and Kisielewski, Bartlomiej and Kittlinger, David and Klose, Daniel and Kodys, Peter and Koffmane, Christian and Konorov, Igor and Krein, Matthäus and Krivokuca, Silvia and Kuhr, Thomas and Kurz, Simon and Kvasnicka, Peter and Lange, Jens Sören and Lautenbach, Klemens and Leis, Ullrich and Leitl, Philipp and Levit, Dmytro and Liemann, Gerhard and Liu, Qingyuan and Liu, Zhen’An and Lück, Thomas and Marinas, Carlos and Mccarney, Sara and Moser, Hans-Günther and Moya, David and Müller, Felix Johannes and Müller, Felix and Niebuhr, Carsten and Ninkovic, Jelena and Paschen, Botho and Paul, Stephan and Peric, Ivan and Pitzl, Daniel and Rabusov, Andrei and Reif, Markus and Reiter, Simon Patrik and Richter, Rainer and Ritter, Martin and Ritzert, Michael and Sanchez, Javier Gonzalez and Scavino, Bianca and Schaller, Gerhard and Schmitz, Jannes and Schnecke, Martina and Schopper, Florian and Schreeck, Harrison and Schwenker, Benjamin and Schwickardi, Marike and Sedlmeyer, Reinhard and Sfienti, Concettina and Simon, Frank and Skambraks, Sebastian and Skorupa, Justin and Soloviev, Yuri and Spruck, Björn and Stefkova, Slavomira and Stever, Reimer and Tafelmayer, Eva and Takahashi, Maiko and Vila, Ivan and Virto, Amparo Lopez and Vogt, Sven and Wang, Chunjie and Wieduwilt, Philipp and Windel, Hendrik and Ye, Hua and Zhao, Jingzhou and Zlebcik, Radek},
	urldate = {2023-09-20},
	date = {2023-05-08},
	langid = {english},
	doi = {10.22323/1.420.0005},
	note = {Conference Name: 10th International Workshop on Semiconductor Pixel Detectors for Particles and Imaging},
}

@article{krohn_global_2020,
	title = {Global Decay Chain Vertex Fitting at B-Factories},
	volume = {976},
	issn = {01689002},
	url = {http://arxiv.org/abs/1901.11198},
	doi = {10.1016/j.nima.2020.164269},
	pages = {164269},
	journaltitle = {Nuclear Instruments and Methods in Physics Research Section A: Accelerators, Spectrometers, Detectors and Associated Equipment},
	shortjournal = {Nuclear Instruments and Methods in Physics Research Section A: Accelerators, Spectrometers, Detectors and Associated Equipment},
	author = {Krohn, J.-F. and Urquijo, P. and Abudinén, F. and Cunliffe, S. and Ferber, T. and Gelb, M. and Gemmler, J. and Goldenzweig, P. and Keck, T. and Komarov, I. and Kuhr, T. and Ligioi, L. and Lubej, M. and Meier, F. and Metzner, F. and Pulvermacher, C. and Ritter, M. and Tamponi, U. and Tenchini, F. and Zupanc, A.},
	urldate = {2023-09-20},
	date = {2020-10},
	eprinttype = {arxiv},
	eprint = {1901.11198 [hep-ex]},
}

@inproceedings{alanazi_survey_2021,
	title = {A survey of machine learning-based physics event generation},
	url = {http://arxiv.org/abs/2106.00643},
	doi = {10.24963/ijcai.2021/588},
	pages = {4286--4293},
	booktitle = {Proceedings of the Thirtieth International Joint Conference on Artificial Intelligence},
	author = {Alanazi, Yasir and Sato, N. and Ambrozewicz, Pawel and Blin, Astrid N. Hiller and Melnitchouk, W. and Battaglieri, Marco and Liu, Tianbo and Li, Yaohang},
	urldate = {2023-09-07},
	date = {2021-08},
	eprinttype = {arxiv},
	eprint = {2106.00643 [hep-ph, physics:nucl-ex]},
}

@misc{hendrycks_gaussian_2023,
	title = {Gaussian Error Linear Units ({GELUs})},
	url = {http://arxiv.org/abs/1606.08415},
	doi = {10.48550/arXiv.1606.08415},
	number = {{arXiv}:1606.08415},
	publisher = {{arXiv}},
	author = {Hendrycks, Dan and Gimpel, Kevin},
	urldate = {2023-09-07},
	date = {2023-06-05},
	eprinttype = {arxiv},
	eprint = {1606.08415 [cs]},
}

@article{eybpoosh_applying_2022,
	title = {Applying inverse stereographic projection to manifold learning and clustering},
	volume = {52},
	issn = {1573-7497},
	url = {https://doi.org/10.1007/s10489-021-02513-0},
	doi = {10.1007/s10489-021-02513-0},
	pages = {4443--4457},
	number = {4},
	journaltitle = {Applied Intelligence},
	shortjournal = {Appl Intell},
	author = {Eybpoosh, Kajal and Rezghi, Mansoor and Heydari, Abbas},
	urldate = {2023-09-07},
	date = {2022-03-01},
	langid = {english},
}

@online{noauthor_robust_nodate,
	title = {Robust Estimation of a Location Parameter on {JSTOR}},
	url = {https://www.jstor.org/stable/2238020},
	urldate = {2023-09-07},
}

@inproceedings{rucco_characterisation_2016,
	location = {Cham},
	title = {Characterisation of the Idiotypic Immune Network Through Persistent Entropy},
	isbn = {978-3-319-29228-1},
	doi = {10.1007/978-3-319-29228-1_11},
	series = {Springer Proceedings in Complexity},
	pages = {117--128},
	booktitle = {Proceedings of {ECCS} 2014},
	publisher = {Springer International Publishing},
	author = {Rucco, Matteo and Castiglione, Filippo and Merelli, Emanuela and Pettini, Marco},
	editor = {Battiston, Stefano and De Pellegrini, Francesco and Caldarelli, Guido and Merelli, Emanuela},
	date = {2016},
	langid = {english},
}

@misc{atienza_persistent_2017,
	title = {Persistent Entropy for Separating Topological Features from Noise in Vietoris-Rips Complexes},
	url = {http://arxiv.org/abs/1701.07857},
	number = {{arXiv}:1701.07857},
	publisher = {{arXiv}},
	author = {Atienza, Nieves and Gonzalez-Diaz, Rocio and Rucco, Matteo},
	urldate = {2023-09-04},
	date = {2017-01-18},
	eprinttype = {arxiv},
	eprint = {1701.07857 [cs]},
}

@misc{friedman_vendi_2023,
	title = {The Vendi Score: A Diversity Evaluation Metric for Machine Learning},
	url = {http://arxiv.org/abs/2210.02410},
	doi = {10.48550/arXiv.2210.02410},
	shorttitle = {The Vendi Score},
	number = {{arXiv}:2210.02410},
	publisher = {{arXiv}},
	author = {Friedman, Dan and Dieng, Adji Bousso},
	urldate = {2023-08-28},
	date = {2023-07-02},
	eprinttype = {arxiv},
	eprint = {2210.02410 [cond-mat, stat]},
}

@misc{zhao_infovae_2018,
	title = {{InfoVAE}: Information Maximizing Variational Autoencoders},
	url = {http://arxiv.org/abs/1706.02262},
	doi = {10.48550/arXiv.1706.02262},
	shorttitle = {{InfoVAE}},
	number = {{arXiv}:1706.02262},
	publisher = {{arXiv}},
	author = {Zhao, Shengjia and Song, Jiaming and Ermon, Stefano},
	urldate = {2023-08-28},
	date = {2018-05-30},
	eprinttype = {arxiv},
	eprint = {1706.02262 [cs, stat]},
}

@inproceedings{gretton_kernel_2006,
	title = {A Kernel Method for the Two-Sample-Problem},
	volume = {19},
	url = {https://proceedings.neurips.cc/paper_files/paper/2006/hash/e9fb2eda3d9c55a0d89c98d6c54b5b3e-Abstract.html},
	booktitle = {Advances in Neural Information Processing Systems},
	publisher = {{MIT} Press},
	author = {Gretton, Arthur and Borgwardt, Karsten and Rasch, Malte and Schölkopf, Bernhard and Smola, Alex},
	urldate = {2023-08-28},
	date = {2006},
}

@article{feydy_geometric_nodate,
	title = {Geometric data analysis, beyond convolutions},
	author = {Feydy, Jean},
	langid = {english},
}

@misc{chen_exploring_2020,
	title = {Exploring Simple Siamese Representation Learning},
	url = {http://arxiv.org/abs/2011.10566},
	doi = {10.48550/arXiv.2011.10566},
	number = {{arXiv}:2011.10566},
	publisher = {{arXiv}},
	author = {Chen, Xinlei and He, Kaiming},
	urldate = {2023-08-25},
	date = {2020-11-20},
	eprinttype = {arxiv},
	eprint = {2011.10566 [cs]},
}

@misc{perez_film_2017,
	title = {{FiLM}: Visual Reasoning with a General Conditioning Layer},
	url = {http://arxiv.org/abs/1709.07871},
	doi = {10.48550/arXiv.1709.07871},
	shorttitle = {{FiLM}},
	number = {{arXiv}:1709.07871},
	publisher = {{arXiv}},
	author = {Perez, Ethan and Strub, Florian and de Vries, Harm and Dumoulin, Vincent and Courville, Aaron},
	urldate = {2023-08-24},
	date = {2017-12-18},
	eprinttype = {arxiv},
	eprint = {1709.07871 [cs, stat]},
}

@misc{corso_principal_2020,
	title = {Principal Neighbourhood Aggregation for Graph Nets},
	url = {http://arxiv.org/abs/2004.05718},
	doi = {10.48550/arXiv.2004.05718},
	number = {{arXiv}:2004.05718},
	publisher = {{arXiv}},
	author = {Corso, Gabriele and Cavalleri, Luca and Beaini, Dominique and Liò, Pietro and Veličković, Petar},
	urldate = {2023-08-21},
	date = {2020-12-31},
	eprinttype = {arxiv},
	eprint = {2004.05718 [cs, stat]},
}

@article{hochreiter_long_1997,
	title = {Long Short-Term Memory},
	volume = {9},
	issn = {0899-7667},
	url = {https://doi.org/10.1162/neco.1997.9.8.1735},
	doi = {10.1162/neco.1997.9.8.1735},
	pages = {1735--1780},
	number = {8},
	journaltitle = {Neural Computation},
	shortjournal = {Neural Comput.},
	author = {Hochreiter, Sepp and Schmidhuber, Jürgen},
	urldate = {2023-08-21},
	date = {1997-11-01},
}

@online{noauthor_long_nodate,
	title = {Long Short-Term Memory {\textbar} Neural Computation},
	url = {https://dl.acm.org/doi/10.1162/neco.1997.9.8.1735},
	urldate = {2023-08-21},
}

@misc{rusch_survey_2023,
	title = {A Survey on Oversmoothing in Graph Neural Networks},
	url = {http://arxiv.org/abs/2303.10993},
	number = {{arXiv}:2303.10993},
	publisher = {{arXiv}},
	author = {Rusch, T. Konstantin and Bronstein, Michael M. and Mishra, Siddhartha},
	urldate = {2023-08-21},
	date = {2023-03-20},
	langid = {english},
	eprinttype = {arxiv},
	eprint = {2303.10993 [cs]},
}

@misc{santurkar_how_2019,
	title = {How Does Batch Normalization Help Optimization?},
	url = {http://arxiv.org/abs/1805.11604},
	doi = {10.48550/arXiv.1805.11604},
	number = {{arXiv}:1805.11604},
	publisher = {{arXiv}},
	author = {Santurkar, Shibani and Tsipras, Dimitris and Ilyas, Andrew and Madry, Aleksander},
	urldate = {2023-08-21},
	date = {2019-04-14},
	eprinttype = {arxiv},
	eprint = {1805.11604 [cs, stat]},
}

@misc{kool_attention_2019,
	title = {Attention, Learn to Solve Routing Problems!},
	url = {http://arxiv.org/abs/1803.08475},
	doi = {10.48550/arXiv.1803.08475},
	number = {{arXiv}:1803.08475},
	publisher = {{arXiv}},
	author = {Kool, Wouter and van Hoof, Herke and Welling, Max},
	urldate = {2023-08-21},
	date = {2019-02-07},
	eprinttype = {arxiv},
	eprint = {1803.08475 [cs, stat]},
}

@misc{ioffe_batch_2015,
	title = {Batch Normalization: Accelerating Deep Network Training by Reducing Internal Covariate Shift},
	url = {http://arxiv.org/abs/1502.03167},
	doi = {10.48550/arXiv.1502.03167},
	shorttitle = {Batch Normalization},
	number = {{arXiv}:1502.03167},
	publisher = {{arXiv}},
	author = {Ioffe, Sergey and Szegedy, Christian},
	urldate = {2023-08-21},
	date = {2015-03-02},
	eprinttype = {arxiv},
	eprint = {1502.03167 [cs]},
}

@article{nguyen_transformers_2019,
	title = {Transformers without Tears: Improving the Normalization of Self-Attention},
	url = {http://arxiv.org/abs/1910.05895},
	doi = {10.5281/zenodo.3525484},
	shorttitle = {Transformers without Tears},
	author = {Nguyen, Toan Q. and Salazar, Julian},
	urldate = {2023-08-21},
	date = {2019-11-02},
	eprinttype = {arxiv},
	eprint = {1910.05895 [cs, stat]},
}

@article{mazur_when_nodate,
	title = {When is one thing equal to some other thing?},
	author = {Mazur, Barry},
	langid = {english},
}

@article{eilenberg_general_nodate,
	title = {{GENERAL} {THEORY} {OF} {NATURAL} {EQUIVALENCES}},
	author = {Eilenberg, Samuel and {MacLANE}, {SAUNDERS}},
	langid = {english},
}

@misc{pang_reward_2023,
	title = {Reward Gaming in Conditional Text Generation},
	url = {http://arxiv.org/abs/2211.08714},
	doi = {10.48550/arXiv.2211.08714},
	number = {{arXiv}:2211.08714},
	publisher = {{arXiv}},
	author = {Pang, Richard Yuanzhe and Padmakumar, Vishakh and Sellam, Thibault and Parikh, Ankur P. and He, He},
	urldate = {2023-08-15},
	date = {2023-06-01},
	eprinttype = {arxiv},
	eprint = {2211.08714 [cs]},
}

@misc{michaud_understanding_2020,
	title = {Understanding Learned Reward Functions},
	url = {http://arxiv.org/abs/2012.05862},
	doi = {10.48550/arXiv.2012.05862},
	number = {{arXiv}:2012.05862},
	publisher = {{arXiv}},
	author = {Michaud, Eric J. and Gleave, Adam and Russell, Stuart},
	urldate = {2023-08-15},
	date = {2020-12-10},
	eprinttype = {arxiv},
	eprint = {2012.05862 [cs]},
}

@misc{amodei_concrete_2016,
	title = {Concrete Problems in {AI} Safety},
	url = {http://arxiv.org/abs/1606.06565},
	doi = {10.48550/arXiv.1606.06565},
	number = {{arXiv}:1606.06565},
	publisher = {{arXiv}},
	author = {Amodei, Dario and Olah, Chris and Steinhardt, Jacob and Christiano, Paul and Schulman, John and Mané, Dan},
	urldate = {2023-08-15},
	date = {2016-07-25},
	eprinttype = {arxiv},
	eprint = {1606.06565 [cs]},
}

@inproceedings{angermueller_model-based_2020,
	title = {Model-based reinforcement learning for biological sequence design},
	url = {https://iclr.cc/virtual_2020/poster_HklxbgBKvr.html},
	eventtitle = {Eighth International Conference on Learning Representations},
	author = {Angermueller, Christof and Dohan, David and Belanger, David and Deshpande, Ramya and Murphy, Kevin and Colwell, Lucy},
	urldate = {2023-08-15},
	date = {2020-04},
	langid = {english},
}

@inproceedings{gong_reinforcement_2019,
	location = {Minneapolis, Minnesota},
	title = {Reinforcement Learning Based Text Style Transfer without Parallel Training Corpus},
	url = {https://aclanthology.org/N19-1320},
	doi = {10.18653/v1/N19-1320},
	eventtitle = {{NAACL}-{HLT} 2019},
	pages = {3168--3180},
	booktitle = {Proceedings of the 2019 Conference of the North American Chapter of the Association for Computational Linguistics: Human Language Technologies, Volume 1 (Long and Short Papers)},
	publisher = {Association for Computational Linguistics},
	author = {Gong, Hongyu and Bhat, Suma and Wu, Lingfei and Xiong, {JinJun} and Hwu, Wen-mei},
	urldate = {2023-08-15},
	date = {2019-06},
}

@online{noauthor_denoising_nodate,
	title = {Denoising Diffusion-based Generative Modeling: Foundations and Applications},
	url = {https://cvpr2022-tutorial-diffusion-models.github.io},
	shorttitle = {Denoising Diffusion-based Generative Modeling},
	titleaddon = {Denoising Diffusion-based Generative Modeling: Foundations and Applications},
	urldate = {2023-08-15},
}

@inproceedings{yang_fudge_2021,
	title = {{FUDGE}: Controlled Text Generation With Future Discriminators},
	url = {http://arxiv.org/abs/2104.05218},
	doi = {10.18653/v1/2021.naacl-main.276},
	shorttitle = {{FUDGE}},
	pages = {3511--3535},
	booktitle = {Proceedings of the 2021 Conference of the North American Chapter of the Association for Computational Linguistics: Human Language Technologies},
	author = {Yang, Kevin and Klein, Dan},
	urldate = {2023-08-15},
	date = {2021},
	eprinttype = {arxiv},
	eprint = {2104.05218 [cs]},
}

@misc{dathathri_plug_2020,
	title = {Plug and Play Language Models: A Simple Approach to Controlled Text Generation},
	url = {http://arxiv.org/abs/1912.02164},
	doi = {10.48550/arXiv.1912.02164},
	shorttitle = {Plug and Play Language Models},
	number = {{arXiv}:1912.02164},
	publisher = {{arXiv}},
	author = {Dathathri, Sumanth and Madotto, Andrea and Lan, Janice and Hung, Jane and Frank, Eric and Molino, Piero and Yosinski, Jason and Liu, Rosanne},
	urldate = {2023-08-15},
	date = {2020-03-03},
	eprinttype = {arxiv},
	eprint = {1912.02164 [cs]},
}

@misc{chan_deep_2021,
	title = {Deep Extrapolation for Attribute-Enhanced Generation},
	url = {http://arxiv.org/abs/2107.02968},
	doi = {10.48550/arXiv.2107.02968},
	number = {{arXiv}:2107.02968},
	publisher = {{arXiv}},
	author = {Chan, Alvin and Madani, Ali and Krause, Ben and Naik, Nikhil},
	urldate = {2023-08-15},
	date = {2021-10-25},
	eprinttype = {arxiv},
	eprint = {2107.02968 [cs, q-bio]},
}

@misc{welleck_generating_2022,
	title = {Generating Sequences by Learning to Self-Correct},
	url = {http://arxiv.org/abs/2211.00053},
	doi = {10.48550/arXiv.2211.00053},
	number = {{arXiv}:2211.00053},
	publisher = {{arXiv}},
	author = {Welleck, Sean and Lu, Ximing and West, Peter and Brahman, Faeze and Shen, Tianxiao and Khashabi, Daniel and Choi, Yejin},
	urldate = {2023-08-15},
	date = {2022-10-31},
	eprinttype = {arxiv},
	eprint = {2211.00053 [cs]},
}

@misc{novak_iterative_2018,
	title = {Iterative Refinement for Machine Translation},
	url = {http://arxiv.org/abs/1610.06602},
	doi = {10.48550/arXiv.1610.06602},
	number = {{arXiv}:1610.06602},
	publisher = {{arXiv}},
	author = {Novak, Roman and Auli, Michael and Grangier, David},
	urldate = {2023-08-15},
	date = {2018-04-13},
	eprinttype = {arxiv},
	eprint = {1610.06602 [cs]},
}

@misc{mallinson_edit5_2022,
	title = {{EdiT}5: Semi-Autoregressive Text-Editing with T5 Warm-Start},
	url = {http://arxiv.org/abs/2205.12209},
	doi = {10.48550/arXiv.2205.12209},
	shorttitle = {{EdiT}5},
	number = {{arXiv}:2205.12209},
	publisher = {{arXiv}},
	author = {Mallinson, Jonathan and Adamek, Jakub and Malmi, Eric and Severyn, Aliaksei},
	urldate = {2023-08-15},
	date = {2022-10-26},
	eprinttype = {arxiv},
	eprint = {2205.12209 [cs]},
}

@misc{guu_generating_2018,
	title = {Generating Sentences by Editing Prototypes},
	url = {http://arxiv.org/abs/1709.08878},
	doi = {10.48550/arXiv.1709.08878},
	number = {{arXiv}:1709.08878},
	publisher = {{arXiv}},
	author = {Guu, Kelvin and Hashimoto, Tatsunori B. and Oren, Yonatan and Liang, Percy},
	urldate = {2023-08-15},
	date = {2018-09-07},
	eprinttype = {arxiv},
	eprint = {1709.08878 [cs, stat]},
}

@article{madani_large_2023,
	title = {Large language models generate functional protein sequences across diverse families},
	volume = {41},
	rights = {2023 The Author(s), under exclusive licence to Springer Nature America, Inc.},
	issn = {1546-1696},
	url = {https://www.nature.com/articles/s41587-022-01618-2},
	doi = {10.1038/s41587-022-01618-2},
	pages = {1099--1106},
	number = {8},
	journaltitle = {Nature Biotechnology},
	shortjournal = {Nat Biotechnol},
	author = {Madani, Ali and Krause, Ben and Greene, Eric R. and Subramanian, Subu and Mohr, Benjamin P. and Holton, James M. and Olmos, Jose Luis and Xiong, Caiming and Sun, Zachary Z. and Socher, Richard and Fraser, James S. and Naik, Nikhil},
	urldate = {2023-08-15},
	date = {2023-08},
	langid = {english},
	note = {Number: 8
Publisher: Nature Publishing Group},
}

@online{noauthor_progen_nodate,
	title = {{ProGen}: Language Modeling for Protein Generation {\textbar} {bioRxiv}},
	url = {https://www.biorxiv.org/content/10.1101/2020.03.07.982272v2},
	urldate = {2023-08-15},
}

@misc{keskar_ctrl_2019,
	title = {{CTRL}: A Conditional Transformer Language Model for Controllable Generation},
	url = {http://arxiv.org/abs/1909.05858},
	doi = {10.48550/arXiv.1909.05858},
	shorttitle = {{CTRL}},
	number = {{arXiv}:1909.05858},
	publisher = {{arXiv}},
	author = {Keskar, Nitish Shirish and {McCann}, Bryan and Varshney, Lav R. and Xiong, Caiming and Socher, Richard},
	urldate = {2023-08-15},
	date = {2019-09-20},
	eprinttype = {arxiv},
	eprint = {1909.05858 [cs]},
}

@misc{padmakumar_extrapolative_2023,
	title = {Extrapolative Controlled Sequence Generation via Iterative Refinement},
	url = {http://arxiv.org/abs/2303.04562},
	doi = {10.48550/arXiv.2303.04562},
	number = {{arXiv}:2303.04562},
	publisher = {{arXiv}},
	author = {Padmakumar, Vishakh and Pang, Richard Yuanzhe and He, He and Parikh, Ankur P.},
	urldate = {2023-08-15},
	date = {2023-06-07},
	eprinttype = {arxiv},
	eprint = {2303.04562 [cs, q-bio]},
}

@article{yeh_novo_2023,
	title = {De novo design of luciferases using deep learning},
	volume = {614},
	rights = {2023 The Author(s)},
	issn = {1476-4687},
	url = {https://www.nature.com/articles/s41586-023-05696-3},
	doi = {10.1038/s41586-023-05696-3},
	pages = {774--780},
	number = {7949},
	journaltitle = {Nature},
	author = {Yeh, Andy Hsien-Wei and Norn, Christoffer and Kipnis, Yakov and Tischer, Doug and Pellock, Samuel J. and Evans, Declan and Ma, Pengchen and Lee, Gyu Rie and Zhang, Jason Z. and Anishchenko, Ivan and Coventry, Brian and Cao, Longxing and Dauparas, Justas and Halabiya, Samer and {DeWitt}, Michelle and Carter, Lauren and Houk, K. N. and Baker, David},
	urldate = {2023-08-15},
	date = {2023-02},
	langid = {english},
	note = {Number: 7949
Publisher: Nature Publishing Group},
}

@article{gainza_novo_2023,
	title = {De novo design of protein interactions with learned surface fingerprints},
	volume = {617},
	rights = {2023 The Author(s)},
	issn = {1476-4687},
	url = {https://www.nature.com/articles/s41586-023-05993-x},
	doi = {10.1038/s41586-023-05993-x},
	pages = {176--184},
	number = {7959},
	journaltitle = {Nature},
	author = {Gainza, Pablo and Wehrle, Sarah and Van Hall-Beauvais, Alexandra and Marchand, Anthony and Scheck, Andreas and Harteveld, Zander and Buckley, Stephen and Ni, Dongchun and Tan, Shuguang and Sverrisson, Freyr and Goverde, Casper and Turelli, Priscilla and Raclot, Charlène and Teslenko, Alexandra and Pacesa, Martin and Rosset, Stéphane and Georgeon, Sandrine and Marsden, Jane and Petruzzella, Aaron and Liu, Kefang and Xu, Zepeng and Chai, Yan and Han, Pu and Gao, George F. and Oricchio, Elisa and Fierz, Beat and Trono, Didier and Stahlberg, Henning and Bronstein, Michael and Correia, Bruno E.},
	urldate = {2023-08-15},
	date = {2023-05},
	langid = {english},
	note = {Number: 7959
Publisher: Nature Publishing Group},
}

@article{freschlin_machine_2022,
	title = {Machine learning to navigate fitness landscapes for protein engineering},
	volume = {75},
	issn = {0958-1669},
	url = {https://www.sciencedirect.com/science/article/pii/S0958166922000465},
	doi = {10.1016/j.copbio.2022.102713},
	pages = {102713},
	journaltitle = {Current Opinion in Biotechnology},
	shortjournal = {Current Opinion in Biotechnology},
	author = {Freschlin, Chase R and Fahlberg, Sarah A and Romero, Philip A},
	urldate = {2023-08-15},
	date = {2022-06-01},
}

@misc{sun_length-extrapolatable_2022,
	title = {A Length-Extrapolatable Transformer},
	url = {http://arxiv.org/abs/2212.10554},
	doi = {10.48550/arXiv.2212.10554},
	number = {{arXiv}:2212.10554},
	publisher = {{arXiv}},
	author = {Sun, Yutao and Dong, Li and Patra, Barun and Ma, Shuming and Huang, Shaohan and Benhaim, Alon and Chaudhary, Vishrav and Song, Xia and Wei, Furu},
	urldate = {2023-08-15},
	date = {2022-12-20},
	eprinttype = {arxiv},
	eprint = {2212.10554 [cs]},
	note = {version: 1},
}

@misc{workshop_bloom_2023,
	title = {{BLOOM}: A 176B-Parameter Open-Access Multilingual Language Model},
	url = {http://arxiv.org/abs/2211.05100},
	doi = {10.48550/arXiv.2211.05100},
	shorttitle = {{BLOOM}},
	number = {{arXiv}:2211.05100},
	publisher = {{arXiv}},
	author = {Workshop, {BigScience} and Scao, Teven Le and Fan, Angela and Akiki, Christopher and Pavlick, Ellie and Ilić, Suzana and Hesslow, Daniel and Castagné, Roman and Luccioni, Alexandra Sasha and Yvon, François and Gallé, Matthias and Tow, Jonathan and Rush, Alexander M. and Biderman, Stella and Webson, Albert and Ammanamanchi, Pawan Sasanka and Wang, Thomas and Sagot, Benoît and Muennighoff, Niklas and del Moral, Albert Villanova and Ruwase, Olatunji and Bawden, Rachel and Bekman, Stas and {McMillan}-Major, Angelina and Beltagy, Iz and Nguyen, Huu and Saulnier, Lucile and Tan, Samson and Suarez, Pedro Ortiz and Sanh, Victor and Laurençon, Hugo and Jernite, Yacine and Launay, Julien and Mitchell, Margaret and Raffel, Colin and Gokaslan, Aaron and Simhi, Adi and Soroa, Aitor and Aji, Alham Fikri and Alfassy, Amit and Rogers, Anna and Nitzav, Ariel Kreisberg and Xu, Canwen and Mou, Chenghao and Emezue, Chris and Klamm, Christopher and Leong, Colin and van Strien, Daniel and Adelani, David Ifeoluwa and Radev, Dragomir and Ponferrada, Eduardo González and Levkovizh, Efrat and Kim, Ethan and Natan, Eyal Bar and De Toni, Francesco and Dupont, Gérard and Kruszewski, Germán and Pistilli, Giada and Elsahar, Hady and Benyamina, Hamza and Tran, Hieu and Yu, Ian and Abdulmumin, Idris and Johnson, Isaac and Gonzalez-Dios, Itziar and de la Rosa, Javier and Chim, Jenny and Dodge, Jesse and Zhu, Jian and Chang, Jonathan and Frohberg, Jörg and Tobing, Joseph and Bhattacharjee, Joydeep and Almubarak, Khalid and Chen, Kimbo and Lo, Kyle and Von Werra, Leandro and Weber, Leon and Phan, Long and allal, Loubna Ben and Tanguy, Ludovic and Dey, Manan and Muñoz, Manuel Romero and Masoud, Maraim and Grandury, María and Šaško, Mario and Huang, Max and Coavoux, Maximin and Singh, Mayank and Jiang, Mike Tian-Jian and Vu, Minh Chien and Jauhar, Mohammad A. and Ghaleb, Mustafa and Subramani, Nishant and Kassner, Nora and Khamis, Nurulaqilla and Nguyen, Olivier and Espejel, Omar and de Gibert, Ona and Villegas, Paulo and Henderson, Peter and Colombo, Pierre and Amuok, Priscilla and Lhoest, Quentin and Harliman, Rheza and Bommasani, Rishi and López, Roberto Luis and Ribeiro, Rui and Osei, Salomey and Pyysalo, Sampo and Nagel, Sebastian and Bose, Shamik and Muhammad, Shamsuddeen Hassan and Sharma, Shanya and Longpre, Shayne and Nikpoor, Somaieh and Silberberg, Stanislav and Pai, Suhas and Zink, Sydney and Torrent, Tiago Timponi and Schick, Timo and Thrush, Tristan and Danchev, Valentin and Nikoulina, Vassilina and Laippala, Veronika and Lepercq, Violette and Prabhu, Vrinda and Alyafeai, Zaid and Talat, Zeerak and Raja, Arun and Heinzerling, Benjamin and Si, Chenglei and Taşar, Davut Emre and Salesky, Elizabeth and Mielke, Sabrina J. and Lee, Wilson Y. and Sharma, Abheesht and Santilli, Andrea and Chaffin, Antoine and Stiegler, Arnaud and Datta, Debajyoti and Szczechla, Eliza and Chhablani, Gunjan and Wang, Han and Pandey, Harshit and Strobelt, Hendrik and Fries, Jason Alan and Rozen, Jos and Gao, Leo and Sutawika, Lintang and Bari, M. Saiful and Al-shaibani, Maged S. and Manica, Matteo and Nayak, Nihal and Teehan, Ryan and Albanie, Samuel and Shen, Sheng and Ben-David, Srulik and Bach, Stephen H. and Kim, Taewoon and Bers, Tali and Fevry, Thibault and Neeraj, Trishala and Thakker, Urmish and Raunak, Vikas and Tang, Xiangru and Yong, Zheng-Xin and Sun, Zhiqing and Brody, Shaked and Uri, Yallow and Tojarieh, Hadar and Roberts, Adam and Chung, Hyung Won and Tae, Jaesung and Phang, Jason and Press, Ofir and Li, Conglong and Narayanan, Deepak and Bourfoune, Hatim and Casper, Jared and Rasley, Jeff and Ryabinin, Max and Mishra, Mayank and Zhang, Minjia and Shoeybi, Mohammad and Peyrounette, Myriam and Patry, Nicolas and Tazi, Nouamane and Sanseviero, Omar and von Platen, Patrick and Cornette, Pierre and Lavallée, Pierre François and Lacroix, Rémi and Rajbhandari, Samyam and Gandhi, Sanchit and Smith, Shaden and Requena, Stéphane and Patil, Suraj and Dettmers, Tim and Baruwa, Ahmed and Singh, Amanpreet and Cheveleva, Anastasia and Ligozat, Anne-Laure and Subramonian, Arjun and Névéol, Aurélie and Lovering, Charles and Garrette, Dan and Tunuguntla, Deepak and Reiter, Ehud and Taktasheva, Ekaterina and Voloshina, Ekaterina and Bogdanov, Eli and Winata, Genta Indra and Schoelkopf, Hailey and Kalo, Jan-Christoph and Novikova, Jekaterina and Forde, Jessica Zosa and Clive, Jordan and Kasai, Jungo and Kawamura, Ken and Hazan, Liam and Carpuat, Marine and Clinciu, Miruna and Kim, Najoung and Cheng, Newton and Serikov, Oleg and Antverg, Omer and van der Wal, Oskar and Zhang, Rui and Zhang, Ruochen and Gehrmann, Sebastian and Mirkin, Shachar and Pais, Shani and Shavrina, Tatiana and Scialom, Thomas and Yun, Tian and Limisiewicz, Tomasz and Rieser, Verena and Protasov, Vitaly and Mikhailov, Vladislav and Pruksachatkun, Yada and Belinkov, Yonatan and Bamberger, Zachary and Kasner, Zdeněk and Rueda, Alice and Pestana, Amanda and Feizpour, Amir and Khan, Ammar and Faranak, Amy and Santos, Ana and Hevia, Anthony and Unldreaj, Antigona and Aghagol, Arash and Abdollahi, Arezoo and Tammour, Aycha and {HajiHosseini}, Azadeh and Behroozi, Bahareh and Ajibade, Benjamin and Saxena, Bharat and Ferrandis, Carlos Muñoz and {McDuff}, Daniel and Contractor, Danish and Lansky, David and David, Davis and Kiela, Douwe and Nguyen, Duong A. and Tan, Edward and Baylor, Emi and Ozoani, Ezinwanne and Mirza, Fatima and Ononiwu, Frankline and Rezanejad, Habib and Jones, Hessie and Bhattacharya, Indrani and Solaiman, Irene and Sedenko, Irina and Nejadgholi, Isar and Passmore, Jesse and Seltzer, Josh and Sanz, Julio Bonis and Dutra, Livia and Samagaio, Mairon and Elbadri, Maraim and Mieskes, Margot and Gerchick, Marissa and Akinlolu, Martha and {McKenna}, Michael and Qiu, Mike and Ghauri, Muhammed and Burynok, Mykola and Abrar, Nafis and Rajani, Nazneen and Elkott, Nour and Fahmy, Nour and Samuel, Olanrewaju and An, Ran and Kromann, Rasmus and Hao, Ryan and Alizadeh, Samira and Shubber, Sarmad and Wang, Silas and Roy, Sourav and Viguier, Sylvain and Le, Thanh and Oyebade, Tobi and Le, Trieu and Yang, Yoyo and Nguyen, Zach and Kashyap, Abhinav Ramesh and Palasciano, Alfredo and Callahan, Alison and Shukla, Anima and Miranda-Escalada, Antonio and Singh, Ayush and Beilharz, Benjamin and Wang, Bo and Brito, Caio and Zhou, Chenxi and Jain, Chirag and Xu, Chuxin and Fourrier, Clémentine and Periñán, Daniel León and Molano, Daniel and Yu, Dian and Manjavacas, Enrique and Barth, Fabio and Fuhrimann, Florian and Altay, Gabriel and Bayrak, Giyaseddin and Burns, Gully and Vrabec, Helena U. and Bello, Imane and Dash, Ishani and Kang, Jihyun and Giorgi, John and Golde, Jonas and Posada, Jose David and Sivaraman, Karthik Rangasai and Bulchandani, Lokesh and Liu, Lu and Shinzato, Luisa and de Bykhovetz, Madeleine Hahn and Takeuchi, Maiko and Pàmies, Marc and Castillo, Maria A. and Nezhurina, Marianna and Sänger, Mario and Samwald, Matthias and Cullan, Michael and Weinberg, Michael and De Wolf, Michiel and Mihaljcic, Mina and Liu, Minna and Freidank, Moritz and Kang, Myungsun and Seelam, Natasha and Dahlberg, Nathan and Broad, Nicholas Michio and Muellner, Nikolaus and Fung, Pascale and Haller, Patrick and Chandrasekhar, Ramya and Eisenberg, Renata and Martin, Robert and Canalli, Rodrigo and Su, Rosaline and Su, Ruisi and Cahyawijaya, Samuel and Garda, Samuele and Deshmukh, Shlok S. and Mishra, Shubhanshu and Kiblawi, Sid and Ott, Simon and Sang-aroonsiri, Sinee and Kumar, Srishti and Schweter, Stefan and Bharati, Sushil and Laud, Tanmay and Gigant, Théo and Kainuma, Tomoya and Kusa, Wojciech and Labrak, Yanis and Bajaj, Yash Shailesh and Venkatraman, Yash and Xu, Yifan and Xu, Yingxin and Xu, Yu and Tan, Zhe and Xie, Zhongli and Ye, Zifan and Bras, Mathilde and Belkada, Younes and Wolf, Thomas},
	urldate = {2023-08-15},
	date = {2023-06-27},
	eprinttype = {arxiv},
	eprint = {2211.05100 [cs]},
}

@misc{su_roformer_2022,
	title = {{RoFormer}: Enhanced Transformer with Rotary Position Embedding},
	url = {http://arxiv.org/abs/2104.09864},
	doi = {10.48550/arXiv.2104.09864},
	shorttitle = {{RoFormer}},
	number = {{arXiv}:2104.09864},
	publisher = {{arXiv}},
	author = {Su, Jianlin and Lu, Yu and Pan, Shengfeng and Murtadha, Ahmed and Wen, Bo and Liu, Yunfeng},
	urldate = {2023-08-15},
	date = {2022-08-08},
	eprinttype = {arxiv},
	eprint = {2104.09864 [cs]},
}

@misc{touvron_llama_2023,
	title = {{LLaMA}: Open and Efficient Foundation Language Models},
	url = {http://arxiv.org/abs/2302.13971},
	doi = {10.48550/arXiv.2302.13971},
	shorttitle = {{LLaMA}},
	number = {{arXiv}:2302.13971},
	publisher = {{arXiv}},
	author = {Touvron, Hugo and Lavril, Thibaut and Izacard, Gautier and Martinet, Xavier and Lachaux, Marie-Anne and Lacroix, Timothée and Rozière, Baptiste and Goyal, Naman and Hambro, Eric and Azhar, Faisal and Rodriguez, Aurelien and Joulin, Armand and Grave, Edouard and Lample, Guillaume},
	urldate = {2023-08-15},
	date = {2023-02-27},
	eprinttype = {arxiv},
	eprint = {2302.13971 [cs]},
}

@misc{chowdhery_palm_2022,
	title = {{PaLM}: Scaling Language Modeling with Pathways},
	url = {http://arxiv.org/abs/2204.02311},
	doi = {10.48550/arXiv.2204.02311},
	shorttitle = {{PaLM}},
	number = {{arXiv}:2204.02311},
	publisher = {{arXiv}},
	author = {Chowdhery, Aakanksha and Narang, Sharan and Devlin, Jacob and Bosma, Maarten and Mishra, Gaurav and Roberts, Adam and Barham, Paul and Chung, Hyung Won and Sutton, Charles and Gehrmann, Sebastian and Schuh, Parker and Shi, Kensen and Tsvyashchenko, Sasha and Maynez, Joshua and Rao, Abhishek and Barnes, Parker and Tay, Yi and Shazeer, Noam and Prabhakaran, Vinodkumar and Reif, Emily and Du, Nan and Hutchinson, Ben and Pope, Reiner and Bradbury, James and Austin, Jacob and Isard, Michael and Gur-Ari, Guy and Yin, Pengcheng and Duke, Toju and Levskaya, Anselm and Ghemawat, Sanjay and Dev, Sunipa and Michalewski, Henryk and Garcia, Xavier and Misra, Vedant and Robinson, Kevin and Fedus, Liam and Zhou, Denny and Ippolito, Daphne and Luan, David and Lim, Hyeontaek and Zoph, Barret and Spiridonov, Alexander and Sepassi, Ryan and Dohan, David and Agrawal, Shivani and Omernick, Mark and Dai, Andrew M. and Pillai, Thanumalayan Sankaranarayana and Pellat, Marie and Lewkowycz, Aitor and Moreira, Erica and Child, Rewon and Polozov, Oleksandr and Lee, Katherine and Zhou, Zongwei and Wang, Xuezhi and Saeta, Brennan and Diaz, Mark and Firat, Orhan and Catasta, Michele and Wei, Jason and Meier-Hellstern, Kathy and Eck, Douglas and Dean, Jeff and Petrov, Slav and Fiedel, Noah},
	urldate = {2023-08-15},
	date = {2022-10-05},
	eprinttype = {arxiv},
	eprint = {2204.02311 [cs]},
}

@misc{raffel_exploring_2020,
	title = {Exploring the Limits of Transfer Learning with a Unified Text-to-Text Transformer},
	url = {http://arxiv.org/abs/1910.10683},
	number = {{arXiv}:1910.10683},
	publisher = {{arXiv}},
	author = {Raffel, Colin and Shazeer, Noam and Roberts, Adam and Lee, Katherine and Narang, Sharan and Matena, Michael and Zhou, Yanqi and Li, Wei and Liu, Peter J.},
	urldate = {2023-08-15},
	date = {2020-07-28},
	eprinttype = {arxiv},
	eprint = {1910.10683 [cs, stat]},
}

@misc{zhang_opt_2022,
	title = {{OPT}: Open Pre-trained Transformer Language Models},
	url = {http://arxiv.org/abs/2205.01068},
	doi = {10.48550/arXiv.2205.01068},
	shorttitle = {{OPT}},
	number = {{arXiv}:2205.01068},
	publisher = {{arXiv}},
	author = {Zhang, Susan and Roller, Stephen and Goyal, Naman and Artetxe, Mikel and Chen, Moya and Chen, Shuohui and Dewan, Christopher and Diab, Mona and Li, Xian and Lin, Xi Victoria and Mihaylov, Todor and Ott, Myle and Shleifer, Sam and Shuster, Kurt and Simig, Daniel and Koura, Punit Singh and Sridhar, Anjali and Wang, Tianlu and Zettlemoyer, Luke},
	urldate = {2023-08-15},
	date = {2022-06-21},
	eprinttype = {arxiv},
	eprint = {2205.01068 [cs]},
}

@misc{brown_language_2020,
	title = {Language Models are Few-Shot Learners},
	url = {http://arxiv.org/abs/2005.14165},
	doi = {10.48550/arXiv.2005.14165},
	number = {{arXiv}:2005.14165},
	publisher = {{arXiv}},
	author = {Brown, Tom B. and Mann, Benjamin and Ryder, Nick and Subbiah, Melanie and Kaplan, Jared and Dhariwal, Prafulla and Neelakantan, Arvind and Shyam, Pranav and Sastry, Girish and Askell, Amanda and Agarwal, Sandhini and Herbert-Voss, Ariel and Krueger, Gretchen and Henighan, Tom and Child, Rewon and Ramesh, Aditya and Ziegler, Daniel M. and Wu, Jeffrey and Winter, Clemens and Hesse, Christopher and Chen, Mark and Sigler, Eric and Litwin, Mateusz and Gray, Scott and Chess, Benjamin and Clark, Jack and Berner, Christopher and {McCandlish}, Sam and Radford, Alec and Sutskever, Ilya and Amodei, Dario},
	urldate = {2023-08-15},
	date = {2020-07-22},
	eprinttype = {arxiv},
	eprint = {2005.14165 [cs]},
}

@inproceedings{ontanon_making_2022,
	location = {Dublin, Ireland},
	title = {Making Transformers Solve Compositional Tasks},
	url = {https://aclanthology.org/2022.acl-long.251},
	doi = {10.18653/v1/2022.acl-long.251},
	eventtitle = {{ACL} 2022},
	pages = {3591--3607},
	booktitle = {Proceedings of the 60th Annual Meeting of the Association for Computational Linguistics (Volume 1: Long Papers)},
	publisher = {Association for Computational Linguistics},
	author = {Ontanon, Santiago and Ainslie, Joshua and Fisher, Zachary and Cvicek, Vaclav},
	urldate = {2023-08-14},
	date = {2022-05},
}

@misc{kim_transformers_2022,
	title = {Transformers Generalize {DeepSets} and Can be Extended to Graphs and Hypergraphs},
	url = {http://arxiv.org/abs/2110.14416},
	doi = {10.48550/arXiv.2110.14416},
	number = {{arXiv}:2110.14416},
	publisher = {{arXiv}},
	author = {Kim, Jinwoo and Oh, Saeyoon and Hong, Seunghoon},
	urldate = {2023-08-14},
	date = {2022-01-21},
	eprinttype = {arxiv},
	eprint = {2110.14416 [cs]},
}

@article{jurewicz_set--sequence_2021,
	title = {Set-to-Sequence Methods in Machine Learning: a Review},
	volume = {71},
	issn = {1076-9757},
	url = {http://arxiv.org/abs/2103.09656},
	doi = {10.1613/jair.1.12839},
	shorttitle = {Set-to-Sequence Methods in Machine Learning},
	pages = {885--924},
	journaltitle = {Journal of Artificial Intelligence Research},
	shortjournal = {jair},
	author = {Jurewicz, Mateusz and Strømberg-Derczynski, Leon},
	urldate = {2023-08-14},
	date = {2021-08-12},
	eprinttype = {arxiv},
	eprint = {2103.09656 [cs]},
}

@software{zhang_deep_2023,
	title = {Deep Set Prediction Networks},
	rights = {{MIT}},
	url = {https://github.com/Cyanogenoid/dspn},
	author = {Zhang, Yan},
	urldate = {2023-08-14},
	date = {2023-06-03},
	note = {original-date: 2019-06-14T06:06:09Z},
}

@misc{bevilacqua_size-invariant_2021,
	title = {Size-Invariant Graph Representations for Graph Classification Extrapolations},
	url = {http://arxiv.org/abs/2103.05045},
	number = {{arXiv}:2103.05045},
	publisher = {{arXiv}},
	author = {Bevilacqua, Beatrice and Zhou, Yangze and Ribeiro, Bruno},
	urldate = {2023-08-14},
	date = {2021-07-06},
	eprinttype = {arxiv},
	eprint = {2103.05045 [cs]},
}

@misc{press_train_2022,
	title = {Train Short, Test Long: Attention with Linear Biases Enables Input Length Extrapolation},
	url = {http://arxiv.org/abs/2108.12409},
	doi = {10.48550/arXiv.2108.12409},
	shorttitle = {Train Short, Test Long},
	number = {{arXiv}:2108.12409},
	publisher = {{arXiv}},
	author = {Press, Ofir and Smith, Noah A. and Lewis, Mike},
	urldate = {2023-08-14},
	date = {2022-04-22},
	eprinttype = {arxiv},
	eprint = {2108.12409 [cs]},
}

@online{noauthor_deepfunc_nodate,
	title = {{DeepFunc}: A Deep Learning Framework for Accurate Prediction of Protein Functions from Protein Sequences and Interactions - Zhang - 2019 - {PROTEOMICS} - Wiley Online Library},
	url = {https://analyticalsciencejournals.onlinelibrary.wiley.com/doi/abs/10.1002/pmic.201900019},
	urldate = {2023-08-14},
}

@misc{zhang_understanding_2021,
	title = {Understanding Failures in Out-of-Distribution Detection with Deep Generative Models},
	url = {http://arxiv.org/abs/2107.06908},
	doi = {10.48550/arXiv.2107.06908},
	number = {{arXiv}:2107.06908},
	publisher = {{arXiv}},
	author = {Zhang, Lily H. and Goldstein, Mark and Ranganath, Rajesh},
	urldate = {2023-08-14},
	date = {2021-07-16},
	eprinttype = {arxiv},
	eprint = {2107.06908 [cs]},
}

@article{krawczuk_gg-gan_2020,
	title = {{GG}-{GAN}: A Geometric Graph Generative Adversarial Network},
	url = {https://openreview.net/forum?id=qiAxL3Xqx1o},
	shorttitle = {{GG}-{GAN}},
	author = {Krawczuk, Igor and Abranches, Pedro and Loukas, Andreas and Cevher, Volkan},
	urldate = {2023-08-14},
	date = {2020-10-02},
	langid = {english},
}

@misc{wang_graphgan_2017,
	title = {{GraphGAN}: Graph Representation Learning with Generative Adversarial Nets},
	url = {http://arxiv.org/abs/1711.08267},
	shorttitle = {{GraphGAN}},
	number = {{arXiv}:1711.08267},
	publisher = {{arXiv}},
	author = {Wang, Hongwei and Wang, Jia and Wang, Jialin and Zhao, Miao and Zhang, Weinan and Zhang, Fuzheng and Xie, Xing and Guo, Minyi},
	urldate = {2023-08-14},
	date = {2017-11-22},
	eprinttype = {arxiv},
	eprint = {1711.08267 [cs, stat]},
}

@misc{luo_diffusion_2021,
	title = {Diffusion Probabilistic Models for 3D Point Cloud Generation},
	url = {http://arxiv.org/abs/2103.01458},
	doi = {10.48550/arXiv.2103.01458},
	number = {{arXiv}:2103.01458},
	publisher = {{arXiv}},
	author = {Luo, Shitong and Hu, Wei},
	urldate = {2023-08-14},
	date = {2021-06-13},
	eprinttype = {arxiv},
	eprint = {2103.01458 [cs]},
}

@misc{kohler_equivariant_2020,
	title = {Equivariant Flows: Exact Likelihood Generative Learning for Symmetric Densities},
	url = {http://arxiv.org/abs/2006.02425},
	doi = {10.48550/arXiv.2006.02425},
	shorttitle = {Equivariant Flows},
	number = {{arXiv}:2006.02425},
	publisher = {{arXiv}},
	author = {Köhler, Jonas and Klein, Leon and Noé, Frank},
	urldate = {2023-08-14},
	date = {2020-10-26},
	eprinttype = {arxiv},
	eprint = {2006.02425 [physics, stat]},
}

@misc{kosiorek_conditional_2020,
	title = {Conditional Set Generation with Transformers},
	url = {http://arxiv.org/abs/2006.16841},
	doi = {10.48550/arXiv.2006.16841},
	number = {{arXiv}:2006.16841},
	publisher = {{arXiv}},
	author = {Kosiorek, Adam R. and Kim, Hyunjik and Rezende, Danilo J.},
	urldate = {2023-08-14},
	date = {2020-07-01},
	eprinttype = {arxiv},
	eprint = {2006.16841 [cs]},
}

@article{mantel_detection_1967,
	title = {The Detection of Disease Clustering and a Generalized Regression Approach},
	volume = {27},
	issn = {0008-5472},
	pages = {209--220},
	number = {2},
	journaltitle = {Cancer Research},
	shortjournal = {Cancer Research},
	author = {Mantel, Nathan},
	date = {1967-02-01},
}

@online{noauthor_biometry_nodate,
	title = {Biometry. The Principles and Practice of Statistics in Biological Research; Statistical Tables {\textbar} Systematic Biology {\textbar} Oxford Academic},
	url = {https://academic.oup.com/sysbio/article-abstract/19/4/391/1648697?redirectedFrom=fulltext},
	urldate = {2023-08-12},
}

@article{kuhr_belle_2019,
	title = {The Belle {II} Core Software},
	volume = {3},
	issn = {2510-2036, 2510-2044},
	url = {http://arxiv.org/abs/1809.04299},
	doi = {10.1007/s41781-018-0017-9},
	pages = {1},
	number = {1},
	journaltitle = {Computing and Software for Big Science},
	shortjournal = {Comput Softw Big Sci},
	author = {Kuhr, T. and Pulvermacher, C. and Ritter, M. and Hauth, T. and Braun, N.},
	urldate = {2023-08-12},
	date = {2019-12},
	eprinttype = {arxiv},
	eprint = {1809.04299 [hep-ex, physics:physics]},
}

@article{hashemi_pixel_2021,
	title = {Pixel Detector Background Generation using Generative Adversarial Networks at Belle {II}},
	volume = {251},
	rights = {© The Authors, published by {EDP} Sciences, 2021},
	issn = {2100-014X},
	url = {https://www.epj-conferences.org/articles/epjconf/abs/2021/05/epjconf_chep2021_03031/epjconf_chep2021_03031.html},
	doi = {10.1051/epjconf/202125103031},
	pages = {03031},
	journaltitle = {{EPJ} Web of Conferences},
	shortjournal = {{EPJ} Web Conf.},
	author = {Hashemi, Hosein and Hartmann, Nikolai and Kuhr, Thomas and Ritter, Martin and Srebre, Matej},
	urldate = {2023-08-12},
	date = {2021},
	langid = {english},
	note = {Publisher: {EDP} Sciences},
}

@misc{binkowski_demystifying_2021,
	title = {Demystifying {MMD} {GANs}},
	url = {http://arxiv.org/abs/1801.01401},
	doi = {10.48550/arXiv.1801.01401},
	number = {{arXiv}:1801.01401},
	publisher = {{arXiv}},
	author = {Bińkowski, Mikołaj and Sutherland, Danica J. and Arbel, Michael and Gretton, Arthur},
	urldate = {2023-08-12},
	date = {2021-01-14},
	eprinttype = {arxiv},
	eprint = {1801.01401 [cs, stat]},
}

@misc{szegedy_rethinking_2015,
	title = {Rethinking the Inception Architecture for Computer Vision},
	url = {http://arxiv.org/abs/1512.00567},
	doi = {10.48550/arXiv.1512.00567},
	number = {{arXiv}:1512.00567},
	publisher = {{arXiv}},
	author = {Szegedy, Christian and Vanhoucke, Vincent and Ioffe, Sergey and Shlens, Jonathon and Wojna, Zbigniew},
	urldate = {2023-08-12},
	date = {2015-12-11},
	eprinttype = {arxiv},
	eprint = {1512.00567 [cs]},
}

@misc{parmar_aliased_2022,
	title = {On Aliased Resizing and Surprising Subtleties in {GAN} Evaluation},
	url = {http://arxiv.org/abs/2104.11222},
	doi = {10.48550/arXiv.2104.11222},
	number = {{arXiv}:2104.11222},
	publisher = {{arXiv}},
	author = {Parmar, Gaurav and Zhang, Richard and Zhu, Jun-Yan},
	urldate = {2023-08-12},
	date = {2022-01-20},
	eprinttype = {arxiv},
	eprint = {2104.11222 [cs]},
}

@online{noauthor_minimum_nodate,
	title = {The minimum of potential energy of a System of point charges},
	url = {https://www.degruyter.com/document/doi/10.1515/dma.1993.3.1.75/html?lang=en},
	urldate = {2023-08-12},
}

@misc{heusel_gans_2018,
	title = {{GANs} Trained by a Two Time-Scale Update Rule Converge to a Local Nash Equilibrium},
	url = {http://arxiv.org/abs/1706.08500},
	doi = {10.48550/arXiv.1706.08500},
	number = {{arXiv}:1706.08500},
	publisher = {{arXiv}},
	author = {Heusel, Martin and Ramsauer, Hubert and Unterthiner, Thomas and Nessler, Bernhard and Hochreiter, Sepp},
	urldate = {2023-08-10},
	date = {2018-01-12},
	eprinttype = {arxiv},
	eprint = {1706.08500 [cs, stat]},
}

@article{kuhr_computing_2011,
	title = {Computing at Belle {II}},
	volume = {331},
	issn = {1742-6596},
	url = {https://dx.doi.org/10.1088/1742-6596/331/7/072021},
	doi = {10.1088/1742-6596/331/7/072021},
	pages = {072021},
	number = {7},
	journaltitle = {Journal of Physics: Conference Series},
	shortjournal = {J. Phys.: Conf. Ser.},
	author = {Kuhr, Thomas},
	urldate = {2023-08-10},
	date = {2011-12},
	langid = {english},
}

@online{noauthor_asymptotics_nodate,
	title = {Asymptotics for Minimal Discrete Energy on the Sphere on {JSTOR}},
	url = {https://www.jstor.org/stable/117605},
	urldate = {2023-08-10},
}

@online{noauthor_xxiv_nodate,
	title = {{XXIV}. On the structure of the atom: an investigation of the stability and periods of oscillation of a number of corpuscles arranged at equal intervals around the circumference of a circle; with application of the results to the theory of atomic structure: The London, Edinburgh, and Dublin Philosophical Magazine and Journal of Science: Vol 7, No 39},
	url = {https://www.tandfonline.com/doi/abs/10.1080/14786440409463107},
	urldate = {2023-08-10},
}

@article{cao_coupling_2015,
	title = {Coupling Learning of Complex Interactions},
	volume = {51},
	issn = {03064573},
	url = {http://arxiv.org/abs/2007.13534},
	doi = {10.1016/j.ipm.2014.08.007},
	pages = {167--186},
	number = {2},
	journaltitle = {Information Processing \& Management},
	shortjournal = {Information Processing \& Management},
	author = {Cao, Longbing},
	urldate = {2023-08-10},
	date = {2015-03},
	eprinttype = {arxiv},
	eprint = {2007.13534 [cs]},
}

@online{noauthor_information_nodate,
	title = {On Information and Sufficiency},
	url = {https://projecteuclid.org/journals/annals-of-mathematical-statistics/volume-22/issue-1/On-Information-and-Sufficiency/10.1214/aoms/1177729694.full},
	urldate = {2023-08-10},
}

@misc{balestriero_cookbook_2023,
	title = {A Cookbook of Self-Supervised Learning},
	url = {http://arxiv.org/abs/2304.12210},
	doi = {10.48550/arXiv.2304.12210},
	number = {{arXiv}:2304.12210},
	publisher = {{arXiv}},
	author = {Balestriero, Randall and Ibrahim, Mark and Sobal, Vlad and Morcos, Ari and Shekhar, Shashank and Goldstein, Tom and Bordes, Florian and Bardes, Adrien and Mialon, Gregoire and Tian, Yuandong and Schwarzschild, Avi and Wilson, Andrew Gordon and Geiping, Jonas and Garrido, Quentin and Fernandez, Pierre and Bar, Amir and Pirsiavash, Hamed and {LeCun}, Yann and Goldblum, Micah},
	urldate = {2023-08-10},
	date = {2023-06-28},
	eprinttype = {arxiv},
	eprint = {2304.12210 [cs]},
}

@online{weng_self-supervised_2019,
	title = {Self-Supervised Representation Learning},
	url = {https://lilianweng.github.io/posts/2019-11-10-self-supervised/},
	author = {Weng, Lilian},
	urldate = {2023-08-10},
	date = {2019-11-10},
	langid = {english},
	note = {Section: posts},
}

@online{noauthor_170301203_nodate,
	title = {[1703.01203] Stochastic Separation Theorems},
	url = {https://arxiv.org/abs/1703.01203},
	urldate = {2023-08-10},
}

@inproceedings{wang_normface_2017,
	title = {{NormFace}: L2 Hypersphere Embedding for Face Verification},
	url = {http://arxiv.org/abs/1704.06369},
	doi = {10.1145/3123266.3123359},
	shorttitle = {{NormFace}},
	pages = {1041--1049},
	booktitle = {Proceedings of the 25th {ACM} international conference on Multimedia},
	author = {Wang, Feng and Xiang, Xiang and Cheng, Jian and Yuille, Alan L.},
	urldate = {2023-08-10},
	date = {2017-10-19},
	eprinttype = {arxiv},
	eprint = {1704.06369 [cs]},
}

@misc{zaheer_deep_2018,
	title = {Deep Sets},
	url = {http://arxiv.org/abs/1703.06114},
	doi = {10.48550/arXiv.1703.06114},
	number = {{arXiv}:1703.06114},
	publisher = {{arXiv}},
	author = {Zaheer, Manzil and Kottur, Satwik and Ravanbakhsh, Siamak and Poczos, Barnabas and Salakhutdinov, Ruslan and Smola, Alexander},
	urldate = {2023-08-10},
	date = {2018-04-14},
	eprinttype = {arxiv},
	eprint = {1703.06114 [cs, stat]},
}

@misc{ravanbakhsh_equivariance_2017,
	title = {Equivariance Through Parameter-Sharing},
	url = {http://arxiv.org/abs/1702.08389},
	doi = {10.48550/arXiv.1702.08389},
	number = {{arXiv}:1702.08389},
	publisher = {{arXiv}},
	author = {Ravanbakhsh, Siamak and Schneider, Jeff and Poczos, Barnabas},
	urldate = {2023-08-10},
	date = {2017-06-13},
	eprinttype = {arxiv},
	eprint = {1702.08389 [cs, stat]},
}

@misc{guttenberg_permutation-equivariant_2016,
	title = {Permutation-equivariant neural networks applied to dynamics prediction},
	url = {http://arxiv.org/abs/1612.04530},
	doi = {10.48550/arXiv.1612.04530},
	number = {{arXiv}:1612.04530},
	publisher = {{arXiv}},
	author = {Guttenberg, Nicholas and Virgo, Nathaniel and Witkowski, Olaf and Aoki, Hidetoshi and Kanai, Ryota},
	urldate = {2023-08-10},
	date = {2016-12-14},
	eprinttype = {arxiv},
	eprint = {1612.04530 [cs, stat]},
}

@misc{yun_are_2020,
	title = {Are Transformers universal approximators of sequence-to-sequence functions?},
	url = {http://arxiv.org/abs/1912.10077},
	doi = {10.48550/arXiv.1912.10077},
	number = {{arXiv}:1912.10077},
	publisher = {{arXiv}},
	author = {Yun, Chulhee and Bhojanapalli, Srinadh and Rawat, Ankit Singh and Reddi, Sashank J. and Kumar, Sanjiv},
	urldate = {2023-08-10},
	date = {2020-02-24},
	eprinttype = {arxiv},
	eprint = {1912.10077 [cs, stat]},
}

@inproceedings{clark_what_2019,
	location = {Florence, Italy},
	title = {What Does {BERT} Look at? An Analysis of {BERT}'s Attention},
	url = {https://aclanthology.org/W19-4828},
	doi = {10.18653/v1/W19-4828},
	shorttitle = {What Does {BERT} Look at?},
	eventtitle = {{BlackboxNLP} 2019},
	pages = {276--286},
	booktitle = {Proceedings of the 2019 {ACL} Workshop {BlackboxNLP}: Analyzing and Interpreting Neural Networks for {NLP}},
	publisher = {Association for Computational Linguistics},
	author = {Clark, Kevin and Khandelwal, Urvashi and Levy, Omer and Manning, Christopher D.},
	urldate = {2023-08-10},
	date = {2019-08},
}

@misc{liu_understanding_2020,
	title = {Understanding the Difficulty of Training Transformers},
	url = {http://arxiv.org/abs/2004.08249},
	doi = {10.48550/arXiv.2004.08249},
	number = {{arXiv}:2004.08249},
	publisher = {{arXiv}},
	author = {Liu, Liyuan and Liu, Xiaodong and Gao, Jianfeng and Chen, Weizhu and Han, Jiawei},
	urldate = {2023-08-10},
	date = {2020-09-18},
	eprinttype = {arxiv},
	eprint = {2004.08249 [cs, stat]},
}

@misc{dosovitskiy_image_2021,
	title = {An Image is Worth 16x16 Words: Transformers for Image Recognition at Scale},
	url = {http://arxiv.org/abs/2010.11929},
	doi = {10.48550/arXiv.2010.11929},
	shorttitle = {An Image is Worth 16x16 Words},
	number = {{arXiv}:2010.11929},
	publisher = {{arXiv}},
	author = {Dosovitskiy, Alexey and Beyer, Lucas and Kolesnikov, Alexander and Weissenborn, Dirk and Zhai, Xiaohua and Unterthiner, Thomas and Dehghani, Mostafa and Minderer, Matthias and Heigold, Georg and Gelly, Sylvain and Uszkoreit, Jakob and Houlsby, Neil},
	urldate = {2023-08-10},
	date = {2021-06-03},
	eprinttype = {arxiv},
	eprint = {2010.11929 [cs]},
}

@misc{jiang_transgan_2021,
	title = {{TransGAN}: Two Pure Transformers Can Make One Strong {GAN}, and That Can Scale Up},
	url = {http://arxiv.org/abs/2102.07074},
	doi = {10.48550/arXiv.2102.07074},
	shorttitle = {{TransGAN}},
	number = {{arXiv}:2102.07074},
	publisher = {{arXiv}},
	author = {Jiang, Yifan and Chang, Shiyu and Wang, Zhangyang},
	urldate = {2023-08-10},
	date = {2021-12-08},
	eprinttype = {arxiv},
	eprint = {2102.07074 [cs]},
}

@inproceedings{hudson_generative_2021,
	title = {Generative Adversarial Transformers},
	url = {https://proceedings.mlr.press/v139/hudson21a.html},
	eventtitle = {International Conference on Machine Learning},
	pages = {4487--4499},
	booktitle = {Proceedings of the 38th International Conference on Machine Learning},
	publisher = {{PMLR}},
	author = {Hudson, Drew A. and Zitnick, Larry},
	urldate = {2023-08-10},
	date = {2021-07-01},
	langid = {english},
	note = {{ISSN}: 2640-3498},
}

@misc{wang_understanding_2022,
	title = {Understanding Contrastive Representation Learning through Alignment and Uniformity on the Hypersphere},
	url = {http://arxiv.org/abs/2005.10242},
	doi = {10.48550/arXiv.2005.10242},
	number = {{arXiv}:2005.10242},
	publisher = {{arXiv}},
	author = {Wang, Tongzhou and Isola, Phillip},
	urldate = {2023-08-10},
	date = {2022-08-15},
	eprinttype = {arxiv},
	eprint = {2005.10242 [cs, stat]},
}

@misc{zhang_word_2018,
	title = {Word Embedding Perturbation for Sentence Classification},
	url = {http://arxiv.org/abs/1804.08166},
	doi = {10.48550/arXiv.1804.08166},
	number = {{arXiv}:1804.08166},
	publisher = {{arXiv}},
	author = {Zhang, Dongxu and Yang, Zhichao},
	urldate = {2023-08-10},
	date = {2018-04-22},
	eprinttype = {arxiv},
	eprint = {1804.08166 [cs]},
}

@misc{devlin_bert_2019,
	title = {{BERT}: Pre-training of Deep Bidirectional Transformers for Language Understanding},
	url = {http://arxiv.org/abs/1810.04805},
	doi = {10.48550/arXiv.1810.04805},
	shorttitle = {{BERT}},
	number = {{arXiv}:1810.04805},
	publisher = {{arXiv}},
	author = {Devlin, Jacob and Chang, Ming-Wei and Lee, Kenton and Toutanova, Kristina},
	urldate = {2023-08-10},
	date = {2019-05-24},
	eprinttype = {arxiv},
	eprint = {1810.04805 [cs]},
}

@misc{sharifzadeh_classification_2020,
	title = {Classification by Attention: Scene Graph Classification with Prior Knowledge},
	url = {http://arxiv.org/abs/2011.10084},
	doi = {10.48550/arXiv.2011.10084},
	shorttitle = {Classification by Attention},
	number = {{arXiv}:2011.10084},
	publisher = {{arXiv}},
	author = {Sharifzadeh, Sahand and Baharlou, Sina Moayed and Tresp, Volker},
	urldate = {2023-08-10},
	date = {2020-12-17},
	eprinttype = {arxiv},
	eprint = {2011.10084 [cs]},
}

@misc{zhang_consistency_2020,
	title = {Consistency Regularization for Generative Adversarial Networks},
	url = {http://arxiv.org/abs/1910.12027},
	number = {{arXiv}:1910.12027},
	publisher = {{arXiv}},
	author = {Zhang, Han and Zhang, Zizhao and Odena, Augustus and Lee, Honglak},
	urldate = {2023-08-09},
	date = {2020-02-18},
	eprinttype = {arxiv},
	eprint = {1910.12027 [cs, stat]},
}

@online{noauthor_visual_nodate,
	title = {Visual and semantic similarity in {ImageNet} {\textbar} Proceedings of the 2011 {IEEE} Conference on Computer Vision and Pattern Recognition},
	url = {https://dl.acm.org/doi/10.1109/CVPR.2011.5995474},
	urldate = {2023-08-03},
}

@misc{rangwani_class_2021,
	title = {Class Balancing {GAN} with a Classifier in the Loop},
	url = {http://arxiv.org/abs/2106.09402},
	doi = {10.48550/arXiv.2106.09402},
	number = {{arXiv}:2106.09402},
	publisher = {{arXiv}},
	author = {Rangwani, Harsh and Mopuri, Konda Reddy and Babu, R. Venkatesh},
	urldate = {2023-08-03},
	date = {2021-06-17},
	eprinttype = {arxiv},
	eprint = {2106.09402 [cs]},
}

@online{noauthor_211106119_nodate,
	title = {[2111.06119] Fine-Grained Image Analysis with Deep Learning: A Survey},
	url = {https://arxiv.org/abs/2111.06119},
	urldate = {2023-08-03},
}

@misc{geng_generative_2020,
	title = {Generative Model without Prior Distribution Matching},
	url = {http://arxiv.org/abs/2009.11016},
	doi = {10.48550/arXiv.2009.11016},
	number = {{arXiv}:2009.11016},
	publisher = {{arXiv}},
	author = {Geng, Cong and Wang, Jia and Chen, Li and Gao, Zhiyong},
	urldate = {2023-07-04},
	date = {2020-09-23},
	eprinttype = {arxiv},
	eprint = {2009.11016 [cs]},
}

@misc{martinez_ld-gan_2023,
	title = {{LD}-{GAN}: Low-Dimensional Generative Adversarial Network for Spectral Image Generation with Variance Regularization},
	url = {http://arxiv.org/abs/2305.00132},
	doi = {10.48550/arXiv.2305.00132},
	shorttitle = {{LD}-{GAN}},
	number = {{arXiv}:2305.00132},
	publisher = {{arXiv}},
	author = {Martinez, Emmanuel and Jacome, Roman and Hernandez-Rojas, Alejandra and Arguello, Henry},
	urldate = {2023-07-04},
	date = {2023-04-28},
	eprinttype = {arxiv},
	eprint = {2305.00132 [cs, eess]},
}

@misc{su_unified_2023,
	title = {A Unified {GAN} Framework Regarding Manifold Alignment for Remote Sensing Images Generation},
	url = {http://arxiv.org/abs/2305.19507},
	doi = {10.48550/arXiv.2305.19507},
	number = {{arXiv}:2305.19507},
	publisher = {{arXiv}},
	author = {Su, Xingzhe and Qiang, Wenwen and Song, Zeen and Gao, Hang and Wu, Fengge and Zheng, Changwen},
	urldate = {2023-07-04},
	date = {2023-05-30},
	eprinttype = {arxiv},
	eprint = {2305.19507 [cs, eess]},
}

@misc{ding_reviving_2023,
	title = {Reviving Shift Equivariance in Vision Transformers},
	url = {http://arxiv.org/abs/2306.07470},
	doi = {10.48550/arXiv.2306.07470},
	number = {{arXiv}:2306.07470},
	publisher = {{arXiv}},
	author = {Ding, Peijian and Soselia, Davit and Armstrong, Thomas and Su, Jiahao and Huang, Furong},
	urldate = {2023-07-04},
	date = {2023-06-12},
	eprinttype = {arxiv},
	eprint = {2306.07470 [cs]},
	note = {version: 1},
}

@misc{zheng_dynamic_2022,
	title = {Dynamic Position Encoding for Transformers},
	url = {http://arxiv.org/abs/2204.08142},
	doi = {10.48550/arXiv.2204.08142},
	number = {{arXiv}:2204.08142},
	publisher = {{arXiv}},
	author = {Zheng, Joyce and Rezagholizadeh, Mehdi and Passban, Peyman},
	urldate = {2023-07-04},
	date = {2022-10-22},
	eprinttype = {arxiv},
	eprint = {2204.08142 [cs]},
	note = {version: 2},
}

@misc{liu_learning_2020,
	title = {Learning to Encode Position for Transformer with Continuous Dynamical Model},
	url = {http://arxiv.org/abs/2003.09229},
	doi = {10.48550/arXiv.2003.09229},
	number = {{arXiv}:2003.09229},
	publisher = {{arXiv}},
	author = {Liu, Xuanqing and Yu, Hsiang-Fu and Dhillon, Inderjit and Hsieh, Cho-Jui},
	urldate = {2023-07-04},
	date = {2020-03-12},
	eprinttype = {arxiv},
	eprint = {2003.09229 [cs, stat]},
	note = {version: 1},
}

@misc{young_dateformer_2023,
	title = {Dateformer: Time-modeling Transformer for Longer-term Series Forecasting},
	url = {http://arxiv.org/abs/2207.05397},
	doi = {10.48550/arXiv.2207.05397},
	shorttitle = {Dateformer},
	number = {{arXiv}:2207.05397},
	publisher = {{arXiv}},
	author = {Young, Julong and Chen, Junhui and Huang, Feihu and Peng, Jian},
	urldate = {2023-07-04},
	date = {2023-02-21},
	eprinttype = {arxiv},
	eprint = {2207.05397 [cs]},
}

@misc{hassani_escaping_2022,
	title = {Escaping the Big Data Paradigm with Compact Transformers},
	url = {http://arxiv.org/abs/2104.05704},
	doi = {10.48550/arXiv.2104.05704},
	number = {{arXiv}:2104.05704},
	publisher = {{arXiv}},
	author = {Hassani, Ali and Walton, Steven and Shah, Nikhil and Abuduweili, Abulikemu and Li, Jiachen and Shi, Humphrey},
	urldate = {2023-07-04},
	date = {2022-06-07},
	eprinttype = {arxiv},
	eprint = {2104.05704 [cs]},
	note = {version: 4},
}

@misc{fan_addressing_2021,
	title = {Addressing Some Limitations of Transformers with Feedback Memory},
	url = {http://arxiv.org/abs/2002.09402},
	doi = {10.48550/arXiv.2002.09402},
	number = {{arXiv}:2002.09402},
	publisher = {{arXiv}},
	author = {Fan, Angela and Lavril, Thibaut and Grave, Edouard and Joulin, Armand and Sukhbaatar, Sainbayar},
	urldate = {2023-07-04},
	date = {2021-01-25},
	eprinttype = {arxiv},
	eprint = {2002.09402 [cs, stat]},
}

@misc{zeng_are_2022,
	title = {Are Transformers Effective for Time Series Forecasting?},
	url = {http://arxiv.org/abs/2205.13504},
	doi = {10.48550/arXiv.2205.13504},
	number = {{arXiv}:2205.13504},
	publisher = {{arXiv}},
	author = {Zeng, Ailing and Chen, Muxi and Zhang, Lei and Xu, Qiang},
	urldate = {2023-07-04},
	date = {2022-08-17},
	eprinttype = {arxiv},
	eprint = {2205.13504 [cs]},
}

@incollection{rumelhart_learning_1987,
	title = {Learning Internal Representations by Error Propagation},
	isbn = {978-0-262-29140-8},
	url = {https://ieeexplore.ieee.org/document/6302929},
	pages = {318--362},
	booktitle = {Parallel Distributed Processing: Explorations in the Microstructure of Cognition: Foundations},
	publisher = {{MIT} Press},
	author = {Rumelhart, David E. and {McClelland}, James L.},
	urldate = {2023-06-30},
	date = {1987},
	note = {Conference Name: Parallel Distributed Processing: Explorations in the Microstructure of Cognition: Foundations},
}

@misc{zhang_how_2022,
	title = {How Does {SimSiam} Avoid Collapse Without Negative Samples? A Unified Understanding with Self-supervised Contrastive Learning},
	url = {http://arxiv.org/abs/2203.16262},
	shorttitle = {How Does {SimSiam} Avoid Collapse Without Negative Samples?},
	number = {{arXiv}:2203.16262},
	publisher = {{arXiv}},
	author = {Zhang, Chaoning and Zhang, Kang and Zhang, Chenshuang and Pham, Trung X. and Yoo, Chang D. and Kweon, In So},
	urldate = {2023-06-28},
	date = {2022-03-30},
	eprinttype = {arxiv},
	eprint = {2203.16262 [cs]},
}

@misc{moon_embedding-dynamic_2022,
	title = {An Embedding-Dynamic Approach to Self-supervised Learning},
	url = {http://arxiv.org/abs/2207.03552},
	number = {{arXiv}:2207.03552},
	publisher = {{arXiv}},
	author = {Moon, Suhong and Buracas, Domas and Park, Seunghyun and Kim, Jinkyu and Canny, John},
	urldate = {2023-06-28},
	date = {2022-07-07},
	eprinttype = {arxiv},
	eprint = {2207.03552 [cs]},
}

@online{noauthor_understanding_nodate,
	title = {Understanding self-supervised and contrastive learning with "Bootstrap Your Own Latent" ({BYOL})},
	url = {https://generallyintelligent.com/research/2020-08-24-understanding-self-supervised-contrastive-learning/},
	urldate = {2023-06-28},
	langid = {american},
}

@misc{caron_emerging_2021,
	title = {Emerging Properties in Self-Supervised Vision Transformers},
	url = {http://arxiv.org/abs/2104.14294},
	doi = {10.48550/arXiv.2104.14294},
	number = {{arXiv}:2104.14294},
	publisher = {{arXiv}},
	author = {Caron, Mathilde and Touvron, Hugo and Misra, Ishan and Jégou, Hervé and Mairal, Julien and Bojanowski, Piotr and Joulin, Armand},
	urldate = {2023-06-27},
	date = {2021-05-24},
	eprinttype = {arxiv},
	eprint = {2104.14294 [cs]},
}

@inproceedings{yim_gift_2017,
	title = {A Gift from Knowledge Distillation: Fast Optimization, Network Minimization and Transfer Learning},
	doi = {10.1109/CVPR.2017.754},
	shorttitle = {A Gift from Knowledge Distillation},
	eventtitle = {2017 {IEEE} Conference on Computer Vision and Pattern Recognition ({CVPR})},
	pages = {7130--7138},
	booktitle = {2017 {IEEE} Conference on Computer Vision and Pattern Recognition ({CVPR})},
	author = {Yim, Junho and Joo, Donggyu and Bae, Jihoon and Kim, Junmo},
	date = {2017-07},
	note = {{ISSN}: 1063-6919},
}

@article{romero_fitnets_2014,
	title = {{FitNets}: Hints for Thin Deep Nets},
	url = {https://www.semanticscholar.org/paper/FitNets%3A-Hints-for-Thin-Deep-Nets-Romero-Ballas/cd85a549add0c7c7def36aca29837efd24b24080},
	shorttitle = {{FitNets}},
	journaltitle = {{CoRR}},
	author = {Romero, Adriana and Ballas, Nicolas and Kahou, S. and Chassang, Antoine and Gatta, C. and Bengio, Yoshua},
	urldate = {2023-06-27},
	date = {2014-12-19},
}

@misc{furlanello_born_2018,
	title = {Born Again Neural Networks},
	url = {http://arxiv.org/abs/1805.04770},
	doi = {10.48550/arXiv.1805.04770},
	number = {{arXiv}:1805.04770},
	publisher = {{arXiv}},
	author = {Furlanello, Tommaso and Lipton, Zachary C. and Tschannen, Michael and Itti, Laurent and Anandkumar, Anima},
	urldate = {2023-06-25},
	date = {2018-06-29},
	eprinttype = {arxiv},
	eprint = {1805.04770 [cs, stat]},
}

@misc{grill_bootstrap_2020,
	title = {Bootstrap your own latent: A new approach to self-supervised Learning},
	url = {http://arxiv.org/abs/2006.07733},
	doi = {10.48550/arXiv.2006.07733},
	shorttitle = {Bootstrap your own latent},
	number = {{arXiv}:2006.07733},
	publisher = {{arXiv}},
	author = {Grill, Jean-Bastien and Strub, Florian and Altché, Florent and Tallec, Corentin and Richemond, Pierre H. and Buchatskaya, Elena and Doersch, Carl and Pires, Bernardo Avila and Guo, Zhaohan Daniel and Azar, Mohammad Gheshlaghi and Piot, Bilal and Kavukcuoglu, Koray and Munos, Rémi and Valko, Michal},
	urldate = {2023-06-26},
	date = {2020-09-10},
	eprinttype = {arxiv},
	eprint = {2006.07733 [cs, stat]},
}

@misc{hinton_distilling_2015,
	title = {Distilling the Knowledge in a Neural Network},
	url = {http://arxiv.org/abs/1503.02531},
	doi = {10.48550/arXiv.1503.02531},
	number = {{arXiv}:1503.02531},
	publisher = {{arXiv}},
	author = {Hinton, Geoffrey and Vinyals, Oriol and Dean, Jeff},
	urldate = {2023-06-25},
	date = {2015-03-09},
	eprinttype = {arxiv},
	eprint = {1503.02531 [cs, stat]},
}

@misc{jing_understanding_2022,
	title = {Understanding Dimensional Collapse in Contrastive Self-supervised Learning},
	url = {http://arxiv.org/abs/2110.09348},
	doi = {10.48550/arXiv.2110.09348},
	number = {{arXiv}:2110.09348},
	publisher = {{arXiv}},
	author = {Jing, Li and Vincent, Pascal and {LeCun}, Yann and Tian, Yuandong},
	urldate = {2023-06-25},
	date = {2022-04-23},
	eprinttype = {arxiv},
	eprint = {2110.09348 [cs]},
}

@misc{chen_simple_2020,
	title = {A Simple Framework for Contrastive Learning of Visual Representations},
	url = {http://arxiv.org/abs/2002.05709},
	doi = {10.48550/arXiv.2002.05709},
	number = {{arXiv}:2002.05709},
	publisher = {{arXiv}},
	author = {Chen, Ting and Kornblith, Simon and Norouzi, Mohammad and Hinton, Geoffrey},
	urldate = {2023-06-25},
	date = {2020-06-30},
	eprinttype = {arxiv},
	eprint = {2002.05709 [cs, stat]},
}

@inproceedings{sohn_improved_2016,
	title = {Improved Deep Metric Learning with Multi-class N-pair Loss Objective},
	volume = {29},
	url = {https://papers.nips.cc/paper_files/paper/2016/hash/6b180037abbebea991d8b1232f8a8ca9-Abstract.html},
	booktitle = {Advances in Neural Information Processing Systems},
	publisher = {Curran Associates, Inc.},
	author = {Sohn, Kihyuk},
	urldate = {2023-06-25},
	date = {2016},
}

@misc{oord_representation_2019,
	title = {Representation Learning with Contrastive Predictive Coding},
	url = {http://arxiv.org/abs/1807.03748},
	doi = {10.48550/arXiv.1807.03748},
	number = {{arXiv}:1807.03748},
	publisher = {{arXiv}},
	author = {Oord, Aaron van den and Li, Yazhe and Vinyals, Oriol},
	urldate = {2023-06-25},
	date = {2019-01-22},
	eprinttype = {arxiv},
	eprint = {1807.03748 [cs, stat]},
}

@article{chechik_large_2010,
	title = {Large Scale Online Learning of Image Similarity Through Ranking},
	volume = {11},
	issn = {1533-7928},
	url = {http://jmlr.org/papers/v11/chechik10a.html},
	pages = {1109--1135},
	number = {36},
	journaltitle = {Journal of Machine Learning Research},
	author = {Chechik, Gal and Sharma, Varun and Shalit, Uri and Bengio, Samy},
	urldate = {2023-06-25},
	date = {2010},
}

@inproceedings{weinberger_distance_2005,
	title = {Distance Metric Learning for Large Margin Nearest Neighbor Classification},
	volume = {18},
	url = {https://proceedings.neurips.cc/paper/2005/hash/a7f592cef8b130a6967a90617db5681b-Abstract.html},
	booktitle = {Advances in Neural Information Processing Systems},
	publisher = {{MIT} Press},
	author = {Weinberger, Kilian Q and Blitzer, John and Saul, Lawrence},
	urldate = {2023-06-25},
	date = {2005},
}

@article{bromley_signature_nodate,
	title = {Signature Verification using a "Siamese" Time Delay Neural Network},
	author = {Bromley, Jane and Guyon, Isabelle and {LeCun}, Yann and Säckinger, Eduard and Shah, Roopak},
	langid = {english},
}

@online{noauthor_self-supervised_nodate,
	title = {Self-supervised learning: The dark matter of intelligence},
	url = {https://ai.facebook.com/blog/self-supervised-learning-the-dark-matter-of-intelligence/},
	urldate = {2023-06-24},
}

@article{le-khac_contrastive_2020,
	title = {Contrastive Representation Learning: A Framework and Review},
	volume = {8},
	issn = {2169-3536},
	url = {http://arxiv.org/abs/2010.05113},
	doi = {10.1109/ACCESS.2020.3031549},
	shorttitle = {Contrastive Representation Learning},
	pages = {193907--193934},
	journaltitle = {{IEEE} Access},
	shortjournal = {{IEEE} Access},
	author = {Le-Khac, Phuc H. and Healy, Graham and Smeaton, Alan F.},
	urldate = {2023-06-24},
	date = {2020},
	eprinttype = {arxiv},
	eprint = {2010.05113 [cs, stat]},
}

@inreference{noauthor_assignment_2023,
	title = {Assignment problem},
	rights = {Creative Commons Attribution-{ShareAlike} License},
	url = {https://en.wikipedia.org/w/index.php?title=Assignment_problem&oldid=1159259304},
	booktitle = {Wikipedia},
	urldate = {2023-06-24},
	date = {2023-06-09},
	langid = {english},
	note = {Page Version {ID}: 1159259304},
}

@article{tomizawa_techniques_1971,
	title = {On some techniques useful for solution of transportation network problems},
	volume = {1},
	rights = {Copyright © 1971 Wiley Periodicals, Inc., A Wiley Company},
	issn = {1097-0037},
	url = {https://onlinelibrary.wiley.com/doi/abs/10.1002/net.3230010206},
	doi = {10.1002/net.3230010206},
	pages = {173--194},
	number = {2},
	journaltitle = {Networks},
	author = {Tomizawa, N.},
	urldate = {2023-06-24},
	date = {1971},
	langid = {english},
	note = {\_eprint: https://onlinelibrary.wiley.com/doi/pdf/10.1002/net.3230010206},
}

@online{noauthor_theoretical_nodate,
	title = {Theoretical Improvements in Algorithmic Efficiency for Network Flow Problems {\textbar} Journal of the {ACM}},
	url = {https://dl.acm.org/doi/10.1145/321694.321699},
	urldate = {2023-06-24},
}

@online{noauthor_hungarian_nodate,
	title = {The Hungarian method for the assignment problem - Kuhn - 1955 - Naval Research Logistics Quarterly - Wiley Online Library},
	url = {https://onlinelibrary.wiley.com/doi/10.1002/nav.3800020109},
	urldate = {2023-06-24},
}

@misc{vignac_top-n_2022,
	title = {Top-N: Equivariant set and graph generation without exchangeability},
	url = {http://arxiv.org/abs/2110.02096},
	doi = {10.48550/arXiv.2110.02096},
	shorttitle = {Top-N},
	number = {{arXiv}:2110.02096},
	publisher = {{arXiv}},
	author = {Vignac, Clement and Frossard, Pascal},
	urldate = {2023-06-23},
	date = {2022-04-01},
	eprinttype = {arxiv},
	eprint = {2110.02096 [cs, stat]},
}

@misc{he_deep_2015,
	title = {Deep Residual Learning for Image Recognition},
	url = {http://arxiv.org/abs/1512.03385},
	doi = {10.48550/arXiv.1512.03385},
	number = {{arXiv}:1512.03385},
	publisher = {{arXiv}},
	author = {He, Kaiming and Zhang, Xiangyu and Ren, Shaoqing and Sun, Jian},
	urldate = {2023-06-22},
	date = {2015-12-10},
	eprinttype = {arxiv},
	eprint = {1512.03385 [cs]},
}

@misc{zhang_self-attention_2019,
	title = {Self-Attention Generative Adversarial Networks},
	url = {http://arxiv.org/abs/1805.08318},
	doi = {10.48550/arXiv.1805.08318},
	number = {{arXiv}:1805.08318},
	publisher = {{arXiv}},
	author = {Zhang, Han and Goodfellow, Ian and Metaxas, Dimitris and Odena, Augustus},
	urldate = {2023-06-22},
	date = {2019-06-14},
	eprinttype = {arxiv},
	eprint = {1805.08318 [cs, stat]},
}

@misc{brock_neural_2017,
	title = {Neural Photo Editing with Introspective Adversarial Networks},
	url = {http://arxiv.org/abs/1609.07093},
	doi = {10.48550/arXiv.1609.07093},
	number = {{arXiv}:1609.07093},
	publisher = {{arXiv}},
	author = {Brock, Andrew and Lim, Theodore and Ritchie, J. M. and Weston, Nick},
	urldate = {2023-06-21},
	date = {2017-02-06},
	eprinttype = {arxiv},
	eprint = {1609.07093 [cs, stat]},
}

@misc{hu_provable_2020,
	title = {Provable Benefit of Orthogonal Initialization in Optimizing Deep Linear Networks},
	url = {http://arxiv.org/abs/2001.05992},
	number = {{arXiv}:2001.05992},
	publisher = {{arXiv}},
	author = {Hu, Wei and Xiao, Lechao and Pennington, Jeffrey},
	urldate = {2023-06-21},
	date = {2020-01-16},
	eprinttype = {arxiv},
	eprint = {2001.05992 [cs, math, stat]},
	note = {version: 1},
}

@misc{miyato_spectral_2018,
	title = {Spectral Normalization for Generative Adversarial Networks},
	url = {http://arxiv.org/abs/1802.05957},
	doi = {10.48550/arXiv.1802.05957},
	number = {{arXiv}:1802.05957},
	publisher = {{arXiv}},
	author = {Miyato, Takeru and Kataoka, Toshiki and Koyama, Masanori and Yoshida, Yuichi},
	urldate = {2023-06-20},
	date = {2018-02-16},
	eprinttype = {arxiv},
	eprint = {1802.05957 [cs, stat]},
}

@misc{de_vries_modulating_2017,
	title = {Modulating early visual processing by language},
	url = {http://arxiv.org/abs/1707.00683},
	doi = {10.48550/arXiv.1707.00683},
	number = {{arXiv}:1707.00683},
	publisher = {{arXiv}},
	author = {de Vries, Harm and Strub, Florian and Mary, Jérémie and Larochelle, Hugo and Pietquin, Olivier and Courville, Aaron},
	urldate = {2023-06-20},
	date = {2017-12-18},
	eprinttype = {arxiv},
	eprint = {1707.00683 [cs]},
	note = {version: 3},
}

@misc{kang_rebooting_2021,
	title = {Rebooting {ACGAN}: Auxiliary Classifier {GANs} with Stable Training},
	url = {http://arxiv.org/abs/2111.01118},
	doi = {10.48550/arXiv.2111.01118},
	shorttitle = {Rebooting {ACGAN}},
	number = {{arXiv}:2111.01118},
	publisher = {{arXiv}},
	author = {Kang, Minguk and Shim, Woohyeon and Cho, Minsu and Park, Jaesik},
	urldate = {2023-06-20},
	date = {2021-11-01},
	eprinttype = {arxiv},
	eprint = {2111.01118 [cs]},
}

@misc{miyato_cgans_2018,
	title = {{cGANs} with Projection Discriminator},
	url = {http://arxiv.org/abs/1802.05637},
	doi = {10.48550/arXiv.1802.05637},
	number = {{arXiv}:1802.05637},
	publisher = {{arXiv}},
	author = {Miyato, Takeru and Koyama, Masanori},
	urldate = {2023-06-20},
	date = {2018-08-14},
	eprinttype = {arxiv},
	eprint = {1802.05637 [cs, stat]},
}

@misc{odena_conditional_2017,
	title = {Conditional Image Synthesis With Auxiliary Classifier {GANs}},
	url = {http://arxiv.org/abs/1610.09585},
	doi = {10.48550/arXiv.1610.09585},
	number = {{arXiv}:1610.09585},
	publisher = {{arXiv}},
	author = {Odena, Augustus and Olah, Christopher and Shlens, Jonathon},
	urldate = {2023-06-20},
	date = {2017-07-20},
	eprinttype = {arxiv},
	eprint = {1610.09585 [cs, stat]},
}

@misc{mirza_conditional_2014,
	title = {Conditional Generative Adversarial Nets},
	url = {http://arxiv.org/abs/1411.1784},
	doi = {10.48550/arXiv.1411.1784},
	number = {{arXiv}:1411.1784},
	publisher = {{arXiv}},
	author = {Mirza, Mehdi and Osindero, Simon},
	urldate = {2023-06-20},
	date = {2014-11-06},
	eprinttype = {arxiv},
	eprint = {1411.1784 [cs, stat]},
}

@book{tomczak_deep_2022,
	location = {Cham},
	title = {Deep Generative Modeling},
	isbn = {978-3-030-93157-5 978-3-030-93158-2},
	url = {https://link.springer.com/10.1007/978-3-030-93158-2},
	publisher = {Springer International Publishing},
	author = {Tomczak, Jakub M.},
	urldate = {2023-06-20},
	date = {2022},
	langid = {english},
	doi = {10.1007/978-3-030-93158-2},
}

@misc{arjovsky_towards_2017,
	title = {Towards Principled Methods for Training Generative Adversarial Networks},
	url = {http://arxiv.org/abs/1701.04862},
	doi = {10.48550/arXiv.1701.04862},
	number = {{arXiv}:1701.04862},
	publisher = {{arXiv}},
	author = {Arjovsky, Martin and Bottou, Léon},
	urldate = {2023-06-20},
	date = {2017-01-17},
	eprinttype = {arxiv},
	eprint = {1701.04862 [cs, stat]},
}

@misc{kang_contragan_2021,
	title = {{ContraGAN}: Contrastive Learning for Conditional Image Generation},
	url = {http://arxiv.org/abs/2006.12681},
	doi = {10.48550/arXiv.2006.12681},
	shorttitle = {{ContraGAN}},
	number = {{arXiv}:2006.12681},
	publisher = {{arXiv}},
	author = {Kang, Minguk and Park, Jaesik},
	urldate = {2023-06-20},
	date = {2021-02-03},
	eprinttype = {arxiv},
	eprint = {2006.12681 [cs]},
}

@misc{arjovsky_wasserstein_2017,
	title = {Wasserstein {GAN}},
	url = {http://arxiv.org/abs/1701.07875},
	doi = {10.48550/arXiv.1701.07875},
	number = {{arXiv}:1701.07875},
	publisher = {{arXiv}},
	author = {Arjovsky, Martin and Chintala, Soumith and Bottou, Léon},
	urldate = {2023-06-20},
	date = {2017-12-06},
	eprinttype = {arxiv},
	eprint = {1701.07875 [cs, stat]},
}

@misc{zhao_energy-based_2017,
	title = {Energy-based Generative Adversarial Network},
	url = {http://arxiv.org/abs/1609.03126},
	doi = {10.48550/arXiv.1609.03126},
	number = {{arXiv}:1609.03126},
	publisher = {{arXiv}},
	author = {Zhao, Junbo and Mathieu, Michael and {LeCun}, Yann},
	urldate = {2023-06-20},
	date = {2017-03-06},
	eprinttype = {arxiv},
	eprint = {1609.03126 [cs, stat]},
}

@misc{lim_geometric_2017,
	title = {Geometric {GAN}},
	url = {http://arxiv.org/abs/1705.02894},
	doi = {10.48550/arXiv.1705.02894},
	number = {{arXiv}:1705.02894},
	publisher = {{arXiv}},
	author = {Lim, Jae Hyun and Ye, Jong Chul},
	urldate = {2023-06-20},
	date = {2017-05-08},
	eprinttype = {arxiv},
	eprint = {1705.02894 [cond-mat, stat]},
}

@misc{unterthiner_coulomb_2018,
	title = {Coulomb {GANs}: Provably Optimal Nash Equilibria via Potential Fields},
	url = {http://arxiv.org/abs/1708.08819},
	shorttitle = {Coulomb {GANs}},
	number = {{arXiv}:1708.08819},
	publisher = {{arXiv}},
	author = {Unterthiner, Thomas and Nessler, Bernhard and Seward, Calvin and Klambauer, Günter and Heusel, Martin and Ramsauer, Hubert and Hochreiter, Sepp},
	urldate = {2023-06-20},
	date = {2018-01-30},
	eprinttype = {arxiv},
	eprint = {1708.08819 [cs, stat]},
}

@inproceedings{salimans_improved_2016,
	title = {Improved Techniques for Training {GANs}},
	volume = {29},
	url = {https://proceedings.neurips.cc/paper_files/paper/2016/hash/8a3363abe792db2d8761d6403605aeb7-Abstract.html},
	booktitle = {Advances in Neural Information Processing Systems},
	publisher = {Curran Associates, Inc.},
	author = {Salimans, Tim and Goodfellow, Ian and Zaremba, Wojciech and Cheung, Vicki and Radford, Alec and Chen, Xi and Chen, Xi},
	urldate = {2023-06-20},
	date = {2016},
}

@misc{brock_large_2019,
	title = {Large Scale {GAN} Training for High Fidelity Natural Image Synthesis},
	url = {http://arxiv.org/abs/1809.11096},
	doi = {10.48550/arXiv.1809.11096},
	number = {{arXiv}:1809.11096},
	publisher = {{arXiv}},
	author = {Brock, Andrew and Donahue, Jeff and Simonyan, Karen},
	urldate = {2023-06-20},
	date = {2019-02-25},
	eprinttype = {arxiv},
	eprint = {1809.11096 [cs, stat]},
}

@misc{grosse_sandwiching_2015,
	title = {Sandwiching the marginal likelihood using bidirectional Monte Carlo},
	url = {http://arxiv.org/abs/1511.02543},
	doi = {10.48550/arXiv.1511.02543},
	number = {{arXiv}:1511.02543},
	publisher = {{arXiv}},
	author = {Grosse, Roger B. and Ghahramani, Zoubin and Adams, Ryan P.},
	urldate = {2023-06-19},
	date = {2015-11-08},
	eprinttype = {arxiv},
	eprint = {1511.02543 [cs, stat]},
}

@inproceedings{hoffman_learning_2017,
	title = {Learning Deep Latent Gaussian Models with Markov Chain Monte Carlo},
	url = {https://proceedings.mlr.press/v70/hoffman17a.html},
	eventtitle = {International Conference on Machine Learning},
	pages = {1510--1519},
	booktitle = {Proceedings of the 34th International Conference on Machine Learning},
	publisher = {{PMLR}},
	author = {Hoffman, Matthew D.},
	urldate = {2023-06-19},
	date = {2017-07-17},
	langid = {english},
	note = {{ISSN}: 2640-3498},
}

@misc{li_approximate_2017,
	title = {Approximate Inference with Amortised {MCMC}},
	url = {http://arxiv.org/abs/1702.08343},
	doi = {10.48550/arXiv.1702.08343},
	number = {{arXiv}:1702.08343},
	publisher = {{arXiv}},
	author = {Li, Yingzhen and Turner, Richard E. and Liu, Qiang},
	urldate = {2023-06-19},
	date = {2017-05-22},
	eprinttype = {arxiv},
	eprint = {1702.08343 [cs, stat]},
}

@inproceedings{zimmermann_nested_2021,
	title = {Nested Variational Inference},
	volume = {34},
	url = {https://proceedings.neurips.cc/paper/2021/hash/ab49b208848abe14418090d95df0d590-Abstract.html},
	pages = {20423--20435},
	booktitle = {Advances in Neural Information Processing Systems},
	publisher = {Curran Associates, Inc.},
	author = {Zimmermann, Heiko and Wu, Hao and Esmaeili, Babak and van de Meent, Jan-Willem},
	urldate = {2023-06-19},
	date = {2021},
}

@inproceedings{zhang_differentiable_2021,
	title = {Differentiable Annealed Importance Sampling and the Perils of Gradient Noise},
	url = {https://openreview.net/forum?id=6rqjgrL7Lq},
	eventtitle = {Advances in Neural Information Processing Systems},
	author = {Zhang, Guodong and Hsu, Kyle and Li, Jianing and Finn, Chelsea and Grosse, Roger Baker},
	urldate = {2023-06-19},
	date = {2021-11-09},
	langid = {english},
}

@article{neal_annealed_2001,
	title = {Annealed importance sampling},
	volume = {11},
	issn = {1573-1375},
	url = {https://doi.org/10.1023/A:1008923215028},
	doi = {10.1023/A:1008923215028},
	pages = {125--139},
	number = {2},
	journaltitle = {Statistics and Computing},
	shortjournal = {Statistics and Computing},
	author = {Neal, Radford M.},
	urldate = {2023-06-19},
	date = {2001-04-01},
	langid = {english},
}

@misc{burda_importance_2016,
	title = {Importance Weighted Autoencoders},
	url = {http://arxiv.org/abs/1509.00519},
	doi = {10.48550/arXiv.1509.00519},
	number = {{arXiv}:1509.00519},
	publisher = {{arXiv}},
	author = {Burda, Yuri and Grosse, Roger and Salakhutdinov, Ruslan},
	urldate = {2023-06-19},
	date = {2016-11-07},
	eprinttype = {arxiv},
	eprint = {1509.00519 [cs, stat]},
}

@misc{ramesh_gatsbi_2022,
	title = {{GATSBI}: Generative Adversarial Training for Simulation-Based Inference},
	url = {http://arxiv.org/abs/2203.06481},
	doi = {10.48550/arXiv.2203.06481},
	shorttitle = {{GATSBI}},
	number = {{arXiv}:2203.06481},
	publisher = {{arXiv}},
	author = {Ramesh, Poornima and Lueckmann, Jan-Matthis and Boelts, Jan and Tejero-Cantero, Álvaro and Greenberg, David S. and Gonçalves, Pedro J. and Macke, Jakob H.},
	urldate = {2023-06-19},
	date = {2022-03-12},
	eprinttype = {arxiv},
	eprint = {2203.06481 [cs, stat]},
}

@misc{rezende_taming_2018,
	title = {Taming {VAEs}},
	url = {http://arxiv.org/abs/1810.00597},
	doi = {10.48550/arXiv.1810.00597},
	number = {{arXiv}:1810.00597},
	publisher = {{arXiv}},
	author = {Rezende, Danilo Jimenez and Viola, Fabio},
	urldate = {2023-06-19},
	date = {2018-10-01},
	eprinttype = {arxiv},
	eprint = {1810.00597 [cs, stat]},
}

@misc{bowman_generating_2016,
	title = {Generating Sentences from a Continuous Space},
	url = {http://arxiv.org/abs/1511.06349},
	doi = {10.48550/arXiv.1511.06349},
	number = {{arXiv}:1511.06349},
	publisher = {{arXiv}},
	author = {Bowman, Samuel R. and Vilnis, Luke and Vinyals, Oriol and Dai, Andrew M. and Jozefowicz, Rafal and Bengio, Samy},
	urldate = {2023-06-19},
	date = {2016-05-12},
	eprinttype = {arxiv},
	eprint = {1511.06349 [cs]},
}

@online{noauthor_random_nodate,
	title = {Random variate generation in one line of code {\textbar} Proceedings of the 28th conference on Winter simulation},
	url = {https://dl.acm.org/doi/10.1145/256562.256623},
	urldate = {2023-06-17},
}

@article{hinton_reducing_2006,
	title = {Reducing the Dimensionality of Data with Neural Networks},
	volume = {313},
	url = {https://www.science.org/doi/10.1126/science.1127647},
	doi = {10.1126/science.1127647},
	pages = {504--507},
	number = {5786},
	journaltitle = {Science},
	author = {Hinton, G. E. and Salakhutdinov, R. R.},
	urldate = {2023-06-17},
	date = {2006-07-28},
	note = {Publisher: American Association for the Advancement of Science},
}

@online{noauthor_introduction_nodate,
	title = {An Introduction to Variational Methods for Graphical Models {\textbar} {SpringerLink}},
	url = {https://link.springer.com/article/10.1023/A:1007665907178},
	urldate = {2023-06-15},
}

@inproceedings{lasserre_principled_2006,
	title = {Principled Hybrids of Generative and Discriminative Models},
	volume = {1},
	doi = {10.1109/CVPR.2006.227},
	eventtitle = {2006 {IEEE} Computer Society Conference on Computer Vision and Pattern Recognition ({CVPR}'06)},
	pages = {87--94},
	booktitle = {2006 {IEEE} Computer Society Conference on Computer Vision and Pattern Recognition ({CVPR}'06)},
	author = {Lasserre, J.A. and Bishop, C.M. and Minka, T.P.},
	date = {2006-06},
	note = {{ISSN}: 1063-6919},
}

@online{noauthor_probabilistic_nodate,
	title = {Probabilistic machine learning and artificial intelligence {\textbar} Nature},
	url = {https://www.nature.com/articles/nature14541},
	urldate = {2023-06-15},
}

@article{wu_comprehensive_2021,
	title = {A Comprehensive Survey on Graph Neural Networks},
	volume = {32},
	issn = {2162-237X, 2162-2388},
	url = {http://arxiv.org/abs/1901.00596},
	doi = {10.1109/TNNLS.2020.2978386},
	pages = {4--24},
	number = {1},
	journaltitle = {{IEEE} Transactions on Neural Networks and Learning Systems},
	shortjournal = {{IEEE} Trans. Neural Netw. Learning Syst.},
	author = {Wu, Zonghan and Pan, Shirui and Chen, Fengwen and Long, Guodong and Zhang, Chengqi and Yu, Philip S.},
	urldate = {2023-06-15},
	date = {2021-01},
	eprinttype = {arxiv},
	eprint = {1901.00596 [cs, stat]},
}

@online{noauthor_improving_nodate,
	title = {Improving language understanding with unsupervised learning},
	url = {https://openai.com/research/language-unsupervised},
	urldate = {2023-06-15},
	langid = {american},
}

@online{noauthor_160706450_nodate,
	title = {[1607.06450] Layer Normalization},
	url = {https://arxiv.org/abs/1607.06450},
	urldate = {2023-06-14},
}

@inproceedings{xu_show_2015,
	title = {Show, Attend and Tell: Neural Image Caption Generation with Visual Attention},
	url = {https://proceedings.mlr.press/v37/xuc15.html},
	shorttitle = {Show, Attend and Tell},
	eventtitle = {International Conference on Machine Learning},
	pages = {2048--2057},
	booktitle = {Proceedings of the 32nd International Conference on Machine Learning},
	publisher = {{PMLR}},
	author = {Xu, Kelvin and Ba, Jimmy and Kiros, Ryan and Cho, Kyunghyun and Courville, Aaron and Salakhudinov, Ruslan and Zemel, Rich and Bengio, Yoshua},
	urldate = {2023-06-14},
	date = {2015-06-01},
	langid = {english},
	note = {{ISSN}: 1938-7228},
}

@online{graves_neural_2014,
	title = {Neural Turing Machines},
	url = {https://arxiv.org/abs/1410.5401v2},
	titleaddon = {{arXiv}.org},
	author = {Graves, Alex and Wayne, Greg and Danihelka, Ivo},
	urldate = {2023-06-14},
	date = {2014-10-20},
	langid = {english},
}

@online{ma_interactive_2017,
	title = {Interactive Attention Networks for Aspect-Level Sentiment Classification},
	url = {https://arxiv.org/abs/1709.00893v1},
	titleaddon = {{arXiv}.org},
	author = {Ma, Dehong and Li, Sujian and Zhang, Xiaodong and Wang, Houfeng},
	urldate = {2023-06-14},
	date = {2017-09-04},
	langid = {english},
}

@online{sordoni_iterative_2016,
	title = {Iterative Alternating Neural Attention for Machine Reading},
	url = {https://arxiv.org/abs/1606.02245v4},
	titleaddon = {{arXiv}.org},
	author = {Sordoni, Alessandro and Bachman, Philip and Trischler, Adam and Bengio, Yoshua},
	urldate = {2023-06-14},
	date = {2016-06-07},
	langid = {english},
}

@inproceedings{luong_effective_2015,
	location = {Lisbon, Portugal},
	title = {Effective Approaches to Attention-based Neural Machine Translation},
	url = {https://aclanthology.org/D15-1166},
	doi = {10.18653/v1/D15-1166},
	eventtitle = {{EMNLP} 2015},
	pages = {1412--1421},
	booktitle = {Proceedings of the 2015 Conference on Empirical Methods in Natural Language Processing},
	publisher = {Association for Computational Linguistics},
	author = {Luong, Thang and Pham, Hieu and Manning, Christopher D.},
	urldate = {2023-06-14},
	date = {2015-09},
}

@online{bahdanau_neural_2014,
	title = {Neural Machine Translation by Jointly Learning to Align and Translate},
	url = {https://arxiv.org/abs/1409.0473v7},
	titleaddon = {{arXiv}.org},
	author = {Bahdanau, Dzmitry and Cho, Kyunghyun and Bengio, Yoshua},
	urldate = {2023-06-14},
	date = {2014-09-01},
	langid = {english},
}

@misc{goodfellow_generative_2014,
	title = {Generative Adversarial Networks},
	url = {http://arxiv.org/abs/1406.2661},
	doi = {10.48550/arXiv.1406.2661},
	number = {{arXiv}:1406.2661},
	publisher = {{arXiv}},
	author = {Goodfellow, Ian J. and Pouget-Abadie, Jean and Mirza, Mehdi and Xu, Bing and Warde-Farley, David and Ozair, Sherjil and Courville, Aaron and Bengio, Yoshua},
	urldate = {2023-06-04},
	date = {2014-06-10},
	eprinttype = {arxiv},
	eprint = {1406.2661 [cs, stat]},
}

@misc{song_how_2021,
	title = {How to Train Your Energy-Based Models},
	url = {http://arxiv.org/abs/2101.03288},
	doi = {10.48550/arXiv.2101.03288},
	number = {{arXiv}:2101.03288},
	publisher = {{arXiv}},
	author = {Song, Yang and Kingma, Diederik P.},
	urldate = {2023-06-04},
	date = {2021-02-17},
	eprinttype = {arxiv},
	eprint = {2101.03288 [cs, stat]},
}

@misc{song_generative_2020,
	title = {Generative Modeling by Estimating Gradients of the Data Distribution},
	url = {http://arxiv.org/abs/1907.05600},
	doi = {10.48550/arXiv.1907.05600},
	number = {{arXiv}:1907.05600},
	publisher = {{arXiv}},
	author = {Song, Yang and Ermon, Stefano},
	urldate = {2023-06-04},
	date = {2020-10-10},
	eprinttype = {arxiv},
	eprint = {1907.05600 [cs, stat]},
}

@misc{ramesh_hierarchical_2022,
	title = {Hierarchical Text-Conditional Image Generation with {CLIP} Latents},
	url = {http://arxiv.org/abs/2204.06125},
	doi = {10.48550/arXiv.2204.06125},
	number = {{arXiv}:2204.06125},
	publisher = {{arXiv}},
	author = {Ramesh, Aditya and Dhariwal, Prafulla and Nichol, Alex and Chu, Casey and Chen, Mark},
	urldate = {2023-06-04},
	date = {2022-04-12},
	eprinttype = {arxiv},
	eprint = {2204.06125 [cs]},
}

@misc{kingma_variational_2023,
	title = {Variational Diffusion Models},
	url = {http://arxiv.org/abs/2107.00630},
	doi = {10.48550/arXiv.2107.00630},
	number = {{arXiv}:2107.00630},
	publisher = {{arXiv}},
	author = {Kingma, Diederik P. and Salimans, Tim and Poole, Ben and Ho, Jonathan},
	urldate = {2023-06-04},
	date = {2023-04-13},
	eprinttype = {arxiv},
	eprint = {2107.00630 [cs, stat]},
}

@misc{ho_denoising_2020,
	title = {Denoising Diffusion Probabilistic Models},
	url = {http://arxiv.org/abs/2006.11239},
	doi = {10.48550/arXiv.2006.11239},
	number = {{arXiv}:2006.11239},
	publisher = {{arXiv}},
	author = {Ho, Jonathan and Jain, Ajay and Abbeel, Pieter},
	urldate = {2023-06-04},
	date = {2020-12-16},
	eprinttype = {arxiv},
	eprint = {2006.11239 [cs, stat]},
}

@inproceedings{kingma_improved_2016,
	title = {Improved Variational Inference with Inverse Autoregressive Flow},
	volume = {29},
	url = {https://papers.nips.cc/paper_files/paper/2016/hash/ddeebdeefdb7e7e7a697e1c3e3d8ef54-Abstract.html},
	booktitle = {Advances in Neural Information Processing Systems},
	publisher = {Curran Associates, Inc.},
	author = {Kingma, Durk P and Salimans, Tim and Jozefowicz, Rafal and Chen, Xi and Sutskever, Ilya and Welling, Max},
	urldate = {2023-06-04},
	date = {2016},
}

@misc{berg_sylvester_2019,
	title = {Sylvester Normalizing Flows for Variational Inference},
	url = {http://arxiv.org/abs/1803.05649},
	doi = {10.48550/arXiv.1803.05649},
	number = {{arXiv}:1803.05649},
	publisher = {{arXiv}},
	author = {Berg, Rianne van den and Hasenclever, Leonard and Tomczak, Jakub M. and Welling, Max},
	urldate = {2023-06-04},
	date = {2019-02-20},
	eprinttype = {arxiv},
	eprint = {1803.05649 [cs, stat]},
}

@misc{rezende_variational_2016,
	title = {Variational Inference with Normalizing Flows},
	url = {http://arxiv.org/abs/1505.05770},
	doi = {10.48550/arXiv.1505.05770},
	number = {{arXiv}:1505.05770},
	publisher = {{arXiv}},
	author = {Rezende, Danilo Jimenez and Mohamed, Shakir},
	urldate = {2023-06-04},
	date = {2016-06-14},
	eprinttype = {arxiv},
	eprint = {1505.05770 [cs, stat]},
}

@misc{kingma_auto-encoding_2022,
	title = {Auto-Encoding Variational Bayes},
	url = {http://arxiv.org/abs/1312.6114},
	doi = {10.48550/arXiv.1312.6114},
	number = {{arXiv}:1312.6114},
	publisher = {{arXiv}},
	author = {Kingma, Diederik P. and Welling, Max},
	urldate = {2023-06-04},
	date = {2022-12-10},
	eprinttype = {arxiv},
	eprint = {1312.6114 [cs, stat]},
}

@misc{rezende_stochastic_2014,
	title = {Stochastic Backpropagation and Approximate Inference in Deep Generative Models},
	url = {http://arxiv.org/abs/1401.4082},
	doi = {10.48550/arXiv.1401.4082},
	number = {{arXiv}:1401.4082},
	publisher = {{arXiv}},
	author = {Rezende, Danilo Jimenez and Mohamed, Shakir and Wierstra, Daan},
	urldate = {2023-06-04},
	date = {2014-05-30},
	eprinttype = {arxiv},
	eprint = {1401.4082 [cs, stat]},
}

@article{goodman_controlled_1950,
	title = {Controlled Selection--A Technique in Probability Sampling},
	volume = {45},
	issn = {0162-1459},
	url = {https://www.jstor.org/stable/2280293},
	doi = {10.2307/2280293},
	pages = {350--372},
	number = {251},
	journaltitle = {Journal of the American Statistical Association},
	author = {Goodman, Roe and Kish, Leslie},
	urldate = {2023-06-04},
	date = {1950},
	note = {Publisher: [American Statistical Association, Taylor \& Francis, Ltd.]},
}

@article{hinton_fast_2006,
	title = {A fast learning algorithm for deep belief nets},
	volume = {18},
	issn = {0899-7667},
	doi = {10.1162/neco.2006.18.7.1527},
	pages = {1527--1554},
	number = {7},
	journaltitle = {Neural Computation},
	shortjournal = {Neural Comput},
	author = {Hinton, Geoffrey E. and Osindero, Simon and Teh, Yee-Whye},
	date = {2006-07},
	pmid = {16764513},
}

@article{hinton_training_2002,
	title = {Training products of experts by minimizing contrastive divergence},
	volume = {14},
	issn = {0899-7667},
	doi = {10.1162/089976602760128018},
	pages = {1771--1800},
	number = {8},
	journaltitle = {Neural Computation},
	shortjournal = {Neural Comput},
	author = {Hinton, Geoffrey E.},
	date = {2002-08},
	pmid = {12180402},
}

@inproceedings{hinton_optimal_1983,
	title = {{OPTIMAL} {PERCEPTUAL} {INFERENCE}},
	url = {https://www.semanticscholar.org/paper/OPTIMAL-PERCEPTUAL-INFERENCE-Hinton-Sejnowski/1718965f492d4e9fe1d98a3fb83efe671a4aed2c},
	author = {Hinton, Geoffrey E. and Sejnowski, J.},
	urldate = {2023-06-04},
	date = {1983},
}

@article{belayneh_calorimetry_2020,
	title = {Calorimetry with deep learning: particle simulation and reconstruction for collider physics},
	volume = {80},
	issn = {1434-6052},
	url = {https://doi.org/10.1140/epjc/s10052-020-8251-9},
	doi = {10.1140/epjc/s10052-020-8251-9},
	shorttitle = {Calorimetry with deep learning},
	pages = {688},
	number = {7},
	journaltitle = {The European Physical Journal C},
	shortjournal = {Eur. Phys. J. C},
	author = {Belayneh, Dawit and Carminati, Federico and Farbin, Amir and Hooberman, Benjamin and Khattak, Gulrukh and Liu, Miaoyuan and Liu, Junze and Olivito, Dominick and Pacela, Vitória Barin and Pierini, Maurizio and Schwing, Alexander and Spiropulu, Maria and Vallecorsa, Sofia and Vlimant, Jean-Roch and Wei, Wei and Zhang, Matt},
	urldate = {2023-05-31},
	date = {2020-07-31},
	langid = {english},
}

@article{brehmer_simulation-based_2021,
	title = {Simulation-based inference in particle physics},
	volume = {3},
	rights = {2021 Springer Nature Limited},
	issn = {2522-5820},
	url = {https://www.nature.com/articles/s42254-021-00305-6},
	doi = {10.1038/s42254-021-00305-6},
	pages = {305--305},
	number = {5},
	journaltitle = {Nature Reviews Physics},
	shortjournal = {Nat Rev Phys},
	author = {Brehmer, Johann},
	urldate = {2023-05-31},
	date = {2021-05},
	langid = {english},
	note = {Number: 5
Publisher: Nature Publishing Group},
}

@article{regadio_synthesis_2022,
	title = {Synthesis of pulses from particle detectors with a Generative Adversarial Network ({GAN})},
	volume = {1033},
	issn = {0168-9002},
	url = {https://www.sciencedirect.com/science/article/pii/S0168900222002108},
	doi = {10.1016/j.nima.2022.166647},
	pages = {166647},
	journaltitle = {Nuclear Instruments and Methods in Physics Research Section A: Accelerators, Spectrometers, Detectors and Associated Equipment},
	shortjournal = {Nuclear Instruments and Methods in Physics Research Section A: Accelerators, Spectrometers, Detectors and Associated Equipment},
	author = {Regadío, Alberto and Esteban, Luis and Sánchez-Prieto, Sebastián},
	urldate = {2023-05-31},
	date = {2022-06-11},
	langid = {english},
}

@article{rajak_autonomous_2021,
	title = {Autonomous reinforcement learning agent for stretchable kirigami design of 2D materials},
	volume = {7},
	rights = {2021 The Author(s)},
	issn = {2057-3960},
	url = {https://www.nature.com/articles/s41524-021-00572-y},
	doi = {10.1038/s41524-021-00572-y},
	pages = {1--8},
	number = {1},
	journaltitle = {npj Computational Materials},
	shortjournal = {npj Comput Mater},
	author = {Rajak, Pankaj and Wang, Beibei and Nomura, Ken-ichi and Luo, Ye and Nakano, Aiichiro and Kalia, Rajiv and Vashishta, Priya},
	urldate = {2023-05-31},
	date = {2021-07-09},
	langid = {english},
	note = {Number: 1
Publisher: Nature Publishing Group},
}

@misc{li_zero-knowledge_2023,
	title = {Zero-Knowledge Zero-Shot Learning for Novel Visual Category Discovery},
	url = {http://arxiv.org/abs/2302.04427},
	doi = {10.48550/arXiv.2302.04427},
	number = {{arXiv}:2302.04427},
	publisher = {{arXiv}},
	author = {Li, Zhaonan and Liu, Hongfu},
	urldate = {2023-05-31},
	date = {2023-02-08},
	eprinttype = {arxiv},
	eprint = {2302.04427 [cs]},
}

@article{anishchenko_novo_2021,
	title = {De novo protein design by deep network hallucination},
	volume = {600},
	rights = {2021 The Author(s), under exclusive licence to Springer Nature Limited},
	issn = {1476-4687},
	url = {https://www.nature.com/articles/s41586-021-04184-w},
	doi = {10.1038/s41586-021-04184-w},
	pages = {547--552},
	number = {7889},
	journaltitle = {Nature},
	author = {Anishchenko, Ivan and Pellock, Samuel J. and Chidyausiku, Tamuka M. and Ramelot, Theresa A. and Ovchinnikov, Sergey and Hao, Jingzhou and Bafna, Khushboo and Norn, Christoffer and Kang, Alex and Bera, Asim K. and {DiMaio}, Frank and Carter, Lauren and Chow, Cameron M. and Montelione, Gaetano T. and Baker, David},
	urldate = {2023-05-31},
	date = {2021-12},
	langid = {english},
	note = {Number: 7889
Publisher: Nature Publishing Group},
}

@misc{shwartz-ziv_compress_2023,
	title = {To Compress or Not to Compress- Self-Supervised Learning and Information Theory: A Review},
	url = {http://arxiv.org/abs/2304.09355},
	doi = {10.48550/arXiv.2304.09355},
	shorttitle = {To Compress or Not to Compress- Self-Supervised Learning and Information Theory},
	number = {{arXiv}:2304.09355},
	publisher = {{arXiv}},
	author = {Shwartz-Ziv, Ravid and {LeCun}, Yann},
	urldate = {2023-05-31},
	date = {2023-05-04},
	eprinttype = {arxiv},
	eprint = {2304.09355 [cs, math]},
}

@misc{haochen_theoretical_2023,
	title = {A Theoretical Study of Inductive Biases in Contrastive Learning},
	url = {http://arxiv.org/abs/2211.14699},
	doi = {10.48550/arXiv.2211.14699},
	number = {{arXiv}:2211.14699},
	publisher = {{arXiv}},
	author = {{HaoChen}, Jeff Z. and Ma, Tengyu},
	urldate = {2023-05-31},
	date = {2023-04-08},
	eprinttype = {arxiv},
	eprint = {2211.14699 [cs, stat]},
}

@misc{battaglia_relational_2018,
	title = {Relational inductive biases, deep learning, and graph networks},
	url = {http://arxiv.org/abs/1806.01261},
	doi = {10.48550/arXiv.1806.01261},
	number = {{arXiv}:1806.01261},
	publisher = {{arXiv}},
	author = {Battaglia, Peter W. and Hamrick, Jessica B. and Bapst, Victor and Sanchez-Gonzalez, Alvaro and Zambaldi, Vinicius and Malinowski, Mateusz and Tacchetti, Andrea and Raposo, David and Santoro, Adam and Faulkner, Ryan and Gulcehre, Caglar and Song, Francis and Ballard, Andrew and Gilmer, Justin and Dahl, George and Vaswani, Ashish and Allen, Kelsey and Nash, Charles and Langston, Victoria and Dyer, Chris and Heess, Nicolas and Wierstra, Daan and Kohli, Pushmeet and Botvinick, Matt and Vinyals, Oriol and Li, Yujia and Pascanu, Razvan},
	urldate = {2023-05-31},
	date = {2018-10-17},
	eprinttype = {arxiv},
	eprint = {1806.01261 [cs, stat]},
}

@article{butter_amplifying_nodate,
	title = {Amplifying Statistics using Generative Models},
	author = {Butter, Anja},
	langid = {english},
}

@misc{buhmann_caloclouds_2023,
	title = {{CaloClouds}: Fast Geometry-Independent Highly-Granular Calorimeter Simulation},
	url = {http://arxiv.org/abs/2305.04847},
	doi = {10.48550/arXiv.2305.04847},
	shorttitle = {{CaloClouds}},
	number = {{arXiv}:2305.04847},
	publisher = {{arXiv}},
	author = {Buhmann, Erik and Diefenbacher, Sascha and Eren, Engin and Gaede, Frank and Kasieczka, Gregor and Korol, Anatolii and Korcari, William and Krüger, Katja and {McKeown}, Peter},
	urldate = {2023-05-30},
	date = {2023-05-08},
	eprinttype = {arxiv},
	eprint = {2305.04847 [hep-ex, physics:hep-ph, physics:physics]},
}

@misc{diefenbacher_l2lflows_2023,
	title = {L2LFlows: Generating High-Fidelity 3D Calorimeter Images},
	url = {http://arxiv.org/abs/2302.11594},
	doi = {10.48550/arXiv.2302.11594},
	shorttitle = {L2LFlows},
	number = {{arXiv}:2302.11594},
	publisher = {{arXiv}},
	author = {Diefenbacher, Sascha and Eren, Engin and Gaede, Frank and Kasieczka, Gregor and Krause, Claudius and Shekhzadeh, Imahn and Shih, David},
	urldate = {2023-05-30},
	date = {2023-02-22},
	eprinttype = {arxiv},
	eprint = {2302.11594 [hep-ex, physics:hep-ph, physics:physics]},
}

@misc{butter_jet_2023,
	title = {Jet Diffusion versus {JetGPT} -- Modern Networks for the {LHC}},
	url = {http://arxiv.org/abs/2305.10475},
	doi = {10.48550/arXiv.2305.10475},
	number = {{arXiv}:2305.10475},
	publisher = {{arXiv}},
	author = {Butter, Anja and Huetsch, Nathan and Schweitzer, Sofia Palacios and Plehn, Tilman and Sorrenson, Peter and Spinner, Jonas},
	urldate = {2023-05-30},
	date = {2023-05-17},
	eprinttype = {arxiv},
	eprint = {2305.10475 [hep-ph]},
}

@misc{leigh_pc-jedi_2023,
	title = {{PC}-{JeDi}: Diffusion for Particle Cloud Generation in High Energy Physics},
	url = {http://arxiv.org/abs/2303.05376},
	shorttitle = {{PC}-{JeDi}},
	number = {{arXiv}:2303.05376},
	publisher = {{arXiv}},
	author = {Leigh, Matthew and Sengupta, Debajyoti and Quétant, Guillaume and Raine, John Andrew and Zoch, Knut and Golling, Tobias},
	urldate = {2023-05-30},
	date = {2023-03-09},
	eprinttype = {arxiv},
	eprint = {2303.05376 [hep-ex, physics:hep-ph]},
}

@misc{buhmann_epic-gan_2023,
	title = {{EPiC}-{GAN}: Equivariant Point Cloud Generation for Particle Jets},
	url = {http://arxiv.org/abs/2301.08128},
	doi = {10.48550/arXiv.2301.08128},
	shorttitle = {{EPiC}-{GAN}},
	number = {{arXiv}:2301.08128},
	publisher = {{arXiv}},
	author = {Buhmann, Erik and Kasieczka, Gregor and Thaler, Jesse},
	urldate = {2023-05-30},
	date = {2023-02-09},
	eprinttype = {arxiv},
	eprint = {2301.08128 [hep-ex, physics:hep-ph, physics:physics]},
}

@misc{cho_learning_2014,
	title = {Learning Phrase Representations using {RNN} Encoder-Decoder for Statistical Machine Translation},
	url = {http://arxiv.org/abs/1406.1078},
	doi = {10.48550/arXiv.1406.1078},
	number = {{arXiv}:1406.1078},
	publisher = {{arXiv}},
	author = {Cho, Kyunghyun and van Merrienboer, Bart and Gulcehre, Caglar and Bahdanau, Dzmitry and Bougares, Fethi and Schwenk, Holger and Bengio, Yoshua},
	urldate = {2023-05-30},
	date = {2014-09-02},
	eprinttype = {arxiv},
	eprint = {1406.1078 [cs, stat]},
}

@misc{axelrod_sample_2019,
	title = {Sample Amplification: Increasing Dataset Size even when Learning is Impossible},
	url = {http://arxiv.org/abs/1904.12053},
	doi = {10.48550/arXiv.1904.12053},
	shorttitle = {Sample Amplification},
	number = {{arXiv}:1904.12053},
	publisher = {{arXiv}},
	author = {Axelrod, Brian and Garg, Shivam and Sharan, Vatsal and Valiant, Gregory},
	urldate = {2023-05-29},
	date = {2019-12-02},
	eprinttype = {arxiv},
	eprint = {1904.12053 [cs, math, stat]},
}

@article{kansal_evaluating_2023,
	title = {Evaluating generative models in high energy physics},
	volume = {107},
	issn = {2470-0010, 2470-0029},
	url = {https://link.aps.org/doi/10.1103/PhysRevD.107.076017},
	doi = {10.1103/PhysRevD.107.076017},
	pages = {076017},
	number = {7},
	journaltitle = {Physical Review D},
	shortjournal = {Phys. Rev. D},
	author = {Kansal, Raghav and Li, Anni and Duarte, Javier and Chernyavskaya, Nadezda and Pierini, Maurizio and Orzari, Breno and Tomei, Thiago},
	urldate = {2023-05-29},
	date = {2023-04-21},
	langid = {english},
}

@article{tombs_method_2022,
	title = {A method to challenge symmetries in data with self-supervised learning},
	volume = {17},
	issn = {1748-0221},
	url = {http://arxiv.org/abs/2111.05442},
	doi = {10.1088/1748-0221/17/08/P08024},
	pages = {P08024},
	number = {8},
	journaltitle = {Journal of Instrumentation},
	shortjournal = {J. Inst.},
	author = {Tombs, Rupert and Lester, Christopher G.},
	urldate = {2023-05-28},
	date = {2022-08-01},
	eprinttype = {arxiv},
	eprint = {2111.05442 [hep-ph, physics:physics]},
}

@misc{hashemi_ultra-high-resolution_2023,
	title = {Ultra-High-Resolution Detector Simulation with Intra-Event Aware {GAN} and Self-Supervised Relational Reasoning},
	url = {http://arxiv.org/abs/2303.08046},
	doi = {10.48550/arXiv.2303.08046},
	number = {{arXiv}:2303.08046},
	publisher = {{arXiv}},
	author = {Hashemi, Hosein and Hartmann, Nikolai and Sharifzadeh, Sahand and Kahn, James and Kuhr, Thomas},
	urldate = {2023-05-28},
	date = {2023-03-07},
	eprinttype = {arxiv},
	eprint = {2303.08046 [hep-ph, physics:physics]},
}

@misc{krippendorf_detecting_2020,
	title = {Detecting Symmetries with Neural Networks},
	url = {http://arxiv.org/abs/2003.13679},
	doi = {10.48550/arXiv.2003.13679},
	number = {{arXiv}:2003.13679},
	publisher = {{arXiv}},
	author = {Krippendorf, Sven and Syvaeri, Marc},
	urldate = {2023-05-28},
	date = {2020-03-30},
	eprinttype = {arxiv},
	eprint = {2003.13679 [hep-th, physics:physics]},
}

@article{barenboim_symmetry_2021,
	title = {Symmetry meets {AI}},
	volume = {11},
	issn = {2542-4653},
	url = {https://scipost.org/10.21468/SciPostPhys.11.1.014},
	doi = {10.21468/SciPostPhys.11.1.014},
	pages = {014},
	number = {1},
	journaltitle = {{SciPost} Physics},
	author = {Barenboim, Gabriela and Hirn, Johannes and Sanz, Veronica},
	urldate = {2023-05-28},
	date = {2021-07-15},
	langid = {english},
}

@misc{dillon_symmetries_2021,
	title = {Symmetries, Safety, and Self-Supervision},
	url = {http://arxiv.org/abs/2108.04253},
	doi = {10.48550/arXiv.2108.04253},
	number = {{arXiv}:2108.04253},
	publisher = {{arXiv}},
	author = {Dillon, Barry M. and Kasieczka, Gregor and Olischlager, Hans and Plehn, Tilman and Sorrenson, Peter and Vogel, Lorenz},
	urldate = {2023-05-28},
	date = {2021-08-09},
	eprinttype = {arxiv},
	eprint = {2108.04253 [hep-ph]},
}

@inproceedings{zietlow_demystifying_2021,
	title = {Demystifying Inductive Biases for (Beta-){VAE} Based Architectures},
	url = {https://proceedings.mlr.press/v139/zietlow21a.html},
	eventtitle = {International Conference on Machine Learning},
	pages = {12945--12954},
	booktitle = {Proceedings of the 38th International Conference on Machine Learning},
	publisher = {{PMLR}},
	author = {Zietlow, Dominik and Rolinek, Michal and Martius, Georg},
	urldate = {2023-05-28},
	date = {2021-07-01},
	langid = {english},
	note = {{ISSN}: 2640-3498},
}

@misc{zhao_bias_2018,
	title = {Bias and Generalization in Deep Generative Models: An Empirical Study},
	url = {http://arxiv.org/abs/1811.03259},
	doi = {10.48550/arXiv.1811.03259},
	shorttitle = {Bias and Generalization in Deep Generative Models},
	number = {{arXiv}:1811.03259},
	publisher = {{arXiv}},
	author = {Zhao, Shengjia and Ren, Hongyu and Yuan, Arianna and Song, Jiaming and Goodman, Noah and Ermon, Stefano},
	urldate = {2023-05-28},
	date = {2018-11-07},
	eprinttype = {arxiv},
	eprint = {1811.03259 [cs, stat]},
}

@misc{hao_data_2019,
	title = {Data Amplification: A Unified and Competitive Approach to Property Estimation},
	url = {http://arxiv.org/abs/1904.00070},
	doi = {10.48550/arXiv.1904.00070},
	shorttitle = {Data Amplification},
	number = {{arXiv}:1904.00070},
	publisher = {{arXiv}},
	author = {Hao, Yi and Orlitsky, Alon and Suresh, Ananda T. and Wu, Yihong},
	urldate = {2023-05-28},
	date = {2019-03-29},
	eprinttype = {arxiv},
	eprint = {1904.00070 [cs, math, stat]},
}

@misc{axelrod_statistical_2022,
	title = {On the Statistical Complexity of Sample Amplification},
	url = {http://arxiv.org/abs/2201.04315},
	doi = {10.48550/arXiv.2201.04315},
	number = {{arXiv}:2201.04315},
	publisher = {{arXiv}},
	author = {Axelrod, Brian and Garg, Shivam and Han, Yanjun and Sharan, Vatsal and Valiant, Gregory},
	urldate = {2023-05-28},
	date = {2022-01-12},
	eprinttype = {arxiv},
	eprint = {2201.04315 [cs, math, stat]},
}

@article{chahrour_comparing_2022,
	title = {Comparing Machine Learning and Interpolation Methods for Loop-Level Calculations},
	volume = {12},
	issn = {2542-4653},
	url = {http://arxiv.org/abs/2111.14788},
	doi = {10.21468/SciPostPhys.12.6.187},
	pages = {187},
	number = {6},
	journaltitle = {{SciPost} Physics},
	shortjournal = {{SciPost} Phys.},
	author = {Chahrour, Ibrahim and Wells, James D.},
	urldate = {2023-05-28},
	date = {2022-06-08},
	eprinttype = {arxiv},
	eprint = {2111.14788 [hep-ph]},
}

@article{touranakou_particle-based_2022,
	title = {Particle-based Fast Jet Simulation at the {LHC} with Variational Autoencoders},
	volume = {3},
	issn = {2632-2153},
	url = {http://arxiv.org/abs/2203.00520},
	doi = {10.1088/2632-2153/ac7c56},
	pages = {035003},
	number = {3},
	journaltitle = {Machine Learning: Science and Technology},
	shortjournal = {Mach. Learn.: Sci. Technol.},
	author = {Touranakou, Mary and Chernyavskaya, Nadezda and Duarte, Javier and Gunopulos, Dimitrios and Kansal, Raghav and Orzari, Breno and Pierini, Maurizio and Tomei, Thiago and Vlimant, Jean-Roch},
	urldate = {2023-05-27},
	date = {2022-09-01},
	eprinttype = {arxiv},
	eprint = {2203.00520 [hep-ex, physics:hep-ph, physics:physics]},
}

@inproceedings{salimans_progressive_2022,
	title = {Progressive Distillation for Fast Sampling of Diffusion Models},
	url = {https://openreview.net/forum?id=TIdIXIpzhoI},
	eventtitle = {International Conference on Learning Representations},
	author = {Salimans, Tim and Ho, Jonathan},
	urldate = {2023-05-25},
	date = {2022-01-28},
	langid = {english},
}

@misc{mikuni_fast_2023,
	title = {Fast Point Cloud Generation with Diffusion Models in High Energy Physics},
	url = {http://arxiv.org/abs/2304.01266},
	doi = {10.48550/arXiv.2304.01266},
	number = {{arXiv}:2304.01266},
	publisher = {{arXiv}},
	author = {Mikuni, Vinicius and Nachman, Benjamin and Pettee, Mariel},
	urldate = {2023-05-25},
	date = {2023-04-03},
	eprinttype = {arxiv},
	eprint = {2304.01266 [hep-ex, physics:hep-ph]},
}

@article{mikuni_score-based_2022,
	title = {Score-based Generative Models for Calorimeter Shower Simulation},
	volume = {106},
	issn = {2470-0010, 2470-0029},
	url = {http://arxiv.org/abs/2206.11898},
	doi = {10.1103/PhysRevD.106.092009},
	pages = {092009},
	number = {9},
	journaltitle = {Physical Review D},
	shortjournal = {Phys. Rev. D},
	author = {Mikuni, Vinicius and Nachman, Benjamin},
	urldate = {2023-05-25},
	date = {2022-11-28},
	eprinttype = {arxiv},
	eprint = {2206.11898 [hep-ex, physics:hep-ph, physics:physics]},
}

@misc{papamakarios_masked_2018,
	title = {Masked Autoregressive Flow for Density Estimation},
	url = {http://arxiv.org/abs/1705.07057},
	doi = {10.48550/arXiv.1705.07057},
	number = {{arXiv}:1705.07057},
	publisher = {{arXiv}},
	author = {Papamakarios, George and Pavlakou, Theo and Murray, Iain},
	urldate = {2023-05-25},
	date = {2018-06-14},
	eprinttype = {arxiv},
	eprint = {1705.07057 [cs, stat]},
}

@misc{xu_generative_2023,
	title = {Generative Machine Learning for Detector Response Modeling with a Conditional Normalizing Flow},
	url = {http://arxiv.org/abs/2303.10148},
	number = {{arXiv}:2303.10148},
	publisher = {{arXiv}},
	author = {Xu, Allison and Han, Shuo and Ju, Xiangyang and Wang, Haichen},
	urldate = {2023-05-22},
	date = {2023-04-27},
	eprinttype = {arxiv},
	eprint = {2303.10148 [hep-ex, physics:physics]},
}

@misc{kingma_improving_2017,
	title = {Improving Variational Inference with Inverse Autoregressive Flow},
	url = {http://arxiv.org/abs/1606.04934},
	number = {{arXiv}:1606.04934},
	publisher = {{arXiv}},
	author = {Kingma, Diederik P. and Salimans, Tim and Jozefowicz, Rafal and Chen, Xi and Sutskever, Ilya and Welling, Max},
	urldate = {2023-05-22},
	date = {2017-01-30},
	eprinttype = {arxiv},
	eprint = {1606.04934 [cs, stat]},
}

@misc{krause_caloflow_2023,
	title = {{CaloFlow} {II}: Even Faster and Still Accurate Generation of Calorimeter Showers with Normalizing Flows},
	url = {http://arxiv.org/abs/2110.11377},
	shorttitle = {{CaloFlow} {II}},
	number = {{arXiv}:2110.11377},
	publisher = {{arXiv}},
	author = {Krause, Claudius and Shih, David},
	urldate = {2023-05-22},
	date = {2023-05-05},
	eprinttype = {arxiv},
	eprint = {2110.11377 [hep-ex, physics:hep-ph, physics:physics]},
}

@misc{liu_generalizing_2023,
	title = {Generalizing to new calorimeter geometries with Geometry-Aware Autoregressive Models ({GAAMs}) for fast calorimeter simulation},
	url = {http://arxiv.org/abs/2305.11531},
	number = {{arXiv}:2305.11531},
	publisher = {{arXiv}},
	author = {Liu, Junze and Ghosh, Aishik and Smith, Dylan and Baldi, Pierre and Whiteson, Daniel},
	urldate = {2023-05-22},
	date = {2023-05-19},
	eprinttype = {arxiv},
	eprint = {2305.11531 [hep-ex, physics:hep-ph, physics:physics]},
}

@misc{durkan_neural_2019,
	title = {Neural Spline Flows},
	url = {http://arxiv.org/abs/1906.04032},
	doi = {10.48550/arXiv.1906.04032},
	number = {{arXiv}:1906.04032},
	publisher = {{arXiv}},
	author = {Durkan, Conor and Bekasov, Artur and Murray, Iain and Papamakarios, George},
	urldate = {2023-05-21},
	date = {2019-12-02},
	eprinttype = {arxiv},
	eprint = {1906.04032 [cs, stat]},
}

@misc{germain_made_2015,
	title = {{MADE}: Masked Autoencoder for Distribution Estimation},
	url = {http://arxiv.org/abs/1502.03509},
	doi = {10.48550/arXiv.1502.03509},
	shorttitle = {{MADE}},
	number = {{arXiv}:1502.03509},
	publisher = {{arXiv}},
	author = {Germain, Mathieu and Gregor, Karol and Murray, Iain and Larochelle, Hugo},
	urldate = {2023-05-21},
	date = {2015-06-05},
	eprinttype = {arxiv},
	eprint = {1502.03509 [cs, stat]},
}

@misc{vaswani_attention_2017,
	title = {Attention Is All You Need},
	url = {http://arxiv.org/abs/1706.03762},
	doi = {10.48550/arXiv.1706.03762},
	number = {{arXiv}:1706.03762},
	publisher = {{arXiv}},
	author = {Vaswani, Ashish and Shazeer, Noam and Parmar, Niki and Uszkoreit, Jakob and Jones, Llion and Gomez, Aidan N. and Kaiser, Lukasz and Polosukhin, Illia},
	urldate = {2023-05-21},
	date = {2017-12-05},
	eprinttype = {arxiv},
	eprint = {1706.03762 [cs]},
}
  \addcontentsline{toc}{chapter}{Bibliography}

  \addcontentsline{toc}{chapter}{\protect Acknowledgements}

\chapter*{Acknowledgements}
This research was supported by the collaborative project IDT-UM~(Innovative Digitale Technologien zur Erforschung von Universum und Materie) funded by the German Federal Ministry of Education and Research~(BMBF) and the Deutsche Forschungsgemeinschaft under Germany's Excellence Strategy – EXC 2094 "ORIGINS" – 390783311.

This thesis is the capstone of four of the most adventurous years of my life. I learned a lot and gathered many unique experiences on this journey. 
None of this would have been possible without the assistance of a myriad of amazing individuals. While I may not be able to thank each and every one of them, I will certainly try.

First of all, I would like to thank Prof. Dr. Thomas Kuhr for his unwavering support and for sharing his expertise in detector physics and the exciting field of flavor physics. 
I vividly recall the day you introduced me to this subject matter. You outlined several potential topics, and when I asked which one was more challenging, you pointed to the PXD project. Thank you sincerely for allowing me the freedom to explore a multitude of different topics, even those outside our designated project.

I would like to extend my sincerest gratitude to Prof. Dr. Lukas Heinrich, Prof. Dr. Volker Tresp, Prof. Dr. Stefen Rullands, Prof. Dr. Daniel Gruen, and Prof. Dr. Lode Pollet for being part of my examination committee.

Dr.~Nikolai Hartmann, thank you for generously dedicating your time to me and offering support at literally any hour of the week. Our discussions have always been vibrant and a synthesis of ideas. Your live coding sessions were truly epic, and I learned a lot from them. 
I'm also deeply grateful for your thorough review and insightful comments on this thesis.

Dr.~Sahand Sharifzadeh, thank you for supporting and believing in me. I have always learned new things from our discussions. Your insights in creating new narratives and out-of-the-box thinking were always inspirational for me.

Dr.~James Kahn, we have not met each other in person even once, but we had numerous online meetings that took hours discussing different ideas in both the LinkFEI and IEA-GAN projects. I learned many tricks and academic hacks from you. 

I thank Dr.~Thomas Lueck for his technical and insightful comments about the PXD problem. Thank you for helping me brush up on the first chapter of this thesis!

I thank my colleagues from the Computational Center for Particle and Astrophysics~(C2PAP), who provided expertise and computation power that greatly assisted the research. 

I also thank David Katheder for his assistance with the GitHub repository preparation and cleaning of my code. 

Dr.~Nikole Hartmann, thank you for imparting your academic experience. Your comments on publications were incredibly helpful and game-changing.

I want to thank my colleagues, Dr. Sviatoslav Bilokin, Yaroslav Kulii, Nathalie Eberlein, Boyang Yu, Pascal Schmolz, Caspar Schmitt, David Koch, Lorenz Gärtner, and Arul Prakash for their constant support and cheerful conversations. 

Achieving this milestone would have been unimaginable without the abundant and heartfelt support from my family and friends. Your unwavering belief in me has been my keystone, making it possible for me to persevere through the most challenging times. Thank you for standing by me.
I look forward to many more years filled with joy, love, and shared triumphs!

To my beloved wife, Setareh, your love has been the alchemy that transformed long nights of despair into hope and self-doubt into self-belief. Your kindness taught me that even in the toughest of times, love could light the way. Your wisdom served as the balance to my often one-track mind, reminding me to see life in all its beautiful complexities. You celebrated my smallest victories and stood by me in my biggest defeats. You were the calm in the academic storm, offering  perspective when I most needed it. I am immensely grateful to have shared this journey with you, and I can't wait to embark on new adventures together.
  \markboth{Acknowledgements}{Acknowledgements}

\end{document}